\documentclass{book}
\usepackage[sfdefault]{carlito} 
\usepackage[headheight=45pt]{geometry}
\usepackage[onehalfspacing]{setspace}
\usepackage{titlesec}
\usepackage[colorlinks,linkcolor=blue,citecolor=blue]{hyperref}
\usepackage{graphicx}
\usepackage{amsmath}
\usepackage{bm}
\usepackage{wrapfig}
\usepackage{xcolor}
\usepackage{authblk}
\usepackage[font=small,labelfont=bf]{caption}
\usepackage{braket}
\usepackage{physics}
\usepackage[version=3]{mhchem}  
\usepackage{xurl}

\usepackage{float}

\usepackage{seqsplit}

\usepackage{subcaption}
\usepackage[export]{adjustbox}
\usepackage{subcaption}
\usepackage{tikz} 
\usetikzlibrary{matrix}

\usepackage{comment}

\usepackage{physics}
\usepackage[version=3]{mhchem}  

\usepackage[normalem]{ulem}
\usepackage[numbers,sort&compress]{natbib}
\usepackage{multirow}
\usepackage{pifont}

\usepackage{smartdiagram}

\usepackage{amssymb,bbm,braket,xspace}

\usepackage{array}

\usepackage[utf8]{inputenc}

\usepackage{dsfont}
\usepackage[capitalise]{cleveref}

\newcommand{\sectionauthor}[2][1]{{\it #2$^\text{#1}$,}}
\newcommand{\sectionlastauthor}[2][1]{{and \it #2$^\text{#1}$}}
\newcommand{\sectionaffil}[2][1]{$^\text{#1}$#2\newline}

\makeatletter
\def\@makechapterhead#1{%
  \vspace*{20\p@}%
  {\parindent \z@ \raggedright \normalfont
    \ifnum \c@secnumdepth >\m@ne
      \if@mainmatter
        \huge\bfseries \thechapter\space\space%
      \fi
    \fi
    \interlinepenalty\@M
    \huge \bfseries #1\par\nobreak
    \vskip 20\p@
  }}
\makeatother

\usepackage{tabularx,booktabs} 
\usepackage[utf8]{inputenc}
\usepackage{amssymb}

\newcolumntype{Y}{>{\centering\arraybackslash}X}

\newcommand{\gi}[1]{\textbf{\textit{\textcolor{gray}{#1}}}}

\newcommand{\rev}[1]{\textcolor{black}{#1}}

\newif\ifshowremoved
\showremovedfalse 

\newcommand{\removed}[1]{%
  \ifshowremoved
    \textcolor{red}{\sout{#1}}%
  \fi
}

\title{Roadmap on Advancements of the \\FHI-aims Software Package}
\date{Feb 17, 2026}

\usepackage{bbm}
\usepackage{fancyhdr}
\begin{document}

\author{}

\maketitle

{\large Authors:\\} 
\begin{onehalfspace}
Joseph W. Abbott,
Carlos Mera Acosta, 
Alaa Akkoush, 
Alberto Ambrosetti,
Viktor Atalla,
Alexej Bagrets,
Jörg Behler,
Daniel Berger,
\rev{Hannah Bertschi}, 
Björn Bieniek,
Jonas Bj\"ork,
{\bf Volker Blum*}, 
Saeed Bohloul,
Connor L. Box,
Nicholas Boyer,
Danilo Simoes Brambila,
Gabriel A. Bramley,
Kyle R. Bryenton, 
María Camarasa-Gó\-mez, 
Christian Carbogno, 
Fabio Caruso,
Sucismita Chutia,
Michele Ceriotti, 
Gábor Csányi,
William Dawson,
Francisco A. Delesma,
Fabio Della Sala,
Bernard Delley,
Robert A. DiStasio Jr.,
Maria Dragoumi,
Sander Driessen,
Marc Dvorak,
Simon Erker,
Ferdinand Evers,
Eduardo Fabiano,
Matthew R. Farrow, 
Florian Fiebig, 
Jakob Filser,
Lucas Foppa,
Lukas Gallandi,
Alberto Garcia,
Ralf Gehrke,
Simiam Ghan, 
Luca M. Ghiringhelli,
Mark Glass,
Stefan Goedecker,
Dorothea Golze,
Matthias Gramzow,
James A. Green,
Andrea Grisafi,
Andreas Gr\"uneis,
Jan G\"unzl,
Stefan Gutzeit, 
Samuel J. Hall, 
Felix Hanke,
Ville Havu,
Xingtao He,
Joscha Hekele,
Olle Hellman,
Uthpala Herath,
Jan Hermann, 
Daniel Hernangómez-Pérez,
Oliver T. Hofmann,
Johannes Hoja,
Simon Hollweger,
Lukas H\"ormann,
Ben Hourahine,
Wei Bin How,
William P. Huhn,
Marcel H\"ulsberg,
Timo Jacob,
Sara Panahian Jand, 
Hong Jiang, 
Erin R. Johnson,
Werner J\"urgens,
J. Matthias Kahk, 
Yosuke Kanai, 
Kisung Kang, 
Petr Karpov,
Elisabeth Keller,
Roman Kempt,
Danish Khan,
Matthias Kick, 
Benedikt P. Klein,
Jan Kloppenburg,
Alexander Knoll,
Florian Knoop,
Franz Knuth,
Simone S. K\"ocher,
Jannis Kockläuner, 
{\bf Sebastian Kokott*}, 
Thomas K\"orzd\"orfer, 
Hagen-Henrik Kowalski,
Peter Kra\-tzer,
Pavel K{\r{u}}s,
Raul Laasner,
Bruno Lang,
Björn Lange,
Marcel F. Langer,
Ask Hjorth Larsen,
Hermann Lederer,
Susi Lehtola, 
Maja-Olivia Lenz-Himmer, 
Moritz Leucke, 
Sergey Levchenko, 
Alan Lewis,
O. Anatole von Lilienfeld,
Konstantin Lion,
Werner Lipsunen,
Johannes Lischner,
Yair Litman, 
Chi Liu,
Qing-Long Liu,
\rev{Songrui Liu}, 
Andrew J. Logsdail, 
Michael Lorke,
Zekun Lou,
Iuliia Mandzhieva,
Andreas Marek, 
Johannes T. Margraf, 
Reinhard J. Maurer,
Tobias Melson,
Florian Merz,
Jörg Meyer,
Georg S. Michelitsch,
Teruyasu Mizoguchi,
Evgeny Moerman, 
Dylan Morgan,
Jack Morgenstein,
Jonathan Moussa,
Akhil S. Nair,
Lydia Nemec,
Harald Oberhofer, 
Alberto Otero-de-la-Roza, 
Ram\'{o}n L. Panadés-Barrueta,
Thanush Patlolla,
Mariia Pogodaeva,
Alexander P\"oppl,
Alastair J. A. Price, 
Thomas A. R. Purcell, 
Jingkai Quan,
Nathaniel Raimbault,
Markus Rampp, 
Karsten Rasim, 
Ronald Redmer, 
Xinguo Ren, 
Karsten Reuter,
Norina A. Richter,
Stefan Ringe, 
Patrick Rinke,
Simon P. Rittmeyer,
Herzain I. Rivera-Arrieta,
Matti Ropo,
{\bf Mariana Rossi*}, 
Victor Ruiz, 
Nikita Rybin,
Andrea Sanfilippo,
{\bf Matthias Scheffler*},
Christoph Scheurer,
Chris\-toph Schober,
Franziska Schubert,
Tonghao Shen, 
Christopher Shepard, 
Honghui Shang,
Kiyou Shibata,
Andrei Sobolev, 
Ruyi Song, 
Aloysius Soon, 
Daniel T. Speckhard, 
Pavel V. Stishenko,
\rev{Elia Stocco}, 
Muhammad \rev{N.} Tahir, 
Izumi Takahara,
Jun Tang,
Zechen Tang,
Thomas Theis,
Franziska Theiss,
Alexandre Tkatchenko, 
Milica Todorović,
George Trenins,
Oliver T. Unke,
Álvaro Vázquez-Mayagoitia,
Oscar van Vuren,
Daniel Waldschmidt,
Han Wang,
Yanyong Wang, 
Jürgen Wieferink,
Jan Wilhelm,
Scott Woodley,
Jianhang Xu, 
Yong Xu, 
Yi Yao,
Yingyu Yao,
Mina Yoon,
Victor Wen-zhe Yu,
Zhenkun Yuan, 
Marios Zacharias, 
Igor Ying Zhang, 
Min-Ye Zhang,
Wentao Zhang,
\rev{Xingchen Zhang}, 
Rundong Zhao, 
Shuo Zhao, 
Ruiyi Zhou, 
Yuanyuan Zhou, 
Tong Zhu \\ \\
{\bf (*)} journal guest editor
\end{onehalfspace}

\tableofcontents

\chapter{Introduction}

\newpage

\section{Introduction and Overview}
\sectionauthor[1,2]{Volker Blum}
\sectionauthor[3,4]{Sebastian Kokott}
\sectionauthor[5]{Mariana Rossi}
\sectionlastauthor[4]{Matthias Scheffler}

\sectionaffil[1]{Thomas Lord Department of Mechanical Engineering and Materials Science, Duke University, Durham, NC 27708, USA}
\sectionaffil[2] {Department of Chemistry, Duke University, Durham, NC 27708, USA}
\sectionaffil[3]{Molecular Simulations from First Principles e.V., D-14195 Berlin, Germany}
\sectionaffil[4]{The NOMAD Laboratory at the Fritz Haber Institute of the Max Planck Society, Faradayweg 4-6, D-14195 Berlin, Germany}
\sectionaffil[5]{Max Planck Institute for the Structure and Dynamics of Matter, 22761 Hamburg, Germany}

\subsection*{Abstract}

Electronic-structure theory is the foundation of the description of materials including multiscale modeling of their properties and functions. Obviously, without sufficient accuracy at the base, reliable predictions are unlikely at any level that follows. The software package FHI-aims has proven to be a game changer for accurate free-energy calculations because of its scalability, numerical precision, and its efficient handling of density functional theory (DFT) with hybrid functionals and van der Waals interactions.  It treats molecules, clusters, and extended systems (solids and liquids) on an equal footing. Besides DFT, FHI-aims also includes quantum-chemistry methods, descriptions for excited states and vibrations, and calculations of various types of transport. Recent advancements address the integration of FHI-aims into an increasing number of workflows and various artificial intelligence (AI) methods. This {\it Roadmap} describes the state-of-the-art of FHI-aims and advancements that are currently ongoing or planned.

\subsection*{Introduction}

The Fritz Haber Institute {\it ab initio} materials simulation (FHI-aims) software project is an accurate and precise electronic-structure and molecular dynamics software that treats molecules, clusters, and extended systems (solids and liquids) on an equal footing. Here the terms "accurate" and "precise" refer to the basic theoretical concepts and employed numerical methods, respectively. The FHI-aims project was conceived in 2003 at the Fritz Haber Institute of the Max Planck Society. FHI-aims has since grown into a large global community. The software is being developed and advanced by numerous researchers without whom it would not exist the way it does. This is exemplified by the 32 topical contributions and the 201 authors featured in this {\it Roadmap}, highlighting the breadth and diversity of expertise within the FHI-aims ecosystem.

This {\it Roadmap} describes the state-of-the-art of the FHI-aims software package and its ongoing advancements and the plans for the near future. \removed{This is not a review paper, and} \rev{D}etailed discussions on the background mathematics and physics are referred to the original publications or to other, topically focused roadmaps~\cite{Blum2024roadmap,Bauer2024,Gavini2023}. In addition to descriptions of the FHI-aims capabilities, most contributions also sketch and provide links to practical tutorials.
In short, the FHI-aims software is described in terms of three topical bullets:
\begin{itemize}
    \item Versatile: FHI-aims serves for the description of molecules, clusters, nanostructures, surfaces, solids, and liquids. It addresses the electronic structure as well as vibrations and (long-time) molecular dynamics. It is also links to state-of-the-art AI tools.
    \item Precise and Accurate: FHI-aims employs DFT with highly efficient implementations of advanced exchange-correlation functionals, many-body methods (GW approximation and Bethe-Salpeter equation) and quantum chemistry methods, such as Møller-Plesset perturbation and coupled-cluster theories. The latter also includes the equation-of-motion approach for excited states.
    \item Scalable: FHI-aims runs on laptops, and for larger systems and long-time molecular dynamics it runs efficiently on highest-performance computers.
\end{itemize}

The {\it Roadmap} covers all these aspects, and more. Besides this 1-Introduction, it is organized into seven further chapters: 2-Basic Concepts and Numerics, 3-Exchange and Correlation, 4-Electronic Excited States, 5-Linear Response Methods, 6-Vibrations and Dynamics, 7-Transport, and 8-Workflows and AI. Each of these chapters contains between 3 and 6 contributions, covering  different subtopics. Let give a brief summary of the various chapters.

Chapter~\ref{ChapBasicConcepts} introduces the foundational methodological concepts employed in FHI-aims. Numerically tabulated atom-centered orbital (NAO) basis sets are employed to mathematically express the electronic structure. The first roadmap segment covers the details of this choice, which is central to the code's ability to achieve high numerical precision while remaining affordable up to very large system sizes. The technical choices (integration grids, electrostatic potential, forces, stresses, structure optimization) necessary to implement electronic structure theory based on NAOs efficiently are covered in a separate segment. Next, approaches to include relativistic effects (scalar, spin-orbit coupled or four-component) are central for the physical accuracy of computational predictions by electronic structure theory. For computational efficiency, linear-algebra based solutions of the self-consistent field eigenvalue problem or the density matrix are essential and are handled by open-source libraries, including the Eigenvalue soLvers for Petaflop Applications (ELPA) library and the ELectronic Structure Infrastructure (ELSI) interface, which integrates this and other approaches into FHI-aims as a dedicated software layer. Finally, the wealth of data acquired in a single electronic structure calculation is immense and appropriate tools to analyze and understand this data are discussed. DFT represents a most important practical concept. However, it only exists in terms of approximations to the exchange and correlation (xc) functional.~\cite{teale2022}

Chapter~\ref{ChapXC} addresses the various levels of approximations, starting from the local-density approximation and Perdew’s generalized gradient approximation, which is the currently widely favored approach, and comes in several flavors. For several applications the self-interaction error of these functionals can be critical. Here the DFT+U method can heal part of the problem, in some cases. A significant advancement of FHI-aims concerns the recently improved implementation of the next level of the treatment of xc, namely hybrid functionals. Here FHI-aims achieves top-notch efficiency. All the mentioned functionals miss the long-range dispersion (van der Waals) interactions which can be critical for a reliable description, e.g. of organic systems. FHI-aims offers several options for adding vdW contributions to the standard xc functionals. The highest level xc treatment in DFT, close to “heaven of xc functionals” is the random-phase approximation (RPA), for which FHI-aims offers total energies and also forces. In fact, FHI-aims also offers quantum-chemistry methods such as coupled-cluster theory. However, this is discussed in the next chapter because it also includes the coupled-cluster equation-of-motion treatment to describe excited states. A special contribution of this chapter deals with explicit and implicit embedding approaches that allows the use of high-level first-principles methods in a reduced region, which is embedded in a coarse-grained environment. This technique is especially relevant for applications in biochemistry and homo- and heterogeneous catalysis.

Chapter~\ref{ChapExcitedStates} reports numerous approaches for calculating electronic excited states. It covers coupled-cluster theory (including the equation of motion), the GW approach, time-dependent DFT (also in its real-time formulation), and the Delta-SCF method. These implementations cover neutral excitations as well as ionization and electron affinities, which means charged excitations. As in essentially all FHI-aims contributions, the methods enable the treatment of core levels and valence bands and they apply to finite systems as well as to extended solids and liquids. Let us just mention real-time time-dependent DFT, because this is a less common concept. Such simulations can address important questions, especially pertaining to non-equilibrium electron dynamic phenomena, including interfacial charge transfer, atom-cluster collisions, topological quantum matter, etc. Furthermore, Ehrenfest dynamics can be performed on top of real-time time-dependent DFT so that non-adiabatic effects of electrons can be taken into account in first-principles molecular dynamics simulations. 

Chapter~\ref{ChapLinearResponse} covers the implementation of methods that are capable of describing the response of the electronic density to external stimuli, in a linear-response regime. Contribution 5.1 summarizes the implementation of density-functional perturbation theory (DFPT) for electric-field and nuclear displacement perturbations. This gives access to polarizabilities and optical dielectric constants of materials, which enter the calculation of several kinds of vibrational spectroscopy (Raman, sum-frequency generation, second-harmonic generation, etc.), as well as the quantities needed for the calculation of electron-phonon and non-adiabatic coupling terms, further detailed in Chapter 6. Contribution 5.2 details the implementation of DFPT for a magnetic-field perturbation, which provides the necessary ingredients for the calculation of NMR shifts. Finally, contribution 5.3 describes the implementation and gives practical guidelines for the calculation of the polarization in periodic systems, including derived quantities such as Born-effective charges and topological invariants. With the methods presented in this chapter, it is possible to assess a wide range of material response properties.

Complementing and enhancing the topics discussed in the previous chapters, Chapter~\ref{ChapVibrationsDynamics} deals with nuclear motion and its impact on electronic properties. Contribution 6.1 centers on the description of FHI-vibes, a package associated with FHI-aims that interfaces to several other packages and workflows for vibrational motion and ab initio molecular dynamics (aiMD). It allows the computation of harmonic, perturbative, and fully anharmonic nuclear motion in several contexts. Contribution 6.2 describes several ways of performing aiMD with FHI-aims and highlights its use in the grand-canonical replica exchange method, which allows one to simulate material phases with fluctuating density or stoichiometry. Contribution 6.3 describes two efficient methods to include the quantum nature of nuclei in atomistic simulations. The nuclear-electronic orbital method, which is implemented natively and can approximate adiabatic and non-adiabatic regimes, and the path-integral molecular dynamics method, which is exact in the adiabatic approximation and can be performed through the connection of FHI-aims with i-PI. Contribution 6.4 looks at the coupling between atomic motion and electronic properties. It describes the computation of electron-phonon coupling and its use in the computation of nuclear motion in metallic environments, where dissipative dynamics that can be described in an electronic friction formalism is possible. In Contribution 6.5, electron-phonon coupling in the adiabatic regime is also addressed by showing how to perform electronic band-structure unfolding with numeric atom-centered orbitals and calculating temperature-dependent band-structures from thermally-displaced structures.

Chapter~\ref{ChapTransport} deals with various aspects and forms of transport: vibrational heat transport, electrical transport of electrons or holes in semiconductors, and transport through single molecules and nanoscale systems where the molecule is positioned between two electrodes. With respect to the description of vibrational heat transport and electrical transport, FHI-aims features the most accurate methods considering the full statistical mechanics and all degrees of anharmonicities. The chapter also discusses perturbative methods that start from the phonon picture and are limited to largely harmonic materials. The more advanced methods employed by FHI-aims (Green-Kubo and Kubo-Greenwood) are particularly relevant for thermal insulators and thermoelectric materials. The method for molecular and nanoscale transport studies enables the treatment of a wide range of quantum phenomena, including quantum interference, non-equilibrium spin-crossover, and much more. Electronic couplings between electronic states localized in different components of a system, often also called transfer-integrals, are a useful tool to understand the efficiency of charge transport. The implementation which allows their calculation is described in the last contribution of this chapter.

Chapter~\ref{ChapWorkflowsAI} covers Workflows and AI. It starts with a description of efficient tools and infrastructure for creating and managing large datasets, visualizing and analyzing diverse data, and assessing the quality of simulation results. A wide array of tools is described, covering developments targeted mostly at FHI-aims, such as the Graphical Interface for Materials Simulations (GIMS), as well as more general community tools that can be efficiently used with FHI-aims. Important recent developments concern workflows that automate the AI guided acquisition of data, that perform AI-guided exploration of materials space, and that offer an AI-guided physico-chemical analysis of data. For example, the creation of machine-learning interatomic potentials is discussed, including active learning strategies to efficiently improve the quality of potentials. Because the structure space can be prohibitively high-dimensional, a strategy to reduce the dimensionality and efficiently sample conformational space is represented by the Bayesian Optimization Structure Search (BOSS) method. To understand structure-property relationships, the AI Sure Independence Screening and Sparsifying Operator (SISSO) method is integrated with FHI-aims. This enables the creation of maps of materials properties and functions to identify regions in materials space where high-performance can be obtained for a certain goal and/or where more data are needed.
Finally, Chapter 8 describes several machine-learning methods that target the electronic-structure directly and aim to learn basic ingredients of DFT, such as the electronic density or the Kohn-Sham Hamiltonian, based solely on a description of the atomic structure. These methods, which are all interfaced with FHI-aims, can be tightly integrated to electronic-structure calculations and offer a route to completely bypass the costly DFT evaluation of all material properties while retaining accuracy. 

In conclusion of this Introduction, let us mention that the lists of authors of the whole paper and of most (but not all) contributions are ordered alphabetically, and the coordinators of the respective contributions are typeset in bold.

We expect that the summaries of features and implementations presented in these contributions will raise the interest of the electronic structure community, and we are happy to receive further questions or comments by email to aims-coordinators@ms1p.org.

\subsection*{Acknowledgment}

We are deeply grateful to Anna Lutz for her dedication in coordinating and incorporating all change requests, and for her thoughtful communication with each contributor to the Roadmap article. Her efforts were instrumental in bringing it all together.

The Roadmap editors and FHI-aims coordinators:\\
Volker Blum, Sebastian Kokott, Mariana Rossi, and Matthias Scheffler

\chapter{Basic Concepts and Numerics}
\label{ChapBasicConcepts}





\section{Numeric Atom-Centered Orbital Basis Sets for DFT and Beyond}
\label{ChapBasisSets}

\sectionauthor[1,2]{J\"org Behler} 
\sectionauthor[3,4]{\textbf{*Volker Blum}}
\sectionauthor[5]{Nicholas Boyer} 
\sectionauthor[6]{Bernard Delley} 
\sectionauthor[7]{Maria Dragoumi} 
\sectionauthor[8]{Stefan Goedecker} 
\sectionauthor[3]{William Huhn} 
\sectionauthor[9,10,11]{Erin R. Johnson} 
\sectionauthor[12]{Elisabeth Keller} 
\sectionauthor[7,13]{\textbf{*Sebastian Kokott}} 
\sectionauthor[3]{Raul Laasner} 
\sectionauthor[14]{Susi Lehtola} 
\sectionauthor[12]{Johannes Margraf} 
\sectionauthor[3]{Jack Morgenstein} 
\sectionauthor[15]{Thanush Patlolla} 
\sectionauthor[16,7,a]{Xinguo Ren} 
\sectionauthor[12]{Karsten Reuter} 
\sectionauthor[17]{\textbf{*Mariana Rossi}}
\sectionauthor[7]{\textbf{*Matthias Scheffler}}
\sectionauthor[3,b]{Yi Yao} 
\sectionauthor[7,18,c]{Igor Ying Zhang} 
\sectionauthor[3,19,d]{Rundong Zhao}
\sectionlastauthor[3]{Tong Zhu} 

\sectionaffil[1]{Lehrstuhl für Theoretische Chemie II, Ruhr-Universität Bochum, 44780 Bochum, Germany}
\sectionaffil[2] {Research Center Chemical Sciences and Sustainability, Research Alliance Ruhr, 44780 Bochum, Germany}
\sectionaffil[3]{Thomas Lord Department of Mechanical Engineering and Materials Science, Duke University, Durham, NC 27708, USA}
\sectionaffil[4]{Department of Chemistry, Duke University, Durham, NC 27708, USA}
\sectionaffil[5]{Department of Chemistry, University of North Carolina at Chapel Hill, Chapel Hill, NC, USA}
\sectionaffil[6]{LTC, Paul Scherrer Institut, Forschungsstrasse 111, 5232 Villigen-PSI, Switzerland}
\sectionaffil[7]{The NOMAD Laboratory at the Fritz Haber Institute of the Max Planck Society, Faradayweg 4-6, D-14195 Berlin, Germany}
\sectionaffil[8]{Department of Physics, University of Basel, Klingelbergstrasse 82, CH-4056 Basel, Switzerland}
\sectionaffil[9]{Department of Chemistry, Dalhousie University, 6243 Alumni Crescent, Halifax, Nova Scotia, B3H 4R2, Canada}
\sectionaffil[10]{Department of Physics and Atmospheric Science, Dalhousie University, 6310 Coburg Road, Halifax, Nova Scotia, B3H 4R2, Canada}
\sectionaffil[11]{Yusuf Hamied Department of Chemistry, University of Cambridge, Lensfield Road, Cambridge, UK, CB2 1EW}
\sectionaffil[12]{Theory Department, Fritz Haber Institute of the Max Planck Society, Faradayweg 4-6, D-14195 Berlin, Germany}
\sectionaffil[13]{Molecular Simulations from First Principles e.V., D-14195 Berlin, Germany}
\sectionaffil[14]{Department of Chemistry, University of Helsinki, P.O. Box 55, FI-00014 University of Helsinki, Finland}
\sectionaffil[15]{William G. Enloe High School, Raleigh, NC 27610, USA}
\sectionaffil[16]{Institute of Physics, Chinese Academy of Sciences, Beijing, 100190, China}
\sectionaffil[17]{Max Planck Institute for the Structure and Dynamics of Matter, 22761 Hamburg, Germany}
\sectionaffil[18]{Department of Chemistry, Fudan University, Shanghai 200433, People’s Republic of China}
\sectionaffil[19]{School of Physics, Beihang University, Beijing 102206, China}

\sectionaffil[]{(*) Coordinator of this contribution}

\rule[0.25ex]{0.35\linewidth}{0.25pt}

\sectionaffil[a]{{\it Current Address:} Institute of Physics, Chinese Academy of Sciences, Beijing, 100190, China}
\sectionaffil[b]{{\it Current Address:} Molecular Simulations from First Principles e.V., D-14195 Berlin, Germany}
\sectionaffil[c]{{\it Current Address:} Department of Chemistry, Fudan University, Shanghai 200433, People’s Republic of China}
\sectionaffil[d]{{\it Current Address:} School of Physics, Beihang University, Beijing 102206, China}




\subsection*{Summary}

Density functional theory (DFT) -- especially semilocal Kohn-Sham DFT and hybrid generalized Kohn-Sham DFT, which are based on an \rev{orthonormal set of} effective single-particle wave function to construct the density -- is arguably the most productive method for quantum-mechanical simulations of materials and molecules. The central numerical choice in implementing DFT is how to discretize the effective single-particle orbitals using a finite set of basis functions. The chosen basis set determines the numerical precision of a particular simulation and also the time to solution. 

FHI-aims employs numerically tabulated atom-centered orbital (NAO) basis functions, exploiting several key advantages for precision and efficiency: (1) The rapidly varying shape of the orbitals and ground-state density towards the nucleus is captured practically exactly by including numerically calculated orbitals of free atoms in the basis set. Thus, accurate simulations including all electrons become straightforward, even for large systems. 
(2) Precise ground-state densities and total energies for molecules and for extended systems can be obtained with rather few additional basis functions beyond the minimal basis, i.e., the overall basis set size remains efficient even for high-precision calculations. 
(3) NAO basis functions can be strictly localized to a spatially limited region around their atomic center, $\mathbf{R}_I$, with a smooth decay to zero at a certain distance. Thus, different regions in space are treated by different subsets of basis functions, allowing one to implement almost all operations in standard DFT in a way that allows the computational effort to scale linearly with system size, $N$, in the limit of large systems. The sole exception is the solution of the matrix-based eigenvalue problem of DFT, which formally scales as $O(N^3)$ and dominates for the largest system sizes, even when employing highly efficient solvers \cite{Marek2014,kokott2024}. \rev{In the limit of a large number of atoms, this cost can often be circumvented by lower-scaling, direct solutions for the density matrix \cite{yu_elsi_2020}. However, the efficiency of these solvers strongly depends on the sparsity patterns (e.g., atom density) in the simulated system. For 3D bulk materials, diagonalization is still often the most efficient (or, at least, most practical) approach even for the largest systems simulated today.}

For ground-state DFT, FHI-aims possesses a complete, easy-to-use and well tested NAO basis set library for elements 1-102, which can be used without any further editing and which allows users to balance the desired precision with computational efficiency. Fast calculations are enabled by small but physically reliable ``light'' basis sets, while progressively larger basis sets (``intermediate'', ``tight'', and beyond) allow one to achieve benchmark-quality numerical precision for targeted observables in production calculations \rev{(though of course dependent on the observable in question)}. The general form of NAO basis functions also allows one to use a wide range of other standard atom-centered orbital basis sets, including pre-constructed NAO basis sets for applications beyond ground-state DFT, as well as Gaussian- or Slater-type orbital (GTO or STO) basis sets.

\subsection*{Current Status of the Implementation}

As noted above, the practical handling of FHI-aims' basis sets is simple in routine simulations. Three well-tested levels of numerical precision, ``light,'' ``intermediate,'' and ``tight,'' cover ground-state DFT and many other tasks across the periodic table. In a practical FHI-aims run, simply copy the desired predefined defaults into FHI-aims' input file \texttt{control.in} prior to running a calculation. 

The remainder of this section reviews the mathematical definition of the basis sets inside the code, as well as the origins of appropriate basis sets for ground-state DFT and beyond. Throughout this section, the term ``precision'' is used to indicate the numerical quality of the result for a given set of physical approximations to the electronic structure problem (e.g., a particular density functional) as opposed to the accuracy of the approximate electronic structure method(s) themselves.

\subsubsection*{\color{gray}Numerical definition of basis functions in FHI-aims}

In electronic structure theory, a set of $N_b$ basis functions $\{\varphi_i(\mathbf{r})\}$ is employed to construct the electron density, $n(\mathbf{r})$, and other observables from effective, single-particle orbitals, $\psi_l(\mathbf{r})$, where $l$ is the orbital index:
\begin{equation}
    \psi_l(\mathbf{r}) = \sum_{i=1}^{N_b} c_{il} \varphi_i(\mathbf{r}) . 
    \label{Eq:KS-Orbitals}
\end{equation}
FHI-aims uses numerically tabulated, atom-centered basis functions of the form \cite{Blum2009}
\begin{equation}
    \varphi_i(\mathbf{r}) = \frac{u_i(r)}{r} \cdot Y_{L}(\Omega) . \label{Eq:NAO}
\end{equation}
Here, $u_i(r)$ is a radial function, usually centered at an atomic position $\mathbf{R}_I$ (but can also be placed on user-defined empty sites). $I$ denotes the center, $Y_{L}(\Omega)$ is a real-valued spherical harmonic function, $r=|\mathbf{r}-\mathbf{R}_I|$ denotes the distance to the atomic position, \rev{and $\Omega=(\theta,\phi)$ are the angles defined by the unit vector $(\vec r - \vec R_I)/r$}. $L$ indicates the combination of angular and magnetic quantum numbers, which are implicitly defined by the overall basis function index $i$ (i.e., $L=L(i)$ in practice). 
The use of numerically tabulated radial functions, $u_i(r)$, in Eq.~(\ref{Eq:NAO}) was pioneered by Averill, Ellis, Zunger, Freeman, Delley, te Velde, Baerends, Koepernik, Eschrig, and others \cite{Averill1973,Zunger1977,Delley1982,Delley1990,teVElde1991,Koepernik1999}. In FHI-aims, the radial functions, $u_i(r)$, are defined via smooth \rev{natural} cubic spline functions on dense, logarithmically spaced grids and, thus, allow for a highly precise representation of any radial function shape. 
Calculations of periodic systems can be performed on exactly the same mathematical footing as non-periodic calculations by using linear combinations of the localized basis functions of Eq.~(\ref{Eq:NAO}) in every unit cell, with unit-cell-dependent phase factors accounting for reciprocal-space sums \cite{Blum2009}.

FHI-aims supports five types of radial functions, $u_i(r)$: 
(i) radial functions of spherical, self-consistent free atoms; 
(ii) radial functions of hydrogen-like atoms (i.e., one-electron atoms with a spherical nuclear potential, $Z/r$, in which $Z$ is an effective parameter; 
(iii) radial functions of spherical, self-consistent free ions; 
(iv) GTO functions; and 
(v) STO functions. 
Radial functions, densities, and potentials for (i) spherical free atoms and (iii) free ions are generated on the fly using the atomic solvers originally included in the fhi98pp code \cite{fuchs1999}, or in the  atom\_sphere \cite{Goedecker2016} or four-component (Dirac) DFTatom codes \cite{Certik2013}, while hydrogen-like functions (ii) can be generated either from their analytical shape or by numerically solving the corresponding radial equation. The NAO radial function types (i-iii) in FHI-aims are strictly localized, i.e., they are non-zero inside a specified radius and they decay smoothly to zero when approaching the boundary. The smooth decay of the radial functions is achieved
by applying a confinement potential during the construction of the basis functions \cite{Blum2009,Delley1990,Junquera2001} (i.e., also for numerically computed, hydrogen-like radial functions). This confinement potential is zero within a given radius, $r_c$, and then smoothly increases to a large value (formally, infinity) over a width, $w$. The intention of this localization is to avoid extended regions in which a basis function is close to, but not quite, zero, while keeping the impact on the overall attainable precision of a calculation minimal. Typical extents of the basis functions are thus still rather large: $r_c+w=5$~{\AA} for light elements and ``light'' computational settings and up to 8~{\AA} for the largest atoms (Cs) and ``tight'' settings. As a result, products of NAO basis functions are only non-zero if they are located on atoms in reasonable proximity to one another, facilitating $O(N)$ scaling of the computational effort with increasing system size (where $N$ is a measure of the system's extent) for calculating Hamiltonian and overlap matrix integrals in semilocal DFT \cite{Havu2009}, the non-local exchange operator \cite{kokott2024}, and the electron density update. \rev{Depending on the exact mathematical context, the measure $N$ might refer to the number of atoms, basis functions, electrons or similar quantities related to the system size.} 

\subsubsection*{\color{gray}Construction principle of the NAO basis set library for ground-state DFT}

Following the same basic principles as other atom-centered basis sets in quantum chemistry, the preconstructed standard basis sets in FHI-aims consist of two parts:
\begin{itemize}
    \item A ``minimal basis'' of atomic basis functions that can accommodate all electrons of each individual atom in a chemical system (non-periodic or periodic). The minimal basis can be constructed to represent the ground-state free-atom limit with high accuracy, including the structure of the orbitals and density near the nucleus, but it is not on its own sufficient to capture properties in extended systems.
    \item Additional radial functions that provide flexibility to represent the orbitals, electron density, and other observables in the full system. Consistent with the experience from other atom-centered orbital approaches (GTOs or STOs), only a relatively small set of specifically selected basis functions in addition to the minimal basis is needed to achieve physically accurate solutions for the orbitals, density, and total ground-state  energy \cite{Delley1990}.
\end{itemize}
The fact that the shape of the radial functions, $u_i(r)$, in Eq.~(\ref{Eq:NAO}) is flexible enables the creation of basis sets that can represent different chemical structures and environments in DFT with high precision, while keeping the overall number of basis functions low \rev{(in the range of single digits to several tens per atom, depending on chemical element and desired precision; see below for some examples)}, i.e., remaining computationally efficient.

The minimal basis in FHI-aims consists of numerically calculated radial functions of self-consistent, spherical and non-spinpolarized free atoms in their ground states. This minimal basis captures the shape of the orbitals and density close to each atomic nucleus nearly exactly in a straightforward way, since the electron-nuclear potential is very large and dominates the solution close to the nucleus. Thus,
shape approximations to the near-nuclear potential, orbitals, or density, such as pseudopotentials, effective core potentials, or projector-augmented waves are not needed. Likewise, the unphysical behavior of the kinetic energy operator that can arise for linear combinations of analytical functions \cite{Zhang2013}, such as \rev{for} contracted Gaussian functions \rev{near a nucleus}, is avoided \rev{(cf. Figure A.2 and associated discussion in Ref. \cite{Zhang2013})}. As an additional advantage, physically correct finite-nucleus potentials for different isotopes can be substituted for the unphysical Coulomb singularity, $-Z_I/r$  \cite{Andrae2000}, if desired (FHI-aims has an implementation of finite-nucleus potentials from experimentally determined charge density distributions \cite{Jager1974,Vries1987,Fricke1995} as compiled in Ref.~\cite{Day2014}), and fully relativistic four-component atomic basis functions can also be incorporated into the formalism in the quasi-four-component (Q4C) approach \cite{Zhao2021}.

Beyond the minimal basis definition, FHI-aims provides a basis-set library of additional radial functions for each atom type (elements 1-102), enabling calculations of increasing precision as more functions are included.
This basis set library was created in an automated fashion, sidestepping any potentially limiting ``human intuition'' in the initial basis set construction steps \cite{Blum2009}. The simple recipe that proved successful was to select specific basis functions one by one from a large pool of candidate functions, using a stepwise variational minimization of non-self-consistently computed total energies of atomic dimers at several different bond distances. The pool of candidate functions included orbitals of hydrogen-like single-electron atoms, as well as doubly charged free ions, considering angular momenta up to $g$ for elements 1-18 and up to $h$ for all elements beyond. 

In practice, the sequences of basis functions resulting from this minimization procedure turn out to be transferable for use as basis functions in more complex structures as well. With few exceptions, the basis functions follow one another in groups of different angular momenta (e.g., a $p$-, $d$- and $s$-type function following the minimal basis for oxygen, see below). These groups are denoted as ``tiers'' or ``levels'' of basis functions for each element. The basis sets in the ``light'', ''intermediate'', and ``tight'' numerical defaults of FHI-aims are constructed by selecting basis functions from these ``tiers'' to achieve basis set convergence commensurate with the respective defaults. \rev{However, we note that convergence of an observable with the basis set size may be dependent on the system type (i.e., free atoms, molecules, surface or bulk systems), the chosen method (i.e.,GGA, meta-GGA, or hybrid functional vs. correlated methods), or the observable itself.}

\begin{figure}[t]
    \centering
    \includegraphics[width=\linewidth]{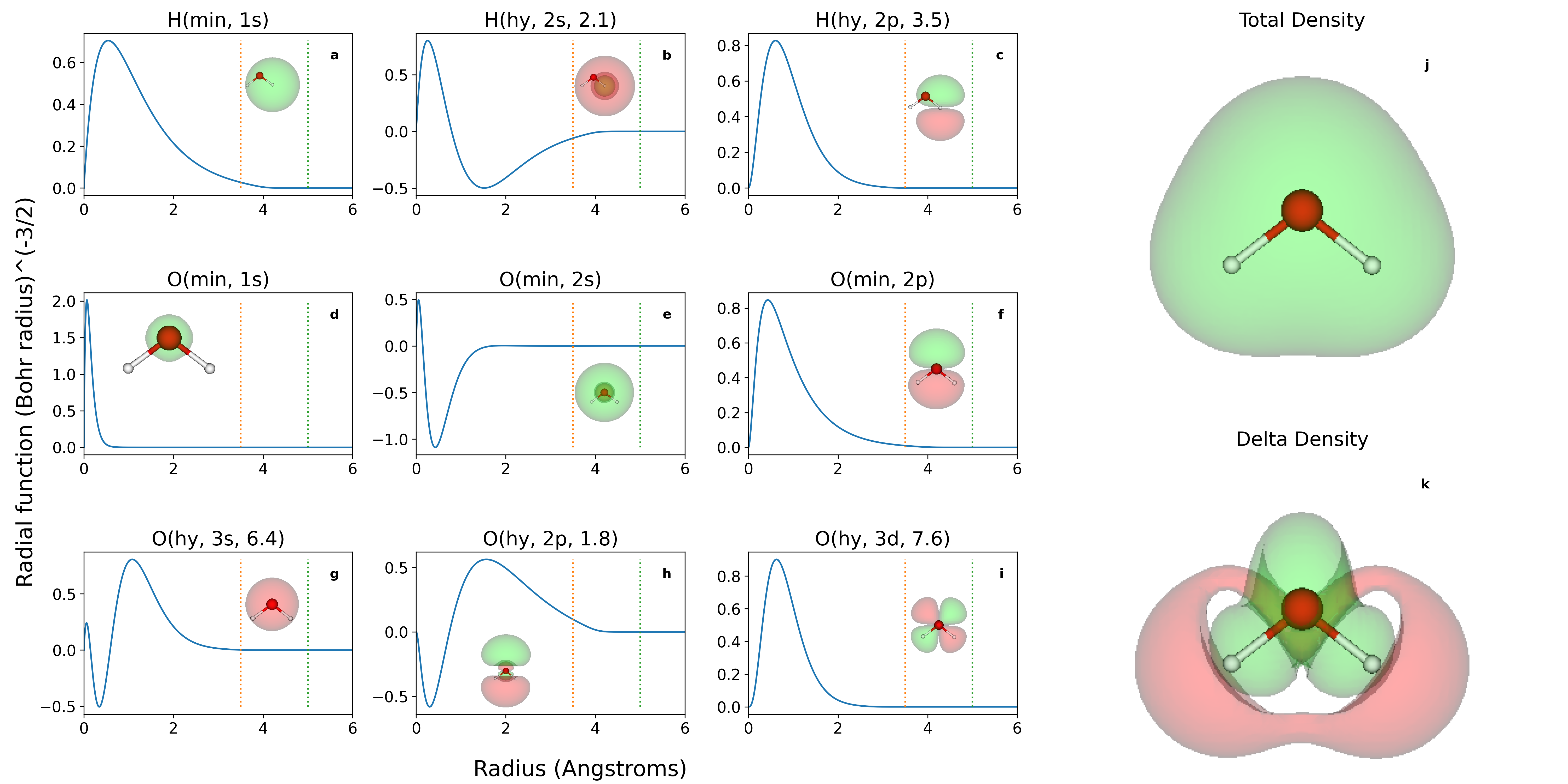}
    \caption{(a-i) Visualization of the NAO basis functions associated with ``light'' settings for a H$_2$O molecule. Each plot shows a radial function, $u_i(r)$, with  an isosurface of the corresponding three-dimensional basis function, $\varphi_i(\mathbf{r})$, as an inset. The numerical definition of each basis function is indicated at the top of each plot. ``min'' indicates a minimal basis function, ``hy'' denotes a basis function from a hydrogen-like atom, for which the final number is the effective nuclear charge, $Z$, used in the construction of that function (e.g., $Z=2.1$ for the hydrogen-like 2$s$ function in b). Vertical red lines show the onset of the confinement potential (3.5~{\AA}) and vertical green lines show the radial distance at which the confinement potential approaches infinity (5~{\AA}). For larger distances, each radial function is strictly zero. (j) Density isosurface of the self-consistent total density of the H$_2$O molecule, computed using FHI-aims' ``light'' settings and the PBE density functional. (k) Difference between the self-consistent density of the H$_2$O molecule and a simple addition of spherical free-atom densities. The isosurfaces of basis functions are shown for a value of $\pm 0.03$ a.u., with green denoting positive and red negative isosurfaces. Similarly, total and difference densities are plotted with positive (green) and negative (red) isosurfaces for values of $\pm 0.03$ a.u.}
    
    \label{fig:basis}
\end{figure}

As an example, Figure \ref{fig:basis} illustrates the basis functions used in ``light'' settings for the H and O atoms in the H$_2$O molecule.
The hydrogen atom has only a single minimal basis function (1$s$, Figure~\ref{fig:basis}a). The next selected basis function is a parameterized, hydrogen-like 2$s$ function (Figure~\ref{fig:basis}b) followed by a 2$p$ hydrogen-like function (Figure~\ref{fig:basis}c). In multi-atom calculations, this initial group of basis functions builds on the atomic solution but provides variational flexibility both regarding the radial character and the angular momentum character of the density around the H atom. The minimal 1$s$, hydrogen-like 2$s$ and 2$p$ functions shown in Figures~\ref{fig:basis}a-c (the first ``tier'') are enabled as part of FHI-aims' ``light'' defaults for H. \rev{Considering that there are three angular momentum functions for a $p$ orbital, this adds up to five radial functions for H with ``light'' settings.}

The light elements B, C, N, O, and F all possess minimal basis sets composed of 1$s$, 2$s$, and 2$p$ functions (shown for O in Figures~\ref{fig:basis}d-e). Consistent with typical STO or GTO basis sets, the next ``tier'' for each of these elements contains one $s$, $p$, and $d$ function, respectively (Figures~\ref{fig:basis}g-i)\rev{, amounting to a total of 14 basis functions per atom}. As in the case of H, this group of functions again provides flexibility both regarding the radial and the angular momentum character of the density around each atom. This first tier of basis functions is used as ``light'' settings in FHI-aims. It yields rather reliable molecular geometries \cite{Blum2009} and closely matches the design principles of so-called ``double zeta plus polarization'' STO and GTO basis sets. 

Figure \ref{fig:basis}j shows the resulting electron density of a H$_2$O molecule for the PBE density functional \cite{Perdew1996}. Figure~\ref{fig:basis}k displays the corresponding difference between the self-consistent electron density, $n(\mathbf{r})$, and a simple superposition of spherical free-atom densities without any chemical bond formation, showing the physically expected shift of charge from the red region near the H atoms to the green region closer to the O atom). As shown, e.g., in Refs.\ \cite{Blum2009,Jensen2017}, the numerical precision can be systematically increased to very high precision \rev{(well below the chemical accuracy threshold of 1kcal/mol for total energies and atomization energies)} as more basis functions are added beyond ``light'' settings. \rev{For example, Figure 2 in Ref.~\cite{Jensen2017} reports 0.8 kcal/mol Median Absolute Deviation (MAD) and 1.4 kcal/mol maximum absolute deviation (max AD) for tight species defaults, and 0.1 kcal/mol MAD and 0.8 kcal/mol max AD for tier4 species defaults for 100+ molecules, per molecule with respect to highly precise values obtained using the multiresolution wavelet code MRCHEM \cite{MRChem2016}. The deviations for atomization energies are correspondingly less, i.e., 0.3 kcal/mol MAD and 1.3 kcal/mol for tight species defaults, and 0.08 kcal/mol MAD and 0.5 kcal/mol max AD for tier4 species defaults, because some errors cancel.}

The computer-generated basis set library in FHI-aims can also safeguard against potential omissions in human intuition. For example, the Zn atom ($Z$=30) has a closed 3$d$ electron shell but not yet a $p$-type valence electron shell in the free atom. The first automatically selected additional basis function is indeed a $p$-type function, reflecting its importance in chemical bonds, then followed by another set of $s$, $p$, $d$ and $f$ functions. This group of 33 basis functions is retained in the the first tier of basis functions for Zn in FHI-aims, corresponding to ``light'' settings.

\rev{In relation to other basis function types than NAOs (e.g., plane waves), it is important to note that comparing just the number of basis functions on their own is not strictly meaningful, since the numerical effort per basis function will vary greatly with the basis function type, desired numerical precision, and other approximations made. Furthermore, precise points of comparison in the literature can be difficult to find. As one literature example, consider the augmented plane wave calculations showcased in Ref.~\cite{Blaha2010}. In this work, for the CO molecule in a periodic box, 1,400 basis functions per atom were used. For a bulk Sb$_2$S$_3$ crystal, 120 basis functions per atom were employed. In contrast, FHI-aims' ``tight'' settings require 39 basis functions per atom for the CO molecule (an advantage brought about because no NAO basis functions are needed to represent the vacuum in a DFT total-energy calculation). In the Sb$_2$S$_3$ case, 44.2 NAO basis functions per atom would be used in FHI-aims (``tight'' settings). Towards the largest systems, the relatively compact basis sets (i.e., relatively low ratio of number of basis functions to number of electrons in the system) will reduce the computational effort for matrix based eigenvalue or density matrix solutions (Sec. \ref{SecELPA}).}

\subsection*{\color{gray}NAO basis sets beyond ground-state DFT}

As with any other type of basis set, basis sets with different characteristics can be required for applications beyond ground-state DFT, e.g., to accurately represent explicit two-electron correlation (not part of ground-state semilocal or hybrid DFT) or to reflect response properties that arise in special spatial regions, such as the response of the density to a core level excitation or to the presence of nuclear spins.\cite{laasner+es2024} For some scenarios, specialized NAO basis sets have been developed and benchmarked, as described in other sections of this review. Examples include:
\begin{itemize}
  \item \textit{Correlated methods for light-element molecules} (e.g., second-order M{\o}ller-Plesset theory, MP2, the random phase approximation, RPA, and $GW$): 
    The valence correlation consistent (VCC) $n$-zeta (nZ) NAO basis sets (``NAO-VCC-$n$Z'') for the elements H-Ar introduced in Ref. \cite{Zhang2013} allow for a systematic extrapolation of ground- and excited-state energies to the complete basis-set limit, using basis sets of increasing order $n$ ($n$=2-5). The principle follows the established scheme for GTO-based correlation-consistent basis sets by Dunning and others \cite{Dunning1989}, but using a minimal basis set of free-atom NAO basis functions and additional hydrogen-like basis functions, all of which are numerically confined to a finite extent as described above.   
  \item \textit{Low-energy neutral excitations in light-element molecules} (e.g., optical absorption and excitons using the Bethe-Salpeter equation (BSE), linear-response or real-time time-dependent DFT): As shown in Ref.~\cite{liu2020all}, numerically precise neutral excitation energies can be achieved by adding a few spatially extended ``augmentation'' basis functions to FHI-aims' standard basis sets. This strategy is analogous to Dunning's augmentation of standard GTO basis sets.\cite{Kendall1992} In practice, Ref.~\cite{liu2020all} shows that adding one $s$-like and one $p$-like augmentation function from Dunning's basis sets to FHI-aims' ``tier 2'' basis sets allows one to compute low-lying excitation energies of small molecules with a numerical precision of a few hundredths of an eV. These basis sets are referred to as ``tier2\_aug2'' in FHI-aims. 
  \item \textit{Neutral excitation energies of light-element molecules involving core levels}: For excitations that promote a core electron to a valence level (e.g., in X-ray absorption spectroscopy), additional basis set flexibility in the near-nuclear region is needed. As shown in Ref. \cite{Yao2022}, precise core-level excitation energies can be obtained by adding a few highly localized STO basis functions to the  ``tier2\_aug2'' basis sets.
  \item \textit{$GW$ calculations of energy band structures of periodic solids:} FHI-aims' standard ``tier 2'' basis sets show qualitatively reliable results \cite{Ren2021}. In line with other basis set types (e.g., plane waves, linearized augmented plane waves or GTOs), key parameters such as the fundamental gap reach a precision of several hundredths or tenths of eV for dense solids. Compared to experiment, the resulting band gaps are drastically improved over values extracted from single-particle eigenvalues of semilocal DFT. As explained in more detail in Ref. \cite{Ren2021}, additional, highly localized STO basis functions can be added to further improve numerical precision. Similarly, optical properties of solids can be computed using the BSE based on $GW$ quasiparticle calculations and FHI-aims' standard ``tier'' basis sets.\cite{Zhou2025}
  \item \textit{Nuclear magnetic resonance observables for light-element molecules:} As shown in Ref.~\cite{laasner+es2024}, nuclear shieldings can be computed with high numerical precision using the NAO-VCC-nZ basis sets. $J$-couplings, i.e., the effective coupling strength between two nuclear spins, require basis sets that include additional, highly localized functions around each nucleus, in line with experience from other basis set types. For $J$-couplings (elements 1-18), FHI-aims offers a series of basis sets called NAO-$J$-$n$. These basis sets combine the NAO-VCC-$n$Z basis sets with highly confined, primitive $s$-type GTOs derived from Jensen's GTO-based pc-J basis sets\cite{Jensen2006,Jensen2010}.
\end{itemize}
In addition to NAO basis sets, FHI-aims can also employ any other basis set prescription based on STOs or GTOs from the standard quantum chemistry literature, since these basis functions have the same numerical form as Eq. (\ref{Eq:NAO}). A broad range of GTO basis sets has been collected at the Molecular Sciences Software Institute (MolSSI) basis set exchange \cite{Pritchard2019} and can be exported in a format suitable for FHI-aims. Thus, direct comparisons and benchmarks with standard quantum-chemistry approaches are possible on equal footing in FHI-aims, e.g., in the GW100 benchmark of molecular excitation energies\cite{Setten2015}.

\subsection*{Usability and Tutorials}

As noted earlier, FHI-aims' standard basis sets are pre-tabulated within ``light'', ``intermediate'', or ``tight'' species defaults. Additionally, a set of ``really\_tight'' species defaults is provided as a starting point for customizable convergence tests, with the same initial basis-set selection as ``tight'' but even more stringent numerical parameters otherwise. The user either appends these defaults for the given elements to the \texttt{control.in} input file of FHI-aims or uses an external package, such as GIMS \cite{Kokott2021} or ASE \cite{larsen2017ase}, to automatically build an input file with the specified defaults.  Their use is documented across all tutorials published for the code (\href{https://fhi-aims-club.gitlab.io/tutorials/tutorials-overview/}{\texttt{https://fhi-aims-club.gitlab.io/tutorials/tutorials-overview/}})

The tabulated defaults that are distributed with the code have been benchmarked for reliable performance. If desired, all defaults can be trivially customized. In addition to other numerical settings (e.g., the real-space grids and the electrostatic potential expansion order\cite{Blum2009}), they contain specifications of the NAO confining parameters, the NAO minimal basis, and the list of all further basis functions to be used in the calculation. This selection can be used as is or (if desired) can be adjusted -- either by simply commenting / uncommenting suggested basis functions or by editing directly, allowing one to add any supported basis-function type with custom parameters. \rev{Thus, the modification of the parameters in the species defaults is straightforward for any user. The meanings of all these parameters are documented in the FHI-aims manual.} Furthermore, standard GTO and STO basis sets from other quantum chemistry codes can be used, e.g., imported from the MolSSI Basis Set Exchange.\cite{Pritchard2019} 

For DFT, the high numerical precision of the pre-tabulated basis sets in FHI-aims has been established in a host of tests against other codes over time. Perhaps most prominently, in the widely cited 2016 ``Delta test'' of computationally predicted equations of state $E(V)$ for 70 elemental solids,\cite{Lejaeghere2016} FHI-aims' ``really\_tight`` settings with the first two angular momentum groups or ``tiers'' enabled, provided essentially indistinguishable numerical precision compared to other benchmark-quality all-electron codes (Wien2k\cite{Blaha2020} and Exciting\cite{gulans2014exciting}). FHI-aims' current ``default\_2020'' light, intermediate, and tight production basis-set choices reflect selections that were assessed using the same Delta test data to ensure reliable precision at each level. A full compilation of the Delta test results is included in Appendix~\ref{fig:DeltaTestSummary} and a set of corresponding input and output files has been deposited at the NOMAD repository.

For total energies, atomization energies and dipole moments of isolated systems, a comparison to high-precision multiresolution wavelet calculations for 211 molecules demonstrated similarly high and systematically increasing precision of FHI-aims' NAO basis sets, down to few-meV total energy differences per molecule on average. \cite{Jensen2017} Energy band structures for 103 elemental and compound solids, covering 66 chemical elements, were later compared at the DFT level between FHI-aims and the high-precision all-electron code Wien2k. \cite{huhn2017} For scalar-relativistic electronic valence states, differences remain below 10 meV on average for the band structures when using FHI-aims' ``tier 2'' NAO basis sets. For conduction bands up to 5~meV above the conduction band minimum or Fermi level, this difference increases to 50~meV for some elements; this is still remarkably close given the limits imposed by finite basis set size on the representation of high-lying unoccupied states. The same comparison was later carried out for self-consistent, spin-orbit-coupled band structures from the Wien2k code, compared to Q4C results from FHI-aims, also including spin-orbit coupling self-consistently. Despite the completely different implementations and overall different methodological choices, both codes agreed to within 50~meV on average even for the largest deviation (metallic Ir, for which FHI-aims' Q4C implementation is expected to be more accurate) for valence energy bands. 

A standard benchmark for main-group quantum chemistry is the GMTKN55 database \cite{gmtkn55}, which consists of 55 individual benchmark sets that can be grouped into five categories, as shown in Table~\ref{t:gmtkn55}. Since the data sets in each category have very different energy scales, their contributions to the total error metrics are weighted by the average magnitude of their component energy differences. The excellent performance of FHI-aims' NAO basis sets is demonstrated by their differences from the GTO values, $|\Delta|$ in Table~\ref{t:gmtkn55}, giving rather close results to those reported in Ref.~\cite{gmtkn55} using quadruple-$\zeta$ GTO basis sets, while including far fewer basis functions. Since this comparison focuses primarily on numerical precision of the different basis set prescriptions, the PBE functional was selected for testing in this work due to its unambiguous definition. As shown by the relatively high values of the weighted mean absolute deviation, compared to the high-level reference data of Ref.\ \cite{gmtkn55}, the physical accuracy of the PBE functional itself is of course still limited due to neglect of dispersion and inherent delocalization error. 

\begin{table}
\caption{Weighted mean absolute deviations (WTMAD2, defined in Ref.~\cite{gmtkn55}, in kcal/mol) from high-level reference data for each category of the GMTKN55 benchmark. Results are shown for the PBE functional using either literature data\cite{gmtkn55} for Gaussian-type orbitals (GTOs) or FHI-aims calculations (NAOs). $|\Delta|$ denotes the difference between the calculated GTO and NAO WTMAD2 values, validating their numerical precision. The GTO calculations used the aug$^\prime$-def2-QZVP basis set, while the NAO calculations used tight species defaults in most cases. The exceptions are for the AHB21, BH76, BH76RC,
G21EA, and IL16 benchmarks, which contain weakly bound anions. Here, the nonstandard ``tier2\_aug2'' basis was used as it includes additional diffuse functions. The other exception is for WATER27 where, due to the large system sizes, tier2\_aug2 was only applied to oxygen atoms for the specific reactions involving anions. The FHI-aims inputs are freely available on github \cite{gmtkn55-inputs}.}
\label{t:gmtkn55}
\centering
\begin{tabular}{l|c|ccc} \hline
\multirow{2}{*}{Dataset description} & Number & \multicolumn{3}{c}{PBE} \\ 
 & of subsets & GTOs & NAOs & $|\Delta|$ \\ \hline
Reaction energies for small systems      & 18 & ~7.93 & ~7.99 & 0.06 \\
Reaction energies for large systems      & ~9 & 16.23 & 16.27 & 0.04 \\
Barrier heights                          & ~7 & 16.72 & 16.42 & 0.30 \\
Intermolecular non-covalent interactions & 12 & 15.68 & 16.01 & 0.33 \\ 
Intramolecular non-covalent interactions & ~9 & 19.70 & 19.93 & 0.23 \\ \hline
Total data set                           & 55 & 13.88 & 13.95 & 0.07 \\ \hline
\end{tabular}
\end{table} 

\subsection*{Future Plans and Challenges}

FHI-aims' standard NAO basis sets for DFT are highly robust and flexible, such that the fundamental basis-set selection has not needed to be changed since its original 2009 publication \cite{Blum2009}. However, several additions are nevertheless desirable and on the way.

Methods that improve over the speed of ``light'' settings yet remain reliable have been a frequently desired wishlist item. The challenge is that the minimal basis of free-atom functions on its own is needed to guarantee the numerical representation of free atoms, but is itself not sufficient to guarantee physically correct results for bonded structures. Recent work by Keller \emph{et al.} \cite{Keller2024} provides a pathway forward, by employing almost minimal basis sets, to which only a single $s$ function is added. Additionally, a parameterized correction energy is introduced that accounts for the some of the missing precision. This correction currently only supports the widely used PBE density functional, but could be extended to other functionals as well. The energy correction results in an average error in unit-cell volume of less than 5\% across a wide range of different materials spanning from perovskites to noble gases and metalloids. In comparison with the commonly used FHI-aims-09 ``light'' and ``tight'' basis sets, this approach was found to offer average speedups of 1.92 and 4.76, respectively, for the cases considered in Ref.\ \cite{Keller2024}. This decrease in computational time makes the simulation of systems up to tens of thousands of atoms, or beyond, significantly more accessible. 

The minimal basis sets of FHI-aims contribute greatly to the precision of electron densities and total energies near each nucleus. One remaining restriction is that numerically exact free-atom solutions in the atomic solver libraries employed in FHI-aims are not available for meta-GGA density functionals or hybrid DFT. Thus, the minimal basis sets for these functionals are currently typically constructed by the local-density approximation or the semilocal PBE functional. While the resulting difference is insignificant for energy differences in chemically bonded systems, a completely consistent treatment of the minimal basis for each density functional would be desirable in order to completely leverage the all-electron precision in FHI-aims across all functionals.

Finally, for speed reasons, the introduction of pseudopotentials in FHI-aims has often been desired, since this change would reduce computational cost to the attainable minimum by outrightly eliminating all core electrons. Indeed, this is the approach taken by the Siesta code \cite{Garcia2020} and others. NAOs are perfectly compatible with pseudopotentials. However, a new optimization of all basis sets, grids, and a complete pseudopotential set would be required. Norm-conserving pseudopotentials are already partially supported in FHI-aims in the context of embedding calculations. This functionality could easily be extended to the fully self-consistent case. In order to increase the speed of calculations even further, full support for norm-conserving pseudopotenials could certainly be envisioned, but has not yet been approached because of the considerable human time investment. 

As an alternative approach to minimize the computational cost associated with core electrons, a frozen-core approximation has been implemented for the eigenvalue solver in the ELSI infrastructure. This approach requires no adjustments to the underlying potential and it allows one to seamlessly re-introduce the core electrons if desired. Carrying the frozen-core approach through the remaining FHI-aims code (beyond the eigensolver stage) is a promising route to achieve similar speedups to an outright pseudopotential variant of FHI-aims.

\subsection*{Acknowledgements}
We acknowledge 
Dr.\ Axel Becke for his assistance with generating the GMTKN55 inputs. 
ERJ thanks the Atlantic Computational Excellence Network (ACENET) for computing resources, the Natural Sciences and Engineering Research Council (NSERC) of Canada for financial support, and the Royal Society for a Wolfson Visiting Fellowship. MS acknowledges support by his TEC1p Advanced Grant (the European Research Council (ERC) Horizon 2020 research and innovation programme, grant agreement No. 740233.





\newpage

\section{Technical Underpinnings for All-Electron Density Functional Theory with Numeric Atom-centered Orbitals }

\sectionauthor[1]{Viktor Atalla}
\sectionauthor[2]{Daniel Berger}
\sectionauthor[1]{Björn Bieniek}
\sectionauthor[3,4]{\textbf{*Volker Blum}}
\sectionauthor[5]{Saeed Bohloul}
\sectionauthor[6,7,a]{Christian Carbogno}
\sectionauthor[1]{Sucismita Chutia}
\sectionauthor[1]{Ralf Gehrke}
\sectionauthor[3]{Mark Glass}
\sectionauthor[8]{Dorothea Golze}
\sectionauthor[8]{Jan Günzl}
\sectionauthor[9]{Ville Havu}
\sectionauthor[10]{Olle Hellman}
\sectionauthor[4]{Uthpala Herath}
\sectionauthor[4]{William Huhn}
\sectionauthor[1,b]{Werner J\"urgens}
\sectionauthor[6,11]{Florian Knoop}
\sectionauthor[6]{Franz Knuth}
\sectionauthor[6,5]{\textbf{*Sebastian Kokott}}
\sectionauthor[3]{Björn Lange}
\sectionauthor[6]{Maja Olivia Lenz-Himmer}
\sectionauthor[6,5]{Konstantin Lion}
\sectionauthor[9]{Werner Lipsunen}
\sectionauthor[12]{Florian Merz}
\sectionauthor[6]{Herzain I. Rivera-Arrieta}
\sectionauthor[1]{Matti Ropo}
\sectionauthor[13]{Mariana Rossi}
\sectionauthor[1]{Andrea Sanfilippo}
\sectionauthor[6]{Matthias Scheffler}
\sectionauthor[1]{Franziska Schubert}
\sectionauthor[14]{Álvaro Vázquez-Mayagoitia}
\sectionauthor[1]{J\"urgen Wieferink}
\sectionauthor[15]{Scott Woodley}
\sectionauthor[4,c]{Yi Yao}
\sectionauthor[1]{Yingyu Yao}
\sectionauthor[16]{Mina Yoon}
\sectionlastauthor[3]{Victor Wen-zhe Yu}

\sectionaffil[1]{Theory Department (since 1/1/2020: The NOMAD Laboratory), Fritz Haber Institute of the Max Planck Society, Faradayweg 4-6, D-14195 Berlin, Germany}
\sectionaffil[2]{Chair for Theoretical Chemistry and Catalysis Research Center, Technische Universit\"at M\"unchen, Lichtenbergstr. 4, D-85747 Garching, Germany}
\sectionaffil[3]{Thomas Lord Department of Mechanical Engineering and Materials Science, Duke University, Durham, NC 27708, USA}
\sectionaffil[4]{Department of Chemistry, Duke University, Durham, NC 27708, USA}
\sectionaffil[5]{Molecular Simulations from First Principles e.V., D-14195 Berlin, Germany}
\sectionaffil[6]{The NOMAD Laboratory at the Fritz Haber Institute of the Max Planck Society, Faradayweg 4-6, D-14195 Berlin, Germany}
\sectionaffil[7]{Theory Department, Fritz Haber Institute of the Max Planck Society, Faradayweg 4-6, D-14195 Berlin, Germany}
\sectionaffil[8]{Faculty of Chemistry and Food Chemistry, Technische Universität Dresden, 01062 Dresden, Germany}
\sectionaffil[9]{Department of Applied Physics, Aalto University, P.O. Box 11000, FI-00076 Aalto, Finland}
\sectionaffil[10]{Weizmann Institute of Science, Rehovot, Israel}
\sectionaffil[11]{Department of Physics, Chemistry and Biology (IFM), Linköping University, SE-581 83, Linköping, Sweden}
\sectionaffil[12]{Lenovo HPC Innovation Center, Stuttgart, Germany}
\sectionaffil[13]{Max Planck Institute for the Structure and Dynamics of Matter, 22761 Hamburg, Germany}
\sectionaffil[14]{Computational Science Division, Argonne National Laboratory, 9700 South Cass Avenue, Lemont, Illinois 60439, United States}
\sectionaffil[15]{University College London, Department of Chemistry, Gower Street, London WC1E 6BT, United Kingdom}
\sectionaffil[16]{Center for Nanophase Materials Sciences, Oak Ridge National Laboratory, Oak Ridge, TN 37830, USA}

\sectionaffil[]{(*) Coordinator of this contribution}
\rule[0.25ex]{0.35\linewidth}{0.25pt}

\sectionaffil[a]{{\it Current Address:} Theory Department, Fritz Haber Institute of the Max Planck Society, Faradayweg 4-6, D-14195 Berlin, Germany}
\sectionaffil[b]{{\it Current Address:} Technische Hochschule Mittelhessen (THM), University of Applied Sciences, Fachbereich MND, Wilhelm-Leuschner-Straße 13, D-61169 Friedberg, Germany}
\sectionaffil[c]{{\it Current Address:} Molecular Simulations from First Principles e.V., D-14195 Berlin, Germany}




\subsection*{Summary}

Numeric atom-centered orbitals (NAOs) are the fundamental discretization choice for the electronic structure problem in FHI-aims. The remaining code is built around this foundation. 
This contribution summarizes the technical choices made to deliver a code that is efficient and convenient to use.
We outline the implementation of the integration scheme and the underlying integration grids, as well as the evaluation of the electrostatic potential using an atom-centered multipole decomposition of the density. The two-electron Coulomb operator is treated in a separate scheme, using an efficient resolution of identity technique. The available choices for the k-grid are introduced. Furthermore, symmetry usage is considered, implemented as an option for periodic systems to reduce the computational effort for real-space and k-space operations. Beyond the basic algorithmic choices, a key element in FHI-aims is the handling of the message-passing interface (MPI) based parallelization. Two schemes exist in distributing the real-space \rev{Hamiltonian} and overlap matrices: a globally sparse index representation and a locally-indexed dense index representation, where the latter is especially efficient for large-scale simulations. From the spatial derivatives of the NAOs, expressions for analytical forces and stresses can be derived, enabling efficient structure relaxation and molecular dynamics simulations. The technical details of the trust-radius-method relaxation algorithm, as well as constrained structure relaxation will be discussed. We lay out how the integration, force and stress evaluation can be further sped up by using graphics processing units (GPUs). Finally, we explain software engineering aspects, including how new developments are integrated into the software project and core functionalities of FHI-aims are continuously regression tested.

\subsection*{Current Status of the Implementation}

\paragraph{Real-space grids.} A central question for a code based on numerically tabulated functions is how to represent integrands for matrix elements, densities, potentials, and other objects relevant to electronic structure theory. A central constraint is that an all-electron method with sharply peaked potentials and rapidly varying orbitals near the nucleus cannot simply be implemented using a uniform, even spaced real-space grid of points in three dimensions -- such a grid would need to be far too dense to resolve each nucleus correctly. Instead, FHI-aims successfully employs the methods of non-uniform, overlapping atom-centered real-space grids pioneered by Becke~\cite{Becke1988} and Delley~\cite{Delley1990}.
In this formalism, real-space integrals are computed numerically on an integration grid that combines atom-centered spheres of grid points around each nucleus~($\mathrm{at}$). Using this integration grid, integrals can be approximated as a summation:
\begin{eqnarray}
\int d^{3} r f(\boldsymbol{r}) & \approx & \sum_{i=1}^{N_{\mathrm{grid}}} w_i f(\boldsymbol{r}_i) \nonumber \\ & = & \sum_{{\mathrm{at}}=1}^{N_\mathrm{at}} \left[ \sum_{s=1}^{N_\mathrm{rad}(\mathrm{at})} w_\mathrm{rad}(\mathrm{at},s) \sum_{t=1}^{N_{\mathrm{ang}}(\mathrm{at},s)} w_{\mathrm{ang}}(\mathrm{at},s,t) \, p_{\mathrm{at}}(\boldsymbol{r}_{\mathrm{at},s,t}) f(\boldsymbol{r}_{\mathrm{at},s,t}) \right] , 
\label{eq:volume_integral}%
\end{eqnarray}
where $w_\mathrm{rad}(\mathrm{at},s)$ represents radial integration grid shell weights. 
The specific choice of radial integration grid shells is detailed in \cite{Blum2009}, Appendix A of \cite{Zhang2013}  and in Appendix D.1 of Knuth et al.~\cite{knuth2015all}, where a variable radial multiplier is used to uniformly increase the number of radial integration grid shells $s$.
The $w_{\mathrm{ang}}(\mathrm{at},s,t)$ are the weights for the spherical integration grid, which typically use the grids by Lebedev and coworkers~\cite{Lebedev1975,Lebedev1976,Lebedev1999} (FHI-aims uses the variants created by Delley \cite{Delley1996}).
The Lebedev grids possesses octahedral, i.e., $O_h$ symmetry. We further provide angular grids with hexagonal symmetry of $D_{6h}$ and, additionally non-symmetric grids ``efficient spherical t-design'' (ESTD) grids~\cite{womersley2018efficient}. They can be triggered by the keyword \verb|force_lebedev d6hgrid| or \verb|force_lebedev estd|, respectively.
The function $p_{\mathrm{at}}(\boldsymbol{r}_{\mathrm{at},s,t})$ is the partition function, which smoothly transitions from zero to one based on its distances to each nucleus.
For the recommended and default partition function used in FHI-aims, we refer to Appendix C in Knuth \textit{et al.}~\cite{knuth2015all}. This partition function is based on the Stratmann's partition function~\cite{Stratmann1996} with our modifications to enhance performance in periodic systems.

For an efficient evaluation of the integrals of the form of Eq.~\eqref{eq:volume_integral}, the set of integration points $P$ is divided into mutually distinct groups or ``batches'' of points $B_\nu \subset P$. The batches $B_\nu$ are distributed across the MPI tasks using a grid adapted cut-plane-method~\cite{Havu2009} demonstrating linear scaling behavior. \rev{As described in the same paper, the density and its gradients are evaluated in a linear-scaling fashion, using the same formalism and the localized nature of NAO basis functions. For periodic systems and for large non-periodic systems, the density is constructed using a density-matrix based formalism rather than by summing over molecular orbitals individually.}

\paragraph{Electrostatic (Hartree) Energy.} The average Coulomb energy, $E_\text{es}$, of electrons and nuclei, approximated by a continuous probability density $n(\mathbf{r})$ for the location of electrons, and fixed positions $\mathbf{R}_I$ for the nuclei $Z_I$ ($I$=1,...,$N_\text{atoms}$), is a key element of any electronic structure calculation. Importantly, and as laid out in Ref. \cite{Blum2009}, this group of terms must be treated together:
\begin{equation}\label{Eq:es}
E_\text{es} = - \int d^3r \left( n(\mathbf{r}) \sum_I \frac{Z_I}{|\mathbf{r} - \mathbf{R}_I|}\right) + \frac{1}{2}\iint d^3r d^3r^{\prime} \left( \frac{n(\mathbf{r}) n(\mathbf{r}^\prime)}{|\mathbf{r} - \mathbf{r}^\prime|} \right) + \frac{1}{2}\sum_{I,J\neq I} \frac{Z_I Z_J}{|\mathbf{R}_J - \mathbf{R}_I|} .
\end{equation}
In the limit of infinitely large or periodic systems, each term in Eq. (\ref{Eq:es}) would diverge towards infinity on its own, even if taken per atom or per unit cell, and even for an overall neutral system, since each term amounts to the interaction of a finite charge with another, overall finite charge. Taken together, the overall system of electrons and nuclei with which each charge interacts is neutral, i.e., a combined treatment as laid out in Ref. \cite{Blum2009} is tractable.

In practice \cite{Blum2009}, FHI-aims first subtracts a sum of spherical free-atom densities from the overall density $n(r)$. The free-atom densities and their electrostatic potentials are large but they can be calculated precisely on dense logarithmic grids, thus removing a large piece from the overall computation of electrostatics.

The remaining density difference between the full density and the sum of free-atom densities, $\delta n(\mathbf{r})$, is then evaluated numerically on the overlapping atom-centered integration grid. To obtain its electrostatic potential, FHI-aims follows Becke's \cite{Becke1988} and Delley's \cite{Delley1990} approach to first partition $\delta n(\mathbf{r})$ into localized density pieces around each atom and then multipole-decomposing these densities. Each multipole component can be treated separately, using the Green's function of the Coulomb operator, to arrive at a corresponding multipole potential component of the electrostatic potential, which can then be summed up to give the overall electrostatic potential $\delta v_\mathrm{es}(\mathbf{r})$ associated with $\delta n(\mathbf{r})$. Some care is required to ensure that each multipole-decomposed density component has the exact same charge or multipole moment when represented on different, dense or sparse integration grids during the procedure. Furthermore, for periodic systems, an Ewald summation technique is applied to each multipole component~\cite{Delley1996}, ensuring that long-range tails of the potential can be summed up efficiently and at once in reciprocal space. 

Charged periodic systems can be treated but require an artificial opposing compensating background charge to give a finite energy per unit cell. This background charge can either be a constant density or it can be distributed as small pieces over the nuclei in the system. In the case of a localized charged defect and a constant background charge, additional techniques such as the Freysoldt-Neugebauer-Vandewalle correction~\cite{Freysoldt2009} can then be employed to approximately restore the energy of an isolated charged defect in an otherwise neutral crystal.

Overall, the electrostatic potential evaluation described above is efficient and, in periodic systems, scales as $O(N) \log N$ with system size $N$. Recently, we redesigned the computation to evaluate the contribution of each multipole component to the Hartree potential across a batch of grid points simultaneously \cite{kokott2024}. This restructuring eliminates branching in the innermost loop, reduces subroutine calls by two orders of magnitude, improves memory access efficiency and cache usage, enables compiler vectorization, and ensures that the Hartree energy evaluation is not a bottleneck even for extremely large periodic computations.

\textbf{Two-electron Coulomb operator} Unlike for the overall density and average electrostatic potential, a piecewise evaluation of the Coulomb operator over orbitals or basis functions cannot be avoided if the exchange operator or explicitly correlated method's ground- or excited-state energies are required. The two-electron Coulomb operator in terms of NAO basis functions $\varphi_i(\mathbf{r})$, 
\begin{equation}
  (ij|kl) = \iint d^3r d^3r^\prime \left( \frac{\varphi_i^*(\mathbf{r}) \varphi_j^*(\mathbf{r^\prime})\varphi_k^*(\mathbf{r^\prime}) \varphi_l(\mathbf{r})}{|\mathbf{r} - \mathbf{r}^\prime|} \right) ,
\end{equation}
poses no particular mathematical difficulty, except that the sheer number of terms initially grows as $N_\text{basis}^4$ in finite systems and, in periodic systems, can never be cut off even for very large distances because of the slow decay of the $\frac{1}{|\mathbf{r} - \mathbf{r}^\prime|}$ potential. In FHI-aims, any techniques that require the actual two-electron Coulomb operator can be treated efficiently using resolution of identity (RI) techniques, which expand products of two basis functions $\varphi_i(\mathbf{r}) \varphi_j(\mathbf{r})$ in terms of simpler individual auxiliary basis function sets $\{P_\mu(\mathbf{r})\}$. The construction of these product basis sets and associated operations for NAOs is described in Ref. \cite{ren2012} \rev{(RI-V)} and, in a localized form, in Ref. \cite{ihrig2015} \rev{(RI-LVL)}. Especially the latter, localized RI technique RI-LVL enables linear-scaling evaluations of the exchange operator even for extreme system sizes and efficient periodic implementations of exact exchange~\cite{levchenko2015,kokott2024} as well as many-body methods \cite{Ren2021,Zhou2025}. \rev{The errors associated with the corresponding RI techniques are discussed in Ref. \cite{ihrig2015}.}

Importantly, the explicit form of the two-electron Coulomb operator is never involved in the overall electrostatic potential calculation in FHI-aims, which rather employs the density-based multipole decomposition formalism described following Eq. (\ref{Eq:es}). \rev{Thus, the precision of the auxiliary basis set and associated RI technique have no impact on the Coulomb potential term, usually the biggest term in complex systems}. The remaining, \rev{usually} smaller exchange or explicitly correlated energy pieces can be computed stably and efficiently with reasonably sized and well benchmarked auxiliary NAO basis sets for RI.

\paragraph{k-point grids.}
By default, for periodic calculations, FHI-aims relies on $\Gamma$-centered regular k-space grids for Brillouin zone integrations, with  fractional coordinates $k_i = (i_1/N_1,i_2/N_2,i_3/N_3)$, where $i_x = 0, ..., N_x -1$ for each reciprocal lattice vector direction $x=1, 2, 3$. The number of k-points $N_1, N_2, N_3$ along their corresponding reciprocal lattice vector can be either specified directly by the keyword \texttt{k\_grid N\_1 N\_2 N\_3} or is computed from the k-point line density $n_k$ \texttt{k\_grid\_density n\_k} via $N_x = n_k\vert\mathbf{L}^\ast_x\vert$, where $\mathbf{L}^\ast_x$ is the reciprocal lattice vector corresponding to lattice vector $\mathbf{L}_x$. 
Additionally, the user can specify a rigid shift of the k-grid by setting 
\texttt{k\_offset d\_1 d\_2 d\_3}, where $(d_1, d_2, d_3)$ is the shift in fractional coordinates. The use of shifted k-grids is only supported for LDA, GGA, and meta-GGA functionals and does not work with any functional that requires exact exchange contributions. Simple sanity checks are performed to determine if the user input is meaningful, e.g., for 1D systems and surface models the corresponding vacuum directions should have only one k-point. For LDA, GGA, and meta-GGA functionals, arbitrary k-points can be generated by a user-defined external list of k-points and weights. 
It is also possible to automatically generate generalized regular k-grids by using the autoGR~\cite{morgan2020generalized} code, which is interfaced with FHI-aims. \rev{In the case of autoGR,} generalized regular k-grids are constructed from rotated Born-von Karman cells that preserve the space-group symmetry of the system. They allow a more efficient sampling of the k-space for oblique crystals. Generalized regular k-grids are supported for all density functionals in FHI-aims.

For any nonrelativistic or scalar-relativistic DFT simulation, the code assumes time-reversal symmetry, which reduces the number of k-point by about half. In addition, it is possible to only compute the irreducible set of k-points based on the space-group symmetry of the input structure by specifying \texttt{symmetry\_reduced\_k\_grid\_spg .true.} in the \texttt{control.in} file. In this case, the symmetry matrices obtained from the library \texttt{spglib}~\cite{Togo2024} are used to reduce the k-grid.

\paragraph{Indexing the real-space matrices and load balancing in real-space operations.} In evaluating Eq. \ref{eq:volume_integral} on an even-spaced grid for a group of matrix elements (e.g, for $N_\text{basis}\times N_\text{basis}$ Hamiltonian matrix elements $h_{ij}$, with $N_\text{basis}\times N_\text{basis}$ corresponding integrands $f(\mathbf{r})$ on each grid point $\mathbf{r}$), nearly ideal, communication-free parallel execution can in principle be achieved by distributing different segments of the integration grid to different MPI tasks. A problem arises with respect to re-assembling the full integrals $h_{ij}$ from the partial integrals calculated on different MPI tasks, since the memory footprint of $h_{ij}$ grows large towards large systems, making it necessary to distribute $h_{ij}$ between different MPI tasks as well. The distribution of non-zero elements of $h_{ij}$, however, will not be the same as the distribution of grid segments in which those matrix elements are computed, leading to a non-trivial communication pattern during execution. 

Optimizing these parallel distributions and the associated communication patterns creates an integration framework that scales efficiently towards extremely large systems and towards extremely large \rev{numbers of} processor cores. We here introduce how real-space matrices like the \rev{Hamiltonian} or overlap matrix are indexed and stored over MPI tasks after the integration is carried out. Since the approach is similar for all kind of real-space matrices, we will focus our notation on the \rev{Hamiltonian} matrix. A detailed explanation can be found in the paper by Blum et al.~\cite{Blum2009} and Huhn et al.~\cite{Huhn2020}.

In brief, the real-space Hamiltonian is evaluated for the disjoint batches of grid points $B_v$ as \cite{Huhn2020}
\begin{equation}
h_{\text{ij}}=\sum_v h_{\text{ij}}\left[B_v\right] \, ,
\label{eq:batch_ham}
\end{equation}
where 
\begin{equation}
h_{\text{ij}}\left[B_v\right]=\sum_{\boldsymbol{r} \in B_v} w(\boldsymbol{r}) \varphi_\text{i}^*(\boldsymbol{r}) \hat{h}_{\text{KS}} \varphi_\text{j}(\boldsymbol{r})
\end{equation}
is the contribution of batch $B_v$ to the real-space Hamiltonian element $h_{\text{ij}}$. 

Once integrals of the real-space Hamiltonian (and overlap) are evaluated for each batch $h_{\text{ij}}\left[B_v\right]$, a local matrix copy of $h_{\text{ij}}$ is updated on each MPI task. Within FHI-aims, different storage formats are used for this local copy depending on the calculation setup (see also the explanation in ~\cite{Huhn2020}). 

For small clusters, every MPI task has a full dense copy of $h_{\text{ij}}$ by default (i.e.~\texttt{packed\_matrix\_format none}). This storage pattern becomes inefficient in terms of memory usage for even moderately sized systems and is thus not used as a default choice for any larger clusters or periodic systems. The default for smaller periodic systems or cluster systems is a globally-indexed sparse format (i.e.\linebreak\texttt{packed\_matrix\_format index}). Here, only the non-zero elements of $h_{\text{ij}}$ are stored in a compressed sparse row~(CSR) format along with two indexing arrays. By default, all matrix elements that could potentially be non-zero according to the extent of the basis functions involved are kept -- the strict localization of the basis functions themselves still leads to a natural sparsity of large matrices that is already sufficient for efficient execution. While the sparse matrix storage in this format is efficient for smaller system sizes, its remaining disadvantage is that the sparse storage of $h_{\text{ij}}$ on each MPI task still scales linearly with system size as $O(N_{\text{atom}})$, independent of the number of MPI tasks being used. This scaling behavior of per-processor memory will thus become the memory bottleneck in truly large-scale calculations.

In order to achieve constant, rather than linear per-processor memory of all matrix storage arrays towards large system sizes and correspondingly large CPU core counts, FHI-aims implements a so-called locally-indexed \rev{sparse or} dense real-space matrix storage format \cite{Huhn2020} (activated by \texttt{use\_local\_index .true.} \rev{or automatically above a certain system size, typically 100 atoms, when supported by all requested observables in a calculation}). In this approach, each processor works on batches \rev{of spatially adjacent grid points and stores only the subset of matrix elements (e.g., $h_{\text{ij}}$) that are non-zero on those grid points}. Thus, individual matrix elements can have non-zero contributions from different processors (i.e., different MPI tasks), but in the limit of large systems, no single MPI task needs to store anything even close to the full matrix size - the actual, locally non-zero matrix segments to be stored are far smaller. The synchronization step to gather the full matrix from its local, distributed pieces is then more complex compared to a single global format. First, the locally-indexed matrix elements are distributed in an overlapping fashion across MPI tasks. Second, the fully integrated matrix is also stored in a distributed format, but normally a dense, block-cyclic format across MPI tasks that is not simply related to the parallelization of the underlying integration grid. 
The distributed format avoids the memory bottleneck of the global format for larger systems, provided that sufficiently many MPI tasks are available, but incurs a small performance overhead. To mitigate this overhead, one can set the keyword \texttt{load\_balancing .true.}\rev{, where the code tries to achieve an even integration load across all cores. Enabling the load balancing also changes the matrix storage format from locally-indexed sparse format to a local dense matrix format (see Ref.~\cite{Huhn2020} for details).}

\rev{In general, computations that use quantities built from basis functions on the real-space grid can take advantage of sparsity in much the same way as for the Hamiltonian and overlap matrices. This includes the evaluation of the density and its derivatives, as well as of LDA, GGA and mGGA exchange-correlation terms, since they are subject to the same locality limitations within each grid-point batch.}

\paragraph{MPI-3 shared memory.} As a next step towards optimizing memory and communication, sharing common arrays between processors located on a physical node in FHI-aims is accomplished following the MPI-3 standard. Compared to the more frequently found OpenMP paradigm, the MPI-3 approach can be quite readily controlled from within a given run, without asking for any specific instructions for shared-memory layout to be provided by a user. At the time of writing, MPI-3 shared memory use is largely restricted to the computation of the exact exchange contribution.\cite{kokott2024} A module for handling MPI-3 internode shared-memory arrays has been added to FHI-aims. In principle, this module makes the use of the shared memory arrays in FHI-aims almost as easy as using non-shared arrays. For reading from such arrays, there is no difference to conventional arrays. Only in case of writing to the shared arrays special care is needed as only a single MPI task is allowed write to an array element at the same time. After writing to the shared arrays, a special synchronisation step is needed. The shared memory modules set up the needed infrastructure, e.g., intranode communicators, which in turn enable the allocation and deallocation, as well as synchronisation of the shared memory arrays.

\paragraph{Forces and stress tensor.} 
In an all-electron formalism, nearly complete basis set convergence can typically be achieved for total energies associated with fixed atomic positions, but the same fixed basis set would not describe a structure completely if the atom in question, but not its basis functions, were moved. As a result, when calculating analytical derivatives of the total energy for a basis set associated with fixed atomic positions, a number of terms arise that must be accounted for. In short (a detailed description is included in Ref. \cite{Blum2009}), in addition to Hellmann-Feynman forces, i.e., the derivative of the Hamiltonian itself, so-called Pulay forces (associated with immobile basis functions), additional derivatives associated with the multipole decoposition of the electrostatic potential, and terms associated with the exchange-correlation functional (generalized gradient functional derivatives, derivatives associated with exact exchange \cite{levchenko2015}, the random phase approximation \cite{Tahir2022}) must be computed. Recently, a correction of the Hellmann-Feynman forces that accounts for residual non-selfconsistency (cf. Eq.~\eqref{eq:rossi-force-correction}) has been implemented, reducing the number of SCF iterations needed for precise forces. In addition, the branching inside of the main loops of the Pulay force evaluation has been simplified and the computation to aggregate or avoid unnecessary computations has been restructured. 

Furthermore, the evaluation of stress,~i.e.,~of the derivatives of the total energy~$E$ under a symmetric strain deformation~$\varepsilon_{\alpha\beta}$ of the unit \rev{cell} with volume~$V$ along the the Cartesian axes~$\alpha,\beta$,
\begin{equation}
\sigma_{\alpha\beta} = \frac{1}{V} \frac{\partial E}{\partial \varepsilon_{\alpha\beta}}
\end{equation}
is supported. A finite-difference approach is available for testing purposes, covering all total-energy methods. For production purposes, all semilocal (LDA, GGA, and mGGA) and hybrid density functionals are supported by analytical stress derivatives. A detailed account of the terms needed for GGA and hybrid functionals is given in Ref.~\cite{knuth2015all}. These implementations enable structural relaxations of lattice degrees of freedom,
as discussed below, enable \rev{\textit{ab initio} molecular dynamics} simulations in isobaric or isostress ensembles (see Contrib.~\ref{ChapMD}), and play a pivotal role in fully anharmonic simulations of heat transport~(see Contrib.~\ref{ChapHeatTransport}).

\paragraph{Structure optimization}

The task of structure optimization is to find a local minimum configuration $x_{\rm min}$ of the DFT total energy $E (x)$, where~\cite{Nocedal1999}
\begin{align}
    x=\left(R_0^x, R_0^y, \ldots, R_N^z ; a_1^x, a_2^x, \ldots, a_3^z\right)~,
    \label{eq:fk.1}
\end{align}
with $3N$ atomic coordinates $R^{x,y,z}_I$, and three lattice vectors $a^{x,y,z}_L$ in the case of periodic solids. The degrees of freedom can be constrained in various ways, leading to a modified total energy function $\tilde{E} (x)$ without implications for the following discussion. For constraining the structure optimization either the space group symmetries can be used or, more generally, parametric constraints can be utilized that maintain local distortions or global symmetries of the structure~\cite{Lenz2019}.
In the quasi-Newton method, a step $s_x$ toward the minimum is predicted by
\begin{align}
    s_x = B^{-1}_{xx'} f_{x'}~.
    \label{eq:fk.5}
\end{align}
where $f_x$ denotes the generalized force, $f_x = - \partial_x E(x)$, and $B$ is an approximation to the true Hessian matrix, $H_{xx'} = \partial_x \partial_{x'} E(x)$. 
The minimum configuration is characterized by $f_x = 0$.
FHI-aims uses a quasi-Newton descent optimization with a hessian approximant $B$ obtained via the Broyden–Fletcher–Gold\-farb-Shanno (BFGS) algorithm~\cite{BROYDEN1970,Fletcher1970,Goldfarb1970,Shanno1970}, and a trust radius method (TRM) to limit the predicted step size. The initial guess for the approximate hessian, $B^0$, is obtained in block-diagonal form via a Lindh model for the atomic degrees of freedom~\cite{Lindh1995}, and
\rev{
\begin{align}
    B_a^0=c (a^{-1})^{T} a^{-1}~
    \label{eq:fk.6}
\end{align}
}
for the lattice degrees of freedom, where $c$ is a numerical constant, $a$ is the lattice matrix defined in Eq.\,\eqref{eq:fk.1}, \rev{and $T$ denotes the matrix transpose}. The choice in Eq.\,\eqref{eq:fk.6} leads to strain-like relaxation steps~\cite{Pfrommer1997,Tadmor1999}, improving the relaxation both in terms of the number of relaxation steps and the final outcome, compared to the original choice of a diagonal matrix $B_a^0$ in Ref.\,\cite{knuth2015all}.
Furthermore, this choice can preserve space-group symmetry when used together with symmetrized forces and stress (cf. discussion below).
More details can be found in appendix D of Ref.\,\cite{KnoopDiss}.

For the case where the predicted step length exceeds the trust radius $r_{\rm trust}$, the step length is adjusted. We note that the adjustment of step length to the trust radius in the default TRM algorithm can lead to symmetry-breaking steps even when symmetric forces and stress are used. This is due to the adjustment
\begin{align}
    s'_x = (B + \lambda)^{-1}_{xx'} f_{x'}~,
    \label{eq:fk.7}
\end{align}
where $\lambda$ is chosen such that $s'_x = r_{\rm trust}$, which can lead to space-group symmetry breaking in some cases, but is usually a good choice in particular for complex potential energy surfaces. To enforce a symmetry-preserving relaxation, the keyword \texttt{LATTICE\_TRM} can be used, where the adjustment is performed via rescaling
\begin{align}
    s'_x = B^{-1}_{xx'} f_{x'} / \lambda~,
    \label{eq:fk.8}
\end{align}
which preserves symmetry. Finally, FHI-aims monitors the consistency of forces, stresses, and total energies during a structure optimization, i.e., the total energy is not allowed to increase beyond a certain target tolerance (typically, 3~meV) over multiple relaxation steps.

In addition to structure optimization to a local mimimum-energy structure, a python package “aimsChain” provides the ability to find transition states and minimum energy paths for chemical reactions. This package supports various flavours of the chain of states methods for finding the minimum energy path. Currently the nudged elastic band method (NEB) \cite{Jonsson1998, Henkelman2000}, the string method \cite{Weinan2007}, and the growing string method \cite{Baron2004} are included. A tutorial is available at \url{https://fhi-aims-club.gitlab.io/tutorials/finding-transition-states-with-aimschain}.

\paragraph{Symmetry for periodic structures.\label{par:symmetry}}

Several ways to use symmetry for the simulation of periodic structures exist in FHI-aims. As noted before, FHI-aims assumes time-reversal symmetry for DFT calculations for the k-space grid in non-relativistic or scalar-relativistic computations by default. In addition, space-group symmetry can be used to further reduce the number of k-space grid points; here, the software package spglib\cite{Togo2024} is used. Space-group symmetry can be employed as well to reduce the number of real-space grid points during the integration of the real-space matrix elements \rev{as well as for the computation of the density and of its gradients}. This is especially beneficial for high-symmetry systems if force or stress evaluations are needed. If space-group symmetry is requested by the user, the input structure either needs to have the numerically correct atomic positions and lattice parameters, or the structure can be refined on the fly to the closest spacegroup that could be found by spglib. If the system has a space-group number larger than 1, the symmetry of the system will be preserved along relaxation trajectories. 

Separately, FHI-aims has a complete framework for parametrically constrained relaxation of certain structures (i.e., user-imposed structural constraints beyond simply preserving the symmetry of a given system).

\paragraph{GPU acceleration.} FHI-aims supports the use of graphics processing units (GPUs) for acceleration of DFT simulations. Currently, the integration of the \rev{Hamiltonian} matrix and overlap matrix, the evaluation of forces and stress components, as well as the solution of the generalized eigenvalue problem can be executed on GPUs~\cite{Huhn2020}. GPUs with support for Nvidia's CUDA and AMD's HIP can be used for the aforementioned parts of the computation. The solution of the generalized eigenvalue problem can be accelerated by the ELPA library (see \textbf{Contribution~2.5}). 

\paragraph{Restarting calculations.}

Often it is useful to restart simulations from already converged density matrices, e.g. for executing post-processing tasks (DOS, band structure, etc.) or to split long calculations into several parts. The current recommended restart functionality is provided by ELSI (see \textbf{Contribution~2.5}) via writing the density matrix to disk. This enables the restart of the calculations for the same number of k-points and basis functions, while not restricting the  number of MPI tasks. 

Recently, a different type of restart from the resolution-of-identity decomposed density has been added~\cite{Lewis2021}.  While this functionality is mostly used for machine-learning-the-density approaches, it can be also used in general for restarting calculations, especially also for different numbers of k-points. More details are given and applications are demonstrated in \textbf{Contribution~8.4}. It is also possible to write checkpoints during the computation of the self-energy for the GW approximation. This feature is introduced in \textbf{Contribution~4.3}.

\begin{figure}[ht]
\centering
\includegraphics[width=0.72\textwidth]{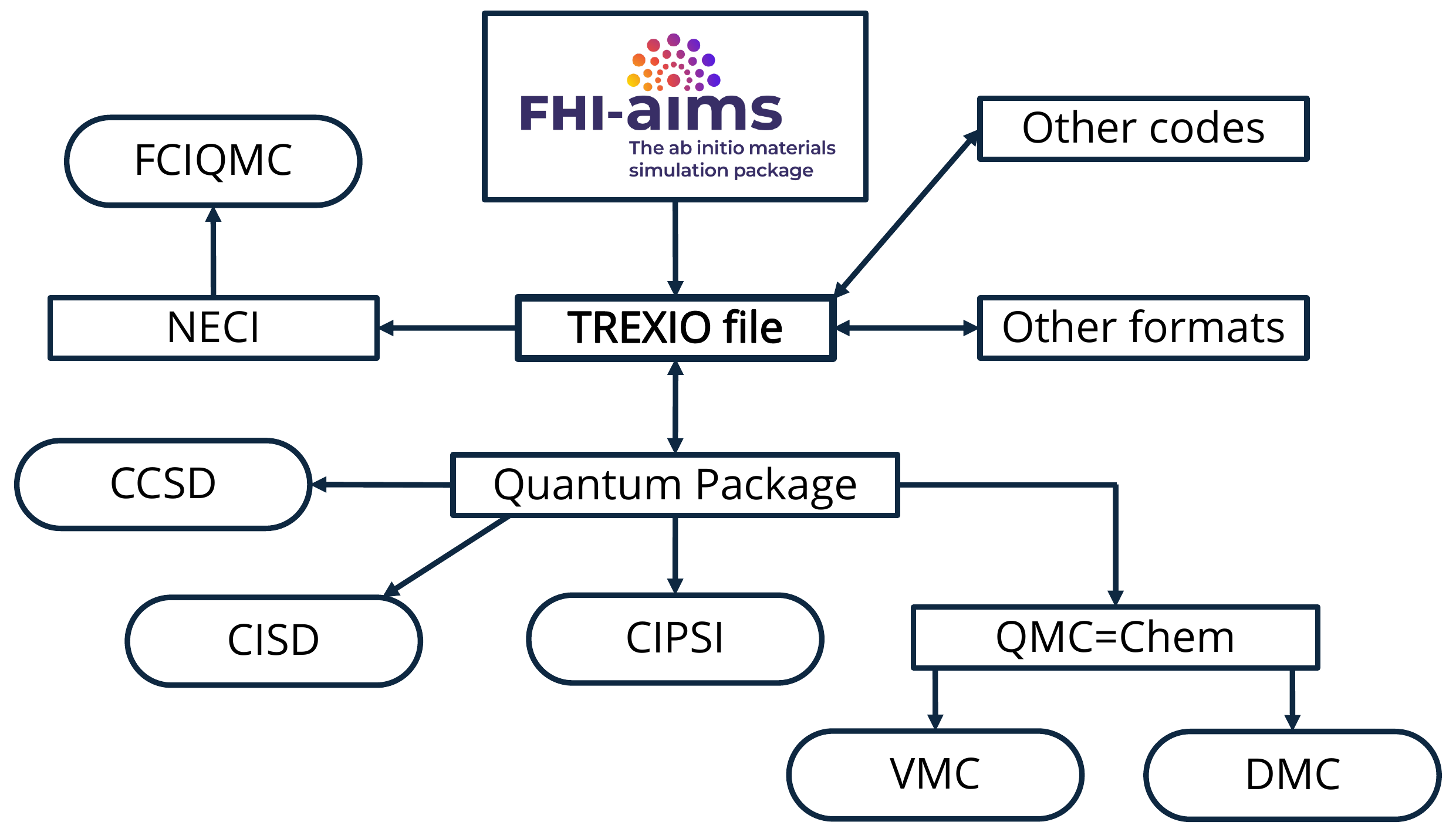}
\caption{Scheme illustrating how an NAO wave function calculated by FHI-aims can be processed by other quantum chemistry codes after being exported to a TREXIO file. Rectangular boxes represent codes, while rounded boxes indicate the possible quantum chemistry methods. Refer to the text for explanations of the abbreviations.} 
\label{fig:trexio}
\end{figure}

\paragraph{Wave function export to TREXIO.}
FHI-aims provides an interface to the TREXIO \cite{posenitskiyTREXIOFileFormat2023} software ecosystem, which allows one to export a wave function as a TREXIO file. TREXIO files store wave functions in a general format, enabling seamless transfer between different quantum chemistry codes.  
Several codes can read wave functions from TREXIO. Figure~\ref{fig:trexio} provides an overview of codes for which interoperability with the NAO-based TREXIO wave function from FHI-aims has been verified. This includes the codes Quantum Package \cite{garnironQuantumPackage202019}, NECI \cite{gutherNECIElectronConfiguration2020} and QMC=Chem \cite{scemamaQuantumMonteCarlo2013, scemamaQuantumMonteCarlo2016}. Quantum Package supports methods like coupled cluster (CCSD), configuration interaction singles and doubles (CISD) or CIPSI (Configuration Interaction using a Perturbative Selection made Iteratively). The NECI software suite enables full configuration interaction quantum Monte Carlo (FCIQMC), while QMC=Chem performs highly accurate variational and diffusion quantum Monte Carlo (VMC, DMC) calculations. Additionally, TREXIO files can be converted into other formats, such as FCIDUMP \cite{knowlesDeterminantBasedFull1989}, to enhance interoperability with other codes.

\paragraph{Software Engineering: Regression testing and continuous integration.}

FHI-aims includes a suite  of regression tests containing over 100 small calculations that test a broad range of FHI-aims' functionality across several different primary code branches (e.g. periodic/non-periodic systems, MPI parallel/serial code execution). To ensure integrity of the code, FHI-aims employs a range of continuous integration pipelines to run the suite of regression tests for different compilers (nvhpc, GNU, Intel oneAPI) on different architectures (CPU and GPU nodes) and for different MPI libraries. The \rev{continuous integration} pipelines are implemented using GitLab \rev{continuous integration/continuous delivery}, through
a \texttt{gitlab-runner} instance installed on a small local cluster. The \rev{continuous integration} process is initiated whenever
a developer pushes code changes to the FHI-aims GitLab repository. The push event creates several \rev{continuous integration} pipelines on the GitLab server, which in turn are executed through the 
\texttt{gitlab-runner} instance on the small local cluster. Every pipeline has a dedicated container (that is, a virtualization of the operation system with pre-installed dependencies). In this container, the current version of FHI-aims is built and
the regression tests are executed. Once the tests have successfully passed, the code changes can be merged into the main line of the code.






\subsection*{Future Plans and Challenges}

FHI-aims in its present form supports some of the most demanding electronic structure calculations ever performed, especially for more demanding hybrid density functional theory calculations~\cite{kokott2024}.
With the advent of exascale computing, the biggest challenge will be to keep FHI-aims current and functional on the most relevant platforms and architectures, which are themselves continually evolving. Even for the existing GPU infrastructure in FHI-aims more opportunities are possible to accelerate the code execution, especially porting the evaluation of the Hartree Potential and the exact exchange contribution to GPUs. Furthermore, new types of GPUs, such as Intel's current generation, are not yet supported at the time of writing and will require developer attention.

\subsection*{Acknowledgements}

The authors acknowledge Arvid Ihrig for numerous contributions to the FHI-aims core functionalities.

We acknowledge the computational resources provided by the Aalto Science-IT project and CSC – IT Center for Science (Finland).

MS acknowledges support by his TEC1p Advanced Grant (the European Research Council (ERC) Horizon 2020 research and innovation programme, grant agreement No. 740233.






\newpage

\section{Relativistic Effects: Scalar Relativity, Spin-Orbit Coupling, and Four-Component Relativity}\label{Sec:Relativity}

\sectionauthor[1,2]{\textbf{*Volker Blum}} 
\sectionauthor[3]{Sander Driessen}
\sectionauthor[4]{Dorothea Golze}
\sectionauthor[5]{Matthias Gramzow}
\sectionauthor[1]{William P. Huhn}
\sectionauthor[5,a]{Timo Jacob}
\sectionauthor[6,7]{Sebastian Kokott}
\sectionauthor[4]{Qing-Long Liu}
\sectionauthor[8]{Karsten Reuter}
\sectionauthor[6]{Matthias Scheffler}
\sectionauthor[9]{Han Wang}
\sectionauthor[1]{Wentao Zhang}
\sectionlastauthor[9]{\textbf{*Rundong Zhao}} 

\sectionaffil[1]{Thomas Lord Department of Mechanical Engineering and Materials Science, Duke University, Durham, NC 27708, USA}
\sectionaffil[2]{Department of Chemistry, Duke University, Durham, NC 27708, USA}
\sectionaffil[3]{Materials Simulation and Modelling, Department of Applied Physics, Eindhoven University of Technology, PO Box 513, 5600 MB
Eindhoven, The Netherlands}
\sectionaffil[4]{Faculty of Chemistry and Food Chemistry, Technische Universität Dresden,  01062 Dresden, Germany}
\sectionaffil[5]{Theory Department (since 1/1/2020: The NOMAD Laboratory), Fritz Haber Institute of the Max Planck Society, Faradayweg 4-6, D-14195 Berlin, Germany}
\sectionaffil[6]{The NOMAD Laboratory at the Fritz Haber Institute of the Max Planck Society, Faradayweg 4-6, D-14195 Berlin, Germany}
\sectionaffil[7]{Molecular Simulations from First Principles e.V., D-14195 Berlin, Germany}
\sectionaffil[8]{Theory Department, Fritz Haber Institute of the Max Planck Society, Faradayweg 4-6, D-14195 Berlin, Germany}
\sectionaffil[9]{School of Physics, Beihang University, Beijing 102206, China}

\sectionaffil[]{(*) Coordinator of this contribution}
\rule[0.25ex]{0.35\linewidth}{0.25pt}

\sectionaffil[a]{{\it Current Address:} Institute of Electrochemistry, Ulm University, 89081 Ulm, Germany}



\subsection*{Summary}
While quantum mechanics is often introduced via the non-relativistic Schr\"odinger equation, this is actually not the correct equation to capture chemical and materials properties across the periodic table. In chemistry and materials science, relativistic effects can have large impacts and must be accounted for. Dirac's equation incorporates relativity more adequately and approximations derived from it are therefore used in practical simulations to ensure qualitatively correct results. The non-relativistic limit is often viewed as sufficient for chemical bonding characteristics of the lightest elements (atomic numbers $Z\le$20), but relativistic effects can have a considerable effect even on the properties of this limited group, e.g., on their core level energies\cite{Keller2020}. The impact of relativity grows rapidly with increasing atomic number $Z$ of the elements involved in a molecule or compound\cite{huhn2017,Zhao2021}. Predicted non-relativistic energy band gaps are off by $\sim$0.3~eV already for the semiconductors ZnO and ZnS ($Z$=30 for Zn). For semiconductors including the heavy 6$p$ main group elements Pb and Bi ($Z$=82 and 83, respectively), energy bands derived from the heavy elements are qualitatively incorrect without including the effects of spin-orbit coupling (SOC). Simply put, our world is a product of relativistic, not non-relativistic underlying interactions, including with electromagnetic fields.

This chapter focuses on relativistic approaches implemented in FHI-aims\cite{Blum2009,huhn2017,Zhao2021}, not a general review of relativistic electronic structure theory, which can be found elsewhere\cite{Grant2007}. NAOs offer a unique opportunity for relativistic calculations since their functional form in the important region near the nucleus can be captured practically exactly, while their computational properties (compact basis sets, localization) make them amenable to treating large, complex systems. FHI-aims can include relativistic effects in density-functional theory (DFT) at four different levels of increasing accuracy, i.e., (1) non-relativistic, (2) scalar-relativistic (scalar Kohn-Sham orbitals with relativistic effects included in the kinetic energy operator)\cite{Blum2009}, (3) scalar-relativistic but with SOC effects included non-selfconsistently in orbitals and band structures \cite{huhn2017}, and (4) four-component orbitals in quasi-four-component (Q4C) Dirac-Kohn-Sham theory for total energies\cite{Zhao2021}. Scalar-relativistic effects are included using the atomic zero-order regular approximation (atomic ZORA) at no extra cost and complexity compared to the non-relativistic case and should therefore be included in all simulations. Non-selfconsistent SOC yields energy band structures that are quantitatively accurate at least up to $Z$=48 (Cd) and cover the properties of heavy-element compounds qualitatively correctly up to Pb and Bi. The Q4C approach provides an excellent foundation for future developments that cover spin, currents, and the interaction of the electrons with external fields.

\subsection*{Current Status of the Implementation}

The Dirac-Kohn-Sham (DKS) DFT equations are built upon single-particle wave functions that are vectors of four scalar functions or ``components'', typically denoted $\left(\phi_1(\mathbf{r}), \phi_2(\mathbf{r}), \chi_1(\mathbf{r}),\chi_2(\mathbf{r})\right)$. As a starting point, we consider the time-independent version of the four-component Dirac equation, 
\begin{equation}\label{Eq:DKS}
\begin{pmatrix}
\begin{pmatrix}
V & 0 \\
0 & V
\end{pmatrix}
& c\boldsymbol{\sigma} \cdot \mathbf{p} \\
c\boldsymbol{\sigma} \cdot \mathbf{p} & 
\begin{pmatrix}
-2mc^2+V & 0 \\
0 & -2mc^2+V
\end{pmatrix}
\end{pmatrix}
\begin{pmatrix}
\phi_1  \\  \phi_2  \\  \chi_1  \\  \chi_2
\end{pmatrix}
=
\epsilon
\begin{pmatrix}
\phi_1  \\  \phi_2  \\  \chi_1  \\  \chi_2
\end{pmatrix} \, .
\end{equation}
$\epsilon$ is the corresponding eigenvalue, $m$ is the electron rest mass, $c$ the speed of light, $\mathbf{p} = - i \hbar \nabla$ is the momentum operator, $V(\mathbf{r})$ is the effective DKS potential and $\boldsymbol{\sigma}$ is a 3-vector with the three individual (2$\times$2) Pauli matrices $\sigma_x$, $\sigma_y$, $\sigma_z$ as its entries. As written, Eq. (\ref{Eq:DKS}) already omits explicit magnetic fields and current terms that appear in the full Dirac equation~\cite{Rehn2020}. Additionally, relativistic corrections to the Coulomb potential itself (which governs the electron-electron and electron-nuclear interactions) are neglected. In this form, Eq. (\ref{Eq:DKS}) provides a practical starting point to connect systematically to the more approximate forms of Schr\"odinger-like scalar-relativistic equations with or without SOC~\cite{huhn2017}. The mathematical steps to derive these expressions are simple and are outlined next.

As formatted above, the matrix equation (\ref{Eq:DKS}) emphasizes its structure in a form of four (2$\times$2) blocks. One commonly combines $\phi_1$ and $\phi_2$ into a two-component vector called the ``large component'' $\phi(\mathbf{r}) = \left(\phi_1(\mathbf{r}), \phi_2(\mathbf{r})\right)$, and $\chi_1$ and $\chi_2$ into another two-component vector called the ``small component'' $\chi(\mathbf{r}) = \left(\chi_1(\mathbf{r}), \chi_2(\mathbf{r})\right)$. With this substitution, the lower two lines of Eq. \ref{Eq:DKS} can be used to express the small component $\chi$ in terms of the large component:
\begin{equation}\label{Eq:small}
    \chi = \frac{\rev{c}}{2mc^2 + \epsilon - V} \, \boldsymbol{\sigma}\cdot\mathbf{p} \phi \, .
\end{equation}
Substituting Eq. \ref{Eq:small} back into the upper two lines of Eq. \ref{Eq:DKS} results in a single expression for the large component $\phi$:

\begin{equation}\label{Eq:large}
 \boldsymbol{\sigma} \cdot \mathbf{p} \frac{c^2}{2mc^2+\epsilon-V} \boldsymbol{\sigma} \cdot \mathbf{p} \phi +  \rev{(V \cdot \mathbbm{1})} \phi = \epsilon\phi \, .
\end{equation}

Eq. (\ref{Eq:large}) is a two-component equation for $\phi$ only, where \rev{$V \cdot \mathbbm{1}$ denotes the scalar potential term $V$ multiplied by the (2$\times$2) unit matrix $\mathbbm{1}$} to preserve the (2$\times$2) structure of the equation. However, unlike Schr\"odinger's non-relativistic equation, Eq. (\ref{Eq:large}) features an $\epsilon$-dependent Hamiltonian on the left hand side and therefore is not solvable as a linear system with orthogonal solutions. Furthermore, based on the original equation (\ref{Eq:DKS}), $\phi$ is not normalized on its own but only together with the small component $\chi$. Nevertheless, Eq. (\ref{Eq:large}) is a practical approximate equation for $\phi$ alone.

Using the explicit form of the Pauli matrices, one may further rewrite the kinetic energy operator in Eq.(\ref{Eq:large}) as the sum of a spin-free part and a spin-containing part,\cite{huhn2017} laying the foundation for the aforementioned scalar-relativistic and SOC approaches:
\begin{equation}
 \left( \mathbf{p}\frac{c^2}{2mc^2+\epsilon-V}\cdot\mathbf{p} + i\mathbf{p}\frac{c^2}{2mc^2+\epsilon-V} \times \mathbf{p}\cdot\boldsymbol{\sigma} \right) \phi +  \rev{(V \cdot \mathbbm{1})}\phi = \epsilon\phi .
  \label{Eq:large-sep}
\end{equation}
By writing down Eqs. (\ref{Eq:DKS})-(\ref{Eq:large-sep}) explicitly, the different levels of relativistic treatments available in FHI-aims can be rationalized straightforwardly. 

\textbf{Non-relativistic limit: Schr\"odinger equation.} In regions of space where $\epsilon - V \ll 2mc^2 \approx $ 1~MeV, $\epsilon-V$ is negligible in the first two terms. By neglecting $\epsilon-V$ indiscriminately in those terms, the second term in Eq. (\ref{Eq:large-sep}) vanishes due to the cross product and the standard non-relativistic Schr\"odinger equation results:
\begin{equation}\label{Eq:Schroedinger}
    \frac{\mathbf{p}^2}{2 m} \phi +  \rev{(V \cdot \mathbbm{1})}\phi = \epsilon\phi \, .
\end{equation}
In spin-polarized Kohn-Sham DFT, $\phi_1$ and $\phi_2$ in $\phi$ assume the roles of spin-up and spin-down electrons, coupled only implicitly through the exchange-correlation potential, which is part of $V$. 
However, the assumption $\epsilon - V \ll 2mc^2$ is never fully justified in all regions of space, since $V$ assumes very large negative values close to a nucleus. As noted initially, the non-relativistic Schr\"odinger equation Eq. (\ref{Eq:Schroedinger}) is nevertheless frequently considered to be sufficient to capture properties associated with the valence electrons, away from the nucleus, in light-element molecules and solids ($Z$=1-20), e.g., chemical bonding energies or optical excitations. However, this assumption is far less appropriate for core electrons, which are localized close to the nucleus. Therefore, the most important use-case of the non-relativistic approximation in FHI-aims are benchmark comparisons of calculated observables to results from other codes in which the exact same, simple kinetic energy operator of Eq. (\ref{Eq:Schroedinger}) is used in both implementations. For production calculations, using scalar-relativistic theory (discussed next) has the same computational cost and is always preferable.

\textbf{Scalar relativistic approach: Atomic ZORA.} Three separate observations allow one to turn Eq. (\ref{Eq:large-sep}) into a Schr\"odinger-like equation that provides significantly broader physical accuracy than the non-relativistic limit. 
\begin{enumerate}
    \item The SOC term (which includes $\boldsymbol{\sigma}$) is primarily relevant in regions in which the gradient of $V$, via the momentum operator $\boldsymbol{p}$ applied to $c^2/(2mc^2+\epsilon - V)$, is large, i.e., near the nucleus. The SOC term is therefore frequently neglected for properties derived from valence electrons.
    \item The potential $V(\mathbf{r})$ is only appreciably large compared to $2 mc^2 + \epsilon$ in regions close to a nucleus. In those regions, $V(\mathbf{r})$ is very similar to the potential associated with a free atom  $A$ containing the same nucleus, $V_A$, which can therefore be used instead of $V$. 
    \item The energies of valence electrons (order $\sim$eV), which are primarily located away from the nucleus, always satisfy $\epsilon \ll 2 m c^2$. The Hamiltonian therefore remains correct for valence electron properties even if setting $\epsilon$ to zero on the left side of Eq. (\ref{Eq:large-sep}). 
\end{enumerate}   
By combining observations (1)-(3), Equation (\ref{Eq:large-sep}) can be transformed into a simpler form\cite{Blum2009}, which is nevertheless accurate for all valence electron properties in the absence of strong SOC\cite{Lejaeghere2016}:
\begin{equation}\label{Eq:at-zora}
\mathbf{p}\frac{c^2}{2mc^2-V_A}\cdot\mathbf{p} \phi +  V\phi = \epsilon\phi
\end{equation}
If the self-consistent $V$ were used in the denominator instead of $V_A$, this would be the original zero-order regular approximation\cite{Lenthe1993}, which suffers from a gauge invariance problem\cite{Lenthe1994}. Using $V_A$, which is determined once and does not change in a multi-atom self-consistent calculation, removes the gauge invariance problem.

In practice, Eq. (\ref{Eq:at-zora}) is never applied in real-space form in FHI-aims, but rather in the form of Hamiltonian matrix elements between basis functions, 
\begin{equation}\label{Eq:h_atzora}
  h_{ij} = \langle \varphi_i | \mathbf{p}\frac{c^2}{2mc^2-V_A}\cdot\mathbf{p} + V | \varphi_j \rangle .
\end{equation}
In this form, the operator can be evaluated to the right, for the free atom potential \rev{$V_A$} associated with the nucleus at which $\varphi_j$ is centered, followed by symmetrization between matrix elements $h_{ij}$ and $h_{ji}$. For a given matrix element, the resulting scalar-relativistic atomic ZORA kinetic energy operator is exactly analogous to the non-relativistic one, since $V_A$ is independent of the three-dimensional structure of the molecule or solid under consideration and there is no dependence of $h_{ij}$ on $\epsilon$. For properties associated with valence electrons, e.g., the structure of solids, the results derived from FHI-aims' atomic ZORA scalar relativity are numerically indistinguishable from scalar-relativistic treatments in other benchmark-quality codes\cite{Lejaeghere2016}. The remaining approximation is restricted to core level energies, for which $\epsilon \ll 2 m c^2$ is not exact. However, this effect can be corrected by a simple correction that depends only on the chemical element in question, not on the structure within which an atom is embedded~\cite{Keller2020}. 

As a note, FHI-aims also implements a second, separate scalar-relativistic approach, ``scaled ZORA'' \cite{Lenthe1994,Blum2009}. Scaled ZORA  addresses the gauge invariance problem of the original ZORA approximately, but not entirely. It is implemented non-selfconsistently, i.e., as a post-processing step. Scaled ZORA can be useful if one is interested in scalar-relativistic core level eigenvalues with absolute positions closer to their spin-orbit averaged positions in the Dirac equation. However, in general the atomic ZORA approach should be used since it is free of any gauge-invariance problems and is broadly supported for practical simulations with FHI-aims.

\textbf{Scalar relativistic approach with second-variational SOC correction: atomic ZORA+SOC.} For single-particle energy levels (i.e., energy band structures), SOC effects can become qualitatively important even if they do not yet affect total energy differences drastically. One widely used approach to restore SOC in energy band structures is therefore to first carry out a self-consistent scalar-relativistic calculation without SOC. In a second step, the SOC operator is evaluated for a subset of the scalar-relativistic, self-consistent orbitals. The resulting, spin-orbit coupled Kohn-Sham single-particle Hamiltonian is then inverted once to obtain approximate spin-orbit coupled eigenvalues $\epsilon$ and two-component orbitals $\phi$. This approach is called a ``second-variational'' SOC correction. The specific Hamiltonian used in FHI-aims for this purpose is \cite{huhn2017}
\begin{equation}
    \left( \mathbf{p}\frac{c^2}{2mc^2+\epsilon-\rev{V_A}}\cdot\mathbf{p} + \frac{i}{4mc^2}\mathbf{p}V \times \mathbf{p}\cdot\boldsymbol{\sigma} \right) \phi +  \rev{(V \cdot \mathbbm{1})}\phi = \epsilon\phi ,
    \label{Eq:SOC}
\end{equation}
where the second term results from a Taylor expansion in $V$ of the earlier SOC term in Eq. \ref{Eq:large-sep}. As discussed in more detail below, this non-selfconsistent approach is remarkably precise for band structures across much of the periodic table \cite{huhn2017} while remaining computationally efficient, since a full two-component self-consistent field cycle is avoided. 

\textbf{Self-consistent four-component relativity: Q4C approach.} For self-consistent relativistic calculations of all four components, which includes self-consistent SOC, FHI-aims implements the quasi-four-component (Q4C) approach\cite{Zhao2021,Liu2007}. This method (closely related to the exact two-component, X2C approach of quantum chemistry\cite{Kutzelnigg2005}) relies on the idea that the small component $\chi$ is most relevant in the near-nuclear regions of a structure, where relativistic effects are strong, so that $\chi$ can be handled by an approximate, fixed relation with the large component $\phi$, which is based on the free atom. In practice, this is accomplished by expanding the four-component wave functions into atom-centered, four-component basis functions instead of expanding each component individually:
\begin{equation}\label{6}
\begin{pmatrix}
\phi_i  \\  \chi_i
\end{pmatrix}
= 
\sum_\mu C_{\mu i} \cdot
\begin{pmatrix}
\varphi^L_\mu  \\  \varphi^S_\mu
\end{pmatrix} 
\, .
\end{equation}
For each of the 4C basis functions here, there is a fixed relation between its large and small components, i.e.,
\begin{equation}\label{7}
\varphi_{\mu}^S = \hat{K} \varphi_{\mu}^L; \quad \hat{K} = \left[ \frac{c}{2mc^2+\epsilon_{A,\mu}-V_A} \right] \left(  \boldsymbol{\sigma}\cdot\mathbf{p} \right) \, .
\end{equation}
Here, $A$ denotes the atom type where this basis function is centered, and $\epsilon_{A,\mu}$ and $V_A$ represent the corresponding orbital energy and atomic radial potential, respectively. For the minimal basis, from which the majority of the density is derived, $\epsilon_{A,\mu}$ is the corresponding eigenvalue of each basis functions, i.e., $\hat{K}$ is the exact free-atom relation Eq. (\ref{Eq:small}) between the atomic $\chi$ and $\phi$. For other basis functions, FHI-aims employs $\epsilon_{A,\mu}$=0 as in the atomic ZORA approximation \cite{Zhao2021}. In FHI-aims, a set of fully-relativistic atomic-orbital basis functions for the minimal basis can be precomputed by solving the radial Dirac equation for free atoms using the open-source atom-solver code DFTATOM~\cite{certik+cpc2013}.

\section*{Usability and Tutorials}

Since different relativistic approaches are implemented directly within FHI-aims, using them is as simple as setting a keyword, \texttt{relativistic}, to an appropriate value. This keyword is used and explained in multiple basic FHI-aims tutorials at \href{https://fhi-aims-club.gitlab.io/tutorials/tutorials-overview/}{https://fhi-aims-club.gitlab.io/tutorials/tutorials-overview}. 
\begin{itemize}
    \item Non-relativistic theory and scalar-relativistic atomic ZORA are supported for essentially all functionality in the code due to their simplicity and mathematically simple structure.
    \item Second-variational, spin-orbit coupled atomic ZORA is supported for energy eigenvalues and energy band structures in semilocal and hybrid DFT, as well as many derived quantities such as element-resolved densities of states, spin textures of energy bands, or absorption spectra.
    \item At the time of writing, the Q4C method in FHI-aims is implemented
    to support system sizes up to and above 100 atoms. It does not yet include support for exchange-correlation beyond non-spinpolarized, semilocal DFT and it does not yet support forces or geometry optimization.
\end{itemize}

As mentioned above, the scalar-relativistic atomic ZORA approach achieves essentially identical results with other codes, indicating that the valence electron treatment in this approach is basically the exact one within the limits of a scalar relativistic approach.\cite{Lejaeghere2016} However, energy band structures from scalar relativistic computations deviate from fully relativistic ones already for compounds containing elements beyond Zn ($Z$=30),\cite{Zhao2021} i.e., SOC is essential for band structure calculations involving even moderately heavy elements.

For energy band structures and (in the case of the Q4C method) total energies, the spin-orbit coupled atomic ZORA and Q4C methods have been extensively benchmarked in Refs.~\cite{huhn2017} and \cite{Zhao2021}. Ref. \cite{huhn2017} defined a comprehensive benchmark of the band structures of 103 metallic, semiconducting and ionic solids for scalar-relativistic and second-variational spin-orbit coupled methods, covering chemical elements from Li to Po. Compared to self-consistent spin-orbit coupled energy band structures derived from the Wien2k code,\cite{Blaha2020} the second-variational atomic ZORA + SOC approach remains quantitatively exact up to at least Cd ($Z$=48). The following 5$p$ and 6$s$ elements (up to Ba) show band structures that agree within 50~meV on average, whereas the energy band structures of 6$p$ elements (up to Po) show deviations up to 0.3~eV. In Ref. \cite{Zhao2021}, the same group of compounds was used to benchmark other methods against the Q4C approach, showing close agreement with Wien2k's self-consistent SOC approach except for the energy band structures of the late 5$d$ metals Os, Ir, and Pt, for which the Q4C approach is presumably more accurate. All corresponding Q4C band structure benchmark data can be accessed via the NOMAD database at 
(\href{https://doi.org/10.17172/NOMAD/2025.03.09-1}{\texttt{https://doi.org/10.17172/NOMAD/2025.03.09-1}}).

\begin{figure}[htbp]
    \centering
    \includegraphics[width=\textwidth]{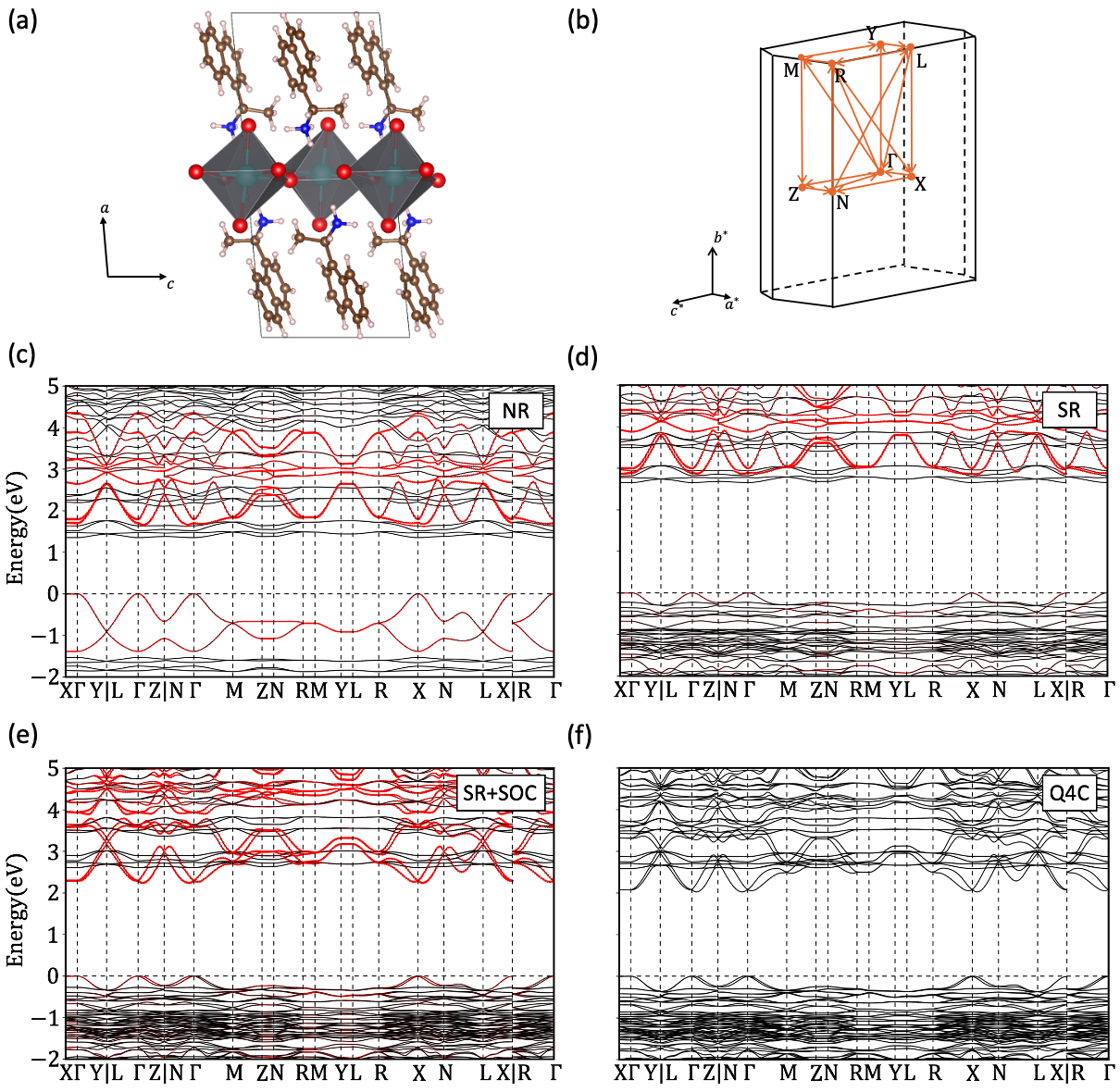}
    \caption{Structural and electronic properties of \rev{the layered hybrid perovskite S-1-(1-napthyl)ethylammonium lead bromide (S-NPB)}.
(a) A representative unit cell of the relaxed S-NPB structure in real space consists of 118 atoms, including 2 lead atoms, 8 bromine atoms, 4 nitrogen atoms,  32 carbon
atoms, and 72 hydrogen atoms in a unit cell.
(b) The corresponding Brillouin zone in reciprocal space, highlighting the k-path employed for band structure calculations.
(c–f) Electronic band structure comparisons under different computational methods: (c) Nonrelativistic (NR), (d) Atomic Zero-Order Regular Approximation Scalar Relativistic (SR), (e) SR with non-selfconsistent Spin-Orbit Coupling (SR+SOC), and (f) Quasi-Four-Component (Q4C) method. The contributions for lead(Pb) in the bands (c-e) are highlighted in red. }
    \label{fig:perovskite}
\end{figure}

Figure \ref{fig:perovskite} shows the impact of relativistic effects on the band structure of an organic-inorganic semiconductor, the chiral hybrid perovskite S-1-(1-napthyl)ethylammonium lead bromide (S-NPB) \cite{jana2020} in a 118-atom structure (Figure \ref{fig:perovskite}a), computed using tight settings along the reciprocal space path shown in Figure \ref{fig:perovskite}b. The density functional used is DFT-PBE, since the objective is a comparison of different relativistic effects on equal footing, rather than a more accurate treatment of the overall band structure using a higher-level approach. Energy bands highlighted in red are those primarily associated with lead ($Z$=82), for which SOC effects are much stronger than for all other elements in the structure. The evolution of energy band structures from (c) non-relativistic to (d) scalar-relativistic (atomic ZORA) shows drastic shifts both of the band gap and of the energy band alignments between Pb and other components in the valence band. Similarly strong effects occur when introducing second-variational SOC (e). Without SOC, a group of flat, molecular-derived bands would form the conduction band minimum, whereas the physically correct inclusion of SOC shifts the conduction bands to the Pb-derived bands. Physically, this shift would have a drastic effect, since the spatial location of carriers in a semiconductor impacts both the electronic and the optical properties of the material. Finally, the transition to (f) with the Q4C approach (including SOC self-consistently, as well as the small component) shows further changes, but much smaller in scale. While the Pb-derived bands are not separately highlighted in this figure (computing a formal decomposition is not yet implemented for the four-component wave function), their character at the conduction band edge is clearly apparent in comparison to (e), with a shift of a few tenths of eV compared to the organic-derived flat bands immediately above. This shift is of the same order as the inherent uncertainty even of higher-level functionals (such as hybrid DFT or the $G_0W_0$ method with different starting points)~\cite{Kim2020}. Compared to Q4C, the band structure features of this closed-shell system are already captured well at the simpler, non-selfconsistent atomic ZORA + SOC approach. However, a higher level of relativistic treatment would be needed for non-collinear spins in open-shell systems or if the $p$-derived conduction bands were partially filled, in which case self-consistency would matter.


\subsection*{Future Plans and Challenges}

For production DFT calculations across most of the periodic table, at least up to heavy elements such as Pb ($Z$=82), relativistic effects are reliably captured by scalar relativity (atomic ZORA) and, for band structures, enhanced by second-variational spin-orbit coupling (atomic ZORA + SOC). 

For phenomena in which relativistic effects near the nucleus or the coupling between spin and orbital degrees of freedom play a particular role, future developments in FHI-aims will build on the Q4C approach, which has both self-consistent spin orbit coupling, the correct four-component wave function near the nucleus, and which can incorporate other effects such as the correct potential of a finite nucleus if needed. At the time of writing, ongoing developments include an extension of Q4C to system sizes well beyond the current reach, by adapting the method to FHI-aims' distributed-parallel, ``locally-indexed'' \cite{Huhn2020} infrastructure for large-scale integrals. Similarly, an extension of the Q4C-supported exchange-correlation functionals to DFT+U and support for spin-polarized exchange-correlation functionals, making non-collinear spin systems accessible in open-shell Q4C, are under active development.

Further important future Q4C developments include forces and stresses for geometry optimization and dynamics, support for hybrid density functionals, as well as extensions of the Q4C approach to many-body methods such as $GW$ or the Bethe-Salpeter equation. Similarly, core-hole spectroscopies, nuclear magnetic resonance observables, and other physical observables which are impacted by nuclear and/or core electron properties are highly desirable targets for a fully relativistic implementation.  
Overall, work towards a complete, NAO based four-component treatment offers much promise for direct simulations of phenomena that are currently not within easy range of standard electronic structure methods.

\subsection*{Acknowledgements}

We thank Dr. Ond\v{r}ej \v{C}ert\'{i}k for his technical support on the open source Dirac atom-solver code DFTATOM he authored. We are grateful to Levi Keller, who would have been a coauthor, for his contributions; Levi Keller passed away on January 28, 2023. For the application of the Q4C approach to large hybrid perovskite systems, V.B. and W.Z. were supported by NSF Award DMR-2323803.



  





\newpage

\section{Solvers for Large-Scale Electronic Structure Theory: ELPA and ELSI}
\label{SecELPA}
\sectionauthor[1]{\textbf{*Petr Karpov}}       
\sectionauthor[1]{\textbf{*Andreas Marek}}    
\sectionauthor[1]{Tobias Melson}     
\sectionauthor[1]{Pavel K{\r{u}}s}         
\sectionauthor[1]{Hermann Lederer}   
\sectionauthor[2]{Bruno Lang}        
\sectionauthor[3]{Alexander P\"oppl} 
\sectionauthor[4]{Jun Tang} 
\sectionauthor[6]{Danilo Simoes Brambila} 
\sectionauthor[5]{Hagen-Henrik Kowalski}  
\sectionauthor[6,a]{Lydia Nemec}            
\sectionauthor[7,8,9,b]{Simone S. K\"ocher}  
\sectionauthor[5,8,c]{Christian Carbogno}   
\sectionauthor[7,8]{Christoph Scheurer}  
\sectionauthor[5]{Matthias Scheffler}     
\sectionauthor[7,8]{Karsten Reuter}      
\sectionauthor[10]{Victor Wen-zhe Yu} 
\sectionauthor[11]{Ben Hourahine}     
\sectionauthor[12]{Alberto Garcia}    
\sectionauthor[13]{William Dawson}    
\sectionauthor[10,14,d]{Yi Yao}         
\sectionauthor[15]{William Huhn}     
\sectionauthor[16]{Jonathan Moussa}  
\sectionauthor[10]{Björn Lange}                 
\sectionauthor[17]{Álvaro Vázquez-Mayagoitia}    
\sectionauthor[18]{Mina Yoon}                    
\sectionauthor[19]{Samuel J. Hall}   
\sectionauthor[19,20,21]{Reinhard J. Maurer}  
\sectionauthor[10]{Uthpala Herath}    
\sectionauthor[5,14]{Konstantin Lion}  
\sectionauthor[5,14]{Sebastian Kokott}  
\sectionauthor[22]{Ville Havu}     
\sectionlastauthor[10,23]{\textbf{*Volker Blum}}    

\sectionaffil[1]{Max Planck Computing and Data Facility, Garching, Germany}
\sectionaffil[2]{University of Wuppertal, Mathematics and Natural Sciences}
\sectionaffil[3]{Intel Corporation, Santa Clara, CA, USA}
\sectionaffil[4]{Annapurna Labs, AWS}
\sectionaffil[5]{The NOMAD Laboratory at the Fritz Haber Institute of the Max Planck Society, Faradayweg 4-6, D-14195 Berlin, Germany}
\sectionaffil[6]{Theory Department (since 1/1/2020: The NOMAD Laboratory), Fritz Haber Institute of the Max Planck Society, Faradayweg 4-6, D-14195 Berlin, Germany}
\sectionaffil[7]{Theoretical Chemistry and Catalysis Research Center, Technische Universit\"at M\"unchen, D-85747 Garching, Germany}
\sectionaffil[8]{Theory Department, Fritz Haber Institute of the Max Planck Society, Faradayweg 4-6, D-14195 Berlin, Germany}
\sectionaffil[9]{Institute of Energy Technologies (IET-1), Forschungszentrum J\"ulich GmbH, Wilhelm-Johnen-Str., 52425 J\"ulich, Germany}
\sectionaffil[10]{Thomas Lord Department of Mechanical Engineering and Materials Science, Duke University, Durham, NC 27708, USA}
\sectionaffil[11]{Department of Physics, SUPA, University of Strathclyde, John Anderson Building, 107 Rottenrow, Glasgow G4 0NG, UK}
\sectionaffil[12]{Institut de Ciència de Materials de Barcelona, ICMAB-CSIC, Campus UAB, 08193 Bellaterra, Spain}
\sectionaffil[13]{RIKEN Center for Computational Science, Kobe 650-0047, Japan}
\sectionaffil[14]{Molecular Simulations from First Principles e.V., D-14195 Berlin, Germany}
\sectionaffil[15]{Intel Corporation, 500 Beaver Brook Road, Boxborough, MA, 01719, USA}
\sectionaffil[16]{Molecular Sciences Software Institute, Blacksburg, Virginia 24060, USA}
\sectionaffil[17]{Computational Science Division, Argonne National Laboratory, 9700 South Cass Avenue, Lemont, Illinois 60439, United States}
\sectionaffil[18]{Center for Nanophase Materials Sciences, Oak Ridge National Laboratory, Oak Ridge, TN 37830, USA}
\sectionaffil[19]{Department of Chemistry, University of Warwick, Gibbet Hill Road, CV4 7AL Coventry, United Kingdom}
\sectionaffil[20]{Department of Physics, University of Warwick, Gibbet Hill Road, CV4 7AL Coventry, United Kingdom}
\sectionaffil[21]{Faculty of Physics, University of Vienna, Vienna A-1090, Austria}
\sectionaffil[22]{Department of Applied Physics, Aalto University, P.O. Box 11000, FI-00076 Aalto, Finland}
\sectionaffil[23]{Department of Chemistry, Duke University, Durham, NC 27708, USA}

\sectionaffil[]{(*) Coordinator of this contribution}
\rule[0.25ex]{0.35\linewidth}{0.25pt}

\sectionaffil[a]{{\it Current Address:} Infineon Technologies AG, Am Campeon 1-15, 85579 Neubiberg, Germany}
\sectionaffil[b]{{\it Current Address:} Institute of Energy Technologies (IET-1), Forschungszentrum J\"ulich GmbH, Wilhelm-Johnen-Str., 52425 J\"ulich, Germany}
\sectionaffil[c]{{\it Current Address:} Theory Department, Fritz Haber Institute of the Max Planck Society, Faradayweg 4-6, D-14195 Berlin, Germany}
\sectionaffil[d]{{\it Current Address:} Molecular Simulations from First Principles e.V., D-14195 Berlin, Germany}




\subsection*{Summary}

In this contribution, we give an overview of the ELPA library and ELSI interface, which are crucial elements for large-scale electronic structure calculations in FHI-aims. ELPA is a key solver library that provides efficient solutions for both standard and generalized eigenproblems, which are central to the Kohn-Sham formalism in density functional theory (DFT). It supports CPU and GPU architectures, with full support for NVIDIA and AMD GPUs, and ongoing development for Intel GPUs. Here we also report the results of recent optimizations, leading to significant improvements in GPU performance for the \textit{generalized} eigenproblem.

ELSI is an open-source software interface layer that creates a well-defined connection between ``user'' electronic structure codes and ``solver'' libraries for the Kohn-Sham problem, abstracting the step between Hamilton and overlap matrices (as input to ELSI and the respective solvers) and eigenvalues and eigenvectors or density matrix solutions (as output to be passed back to the ``user'' electronic structure code). In addition to ELPA, ELSI supports solvers including LAPACK and MAGMA, the PEXSI and NTPoly libraries (which bypass an explicit eigenvalue solution), and several others. 

ELSI, ELPA, and other solver libraries supported in ELSI are mature, well-tested software, ensuring efficient support for large-scale simulations on current HPC systems. Future plans for ELPA include further optimization of GPU routines, particularly integrating the GPU collective communication libraries (NVIDIA's NCCL and AMD's RCCL) into the solution of tridiagonal the matrix problem and backtransformation of the eigenvectors in standard eigenproblem solvers. Similarly expanding support for Intel GPUs through deeper integration with Intel's oneAPI libraries, specifically oneCCL, is planned. These advancements aim to ensure that ELPA, accessible either with its own Fortran/C/C++ APIs or via ELSI, remains at the forefront of large-scale computations, offering researchers powerful tools for addressing increasingly complex systems in electronic structure simulations. Future developments in ELSI will also continue to focus on GPU support and exascale readiness, continued support for a broad range of solvers, as well as extensions to solvers, e.g., for constrained density functional theory.

\subsection*{Current Status of the Implementation}
\gi{Introduction: From effective single-particle equations to generalized eigenproblem and density}

Using the Kohn-Sham (KS) \cite{Kohn1965} or generalized Kohn-Sham (gKS) \cite{Seidl1997} formalism, a full problem of $n_\mathrm{el}$ interacting electrons can be transformed to $n_\mathrm{el}$ or more auxiliary single-particle equations, $\hat{H}\psi_l = \varepsilon_l\psi_l$, e.g., for the scalar-relativistic or non-relativistic kinetic energy operator $\hat{t}$:
\begin{equation}
\left( \hat{t} + V_{\textrm{ext}}(\mathbf{r}) + \int \frac{n(\mathbf{r}')}{|\mathbf{r}-\mathbf{r'}|}d\mathbf{r}' + v_{\mathrm{xc}}(\mathbf{r}))\right) \psi_l (\mathbf{r}) =
\varepsilon_l \psi_l (\mathbf{r}) ,
\label{full_schr_eq}
\end{equation}
\rev{where $V_{\textrm{ext}}(\mathbf{r})$ is the external potential, $n(\mathbf{r})=\sum_{l=1}^N f_l |\psi_l(\mathbf{r})|^2$ is the electron density, $\psi_l$ are the effective single-particle orbitals of DFT with occupation numbers $f_l$ and eigenvalues $\varepsilon_l$. $N$ here denotes the number of single-particle states considered in the calculation.}
The exchange-correlation potential $v_{\mathrm{xc}}$ is the only unknown term and can be treated using various density functional approximations \cite{Blum2024roadmap}. To solve the equations (\ref{full_schr_eq}) numerically, one has to choose a convenient set of $N$ 
basis functions and approximate the KS state $\psi_l(\mathbf{r}) = \sum_{i=1}^N c_{li} \varphi_i (\mathbf{r})$. The set of equations (\ref{full_schr_eq}) then transforms to the generalized eigenvalue problem
\begin{equation}
H C = S C \varepsilon
\label{generalized_eigenproblem}
\end{equation}
where $H$ and $S$ are the Hamiltonian and the overlap matrices, respectively, with elements $h_{ij} = \int \varphi_i (\mathbf{r})^* \hat{H} \varphi_j(\mathbf{r}) d\mathbf{r}$, $s_{ij} = \int \varphi_i (\mathbf{r})^* \varphi_j(\mathbf{r}) d\mathbf{r}$; $C$ is the matrix of eigenvectors (stored as matrix columns); and $\varepsilon=\mathrm{diag}(\varepsilon_1,...,\varepsilon_N)$. Given $C$, one may obtain the total energy $E[n(\mathbf{r})]$ via the ground-state density $n(\mathbf{r})$.

\gi{Solving the generalized eigenproblem}

Since the basis functions $\{\varphi_i({\mathbf{r}})\}$ are not orthonormal, $S$ is not an identity matrix. In order to solve the generalized eigenproblem, $H C = \varepsilon S C$, for real symmetric/complex Hermitian matrices $H$, $S$, and positive-definite matrix $S$, one usually reduces it to the standard eigenproblem via the following steps:
\begin{samepage} 
\begin{enumerate}
\item
Cholesky decomposition of $S = U^T U$, where $U$ is an upper-triangular matrix.

\item
Inversion of the upper triangular matrix $U \rightarrow U^{-1}$.

\item
Calculation of two matrix products: $\tilde{H} = (U^{-1})^T H U^{-1}$.
\end{enumerate}
\end{samepage} 
Here, $A^T$ corresponds to the transpose or conjugate-transpose of a real- or complex-valued matrix $A$, respectively. This reduces the \textit{generalized} eigenproblem to the \textit{standard} one, $\tilde{H} \tilde{C} = \varepsilon \tilde{C}$, which has the same set of eigenvalues. If the eigenvectors are needed, one has to perform the backtransformation, which requires another matrix-matrix multiplication: $C = U^{-1} \tilde{C}$.

A strict condition for the applicability of steps 1 to 3 is that the positive-definite matrix $S$ must not be numerically singular, i.e., the overlap matrix $S$ itself must not have eigenvalues that are zero; otherwise, Eq. (\ref{generalized_eigenproblem}) could have multiple different solutions $C$ that correspond to the same physical eigenfunctions $\{\psi_l(\mathbf{r})\}$. As a consequence, in practical computations, $S$ should not be ill-conditioned, i.e., the ratio between the largest and the smallest eigenvalue of $S$ should normally not exceed the numerical range (around $10^{12}$) accessible by double precision numbers. 

The standard FHI-aims basis sets are compact enough to lead to well-conditioned overlap matrices $S$ even for well-converged DFT calculations and steps 1 to 3 are therefore routinely used. Prior to any calculation, FHI-aims computes the eigenvalues $\sigma_i$ of the overlap matrix and alerts the user if a dense basis set with a high condition number of the overlap matrix is detected. In the latter case (high condition number of $S$), electronic structure calculations can then be carried out by transforming the eigenproblem (\ref{generalized_eigenproblem}) to the basis of eigenvectors $D_i$ of the overlap matrix, but omitting any $D_i$ for which $\sigma_i$ is smaller than a small positive number $\delta$ (typically $\delta \le 10^{-5}$). This procedure is standard in the community and is implemented in ELSI.

\gi{Circumventing the eigenvalue problem: Density-matrix based solutions}

In the limit of large systems, the computational effort for solving the eigenproblem Eq. (\ref{generalized_eigenproblem}) scales cubically with system size and becomes the computational bottleneck for Kohn-Sham DFT. Solving the eigenvalue problem (\ref{generalized_eigenproblem}) is one option to obtain $n(\mathbf{r})$. However, since the actual targets of DFT are $n(\mathbf{r})$ and $E[n(\mathbf{r})]$, the eigenvalues and eigenvectors are technically not required. Alternative strategies therefore target finding the single-particle density matrix $P_{ij} = \sum_{l=1}^N f_l c^*_{li} c_{lj}$ without an explicit eigenvalue solution. The density can then be computed as 
  \begin{equation}
  n(\mathbf{r}) = \sum_{ij} \int \varphi_i(\mathbf{r})^* P_{ij} \varphi_j(\mathbf{r}) d\mathbf{r} . \label{density_dm}
  \end{equation}
Especially for sparse, localized basis sets, the computational effort for solvers that target the density matrix without an explicit eigenvalue/eigenvector solution can scale quadratically or linearly with the system size. For sufficiently large systems, such solvers can therefore outperform the cubic-scaling eigenvalue solution. The crossover point, i.e., the system size beyond which the eigenvector-free solution becomes more favorable, depends on the specifics of the system, level of theory, and computer hardware used. Thus, maintaining access to different solver types within a single electronic structure code is desirable.

\begin{figure}[H]
    \centering
    \includegraphics[width=0.7\linewidth]{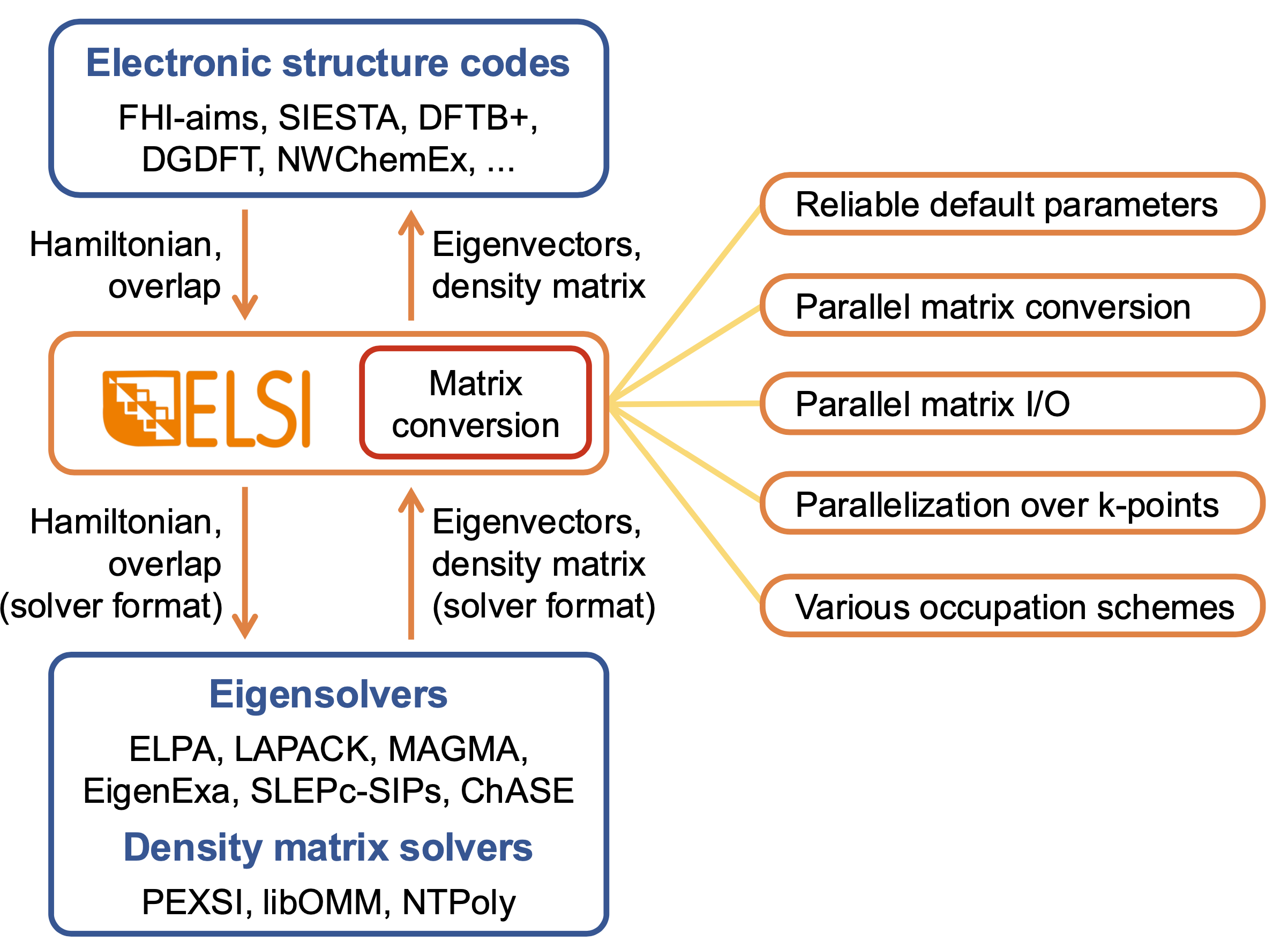}
    \caption{Tasks performed by the ELSI interface software, connecting different eigenvalue and density matrix solvers to electronic structure codes including FHI-aims, Siesta, DFTB+, NWChemEx and others. ELSI provides a uniform interface that is callable in Fortran, C, C++, and Python, and handles matrix format conversion between user codes and solver libraries. The solvers presented here include eigenvalue and density matrix solvers connected to ELSI at the time of writing. The ELSI interface provides functionality for tasks common to eigenvalue and density matrix operations, such as calculating occupation numbers or trivially parallel handling of independent eigenvalue problems for different $k$-points and spin channels.}
    \label{fig:ELSI}
\end{figure}

\gi{ELSI}

The ELSI interface software \cite{elsi_webpage,yu_elsi_2018,yu_elsi_2020} (Figure \ref{fig:ELSI}) provides a uniform code layer that handles eigenvalue problems or density matrix calculations, supporting several electronic structure codes (FHI-aims \cite{Blum2009}, DFTB+ \cite{dftb_plus}, Siesta \cite{Garcia2020}, NWChemEx \cite{NWChemEx}) and integrating ten different solvers.
The solvers have different APIs, making their direct integration into an electronic structure software a non-trivial task. 
Moreover, different electronic structure codes would all need to reimplement essentially the same connections to different solvers.
ELSI simplifies the integration and use of multiple solver libraries by providing a unified interface, allowing users to access various eigensolvers and density matrix solvers optimized for various problem sizes and types. The interface is designed around a derived type, the \texttt{elsi\_handle}, to pass data between the user code that calls ELSI, the ELSI interface layer itself, and the solver codes. This construct enables programs written in Fortran, C, C++ or Python to connect to the ELSI interface with equal ease. Additionally, conversions between different distributed-parallel matrix formats are handled efficiently in ELSI. 

ELSI supports cubic scaling eigensolvers like LAPACK \cite{netlib_lapack,anderson:1999} and ELPA \cite{elpa_webpage,Marek2014,marek:2024}, as well as reduced scaling methods such as PEXSI \cite{PEXSI-1,PEXSI-2} or the linear-scaling NTPoly \cite{NTPoly} density matrix solvers. Further supported solvers include the dense eigensolvers EigenExa \cite{EigenExa} and MAGMA \cite{MAGMA}, the iterative eigensolvers ChASE \cite{ChaSE1, ChaSE2}, SLEPc-SIPS \cite{SLEPc-SIPS}, and libOMM \cite{libOMM}, as well as BSEPACK \cite{BSEpack} and DLA-Future~\cite{DLA-Future,Solca:2024}. Although implemented in ELSI, ChASE and DLA-Future are not currently supported in FHI-aims. \rev{Given the plethora of solvers supported, we refer to the respective original publications for their in-depth mathematical definitions.} 
ELSI returns either eigenvalues and eigenvectors or the density matrix and its energy-weighted counterpart. Since the density matrix includes the occupation numbers of the Kohn-Sham orbitals, ELSI also provides an implementation of numerically precise occupation numbers. For FHI-aims, ELSI is the designated source of standardized computed occupation numbers, $f_l$, including several non-Aufbau occupation approaches to perform occupation-constrained DFT calculations (e.g., the $\Delta$SCF method).

\gi{ELPA}

ELPA is an open-source massively parallel direct eigensolver library for real symmetric or complex Hermitian eigenvalue problems \cite{elpa_webpage,Marek2014,marek:2024}
that is supported by most major DFT software packages \cite{abinit,Deslippe2012,cp2k,cpmd,dftb_plus,Blum2009,gpaw,nwchem,openMX,quantumATK,quantum_espresso_1,quantum_espresso_2,Garcia2020,vasp,Blaha2020}.
ELPA is capable of solving both \textit{standard} and \textit{generalized} eigenproblems: users are not required to perform the transformation from the \textit{generalized} problem manually, although interfaces for each individual transformation step are also provided.
Moreover, ELPA is ported to GPU accelerators (including NVIDIA, AMD, and Intel) \cite{{kus:2019_gpu,Yu:2021}}. 
To the best of our knowledge, ELPA has recently set a new record \cite{wlazlowski:2024} of a $3.2 \cdot 10^6 \times 3.2 \cdot 10^6$ dense matrix diagonalization for all eigenvalues and eigenvectors of the standard real-valued eigenproblem on the LUMI supercomputer, which is equipped with AMD MI250X GPUs.

\subsection*{Usability and Tutorials}

\gi{Solver selection in FHI-aims}

The numerical atomic orbital (NAO) basis sets in FHI-aims \cite{Blum2009} are rather compact, i.e., the number of basis functions $N$ required to represent the single-particle states with high precision is typically not larger than a single-digit multiple of $n_\mathrm{el}$. Thus, since the number of basis functions $N$ determines the dimension of $H$ and $S$, typically a large fraction of their eigenpairs is needed.
For this scenario and for small and mid-sized systems, direct eigensolvers, such as LAPACK, ScaLAPACK \cite{netlib_scalapack,blackford:1997} or ELPA are the solvers of choice. 

One advantage of ELSI is its ability to enable direct, comparative benchmarks of different solvers within a single software framework, allowing one to make a choice that is specific to system and hardware characteristics of a particular computational campaign.
Beyond the default serial (LAPACK) and parallel (ELPA) solvers available in FHI-aims, a range of additional solvers can be accessed through ELSI \rev{and are supported by FHI-aims directly by choosing appropriate keywords}. At the time of writing, these include PEXSI, NTPoly, EigenExa, SLEPc-SIPS, libOMM, and MAGMA.

Past results for the NAO basis sets of FHI-aims showed that the crossover point beyond which different density-matrix-based solvers outperform ELPA lies beyond several hundred to several thousand atoms \cite{yu_elsi_2020}. On parallel and/or GPU hardware, performing a direct eigenvalue solution using ELPA is therefore preferable for the majority of current production calculations in FHI-aims. For larger systems and for semilocal DFT, the solution effort for the pole expansion and selected inversion (PEXSI) \cite{PEXSI-1,PEXSI-2} method scales at most quadratically with system size for three-dimensional systems, and even subquadratically for two- or one-dimensional geometries. Benchmarks in Ref. \cite{yu_elsi_2020} showed that, for 1D systems, PEXSI outperformed a direct diagonalization for as few as 400 atoms. For 2D systems, the crossover was found around 1,000 atoms. Alternatively, for non-metallic systems, density matrix purification allows one to directly solve for the density matrix with a solution effort that grows linearly with the system size, independent of the density functional, as implemented, e.g., in the NTPoly \cite{NTPoly} library for sparse matrix function calculations. \rev{Benchmarks in Ref.~\cite{yu_elsi_2020}, for the DFTB+ code with significantly more sparse matrices than FHI-aims, show that the performance crossover depends heavily on the system type and matrix sparsity, and should be tested individually prior to production calculations for the largest systems.}

\gi{ELPA - Usability}

Within FHI-aims, ELSI and ELPA (as well as LAPACK calls for serial linear algebra steps) are automatically employed with parallelization strategies that leverage trivial parallelism (e.g., for separate $k$-points or spin channels) whenever possible. Since the performance of solvers other than ELPA depends on system, problem and hardware characteristics, their selection is a user choice by setting specific keywords.

One key consideration pertains to high-performance hardware beyond CPUs. Specifically, utilizing GPUs is essential for leveraging the computational power of modern supercomputers.
This not only accelerates matrix diagonalization but also allows for the handling of larger matrices,
enabling the investigation of physical systems with more atoms.

Much effort has been spent porting ELPA to GPUs. Nowadays ELPA fully supports both NVIDIA and AMD GPUs, including support for their respective GPU collective communication libraries, NCCL \cite{nccl} and RCCL \cite{rccl}, respectively, for the most computationally intensive parts of the algorithm.
Partial support is available for Intel GPUs using SYCL \cite{intelsycl} and oneMKL \cite{intelonemkl}, though this is currently limited to solving the standard eigenproblem.

For the core solver of the \textit{standard} eigenproblem, ELPA provides implementations of the conventional one-stage diagonalization method  (``ELPA1'') and the two-stage diagonalization (``ELPA2'') \cite{lang:1993, lang:1994, auckenthaler:2011}.
On CPUs, as a rule of thumb, the ELPA2 solver is preferable and it is also typically up to 1.5-2 times faster than MKL's ScaLAPACK \cite{kus:2019}.
However, on GPUs, the ELPA1 solver is typically faster, providing a speedup of three times over the best setting on the CPU-only nodes if the local matrices per GPU are not extremely small \cite{kus:2019_gpu}. If only a subset of all eigenvectors is computed (as is often the case in DFT), then ELPA2-GPU becomes more competitive compared to ELPA1-GPU as the number of targeted eigenvectors decreases.

For the \textit{generalized} eigenproblem, until recently ELPA-GPU faced a significant bottleneck in the matrix-matrix multiplication step.
ELPA relied on ScaLAPACK or its own implementations of the SUMMA \cite{de_geijn:1997,chtchelkanova:1997} and Cannon's \cite{cannon:1969} algorithms for parallel matrix multiplication, none of which were fully GPU-ported until the 2024.05 release, leaving a substantial part of the computation to CPUs.
Table~\ref{table:elpa-generalized} compares the ELPA 2023.11 and 2024.05 releases, with the latter including a full GPU port for the matrix-matrix multiplication (`Multiply'). This is especially important for `NCCL' setup (1 MPI task per GPU), where the total `Multiply' time for forward and backward transformation is reduced from 559 to 25.6 seconds. 

GPU porting of the matrix-matrix multiplication brings the performance of the ELPA-GPU \textit{generalized} eigensolver in line with that of the core \textit{standard} eigenvalue solver, now achieving a similar speedup of approximately three to four times over the best-performing ELPA-CPU configuration. It is important to note that the CPU side of the comparison (last row of Table~\ref{table:elpa-generalized}) involves all CPU cores on a given computational node. For instance, as shown in Table~\ref{table:elpa-generalized}, we compare a GPU setup with 4 GPUs and tuned from 1 to 18 CPU cores per GPU (using ELPA1) against the CPU setup with all 72 CPU cores (using ELPA2). At the moment, FHI-aims is best tested with NVIDIA MPS \cite{mps}, as it allows multiple MPI processes per GPU to be used efficiently, which can be particularly beneficial for steps like `Solve'. NVIDIA MPS using all CPU cores of a node seems still to provide a very good speedup compared to the CPU-only ELPA (cf.~Table~\ref{table:elpa-generalized}). In cases where ELPA dominates the FHI-aims runtime, it is worthwhile to tune for the best MPI process per GPU ratio. For the future development of ELPA, as more of its parts being ported to GPU over time, we expect that the `NCCL' code path will become more performant than the `MPS' one.
This is because NCCL enables direct data transfer between GPU devices, eliminating the need for costly GPU-CPU memory copies required for MPI operations.

Overall, we recommend using ELPA-GPU when possible: it can provide up to a 4x speedup for the complete solution of standard and generalized eigenproblems, and even up to a 10x speedup for individual solution steps.
Despite this drastic improvement provided by the ELPA 2024.05 release, the `Multiply' step still dominates the time for forward and back transformations from \textit{generalized} to \textit{standard} eigenproblem, motivating its further optimization.

ELSI supports ELPA-GPU (NVIDIA only) directly in ELPA's 2020 release, and as an externally compiled library for later versions and other GPU types. Since newer ELPA versions offer significant performance enhancements on GPUs, it is recommended to use ELSI with externally linked ELPA.

\begin{table}[t]
\centering
\begin{tabularx}{\textwidth}{c || *{7}{Y} || Y}
\toprule
& \multicolumn{3}{c}{Forward transformation} & \multicolumn{3}{c}{Standard EVP} & Backtr. &  \\
\cmidrule(lr){2-4} \cmidrule(lr){5-7} \cmidrule(lr){8-8} 
ELPA version               & Choles.  & Invert  & Multiply        & Tridiag.  & Solve  & Back     & Mult.          &  Total \\
\midrule
2023.11, MPS, 18 MPI per GPU  & $6.5$ & $5.7$   & $\mathbf{34.7}$ & $55.3$    & $11.7$ & $18.0$   & $\mathbf{24.5}$& $156.4$ \\
2024.05, MPS, 18 MPI per GPU  & $6.1$ & $4.9$   & $21.4$          & $54.5$    & $12.1$ & $17.6$   & $12.1$         & $128.9$ \\
2024.05, MPS, 4  MPI per GPU  & $5.9$ & $5.1$   & $8.0$           & $51.3$    & $11.3$ & $12.6$   & $3.3 $         & $97.8$ \\
\midrule
2023.11, NCCL, 1 MPI per GPU  & $2.8$ & $2.1$   & $\mathbf{294}$  & $49.7$    & $40.4$ & $15.3$   & $\mathbf{265}$ & $669$ \\
2024.05, NCCL, 1 MPI per GPU  & $2.3$ & $1.8$   & $18.8$          & $49.5$    & $40.1$ & $14.6$   & $6.8 $         & $134$ \\
\midrule
2024.05, CPU & $27.3$ & $23.4$   & $72.7$         & $54.4$    & $42.9$ & $141$    & $30.3$         & $393$ \\
\bottomrule
\end{tabularx}
\caption{Comparison of the wall-clock times (in seconds) of the ELPA 2023.11 vs 2024.05 releases in different GPU and CPU setups for the individual steps of the \textit{generalized} eigenproblem.
The matrix size is $40960\times40960$, and the diagonalization is performed to obtain all eigenvalues and eigenvectors on a single node on MPCDF's system \textit{Raven} \cite{raven}.
For the GPU calculations the ELPA1 solver with four NVIDIA A100 (40GB, SXM) GPUs was used. The `MPS' configuration uses NVIDIA's Multi-Process Service \cite{mps} and several MPI processes per GPU. 
For the ELPA 2023.11 release, 18 MPI processes per GPU (full Raven node) show the best performance, while for ELPA 2024.05, the best performance is achieved with 4 MPI processes per GPU.
The `NCCL' configuration uses the NVIDIA Collective Communication Library \cite{nccl}, which is constrained to 1 MPI process per GPU.
For the CPU calculations the ELPA2 solver with 72 MPI processes on two 36-core Intel Xeon Platinum 8360Y processors was used. 
}
\label{table:elpa-generalized}
\end{table}

\gi{Tutorials}

Within the FHI-aims ecosystem, ELPA and ELSI are automatically used in standard tutorials, however, ELSI itself comes with its own extensive manual in its repository. Within the standard build process of FHI-aims, the 2020.05.001 version of ELPA is included as a default build step. A newer version of ELPA can be included first by compiling an external version of the ELPA library and then linking to this version in a subsequent build of FHI-aims; this process is described as a dedicated tutorial accessible at \url{https://fhi-aims-club.gitlab.io/tutorials/tutorials-overview/}.

Since the 2024.05 release, ELPA provides a comprehensive and self-contained \textit{ELPA Manual: User's Guide and Best Practices} \cite{marek:2024}. This manual offers ELPA's quick-start guide and code examples, as well as detailed instructions on installation, how to use ELPA in applications, and troubleshooting.
It is an essential resource for both new and experienced users, ensuring optimal performance and integration of ELPA in applications.
Additionally, we provide a GitHub repository containing teaching materials from recent tutorials on ScaLAPACK and ELPA \cite{karpov:2023}. The repository includes presentation slides and various code examples that demonstrate the practical use of ELPA, covering both the CPU and GPU versions of the library.


\subsection*{Future Plans and Challenges}

Several key optimizations are planned for ELPA to further improve its performance on modern HPC systems.
The primary focus is on enhancing the GPU solvers for the standard eigenproblem, particularly focusing on the ELPA2 tridiagonalization, the solve step of the tridiagonal matrix, and the backtransformation steps, which are not yet utilizing the GPU collective-communication libraries. This should allow the code to keep all the data on the GPU memory, avoiding the costly data transfers between the CPU and GPU. Overall, we expect that with more and more parts of ELPA being ported to GPU, using NCCL/RCCL/oneCCL collective communication libraries allowing direct GPU-GPU communication will become more advantageous; in particular, we expect that ELPA's `NCCL' codebranch will outperform the `MPS' one in the near future. However, the use of NCCL and alike will require a reorganization of the management of the GPU use in FHI-aims: it is necessary to set up the dense matrices directly in the GPU memory corresponding to a given MPI task, instead of first distributing them in parallel across all CPUs.

Additionally, work is underway to optimize ELPA's parallel matrix-matrix multiplication routine, which is used for the transformation of the generalized to the standard eigenproblem. Although the GPU implementation has already reduced the bottleneck as we reported in this contribution, the matrix multiplication remains the most time-consuming part of the process.

Expanding support for Intel GPUs is another major goal. Currently, ELPA's functionality on Intel GPU hardware is limited to solving the standard eigenproblem. We plan to utilize oneCCL and further oneMKL operations to achieve feature parity with the other GPU implementations in ELPA.

Generally, ELPA aims to support and be optimized for all modern and emerging hardware architectures, including Accelerated Processing Units (APUs) \cite{apu} and Vector Processing Units (VPUs) \cite{vpu}. Notably, there is an ongoing effort to port ELPA kernels to the RISC-V architecture \cite{risc-V-elpa}.

ELSI will need similar architecture-specific optimizations. Additionally, support for a broader ecosystem of new and emerging solvers is a constant target, with recent, community-supported additions to ELSI of the Chebyshev eigensolver ChaSE \cite{ChaSE1,ChaSE2} and a new, distributed linear algebra eigensolver DLA-Future \cite{DLA-Future,Solca:2024} require implementation and testing in FHI-aims. For example, demonstration calculations of extremely large DFT calculations were performed on Google's proprietary Tensor Processing Units (TPUs) \cite{Pederson:2022}, based on an adaptation of the FHI-aims/ELSI software stack. This adaptation also made use of an efficient combination of single- and double-precision solutions during the self-consistent field cycle. Similar use of faster single-precision solutions in ELPA is possible using a reduced, frozen-core eigenvalue solution via ELSI and FHI-aims, offering further optimization potential in the future \cite{yu2021}.

\subsection*{Acknowledgements}
We are grateful to Marc Torrent for sharing his preliminary ELPA benchmarks with ABINIT and for fruitful discussions, which have triggered the work on optimizing the generalized eigenproblem solver in ELPA-GPU.
We gratefully acknowledge the support provided by Markus Hrywniak from NVIDIA.
We also thank Markus Rampp for the useful comments on the initial draft of the manuscript.
This project was supported by NOMAD Center of Excellence
(European Union's Horizon 2020 research and innovation program, Grant Agreement No. 951786) and the embedded CSE programme of the ARCHER2 UK National Supercomputing Service (http://www.archer2.ac.uk).
We acknowledge CSC for awarding us access to LUMI supercomputer (Kajaani, Finland).
ELSI was supported by the National Science Foundation (NSF), USA under Award No. 1450280. This research used resources of the Argonne Leadership Computing Facility, a U.S. Department of Energy (DOE) Office of Science user facility at Argonne National Laboratory and is based on research supported by the U.S. DOE Office of Science-Advanced Scientific Computing Research Program, under Contract No. DE-AC02-06CH11357.
We also thank all the colleagues who contributed to ELSI/ELPA in the past and/or in the context of other projects, specifically: Jianfeng Lu, Lin Lin, Fabiano Corsetti, Weile Jia, Raul Laasner, Yingzhou Li, David B. Williams-Young, Xinzhe Wu, Edoardo di Napoli, Rocco Meli, Carolin Penke.
MS acknowledges support by his TEC1p Advanced Grant (the European Research Council (ERC) Horizon 2020 research and innovation programme, grant agreement No. 740233.



  





\newpage

\section{Electronic-Structure Analysis Tools in Numeric Atomic Orbital Framework}
\label{ChapBSDOS}

\sectionauthor[1,2]{Volker Blum}
\sectionauthor[3]{Simon Erker}
\sectionauthor[4]{James A. Green}
\sectionauthor[3,5,a]{Oliver T. Hofmann}
\sectionauthor[3]{Simon Hollweger}
\sectionauthor[7]{Lukas H\"{o}rmann}
\sectionauthor[4,6]{\textbf{ *Sebastian Kokott}}
\sectionauthor[6]{Hagen-Henrik Kowalski}
\sectionauthor[5,8,b]{Sergey V. Levchenko}
\sectionauthor[2]{Chi Liu}
\sectionauthor[9]{Teruyasu Mizoguchi}
\sectionauthor[8]{Mariia Pogodaeva}
\sectionauthor[5]{Norina A. Richter}
\sectionauthor[10]{Mariana Rossi}
\sectionauthor[6]{Matthias Scheffler}
\sectionauthor[9]{Kiyou Shibata}
\sectionauthor[2]{Ruyi Song}
\sectionauthor[5,c]{Aloysius Soon}
\sectionauthor[9]{Izumi Takahara}
\sectionauthor[1,d]{Yi Yao}
\sectionlastauthor[11]{Rundong Zhao}

\sectionaffil[1]{Thomas Lord Department of Mechanical Engineering and Materials Science, Duke University, Durham, NC 27708, USA}
\sectionaffil[2]{Department of Chemistry, Duke University, Durham, NC 27708, USA}
\sectionaffil[3]{Institute of Solid State Physics, Graz University of Technology, Petersgasse 16/II, 8010 Graz, Austria}
\sectionaffil[4]{Molecular Simulations from First Principles e.V., D-14195 Berlin, Germany}
\sectionaffil[5]{Theory Department (since 1/1/2020: The NOMAD Laboratory), Fritz Haber Institute of the Max Planck Society, Faradayweg 4-6, D-14195 Berlin, Germany}
\sectionaffil[6]{The NOMAD Laboratory at the Fritz Haber Institute of the Max Planck Society, Faradayweg 4-6, D-14195 Berlin, Germany}
\sectionaffil[7]{Department of Chemistry and Department of Physics, University of Warwick, Coventry, CV3 5AB, UK}
\sectionaffil[8]{Skolkovo Institute of Science and Technology, Moscow 121205, Russia}
\sectionaffil[9]{Institute of Industrial Science, The University of Tokyo, Tokyo 153-8505, Japan}
\sectionaffil[10]{Max Planck Institute for the Structure and Dynamics of Matter, 22761 Hamburg, Germany}
\sectionaffil[11]{School of Physics, Beihang University, Beijing 102206, China}


\sectionaffil[*]{Coordinator of this contribution.}
\rule[0.25ex]{0.35\linewidth}{0.25pt}

\sectionaffil[a]{{\it Current Address:} Institute of Solid State Physics, Graz University of Technology, Petersgasse 16/II, 8010 Graz, Austria}
\sectionaffil[b]{{\it Current Address:} Skolkovo Institute of Science and Technology, Moscow 121205, Russia}
\sectionaffil[c]{{\it Current Address:} Department of Materials Science \& Engineering, Yonsei University, Seoul 03722, Republic of Korea}
\sectionaffil[d]{{\it Current Address:} Molecular Simulations from First Principles e.V., D-14195 Berlin, Germany}




\subsection*{Summary}
Modern first-principles electronic-structure methods allow us to calculate the \rev{wavefunctions and energies for ground and excited electronic states.} In principle, this provides a complete description of system's physical and chemical properties in a wide range of temperature and pressure conditions. However, the information contained in a many-electron wavefunction, or even in the set of one-electron states and energies obtained from effective-potential theories (such as Kohn-Sham density-functional theory or the Hartree-Fock approximation) is too vast and redundant to be useful for physical understanding without further processing. Therefore, a set of electronic-structure analysis tools have been developed, including density of states (total and projected), band structure, charge and spin partitioning, visualization of electronic density and density differences, isosurfaces of one-electron states and their densities, and electron-localization functions. In this Chapter, these tools and some particularities of their implementation in a numeric atomic orbital (NAO) framework are described.


\subsection*{Current Status of the Implementation}
\subsubsection{Density of States (DOS) and Band Structure (BS)}

Total DOS is calculated from one-electron energies $\epsilon_{n{\bf k}}$ as follows:
\begin{equation}
    g(\epsilon) = \frac{1}{F}\sum_{n{\bf k}} w_{\bf k} \delta(\epsilon-\epsilon_{n{\bf k}}) \, ,
    \label{eq:dos}
\end{equation}
\rev{where the ${\bf k}$-point grid is taken in the first Brillouin zone (BZ), $w_{\bf k}$ are ${\bf k}$-point weights (without any spatial symmetry $w_{\bf k}=1/N$, where $N$ is the total number of ${\bf k}$-points in the grid)}, and \rev{$F$} is a normalization factor. \rev{This factor reflects physical scaling of the number of states per energy unit with the system size.} By default $F$ is set to 1, implying normalization per unit cell. When comparing DOS for different unit cells, however, it should be set to the cell volume or another factor that scales with the cell size. 

Projected DOS (PDOS) in FHI-aims is based on Mulliken partitioning, and is calculated as follows:
\begin{equation}
    g_\nu (\epsilon) = \frac{1}{F}\sum_{\mu}\sum_{n{\bf k}} w_{\bf k} \langle \psi_{n{\bf k}} | \phi_\mu \rangle S_{\mu \nu }^{-1}({\bf k})\langle \phi_\nu|\psi_{n{\bf k}}\rangle \delta(\epsilon-\epsilon_{n{\bf k}}) \, ,
    \label{eq:pdos}
\end{equation}
where $\psi_{n{\bf k}}$ are one-electron wavefunctions, $\phi$ are a set of localized functions, $\nu$, $\mu$ are indices of localized orbitals, and $S_{\mu \nu }({\bf k})$ is the overlap matrix. \rev{By construction, $\sum_\nu g_\nu (\epsilon) = g(\epsilon)$.} In FHI-aims, NAOs are used to calculate the projections $\langle \phi_A|\psi_{n{\bf k}}\rangle$. These orbitals are in general non-orthogonal, and interpretation of PDOS in terms of NAO contributions becomes more ambiguous with increasing basis set size. Therefore, PDOS should be used only as a qualitative analysis tool. \rev{In practice, the equation \ref{eq:pdos} is evaluated via Kohn-Sham coefficients $c_{n{\bf k}}^\alpha$ in wavefunctions $\psi_{n{\bf k}}=\sum_\alpha c_{n{\bf k}}^\alpha \sum_{\bf R} \phi_\alpha ({\bf r}+{\bf R})e^{i{\bf kR}}$ (${\bf R}$ is a lattice vector):
\begin{equation}
g_\nu (\epsilon) = \frac{1}{F}\sum_{\mu}\sum_{n{\bf k}} w_{\bf k} (c_{n{\bf k}}^\nu)^* S_{\nu \mu}({\bf k})c_{n{\bf k}}^\mu \delta(\epsilon-\epsilon_{n{\bf k}}) \, ,
\end{equation}
} One can also use molecular or atomic-cluster orbitals as the projection functions $\phi$, resulting in molecular orbital (MO) projected DOS (MODOS). 

In crystal/MO overlap population (COOP/MOOP) analysis\cite{hughbanks1983}, orbital overlap is quantified to characterize chemical bonds (bonding, antibonding, or non-bonding). This is achieved by omitting the sum over basis functions $\mu$ in equation \ref{eq:pdos} and analyzing the obtained pair-contributions to DOS\cite{takahara2024}. 

By convention, BS (the dependence of $\epsilon_{n{\bf k}}$ on ${\bf k}$) is plotted along straight lines connecting special points in the first BZ. The special points depend on the crystal symmetry, and are tabulated for most common crystal structures\cite{hinuma2017}. Similar to DOS, BS can be resolved in terms of NAO contributions by projecting $\epsilon_{n{\bf k}}$ onto NAOs. The projection is part of a Mulliken charge partitioning\cite{mulliken1955,mulliken1955-2}.

In supercell calculations, for example for modelling defects or thermally perturbed crystal structures, bands with a $\bf k$-vector from the larger first BZ of the primitive cell fold into the smaller BZ of the supercell. A special procedure called band unfolding is required to recover important information about topology of electronic bands. Band unfolding is implemented in FHI-aims and is discussed in detail in Chapter 6.5.

\subsubsection{Charge Partitioning (CP)}

CP allows one to assign a certain charge or spin to a particular orbital and/or atom. This is a useful tool for analyzing charge/spin redistribution in a system compared to, for example, free atoms or another system. However, there is no unique way to perform charge partitioning, since electrons in a many-electron system are often shared by several atoms. \rev{One can distinguish orbital-based (Mulliken\cite{mulliken1955,mulliken1955-2}), density-based (Hirshfeld\cite{hirshfeld1977,bultinck2007} and Bader\cite{bader1990}), and potential-based (by fitting an electrostatic potential, ESP) partitioning schemes. The ESP charges can be quite useful as they correctly reproduce electrostatics of the system. All of these methods except Bader partitioning are implemented in FHI-aims for both periodic and non-periodic systems. Bader partitioning and electron localizability indicator (ELI-D, a flavor of electron localization function, ELF, discussed in the next section) are accessible via the FHI-aims interface with the DGrid program package\cite{kohout2021}.}

Berry-phase polarization\cite{resta1994} allows one to calculate (up to an integer number of quanta) polarization ${\bf P}$ of an infinite periodic crystal with a band gap. Polarization can be used to calculate Born effective charges for any atom $A$ as $Z_{A,ij}=\frac{V}{e}\frac{\partial P_{i}}{\partial R_{A,j}}$, where $V$ is the unit cell volume, $P_i$ is $i$th component of vector ${\bf P}$, and $R_{A,j}$ is the $j$th coordinate of atom $A$. Berry-phase polarization and Born effective charges are useful tools for understanding properties of systems with broken inversion symmetry, such as ferroelectrics, or polarization changes during dynamics. \rev{The details of implementation in FHI-aims can be found in Ref.\cite{carbognob2025}.}

\subsubsection{Electron Localization}

Solution of the electronic problem in a mean-field approximation usually results in delocalized one-electron states. However, observables such as total energy or electron density are invariant with respect to any unitary transformation among states, provided occupied and unoccupied states are not mixed. This freedom is used in FHI-aims to find MOs that are more localized and therefore easier to interpret in the context of chemical bonding. 

An interesting way to analyze localization of electrons is the ELF\cite{becke1990}. \rev{Different from the orbital transformations, which can be cast into transformations of a one-particle density matrix, ELF is a way of understanding a two-particle density matrix.} Topological analysis of the ELF yields regions of local pairing of electrons, where the ELF is maximal. ELF gives additional information to the widely used Bader topological analysis\cite{bader1990}, as demonstrated in \cite{becke1990} by revealing the shell structure of heavy atoms using the ELF. Flavors\cite{savin1996,kohout1996} of the original formulation\cite{becke1990} are also available in FHI-aims. Since the ELF is a function in three dimensions, software that can plot volumetric data is necessary to visualize it. FHI-aims can output a volumetric data file (in CUBE format\cite{cube}) for such a visualization.

\subsubsection{Volumetric Data (electron density, electrostatic potential, etc.)}
\label{sec:vol}

\begin{figure}[hbtp] 
   \centering
   \includegraphics[width=10cm]{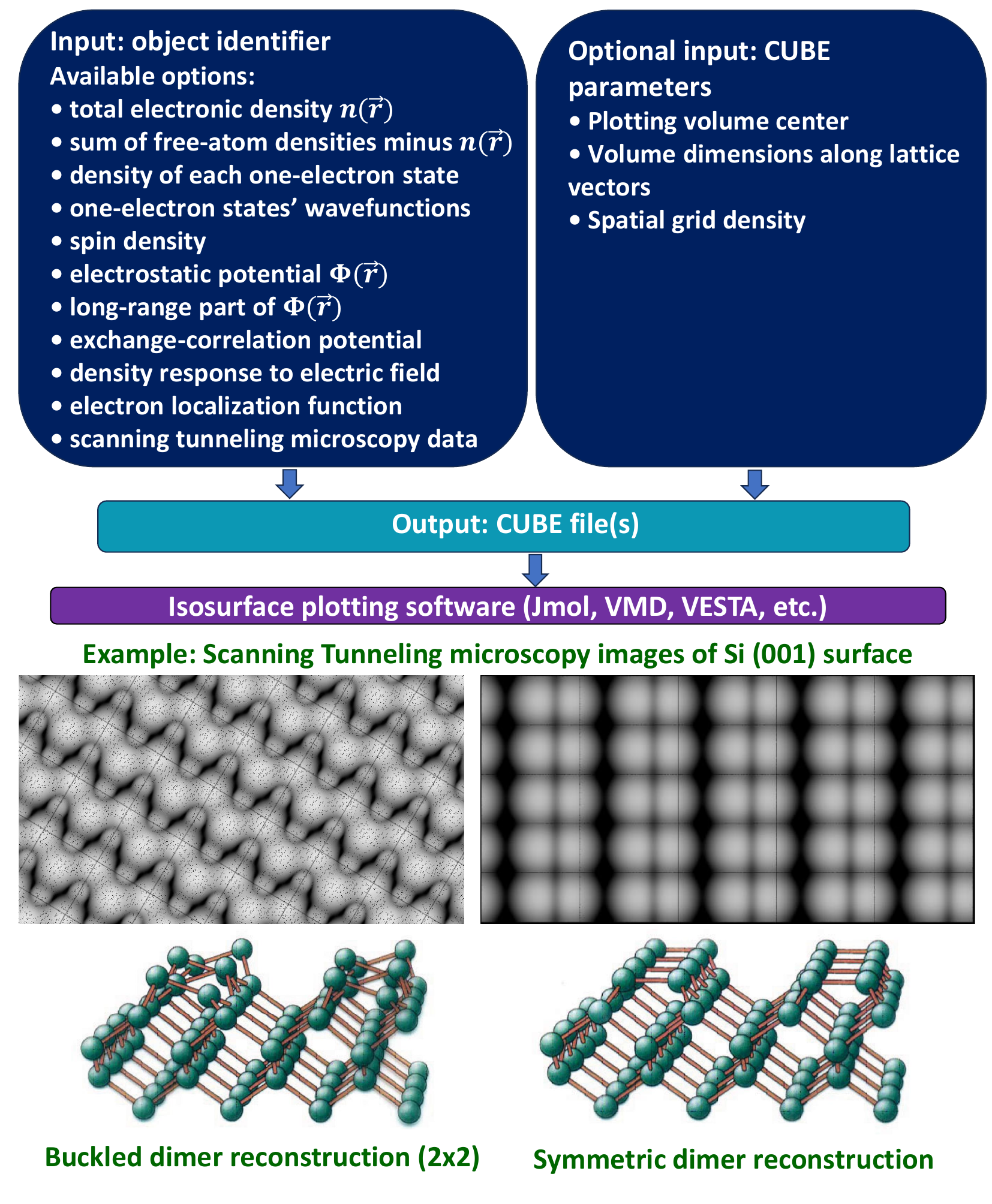}
    \caption{CUBE\cite{cube} output infrastructure in FHI-aims. The density of each one-electron state can be plotted for different ${\bf k}$-points and spin-channels, when relevant. Both real and imaginary parts of electronic wavefunctions can be plotted, if available. Spin density is defined as the difference between spin-majority and spin-minority densities. The long-range part of the electrostatic potential is defined by Ewald partitioning. Density response to an applied electric field needs a perturbation theory calculation. ELF can be calculated in different flavors\cite{becke1990,savin1996,kohout1996}. Scanning tunneling microscopy (STM) data are calculated in the Tersoff-Hamann approximation\cite{tersoff1985}. Simulated STM images of the Si(001) surface with (left) and without (right) buckled dimer reconstruction, calculated with DFT-PBE functional, are shown as an example. The STM images were produced with the VMD software package (see \url{https://fhi-aims-club.gitlab.io/tutorials/stm_visualization/} for a tutorial).}
   \label{fig:cubeout}
\end{figure}

Three-dimensional visualization can provide vast amounts of information for understanding electronic structure. Currently FHI-aims can output data for visualizing objects listed in Fig. \ref{fig:cubeout}. 

\subsection*{Usability and Tutorials}
\begin{figure}[hbtp]
   \centering
   \includegraphics[width=12cm]{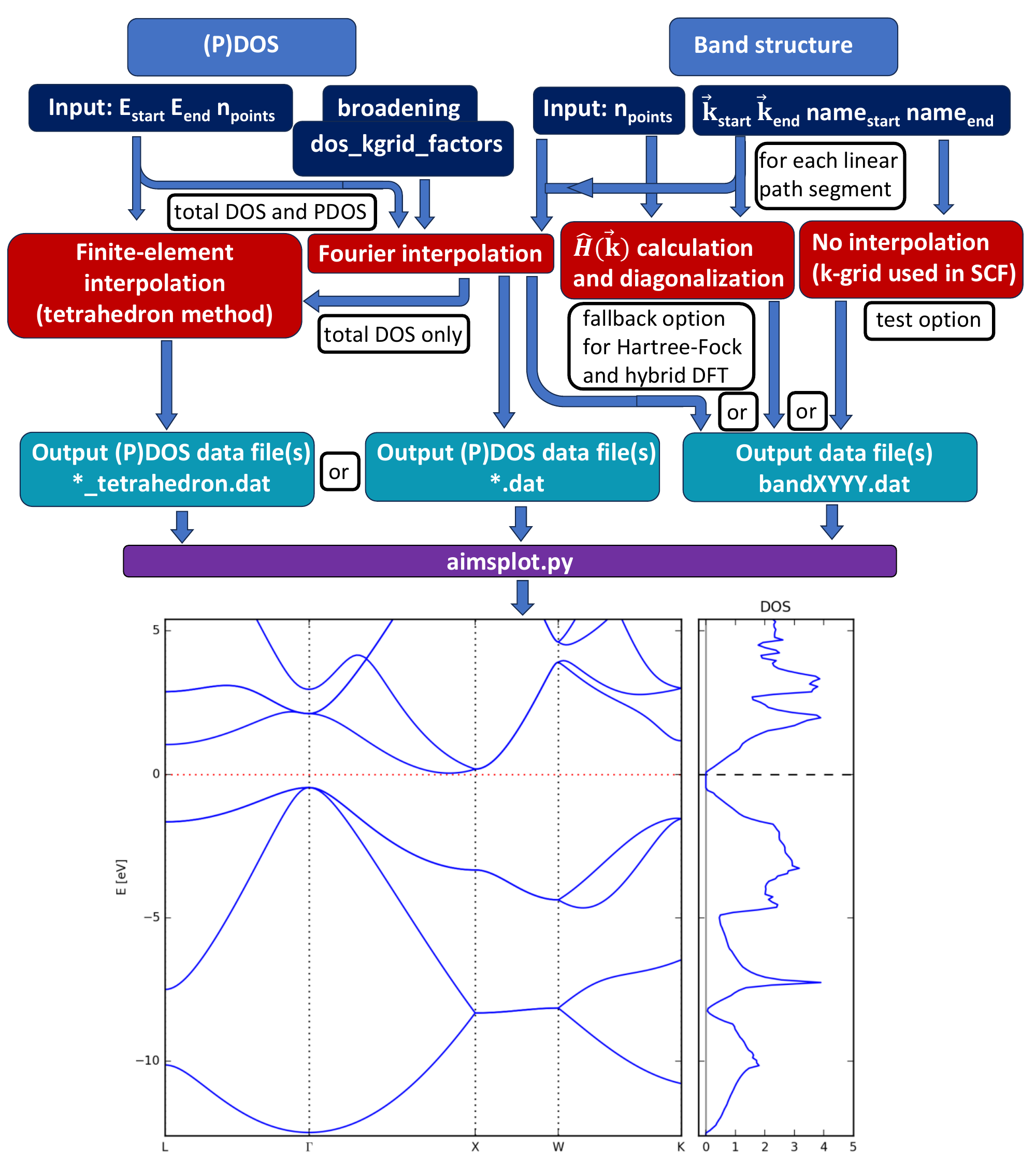}
    \caption{Available functionality for DOS and band structure calculations and plotting. E$_{\rm start}$ and E$_{\rm end}$ are starting and ending points, respectively, on the energy axis for DOS plotting (in eV), n$_{\rm points}$ is the number of points on the energy axis for DOS and along a linear path segment in ${\bf k}$-space for band structure plotting. $\vec{\rm k}_{\rm start}$ and $\vec{\rm k}_{\rm end}$ are coordinates of starting and ending points of a linear path segment in ${\bf k}$-space in the basis of reciprocal lattice vectors. Names of starting and ending points of each path segment are given by name$_{\rm start}$ and name$_{\rm end}$, respectively. An example combined band-structure/DOS plot obtained with python script aimsplot.py is also shown. The script is part of FHI-aims distribution.}
   \label{fig:dosband}
\end{figure}

\begin{figure}[hbtp]
   \centering
   \includegraphics[width=12cm]{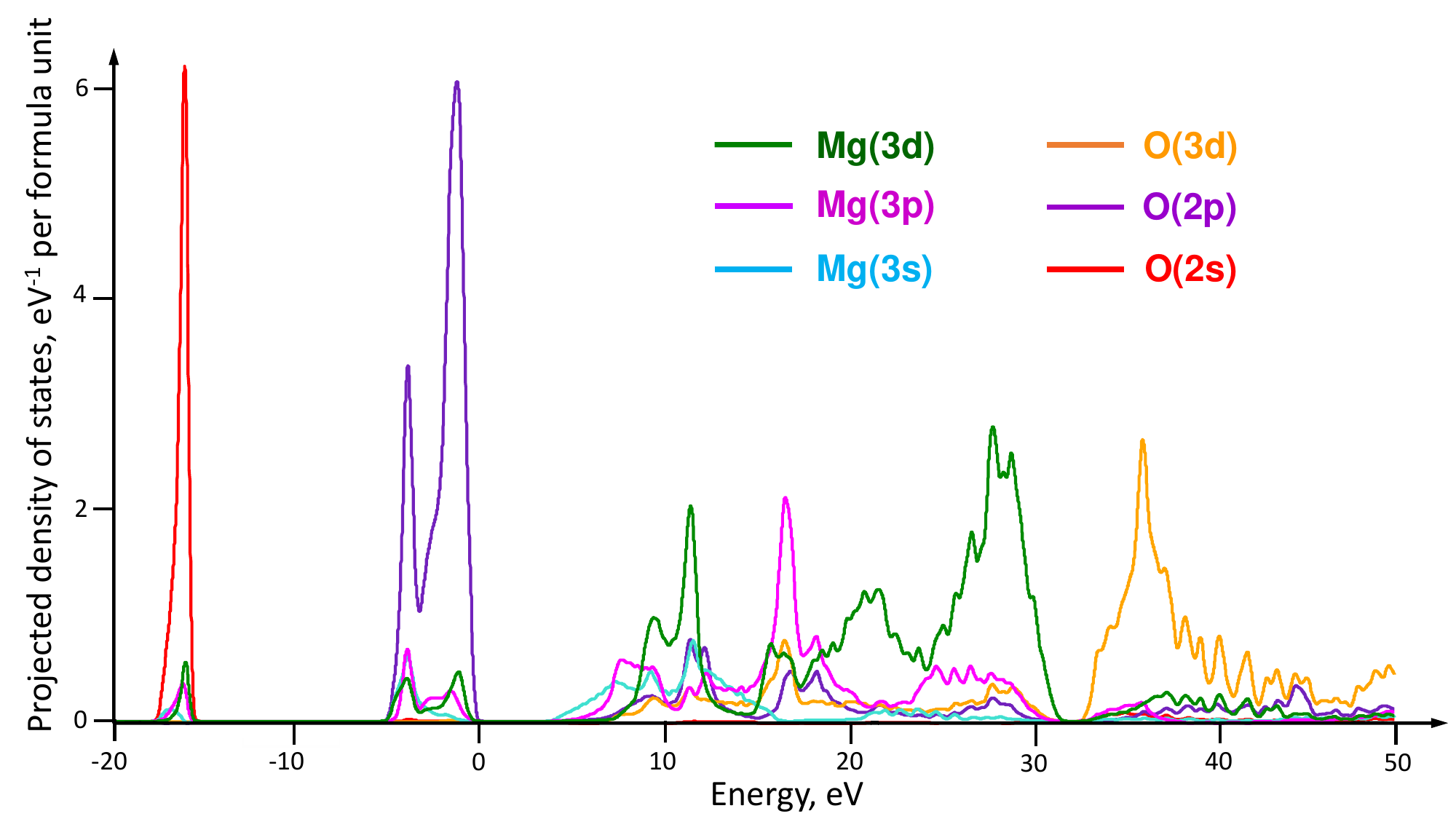}
    \caption{Species PDOS for MgO calculated with DFT-PBE, tight settings. }
   \label{fig:pdos}
\end{figure}

The necessary input parameters and methods for plotting (P)DOS and band structure are shown in Fig. \ref{fig:dosband}. 
In practice, the $\delta$-function in equations \ref{eq:dos} and \ref{eq:pdos} is replaced by a Gaussian function at each ${\bf k}$-point. Each Gaussian is normalized to 1, and its width is a parameter (denoted "broadening" in Fig. \ref{fig:dosband}). To get a converged DOS, one needs to increase the ${\bf k}$-grid density and decrease the width. In order to achieve this in a reasonable computational time, an interpolation can be used to calculate $\epsilon_{n{\bf k}}$ on a ${\bf k}$-grid denser than the one used in the actual SCF calculation. One such interpolation scheme is the tetrahedron method\cite{macdonald1979,zaharioudakis2004}. With this method, the Gaussian width is not requested, since it is calculated automatically. Another interpolation scheme exploits the locality of the NAO basis set. In this basis set, one can calculate the Hamiltonian matrix in real space for a set of lattice vectors ${\bf R}$. Then the Hamiltonian matrix on a dense ${\bf k}$-point grid (defined by the parameter ``\texttt{dos\_kgrid\_factors}`` in Fig. \ref{fig:dosband})) can be calculated by a Fourier transform of the real-space matrix (Fourier interpolation), and then diagonalized to yield new $\epsilon_{n{\bf k}}$ without self-consistency. This approach is sometimes called perturbative. Both the tetrahedron (the recommended way to calculate (P)DOS) and perturbative approaches are implemented in FHI-aims. Moreover, they can be combined to improve total DOS (but currently not PDOS) convergence even further. For PDOS, there are three choices for the sets of $\phi_A$ is FHI-aims: atom, species, and MO. For atom sets, PDOS is output for every atom separately, while for species sets, sum of PDOS for all atoms of the same species are calculated. An example of species PDOS is shown in Fig. \ref{fig:pdos}.

In the case of spin-polarized calculations, DOS for the different spin-channels can be calculated and analyzed separately. Moreover, DOS can be calculated with spin-orbit coupling (SOC) post-SCF correction\cite{huhn2017}. For calculations without periodic boundary conditions, DOS becomes a set of separate states (some of which can be degenerate), since there are no ${\bf k}$-points and no bands. PDOS/MODOS can be used to analyze the orbital character of states in such systems as well.

Just as for DOS, a ${\bf k}$-space interpolation scheme is needed to converge BS. In FHI-aims this can be done with the perturbative approach. Bands can be also calculated with SOC post-SCF correction~\cite{huhn2017}. FHI-aims allows one to characterize bands in terms of spin texture, i.e., expectation values of Pauli matrices for spinors at every $n{\bf k}$\cite{jana2020,song2024} in calculations with SOC. 

There is a difficulty in perturbative interpolation from real to reciprocal space of electronic structure when using non-local potentials, as, for example, is the case for exact exchange in hybrid functionals, or the Hartree-Fock approximation. For a non-local potential the Hamiltonian decays slowly with distance. For a given ${\bf k}$-grid the electronic problem for non-local potential can be solved efficiently by folding in the long-range effects into a ${\bf R}$-dependent matrix for a moderate number of lattice vectors ${\bf R}$ (forming a so-called Born-von Karman supercell). However, this matrix may not be good enough for the Fourier interpolation on a denser ${\bf k}$-grid. This problem can be addressed by either increasing the ${\bf k}$-grid density in the SCF calculation, or by using a much less computationally efficient ${\bf k}$-space implementation of exact exchange. This is relevant for both DOS and BS calculations.

Tutorials are freely available online on how to compute BS and (P)DOS in FHI-aims. In particular, Part 3 and the Appendix of the ``Basics of Running FHI-aims'' (\url{https://fhi-aims-club.gitlab.io/tutorials/basics-of-running-fhi-aims/}) details the keywords required, output produced, and plotting tools available in FHI-aims, and compares the interpolative approaches to computing the total DOS.

For molecular systems, ESP charges are found by minimizing the difference between the actual ESP (from the electronic-structure calculation) and the ESP due to charges, under the condition that the sum of all charges is equal to the total charge of the system. The minimization is performed on a grid of points in space that fit between spheres around each atom, with the radii of these spheres a multiple of van der Walls radii for corresponding atoms. For periodic systems, two different methods are implemented\cite{campana2009,chen2010}. For the method in Ref.\cite{campana2009} one needs to provide electronegativity, self-Coulomb interaction, and weighting factors for each atom to ensure physically reasonable resulting charges. In the second method\cite{chen2010} charges for constraining ESP charges have to be provided directly, for example from Mulliken or Hirshfeld partitioning schemes.

For localization of MOs in non-periodic systems, the unitary optimization of localized MOs, developed by Lehtola and J\'{o}nsson\cite{lehtola2013}, with the Pipek-Mizey localization method\cite{pipek1989} is implemented in FHI-aims. The localization method requires a state-projected population matrix on each atom, and thus relies on a partitioning scheme. Both Mulliken and Hirshfeld partitioning schemes can be used in the localization procedure in FHI-aims. In addition, maximally localized Boys MOs can be calculated, following the procedure of finding the localizing unitary transformation described in Ref.~\cite{guo2011}.

\begin{figure}[hbtp]
   \centering
   \includegraphics[width=12cm]{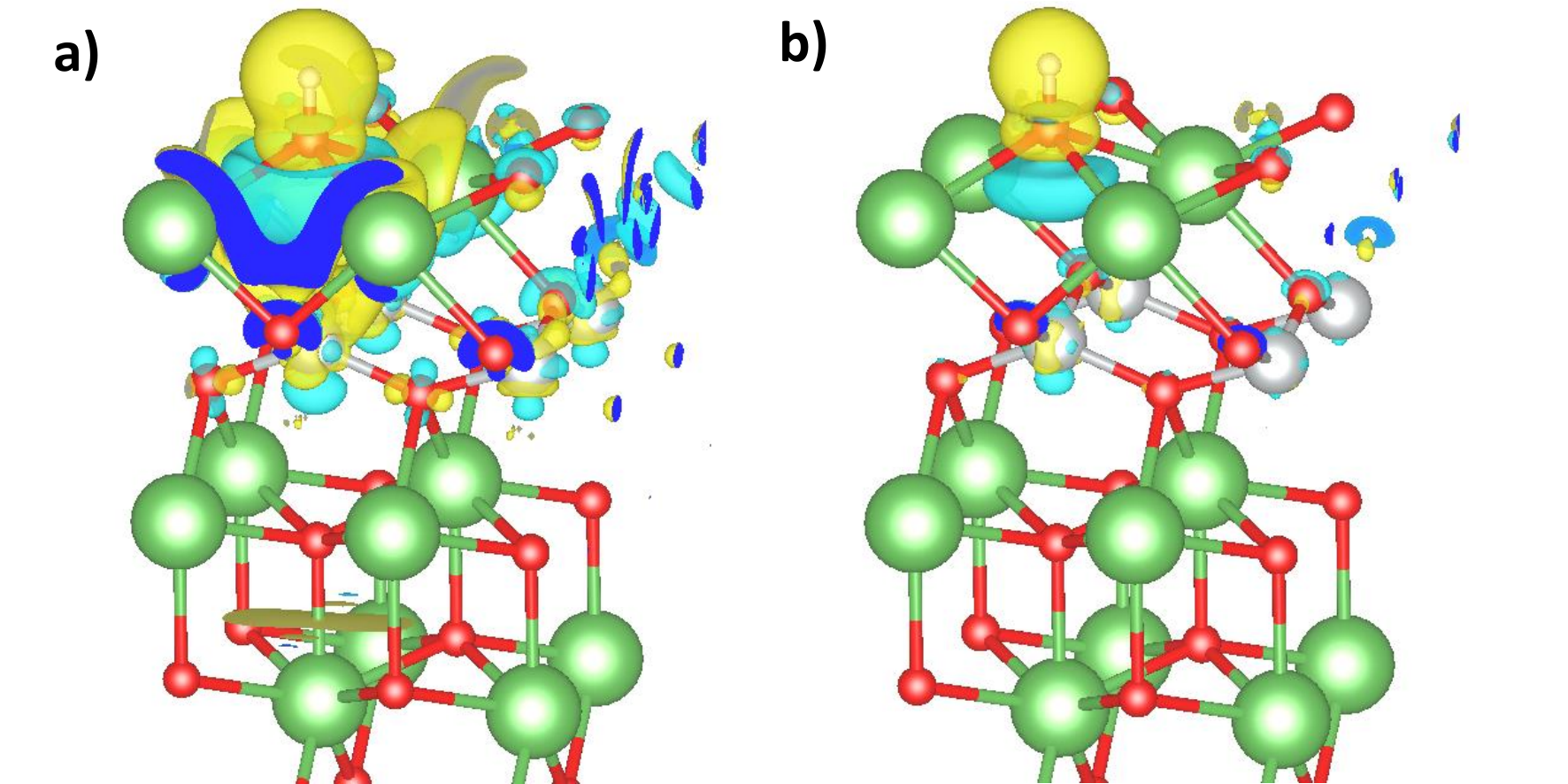}
    \caption{Electron density change upon adsorption of a hydrogen atom on a (001) surface (2$\times$2 surface supercell) of La$_2$NiO$_4$ Ruddlesden-Popper oxide, calculated with the DFT-RPBE functional, tight settings. Oxygen atoms are shown as red spheres, Ni -- grey, La -- green, H -- white. The yellow-colored isosurface shows electron depletion, cyan -- electron accumulation. The isosurface value is 0.01 $|e|$/\AA$^3$ in panel a), and 0.03 $|e|$/\AA$^3$ in panel b). The image was obtained using the VESTA program and a CUBE file, obtained from FHI-aims-generated CUBE files for a surface with adsorbed H, a clean surface, and a free H atom. The positions of atoms in all systems are the same as in the relaxed surface with adsorbed H.}
   \label{fig:densdiff}
\end{figure}

FHI-aims can be requested to output volumetric data for any of the quantities listed in Fig. \ref{fig:cubeout} in the Gaussian CUBE file format \cite{cube}. Several visualization software packages (e.g., Jmol, VMD, VESTA) can plot isosurfaces using the volumetric data from CUBE files. An example of such a plot is shown in Fig. \ref{fig:densdiff}.


\subsection*{Future Plans and Challenges}
As was demonstrated above, the NAO basis and its efficient implementation in FHI-aims provides a powerful platform for electronic-structure analysis. There are many more possibilities for implementation of useful analysis tools in the future. Among ongoing developments are Wannier functions for periodic systems, Bader charge and magnetic moment analysis, and a combined perturbative/tetrahedron interpolation scheme for PDOS. An improved version of the tetrahedron method\cite{blochl1994} is also desired. 

In general, the FHI-aims development team is listening attentively to the requests from community regarding electronic-structure analysis tools. Do not hesitate to contact the team if you need a particular functionality, or if you are willing to implement the functionality in FHI-aims yourself. The only challenges here are the labor and computational costs. The latter is mainly associated with the computational cost of the underlying electronic-structure theory. For example, as discussed above, band-structure calculations can be quite expensive with hybrid functionals. Beyond-DFT approaches, such as $GW$ and other many-body perturbation theory methods, are even more expensive, and the automatic functionality for DOS and band-structure calculations with these methods was not implemented for this reason. Development of robust interpolation schemes, as well as increasing efficiency of the methods themselves, e.g., by using graphical accelerators, will resolve these issues.

\subsection*{Acknowledgements}
We acknowledge contributions from Frank Wagner and Alim Ormeci to the development of the interface between FHI-aims and DGrid. MS acknowledges support by his TEC1p Advanced Grant (the European Research Council (ERC) Horizon 2020 research and innovation programme, grant agreement No. 740233.

\newpage

\section{Explicit and Implicit Embedding Approaches}
\sectionauthor[1]{Daniel Berger}
\sectionauthor[2,3]{Volker Blum} 
\sectionauthor[4]{Gabriel A. Bramley} 
\sectionauthor[4]{Matthew R. Farrow} 
\sectionauthor[5]{Jakob Filser} 
\sectionauthor[5]{Matthias Kick} 
\sectionauthor[6]{\textbf{ *Andrew J. Logsdail}} 
\sectionauthor[7]{Harald Oberhofer} 
\sectionauthor[5]{Karsten Reuter} 
\sectionauthor[8]{Stefan Ringe} 
\sectionauthor[6]{Pavel V. Stishenko} 
\sectionauthor[6]{Oscar van Vuren} 
\sectionauthor[9]{Daniel Waldschmidt} 
\sectionlastauthor[2,a]{Yi Yao} 

\sectionaffil[1]{Chair for Theoretical Chemistry and Catalysis Research Center, Technische Universit\"at M\"unchen, Lichtenbergstr. 4, D-85747 Garching, Germany}
\sectionaffil[2]{Thomas Lord Department of Mechanical Engineering and Materials Science, Duke University, Durham, NC 27708, USA}
\sectionaffil[3]{Department of Chemistry, Duke University, Durham, NC 27708, USA
}
\sectionaffil[4]{Department of Chemistry, Kathleen Lonsdale Materials Chemistry, University College London, London, United Kingdom}
\sectionaffil[5]{Theory Department, Fritz Haber Institute of the Max Planck Society, Faradayweg 4-6, D-14195 Berlin, Germany}
\sectionaffil[6]{Cardiff Catalysis Institute, School of Chemistry, Cardiff University, Park Place, Cardiff CF10 3AT, United Kingdom}
\sectionaffil[7]{Department of Physics and Bavarian Center for Battery Technologies, University of Bayreuth, Bayreuth, Germany}
\sectionaffil[8]{Department of Chemistry, Korea University, Seoul 02841, Republic of Korea}
\sectionaffil[9]{Technische Universität München (TUM), Munich, Germany}

\sectionaffil[*]{Coordinator of this contribution.}
\rule[0.25ex]{0.35\linewidth}{0.25pt}

\sectionaffil[a]{{\it Current Address:} Molecular Simulations from First Principles e.V., D-14195 Berlin, Germany}





\subsection*{Summary}

Applied electronic structure calculations often benefit from approaches that \rev{simplify the system of interest, as approximate models reduce the number of atoms and electrons in a system to make the calculations more tractable; for example, the simulation of molecular adsorption on an extended surface is often performed with fixed-size periodic cells, finite-thickness surface slab models, a vacuum environment, and without thermal effects, which are all simplifications of the laboratory reality}. However, models of real chemical systems can become inaccurate with such an approach, and this motivates efforts towards hierarchical representations where the all-electron site or species of interest is embedded within a coarser representation of the environment. Such multiscale embedding approaches typically ensure that the highest accuracy is maintained on the sites or species of interest, whilst the environment response is sufficiently captured and its energy contributions are included. The most common field of application for multiscale embedding is biological chemistry \cite{warshel1976, senn2009}, though there is demonstrable value for translation to homogeneous and heterogeneous \rev{systems, and application in specific fields such as electrochemistry,} where long-range interactions can play a crucial role \cite{ringe2022, csizi2023, bramley2023}.

\begin{figure}[ht]
    \centering
    \includegraphics[width=0.5\textwidth]{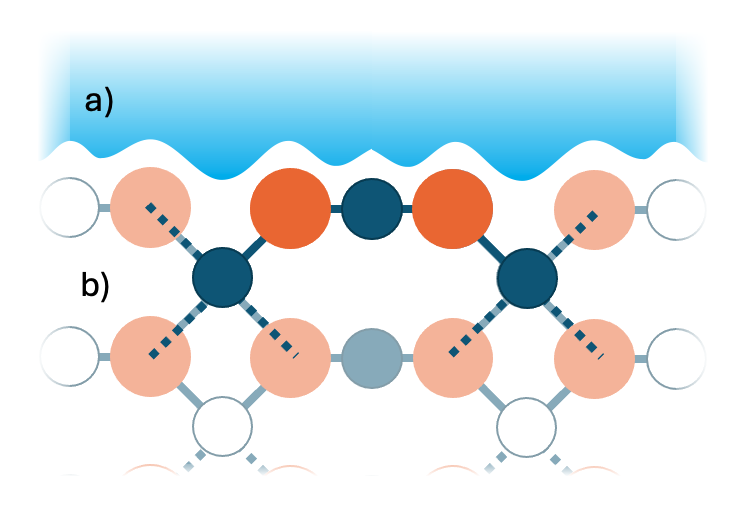}
    \caption{Schematic representations of: a) \textbf{Implicit embedding} environment; and b) \textbf{Explicit embedding} environment. The fully visible species represent quantum mechanical atoms of interest, with blue and orange spheres representing cation and anion species, respectively. For implicit embedding, the blue medium represents the surrounding environment with a dielectric, $\epsilon$; in the explicit embedding model, the partially transparent species represent pseudoised near-neighbours, and white spheres represent the long-range embedding environment that may be included as, \textit{e.g.}, multipolar charges.}
    \label{fig:schematic_of_implicit_and_explicit_embedding}
\end{figure}

Embedding approaches apply either an explicit or implicit effective representation of the environment, as schematically shown in Fig.~\ref{fig:schematic_of_implicit_and_explicit_embedding}. In \textbf{explicit embedding}, basis centres (typically atomic sites) in the embedding environment can be coarsened, by removing electronic degrees of freedom and instead centering an effective embedding potential at the same point in the form of a monopole, dipole, or higher order multipole. Alternatively, hierarchical basis representations and/or approximate treatment of core electrons (such as freezing or pseudoising of electrons) can be applied in the embedding environment, providing greater subtlety in the form of the applied embedding potential \cite{berger2014, yu2021}. The latter approaches are particularly important where directed bonding interactions influence the active site. In contrast, \textbf{implicit embedding} considers the holistic dielectric response to/from an encapsulating medium for the active site or species. Often, the embedding medium is considered to be liquid, and response to any stimulus is represented in the all-electron calculation; directional bonding is therefore not considered, but rather the correct energetics of the system in its surroundings are approximated \cite{ringe2022}.

\subsection*{Current Status of the Implementation}
FHI-aims has infrastructure that supports both implicit and explicit embedding approaches, including connectivity to external packages that can act as drivers for calculation workflows \cite{lu2018}. The FHI-aims functionality allows for the energy 
and forces 
of a system to be calculated self-consistently under the influence of an 
embedding environment.

\textbf{Explicit embedding} is built on the all-electron approach used in a standard FHI-aims calculation, with the effective embedding potential from the surrounding environment captured through surrogate models centred at atomic sites. The embedding environments can take a coarse multipolar form, which is trivially included in the one-electron contributions to the Fock matrix in a manner similar to nuclei; however, this representation typically lacks the subtlety of the density distribution on a real atomic species, and therefore pseudopotential-type embedding is also available. The pseudopotential implementation in FHI-aims takes a Kleinman-Bylander form \cite{berger2014}, which is separable into local and non-local components with additional terms for the non-linear core correction. The pseudopotentials can be particularly valuable for simulations of solid materials with ionic bonding \cite{richter2013, stecher2016, Kick2019}. For effective management of explicit embedding simulations, coupling FHI-aims with external packages that are designed for multiscale simulations is valuable for clear partitioning of the subsystems that require all-electron representation. FHI-aims presently supports an interface to the QM/MM wrapper, ChemShell \cite{lu2018, lu2023multiscale}, which handles the appropriate labelling and geometry specification of pseudopotential centres (when close to the active site of interest), and/or multipolar charges (when at further distance). The infrastructure in FHI-aims is currently suitable for molecular calculations of energies, forces, and other observables.

\textbf{Implicit embedding} takes a contrasting holistic approach by representing the surrounding solvent environment as an effective embedding medium. From a classical electrostatic perspective, the integration of the environment's degrees of freedom is achieved by modifying the Poisson equation to the generalised form:
\begin{equation}\label{eq:GPE}
    \nabla [\epsilon_0 \epsilon(\textbf{r}) \nabla \Phi(\textbf{r})] = - 4 \pi \left(n(\textbf{r}) +n_{\rm ion}[n(\textbf{r})]\right)\quad. 
\end{equation}

The solution of \cref{eq:GPE} provides the electrostatic potential, $\Phi (\textbf{r})$, given the all-electron density representation of the solute, $n (\textbf{r})$, the possible charge density of solvent ions, $n_{\rm ion}$, and the dielectric response of the solvent, $\epsilon (\textbf{r})$, which is commonly functionalised based on the electron density tail of the solute \cite{ringe2022}.


FHI-aims offers three implicit embedding approaches based on the generalised Poisson equation. An additional interface with the Environ package provides further diversity of options for implicit embedding \cite{andreussi2012,filser25}.

The first approach is the \textit{multipole-expansion} (MPE)\cite{filser2021} method, where the dielectric function is represented by a sharp step function of the electron density, leading to a discrete interface between the environment and the embedded system.
The sharp interface in MPE reduces \cref{eq:GPE} to a
problem that can be cast into an overdetermined system of linear equations (SLE), and 
the least squares solution can be solved efficiently using standard algorithms. 
The electrostatic potential in MPE is represented in a finite basis-set expansion, and computational efficiencies can be obtained by separating the regions in the solute 'cavity' space into atom-centered sub-regions to create multiple sub-cavities (MPE-$n$c), limiting the number of basis functions inside each sub-cavity to a size-independent finite number. The outcome is greatly reduced cost when preparing the SLE, as well as opportunity to apply a sparse solver algorithm.

The second implicit embedding approach is the \textit{Stern layer modified Poisson-Boltzmann} (MPB) method, where a smooth dielectric permittivity function of the electron density is applied, which then requires \cref{eq:GPE} to be solved over the whole computational domain. MPB allows for higher flexibility in the model, and also allows the introduction of arbitrary forms for the ionic charge density in the solvent, $n_{\rm ion}$, parameterised as a function of the electron density. The increased complexity of the interface leads inherently to 
higher computational cost, with MPE requiring negligible computation time compared to MPB. 

The third implicit embedding approach is the conductor-like screening model (COSMO).\cite{klamt1993cosmo}. Instead of solving the Poisson equation with a modified $\epsilon (\textbf{r})$, COSMO uses $\epsilon = \infty$ in the medium, independent of its dielectric constant. This choice of boundary conditiion enables the solution of the Poisson equation through a more tractable boundary value problem of electric field inside a conductor. To account for the effect of a finite dielectric constant, an empirical scaling function is used to adjust the surface charge, given by:
\begin{equation}\label{eq:COSMO_scaling}
f(\epsilon)=\frac{\epsilon-1}{\epsilon+k},
\end{equation}
where $k$ is an empirical parameter, typically set to 0.5 or 0. We implemented the smooth version of COSMO by York and Karplus\cite{york1999smooth}, which eliminates discontinuities with respect to the atomic positions present in the original COSMO formulation.

\subsection*{Usability and Tutorials}
\textbf{Explicit embedding} approaches using multipolar charges and pseudoised atoms are realised \textit{via} inclusion in the model geometry file (\texttt{geometry.in}) using the definitions \texttt{multipole} or \texttt{pseudocore}, respectively. For the \texttt{pseudocode} species, a standard species definition is also required with the calculation settings, including integration grids and a minimal basis, and with specific direction to a suitable \texttt{.cpi} format file containing the pseudopotential details \cite{fuchs1999, opium2024}. If the pseudopotentials are being used as a replacement for an all-electron species, the basis functions must be removed for subsequent applied calculations and a representative charge must be set appropriately, as detailed in the software manual. 
FHI-aims also supports the non-linear core correction, which remedies the non-phyiscal linearisation of the exchange-correlation density functional \cite{berger2014}.
Multipole species are simpler to define, requiring definition of location, multipole order, and charge in the system geometry, with no contribution in the \texttt{control.in} file. Forces are available \textit{via} the Hellmann-Feynman formalism, with the \texttt{qmmm} keyword necessary in the \texttt{control.in} file for forces on multipoles. For all-electron species, spurious Coulomb singularities have been observed when multipoles spatially overlap with an integration grid point (\textit{i.e.}, around real atoms); these singularities can be avoided by ensuring reasonable distance separation ($>5~\textrm{\AA}$) between real and embedding species. A full tutorial showcasing how explict embedding calculations are performed with FHI-aims and ChemShell is given in the software tutorials \cite{qmmm2024}.

\textbf{Implicit embedding} is available \textit{via} a range of input keywords that manage the solvent model, and respective subsettings. The \texttt{solvent} keyword selects between the different methods. The options available for MPE include single cavity, piece-wise cavity (MPE-$n$c), and piece-wise solvent representation (MPE-$n$cps), with the latter supporting both molecular and periodic system representations \cite{filser2021,stishenko2024}.
For MPE-$n$c and MPE-$n$cps, default settings are optimised for water as a solvent, but physical model parameters can also be set by the user. These parameters include the solvent dielectric constant, $\epsilon$; the choice of the isosurface that defines the interface between the solvent and solute; and whether non-electrostatic contributions (such as surface tension) are required in the total energy.

For the MPB approach, energies and analytical forces are supported for non-periodic systems \cite{ringe2017phd}. After the appropriate choice for \texttt{solvent} in the calculation input, the dielectric function must be specified. At the current stage, only the dielectric function from the self-consistent continuum solvation (SCCS) method\cite{andreussi2012} has been extensively tested with the MPB package. The SCCS dielectric function has been popular in the community, and parameters for ionic\cite{dupont2013} and neutral molecular solutes in aqueous\cite{andreussi2012}, and non-aqueous\cite{hille2019} solutions (with generalization to any non-aqueous solvent), have been proposed. Implicit embedding methods rely critically on parameterizations, and therefore careful review of these references is encouraged for users. A notable additional feature of the MPB implementation is the possibility for modelling additional continuum charge distributions in the embedded region, which could for example represent salts in electrolytes. This is commonly achieved by modelling the ions as an ideal gas (Poisson-Boltzmann equation) with possibility for additional repulsion between the ions, to account for hydrated ion-ion interactions, and the ion-solute interactions that form a Stern layer \cite{ringe2022,ringe2017phd,ringe2016}. Parameterisations for various dissolved salts are available and have been well-tested for molecular solutes \cite{ringe2017}. Full details of all settings are available in the FHI-aims manual.

For the COSMO approach, both energies and analytical forces are supported for non-periodic systems. To enable a COSMO calculation, one must add the command \texttt{solvent cosmo} into the \texttt{control.in} file. Two additional inputs are required: the dielectric constant of the medium, specified with \texttt{cosmo\_epsilon $<$value$>$}, and a file that represents the grid points on the dielectric surface. Full details on the format of the grid points file can be found in the FHI-aims manual.

The external package Environ offers a wide range of cavity definitions, electrolyte models and additional energy contributions. Its interface with FHI-aims is kept as generic as possible, thereby making new developments in the Environ package immediately available to users of FHI-aims.\cite{andreussi2012,filser25}

\subsection*{Future Plans and Challenges}
The development and deployment of embedding approaches remains an active domain, with increasing need to understand chemical systems at large scale. Explicit atomistic and electronic embedding can therefore benefit from additional options for defining the environment, \textit{via} either density \cite{huang2011} or wavefunction-based embedding \cite{lee2019}, and these are fields of active on-going research. These developments will allow better representation of short-range interatomic interactions, and the coarser point charge and pseudopotential representations can be deployed at a greater distance from the active centre. Care will be required to manage the calculation model and coupling of energy landscapes accurately and coherently, and work is on-going in this domain within FHI-aims and complementary software packages, such as ChemShell. Furthermore, accessibility is as desirable as functionality, with a need for simpler workflows and transferability of approaches across the physical, chemical, biological, and material domains. Complementary to these aspects is the opportunity to integrate machine-learning representations of surrogate landscapes, as discussed elsewhere in this Roadmap. 

Implicit environment embedding is equally important in future workflows, and will benefit from the forthcoming implementation of forces for the MPE model. These forces are in active development for molecular models but further ambitions exist for this functionality in periodic modelling approaches. 
The MPB model is complementary and offers further a high degree of flexibility for future implementations of advanced solvation techniques, such as non-linear or non-local dielectric response \cite{ringe2022}, and similar goals exist for extension to periodic modelling.
The combined outcome will be access to models for solid/liquid interfaces relevant to the most timely challenges in environmentally-relevant chemistry, such as renewable energy and green catalysis.

Work also continues overall to improve the scalability of the implicit embedding approaches, by systematically constructing basis sets of limited size for piecewise solvent sub-regions, and seeking and implementing efficiently scaling sparse solvers such that hierarchical models are not a bottleneck when deployed \cite{stishenko2024}. 
Aspirations to couple implicit and explicit embedding approaches exist, but further development of the underlying framework is required before any attempts to realise this complete embedding model.

\subsection*{Acknowledgements}
We acknowledge fruitful discussions with Oliviero Andreussi, Reinhard J. Maurer, Christoph Scheurer, Alexey Sokol and Scott Woodley. 
AJL, PVS, and GAB acknowledge funding by the UKRI Future Leaders Fellowship program (MR/T018372/1, MR/Y034279/1). 
OvV acknowledges funding of a PhD scholarship by Cardiff University.
AJL, JF, HO, GAB, and PVS acknowledge funding from the ARCHER2 eCSE Programme (eCSE08-13).
HO acknowledges support from the German Science Foundation (DFG) under grant number OB 425/9-1.
SR additionally acknowledges financial support from the National Research Foundation of Korea (NRF), funded by the Ministry of Science and ICT (grant no. 2021R1C1C1008776).

%
%
%


\chapter{Exchange and Correlation}
\label{ChapXC}





\newpage

\section{Exchange and Correlation for Density Functional Theory and the DFT+{\it U} Approach}

\sectionauthor[1,2]{Volker Blum} 
\sectionauthor[3,4]{Fabio Della Sala}
\sectionauthor[3,4]{Eduardo Fabiano}
\sectionauthor[5,6,a]{Jan Hermann} 
\sectionauthor[7]{J. Matthias Kahk} 
\sectionauthor[8]{Matthias Kick} 
\sectionauthor[5,9]{Sebastian Kokott} 
\sectionauthor[10]{Susi Lehtola} 
\sectionauthor[11]{Johannes Lischner} 
\sectionauthor[12]{\textbf{ *Andrew J. Logsdail}} 
\sectionauthor[13]{Harald Oberhofer} 
\sectionauthor[8]{Karsten Reuter} 
\sectionauthor[14]{Mariana Rossi} 
\sectionauthor[5]{\textbf{ *Matthias Scheffler}} 
\sectionlastauthor[6]{Alexandre Tkatchenko} 


\sectionaffil[1]{Thomas Lord Department of Mechanical Engineering and Materials Science, Duke University, Durham, NC 27708, USA}
\sectionaffil[2]{Department of Chemistry, Duke University, Durham, NC 27708, USA}
\sectionaffil[3]{Institute for Microelectronics and Microsystems (CNR-IMM), Via Monteroni, Campus Unisalento, 73100 Lecce, Italy}
\sectionaffil[4]{Center for Biomolecular Nanotechnologies, Istituto Italiano di Tecnologia, Via Barsanti 14, 73010 Arnesano (LE), Italy}
\sectionaffil[5]{The NOMAD Laboratory at the Fritz Haber Institute of the Max Planck Society, Faradayweg 4-6, D-14195 Berlin, Germany}
\sectionaffil[6]{Department of Physics and Materials Science, University of Luxembourg, L-1511 Luxembourg City, Luxembourg}
\sectionaffil[7]{Institute of Physics, University of Tartu, W. Ostwaldi 1, 50411 Tartu, Estonia}
\sectionaffil[8]{Theory Department, Fritz Haber Institute of the Max Planck Society, Faradayweg 4-6, D-14195 Berlin, Germany}
\sectionaffil[9]{Molecular Simulations from First Principles e.V., D-14195 Berlin, Germany}
\sectionaffil[10]{Department of Chemistry, University of Helsinki, P.O. Box 55 (A. I. Virtasen aukio 1),
FI-00014 University of Helsinki, Finland}
\sectionaffil[11]{Department of Physics, Department of Materials, and the Thomas Young Centre for Theory and Simulation of Materials, Imperial College London, London SW7 2AZ, United Kingdom}
\sectionaffil[12]{Cardiff Catalysis Institute, School of Chemistry, Cardiff University, Park Place, Cardiff CF10 3AT, United Kingdom}
\sectionaffil[13]{Department of Physics and Bavarian Center for Battery Technologies, University of Bayreuth, Bayreuth, Germany}
\sectionaffil[14]{Max Planck Institute for the Structure and Dynamics of Matter, 22761 Hamburg, Germany}



\sectionaffil[*]{Coordinator of this contribution.}
\rule[0.25ex]{0.35\linewidth}{0.25pt}

\sectionaffil[a]{{\it Current Address:} Microsoft Research AI for Science, Karl-Liebknecht-Str 32, 10178 Berlin, Germany}



\subsection*{Summary}
In density functional theory (DFT), the ground state total energy is written as a function of the density, \textit{n}, as \cite{hohenberg1964, Kohn1965, Blum2009, teale2022}:
\rev{
\begin{equation}
    \label{eq:Edft}
    E_{\textrm{DFT}} [n] = 
    T_\textrm{s} [n]
    + \int d^{3}\textbf{r}~v_{\textrm{ext}}(\textbf{r}) n(\textbf{r})
    + E_\textrm{H} [n]
    + E_{\textrm{xc}} [n],
\end{equation}
where \textbf{r} is a spatial coordinate.} In Eq.~\eqref{eq:Edft}, the first term (\rev{$T_\textrm{s} [n]$}) is the kinetic energy of non-interacting electrons; the second term is the interaction energy of the electron density with an external potential, $v_{\textrm{ext}}$, generated by, \textit{e.g.}, the nuclei; the third term (\rev{$E_\textrm{H} [n]$}) is the Hartree energy, which describes the electrostatic electron-electron interaction; and the fourth term (\rev{$E_{\textrm{xc}} [n]$}) accounts for the many-electron exchange and correlation, also correcting the spurious self-interaction contained in the Hartree term. 

To calculate 
\rev{$E_{\textrm{DFT}} [n]$}, 
\rev{$T_\textrm{s} [n]$} 
can be evaluated exactly only when expressed in terms of single-particle wave functions. Furthermore, there is no proof that the functional for evaluating 
\rev{$E_{\textrm{xc}} [n]$}
can be written down in terms of a closed mathematical form; indeed, 
\rev{$E_{\textrm{xc}} [n]$} 
is likely just an algorithm that, in its exact implementation, requires the solution of the many-body Schr\"{o}dinger equation \cite{teale2022}. Importantly, Kohn-Sham DFT enables a route towards approximate functionals that can be solved. 


Approximations to $E_{\textrm{xc}}$ 
are central to DFT and they may be complex and intricate. The increasing intricacy is captured in 
the so-called Jacob’s ladder \cite{perdew2001}, which stands on the ground (The Hartree approximation) and reaches towards heaven (The exact solution). The first rung of the ladder is the local-density approximation (LDA) where the density functional approximation (DFA) depends only on the electron density, 
$n$,
and the exact solution is known numerically. The second rung of the ladder refers to the family of generalised gradient approximations (GGAs), which depend on 
$n$ and $\nabla n$.
Then, on the third rung, meta-GGAs are introduced with dependence on the kinetic energy density 
($\tau$)
or higher order derivatives of the density (\textit{e.g.}, 
$\nabla^{2} n$).
On rungs further up the ladder, DFAs also consider occupied Kohn-Sham states (\textit{e.g.} hybrid-DFT) and then even unoccupied Kohn-Sham states. In principle, higher rungs of the ladder, and therefore more complex DFAs, offer greater mathematical flexibility and improved accuracy. For example, a recent development direction is represented by
the Hartree-Fock adiabatic connection (HFAC) methods \cite{seidl18,daas21,constantin25}, which can be considered as
non-linear double-hybrid functionals.
HFAC functionals merge the MP2 term with density functionals from the strong-correlation limit \cite{daas22}, so that
they can be applied also to systems with vanishing gap, and with very accurate results for non-covalent interactions \cite{daas21,constantin25}. \rev{The reader is directed to the wider literature for more complete documentation of DFA formalisms and discussion of applicability for different chemical systems \cite{perdew2001, lehtola2018, koch2015chemist, sholl2022density}.}

Alongside DFA choices, corrective approaches also exist to improve system representation.
DFT+$U$ is particularly successful at addressing the self-interaction error (SIE) for extended transition metal oxides, lanthanides, and actinide compounds with partially filled d- or f-shells, with outcomes similar to hybrid-DFT after careful parameterisation \cite{Agapito2015}. 
The Hubbard-corrected energy, $E_{\textrm{DFT}+U}$, is obtained by applying corrections to states that show a high degree of electron localization, described as correlated subspace, where the SIE can be large \cite{Kick2019}: 
\begin{equation}
    E_{\textrm{DFT}+U} [n] =
    E_{\textrm{DFT}} [n]
    + E_U [\textbf{n}_I]
    - E_\textrm{dc} [\textbf{n}_I].
\end{equation}
Here, $\textbf{n}_I$ refers to the occupation number matrix of a correlated state for atom site $I$, and $E_{\textrm{dc}}$ addresses double-counting of the Coulomb interaction. The DFT+$U$ formalism is simple and can provide improved accuracy with limited additional computational cost on the underlying DFAs.


In this Section of the Roadmap, the 
LDA, GGA, and meta-GGA \textbf{exchange-correlation DFAs} are considered, and we will also consider the \textbf{Hubbard-corrected DFT+$\boldsymbol{U}$} \cite{Anisimov1991,Dudarev1998,Kick2019}. More complex DFAs that depend on the single-particle wave function, and methods for including correlated effects will be discussed \rev{in upcoming Sections. Importantly, these methods frequently build upon the basic DFAs discussed in the present section. Hybrid DFAs, which add a fraction of nonlocal exchange to $E_\textrm{xc}$, will be covered in Section ~\ref{ChapHybrids}. Van der Waals corrections, a correlation effect within $E_\textrm{xc}$, will be addressed in Section ~\ref{ChapDispInteract}. The fifth-rung random phase approximation (RPA), discussed in Section ~\ref{ChapRPA}, also includes van der Waals interactions, along with other aspects of correlation in chemical bonding. Finally,} coupled cluster theory for ground and excited states is addressed in Section ~\ref{ChapCoupledCluster}.

\subsection*{Current Status of the Implementation}
\subsubsection*{Exchange-correlation DFAs}

FHI-aims contains internal implementations of the energy density and its derivatives for various DFAs, which are accessible as separate subroutines. 
Complementary to the bundled implementations, FHI-aims can also interface to two libraries that provide auto-generated code for DFAs, substantially extending the list of available DFAs, and enabling access to high order derivatives. 
An interface to the LibXC library \cite{lehtola2018}, which has become the standard implementation in $>$40 electronic structure packages, provides FHI-aims with $>$600 fully tested DFAs. FHI-aims is currently distributed with LibXC version 6.1.0.
Access to such a standard implementation is beneficial for the reproducibility of computational results \cite{Ghiringhelli2023_SD_626}; note that differences in bundled DFA implementations can lead to different programs giving vastly different results for nominally the same functional \cite{Brakestad2021_JCP_214302, Lehtola2023_JCP_114116}.
Additionally, an interface to the dfauto package \cite{strange2001} provides a numerically robust framework for developing and testing new DFAs, noting that new implementations require developer interaction with the FHI-aims codebase. 
A functional form for dfauto is specified in the Maple language, and the program then automatically derives the functional derivatives and generates the Fortran code.

Both LibXC and dfauto have been extensively tested against the internal implementations of various DFAs, and the libraries have been found to be numerically accurate for the LDA, GGA, and meta-GGA methods outlined.
The bundled implementations, as well as the DFAs from LibXC and dfauto, are also compatible with hybrid-DFT methods. For dispersion-corrected DFT, correlation energy is typically evaluated as an additive term to the total energy, as discussed in Section 3.3 in this Roadmap, although an implementation also exists for the dispersion-corrected DFA of Dion \textit{et al.} that is evaluated during the SCF cycle \cite{dion2004, gulans2009}.

The energy density is evaluated discretely over a radial grid in real space, allowing batch-wise parallelisation on distributed computing architectures. 
The quantities that are necessary for estimating the energy at a given spatial point can include the density ($n$), gradient of the density ($\nabla n$), and the local kinetic energy density ($\tau$), for which the respective DFA families are called the local density approximation (LDA), generalised gradient approximation (GGA), and meta-GGA approaches. 
The implementations calculate and return partial derivatives, allowing for calculation of a potential for the self-consistent field (SCF) cycle. For the GGA, the potential is calculated as \cite{Blum2009}:
\begin{equation}
\label{eqn:vxc_spin_pair}
v_{\textrm{xc}}^{\textrm{GGA}} = \frac{\delta E_{\textrm{xc}} [n,\gamma]}{\delta n} = \frac{\partial f_{\textrm{xc}}}{\partial n} - 2 \nabla \cdot \Bigg[ \frac{\partial f_{\textrm{xc}}}{\partial \gamma} \nabla n \Bigg]
\end{equation}
where $\gamma$ is the scalar product of the spin-density gradients, \textit{i.e.}, $|\nabla n|^{2}$, and $f_{\textrm{xc}}$ is the spatially dependent exchange correlation energy density; \rev{a rearrangement is applied in the implementation of this term, in order to avoid an expensive evaluation of $\nabla^2 n$~\cite{ Blum2009,Pople1992}.}
For the meta-GGA approach, \rev{the required derivative ${\delta E_{\textrm{xc}} [n,\gamma,\tau]}/{\delta n}$ cannot be easily obtained as the derivative requires knowledge of ${\delta \tau}/{\delta n}$ but $\tau$ is not an explicit functional of the electron density, $n$; it is a functional of the occupied electronic states:
\begin{equation}
\label{eqn:kinetic-energy-density}
\tau = \frac{1}{2} \sum_{i} f_{i} | \nabla \psi_{i} | ^{2}
\end{equation}
where $\psi_i$ is the $i$th molecular orbital and $f_i$ is the occupation of this orbital. 
It is possible to calculate ${\delta \tau}/{\delta n}$ \textit{via} the optimised effective potential (OEP) method as implemented by Sharp and Horton \cite{sharp1953variational} or in the Krieger-Li-Iafrate approximation \cite{krieger1990derivation}, but this approach is cumbersome and expensive, making it restrictive for practical applications \cite{arbuznikov2003self}.}
\rev{Therefore,} the potential is calculated by evaluating the functional derivative with respect to $\psi$, as first suggested by Neumann, Nobes and Handy (NNH) \cite{neumann1996}:
\begin{equation}
v_{\textrm{xc}}^{\textrm{mGGA}} = \frac{\delta E_{\textrm{xc}} [n,\gamma,\tau]}{\delta n} \leftarrow \frac{1}{2} \frac{\delta E_{\textrm{xc}} [n,\gamma,\tau]}{\delta \psi}
\end{equation}
and thus we can evaluate the derivatives with respect to the orbitals instead of the density \cite{sun2011}:
\begin{equation}
\label{eqn:v_xc_mgga}
v_{\textrm{xc}}^{\textrm{mGGA}} \psi = v_{\textrm{xc}}^{\textrm{GGA}} \psi - \frac{1}{2} \nabla \cdot \Bigg[ \frac{\partial f_{\textrm{xc}}}{\partial \tau} \nabla \Bigg] \psi.
\end{equation}

\begin{figure}[ht!]
    \centering
    \includegraphics[width=\textwidth]{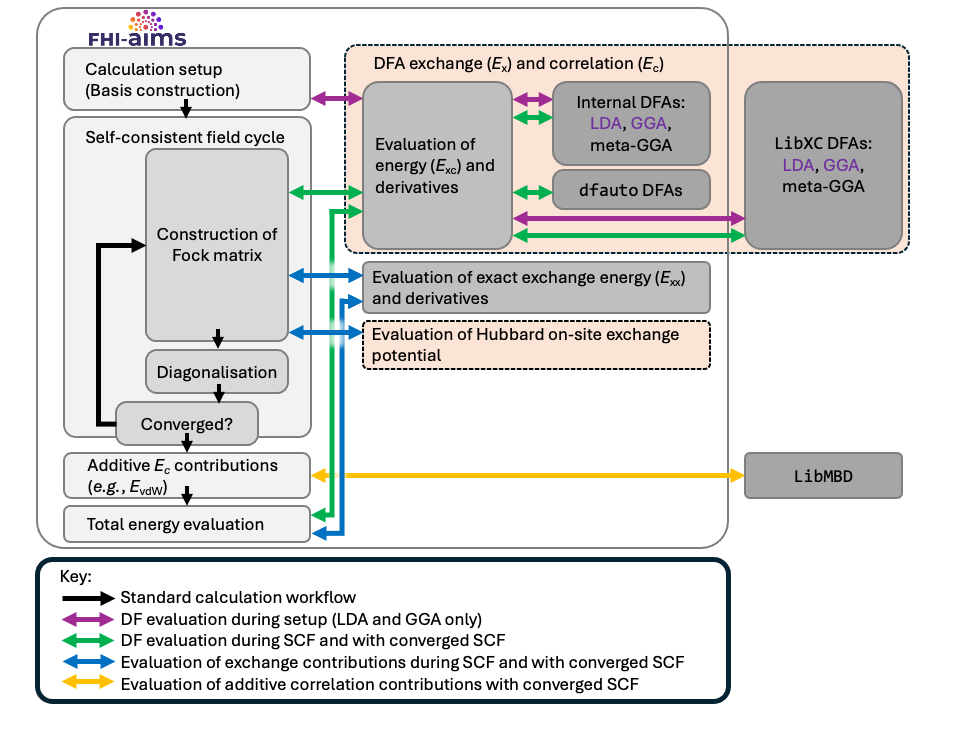}
    \caption{Schematic representation of the FHI-aims infrastructure for evaluation of exchange and correlation contributions. Highlighted in dashed boxes with an orange background are the aspects discussed in this contribution, namely the evaluation of density functional approximations (DFAs) and the evaluation of Hubbard on-site potentials. Additional energy components discussed elsewhere in this Roadmap include exact exchange, perturbative correlation, and additive van der Waals corrections. A key is provided for the \rev{calculation} workflows during calculation setup, during the self-consistent field (SCF) cycle, and during post-convergence energy evaluation.
    }
    \label{fig:exchange_correlation_infrastructure}
\end{figure}

The SCF cycle should be converged to obtain ground-state energies (Fig.~\ref{fig:exchange_correlation_infrastructure}). 
The framework is robust and computationally highly efficient, capable of performing DFT calculations on $>$10,000 cores \cite{bungartz2020}. \rev{Details of the configuration and convergence of the SCF cycle are available in Section 3.10 of the FHI-aims manual \cite{fhi-aims-manual}.}

\subsubsection*{Hubbard-corrected DFT+$\boldsymbol{U}$}

The popular spherically averaged form of DFT+$U$ is available in FHI-aims for addressing SIE in DFAs \cite{Dudarev1998}.
A scalar and site-dependent Hubbard parameter
($U_I$) is applied to determine the interaction strength of electrons localised at atom site $I$. 
The user is required to determine $U$ \textit{via} careful parameterisation to yield an experimentally known observable, such as a band structure \cite{himmetoglu2014}. 
The $U$ parameters depend on the atom sites and their surroundings; the specifics of the projection operator used to determine the occupation of the correlated subspace; and on the employed basis set to form the projection operator.
The projector functions ($\Phi$) for an atomic site $I$ and a magnetic quantum number $m$ are written in FHI-aims as a linear combination of the numerical atomic orbital (NAO) basis functions, $\Phi_{Im} = \sum_i c^i\phi^i_{Im}$, where $c^i$ corresponds to the coefficient of an individual NAO basis function $\phi^i$. 
NAOs are advantageous for this task, because they are the DFT solutions of non-spin-polarised free atoms and thus reflect the character of localised states well, especially when a minimal basis is used.
The NAO formalism \rev{can} result in equivalent charge localization \rev{with $U$ values that are slightly lower, by roughly 1-2 eV, than those reported in other formalisms~\cite{Kick2019}}.

\subsection*{Usability and Tutorials}
\subsubsection*{Exchange-correlation DFAs}
FHI-aims currently provides access to $\sim$40 LDA, GGA, and meta-GGA DFAs, as well as hybrid-DFT, \textit{via} bundled implementations; full details of the available choices are in the software manual \cite{fhi-aims-manual} \rev{, and the reader is directed to wider literature for discussion of most appropriate DFAs for specific chemical systems \cite{koch2015chemist, sholl2022density}.}
The implementations support spin-paired and spin-unpaired calculations. 
Choosing a DFA for a DFT calculation requires a simple input line: \texttt{xc $<$name$>$}, where \texttt{$<$name$>$} is an alphanumeric label as defined in the manual.
For the LibXC and dfauto implementations, an additional keyword on the input line defines that these packages are being used: \texttt{xc libxc $<$name$>$} and \texttt{xc dfauto $<$name$>$}, respectively. 
For the LibXC DFAs, \texttt{$<$name$>$} is as listed in the LibXC documentation \cite{libxc2024}, with capacity to couple distinct exchange and correlation options. 

The flexible framework developed for DFA implementation means that
different DFAs can be used
at different points of the 
calculation workflow. 
Any DFA can be used to perform $x$ initial SCF iterations, for instance, to quickly determine a qualitative electronic structure before switching to a more expensive (or harder to converge) DFA, with the command \texttt{xc$\_$pre $<$$x$$>$ $<$name$>$}. 
A different DFA can then be used to converge the SCF cycle.
Any DFA can also be used for estimating the total energy on top of a converged wave function from another method (\textit{e.g.}, density-corrected methods \cite{song2022}, for example), using the command \texttt{total\_energy\_method $<$name$>$}.
The same command also provides access to contemporary double-hybrids \cite{su2016, wang2021doubly}.

Any LDA and GGA DFAs in LibXC, or a subset of internally implemented LDA and GGA DFAs, can be used for free-atom calculations during the initialisation phase of FHI-aims, which defines the free-atom minimal basis set. 
The default DFA for free atom calculations has been well-tested, but it can also be changed in the input; this impacts absolute energies, but changes in relative energies are typically minimal.

Meta-GGA DFAs are known to require the use of larger radial grids than LDAs or GGAs, and a more thorough radial sampling can be necessary to obtain accurate derivatives \rev{ \cite{wheeler2010integration,yao2017,bartok2019regularized,Lehtola2022_JCP_174114}.}
\rev{The radial grids are a numerically convergent parameter within a DFT software package, and in FHI-aims}
a denser radial grid can be obtained  by changing the overall FHI-aims basis accuracy option (\textit{e.g.}, ``light'' $\rightarrow$ ``intermediate''), or by increasing the values of flags that  control the number of radial grid points, such as the \texttt{radial$\_$multiplier} parameter.

\subsubsection*{Hubbard-corrected DFT+$\boldsymbol{U}$}
The DFT+$U$ functionality is compatible with all 
major 
FHI-aims features and can be easily evoked by defining species-specific $U$ values for the desired angular momentum channel, as per the FHI-aims software manual.
Default settings are optimised for systems with rather localised electrons, and there is high flexibility in the choice of double-counting corrections and the calculation of the occupation matrix \cite{Kick2019}; for the latter, the options are a pure on-site electron counting or a Mulliken-like description.
The features for occupation matrix control \cite{Allen2014} allow fixing of the occupation matrix during an SCF cycle.
The constraint acts like a bias potential, enabling convergence of the electron density towards a desired configuration, such as excess electrons localised on specific sites, a specific magnetic ordering, or a transition state \cite{PhysRevB.79.235125,Allen2014}. 
The subsequent electronic solution can be used to restart an unconstrained self-consistent calculation. 
The occupation matrix control shows great promise for challenging electronic structure situations and hard-to-converge systems.\cite{Kick2019,Kick2020,Kick2020_2,Kick2021}


\subsection*{Future Plans and Challenges}
The DFT framework in FHI-aims is robust and complete, allowing one to evaluate the total energy and perform self-consistent field calculations for a range of LDA, GGA, and meta-GGA approaches to determine the ground state electronic structure.
A thorough and extensive library of contemporary DFAs is obtained by interfacing with external packages, particularly LibXC.
The LibXC interface in FHI-aims currently does not support meta-GGAs that depend on the Laplacian of the density ($\nabla^2 n$); such support will hopefully be added in the future.
We also aim to ensure the completeness of implementations that depend on an external evaluation of correlation (\textit{i.e.}, dispersion-corrected methods). 
These challenges are within FHI-aims, where the specific system observable and/or energy contribution must be calculated, and not in LibXC. 

With respect to future opportunities, work is on-going towards orbital-free approaches for evaluation of the kinetic energy density, including contemporary orbital-free meta-GGAs \cite{mejia2018}, as well as more established approaches that use kinetic energy functionals to address non-additive errors in hierarchical density embedding \cite{wesolowski2015}.
Future developments of the DFT+$U$ method in FHI-aims will focus on implementing the linear response method of computing DFT+$U$ \cite{Cococcioni2005}, the site-to-site coupling term generally denoted as DFT+$U$+$V$ \cite{LeiriaCampo2010}, and further optimisation of the employed projectors.


\subsection*{Acknowledgements}
We acknowledge fruitful discussions with Albert P. Bart\'{o}k and Sheng Bi. 
AJL acknowledges funding by the UKRI Future Leaders Fellowship program (MR/T018372/1, MR/Y034279/1).
SL thanks the Academy of Finland for financial support under project numbers 350282 and 353749.
HO acknowledges support from the German Science Foundation (DFG) under grant numbers OB 425/9-1 and, together with MK, OB 425/3-2.
MS acknowledges support by his TEC1p Advanced Grant (the European Research Council (ERC) Horizon 2020 research and innovation programme, grant agreement No. 740233.

%
%
%






\newpage
\section{Hybrid Density Functionals for Large Scale Simulations}
\label{ChapHybrids}

\sectionauthor[1,2]{Volker Blum}
\sectionauthor[10,3,a]{Christian Carbogno}
\sectionauthor[4]{Lukas Gallandi}
\sectionauthor[5]{James A. Green}
\sectionauthor[6,7]{Stefan Gutzeit}
\sectionauthor[7,8,b]{Felix Hanke}
\sectionauthor[9]{Danish Khan}
\sectionauthor[9]{O. Anatole von Lilienfeld}
\sectionauthor[5,10]{\textbf{ *Sebastian Kokott}}
\sectionauthor[4]{Thomas K\"orzd\"orfer}
\sectionauthor[7,11,c]{Sergey Levchenko}
\sectionauthor[12]{Michael Lorke}
\sectionauthor[13]{Florian Merz}
\sectionauthor[9]{Alastair J. A. Price}
\sectionauthor[14]{Markus Rampp}
\sectionauthor[6]{Ronald Redmer}
\sectionauthor[15,10,d]{Xinguo Ren}
\sectionauthor[16]{Mariana Rossi}
\sectionauthor[10]{Matthias Scheffler}
\sectionauthor[5,10]{Andrei Sobolev}
\sectionauthor[7]{J\"urgen Wieferink}
\sectionlastauthor[5,10,e]{Yi Yao}

\sectionaffil[1]{Thomas Lord Department of Mechanical Engineering and Materials Science, Duke University, Durham, North Carolina 27708, USA}
\sectionaffil[2]{Department of Chemistry, Duke University, Durham, NC 27708, USA}
\sectionaffil[3]{Theory Department, Fritz Haber Institute of the Max Planck Society, Faradayweg 4-6, D-14195 Berlin, Germany}
\sectionaffil[4]{Institut für Chemie, Universit\"at Potsdam, D-14476 Potsdam, Germany}
\sectionaffil[5]{Molecular Simulations from First Principles e.V., D-14195 Berlin, Germany}
\sectionaffil[6]{Institut für Physik, Universit\"at Rostock, D-18051 Rostock, Germany}
\sectionaffil[7]{Theory Department (since 1/1/2020: The NOMAD Laboratory), Fritz Haber Institute of the Max Planck Society, Faradayweg 4-6, D-14195 Berlin, Germany}
\sectionaffil[8]{Surface Science Research Centre and Department of Chemistry, University of Liverpool, Oxford Street, Liverpool L69 3BX, United Kingdom}
\sectionaffil[9]{Chemical Physics Theory Group, Department of Chemistry, University of Toronto, St. George Campus, Toronto, ON, Canada.}
\sectionaffil[10]{The NOMAD Laboratory at the Fritz Haber Institute of the Max Planck Society, Faradayweg 4-6, D-14195 Berlin, Germany}
\sectionaffil[11]{Skolkovo Institute of Science and Technology, Moscow 121205, Russia}
\sectionaffil[12]{Faculty of Physics, University of Duisburg-Essen,D-47057 Duisburg, Germany}
\sectionaffil[13]{Lenovo HPC Innovation Center, Stuttgart, Germany}
\sectionaffil[14]{Max Planck Computing and Data Facility, D-85748 Garching, Germany}
\sectionaffil[15]{Institute of Physics, Chinese Academy of Sciences, Beijing, 100190, China}
\sectionaffil[16]{Max Planck Institute for the Structure and Dynamics of Matter, 22761 Hamburg, Germany}

\sectionaffil[*]{Coordinator of this contribution.}
\rule[0.25ex]{0.35\linewidth}{0.25pt}

\sectionaffil[a]{{\it Current Address:} Theory Department, Fritz Haber Institute of the Max Planck Society, Faradayweg 4-6, D-14195 Berlin, Germany}
\sectionaffil[b]{{\it Current Address:} CuspAI Ltd, 20 Station Road, Cambridge, CB1 2JD, United Kingdom}
\sectionaffil[c]{{\it Current Address:} Skolkovo Institute of Science and Technology, Moscow 121205, Russia}
\sectionaffil[d]{{\it Current Address:} Institute of Physics, Chinese Academy of Sciences, Beijing, 100190, China}
\sectionaffil[e]{{\it Current Address:} Molecular Simulations from First Principles e.V., D-14195 Berlin, Germany}


\subsection*{Summary}

\begin{figure}
    \centering
    \includegraphics[width=\textwidth]{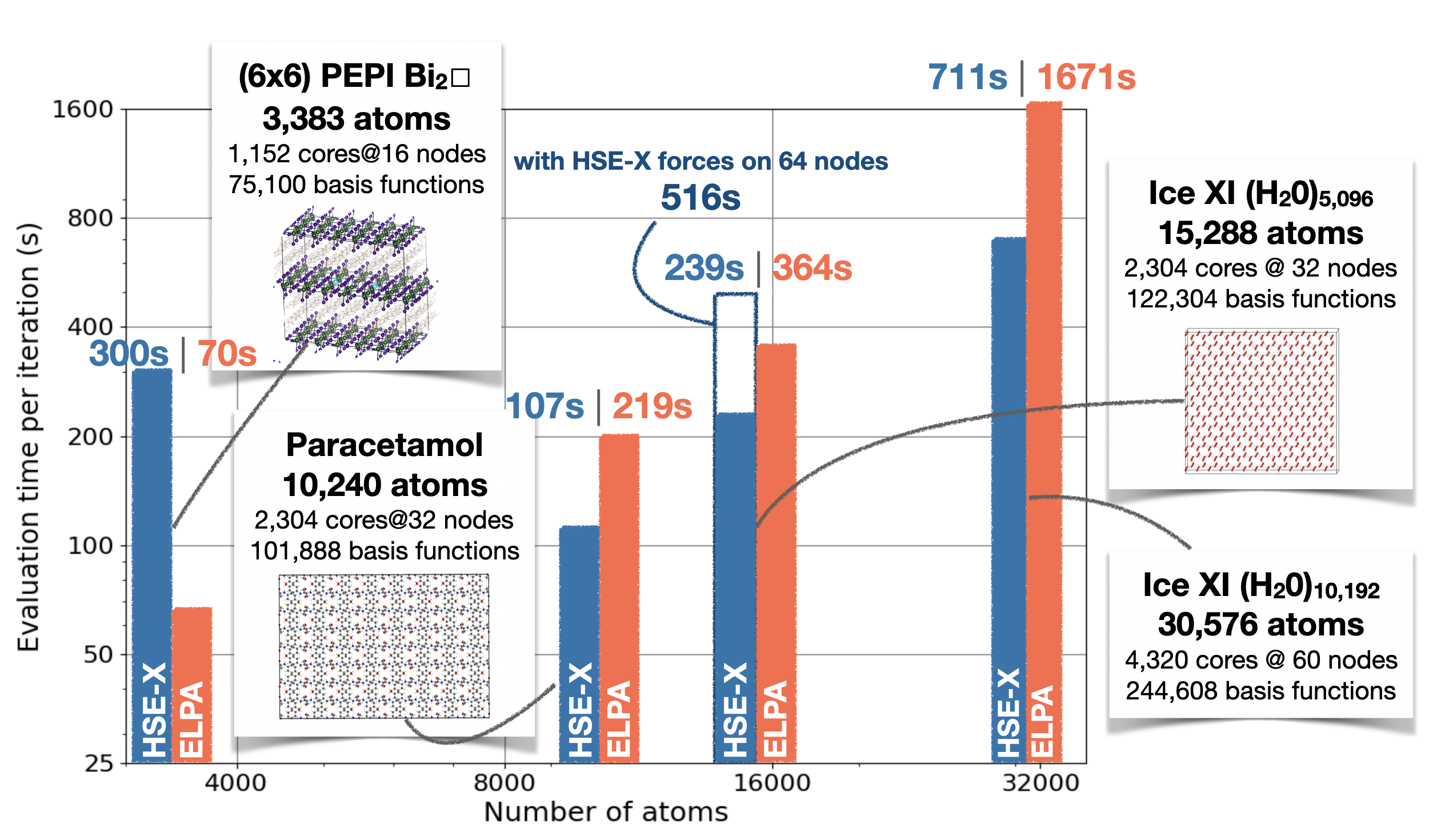}
    \caption{Benchmark results for the largest periodic structures analyzed in Ref.~\cite{kokott2024}. Average runtimes per self-consistent field iteration are shown for the HSE06 exchange operator (blue bars) and the ELPA two-stage eigenvalue solver (red bars). Simulations were performed using the HSE06 hybrid functional for the following systems (from left to right): phenylethylammonium lead iodide (PEPI) with a defect complex (indicated by a square in the chemical formula)~\cite{Lu2023}, a 4$\times$4$\times$4 paracetamol supercell, a 15,288-atom Ice XI supercell (including force evaluation), and a 30,576-atom Ice XI supercell. Calculations were conducted on the Raven HPC cluster at MPCDF, using Intel Xeon IceLake (Platinum 8360Y) nodes with 72 cores per node. Figure adapted from Ref.~\cite{kokott2024}.}
    \label{fig:benchmark-overview}
\end{figure}

Hybrid density functionals (hybrids)\rev{, which incorporate a fraction of nonlocal exchange into $E_\textrm{xc}$ of Eq. (\ref{eq:Edft}),} have emerged as a practical reference method for {\it ab initio} electronic-structure-based simulations. They address several accuracy issues inherent in lower levels of density functional approximations (DFAs) while remaining computationally feasible on current high-performance computers. The main computational bottleneck for atomistic simulations using hybrid density functional theory (DFT) is the evaluation of the non-local exact exchange (EXX). The localized resolution-of-identity-based real-space implementation RI-LVL~\cite{ren2012,ihrig2015} of the exact exchange algorithm in FHI-aims~\cite{levchenko2015} enables the computation of the EXX operator with linear scaling by the number of atoms in the system. The RI-LVL algorithm was recently optimized to allow for much improved exploitation of sparsity and load balancing across ten thousands of parallel computational tasks. All technical details are summarized in Ref.~\cite{kokott2024}. The results demonstrate drastically improved memory and runtime performance, scalability, and workload distribution on CPU clusters. The current reach of the hybrids is documented by run times and scaling of hybrid DFT simulations for several challenging materials, including hybrid organic/inorganic perovskites~\cite{Lu2023} and organic crystals, with up to 30,576 atoms (101,920 electrons described by 244,608 basis functions) in the simulation cell~\cite{kokott2024} (cf. Figure~\ref{fig:benchmark-overview}). Remarkably, despite the scale of these systems, the simulations can be conducted with modest computational resources. Finally, optimizations have been implemented in the computation of band structures and density of states for periodic hybrid functional simulations, alongside the support for new hybrid functionals.

\subsection*{Current Status of the Implementation}

The key concept for the efficient evaluation of the exact exchange is the RI-LVL real space formalism.~\cite{ren2012,ihrig2015} The EXX operator $X_{ij}(\mathbf{R})$ in real space then reads:
\begin{align}
    X_{i j,\sigma}(\mathbf{R})=& \sum_{k \mathbf{R}^{\prime}} \sum_{\mathbf{R}^{\prime \prime}} \sum_{\mu \mathbf{Q}^{\prime}} \sum_{\nu \mathbf{Q}^{\prime \prime}} \tilde{C}_{i k\left(\mathbf{R}^{\prime}\right)}^{\mu(\mathbf{Q}^{\prime})} V_{\mu \nu\left(\mathbf{R}+\mathbf{Q}^{\prime \prime}-\mathbf{Q}^{\prime}\right)} \tilde{C}_{j l\left(\mathbf{R}^{\prime \prime}\right)}^{\nu(\mathbf{Q}^{\prime \prime})} \nonumber \\
    &\times D_{k l,\sigma}\left(\mathbf{R}+\mathbf{R}^{\prime \prime}-\mathbf{R}^{\prime}\right)
    \label{eq:realspace_exchange}
\end{align}
where $\tilde{C}_{i k\left(\mathbf{R}\right)}^{\mu(\mathbf{Q})}$ are the RI expansion coefficients in the flavor of the RI-LVL~\cite{ihrig2015}, $V_{\mu \nu\left(\mathbf{R}\right)}$ the Coulomb matrix (cf. Eq.~\eqref{eq:Coulomb_Matrix}), and
\begin{align}
    D_{k l,\sigma}(\mathbf{R})=\frac{1}{N_\mathbf{k}}\sum_\mathbf{k}\sum_{m} f_{m , \sigma}(\mathbf{k}) c_{k m , \sigma}(\mathbf{k}) c_{l m, \sigma}^{*}(\mathbf{k}) e^{i\mathbf{kR}}
    \label{eq:realspace_dm}
\end{align}
is the Fourier transform of the density matrix, with the occupation numbers $f_{m , \sigma}(\mathbf{k})$, the Kohn-Sham eigenvectors $c_{k m ,\sigma}(\mathbf{k})$ from the solution of KS eigenvalue problem of the previous SCF iteration, and the number of k-points in the Brillouin zone $N_\mathbf{k}$. $\mathbf{R}$ and $\mathbf{Q}$ denote linear combinations of lattice vectors in the Born-von Karman cell; the sum over them is not restricted to the extent of the Born-von Karman cell, but solely by the overlap of the basis functions. The Latin symbols $i$, $j$, $k$, and $l$ denote indices of the numerical atomic orbital (NAO) basis functions, $m$ the eigenstate index,  the Greek symbols $\mu$ and $\nu$ are the indices of the product basis functions, and $\sigma$ is the spin index. All basis functions and auxiliary basis functions are labeled with a real-space lattice vector. Dropping that index refers to basis functions at an atom in the cell $\mathbf{R} = \mathbf{0}$, e.g., $i=i(\mathbf{0})$. \rev{The RI-LVL expansion is constrained so that products of basis functions $\phi_i(r-R)$ on atom $I$ and $\phi_j(r-R^\prime)$ on atom $J$ are expressed using auxiliary basis functions $P_\mu(r-R)$, which must be assigned to the same atom pair $\mathcal{P}(I J)$,
\begin{align}
    \phi_{i}(\mathbf{r} - \mathbf{R}) \phi_{j}(\mathbf{r}- \mathbf{R}^\prime)=
    \sum_{\mu(\mathbf{R})\in \mathcal{P}(I J)} \left( \tilde{C}_{i(\mathbf{R}) j(\mathbf{R}^\prime)}^{\mu(\mathbf{R})} P_{\mu}(\mathbf{r}- \mathbf{R}) + \tilde{C}_{i(\mathbf{R}) j(\mathbf{R}^\prime)}^{\mu(\mathbf{R^\prime})} P_{\mu}(\mathbf{r}- \mathbf{R}^\prime)\right).
\end{align}
All detailed expressions involved in Eq.~\eqref{eq:realspace_exchange} are given in Ref.~\cite{levchenko2015}.}

The optimal distribution and storage of the RI coefficients $\mathbf{\tilde{C}}$, the Coulomb matrix $\mathbf{V}$, and the density matrix $\mathbf{D}$, as well as intermediate matrices of the multiplications in Eq.~\eqref{eq:realspace_exchange} is a hard task and strongly depends on the simulated system. The code tries to utilize the resources in an optimal fashion: Several rows of $X_{ij,\sigma}$ are aggregated into batches (so called Fock matrix blocks). The aggregation of rows of $X_{ij,\sigma}$ into larger blocks enables efficient CPU cache usage. If enough memory per node is available, the code duplicates all the data (that is, $\mathbf{\tilde{C}}$, $\mathbf{V}$, $\mathbf{D}$, index lists storing sparsity patterns, and communication patterns) that is needed to compute a row of the real-space Fock matrix $X_{ij,\sigma}$ (so called instances). This duplication allows a significant reduction of the Message Passing Interface (MPI) communication and enables almost perfect strong scaling for a wide range of core counts.\cite{kokott2024}  Both $\mathbf{\tilde{C}}$ and $\mathbf{V}$ are sparse matrices and only computed and compressed once per self-consistent-field (SCF) cycle. The compression scheme removes columns and rows of the corresponding matrix, wherever the norm of the column and row is below $10^{-10}$. $\mathbf{D}$ is computed every SCF iteration and, formally, is the only input quantity for an EXX SCF evaluation. The real-space EXX operator $X_{ij}(\mathbf{r})$ is linked to the corresponding k-space operator $K_{ij,\sigma} (\mathbf{k})$ via the following Fourier transformation
\begin{align}
    K_{ij,\sigma} (\mathbf{k}) = \sum_\mathbf{R} e^{i\mathbf{k}\cdot\mathbf{R}} X_{ij,\sigma}(\mathbf{R}).
    \label{eq:fock_operator_k}
\end{align}
During a SCF iteration, $K_{ij,\sigma} (\mathbf{k})$ undergoes the same mixing scheme as used for the the density matrix. By default in FHI-aims, this is the Pulay mixing, which ensures a faster convergence of the SCF cycle. Then, the mixed EXX operator $\mathbf{K}_\text{mix}(\mathbf{k)}$ (or a fraction thereof - depending on the used hybrid functional) is added to the Hamiltonian for solving the generalized eigenvalue problem.

The EXX contribution $K_{ij,\sigma} (\mathbf{k})$ can be computed also on generalized regular k-grids~\cite{moreno1992optimal}, which are k-grids constructed from rotated supercells to allow for a more efficient sampling of the Brillouin zone. This helps to reduce the needed compute resources for systems with a skewed cell, e.g. trigonal crystals. The generalized regular k-grid can be either specified by manual user input or automatically generated by autoGR~\cite{morgan2020generalized}. Additionally, $K_{ij,\sigma} (\mathbf{k})$ can be computed at an arbitrary k-point in the Brillouin zone by a Fourier interpolation, e.g., needed for the computation of the band structure or the density of states. The Fourier interpolation requires a sufficiently dense k-grid. \rev{Empirically, our tests show that the Born-von Karman (BvK) cell should be large enough to contain all unique pairs of basis functions which have at least 25\% of overlap measured by the maximal extent of the basis functions.} The code checks at runtime whether this requirement is fulfilled. If the density of the k-grid is too low, the code stops and suggests a minimal needed k-grid.


\subsection*{Usability and Tutorials}

\begin{figure}
    \centering
    \includegraphics[width=\textwidth]{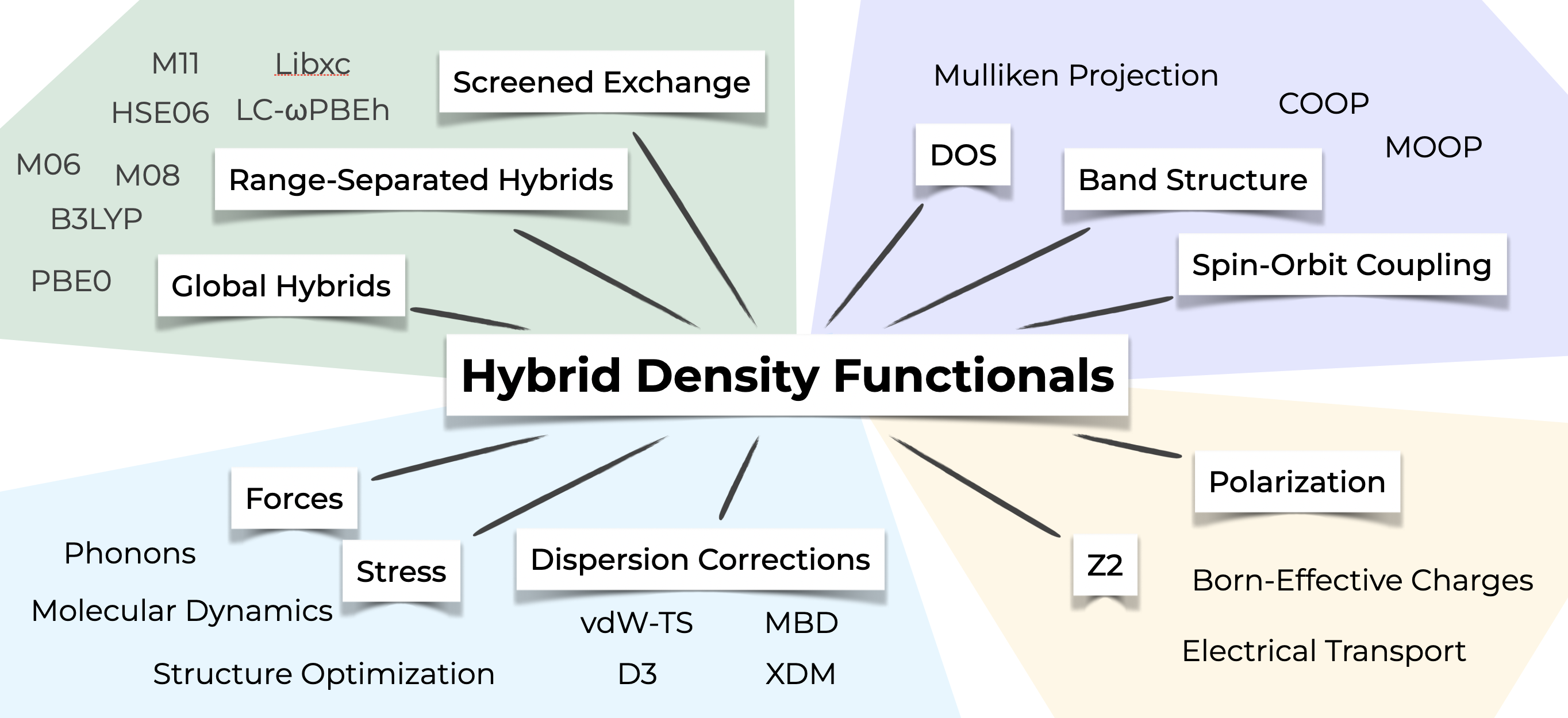}
    \caption{Overview of features implemented in FHI-aims with support for hybrid density functionals.}
    \label{fig:feature-overview}
\end{figure}

The exact exchange code implementation has support for a large number of features in FHI-aims as summarized in Figure~\ref{fig:feature-overview}. It can perform computation of energy and forces for periodic and non-periodic systems, and, in addition, stress~\cite{knuth2015all} for periodic systems. The key concept of any hybrid functional is the mixing of a fraction of exact exchange with fraction(s) of LDA, GGA, or meta-GGA exchange. Often, a range-separation function is introduced to divide the Coulomb potential $v$ into long- and short-range parts. In FHI-aims, the error function (erf) is available as a range-separation function: 
\begin{align}
    v(r) = \underbrace{\frac{1-\text{erf}(\omega r)}{r}}_{v_\text{SR}(r)} + \underbrace{\frac{\text{erf}(\omega r)}{r}}_{v_\text{LR}(r)} \, ,
    \label{eq:coulomb_kernel}
\end{align}
where $r = \vert\mathbf{r}-\mathbf{r}^{\prime}\vert$, $\omega$ (an adjustable inverse length) is the range-separation parameter, and $v_\text{SR}(r)$ and $v_\text{LR}(r)$ are the short- and long-range Coulomb potential, respectively. \rev{As an alternative choice for the range-separation function, the exponential function $\text{exp}(-\omega r)$ was enabled, recently.} In general, the following contributions to the exchange energy $E_x$ are obtained:
\begin{align}
    E_x(\alpha,\beta,\omega) & =  \alpha E_\text{EXX} + \beta E^\text{SR}_\text{EXX}(\omega)  + (1-\alpha) E_\text{x-DFA} - \beta E^\text{SR}_\text{x-DFA}(\omega) \, ,
\end{align}
where $\alpha + \beta \leq 1$ is required. $E_\text{EXX}$ is the EXX energy using the full Coulomb potential and $E^\text{SR}_\text{EXX}(\omega)$ is the short-range EXX energy\rev{, where the short-range Coulomb potential $\frac{1-\text{erf}(\omega r)}{r}$ of Eq.~\eqref{eq:coulomb_kernel} is used. Breaking down the EXX contributions into long- and short-range part, we find that the parameter $\alpha + \beta$ governs the short-range EXX contribution and $\alpha$ the long-range EXX contribution}. Similarly, $E_\text{x-DFA}$ is the semilocal density functional approximation (DFA) exchange energy for the full-range Coulomb operator and $E^\text{SR}_\text{x-DFA}(\omega)$ is the short-range semilocal DFA exchange energy. Choosing a functional via the \texttt{xc} flag in the \texttt{control.in} file will automatically set the default mixing parameters. An exception is the parameter  
$\omega$, which the user needs to specify for range-separated hybrids directly implemented in FHI-aims, since $\omega$ can vary depending on the chosen reference publication. \rev{However, the most common choices for $\omega$ can be used through LibXC~\cite{lehtola2018} by specifying the corresponding LibXC flag for the hybrid functional.} The Hartree-Fock method and different hybrid functionals are implemented in FHI-aims, such as, global hybrids (PBE0~\cite{ernzerhof1999,adamo1999}, PBEsol0~\cite{del2012non}, B3LYP~\cite{stephens1994ab}), range-separated hybrids (HSE06~\cite{heyd2003,heyd2006}, LC-$\omega$PBEh~\cite{vydrov2006assessment}), hybrid meta-generalized gradient functionals (M06 suite~\cite{zhao2008m06}, M08 suite\cite{zhao2008exploring}, M11~\cite{peverati2011improving}), and the Koopmans-com\-pliant screened exchange~\cite{lorke2020koopmans}. Many more hybrid functionals are supported through the XC functional library LibXC~\cite{lehtola2018}. In addition, for materials including heavy elements, perturbative spin-orbit coupling can be combined with the hybrid functionals~\cite{huhn2017}. Moreover, the hybrid functionals can be supplemented by dispersion corrections vdw-TS, D3, MBD, and XDM, and optimized dispersion parameters are provide alongside. 

FHI-aims has been extended to support adaptive hybrid functionals through the methodology of aPBE0~\cite{aPBE0} using on-the-fly machine learning models to estimate EXX admixtures for improved accuracy. A tutorial is provided in Ref.~\cite{tutorialadaptivehybrids} on how to get the optimal fraction of EXX using the Atomic Simulation Environment (ASE). When applicable (i.e. query system is sufficiently similar to training instances), this integration allows for enhanced predictions of energetics, electron densities, and spin-state properties while maintaining computational efficiency. When not applicable, aPBE0 will revert to the default PBE0 admixture ratio of $0.25$.

The current EXX code is production-ready and has been tested on several high-performance compute platforms and CPU architectures, i.e., x86\_64 and arm64 architectures. When a hybrid functional is specified in the \texttt{control.in} file, the code tries to find the best setup for the resources specified by the user with the primary goal to avoid out-of-memory scenarios. However, it is also possible to manually tune the setup via keywords in the \texttt{control.in} file. The manual tuning sometimes enables additional speed-ups of 20-30\%. \rev{The most important parameters used for the memory estimation are printed in the main output file and can be used by the user to find a better memory distribution.} The performance of the code also varies for different hybrid functionals: Hybrid functionals that use a screened Coulomb potential (e.g, HSE06) are significantly faster (especially for dense systems, like bulk materials) than functionals that use a bare Coulomb potential, e.g., runtimes and required CPU resources increase by a factor of six when using the PBE0 functional instead of the HSE06 functional for the hematite crystal. Recently, a screening for the Coulomb potential has been implemented approximately accounting for long-range exact exchange contributions~\cite{kokott2025almostpbe0}. Using this screening function for the PBE0 functional gave results close to the original PBE0 functional, but come with significantly reduced computational cost and memory.

\subsection*{Future Plans and Challenges}

The above-described improvements have become part of the standalone, open-source library libxc-X~\cite{libxcX}. Writing and maintaining efficient HPC software, will be a big challenge in the future. Thus, making the code implementation open source will enable future research and development within the scientific community.

Graphics processing unit (GPU) strategies will be needed for the exact exchange algorithm, but are not yet exploited, as the porting of CPU code to GPU architecture is not at all straightforward. In the CPU implementation, the inherent sparsity of real-space approach keeps the size of matrices used for dense matrix-matrix operations moderate. \rev{The main problem is that RI coefficients and the Coulomb matrix are cast into a compressed format, but have to be expanded again for later matrix multiplications. This scheme only allows to carry out dense matrix algebra for small or medium sized batches.} Thus, with the current algorithm the full capabilities of GPUs cannot be used, and speedups would be limited by communication. An overhaul of the algorithm, and GPU-specific storage and communication patterns will be needed to make it amenable for heterogeneous, GPU-accelerated architectures.

From a technological standpoint, the availability of sufficiently large memory per compute node and task will be crucial for any advanced electronic structure methods, particularly since non-local operators, which often require a careful balance between data locality and communication across nodes and tasks, are typically evaluated. The closer integration of accelerators within HPC nodes, as newly introduced with the Nvidia Grace-Hopper and AMD MI300A processors, shows great promise. However, communication bottlenecks between GPU and CPU tend to vary significantly across different platforms. In general, library APIs for solving mathematical and physical problems are often either vendor-specific or not optimized for performance across various platforms. Addressing these issues by creating vendor-agnostic, high-level APIs that perform efficiently on different architectures would extend the applicability of scientific code, reduce the need for code duplication, and significantly lower research costs. Notably, the ELPA eigensolver has such a vendor-agnostic infrastructure API already implemented. Nevertheless, optimizing communication patterns between CPUs and GPUs for specific architectures remains a complex task. Additionally, new workload distribution models may be required to fully utilize all available resources, such as enabling simultaneous computation on both GPUs and CPUs, as current practices often leave CPUs idle while GPUs perform most of the work.

\subsection*{Acknowledgements}

The authors would like to acknowledge the contribution of Rainer Johanni, who implemented important infrastructure for the hybrid DFT code part in FHI-aims, but passed away much too early. We also acknowledge Arvid Ihrig for numerous contributions to the RI-LVL implementation.

MS acknowledges support by his TEC1p Advanced Grant (the European Research Council (ERC) Horizon 2020 research and innovation programme, grant agreement No. 740233.

\newpage

\section{Density-Functional Based Methods for Dispersion Interactions}
\label{ChapDispInteract}
\sectionauthor[1]{Alberto Ambrosetti}
\sectionauthor[2]{Kyle R. Bryenton}
\sectionauthor[3]{Robert A. DiStasio Jr.} 
\sectionauthor[4]{Simiam Ghan}
\sectionauthor[5]{James A. Green}
\sectionauthor[6,7,a]{Jan Hermann}
\sectionauthor[8]{Johannes Hoja}
\sectionauthor[2,9,10]{\textbf{ *Erin R. Johnson}} 
\sectionauthor[11,12,13]{Reinhard J. Maurer}
\sectionauthor[14]{Alberto Otero-de-la-Roza}
\sectionauthor[15,16]{Alastair J. A. Price}
\sectionauthor[17]{Victor Ruiz}
\sectionauthor[6]{Matthias Scheffler}
\sectionlastauthor[7,18]{\textbf{ *Alexandre Tkatchenko}} 

\sectionaffil[1]{Universit\`{a} degli Studi di Padova}
\sectionaffil[2]{Department of Physics and Atmospheric Science, Dalhousie University, 6310 Coburg Road, Halifax, Nova Scotia, B3H 4R2, Canada}
\sectionaffil[3]{Department of Chemistry and Chemical Biology, Cornell University, Ithaca, New York 14853, USA}
\sectionaffil[4]{Department of Physics, Technical University of Denmark, Fysikvej,
313, 524, 2800 Kgs.\ Lyngby, Denmark}
\sectionaffil[5]{Molecular Simulations from First Principles e.V., D-14195 Berlin, Germany}
\sectionaffil[6]{The NOMAD Laboratory at the Fritz Haber Institute of the Max Planck Society, Faradayweg 4-6, D-14195 Berlin, Germany}
\sectionaffil[7]{Department of Physics and Materials Science, University of Luxembourg, L-1511 Luxembourg City, Luxembourg}
\sectionaffil[8]{Department of Chemistry, University of Graz, Heinrichstra{\ss}e 28, 8010 Graz, Austria}
\sectionaffil[9]{Department of Chemistry, Dalhousie University, 6243 Alumni Crescent, Halifax, Nova Scotia, B3H 4R2, Canada}
\sectionaffil[10]{Yusuf Hamied Department of Chemistry, University of Cambridge, Lensfield Road, Cambridge, CB2 1EW, United Kingdom}
\sectionaffil[11]{Department of Chemistry, University of Warwick, Gibbet Hill Road, CV4 7AL Coventry, United Kingdom}
\sectionaffil[12]{Department of Physics, University of Warwick, Gibbet Hill Road, CV4 7AL Coventry, United Kingdom}
\sectionaffil[13]{Faculty of Physics, University of Vienna, Vienna A-1090, Austria}
\sectionaffil[14]{Departamento de Qu\'{i}mica F\'{i}sica y Anal\'{i}tica and MALTA-Consolider Team, Facultad de Qu\'{i}mica, Universidad de Oviedo, 33006 Oviedo, Spain}
\sectionaffil[15]{Department of Chemistry, University of Toronto, St.\ George Campus, Toronto, ON, Canada}
\sectionaffil[16]{Acceleration Consortium, University of Toronto. 80 St.\ George Street, Toronto, ON M5S 3H6}
\sectionaffil[17]{Theory Department (since 1/1/2020: The NOMAD Laboratory), Fritz Haber Institute of the Max Planck Society, Faradayweg 4-6, D-14195 Berlin, Germany}
\sectionaffil[18]{Institute for Advanced Studies, University of Luxembourg, L-1511 Luxembourg, Luxembourg}

\sectionaffil[*]{Coordinator of this contribution.}
\rule[0.25ex]{0.35\linewidth}{0.25pt}

\sectionaffil[a]{{\it Current Address:} Microsoft Research AI for Science, Karl-Liebknecht-Str 32, 10178 Berlin, Germany}




\subsection*{Summary}
Density-functional theory (DFT) is widely recognized as the standard computational technique for modeling the electronic structure of molecules and materials with up to 10,000 atoms. Despite its proven capabilities, achieving reliability and high accuracy in DFT simulations of intermolecular interactions necessitates the inclusion of London dispersion or van der Waals (vdW) forces. \rev{While they are formally a part of $E_\textrm{xc}$ in Eq. (\ref{eq:Edft}),} such forces result from highly non-local electron correlations that are not captured by local, semi-local, or hybrid exchange-correlation functionals. Numerous post-SCF (self-consistent field) and full SCF vdW/dispersion methods have been developed to pair with popular DFT functionals, demonstrating excellent performance at low computational cost. Here, we describe the vdW/dispersion methods currently implemented in the Fritz Haber Institute \textit{ab initio} materials simulation (FHI-aims) program \cite{Blum2009}.

\subsection*{Current Status of the Implementation}
Currently, there are several post-SCF and full SCF vdW/dispersion methods available in FHI-aims from which to choose. These include the Tkatchenko-Scheffler (TS) model \cite{ts}; two variants of the many-body dispersion (MBD) method—the range-separated, self-consistently screened version (MBD@rsSCS) \cite{Tkatchenko2012,rsscs} and the nonlocal version (MBD-NL) \cite{mbdnl}; and the exchange-hole dipole moment (XDM) method \cite{c6810posthf,xdmchapter,xdmaims}. TS, MBD@rsSCS, and MBD-NL are implemented within the libmbd library, while XDM is implemented in the \texttt{xdm.f90} routine. With each of these methods, the dispersion energy is computed using the self-consistent (SC) electron density from a DFT calculation and, in some cases, other density-dependent properties. It is then added to the self-consistent energy to give the total DFT energy:
\begin{equation}
E_\mathrm{DFT} = E_\mathrm{SC} + E_\mathrm{disp}\,.
\end{equation}
As a result, all of the vdW/dispersion methods listed above have 1-2 empirical parameters within their formulation that depend on the choice of density functional. Similarly, the forces and stresses are derived from the dispersion energy and can be added to the DFT forces and stresses. 

Both TS and MBD@rsSCS are also implemented with full self consistency~\cite{libmbd,Ferri-PRL,Ferri-PRM}, meaning that the vdW/dispersion interactions not only affect the energy (and forces/stresses), but also the electron density, orbitals, and all the properties of interest derived from the orbitals and electron density. These effects can be important for large and/or polarizable systems~\cite{Ferri-PRL,Ferri-PRM}.

Because the evaluation of the pairwise dispersion energy depends on a summation over all atomic pairs, it formally scales as $\mathcal{O}(N^2)$, where $N$ is the number of atoms. On the other hand, the MBD Hamiltonian requires diagonalization of the oscillator coupling matrix, hence MBD/MBD-NL methods scale as $\mathcal{O}(N^3)$. Typically, both pairwise (TS/XDM) and MBD dispersion methods do not increase the cost of the underlying semilocal or hybrid DFT calculation. This is because the pairwise energy expression can be evaluated in a linearly scaling fashion by employing suitable and controlled cutoffs. The MBD expression is efficiently implemented with Ewald summation and employs modern diagonalization routines, whereas the size of the MBD matrix is small compared to the Kohn-Sham matrix. \rev{Hence, the cost of non-self-consistent dispersion calculations is negligible compared to the underlying Kohn-Sham calculation.} In contrast, the self-consistent TS/MBD calculations can increase the computational cost by a factor of 2-3 or higher depending on the system size. This is because the Hirshfeld partitioning and the calculation of the dispersion potential is done at every SCF step of the DFT calculation. \rev{This computational overhead is essentially independent of the specific method used for dispersion interactions.} 

In addition to the four dispersion methods discussed above, the non-local van der Waals density functional of Dion and coworkers \cite{dion2004,gulans2009} is also present within FHI-aims, although it has not been extensively tested and it is still considered experimental. The D3 dispersion correction by Grimme et. al.~\cite{Grimme2010,Grimme2011} is also implemented in FHI-aims via the s-dftd3 library~\cite{sdftd3,Ehlert2024_s-dftd3}, and is capable of computing energies, forces and stresses.

\subsection*{Usability and Tutorials}

The TS, MBD@rsSCS, MBD-NL, and XDM methods can easily be used in FHI-aims by specifying their keywords in the \texttt{control.in} file. These are:

\begin{center}
\begin{tabular}{ll}
TS & \verb+vdw_ts+ \\
MBD@rsSCS & \verb+many_body_dispersion+  \\
MBD-NL & \verb+many_body_dispersion_nl+ \\
XDM & \verb+xdm [basis] or xdm [a1 a2]+ \\
\end{tabular}
\end{center}


Note that an older implementation of the TS method (not included with libmbd) can also be called via the \texttt{vdw\_correction\_hirshfeld} keyword.
While additional options can be called with each keyword, 
the default settings are recommended for most applications. The options for the TS, MBD@rsSCS, and MBD-NL methods, specified on the same line as the main keyword, are:
\begin{center}
\begin{tabular}{ll}
\verb+self_consistent+ & Add MBD or TS contribution to the XC potential \\
\verb+vdw_params_kind+ & \verb+"ts"+ for standard or \verb+"tssurf"+ for surface-adjusted vdW parameters \\
\verb+beta+ & Damping parameter $\beta$ for MBD  \\
\verb+sr+ & Damping parameter $s_R$ for TS \\
\verb+k_grid+ & Integration $k$-grid for the MBD energy of periodic system \\
\verb+zero_negative+ & Zero out negative MBD eigenvalues \\
\verb+do_rpa+ & Calculate MBD energy via the imaginary-frequency integral \\
\verb+rpa_rescale_eigs+ & Rescale negative eigenvalues in the imaginary-frequency integral \\
\end{tabular}
\end{center}

\rev{We suggest employing "tssurf" parameter set only when calculating adsorption processes on surfaces and using TS or MBD@rsSCS methods. The damping parameter for TS, MBD@rsSCS, and MBD-NL methods depend only weakly on the basis set in FHI-aims from "light" to "really tight" settings. The fitting of the damping parameter was done using the "tight" settings.}

For XDM, \rev{the damping parameters depend on the basis set. Hence,} additional options \rev{must} be specified on the same line as the main keyword to inform the code what Becke-Johnson damping parameters to use. As of FHI-aims release 240920, two keyword options are available for this purpose:
\begin{center}
\begin{tabular}{ll}
\verb+xdm [basis]+ & If using a default basis from \texttt{species\_defaults}, specify its name here\\
\hspace{3em} or \\
\verb+xdm [a1 a2]+ & Set the damping parameters, $a_1$ (unitless) and $a_2$ (in \AA), manually \\
\end{tabular}
\end{center}
The exchange-correlation functional is detected from the \texttt{control.in} and, when the \texttt{[basis]} optional parameter is specified, the XDM routine automatically sets optimal damping parameters for that combination. Most common basis sets are supported by this automatic parameter setting, including \texttt{light}, \texttt{intermediate}, \texttt{tight}, and \texttt{lightdenser}\rev{; the calculation will terminate with an error if the basis-set name is not recognised}. The FHI-aims manual provides a comprehensive list of basis defaults and functionals for which XDM damping parameters are available. If your functional/basis combination is not natively supported by XDM, you may manually enter \texttt{[a1 a2]} as two floating-point numbers ($a_2$ in \AA ngstrom units). Detailed instructions on fitting XDM's damping parameters may be found in the XDM tutorial~\cite{XDMTutorial}.

Dispersion/vdW methods are not recommended for use with the local density approximation (LDA) due to its well-established systematic overbinding tendency \cite{bjorkman}. However, damping parameters for TS, MBD@rsSCS, MBD-NL, and XDM are available for use with the popular GGA and hybrid functionals, PBE, HSE06, and PBE0. Specifically for XDM, slightly better performance can be obtained using B86b rather than PBE exchange \cite{requirements}, as in the B86bPBE GGA and the B86bPBE0 hybrid functional \cite{xdmaims}, which is an analogue of PBE0.

The D3 dispersion correction can be activated via the \texttt{d3} keyword in the \texttt{control.in} file. Both the zero ~\cite{Grimme2010} and Becke-Johnson ~\cite{Grimme2011} damping options are possible, and may be selected via the \texttt{damping} option on the same line as the \texttt{d3} keyword. After detection of the exchange-correlation functional in \texttt{control.in}, automatic selection of the D3 parameters is supported for over 100 functionals, comprising both internally implemented functionals, and those from LibXC. Alternatively, user specific parameters may be entered on the same line as the \texttt{d3} keyword, with all options documented in the FHI-aims manual.



\section*{Future Plans and Challenges}
An extension of MBD (MBD+C) is currently being developed~\cite{mbd+c} to explicitly account for the dynamical response of metallic electrons. With respect to MBD@rsSCS and MBD-NL, that rely on harmonically bound charges, MBD+C additionally includes interatomic electron-hopping terms, in analogy to tight-binding models.  MBD+C aims to provide a unified many-body description of bound and delocalized  electrons, which is necessary in order to capture correct power-law scalings and to automatically include plasmon contributions and metallic screening.

For XDM, a key challenge is the inclusion of additional force and stress terms that are currently neglected. The present implementation assumes that, when calculating the atomic forces or stress tensor, the dispersion coefficients do not change with atomic positions or unit-cell lengths. While this is a fair assumption for isolated molecules and molecular crystals, it is poor\rev{er} for harder solids\rev{,} where geometry optimizations can result in overly compact unit-cell volumes \rev{(by $\le 1$\%) and} higher energies \rev{(by $\le 0.04$ eV)} than the true potential-energy minimum \cite{requirements}. While \rev{the additional stress and force} terms are almost always neglected in all dispersion corrections, their omission disproportionately affects XDM due to its highly varying dispersion coefficients near equilibrium geometries \cite{manyworries,oscillators}. Accounting for these missing force and stress terms is an area of active work, and ultimately depends on implementing the derivatives of the Hirshfeld weights that partition the total electron density into atomic pieces with respect to atomic coordinates.

Additional work seeks to improve the atomic polarizabilities used with XDM by replacing the simple scaling of free-atomic values using ratios of Hirshfeld atomic volumes. Options to address this include implementing an atomic polarizabilty functional such as that used by MBD-NL \cite{mbdnl}, or replacing the current simple Hirshfeld partitioning with a more sophistical iterative Hirshfeld \cite{hi} or fractionally ionic \cite{fi} scheme. While iterative Hirshfeld partitioning is already implemented in FHI-aims, it currently only works reliably for the aperiodic case. Lastly, a manuscript is in preparation where \rev{the XDM dispersion energy is damped by a different, 1-parameter function depending on atomic numbers \cite{becke2024}; this method will soon be available in FHI-aims.}

Finally, it is also possible to include beyond-dipole dispersion terms in the TS and MBD methods~\cite{OQDO, Alberto-highorder}. To determine the multipolar parameters, it is possible to use scaling laws based on quantum Drude oscillators and/or their ratios determined by the XDM approach~\cite{c6810posthf}.

\section*{Acknowledgements}
The authors gratefully acknowledge contributions from Nicola Ferri and Ville Havu. RAD gratefully acknowledges financial support from the National Science Foundation under Grant No.\ CHE-1945676 and an Alfred P.\ Sloan Research Fellowship. ERJ thanks the Natural Sciences and Engineering Research Council (NSERC) of Canada for financial support, and the Royal Society for a Wolfson Visiting Fellowship. 
AOR thanks the Spanish Ministerio de Ciencia e Innovaci\'on and the Agencia Estatal de Investigaci\'on, projects PGC2021-125518NB-I00 and RED2022-134388-T co-financed by EU FEDER funds; the Spanish MCIN/AEI/10.13039/501100011033 and European Union NextGenerationEU/PRTR for grants TED2021-129457B-I00 and CNS2023-144958.
AT acknowledges the European Research Council (ERC AdG FITMOL) for funding.
MS acknowledges support by his TEC1p Advanced Grant (the European Research Council (ERC) Horizon 2020 research and innovation programme, grant agreement No. 740233.

\newpage

\section{Random Phase Approximation and Beyond for Total Energies and Forces}
\label{ChapRPA}

\sectionauthor[1,2]{Volker Blum}
\sectionauthor[3]{Francisco A. Delesma}
\sectionauthor[3]{Dorothea Golze}
\sectionauthor[4]{Florian Merz}
\sectionauthor[5,10,a]{\textbf{ *Xinguo Ren}}
\sectionauthor[6,7,8,9]{Patrick Rinke}
\sectionauthor[10]{Matthias Scheffler}
\sectionauthor[5]{Muhammad \rev{N.} Tahir}
\sectionauthor[11]{Alexandre Tkatchenko}
\sectionauthor[10,12,b]{Igor Ying Zhang}
\sectionlastauthor[13]{Tong Zhu}

\sectionaffil[1]{Thomas Lord Department of Mechanical Engineering and Materials Science, Duke University, Durham, NC 27708, USA}
\sectionaffil[2]{Department of Chemistry, Duke University, Durham, NC 27708, USA
}
\sectionaffil[3]{Faculty of Chemistry and Food Chemistry, Technische Universit\"at Dresden, 01062 Dresden, Germany}
\sectionaffil[4]{Lenovo HPC Innovation Center, Meitnerstr. 9, D-70563 Stuttgart, Germany}
\sectionaffil[5]{Institute of Physics, Chinese Academy of Sciences, 3rd South Str. 8, Beijing 100190, China}
\sectionaffil[6]{Department of Applied Physics, Aalto University, P.O. Box 11000, FI-00076 Aalto, Finland}
\sectionaffil[7]{Physics Department, TUM School of Natural Sciences, Technical University of Munich, Garching, Germany}
\sectionaffil[8]{Atomistic Modelling Center, Munich Data Science Institute, Technical University of Munich, Garching, Germany}
\sectionaffil[9]{Munich Center for Machine Learning (MCML), Munich, Germany}
\sectionaffil[10]{The NOMAD Laboratory at the Fritz Haber Institute of the Max Planck Society, Faradayweg 4-6, D-14195 Berlin, Germany}
\sectionaffil[11]{Department of Physics and Materials Science, University of Luxembourg, L-1511, Luxembourg City, Luxembourg}
\sectionaffil[12]{Collaborative Innovation Center of Chemistry for Energy Materials, Shanghai, Key Laboratory of Molecular Catalysis and Innovative Materials, MOE Key Laboratory of Computational Physical Sciences, Department of Chemistry, Fudan University, Shanghai 200433, China}
\sectionaffil[13]{Department of Electrical and Computer Engineering, University of Toronto, Toronto, Ontario M5S 1A4, Canada }

\sectionaffil[*]{Coordinator of this contribution.}
\rule[0.25ex]{0.35\linewidth}{0.25pt}

\sectionaffil[a]{{\it Current Address:} Institute of Physics, Chinese Academy of Sciences, Beijing, 100190, China}
\sectionaffil[b]{{\it Current Address:} Collaborative Innovation Center of Chemistry for Energy Materials, Shanghai, Key Laboratory of Molecular Catalysis and Innovative Materials, MOE Key Laboratory of Computational Physical Sciences, Department of Chemistry, Fudan University, Shanghai 200433, China}




\subsection*{Summary}
The random phase approximation (RPA) is classified under the fifth rung of Jacob's ladder, which organizes exchange-correlation (XC) functionals in Kohn-Sham (KS) density functional theory \cite{perdew2001}. Over the past two decades, RPA has gained recognition in computational chemistry, physics, and materials science \cite{Ren2012b,Eshuis2012} for its ability to accurately capture non-local electron correlations and effectively describe subtle energy differences, such as adsorption energies of molecules on surfaces and energy differences between crystal polymorphs. The RPA concept also led to the development of the many-body dispersion method \cite{Tkatchenko2012}, which addresses van der Waals interactions beyond atomic pairwise summations. While the standard RPA approach tends to underestimate the atomization energies of molecules and the cohesive energies of solids, this issue can be mitigated by incorporating corrections from renormalized single excitations (rSE) and second-order screened exchange contributions \cite{Ren2011,Grueneis2009,Ren2013}. Since 2009, the FHI-aims code has included an RPA implementation for molecular geometries using numeric atomic-centered orbitals \cite{Blum2009,Ren2009,ren2012}. Recently, this implementation has been extended to periodic systems and force calculations for molecules \cite{Tahir2022}. 
\rev{The RPA energy expression in FHI-aims is applied as a post-processing correction, i.e., it is computed following a self-consistent calculation with a different DFA and then effectively replaces the $E_\textrm{xc}$ term in Eq. (\ref{eq:Edft}).}
\rev{One prominent feature of RPA-based methods is that they can describe vdW interactions seamlessly and automatically capture the non-additive and non-anisotropic vdW effects. However, compared to the dedicated vdW-inclusive approaches presented in Sec.~\ref{ChapDispInteract}, RPA is computationally much more demanding. Nevertheless, in complex chemical environments and when there are strong variations among the results obtained by the empirical/semi-empirical vdW-inclusive approaches, RPA can be used as a benchmark method.}

The implementation of RPA in FHI-aims is based on the resolution of identity (RI) approximation, where the Kohn-Sham density response function and Coulomb operator are represented in terms of a set of auxiliary basis functions (ABFs). Different RI flavors are implemented, including a global RI based on the Coulomb-metric (RI-V) \cite{ren2012} and a localized RI based on a two-center approximation (LRI, also known as RI-LVL) \cite{ihrig2015}. The single-point RPA calculations for  finite systems (molecular geometries) can be performed using both RI-V and RI-LVL, while the geometry relaxations and single-point RPA energy calculations for periodic systems are exclusively based on RI-LVL. Currently, RPA calculations can be done for molecules with over 100  atoms, and RPA (and RPA+rSE) geometry relaxation can be performed for systems with a few tens of atoms \cite{Tahir2024}. The current periodic RPA implementation in FHI-aims enables the computation of relatively simple crystal structures with finite $\mathbf{k}$-point sampling. \rev{However, the code infrastructure for RPA force calculations for periodic systems is still under development.} Large-scale periodic calculations can be done by interfacing FHI-aims with LibRPA \cite{Shi2024}, a standalone package for evaluating RPA correlation energy based on low-scaling, real-space algorithm.

\subsection*{Current Status of the Implementation}
In standard RPA calculations, the total energy is given by $E^\text{RPA} = E^\text{HF}\left[\{\psi_i^\text{KS}\}\right] + E_c^\text{RPA}$, where $E^\text{HF}\left[\{\psi_i^\text{KS}\}\right]$ is the Hartree-Fock total energy evaluated using the KS orbitals and $E_c^\text{RPA}$ is the RPA
correlation energy. A convenient expression to evaluate $E_c^\text{RPA}$ is 
\begin{equation}
        E^\text{RPA}_c=\frac{1}{2\pi}\int^{\infty}_0 \mathrm{d}\omega\, \text{Tr}\left[\ln(1-\chi^0(i\omega)V)+\chi^0(i\omega)V\right]\, 
    \label{eq:E_c_RPA}
\end{equation}
where $\chi^0$ and $V$ can be understood as matrices in the ABF basis. Specifically,
\begin{equation}
   \chi^0_{\mu\nu} (i\omega)=\sum_{m,n,\sigma}\frac{(f_{m,\sigma}-f_{n,\sigma})C_{m,n,\sigma}^\mu C_{n,m,\sigma}^\nu}{\epsilon_{m,\sigma}-\epsilon_{n,\sigma}-i\omega}\, 
   \label{eq:chi0_matrix}
\end{equation}
where $C_{m,n,\sigma}^\mu$ are the RI expansion coefficients of products of KS spin-orbitals in terms of ABFs $(\{P_\mu(\mathbf{r}\})$,
\begin{equation}
    \psi_{m,\sigma}(\mathbf{r})\psi_{n,\sigma}(\mathbf{r}) = \sum_{\mu} C_{m,n,\sigma}^\mu P_\mu(\mathbf{r})
    \label{eq:RI_expan}
\end{equation}
and $V$ is the Coulomb matrix
\begin{equation}
    V_{\mu\nu} = \int d\mathbf{r} d\mathbf{r'} \frac{P_\mu(\mathbf{r})P_\nu(\mathbf{r'})}{|\mathbf{r}-\mathbf{r'}|}\, .
    \label{eq:Coulomb_Matrix}
\end{equation}
\rev{For an in-depth discussion of the theoretical basis and algorithmic formulation behind these equations, the readers are referred to Refs.~\cite{Ren2012b,IgorZhang2025}.}
In FHI-aims, we use an ``on-the-fly" procedure to generate suitable atom-centered ABFs for a given single-particle NAO basis set \cite{ren2012,ihrig2015}. The $C_{m,n,\sigma}^\mu$ expansion coefficients in Eq.~\ref{eq:RI_expan} are obtained by first computing the RI
coefficients $\tilde{C}_{i,j}^\mu$ that expand pair products of atomic basis functions $\phi_i\phi_j$ in terms of ABFs, and then transform $\tilde{C}_{i,j}^\mu$ into $C_{m,n,\sigma}^\mu$ by multiplying with the
KS eigenvectors $c_{n\sigma}^i$. The different methods for determining $\tilde{C}_{i,j}^\mu$ distinguish the various flavors of RI, such as RI-V and RI-LVL. The primary challenge in RPA correlation energy calculations is evaluating the $\chi^0$ matrix as described in Eq.~\ref{eq:chi0_matrix}, which scales with the system size as $O(N^4)$.

Extending RPA energy calculations from finite systems to periodic systems with finite ${\mathbf{k}}$-point sampling presents a significant challenge because the KS orbitals carry a Bloch wavevector. Consequently, $C_{m,n,\sigma }^{\mu}$, $\chi^0_{\mu\nu}$, and $V_{\mu\nu}$ all become momentum-dependent, i.e., $C_{m,n,\sigma}^{\mu} \rightarrow C_{m,n,\sigma}^{\mu}(\mathbf{k},\mathbf{q})$, $\chi^0_{\mu\nu} \rightarrow \chi^0_{\mu\nu}(\mathbf{q})$, and
$V_{\mu\nu} \rightarrow V_{\mu\nu}(\mathbf{q})$ and additional 
sums over $\mathbf{q}$ (or $\mathbf{k}$) points are required for the trace or summation operations in Eq.~\ref{eq:E_c_RPA} and \ref{eq:chi0_matrix}. The extension to reciprocal space  is straightforward except for $\mathbf{q}=0$ where certain elements
of $V_{\mu\nu}(\mathbf{q}=\mathbf{0})$ diverge. In our implementation, the truncated Coulomb operator is used at $\mathbf{q}=0$ to avoid any such
divergences, while the full Coulomb operator is used elsewhere in the Brillouin zone (BZ).

\begin{figure}[ht!]
    \centering
    \includegraphics[width=0.85\textwidth]{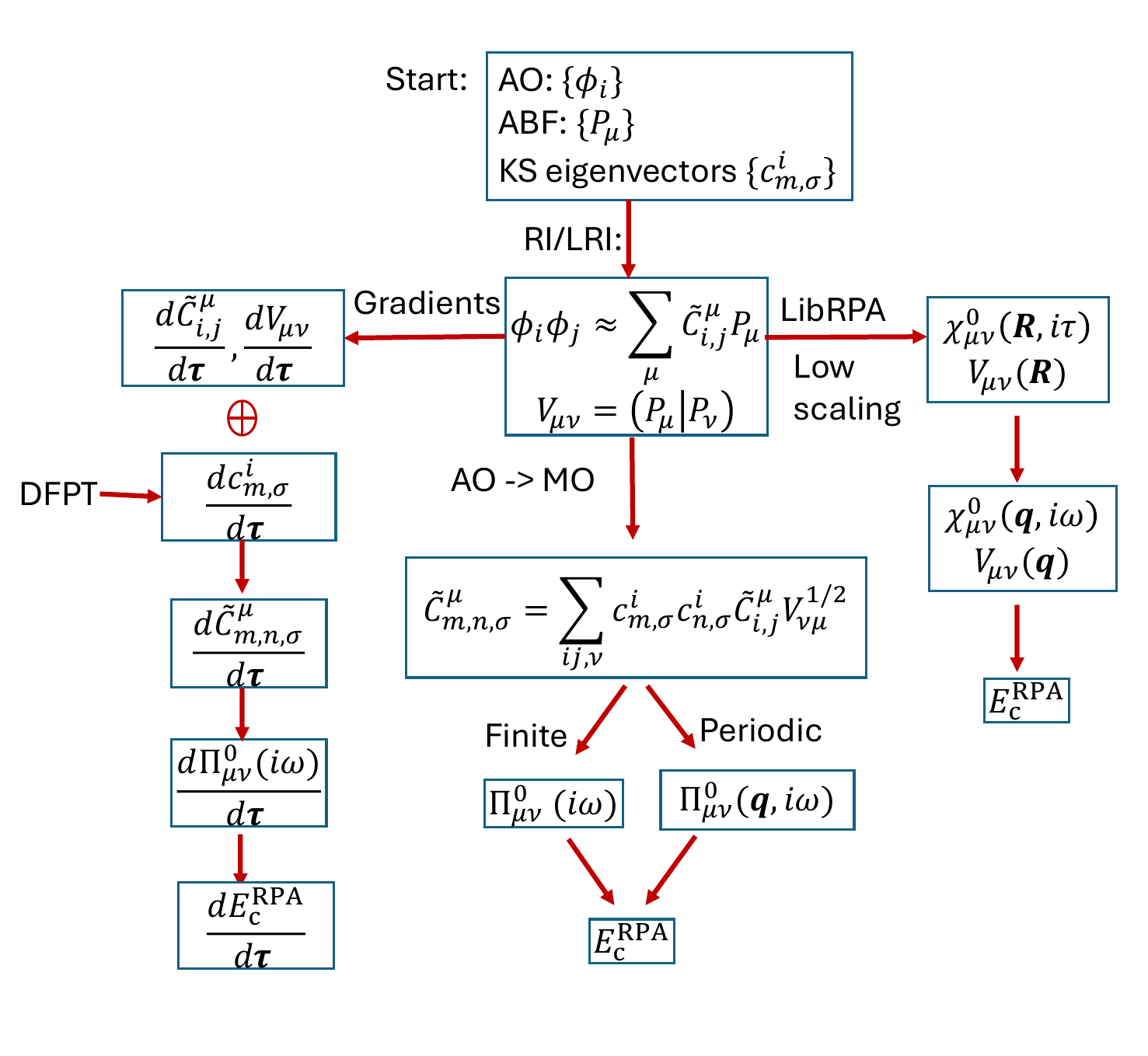}
    \caption{Illustration of the major computational steps in FHI-aims for evaluating the correlation part of the RPA ground-state energies and forces. Here AO denotes atomic orbitals and MO denote KS molecular orbitals. Note that $\tilde{C}_{m,n,\sigma}^\mu$ here is related to ${C}_{m,n,\sigma}^\mu$ in Eq.~\ref{eq:RI_expan} by $\tilde{C}_{m,n,\sigma}^\mu=\sum_{\nu}{C}_{m,n,\sigma}^\nu V^{1/2}_{\nu,\mu}$, and similarly $\Pi^0$ is related to $\chi^0$ by $\Pi^0 = V^{1/2} \chi^0 V^{1/2}$.}
    \label{fig:RPA_flowchart}
\end{figure}

As previously noted, FHI-aims also supplies gradients of the RPA total energy with respect to nuclear displacements, facilitating structure relaxations based on RPA forces. In our implementation, the total RPA force on atom $I$ located at spatial position $\mathbf{\bm \tau}_I$ is decomposed into the following terms
\begin{equation}
     \label{eq:RPA_force_total}
 \begin{split}
 \mathbf{F}_{I}^{\text{RPA}}&=-\dfrac{dE^\text{RPA}}{d\mathbf{\bm \tau}_{I}}\\&
 =-\dfrac{dE^{\text{DFA}}}{d\mathbf{\bm \tau}_{I}}+\dfrac{dE^{\text{DFA}}_{xc}}{d\mathbf{\bm \tau}_{I}}-\dfrac{dE^{\text{EX}}_{x}}{d\mathbf{\bm \tau}_{I}}-\dfrac{dE^{\text{RPA}}_{c}}{d\mathbf{\bm \tau}_{I}}\, 
 \end{split}
\end{equation}
where $-\dfrac{dE^{\text{DFA}}}{d\mathbf{\bm \tau}_{I}}$ is the gradient of a local or semilocal density functional approximations (DFA) and has been long available in FHI-aims. The RPA gradient is obtained by subtracting the contribution of the local/semilocal XC functional from the DFA force, and adding the exact exchange and RPA correlation terms instead. To calculate the latter two, the derivatives of the KS eigenvectors with respect to the nuclear displacements are calculated with density functional perturbation theory (DFPT) \cite{Shang2017}. Further details can be found in Ref.~\cite{Tahir2022}. Currently, the RPA force calculation functionality in FHI-aims is fully developed for molecular geometries, while the implementation for periodic systems is still in progress. The main computational steps for the correlation part of the RPA energy and force calculations, as implemented in FHI-aims (and LibRPA), are shown in Fig.~\ref{fig:RPA_flowchart}.

\subsection*{Usability and Tutorials}
FHI-aims RPA total energy and force calculations are well-established by now. Here, we provide a brief overview of how to perform these calculations using FHI-aims.

\textit{Total Energies:} All functionalities for RPA-based total energy calculations are now accessible through simple keywords in the \texttt{control.in} file, as documented in the FHI-aims manual. Currently, RPA-type calculations are carried out in a post-processing manner, and to invoke RPA-based energy calculations, the keywords 
\begin{center}
  \begin{verbatim}
      xc dfa
      total_energy_method type
  \end{verbatim}
\end{center}
need to be set in the \texttt{control.in} file.  While \texttt{dfa} specifies the prior density functional approximation (dfa) that generates input orbitals and orbital energies for RPA-type calculations, \texttt{type} is a keyword (string) that specifies a chosen post-processing RPA or beyond-RPA method. For example, for a standard RPA@PBE calculation, \texttt{dfa} should be \texttt{PBE} and \texttt{type} should be \texttt{RPA}. In addition to RPA, other choices like \texttt{rpa+sosex} and  \texttt{rpt2} are also legitimate,
which yields RPA+SOSEX and the renormalized 2nd-order perturbation theory (rPT2) total energies, respectively.
Such beyond-RPA methods usually give improved results when the standard RPA shows pronounced underbinding behavior. However, RPA+SOSEX and rPT2 are substantially more expensive than RPA. On the other hand, the rSE correction 
is very cheap, and this term is automatically calculated whenever one performs RPA calculations. RPA+rSE typically describes van der Waals bonded systems more accurately than the standard RPA. Further details about 
the performance of RPA and beyond-RPA methodologies can be found in Refs.~\cite{Ren2012b,Ren2013}.
Periodic RPA energy calculations using FHI-aims can be done as usual by setting the \texttt{k\_grid} 
in \texttt{control.in} and \texttt{lattice\_vector} in the \texttt{geometry.in} file. An example of such calculations for the
Ar crystal was described in Ref.~\cite{Yang2022}. Periodic RPA calculations can be efficiently performed using the LibRPA library, which implements a low-scaling real-space RPA algorithm \cite{Shi2024}. These low-scaling RPA calculations utilize the minimax imaginary time-frequency grids provided by the GreenX library \cite{Azizi2023}, accessible in FHI-aims with the keyword \texttt{freq\_grid\_type minimax}. Recent studies have shown that minimax imaginary frequency grids converge approximately 3.5 times faster than modified Gauss-Legendre grids for integrating the RPA correlation energy, Eq. \eqref{eq:E_c_RPA}. This enhancement significantly reduces the computational prefactor in canonical RPA calculations for both solids and molecules \cite{Azizi2024}.  \rev{At present, the coupling between FHI-aims and LibRPA is implemented via a file-based interface. To prepare the data, the user only needs to add the keyword “\texttt{output librpa}” to the FHI-aims input, which triggers the export of all quantities required by LibRPA in the expected format. No additional manual post-processing of the generated files is needed. With this exported dataset and a separate LibRPA input file specifying the desired calculation settings, the user can directly run the low-scaling RPA calculation using LibRPA. For more details, please refer to Ref.~\cite{Shi2025}.}


\textit{Forces:} 
The force calculations can be done for molecules at the level of RPA and RPA+rSE using FHI-aims. To run
such calculations, additional keywords related to DFPT and choice of the RPA force method also need to be set.
For example, structure relaxations based on RPA@PBE gradients can be performed using the following set of keywords in \texttt{control.in} file
\begin{center}
\begin{verbatim}
  xc  pbe
  relativistic  none 
  occupation_type  gaussian  0.001
  RI_method lvl
  DFPT  vibration_reduce_memory 
  relax_geometry  bfgs  1e-2
  sc_accuracy_forces  1e-2
  least_memory_4  .true.
  rpa_force   freq_formula_method
  frequency_points  16
  my_prodbas_threshold  1.e-12 
\end{verbatim}
\end{center}
Both RPA energy and force implementations in FHI-aims are MPI-parallelized. Depending on the system size, parallel calculations can scale up to hundreds to thousands of CPU cores. A recent performance benchmark and applications to water clusters of the RPA force implementation can be found in Ref.~\cite{Tahir2024}.
\rev{Regarding the timings, the current state of the art of computational efficiency of molecular RPA and RPA+rSE force calculations using FHI-aims can be found in Ref.~\cite{Tahir2025}. As for the improvement from the canonical scaling in FHI-aims to the sub-quadratic scaling in LibRPA for periodic RPA calculations, detailed timings together with the scaling behavior has been reported in Ref.~\cite{Shi2024}.}
The basic functionalities of RPA and beyond-RPA energy and force calculations are detailed in an online tutorial \cite{tutorial}.

An important issue in RPA calculations is the choice of the single-particle basis set. The standard FHI-aims basis sets (\textit{tier's}) can be used, but it is mandatory to perform a counterpoise (CP) correction to the basis set superposition error (BSSE). Furthermore, it is not possible to extrapolate to the complete basis set (CBS) limit
using \textit{tiers}. A recommended choice of basis sets for RPA calculations using FHI-aims are NAO-VCC-$n$Z \cite{Zhang2013} or its localized variants. In practice, it is often a good idea to perform 
a CP correction and compare the CP-corrected and uncorrected results. For elements where NAO-VCC-$n$Z is not
available, correlation-consistent Gaussian basis sets provide a viable alternative.


\subsection*{Future Plans and Challenges}
RPA calculations based on NAO basis sets encounter several challenges that we aim to address in future work. Firstly, the RPA correlation energy converges slowly with respect to single-particle basis sets, often necessitating extrapolation to the CBS limit to achieve high-quality results. In FHI-aims, such ``correlation consistent'' NAO-VCC-$n$Z \cite{Zhang2013} that facilitate CBS extrapolation are officially available only for light elements up to Ar. Recent efforts have focused on developing NAO-VCC-nZ basis sets beyond Ar, but so far, success has been limited to a few selected main-group elements \cite{Yang/Zhang/Ren:2024}. Additionally, BSSE is significant in NAO-based RPA calculations. While BSSE can be corrected using the counterpoise scheme for binding energies of finite systems, counterpoise corrections are not straightforward for periodic systems. Practical periodic RPA calculations depend on the relatively small BSSE of the newly developed NAO-VCC-nZ basis sets and the cancellation between BSSE and basis set incompleteness errors. However, the remaining uncertainty is often difficult to estimate. Therefore, a better strategy is urgently needed to address the basis set issues for RPA calculations with NAOs.

The second issue concerns the handling of the Coulomb singularity at the $\Gamma$ point in periodic RPA calculations. Ideally, the RPA integrand does not exhibit diverging behavior as $\mathbf{q} \rightarrow 0$ in the Brillouin zone (BZ), at least for insulating systems, because the $1/q^2$ divergence in the Coulomb operator is offset by the $q^2$ factor in the KS response function $\chi^0(\mathbf{q})$. However, in the matrix representation using auxiliary basis functions (ABFs), this cancellation is incomplete, necessitating special numerical treatment. Practically, a mixed scheme—using the truncated Coulomb operator for $\mathbf{q=0}$ and the full Coulomb operator elsewhere in the BZ—works quite well. Nonetheless, the performance of this mixed scheme degrades for very sparse $\mathbf{k}$ grids or $\Gamma$-point-only calculations. Developing a more consistent $\Gamma$-point treatment scheme for periodic RPA calculations remains a goal for future development in FHI-aims. \rev{A related Coulomb singularity issue in the periodic $GW$ calculations and how it was addressed in FHI-aims can be found in Ref.~\cite{Ren2021}.}

The third issue pertains to fractional occupations in RPA calculations. These fractional occupations frequently arise in preceding KS-DFT calculations, particularly in systems with open-shell transition metal atoms, defects, or radicals. While the RPA formalism can theoretically accommodate fractional occupations in the KS response function, the practical results are highly sensitive to these occupation numbers, often leading to a decline in RPA performance. Consequently, there is an urgent need to advance beyond the standard RPA to effectively handle such cases. This challenge is also closely linked to the broader goal of extending the RPA formalism to address systems with strong multi-configurational characteristics and finite temperatures.

\subsection*{Acknowledgements}
We acknowledge the funding support from the National Natural Science Foundation of China (Grant Nos 12374067 and 12188101) and the National Key Research and Development Program of China (Grant Nos 2022YFA1403800 and 2023YFA1507004). The authors acknowledge the European Union’s Horizon 2020 research and innovation program for financial support under the grant number 951786 (Nomad Center of Excellence) and the Horizon Europe MSCA Doctoral network grant n.101073486 (EUSpecLab). D. G. acknowledges funding by the Emmy Noether Program of the German Research Foundation (project number 453275048). We also wish to acknowledge CSC – IT Center for Science (Finland), the Jülich Supercomputer Computer Center (Germany) and Aalto Science-IT project for computational resources. MS acknowledges support by his TEC1p Advanced Grant (the European Research Council (ERC) Horizon 2020 research and innovation programme, grant agreement No. 740233.

\chapter{Electronic Excited States}
\label{ChapExcitedStates}






\newpage

\section{Coupled-Cluster Theory for the Ground State and for Excitations}
\label{ChapCoupledCluster}

\sectionauthor[1]{Andreas Gr\"uneis}
\sectionauthor[2]{\textbf{*Evgeny Moerman}}
\sectionauthor[2]{\textbf{*Matthias Scheffler}}
\sectionauthor[2,a]{Tonghao Shen}
\sectionlastauthor[2,a]{Igor Ying Zhang}

\sectionaffil[1]{Institute for Theoretical Physics, TU Wien, Wiedner Hauptstra{\ss}e 8--10/136, 1040 Vienna, Austria}
\sectionaffil[2]{The NOMAD Laboratory at the Fritz Haber Institute of the Max Planck Society, Faradayweg 4-6, D-14195 Berlin, Germany}

\sectionaffil[*]{Coordinator of this contribution.}
\rule[0.25ex]{0.35\linewidth}{0.25pt}

\sectionaffil[a]{{\it Current Address:} Department of Chemistry, Fudan University, Shanghai 200433, People’s Republic of China}



\newlength{\tempdima}
\newcommand{\rowname}[1]
{\rotatebox{90}{\makebox[\tempdima][c]{\textbf{#1}}}}




\subsection*{Summary}

In the molecular quantum chemistry community, coupled-cluster (CC) methods are well-recognized for
their systematic convergence and reliability. The extension of 
the theory to extended systems has been comparably recent~\cite{zhang2019coupled}, so that developments and studies of periodic CC methods for both the ground-state and for excited states are still active fields 
of research and provide valuable benchmark data when the
reliability of density functional approximations is questionable. In this contribution we describe the CC-aims interface between
the FHI-aims and the Cc4s software packages. 
This linkage makes a variety of correlated wave function-based ground-state methods including 
M\o ller-Plesset
perturbation theory (MP2), the random-phase
approximation (RPA) and the
\emph{gold-standard of quantum chemistry} CCSD(T)
method for both molecular and periodic applications accessible.
This contribution discusses these ground-state methods for clusters and molecules, as 
well as for periodic systems. In particular, we discuss recent advancements and the
implementation of the equation-of-motion CC method
for the calculation of ionization (IP-EOM-CCSD) and
electron attachment (EA-EOM-CCSD) processes.
Open questions and routes to solutions are discussed as well.


\subsection*{Current Status of the Implementation}

\begin{figure}
    \centering
    \includegraphics[width=\linewidth]{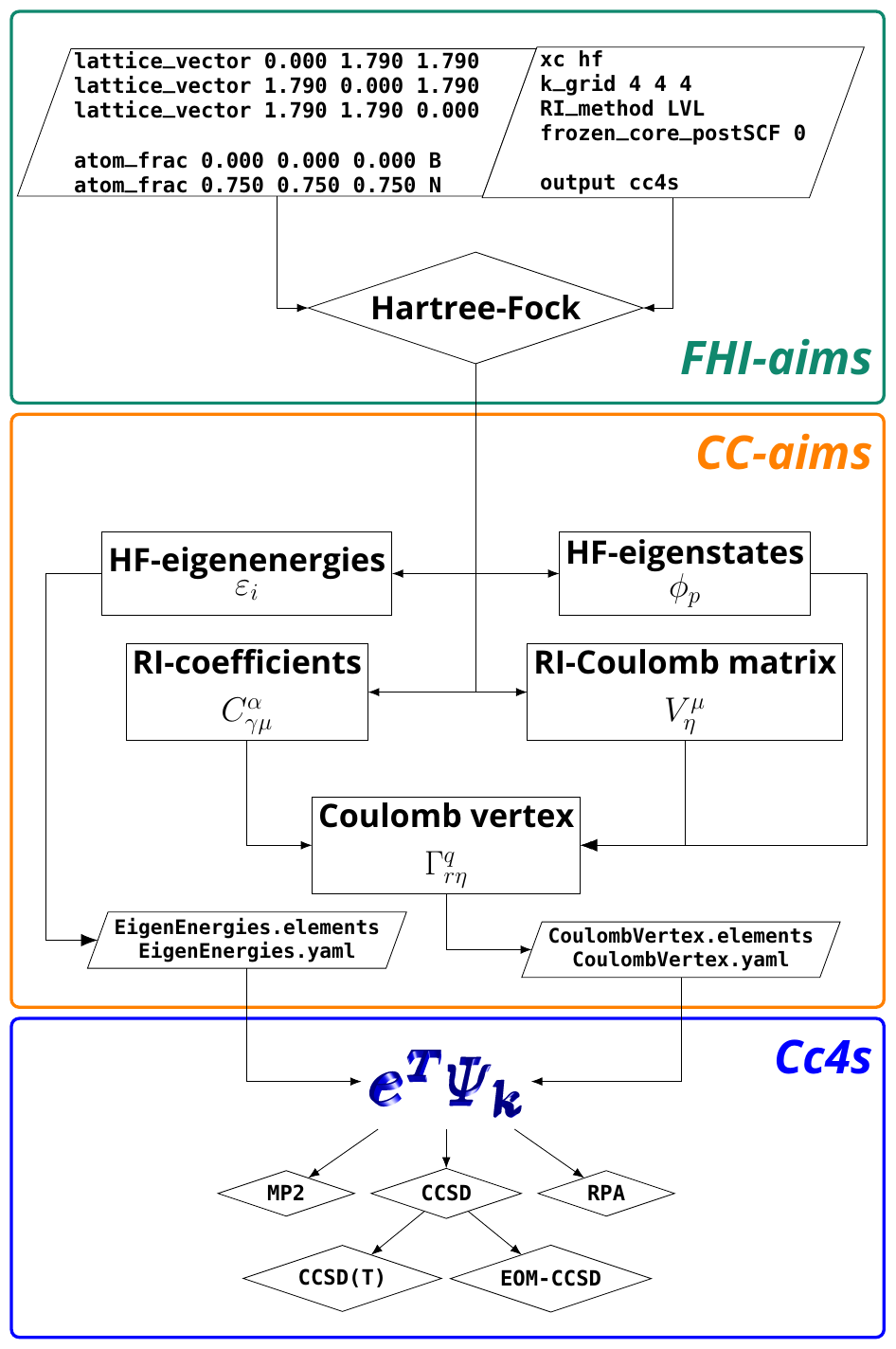}
    \caption{The Cc4s@FHI-aims workflow}
    \label{fig:workflow}
\end{figure}

The Cc4s code constitutes an open-source quantum chemistry software package, which features several correlated wave function methods. As a post-SCF code, Cc4s requires single-particle
eigenergies and wave functions from a mean-field
calculation, which another electronic structure
package must provide. Using the CC-aims interface~\cite{moerman2022interface}
the relevant quantities are conveniently obtained from 
a FHI-aims Hartree-Fock calculation. CC-aims, then, parses and converts the
needed quantities
(e.g the single-particle eigenergies) 
to a Cc4s-compatible format and
computes additional quantities 
(e.g the Coulomb vertex). The files generated by CC-aims
are then used to launch a Cc4s calculation with any of the therein implemented
wave-function methods. This workflow is illustrated in Figure \ref{fig:workflow}.

Below, we discuss the coupled-cluster approach for the electronic ground-state
and for excited states for molecules and clusters and for periodic solids.

In practice, the limiting factor of CC methods
is the substantial memory requirement, which results from the size of the Coulomb integral tensor

\begin{equation}
     V^{pq}_{rs} = \int d\bm{r}\bm{r}'\, 
    \frac{\phi_p^{*}({\bm{r}})\phi_q^{*}(\bm{r'})\phi_r({\bm{r}})\phi_s(\bm{r'})}{|\bm{r} - \bm{r'}|}\text{,}
\end{equation}

where $\phi_p$, $\phi_q$, $\phi_r$ and $\phi_s$
denote four single-particle wave functions.
As a consequence, the memory to store
this tensor scales with the fourth power of 
the basis set size. By performing a low-rank
decomposition of $V^{pq}_{rs}$, the Coulomb vertex
$\Gamma_{r}^{p,\eta}$,
a rank three tensor with a significantly smaller memory footprint~\cite{hummel2017low} is defined

\begin{equation}
    V^{pq}_{rs} \approx
    \sum_{\eta}
    \Gamma_{r}^{p\eta} \Gamma^{q*}_{s\eta}\text{,}
\end{equation}

where $\eta$ indexes the basis functions of an
auxiliary basis. As a memory-saving measure,
CC-aims computes the Coulomb vertex. The auxiliary basis is that of the resolution-of-the-identity (RI) employed in FHI-aims. In combination with CC-aims, one can choose between the RI-V scheme for molecular
applications and its localized approximation, the RI-LVL scheme for periodic
ones. While the former is deemed very
accurate, the latter is more memory-efficient
but in general of insufficient accuracy~\cite{ihrig2015}. 
However, the incompleteness of the RI-LVL auxiliary basis for Hartree-Fock, MP2 and RPA methods can be satisfactorily resolved by manually adding a few auxiliary $f$ -, $g$- and $h$-type basis functions with small eﬀective charges~\cite{ihrig2015}. This approach is also applicable for CC calculations.
The memory footprint of the Coulomb vertex can be further reduced by performing 
a principal component analysis, with which
up to 70\% of the Coulomb vertex can be discarded
in many applications~\cite{hummel2017low}.

Currently, the periodic CC infrastructure of FHI-aims does not allow to perform spin-polarized calculations and the starting point needs to be canonical Hartree-Fock orbitals. For periodic applications, the most important constraint is given by the inability of 
Cc4s to perform an explicit $k$-point summation. 
Instead, Cc4s requires a super cell based
treatment of a periodic system. 
As a consequence, Cc4s does not make use of the
translational symmetry of crystalline systems,
which would reduce 
the memory scaling by a factor of $N_k$
and the computational scaling by a factor of $N_k^2$, with $N_k$ being the number of $k$-points.

\subsection*{Usability and Tutorials}

\begin{figure}[htbp]
\centering
 \begin{tikzpicture}   
 \matrix (fig) [matrix of nodes]{
 \includegraphics[width=0.47\linewidth]{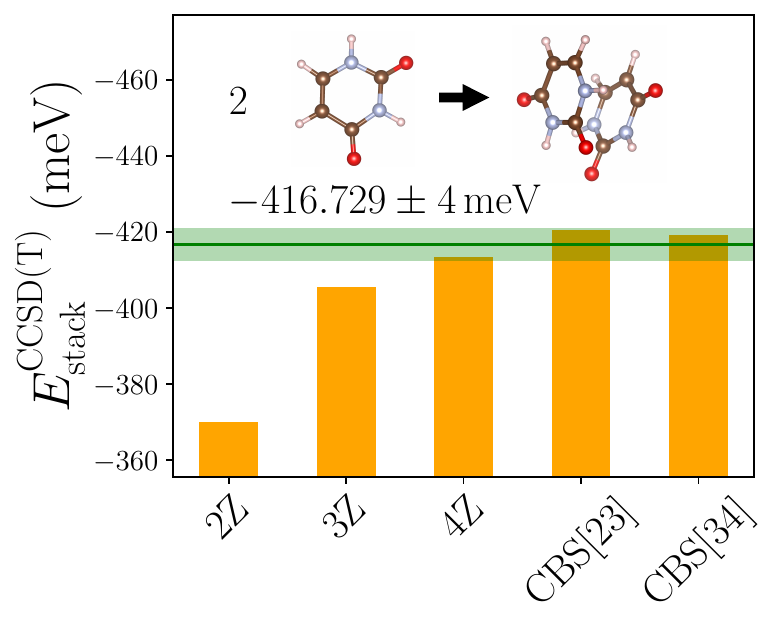}&
 \vspace{1cm}
 \raisebox{0.46cm}{\includegraphics[width=0.47\linewidth]
 {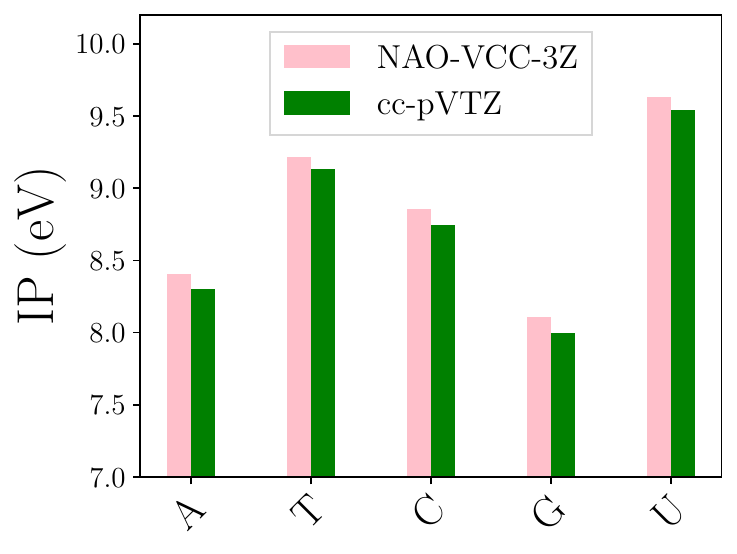}}
 \\[0cm] 
 |[text width=1.85in]| {\subcaption{}\label{fig:1a}}
 &
 |[text width=1.85in]| {\subcaption{}\label{fig:1b}}
 \\[0.0cm] 
 \includegraphics[width=0.47\linewidth]
 {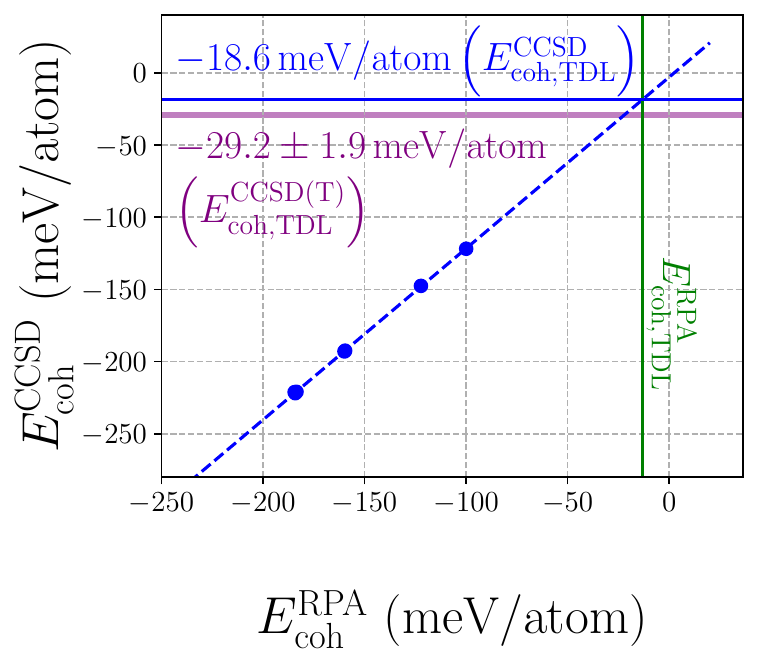}
 &
 \includegraphics[width=0.47\linewidth]{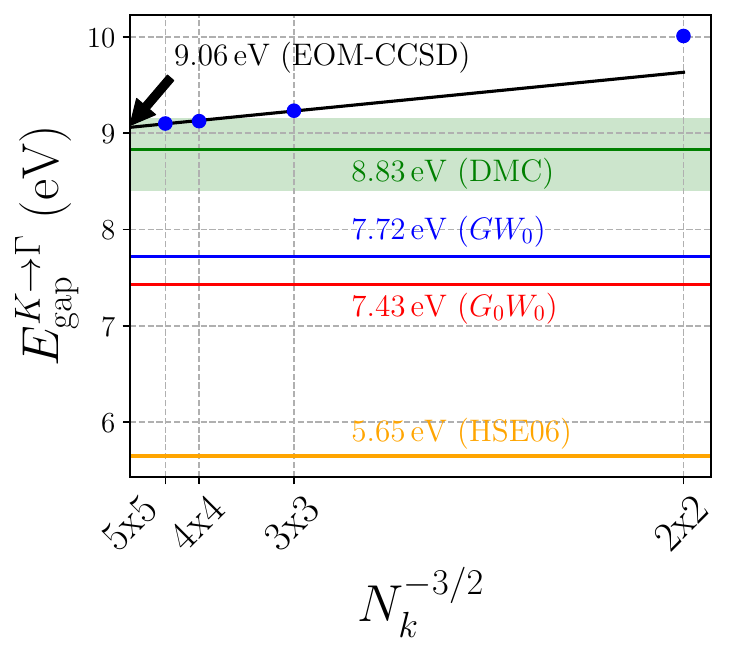}
 \\[0cm] 
 |[text width=1.85in]| {\subcaption{}\label{fig:1c}}
 &
 |[text width=1.85in]| {\subcaption{}\label{fig:1d}}
 \\
  };
  
  \draw[-latex, white] ([xshift=-2mm,yshift=2mm]fig-1-1.south west)  -- ([xshift=-2mm,yshift=-2mm]fig-1-1.north west) node[midway,above,sloped, black]{Molecules};

 \draw[-latex, white] ([xshift=-2mm,yshift=2mm]fig-3-1.south west)  -- ([xshift=-2mm,yshift=-2mm]fig-3-1.north west) node[midway,above,sloped, black]{Extended solids};
 \draw[-latex, white] ([yshift=2mm,xshift=2mm]fig-1-1.north west)  -- ([yshift=2mm,xshift=-2mm]fig-1-1.north east) node[midway,above, black]{Ground state};
 \draw[-latex, white] ([yshift=7.5mm,xshift=2mm]fig-1-2.north west)  -- ([yshift=7.5mm,xshift=-2mm]fig-1-2.north east) node[midway,below, black]{Excited state};
 \end{tikzpicture}
    \caption{A selection of molecular and periodic
CC calculations with the NAO-VCC-$n$Z basis sets:
Figure \ref{fig:1a} 
stacking energy of a uracil dimer on the CCSD(T) level of theory compared to Reference~\cite{al2021interactions},
Figure \ref{fig:1b} 
vertical ionization potentials of adenine (A), thymine (T), cytosine (C), guanine (G)
and uracil (U) via IP-EOM-CCSD compared to Reference~\cite{tripathi2022performance}, 
Figure \ref{fig:1c} 
Finite-size convergence of the cohesive energy
of Neon using the CCSD method plotted against the RPA results.
The (T)-correction has been found to be mostly system-size-independent
and has been added to the extrapolated CCSD result.
Figure \ref{fig:1d} 
Electronic band gap of two-dimensional boron nitride via IP- and EA-EOM-CCSD compared to HSE06 and higher-level correlated methods in Reference~\cite{hunt2020diffusion}. The opaque green area illustrates
the uncertainty of the DMC result.}
    \label{fig:cc-use-cases}
\end{figure}

A variety of FHI-aims based calculations with the CCSD, CCSD(T) and the EOM-CCSD implementations in Cc4s have already 
been performed for small and medium-sized molecules
and for crystalline insulators and semiconductors.
Tutorials for using the workflow involving
FHI-aims, CC-aims and Cc4s for both molecules
and crystalline materials are available on
the FHI-aims platform~\cite{aimscctutorial}.
These tutorials include a detailed step-by-step guide on the installation of the
necessary software (i.e CC-aims and Cc4s) and the calculation of
ground-state and excitation energies. Currently, the tutorial consists of 
three parts. The first one introduces the FHI-aims/CC-aims/Cc4s workflow
by an exemplary calculation of the MP2 correlation energy of the paracetamol
molecule. The second tutorial focuses on a periodic application and demonstrates
the importance of a careful consideration of the system-size dependent error
for correlated wave function methods. To that end, the necessary steps involved in
the calculation of the CCSD cohesive energy of Neon are presented, the results
of which are also shown in Figure \ref{fig:1c}. In the third tutorial, 
the additional steps to compute quasi-particle energies via the 
IP- and EA-EOM-CCSD method are described in detail, for which the
ionization potential of uracil (see Figure \ref{fig:1b})
and the band gap of \textit{trans}-polyacetylene is computed~\cite{moerman2024finite}.

Figure \ref{fig:cc-use-cases} shows four typical 
applications of CC theory using FHI-aims for molecular
and extended system involving both the ground state
and charged excited states. Figure \ref{fig:1a}
and \ref{fig:1b} confirm the accuracy of both ground state 
and excited state properties for molecules using our 
valence-correlation
consistent NAO-VCC-$n$Z basis sets~\cite{Zhang2013}. Via extrapolation to the complete basis set (CBS) limit, we are able to reproduce the stacking energy of uracil (Figure \ref{fig:1a}) within a few
meV. 
Going beyond the ground-state, in Figure \ref{fig:1b} we show results 
of the IP-EOM-CCSD method to obtain the ionization potentials
of the five nucleobases, which are in good agreement (within $\approx 100\,\text{meV}$) with previously reported
values by Tripathi et al.~\cite{tripathi2022performance}. 

The remaining deviation stems from the use of different finite basis sets: cc-pVTZ in Reference~\cite{tripathi2022performance} and NAO-VCC-3Z in our study. For ground-state properties, we have found that the NAO-VCC-$n$Z
basis set performs better than the cc-pV$n$Z  basis set for advanced correlation methods, including MP2, RPA, and CCSD(T)~\cite{Zhang2013, shen2019massive}. This discrepancy can be further minimized by extrapolating to the CBS limit.

The remaining deviation is attributable to the utilization of a 3Z basis set, for which 
a remaining difference of 
that magnitude to the CBS limit is expected.

Figure \ref{fig:1d} demonstrates the 
applicability of the  EOM-CCSD method to obtain
quasi-particle band gaps. For that we studied the
fundamental, indirect $K\to\Gamma$ band gap of two-dimensional
hexagonal boron nitride (hBN), for which different state-of-the-art
methods like generalized KS-DFT and the $GW$ approximation, and highly accurate methods like
Diffusion Monte Carlo (DMC) yielded very different results. Our analysis
shows good agreement with the DMC result by 
Hunt et al.~\cite{hunt2020diffusion}, confirming that an accurate
treatment of electronic correlation is crucial in this system.
Figure \ref{fig:1c} shows the cohesive energy of the Neon crystal.
The stability of this crystal is almost exclusively determined by 
van-der-Waals interactions, which CC methods are known to precisely
capture. 
With Cc4s@FHI-aims we obtain a cohesive energy of 
$-17.9\,\text{meV/atom}$ at the CCSD level. 
The experimental value, already adjusted for zero-point fluctuations, is $-27\,\text{meV/atom}$, with an estimated zero-point contribution of $7\,\text{meV/atom}$~\cite{rosciszewski1999ab}.
For these calculations the NAO-VCC-2Z and -3Z basis sets were employed. Performing a conventional two-point extrapolation 
to the CBS limit yields the results in Figure \ref{fig:1c} 
with a remaining difference to the CBS limit of $8\,\text{meV/atom}$, which was added to all
of the data points. A more accurate cohesive energy was obtained by accounting for triple excitations in a 
perturbative manner. That triple (T)-correction was found
to be mostly independent of the system size, and
determined to lie between $-8.7\,\text{meV/atom}$
and $-12.6\,\text{meV/atom}$. By adding these values
to the extrapolated CCSD cohesive energy, a CCSD(T) 
result of $-29.2\pm1.9\,\text{meV/atom}$ was found,
which is in excellent agreement with the experimental 
finding.

\subsection*{Future Plans and Challenges}

As mentioned before, currently the biggest limitation in performing
periodic CC calculations via Cc4s is the lack of an explicit $k$-point
summation. 
The translational symmetry of a periodic system is reflected in the
block-sparsity of the CC tensors. A block-sparse implementation of the
tensor contraction engine in Cc4s is currently under development and is
expected to severely reduce the computational and memory requirements of 
CC calculations.
Further future plans involve the extension of the Cc4s functionalities to include spin-polarized and non-HF based calculations. 
The convergence of CC ground and excited state properties with respect to system size can be very slow. As has been in part demonstrated in Figure \ref{fig:1c}, one can correct the corresponding finite-size errors by additionally employing another, computationally cheaper non-CC method (e.g the RPA for the ground-state or the $G_0W_0$ method for quasi-particles~\cite{moerman2025}), which exhibits a similar system size convergence. However, via the implementation of a block-sparse tensor treatment in Cc4s, more work has to be done to verify the accuracy and the limitations of this methodology.
Development of new strategies to reduce the basis set incompleteness error is the second, very important future goal,
to improve the precision of CC calculations. In ground-state CC applications in particular, where the quantity of interest itself is sometimes
on the order of a few meV, one is forced to use very big basis
sets, even if extrapolation schemes are accessible to achieve the necessary precision. 
One solution is to use the much more compact natural orbital basis instead of the 
canonical HF basis~\cite{lowdin1955quantum}. One way to construct natural orbitals for correlated wave function methods is to construct the one-electron reduced density matrix at the MP2 level~\cite{gruneis2011natural}. Diagonalization of the density matrix yields 
the natural orbitals and selection of only those with a sufficiently high
occupation numbers allows to significantly reduce the number of 
single-particle states involved in the subsequent CC calculation,
thus reducing the overall computational cost. The possibility to construct natural orbitals in FHI-aims or CC-aims is therefore a very desirable
feature. 
\rev{
Alternative approaches to reduce the computational cost of CC methods for larger molecules and periodic structures include embedding methods and local CC schemes. Embedding approaches are particularly valuable when a small region requiring highly accurate CC treatment can be identified, while remaining atoms are treated with less expensive methods such as DFT~\cite{libisch2014embedded}, MP2~\cite{shi2025accurate}, or RPA~\cite{schafer2021surface}.
Local CC methods, including local natural orbital (LNO)~\cite{nagy2018optimization} and the domain based local pair natural orbital (DLPNO)~\cite{liakos2015exploring}
CC approaches directly address the steep computational scaling by localizing molecular orbitals and considering only those contributing most significantly 
to the correlation energy. By approximating electronic correlation as local, these methods achieve nearly linear system-size scaling. Recently, the
LNO-CC framework has been expanded to periodic solids~\cite{ye2024periodic}, paving the way to broader application of coupled-cluster methods in more complex materials. While these 
frameworks enable treatment of systems inaccessible to canonical CC 
methods
~\cite{al2021interactions,nagy2019approaching,sandler2021accuracy,szabo2023linear}
, errors from the local approximation can become 
problematic for larger systems where long-range interactions are significant.
Despite these advances in approximate methods, canonical CC implementations remain essential for rigorous benchmarking and method development. 
The systematic hierarchy of CC approximations provides well-defined reference data that are crucial for assessing the accuracy of emerging electronic 
structure methods, including local and embedding approaches, as well as machine learning potentials and density functional approximations. 
The development and optimization of canonical CC methods for both molecules and solids, as presented in this work and now 
accessible to all FHI-aims users, therefore continues to play a vital role in advancing the field of computational electronic structure theory.
}

\subsection*{Acknowledgements}
MS acknowledges support by his TEC1p Advanced Grant (the European Research Council (ERC) Horizon 2020 research and innovation programme, grant agreement No. 740233.

\newpage

\section{All-Electron \textit{GW} for Finite Systems}
\sectionauthor[1,2]{Volker Blum}
\sectionauthor[3,4,a]{Fabio Caruso}
\sectionauthor[5]{\textbf{ *Dorothea Golze}}
\sectionauthor[5]{Jannis Kockl\"auner}
\sectionauthor[5]{Moritz Leucke}
\sectionauthor[5]{Qinglong Liu}
\sectionauthor[5]{Ram\'{o}n L. Panadés-Barrueta}
\sectionauthor[6,7,b]{\textbf{ *Xinguo Ren}}
\sectionauthor[8,9,10,11]{Patrick Rinke}
\sectionauthor[12]{Yanyong Wang}
\sectionlastauthor[13]{Tong Zhu}

\sectionaffil[1]{Thomas Lord Department of Mechanical Engineering and Materials Science, Duke University, Durham, NC 27708, USA}
\sectionaffil[2]{Department of Chemistry, Duke University, Durham, NC 27708, USA}
\sectionaffil[3]{Theory Department (since 1/1/2020: The NOMAD Laboratory), Fritz Haber Institute of the Max Planck Society, Faradayweg 4-6, D-14195 Berlin, Germany}
\sectionaffil[4]{Christian-Albrechts-Universit\"at zu Kiel, 24118 Kiel, Germany}
\sectionaffil[5]{Faculty of Chemistry and Food Chemistry, Technische Universit\"at Dresden, 01062 Dresden, Germany}
\sectionaffil[6]{Institute of Physics, Chinese Academy of Sciences, 100190 Beijing, China}
\sectionaffil[7]{The NOMAD Laboratory at the Fritz Haber Institute of the Max Planck Society, Faradayweg 4-6, D-14195 Berlin, Germany}
\sectionaffil[8]{Department of Applied Physics, Aalto University, P.O. Box 11000, FI-00076 Aalto, Finland}
\sectionaffil[9]{Physics Department, TUM School of Natural Sciences, Technical University of Munich, Garching, Germany}
\sectionaffil[10]{Atomistic Modelling Center, Munich Data Science Institute, Technical University of Munich, Garching, Germany}
\sectionaffil[11]{Munich Center for Machine Learning (MCML), Munich, Germany}
\sectionaffil[12]{CAS Key Laboratory of Quantum
Information, University of Science and Technology of China,
Hefei, Anhui 230026, China}
\sectionaffil[13]{University of Toronto, Toronto, Ontario M5S 1A4, Canada}

\sectionaffil[*]{Coordinator of this contribution.}
\rule[0.25ex]{0.35\linewidth}{0.25pt}

\sectionaffil[a]{{\it Current Address:} Christian-Albrechts-Universit\"at zu Kiel, 24118 Kiel, Germany}
\sectionaffil[b]{{\it Current Address:} Institute of Physics, Chinese Academy of Sciences, Beijing, 100190, China}




\subsection*{Summary}
\label{SecGWFinite}
\setlength{\intextsep}{0.0pt}%
\setlength{\columnsep}{8pt}%

\begin{figure}[t]
    \centering
    \includegraphics[width=0.38\linewidth]{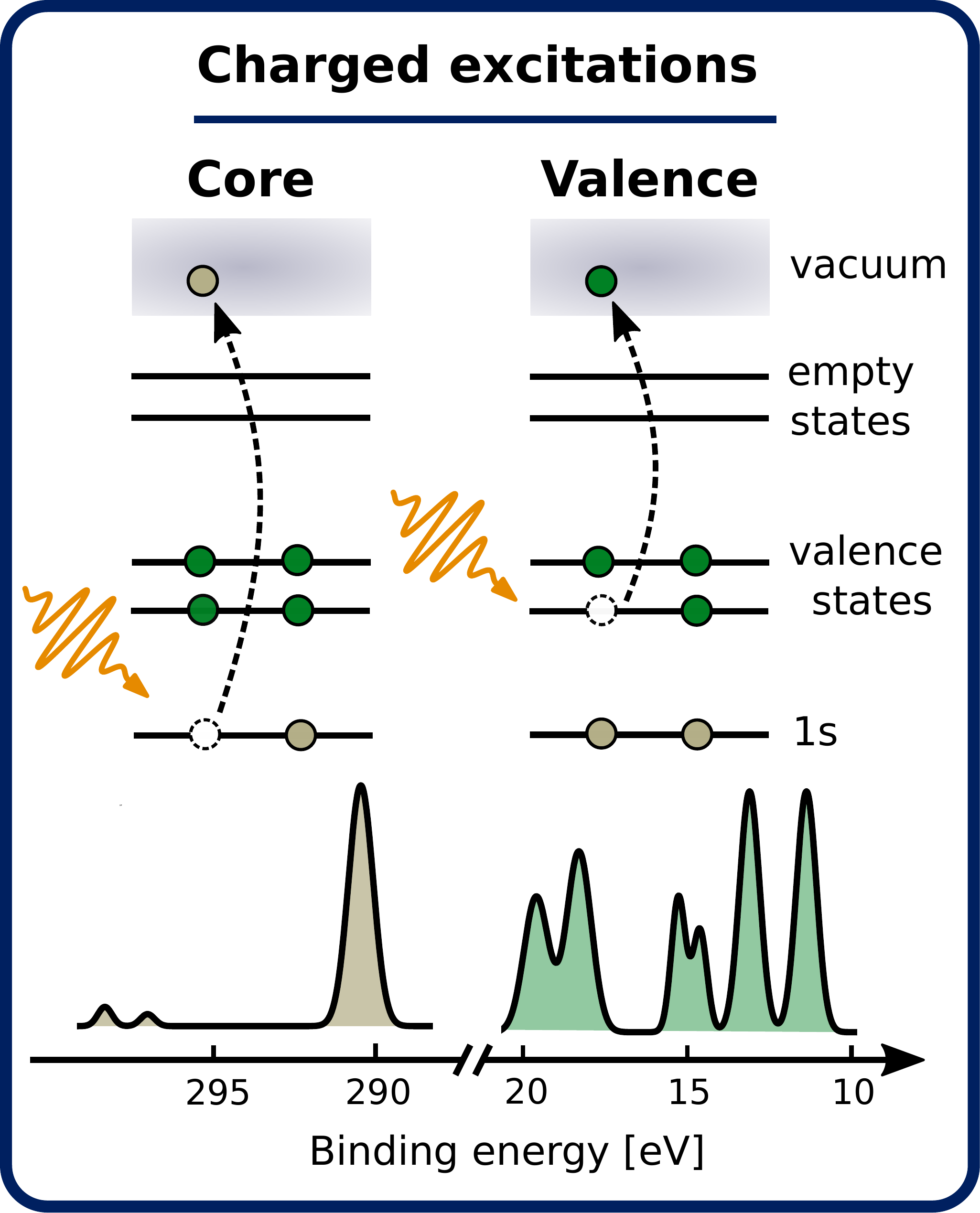}
    \caption{The $GW$ implementation in FHI-aims for finite systems spans the entire energy spectrum, from deep core to valence excitations.} \label{fig:toc_gw}
\end{figure}

The $GW$ approximation~\cite{Hedin1965} to Hedin's equations is a widely used method for predicting charged excitations, as measured by direct or inverse photoemission spectroscopy. While $GW$ was initially applied to compute the band structures of materials, its use has since expanded to finite systems~\cite{Golze2019}. For atoms, molecules or clusters, our all-electron $GW$ implementation is suitable for the computation of deep core excitations with energies larger than 100~eV~\cite{Golze2018,Golze2020,Li2022}, semi-core and valence excitations as well as electron affinities and fundamental gaps, see Figure~\ref{fig:toc_gw}. Additionally, the total spectral function can be computed, including both quasiparticle excitations and satellite features.

FHI-aims offers different $GW$ flavors for finite systems, including single-shot $G_0W_0$ with and without a Hedin shift in the Green's function, fully self-consistent, partially self-consistent, and eigenvalue self-consis\-tent schemes. Scalar relativistic effects and spin-orbit coupling can both be taken into account. The frequency integration of the self-energy is performed with one of two full-frequency methods: imaginary frequencies and analytic continuation or contour deformation. Our implementation scales $O(N^4)$ with respect to system size $N$, enabling calculations up to 200-250 atoms on current high-performance computing (HPC) platforms. Furthermore, FHI-aims offers the following vertex corrections to go beyond $GW$: second-order screened exchange and full second-order self-energy corrections and cumulant expansion of the Green's function.

\subsection*{Current Status of the Implementation}

\subsubsection{\textit{GW} Quantities: Self-Energy, Spectral Function and Quasiparticle Energies}

The self-energy $\Sigma$ is the key quantity in $GW$, encapsulating quantum mechanical effects of correlation and exchange between the excited electron or hole and the surrounding electrons. In practice, $GW$ is often performed as a first-order perturbation (denoted $G_0W_0$), where the self-energy is given by
\begin{equation}
  \Sigma(\mathbf{r},\mathbf{r}',\omega)=
\frac{i}{2\pi}\int  d\omega'  e^{i\omega'\eta} 
G_0(\mathbf{r},\mathbf{r}',\omega+\omega')W_0(\mathbf{r},\mathbf{r}',\omega').
\label{eq:selfenergy}
\end{equation}
$G_0$ denotes the mean-field Green's function and $W_0$ is the screened Coulomb interaction in the random phase approximation (RPA), and $\eta$ a small positive infinitesimal. $G_0$ and $W_0$ are computed from the mean-field single-particle orbitals $\{\psi_n\}$ and corresponding eigenvalues $\{\varepsilon_n\}$, typically obtained from Kohn-Sham density functional (KS-DFT)\footnote{A less common starting point alternative is Hartree-Fock theory.}. The poles of the full propagator $G=G_0 + G_0\Sigma G$ correspond to the electron removal and addition energies measured in direct and inverse photoemission spectroscopy, respectively. For more details on the $GW$ theory and basic equations, we refer the reader to a recent review article~\cite{Golze2019}. 

\begin{figure}[t]
    \centering
    \includegraphics[width=0.99\linewidth]{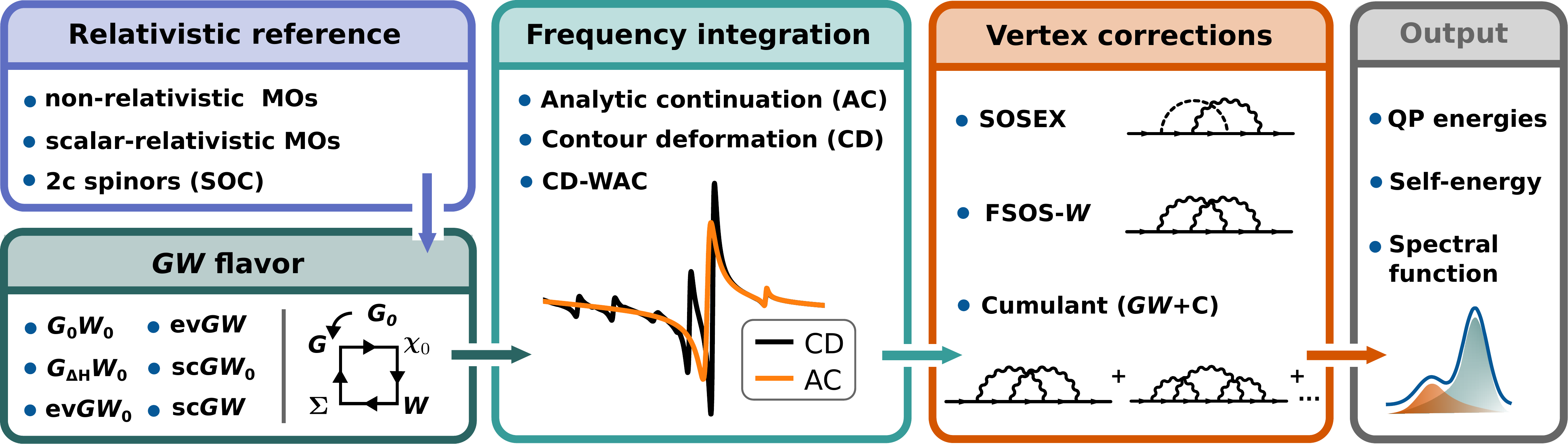}
    \caption{Workflow and choices available for \textit{GW} calculations of finite systems with FHI-aims.}
    \label{fig:implementation}
\end{figure}
A link to the measured photocurrent can be approximately established via the spectral function $A(\omega)$  
\begin{equation}
   A(\omega) = \sum_m \frac{1}{\pi}\left|\mathrm{Im}\bra{\psi_m}G(\omega)\ket{\psi_m}\right| = \frac{1}{\pi}\sum_m\frac{\left|\mathrm{Im}\Sigma_m(\omega)\right|}{\left[\omega-\varepsilon_m-\left(\mathrm{Re}\Sigma_m(\omega)-v_m^{xc}\right)\right]^2+\left[\mathrm{Im}\Sigma_m(\omega)\right]^2}
   \label{eq:spectral}
\end{equation}
where $m$ runs over all states and where $v^{xc}_m=\Braket{\psi_m|v^{xc}|\psi_m}$ is the exchange-correlation potential from DFT and $\Sigma_m=\Braket{\psi_m|
\Sigma(\omega)|\psi_m}$ denotes the projection into the KS orbital basis. The main peaks in the spectral function are the quasiparticle (QP) energies $\varepsilon_{n}^{\mathrm{QP}}$. Additionally, the spectral function includes satellite features, which arise from many-body effects such as plasmon or shake-up processes. 

For QP energies, solutions to Equation~\eqref{eq:spectral} simplify to the quasiparticle equation:
\begin{equation}
\label{eq:qpe}
  \varepsilon_{n}^{\mathrm{QP}}=\varepsilon_{n}+\mathrm{Re}\Sigma_n\left(\varepsilon_{n}^{\mathrm{QP}}\right)-v^{\rm xc}_n.
\end{equation}
Equation~\eqref{eq:qpe} is a non-linear equation, which is solved iteratively in FHI-aims. To avoid recalculating the self-energy in each iteration of Equation~\eqref{eq:qpe}, a linearized approach, referred to as the $Z$-shot method, is also available~\cite
{Golze2019}. Additionally, the output of the self-energy and the spectral function in a specified frequency range can be requested. The workflow and the available choices are summarized in Figure~\ref{fig:implementation}. 

\subsubsection{Relativistic Reference}

Relativistic effects are introduced into the $GW$ equations through the reference state. In FHI-aims, users can choose between non-relativistic or scalar-relativistic molecular orbitals (MOs) and 2-component (2c) spinors. The atomic zeroth-order regular approximation (ZORA) is the default scalar-relativistic method. For non-relativistic calculations and those using atomic ZORA in preceding DFT steps, relativistic corrections can be applied as a post-processing step for 1s core-level excitations, referencing fully relativistic calculations~\cite{Keller2020}. To include spin-orbit coupling (SOC), we recently implemented a 2c-$GW$ scheme~\cite{Aryasetiawan2008}. Our 2c-$GW$ method~\cite{Liu2024} supports 2c spinors from different levels of theory, with the perturbative second-variational SOC implementation\cite{huhn2017} currently used by default. 

\subsubsection{\textit{GW} Flavors}

FHI-aims offers various $GW$ flavors beyond the $G_0W_0$ default. In eigenvalue self-consistency, the eigenvalues are iteratively updated in $G$ (ev$GW_0$) or in both $G$ and $W$ (ev$GW$). Eigenvalue self-consistency increases the computational cost relative to $G_0W_0$. To address this, we recently implemented a Hedin shift in the Green's function ($G_{\Delta\mathrm{H}}W_0$)~\cite{Li2022}, which approximates the ev$GW_0$ scheme with minimal additional computational cost compared to $G_0W_0$. The most conceptually rigorous and computationally demanding approach is fully self-consistent $GW$, denoted as sc$GW$~\cite{Caruso2012,Caruso2013}. In sc$GW$, all four quantities -- $G$, the polarizability $\chi_0$, $W$, and $\Sigma$ -- are iterated until self-consistency in the Green's function is achieved, as shown in the dark-green box in Figure~\ref{fig:implementation}. Additionally, a partial self-consistent scheme (sc$GW_0$) is available, where $W$ is fixed at the $W_0$ level, and only $G$ is determined self-consistently. The approximate self-consistent schemes reduce the dependence on the mean-field starting point, but only sc$GW$ removes it fully~\cite{Caruso2012}. In addition to the QP energies, sc$GW_0$ and sc$GW$ provide also access to the total energies~\cite{Caruso2012}. \rev{Further details on differences, advantages and disadvantages of the various $GW$ flavors can be found in Ref.~\cite{Golze2019}.}

\subsubsection{Frequency Integration}

FHI-aims offers two techniques, \rev{both with canonical scaling with respect to system size,} to solve the frequency integral in Equation~\eqref{eq:selfenergy}: the analytic continuation (AC) of the self-energy~\cite{ren2012} and the contour-deformation (CD) technique~\cite{Golze2018}. For a comprehensive comparison of both methods, we refer the reader to Ref.~\cite{Golze2019}, restricting the description here to the key ideas. The AC scheme is the default method~\cite{ren2012}, available for all $GW$ flavors. In this approach, the self-energy $\Sigma$ is computed on the imaginary frequency axis and then analytically continued to the real-frequency axis by fitting the matrix elements $\Sigma_n(i\omega)$ to a multipole model function. Two common models are available for this purpose: a 2-pole model and a Pad\'{e} approximant.

Conversely, in the CD approach the integrand in Equation~\eqref{eq:selfenergy} is expanded in the complex plane. The self-energy matrix elements $\Sigma_n(\omega)$ are obtained as the sum of two terms: an integral term along the imaginary frequency axis and a residue term that depends on real-frequency arguments. The calculation of the latter typically increases the computational cost compared to the AC approach. To mitigate the computational demands, we recently implemented the CD-WAC method~\cite{Panades2023} (CD with $W$ analytic continuation), which applies an AC for the screened Coulomb interaction $W$ in the residue term. CD and CD-WAC are only implemented for $G_0W_0$, $G_{\Delta\mathrm{H}}W_0$, ev$GW_0$ and ev$GW$ and are currently only available for non- and scalar-relativistic references. 

\rev{Most recently~\cite{Kocklaeuner2025}, a third method has been introduced: an exact frequency treatment in which the frequency convolution is performed analytically, yielding a sum-over-states expression for the correlation part of the self-energy. The RPA excitation energies and transition moments entering this expression are obtained by solving Casida-like equations. This approach is substantially more computationally demanding than AC, CD, or CD-WAC and is therefore not intended for routine calculations. Instead, it serves as a high-accuracy reference for validating more scalable frequency-integration techniques.}

\subsubsection{Algorithm, Performance and Numerical Accuracy}

The AC~\cite{ren2012}, CD~\cite{Golze2018} and CD-WAC~\cite{Panades2023} implementations are based on numeric atom centered orbitals (NAOs) and on the resolution-of-the-identity (RI) approach, which reformulates the four-center two-electron Coulomb integrals as product of two and three-center integrals. Our implementation employs the Coulomb fitting scheme (RI-V)~\cite{Vahtras1993}, which is the most widely used and most accurate RI technique. The AC and CD-WAC schemes scale $O(N^4)$ with respect to system size $N$, while the scaling of CD depends on the energy range. Specifically, CD scales as $O(N^4)$ for valence states, but $O(N^5)$ for deep core-levels~\cite{Golze2018}. The AC, CD and CD-WAC implementations are parallelized using a standard message passing interface (MPI) and scale well up to 2000 - 3000 CPU cores. More details on the parallel performance can be found in Ref.~\cite{Golze2018}. With the CD method, core-level calculations are feasible for systems with up to 100–120 atoms on current HPC platforms. With CD-WAC and AC, both core and valence level calculations can be performed for systems with up to 200–250 atoms. \rev{We refer the reader to Refs.~\cite{Golze2018,Panades2023} for performance benchmarks.}

Among the three implementations, the CD method is the most accurate, reproducing all features of the self-energy with deviations in the sub-millielectronvolt range when compared to an exact frequency treatment, i.e., the full diagonalization of Casida-like equations at the RPA level~\cite{Golze2018}. At the $G_0W_0$ level, CD-WAC approximates the CD results with errors $<1$ meV for valence states and $<5$ meV for 1s core levels~\cite{Panades2023}.
While AC is computationally most efficient and has the smallest computational prefactor, it fails to accurately describe the self-energy for deep core and semi-core levels~\cite{Golze2018}. For frontier orbital excitations, however, AC with a suitable Pad\'{e} model reproduces exact frequency calculations to within 3~meV on average~\cite{Setten2015}.

\subsubsection{Beyond-\textit{GW}: Vertex Corrections}

FHI-aims also offers vertex corrections to the self-energy, specifically: second-order screened exchange (SOSEX)~\cite{Ren2015} and an approximate full second-order self-energy in terms of $W$ (FSOS-$W$)~\cite{Wang2021}. Additionally, it includes propagator corrections, such as the cumulant expansion of the Green's function $G$~\cite{Kocklaeuner2024}. The corresponding Feynman diagrams are displayed in Figure~\ref{fig:implementation}. Diagrammatically, SOSEX adds a second-order exchange diagram to the $GW$ self-energy, while $W$ is still evaluated at the RPA level. $G_0W_0+\mathrm{SOSEX}$ mitigates the dependence on the mean-field reference compared to $G_0W_0$ and can improve the relative energy positions in molecular valence photoemission spectra~\cite{Ren2015}. FSOS-$W$ can be considered as a dynamic extension of SOSEX, replacing the bare Coulomb interaction (dashed line in Figure~\ref{fig:implementation}) in the diagram by the frequency-dependent screened Coulomb interaction $W$ (wiggly line). The $GW$+FSOS-$W$ self-energy was also referred to as $GW\Gamma^{(1)}$ in Ref.~\cite{Wang2021} or $GW$+$G_3W_2$ in Ref.~\cite{Bruneval2024}. Currently, FHI-aims implements an approximate single-shot version of $GW\Gamma^{(1)}$ ($G_0W_0\Gamma_0^{(1)}$) and neglects all terms in the self-energy that involve three occupied or three virtual MOs~\cite{Bruneval2024}. $G_0W_0\Gamma_0^{(1)}$ is very similar in performance to $G_0W_0$+SOSEX. Both schemes are only implemented with AC and exhibit an $O(N^5)$ scaling.

The cumulant expansion improves the prediction of  satellite positions and intensities in the spectral function~\cite{Aryasetiawan1996,Guzzo2011,Gatti2013,Caruso2015,Zhou2015,Zhou2018}, which are poorly described in $GW$. The $GW+C$ approach is based on an exponential
ansatz for the Green's function, which adds an infinite number of bosonic diagrams to the propagator $G$, while $W$ remains again fixed at the RPA level. The FHI-aims $G_0W_0+C$ implementation~\cite{Kocklaeuner2024,Kocklaeuner2025} employs the spectral representation of the self-energy, as previously suggested~\cite{Aryasetiawan1996,Zhou2015}. In this approach, the cumulant expansion is performed as a post-processing step after the computation of the $GW$ self-energy, without significantly increasing the computational cost. The scaling of $G_0W_0+C$ is determined by the $GW$ implementation. $G_0W_0+C$ is implemented exclusively with CD and CD-WAC because satellites are related to pole-features in the self-energy, which are inadequately captured with AC.

\subsection*{Usability and Tutorials}
Comprehensive tutorials on how to perform $GW$ calculations for finite systems with FHI-aims are available at \url{https://fhi-aims-club.gitlab.io/tutorials/rpa-and-gw-for-molecules-and-solids}, covering general aspects~\cite{tutorialgeneral} and core-level spectroscopy specifically~\cite{tutorialcore}. The finite-system $GW$ implementation in FHI-aims is widely used and has contributed to the development of well-established molecular benchmark sets, such as GW100 \cite{Setten2015} and CORE65 \cite{Golze2018}. Additionally, it has been employed to create comprehensive computational databases for machine learning applications. One example is the GW5000 dataset, which includes frontier orbital excitations of molecules with up to 100 atoms \cite{Stuke2020}. Another example is the 1s core-level database, containing more than 15,000 entries for CHO-containing molecules and clusters with up to $\sim$110 atoms \cite{Golze2022}.

\begin{wrapfigure}[24]{r}{0.45\textwidth}
    \centering
    \includegraphics[width=\linewidth]{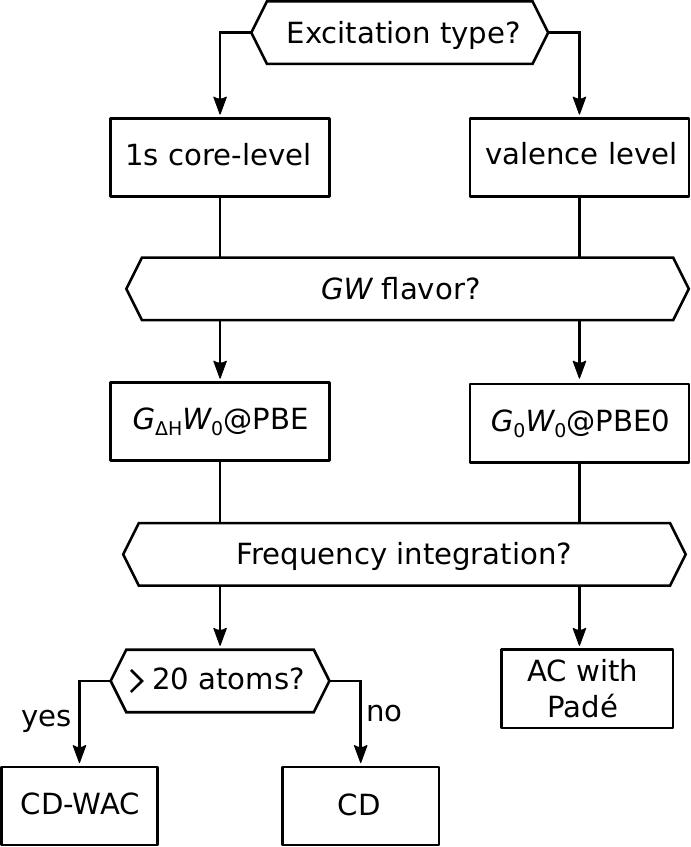}
    \caption{Choices regarding $GW$ flavor and frequency integration depending on the excitation type for FHI-aims calculations.}
    \label{fig:useability}
\end{wrapfigure}

For QP energies, the vertex-corrected $G_0W_0+$SOSEX and $G_0W_0\Gamma_0^{(1)}$ schemes outperform $G_0W_0$ when using conventional Perdew-Burke-Ernzerhof~\cite{Perdew1996} (PBE) and PBE0~\cite{adamo1999,ernzerhof1999} starting points. However, $G_0W_0$ with judiciously chosen starting points provides a similarly accurate, yet computationally more efficient alternative~\cite{Ren2015,Wang2021,Bruneval2024}. For satellites, we recommend $G_0W_0+C$, though we note it yields QP energies at the $G_{\Delta \text{H}}W_0$ level. To reduce the starting point dependence, we recommend the partially self-consistent schemes ev$GW_0$ or sc$GW_0$~\cite{Golze2019}, as ev$GW$ and sc$GW$ underscreen and thus overestimate excitation energies~\cite{Marom2012,Caruso2016,Li2022}. 

Safe recommendations for finite-system $GW$ calculations in FHI-aims, considering both core and valence levels, are summarized in Figure~\ref{fig:useability}. Starting with the valence excitations, we suggest to perform $G_0W_0$ on-top of KS-DFT calculations with the PBE0 functional, which we denote as $G_0W_0@$PBE0. The expected average deviation from experiment is around 0.2~eV~\cite{Caruso2016,Ren2015,Golze2019}. Similar or even better performance is expected when using ev$GW_0$@PBE~\cite{Schambeck2024}, albeit at the cost of increased computational cost. Using AC with a Pad\'{e} model with 16 parameters is usually sufficient for frontier orbitals~\cite{Setten2015} and deeper valence states with binding energies $<20$~eV~\cite{Wilhelm2021}.  Alternatively, CD can be used, which has a slightly higher computational prefactor for valence states.

Turning to 1s core-levels, $G_{\Delta\mathrm{H}}W_0@$PBE is the optimal choice, yielding deviations of 0.3~eV and 0.2~eV for absolute and relative 1s core-level excitations, respectively~\cite{Li2022}. ev$GW_0$@PBE performs equally well, but is at least an order of magnitude more expensive for core-level calculations. Alternatively, we proposed a $G_0W_0@$PBEh$(\alpha=0.45)$ approach~\cite{Golze2020}, using a hybrid functional with 45\% of exact exchange. However, the optimal amount of exact exchange varies with atomic number~\cite{Golze2020} and was specifically optimized for second-row elements, making it unsuitable for heavier elements. We thus recommend $G_{\Delta\mathrm{H}}W_0@$PBE over $G_0W_0@$PBEh$(\alpha=0.45)$. The CD method is mandatory for deep core states. For systems larger than 20 atoms, the CD-WAC approach will be computationally more efficient than CD. 

Another important consideration is the choice of the basis set. As with other correlated methods, $GW$ converges slowly with respect to basis set size. The results, even with the largest localized basis sets, are not fully converged, as demonstrated in Ref.~\cite{Golze2019}. A basis set extrapolation procedure must be employed, which is detailed in the tutorials~\cite{tutorialgeneral,tutorialcore}. For valence states, the NAO-VCC-$n$Z basis set family~\cite{Zhang2013} can be used. Alternatively, Gaussian basis sets, which are treated numerically in FHI-aims, such as the Dunning basis sets cc-pV$n$Z~\cite{Dunning1989, Wilson1996} or the def2-TZPV/def2-QZVP basis sets~\cite{Weigend2005} extrapolate well in our experience~\cite{Setten2015,Golze2019}. For 1s states, the cc-pV$n$Z series works sufficiently~\cite{Golze2020}, but core-rich Gaussian basis sets such as ccX-$n$Z~\cite{Ambroise2021, Mejia2022} have shown a better convergence behavior~\cite{Mejia2022}. If basis set extrapolation cannot be performed due to computational cost, or if the focus is on the spectral function rather than QP energies, we recommend using NAO-VCC-$5$Z for valence states and tier2+STO3 for deep core states. The latter combines NAOs with Slater-type orbitals (STOs), as described in the Supporting Information of Ref.~\cite{Yao2022}. These basis sets typically reproduce extrapolated results within 0.1 eV.


\subsection*{Future Plans and Challenges}
Future developments will improve the computational efficiency and add new features. On the feature side, core-level calculations are currently restricted to 1s excitations, and we plan to extend the implementation to SO-coupled $p$, $d$ and $f$ excitations in the near future. This will require adding CD and CD-WAC to the 2c-$GW$ implementation, \rev{with an experimental 2c-CD version already available.} Additionally, we plan to enable the usage of 2c-$GW$ in combination with the quasi-four-component (Q4C)~\cite{Zhao2021} scheme, which is also available in FHI-aims. The Q4C spinors should provide a more accurate relativistic reference compared to the second-variational SOC approach.  Furthermore, we plan to extend the Sternheimer RPA implementation~\cite{Peng2023} to $GW$ for benchmarking purposes.  

On the performance side, work on low-scaling $GW$ algorithms is on-going. Low-scaling RPA is already functional and the extension to $GW$ is in progress. Our low-scaling RPA and $GW$ algorithms rely on the separable RI scheme, which was recently implemented~\cite{Delesma2024}. The separable RI coefficients are currently obtained as fit to the RI-V coefficients. To reduce the computational prefactor, we plan to replace the Coulomb metric in RI-V with the similarly accurate truncated Coulomb metric. Furthermore, the GPU acceleration for all steps in the low-scaling algorithms is currently ongoing.

\subsection*{Acknowledgements}
D. Golze acknowledges financial support by the Deutsche Forschungsgemeinschaft (DFG, German Research Foundation) for funding via the Emmy Noether Programme (project number~453275048) and for funding within the CRC 1415 (Chemistry of Synthetic Two-Dimensional Materials, project number 417590517) as well as within the RTG 2861 (Research Training Group, project number 491865171). D.G. also acknowledges support by the Innovation Study Exa4GW. 
The Innovation Study Exa4GW has received funding through the Inno4scale project, which is funded by the European High-Performance Computing Joint Undertaking (JU) under Grant Agreement No 101118139. The JU receives support from the European Union's Horizon Europe Programme.
P. Rinke acknowledges the European Union’s Horizon 2020 Research and Innovation Program for financial support under Grant No. 951786 (Nomad Center of Excellence). We further acknowledge CSC–IT Center for Science (Finland), the J\"ulich Supercomputer Center (Germany), the Paderborn Center For Parallel Computing (PC2, Germany) and the Aalto Science-IT project for computational resources. X. Ren acknowledges the funding support by the National Natural Science Foundation of China (Grant Nos 12374067 and 12188101) and the National Key Research and Development Program of China (Grant Nos 2022YFA1403800 and 2023YFA1507004).




  


\newcommand{\bk}{\mathbf{k}}
\newcommand{\bq}{\mathbf{q}}
\newcommand{\XR}[1]{\color{red}{\bf #1 }}
\newcommand{\MYZ}[1]{\textcolor{blue}{\bf #1}}




\newpage

\section{{\it GW} Calculations for Periodic Systems}

\sectionauthor[1,2]{Volker Blum}
\sectionauthor[3]{Francisco A. Delesma}
\sectionauthor[1]{Uthpala Herath}
\sectionauthor[4]{Hermann Lederer} 
\sectionauthor[5]{Florian Merz}
\sectionauthor[6,11,a]{\textbf{ *Xinguo Ren}}
\sectionauthor[7,8,9,10]{Patrick Rinke}
\sectionauthor[11]{Matthias Scheffler}
\sectionauthor[11,b]{Yi Yao}
\sectionlastauthor[6,11]{Min-Ye Zhang}

\sectionaffil[1]{Thomas Lord Department of Mechanical Engineering and Materials Science, Duke University, Durham, NC 27708, USA}
\sectionaffil[2]{Department of Chemistry, Duke University, Durham, NC 27708, USA}
\sectionaffil[3]{Faculty of Chemistry and Food Chemistry, Technische Universit\"at Dresden, 01062 Dresden, Germany}
\sectionaffil[4]{Max Planck Computing and Data Facility, Giessenbachstrasse 2, D-85748 Garching, Germany}
\sectionaffil[5]{Lenovo HPC Innovation Center, Meitnerstr. 9, D-70563 Stuttgart, Germany}
\sectionaffil[6]{Institute of Physics, Chinese Academy of Sciences, 3rd South Str. 8, Beijing 100190, China}
\sectionaffil[7]{Department of Applied Physics, Aalto University, P.O. Box 11000, FI-00076 Aalto, Finland}
\sectionaffil[8]{Physics Department, TUM School of Natural Sciences, Technical University of Munich, Garching, Germany}
\sectionaffil[9]{Atomistic Modelling Center, Munich Data Science Institute, Technical University of Munich, Garching, Germany}
\sectionaffil[10]{Munich Center for Machine Learning (MCML), Munich, Germany}
\sectionaffil[11]{The NOMAD Laboratory at the Fritz Haber Institute of the Max Planck Society, Faradayweg 4-6, D-14195 Berlin, Germany}

\sectionaffil[*]{Coordinator of this contribution.}
\rule[0.25ex]{0.35\linewidth}{0.25pt}

\sectionaffil[a]{{\it Current Address:} Institute of Physics, Chinese Academy of Sciences, Beijing, 100190, China}
\sectionaffil[b]{{\it Current Address:} Molecular Simulations from First Principles e.V., D-14195 Berlin, Germany}




\subsection*{Summary}
The $GW$ approximation \cite{Hedin1965} is a state-of-the-art approach for computing single-electron excitation energies of real materials from first principles \cite{Hybertsen1986}. It can be applied to a large variety of chemical systems across different characteristics and dimensionalities \rev{(for a review of the underlying theory, see Refs.\cite{Reining2018,Golze2019})}.  Compared to the (generalized) Kohn-Sham density functional theory (KS-DFT), $GW$  has a much more rigorous foundation as an approach for determining single-election excitation energies and usually yields much better electronic band structures, particularly the band gaps. However, the improvement of $GW$ over KS-DFT comes with a significantly higher computational cost, which stimulates continuing efforts to improve the algorithms for $GW$ calculations, aiming at more efficient (and sufficiently accurate) implementations. Since its first application to real materials nearly four decades ago \cite{Hybertsen1986}, various $GW$ codes have been developed based on different numerical frameworks 
\cite{Rojas1995,Deslippe2012,Shishkin2006,Friedrich2010,Jiang2013,Zhu2021}. 
These implementations largely fall into those based on (augmented) plane waves and those based on local atomic orbitals. Typically, the plane-wave based implementations are more suitable for simulating periodic systems whereas atomic-orbital based implementations are preferred for molecular systems. 
The $GW$ method was implemented in FHI-aims \cite{Blum2009} at the early stage of its development \cite{ren2012}. The initial implementation was for finite systems \cite{ren2012}, as described in the previous Contribution \ref{SecGWFinite}. Here, we focus on the extension of the $GW$ implementation to periodic systems \cite{Ren2021}. The periodic implementation of $GW$ in the FHI-aims is based on the localized resolution of identity (termed RI-LVL) \cite{ihrig2015}. This significantly reduces the amount of integrals to evaluate and store, and makes the periodic $GW$ implementation with NAO basis sets feasible. However, there is a higher demand for the size and quality of the auxiliary basis sets to ensure the adequate accuracy of RI-LVL for periodic calculations. Furthermore, special care must be taken to deal with the singularity of the screened Coulomb interaction at the $\Gamma$ point. Currently, the periodic $GW$ implementation in FHI-aims allows us to compute the quasiparticle energies for extended systems with small or medium-sized unit cells with finite $\mathbf{k}$ points, and plot the band structures across the Brillouin zone (BZ).


\subsection*{Current Status of the Implementation}
The periodic $GW$ method in FHI-aims employs the canonical algorithm operating solely in the reciprocal space and (imaginary) frequency domain. For KS spin state $\psi^{\bk}_{n\sigma}$ with orbital energy $\epsilon_{n\sigma}^{\bk}$, the correlation self-energy from one-shot $G_0W_0$ is expressed as \cite{Ren2021}
\begin{equation}\label{eq:pgw-sigma-diagonal}
\Sigma^{\mathrm{c}}_{n}(\bk\sigma,i\omega) = -\frac{1}{2 \pi} \sum_{m,\bq} \sum_{\mu, \nu}
\int_{-\infty}^{\infty} \dd{\omega'}
\frac{C_{n,m,\sigma}^\mu(\bk, \bk-\bq) W^{\mathrm{c}}_{0, \mu\nu} \left(\bq, i \omega'\right) C_{m,n,\sigma}^\nu(\bk-\bq, \bk)}
{i \omega-i\omega'+E_{\mathrm{F}}-\epsilon_{m, \sigma}^{\bk-\bq}}
\end{equation}
where $C_{n,m,\sigma}^\mu$ are the RI decomposition coefficients in KS space and $W^{\mathrm{c}}_{0, \mu\nu}=W_{0, \mu\nu} - V_{\mu\nu}$ is the correlation part of screened Coulomb interaction expanded by auxiliary basis functions (ABFs). The summation runs over auxiliary basis indices $\mu, \nu$, band index $m$ and wave-vector $\bq$ sampled in the first BZ.
The frequency integral is evaluated numerically using standard quadrature grids.
The self-energy at real frequency is then obtained by analytically continuing the data on the imaginary axis and used to compute the quasi-particle (QP) energy $\epsilon^{\mathrm{QP}}$ of interest by solving the QP equation.

In the current implementation, RI-LVL plays an important role in reducing the time and memory cost for the RI triple coefficients in KS space, which are obtained by transforming those in atomic orbital (AO) basis with wave function expansion coefficients $\{c^{i}_{n\sigma}(\bk)\}$
\begin{equation}\label{eq:pgw-ricoeff-nm-k}
C_{n,m,\sigma}^\mu(\bk, \bk-\bq) = \sum_{ij} c^{i*}_{n\sigma}(\bk) c^{j}_{m\sigma}(\bk-\bq) \tilde{C}_{i,j}^\mu(\bk, \bk-\bq).
\end{equation}
The AO coefficients $\tilde{C}_{i,j}^\mu(\bk, \bk-\bq)$ are computed by Fourier transforming their real-space counterparts. In global RI, indices $i,j$ have to run over all the atomic basis with non-negligible Coulomb interaction with specific $\mu$, and the Fourier transformations with respect to $\bk$ and $\bk-\bq$ are coupled. The local approximation forces either $i$ or $j$ basis to be located at the same atom and unit cell as the auxiliary function $\mu$ in the real space expansion. As a result, the transformation involves only two nonzero sectors, each dependent on a single wave vector. This makes the storage of real-space triple coefficients affordable and the entire transformation efficient, hence allowing on-the-fly evaluation of $C_{n,m,\sigma}^\mu$ to avoid its expensive storage.

To handle the finite size effect and achieve fast convergence of $\bq$-point sampling in periodic $GW$, the singular nature of the Coulomb interaction near the $\Gamma$ point has to be carefully addressed. This singularity is involved in both calculations of the screened Coulomb interaction and the self-energy. In FHI-aims, the first case is approached by computing the dielectric matrix at the $\Gamma$ point in Coulomb eigenvector representation using relevant terms from $\bk\cdot\mathbf{p}$ theory. For the self-energy calculation, the divergence of screened Coulomb is circumvented by synthesizing it with a truncated Coulomb interaction as used in the Hartree-Fock and hybrid functional calculations.

\begin{figure}[ht]
	\centering
	\includegraphics[width=0.6\textwidth]{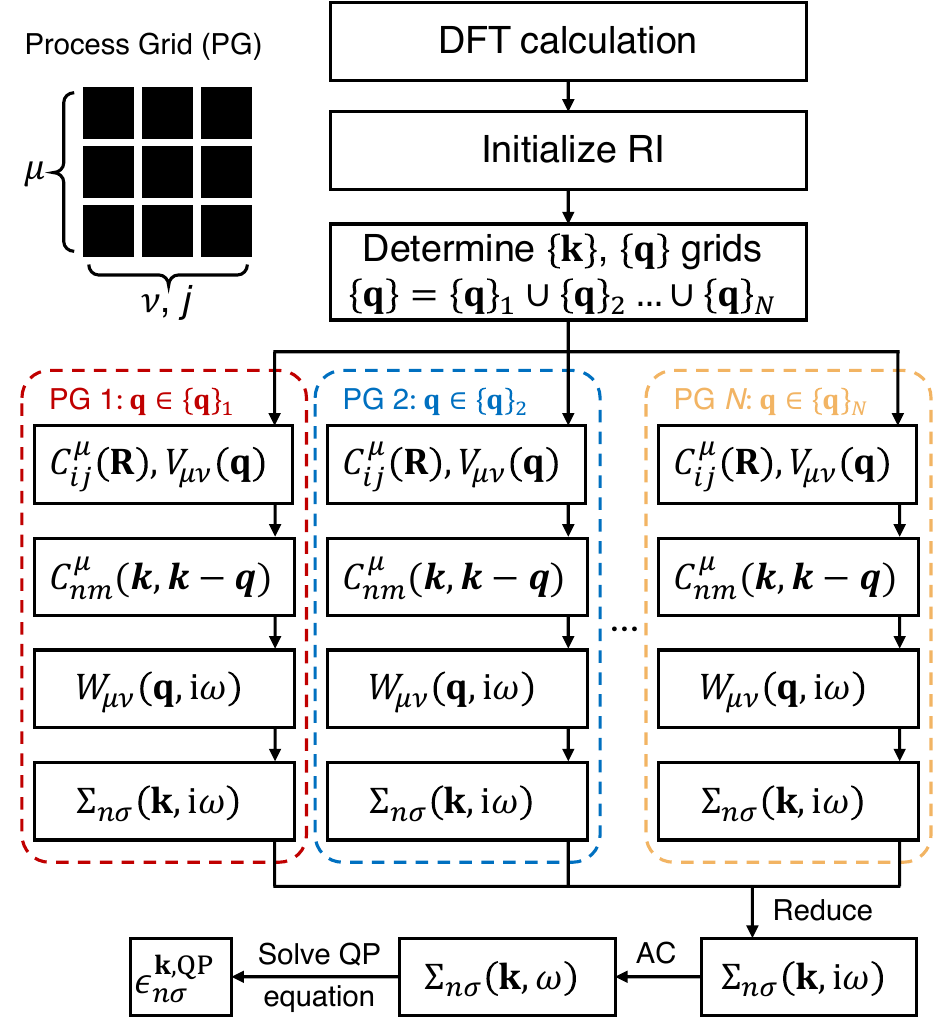}
	\caption{Workflow of canonical periodic $G_0W_0$ algorithm in FHI-aims}
	\label{fig:pgw-workflow}
\end{figure}

The overall cost for the self-energy evaluation of Eq.\eqref{eq:pgw-sigma-diagonal} scales as $\mathcal{O}(N^4 N_{\omega} N_{\bk} N_{\bq})$, with $N$ being the system size of the unit cell, and has a large prefactor due to the use of all unoccupied states. This high computational load necessitates an efficient and scalable parallelization strategy to fully leverage modern massively parallel computers. In FHI-aims, we have implemented a two-level parallelization scheme based on message passing interface (MPI) and its workflow is sketched in Figure \ref{fig:pgw-workflow}. Upon a converged DFT calculation and RI initialization, $\bk$- and $\bq$-grids required for $GW$ are determined. The $\bq$-grids are further grouped and distributed to a batch of process grids, each consisting of MPI tasks mapped to a square array. Within each grid, row ABF indices are distributed across process rows, while column ABFs and second AO indices of RI coefficients are distributed across process columns. The calculation of $W$ and contributions to $\Sigma$ is then performed in an embarassingly parallel manner among process grids. Finally, the self-energy operator is reduced from all tasks for analytic continuation and solution of QP equation. We note that the MPI tasks are scheduled to favor more batches of process grids while minimizing the load imbalance. This scheme allows FHI-aims to effectively handle both small systems using dense $\bq$ sampling and large ones with only a few $\bq$ points.

\subsection*{Usability and Tutorials}

To ensure periodic $GW$ is a reliable feature in FHI-aims, its accuracy and efficiency requires careful investigation.
The accuracy of our periodic $GW$ implementation has been benchmarked in a previous publication \cite{Ren2021} against the all-electron LAPW-based $GW$ code FHI-gap \cite{Jiang2013,Jiang2016}. \rev{It has also been compared recently with implementations in ABINIT, GPAW, and exciting for several representative systems with diverse characteristics \cite{Azizi2025}}. Among many aspects, the efficacy of RI-LVL, the BZ sampling convergence and the impact of one-electron basis are demonstrated. It is found that additional basis functions to the one-electron basis set for on-the-fly construction of ABFs are required to mitigate the error due to the local approximation to RI and obtain accurate band gaps. On top of the tier2 species defaults, FHI-aims is shown to deliver $G_0W_0$ band gaps for semiconductors, in similar accuracy ($\sim0.2$ eV) to those with local-orbital-enhanced LAPW basis from FHI-gap, except for challenging systems like ZnO and LiF. The agreement can be further improved by including Slater-type orbitals in the one-electron basis. We also note that the $\bk$-points on the regular $\bk$-grid of a self-consistent calculation and band paths are treated on equal footing through an on-the-fly Fourier transform for necessary RI coefficients, so that the band structure calculation is free of error arising from self-energy interpolation.


In order to obtain accurate $G_0W_0$ band gap results and band structures, it is essential to minimize the RI error due to the local approximation in RI-LVL. Although more basis functions in the ABFs construction can lead to better accuracy, it will considerably slow down the calculation. Therefore, it is of great interest to identify the optimal additional functions that lead to adequate accuracy gain, but with as few extra ABFs as possible. For this purpose, we have benchmarked the impact on $G_0W_0$ band structure due to different combinations of additional hydrogen-like functions in a few systems, from which promising candidates are selected and tested on a larger set of materials. It results in improved species defaults for periodic $GW$ calculations with optimal additional functions on top of the DFT defaults. They are available in the FHI-aims species collection under ``\texttt{\_gw}''-suffixed folders.

\begin{figure}[ht]
	\centering
	\includegraphics[width=0.8\textwidth]{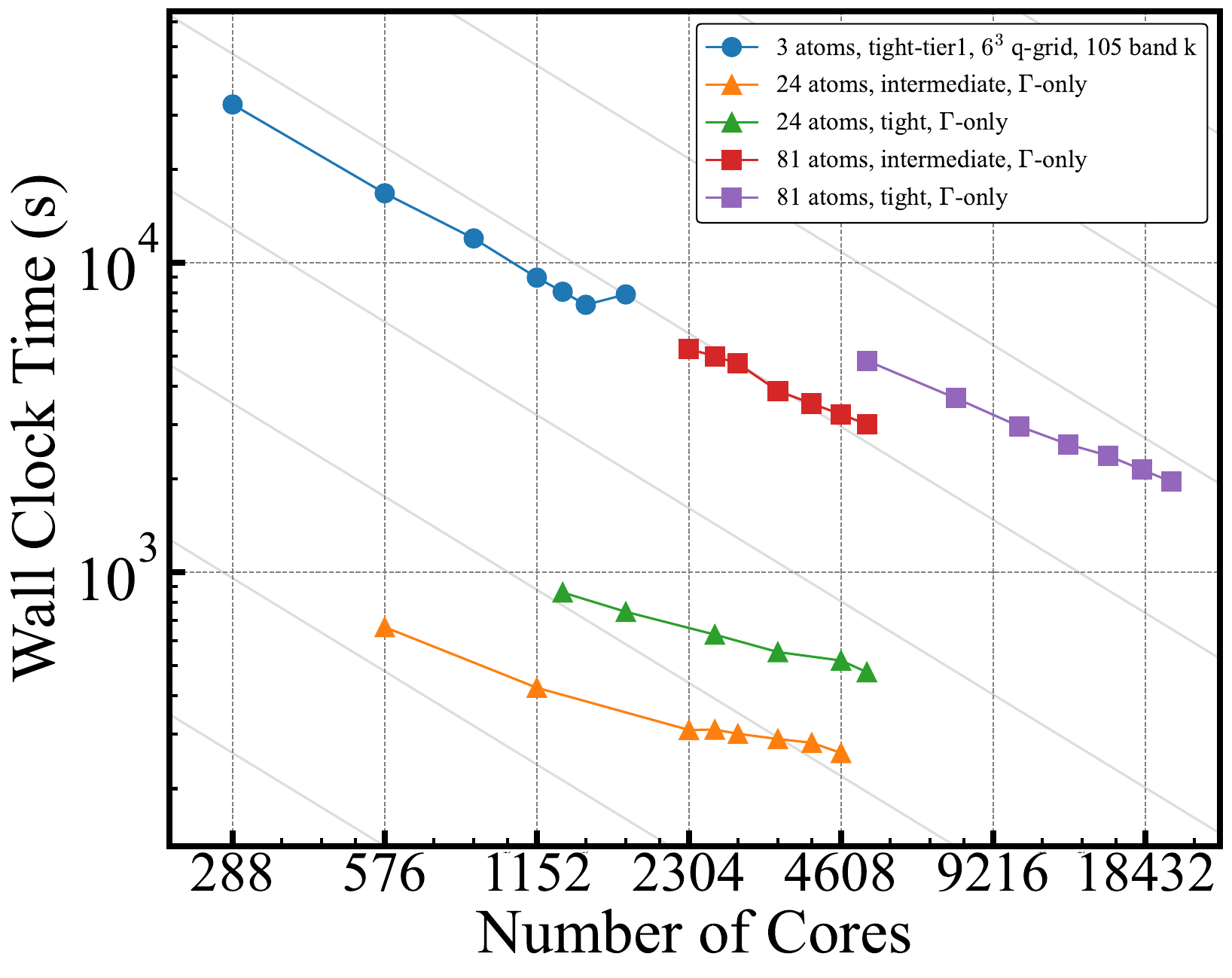}
	\caption{Strong scaling test of FHI-aims periodic $G_0W_0$ calculations for \ce{ZrO2} unit cell (3 atoms) and different super cells. The test is conducted on the Raven system (MPCDF, Germany). Each computer node is equipped with 72 cores (Intel Platinum 8360Y processors) and 256 GB memory.}
	\label{fig:pgw-zro2-strong-scaling}
\end{figure}

In terms of performance, the aforementioned MPI parallelization scheme is designed to run FHI-aims periodic $GW$ efficiently on massive cores for both small- and large-size systems. We have conducted a strong scaling test for the implementation using different sizes of \ce{ZrO2} cell, as shown in Figure \ref{fig:pgw-zro2-strong-scaling}. For the 3-atom unit cell with tight-tier1 basis (77 AOs, 2220 ABFs) and 105 band k-points, we observe nearly perfect scaling up to 1440 MPI tasks (20 nodes). Calculation of an 81-atom super cell with tight basis (3429 AOs, 31617 ABFs) and only $\Gamma$ point scales reasonably well up to 20736 MPI tasks (288 nodes).

To facilitate long calculation times for systems with many $\bq$-points, for example for when there is time-limit restrictions on HPC resources, a restart functionality has been implemented for periodic $GW$ calculations, controlled by the \texttt{restart\_periodic\_gw} keyword.
When activated, each MPI task generates a separate checkpoint file, which records the number of $\bq$ points processed locally by the task (see Figure \ref{fig:pgw-workflow}), along with the accumulated contribution to the correlation self-energy. If a job is interrupted, restarting the calculation with the same number of MPI tasks allows the checkpoint files to be loaded, enabling the calculation to resume from the next unprocessed $\bq$-point.

An informative tutorial for periodic $GW$ is available on the official site as part of the tutorial series ``GW and BSE for molecules and solids'' \cite{tutorial}. It covers a comprehensive list of technical factors mentioned above for converging $G_0W_0$ calculations, and walks through the procedure in great detail. \rev{In particular, we include a worked example for monolayer MoS$_2$, a prototypical two-dimensional material, to illustrate how the $GW$ approach can be applied in the context of low-dimensional functional materials.}


\subsection*{Future Plans and Challenges}
Several directions can be explored for the further development of the periodic $GW$ functionalities in FHI-aims, which will be briefly discussed below.

First, the current periodic $GW$ implementation is based on the $\mathbf{k}$-space formalism, and its $O(N^4)$ scaling limits its application to relatively small systems. \rev{To treat real-world materials problems such as defects and interfaces, where large supercells are typically required, efficient low-scaling implementations are highly desirable, analogous to the low-scaling strategies developed for hybrid functionals (see Contrib.~\ref{ChapHybrids}).} In principle, interfacing FHI-aims with LibRPA \cite{Shi2024}, where the low-scaling real-space $GW$ algorithm has been implemented, enables one to treat large systems. However, in this case, the high memory cost for storing the RI coefficients and Coulomb matrix often becomes the computational bottleneck. Thus, for further improvement, it is necessary to develop a more efficient strategy for memory usage and distributions for large-scale periodic $GW$ calculations.

Second, for periodic $GW$, currently only the one-shot $G_0W_0$ scheme with (semi-)local density functional approximation is implemented in FHI-aims. It would be highly desirable to implement a certain self-consistent $GW$ scheme, particularly those that can deliver wavefunctions. Such quasiparticle wavefunctions are necessary to describe, e.g., topological materials at a theoretical level that goes beyond KS-DFT.

Third, the current periodic implementation of $G_0W_0$ in FHI-aims can only be used to reliably calculate the band structures of the insulating materials. To deal with metallic systems reliably, additional implementation work is in order, including the incorporation of intraband transitions in dielectric function calculations and accurate treatment of the Fermi surface in the BZ integration. Such an extension is a prerequisite for applying FHI-aims $GW$ to correlated metallic systems, such as transition metals and doped cuprates.

Fourth, there has been considerable interest in combining electronic self-energy and vibrational self-energy from electron-phonon coupling in a unified framework \cite{Giustino2017}. The quasiparticle energies determined from such a combined self-energy will naturally capture the temperature effect on the electronic band structures.  Additionally, such a combined self-energy can also be used to calculate the transition temperature of superconductors via the Eliashberg theory \cite{Pellegrini2024}.
It is also of interest to extend the combined self-energy beyond the phonon picture by integrating $G_0W_0$ with molecular dynamics and the band unfolding technique (see Contrib.~\ref{ChapUnfolding}).

Lastly, when it comes to studying properties of materials consisting of heavy elements, relativistic effects such as spin-orbit coupling (SOC) need to be taken into account. FHI-aims provides spin-orbit coupled energy corrections through a second-variational method \cite{huhn2017} for (semi-)local and hybrid DFT calculations. It would be of great interest to incorporate this into periodic GW calculations by calculating the correlation self-energy shown in Eq.~\ref{eq:pgw-sigma-diagonal} using the SOC corrected wave-functions leading to a QP equation with relativistic corrections.

\subsection*{Acknowledgements}
 We acknowledge the funding support from the National Natural Science Foundation of China (Grant Nos 12374067 and 12188101) and the National Key Research and Development Program of China (Grant Nos 2022YFA1403800 and 2023YFA1507004). MS acknowledges support by his TEC1p Advanced Grant (the European Research Council (ERC) Horizon 2020 research and innovation programme, grant agreement No. 740233.

\newpage

\section{Neutral Excited States: Casida Equation for Linear-Response Time-Dependent Density Functional Theory and Bethe-Salpeter Equation}
\label{neutralexcitation}

\sectionauthor[1,2]{\textbf{*Yosuke Kanai}}
\sectionauthor[3]{\rev{Matthias Kick}}
\sectionauthor[1]{Ruiyi Zhou}
\sectionauthor[4]{Chi Liu}
\sectionauthor[5]{Jan Kloppenburg}
\sectionauthor[1,6,a]{Yi Yao}
\sectionauthor[7]{Dorothea Golze}
\sectionauthor[8]{Marc Dvorak}
\sectionauthor[8,9,10,11]{Patrick Rinke}
\sectionauthor[12,13,b]{Xinguo Ren}
\sectionlastauthor[4,6]{Volker Blum}

\sectionaffil[1]{Department of Chemistry, University of North Carolina at Chapel Hill}
\sectionaffil[2]{Department of Physics and Astronomy, University of North Carolina at Chapel Hill}
\sectionaffil[3]{Theory Department, Fritz Haber Institute of the Max Planck Society, Faradayweg 4-6, D-14195 Berlin, Germany}
\sectionaffil[4]{Department of Chemistry, Duke University, Durham, NC 27708, USA}
\sectionaffil[5]{Department of Chemistry and Materials Science, Aalto University, Finland}
\sectionaffil[6]{Thomas Lord Department of Mechanical Engineering and Materials Science, Duke University, Durham, NC 27708, USA}
\sectionaffil[7]{Faculty of Chemistry and Food Chemistry, Technische Universität Dresden, 01062 Dresden, Germany}
\sectionaffil[8]{Department of Applied Physics, Aalto University, P.O. Box 11000, FI-00076 Aalto, Finland}
\sectionaffil[9]{Physics Department, TUM School of Natural Sciences, Technical University of Munich, Garching, Germany}
\sectionaffil[10]{Atomistic Modelling Center, Munich Data Science Institute, Technical University of Munich, Garching, Germany}
\sectionaffil[11]{Munich Center for Machine Learning (MCML), Munich, Germany}
\sectionaffil[12]{Institute of Physics, Chinese Academy of Sciences, 100190 Beijing, China}
\sectionaffil[13]{The NOMAD Laboratory at the Fritz Haber Institute of the Max Planck Society, Faradayweg 4-6, D-14195 Berlin, Germany}

\sectionaffil[]{(*) Coordinator of this contribution}
\rule[0.25ex]{0.35\linewidth}{0.25pt}

\sectionaffil[a]{{\it Current Address:} Molecular Simulations from First Principles e.V., D-14195 Berlin, Germany}
\sectionaffil[b]{{\it Current Address:} Institute of Physics, Chinese Academy of Sciences, Beijing, 100190, China}




\subsection*{Summary}

\begin{figure}[ht]
    \centering
    \includegraphics[width=0.3\textwidth]{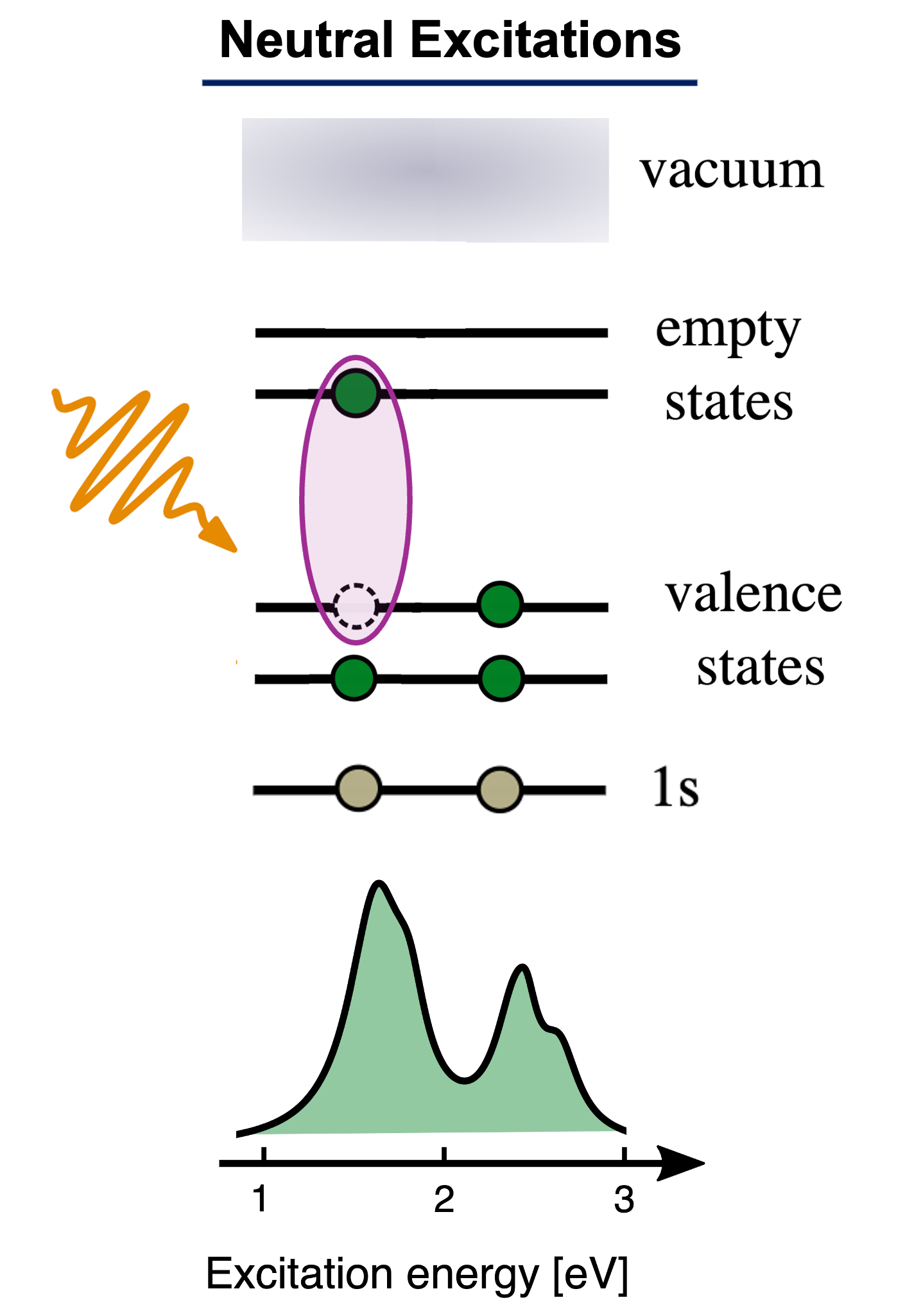}

    \caption{A schematic representation of a neutral excitation (i.e. exciton) modeled using BSE@$GW$ or LR-TDDFT via Casida equation. In this simplified picture, a valence electron is promoted to an empty state by optical absorption but retains an interaction (pink ellipse) with the resulting hole via a screened Coulomb interaction, which incorporates an effective response from all other particles in the system. Experimentally measurable properties such as the optical absorption spectrum can be computed as a function of the excitation energy using BSE@$GW$ or LR-TDDFT.}
    \label{fig:excitation}
\end{figure}

By transforming to the frequency domain, the time-dependent electronic structure theory of a system can be straightforwardly cast within the linear response theory. First-order responses to external stimuli such as electromagnetic fields are largely responsible for experimentally measurable quantities like optical or X-ray absorption spectra. 
These response properties can be written in the Lehmann representation, and computing resonance frequencies from corresponding eigenstates using this approach is of great practical interest.
In the quasi-particle picture, neutral excitations, especially bound states (``excitons''), include the screened interaction between the excited electron and hole as schematically depicted in Figure \ref{fig:excitation}.
In FHI-aims, the Bethe-Salpeter equation (BSE) based on the $GW$ method (Section~\ref{SecGWFinite}), i.e., BSE@$GW$, and Casida's equation of linear-response time-dependent density functional theory (LR-TDDFT) are implemented for calculating electronic excited states using ground-state DFT as the starting point. 
While these two methods derive from very different theoretical formalisms (i.e. BSE from the Green's function theory and the Casida's equation for TDDFT), the practical computational algorithms share close similarities and are thus convenient for software development.
In the FHI-aims code, the BSE@$GW$ and LR-TDDFT approaches are implemented for molecular systems, based on scalar-relativistic density functional approximations, spanning both valence excitations as well as core-level excitations. The BSE@$GW$ approach to compute absorption spectra is also available for extended periodic solids.

\subsection*{Current Status of the Implementation}


\subsubsection{Bethe-Salpeter Equation with $GW$ method}

Based on the quasi-particle description obtained using the $GW$ method as discussed in Section~\ref{SecGWFinite}, the BSE takes advantage of the two-particle Green's function to model the screened two-particle interaction between the hole and excited electron. Therefore, the approximations made at the $GW$ calculation carries over to the BSE calculation as well. In particular, the $GW$ approximation to the self-energy is the basis of the BSE implementation in the FHI-aims code \cite{liu2020all,zhou2024all}. Additionally, as done in most cases, the frequency dependence of the screened interaction, $W$, is neglected at the BSE stage so that the BSE can be solved efficiently as a linear problem. 
Within linear response theory, the BSE is formulated as solving the generalized eigenvalue equation
\cite{rohlfing2000electron,strinati1984effects}
\begin{equation}\label{eq:bse}
    \begin{bmatrix}
    \mathbf{A} & \mathbf{B}\\ 
        \mathbf{B^*} & \mathbf{A^*}
    \end{bmatrix}
    \begin{bmatrix}
        \mathbf{X} \\
        \mathbf{Y}
    \end{bmatrix}
    = \Omega
    \begin{bmatrix}
        \mathbf{I} & \mathbf{0} \\
        \mathbf{0} & \mathbf{-I}
    \end{bmatrix}
    \begin{bmatrix}
        \mathbf{X} \\
        \mathbf{Y}
    \end{bmatrix}
\end{equation}
where $\Omega$ is the excitation energy,
$\mathbf{X}$ and $\mathbf{Y}$ are the transition amplitudes in a product basis of occupied single-quasiparticle orbitals $i$ and unoccupied single-quasiparticle orbitals $a$.
More specifically, in Eq.\ref{eq:bse}, the $\mathbf{A}$, $\mathbf{B}$ matrices are defined as
\begin{align}
    A_{ia,jb} &= \delta_{ij} \delta_{ab} (\epsilon_a^{\text{QP}}-\epsilon_i^{\text{QP}}) + v_{ia,jb} - W_{ij,ab}(\omega=0) \\
    B_{ia,jb} &= v_{ia,bj} - W_{ib,aj}(\omega=0) 
\end{align}
where $\epsilon_{i/a}^{\text{QP}}$ is the quasi-particle energy which is typically obtained from a separate $GW$ calculation as discussed in Section~\ref{SecGWFinite}.   
$v$ is the bare Coulomb interaction, and $W(\omega=0)$ denotes that the frequency-dependence of the screened interaction is typically neglected as also done in the FHI-aims implementation. Refs. \cite{liu2020all,zhou2024all} focused on optical excitations, i.e., from valence bands to low-lying conduction bands. However, as an all-electron implementation, FHI-aims also allows one to compute core-electron excitations as relevant, e.g., for X-ray absorption spectroscopy \cite{Yao2022}. High numerical precision for the 1$s$ excitation of a simple molecule (H$_2$O) with available experimental reference data was  demonstrated in Ref. \cite{Yao2022}, including an extension of the applicable basis sets by tight Slater-type orbitals to capture the core orbitals' response to the excitation.
In particular, the efficacy of the BSE@$GW$ method was compared to the equation-of-motion coupled-cluster theory methods and experiments.

Building on the BSE implementation for isolated systems \cite{liu2020all} and $GW$ method for extended systems \cite{Ren2021}, our all-electron NAO-based implementation of the BSE method was recently generalized  also for extended periodic systems with the Brillouin zone sampling by Zhou \textit{et al.}\cite{zhou2024all}. Currently, only the Tamm-Dancoff approximation (TDA), which amounts to neglecting $\mathbf{B}$ in Eq. \ref{eq:bse}, is implemented for the periodic version. 

Building on the non-periodic BSE@$GW$ implementations, FHI-aims includes a quantum embedding theory called dynamical configuration interaction (DCI)  for finite systems. The foundations of the theory are covered in Refs.~\cite{Dvorak2019a,Dvorak2019b}. DCI is an active-space method, in which a fixed active space of strongly correlated orbitals is treated with configuration interaction (CI). This active space is then embedded in a high-energy, dynamically correlated BSE@$GW$-like bath. By combining CI and BSE@$GW$, DCI aims to leverage the strengths of both theories, providing a balanced treatment of static and dynamic correlation.
DCI offers access to total energies of both ground- and excited-state potential energy surfaces. Neutral excitation energies are obtained as the difference between these total energies.

\subsubsection{Casida Equation: Linear-Response Formulation of Time-Dependent DFT} 

The Casida equation (i.e., LR-TDDFT) is derived by formulating the polarizability 
as a sum over states of the many-body system. In 1995, Casida showed 
that the square of excitation energies can be obtained as an eigenvalue problem in the matrix form:
\begin{equation}\label{eq:casida}
\mathbf{C} \mathbf{F_s}=\Omega^2 \mathbf{F_s}.
\end{equation}

Here, the $\mathbf{C}$ is the so-called Casida matrix, which has the same dimension of $\mathbf{A}$ and $\mathbf{B}$ in the BSE Eq.\ref{eq:bse}.
$\Omega$ are the neutral many-body excitation energies. $\mathbf{F_s}$ are the associated eigenvectors and can be related to the oscillator strengths via the dipole operator\cite{casida_LR}. In Eq. (\ref{eq:casida}), the TDA was adapted, which is widely used for LR-TDDFT.
The $\mathbf{C}$ matrix can be written in the basis of product of the (g)KS orbitals
\begin{equation}\label{eq:casida_omega}
C_{i a, j b}(\omega)=\delta_{i, j} \delta_{a, b}\left(\epsilon_a-\epsilon_i\right)^2+2 \sqrt{\left(\epsilon_a-\epsilon_i\right)} K_{i a, j b}(\omega) \sqrt{\left(\epsilon_b-\epsilon_j\right)},
\end{equation}
where $\epsilon_a$, $\epsilon_i$ are KS eigenvalues. The kernel $K_{i a, j b}$ is defined as
\begin{equation}\label{eq:casida_kernel}
K_{i a, j b}(\omega)= \iint \psi_i^{\dagger}(\mathbf{r}) \psi_a(\mathbf{r})\left[ \frac{1}{\left|\mathbf{r}-\mathbf{r}^{\prime}\right|}
+f_{\mathrm{xc}}\left[n_0\right]\left(\mathbf{r}, \mathbf{r}^{\prime}, \omega\right)\right] \psi_j\left(\mathbf{r}^{\prime}\right) \psi_b^{\dagger}\left(\mathbf{r}^{\prime}\right) d \mathbf{r} d \mathbf{r}^{\prime}
\end{equation}
where $n_0$ is the (g)KS ground state electron density and $f_{xc}[n_0]$ is the exchange-correlation (XC) kernel, which is a functional derivative that describes how the exchange-correlation potential changes with respect to the electron density~\cite{onida2002electronic}. FHI-aims currently supports the LDA XC kernel, implemented within the adiabatic approximation (i.e. $f_{\mathrm{xc}}\left(\mathbf{r}, \mathbf{r}^{\prime}, \omega\right) = f_{\mathrm{xc}}\left(\mathbf{r}, \mathbf{r}^{\prime}\right)$) for isolated (non-periodic) systems.  

\begin{figure}[ht]
    \centering
    \includegraphics[width=0.48\textwidth]{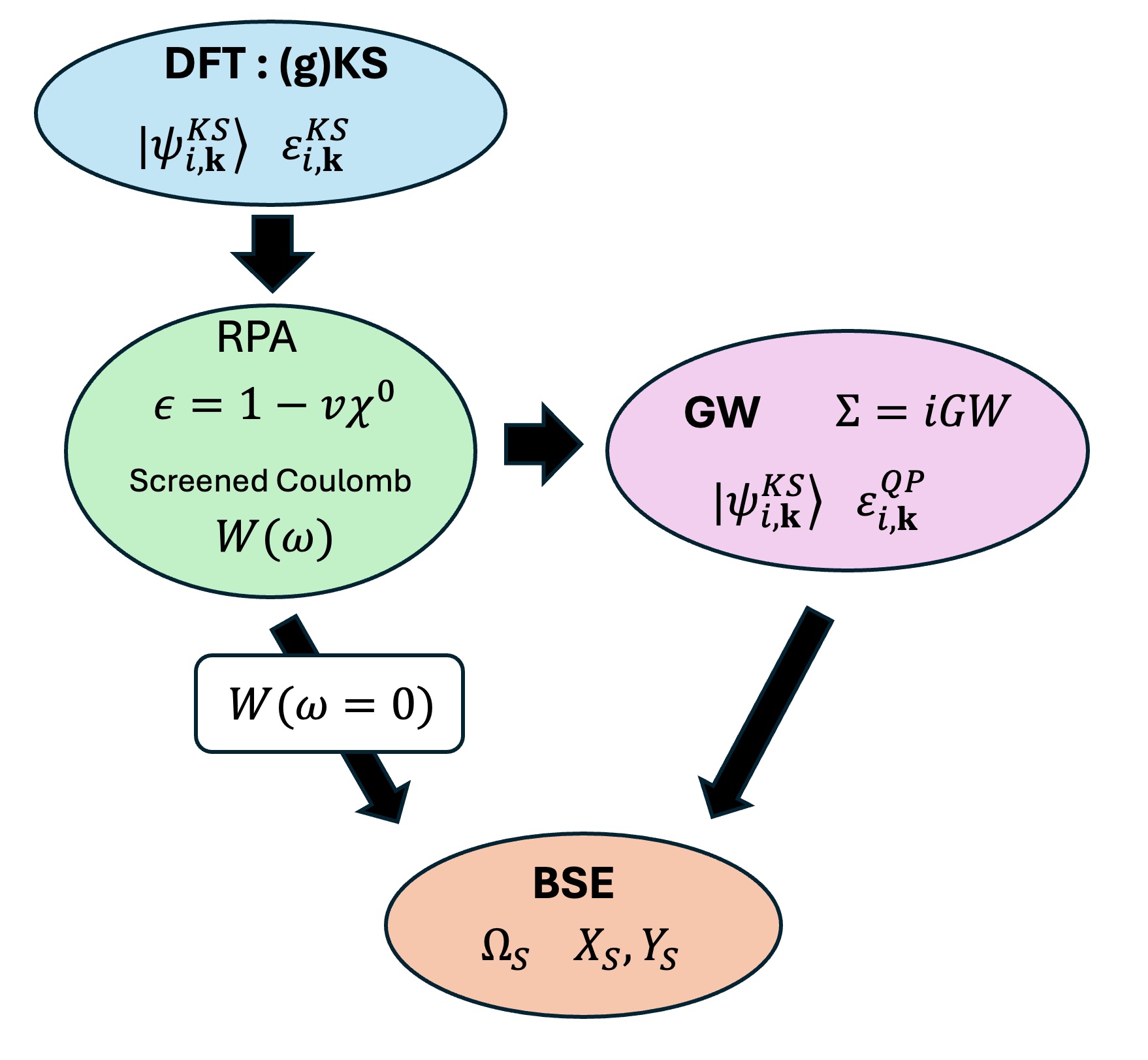}
    \includegraphics[width=0.28\textwidth]{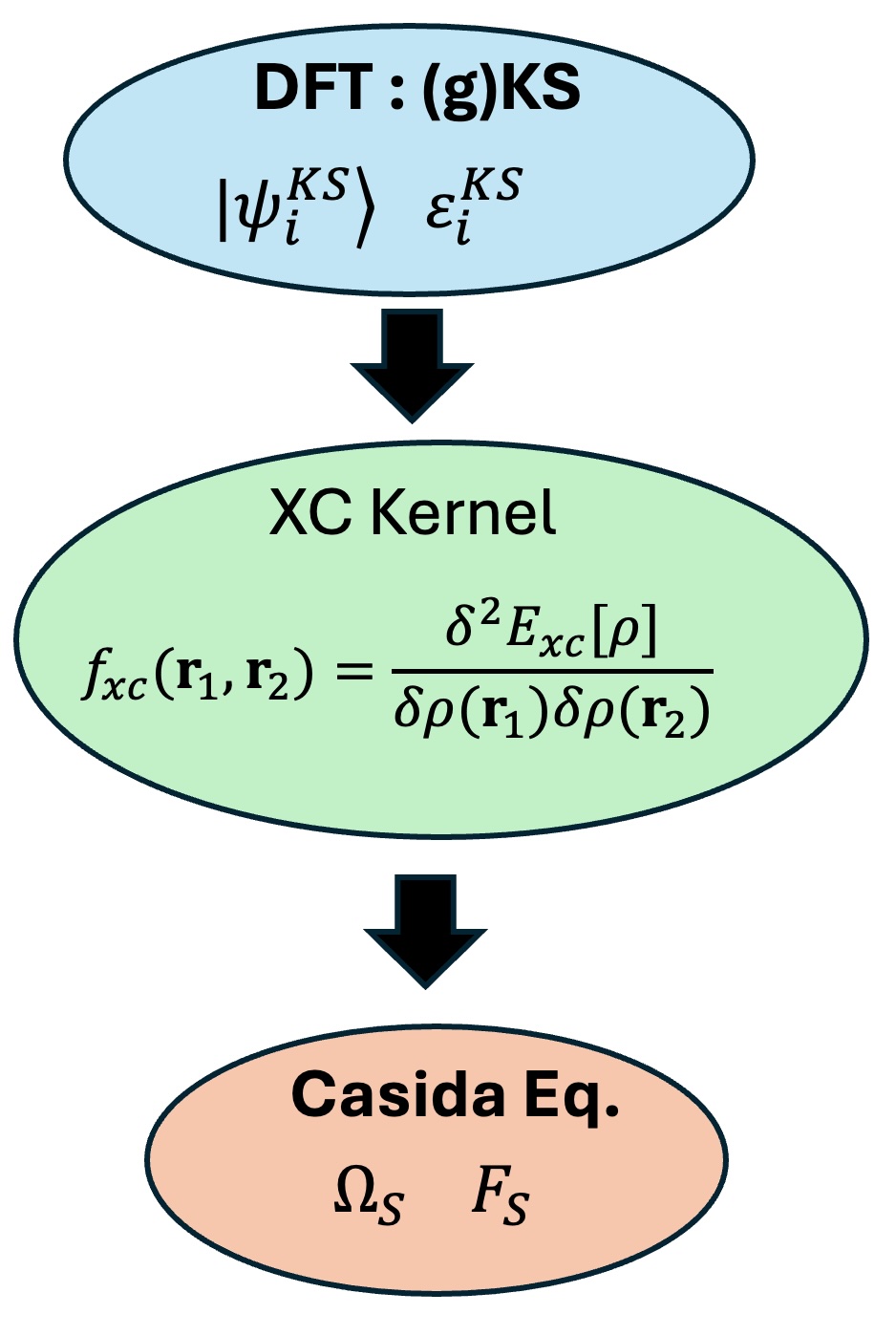}

    \caption{A schematic representation of the flow of the BSE and Casida Equations in the FHI-aims code. Both the BSE and the Casida equation start from performing a ground-state DFT calculation, with periodic boundary conditions supported for the BSE. The BSE additionally relies on the $GW$ method for calculation of the quasi-particles energies. The screened Coulomb interaction is computed within the random phase approximation (RPA). Casida's equation, on the other hand, does not involve an extra step for the single-particle energies and orbitals. However, the exchange-correlation (XC) kernel needs to be approximated. Currently the adiabatic LDA kernel is implemented and the implementation works only for isolated (non-periodic) systems.
    }
    \label{fig:placeholder}
\end{figure}


\subsection*{Usability and Tutorials}

The BSE tutorial for FHI-aims,\footnote{\url{https://fhi-aims-club.gitlab.io/tutorials/rpa-and-gw-for-molecules-and-solids/}}, offers a concise introduction to the BSE@$GW$ method. It includes several straightforward examples for running BSE calculations in molecular systems and extended periodic systems using the FHI-aims code. This tutorial covers the fundamentals of performing neutral excitation calculations with the BSE@$GW$ method. Importantly, it also delves into numerical topics such as basis set convergence and various levels of approximation and algorithms used in the $GW$ method, which computes the quasi-particle energies needed as inputs for BSE calculations.

The BSE@$GW$ implementation in FHI-aims is suitable for both molecular valence excitations, as studied in UV-vis spectroscopy~\cite{liu2020all}, and molecular $K$-edge core-electron excitations, as measured in X-ray absorption spectroscopy~\cite{Yao2022}. For molecular valence excitations, reasonable accuracy is often achieved by using BSE@$G_0W_0$ on top of DFT calculations using a hybrid density functional with 20-30\% of exact exchange~\cite{BSE_review}, such as PBE0~\cite{adamo1999,ernzerhof1999}. In the preceding $G_0W_0$ step, employing an analytic continuation of the self-energy with a Pad\'e model containing 16 parameters is sufficient~\cite{liu2020all}. For fast basis set convergence of low-lying excitation energies in small molecules, augmenting standard NAO basis sets with additional extended, diffuse functions is desirable, in line with the broader experience of quantum chemistry. Benchmarks by Liu \textit{et al.}\cite{liu2020all}, show that a basis set ``tier2+aug2'', where aug2 represents two augmentation functions from the aug-cc-pV$n$Z Gaussian basis sets\cite{Kendall1992,Woon1994}, facilitates excellent numerical convergence of low-lying excitation energies. The tier2+aug2 basis sets agree with results from the much larger, GTO-based aug-cc-pV5Z basis sets within 0.1 eV~\cite{liu2020all}.

For molecular $K$-edge transitions in light elements, Ref. \cite{Yao2022} demonstrated good performance when executing BSE@$G_0W_0$ on top of the PBEh($\alpha$) hybrid DFT functional with at least 45\% exact exchange ($\alpha$). The value $\alpha = 0.45$ is optimal for second-row elements, but should be gradually increased for heavier elements. We consider the $G_0W_0@$PBEh scheme superior to partially self-consistent schemes, which we recommend for $GW$-based charged excitations, as the calculation of $W$ in BSE requires high-quality underlying KS orbitals. At the DFT level, $K$-edge calculations should be performed using the atomic zeroth-order regular approximation (ZORA), with the scalar-relativistic correction from Ref.\cite{Keller2020} applied to the $GW$ quasiparticle energies. The contour deformation technique should be used to compute the $GW$ self-energy. For $K$-edge calculations, we recommend using the tier2+aug2+STO3 basis set, where STO3 refers to the set of Slater-type orbitals provided in the Supporting Information of Ref.\cite{Yao2022}. With this setup, mean absolute errors of 0.63 eV relative to experiment were achieved~\cite{Yao2022}. 

The option for periodic systems is a new addition for BSE@$G_0W_0$ calculations in the FHI-aims code at the time of writing. Benchmark calculations for Si and MgO are reported in Ref ~\cite{zhou2024all}. Here, the localized RI framework of Ref. \cite{ihrig2015} makes the expansion of the Coulomb operator tractable for periodic systems with demonstrated high precision.
%
%
%
In the preceding $G_0W_0$ calculation, one may employ analytic continuation of the self-energy either with a two-pole model or Pad\'e model \cite{Ren2021}. Based on the work in Ref. ~\cite{zhou2024all}, we recommend the standard FHI-aims tier2 basis sets, which converge to 0.2 eV or better. For the Brillouin Zone (BZ) sampling, only the $\Gamma$-centered sampling is currently implemented in the FHI-aims code. For the absorption spectrum of some semiconductor materials, 
a large number of k points is required and a systematic convergence test for BZ sampling is necessary \cite{zhou2024all}. 


The tutorial on LR-TDDFT for molecules, using the Casida equation\footnote{\url{https://fhi-aims-club.gitlab.io/tutorials/rpa-and-gw-for-molecules-and-solids/Part-1/LRTDDFT/}} provides a concise introduction of how to use FHI-aims code to compute excited-state energies for simple molecules, using the LDA XC kernel. For molecular valence excitations, we recommended the same tier2+aug2 basis set as for the BSE@$G_0W_0$ calculations, which reproduces aug-cc-pV5Z results within 0.1 eV\cite{liu2020all}.

DCI generally targets multireference problems that BSE@$GW$ alone struggles to address. The study of double excitations and complicated spin multiplets lies squarely in DCI's domain. The method is well-suited for systems in which a small number of strongly interacting electrons (on the order of 2–10) is surrounded by a bath of weakly interacting electrons. Frontier orbitals are a natural choice for the active space, making defect and impurity systems prime candidates for DCI. Active spaces of up to eight levels and a bath of approximately 300 orbitals are recommended. A comprehensive documentation can be found in the FHI-aims manual. The largest system calculated with DCI so far was gas-phase free-base and Mg-porphyrin~\cite{Dvorak2020}.

\subsection*{Future Plans and Challenges}

The so-called ``starting-point dependence'' of the BSE as well as $GW$ methods (i.e., the choice of the density functional on which the initial self-consistent ground state DFT calculations of orbitals and polarizabilities is based) continues to be an active area of research for practical application of Green's function theory in electronic structure calculations\cite{Golze2019}. Using the FHI-aims code, some progress has been made for molecular systems with the renormalized singles method\cite{Bhattacharya_JPCA_2024}. 
Practical applications of the periodic BSE method are still limited by the necessarily dense Brillouin Zone (BZ) sampling. $\Gamma$-centered sampling is currently implemented in the FHI-aims code.  \rev{ Optical absorption spectra of extended systems are known to converge slowly with respect to BZ sampling. However, explicitly calculating the screened Coulomb interaction matrix, $W$, on a dense grid is often computationally intractable due to severe memory and CPU constraints. To address this, future implementations will incorporate interpolation strategies, such as the double-grid technique (calculating $W$ on a coarse grid and interpolating to a fine grid) \cite{kammerlander_prb_2012,sander2015beyond} or Fourier interpolation methods \cite{deslippe2012berkeleygw}.} \rev{As we make further progress with the Brillouin Zone sampling, the transitions such as those mediated by phonons with a finite momentum can be computed. This will permit further usage of this many-body method for computing the exciton-phonon coupling in various applications.}

In principle, the implementation of Casida's equation in FHI-aims can be straightforwardly extended for periodic systems by utilizing the periodic BSE implementation. At the same time, the proper description of XC kernel for extended systems is necessary. 
In particular, XC kernel of the adiabatic LDA fails to behave correctly in the long-wavelength limit\cite{PhysRevB.56.12811,Byun_2020}. Several remedies have been proposed in the literature, including the use of screened exact exchange\cite{PhysRevB.92.035202, PhysRevMaterials.3.064603} and corrections motivated by many-body approaches like BSE\cite{PhysRevLett.88.066404}. We intend to examine and incorporate some of these recent advancements in the future.
\rev{
On the technical side of software development, a matrix-free version of the Casida equations is also planned, employing iterative eigenvalue solver techniques. This approach will enable simulations of larger systems by avoiding the storage of the full Casida matrix, while also supporting hybrid XC functionals.}

\subsection*{Acknowledgments}
D. Golze acknowledges financial support by the Deutsche Forschungsgemeinschaft (DFG, German Research Foundation) for funding via the Emmy Noether Programme (project number~453275048). We are grateful to Dr. Heiko Appel for his insights and support with the LR-TDDFT implementation in FHI-aims.



  





\newpage

\section{Real-Time Time-Dependent Density Functional Theory}
\label{SecRTTDDFT}

\sectionauthor[1]{\rev{Hannah Bertschi}} 
\sectionauthor[2]{Joscha Hekele}
\sectionauthor[3,4]{\textbf{ *Yosuke Kanai}}
\sectionauthor[2]{\textbf{ *Peter Kratzer}}
\sectionauthor[3]{Christopher Shepard}
\sectionauthor[3]{Jianhang Xu}
\sectionlastauthor[5,6,a]{Yi Yao}

\sectionaffil[1]{Max Planck Institute for the Structure and Dynamics of Matter, D-22761 Hamburg, 
Germany}
\sectionaffil[2]{Faculty of Physics, University of Duisburg-Essen, D-47057 Duisburg, Germany}
\sectionaffil[3]{Department of Chemistry, University of North Carolina at Chapel Hill}
\sectionaffil[4]{Department of Physics and Astronomy, University of North Carolina at Chapel Hill}
\sectionaffil[5]{The NOMAD Laboratory at the FHI of the Max Planck Society, D-14195 Berlin, Germany}
\sectionaffil[6]{Thomas Lord Department of Mechanical Engineering and Materials Science, Duke University,
Durham, North Carolina 27708, USA}

\sectionaffil[*]{Coordinator of this contribution.}
\rule[0.25ex]{0.35\linewidth}{0.25pt}

\sectionaffil[a]{{\it Current Address:} Molecular Simulations from First Principles e.V., D-14195 Berlin, Germany}




\subsection*{Summary}
The explicit real-time propagation approach for time-dependent density functional theory (RT-TDDFT) has become an increasingly popular first-principles computational method for modeling various time-dependent electronic properties of complex matter. Readers are referred to the comprehensive review by Li {\it et al.}~\cite{rttdest_review_xiaosong} on the development of the real-time propagation approaches for electronic structure theory methods in general. 
The theoretical formalism of TDDFT is based on the Runge-Gross theorem~\cite{PhysRevLett.52.997}, and TDDFT is widely used within the linear response theory approach as has been discussed in Section~\ref{neutralexcitation}.
The real-time propagation approach allows us to further expand the scope of the applicability beyond the linear response or the traditional perturbative regimes, and RT-TDDFT simulations are increasingly employed to help answer various outstanding questions, especially pertaining to non-equilibrium electron dynamic phenomena, including interfacial charge transfer, atom-cluster collisions, topological quantum matter, etc.~\cite{xu2023jacs,JCP_2021,Schleife2022MRS}. 
In advantage over linear response theory based on the Casida equation \cite{casida_LR}, RT-TDDFT solely requires the application of the system's Hamiltonian to some starting wavefunction; hence its computational efforts scales linearly with the system size. 
Excited-state energies, including many-body interactions, can be obtained via Fourier transformation of time series. 
However, as becomes apparent from the Fourier sampling theorem, determining highly resolved excitation spectra will require long propagation times, and hence efficient propagation methods.
Furthermore, Ehrenfest dynamics \cite{Ehrenfest} can be performed on top of RT-TDDFT so that non-adiabatic effects of electrons 
can be taken into account in first-principles molecular dynamics simulations. 
Here, FHI-aims \cite{Blum2009} with its atom-centered basis functions has the advantage that basis functions are 'carried around' with the atoms as they move, resulting in an accurate description already with modest computational effort. In contrast to plane-wave basis sets, Pulay forces due to the continuously occurring basis set transformations need to be taken into account.

\subsection*{Current Status of the Implementation}
\begin{figure}[ht]
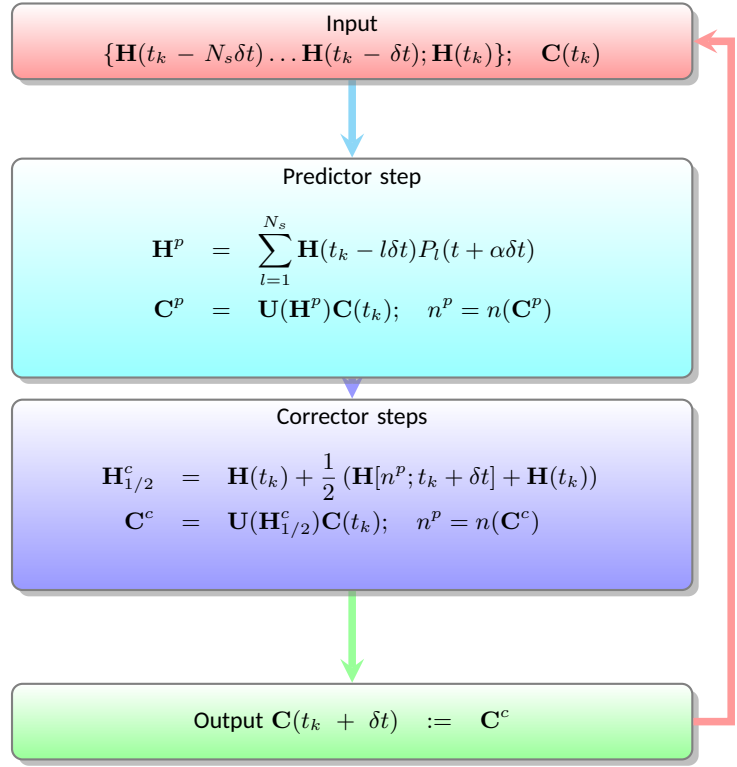

\begin{center}
\smartdiagramset{module minimum width=9cm, 
module y sep=3,
text width=8cm
}
\smartdiagram[flow diagram]{
\centerline{Input}
$
\{ \mathbf{H}(t_k - N_s \delta t) \ldots  \mathbf{H}(t_k - \delta t); \mathbf{H}(t_k) \}; \quad \mathbf{C}(t_k)
$
,
\centerline{Predictor step}
\begin{eqnarray*}
\mathbf{H}^p &=& \sum_{l=1}^{N_s} \mathbf{H}(t_k - l \delta t) P_l(t+ \alpha \delta t)  \\
\mathbf{C}^p &=& \mathbf{U}(\mathbf{H}^p) \mathbf{C}(t_k); \quad
n^p = n( \mathbf{C}^p )
\end{eqnarray*},
\centerline{Corrector steps} 
\begin{eqnarray*}
\mathbf{H}^c_{1/2} &=& \mathbf{H}(t_k) + \frac{1}{2} \left(\mathbf{H}[n^p; t_k + \delta t] + \mathbf{H}(t_k) \right) \\
\mathbf{C}^c &=& \mathbf{U}(\mathbf{H}^c_{1/2}) \mathbf{C}(t_k); \quad
n^p = n( \mathbf{C}^c )
\end{eqnarray*}
,
Output $\mathbf{C}(t_k + \delta t) := \mathbf{C}^c$
}
\end{center}
\caption{Flowchart of a predictor-corrector scheme with time step $\delta t$ and a unitary propagator $\mathbf{U}$ applied to expansion coefficients $\mathbf{C}$ in Eq.~\ref{eq:ksdiff}, Charge density $n^p$ is predicted by polynomial expansion with polynomials $P_l$ from $N_s$ previous Hamiltonians $\mathbf{H}$. The corrector step works with a mid-point Hamiltonian $\mathbf{H}^c_{1/2}$ and may require iteration to converge $n^p$.}
    \label{fig:flowchart}
\end{figure}

\textbf{RT-TDDFT}
The expansion coefficients $c_{in}(t)\in\mathbf{C}$ of the KS orbitals here contain the time-dependence of the electronic system as a matrix $\mathbf{C}\in\mathbf{C}^{N_\mathrm{basis}\times N_\mathrm{occ}}$. Only the $N_\mathrm{occ}$ initially occupied orbitals are evolved in time. The time-dependent Kohn-Sham (TDKS) equation (also used for the NAO implementation in the FHI-aims code, see below) is
\begin{align}
\label{eq:ksdiff}
\frac{d}{dt}\mathbf{C}(t) &= -i\mathbf{S}^{-1}\mathbf{H}(t)\mathbf{C}(t)
\end{align} 
with the overlap matrix $\mathrm{S}_{ij}=\langle\phi_i|\phi_j\rangle$ and the Hamiltonian matrix $\mathrm{H}_{ij}=\langle\phi_i|\mathcal{H}^\mathrm{KS}|\phi_j\rangle$ is then solved to describe electron dynamics. Note the implicit time-dependence via the electron density (i.e. $\mathbf{H}(t) \equiv \mathbf{H}[n(t),t]$ )
The efficient and accurate solution of this equation is the key functionality in every RT-TDDFT code. 
Starting with release 210716, the
FHI-aims code has several options for integrating the TDKS equations. These include 
the Crank-Nicolson scheme \cite{crank1947practical} to variable order, the exponential midpoint rule (including its variant with enforced time-reversal symmetry), but also more complicated schemes such as the commutator-free Magnus expansion of order four, and many more. 
The latter schemes require the computation of an exponential of the Hamiltonian matrix; for this task the user can choose between a straightforward eigen decomposition or a 'scaling and squaring' method that is often computationally more affordable since only matrix products are required.
Maintaining self-consistency between the charge density and the Hamiltonian is crucial in any integration of the TDKS equations and often limits the time step $\delta t$ that can be used for integrating Eq.~\ref{eq:ksdiff}. This problem can be mitigated by using one of the implemented predictor-corrector schemes \cite{qchem_tddft_2} for the propagation, see Fig.~\ref{fig:flowchart}.  
All options are documented in the FHI-aims manual. 
 
The fundamental approximation in most RT-TDDFT simulations is the so-called adiabatic approximation to 
the XC potential that ignores the (in principle mandated) time-non-local character of the XC potential ('history dependence'). 
Likewise, our present implementation does not consider the dependence of the functional on the current as required in TD-current-DFT. 

In RT-TDDFT, we are generally interested in the time-dependent response of the system to some external perturbation. 
Interaction with the light is often modeled using the classical electromagnetic field.
Within the dipole (or long-wavelength) approximation for the electromagnetic field (i.e. neglecting any spatial dependence), the external potential  is given in the velocity gauge as $\mathcal{V}_\mathrm{ext}(t) = -i\mathbf{A}(t)\cdot\nabla + \mathbf{A}^2$
or in the length gauge as
$ \mathcal{V}_\mathrm{ext}(t) = \hat{\mathbf{r}}\cdot\mathbf{E}(t)$
as connected to each other by a gauge transformation. $\bf{A}(t)$ is the vector potential and $\bf{E}(t) = -\partial_t\bf{A}(t)$ is the electric field. 

\textbf{Ehrenfest Dynamics}
When the external perturbation is slowly varying, the electrons remain in the equilibrium ground state for a given a set of atomic nuclear positions. In this adiabatic limit, the system's response can be studied by solving for the instantaneous eigenstate of the time-dependent Hamiltonian. 
First-principles molecular dynamics (FPMD) has found great success in various areas of chemistry and condensed matter physics. 
When the external perturbation is fast-varying, however, the time evolution of the quantum state of electrons needs to be explicitly modeled instead of assuming them to remain in the ground state. In the Ehrenfest dynamics approach, the force on classical atomic nuclei can be obtained from the electrostatic interaction with the time-dependent electron density, which can be described using RT-TDDFT.
Electrons are described using the TDKS equations with a non-adiabatic term $\mathbf{G}$:
\begin{equation}
\begin{aligned}
\label{eq:EF-e}
\frac{d}{dt}\mathbf{C}(t) &= -i\mathbf{S}^{-1}(\mathbf{H}+\mathbf{G})\mathbf{C}(t), \\
G_{ij} &= -i\sum_I\dot{\vec{R}}_I\cdot\vec{\mathbf{B}}_I,
\end{aligned} 
\end{equation}
where $\vec R_I$ is the coordinate of nucleus $I$, and $\vec{\mathbf{B}}_I = \langle\phi_i|\nabla_I|\phi_j\rangle$.
The nuclei are propagated using the conventional energy gradient with non-adiabatic contributions,
\begin{equation}
\begin{aligned}
\label{eq:EF-i}
\vec F_I &= \vec{F}_I^\text{HF} + \vec{F}_I^\text{MP} + \vec{F}_I^\text{XC} + \vec{F}_I^\text{DBC}+ \vec{F}_I^\text{NC}, \\
\vec{F}_I^\text{DBC} & = -\sum_n^{N_\text{occ}}\sum_{ij}^{N_\text{basis}}f_nc^*_{in}c_{jn}\left[
\langle\nabla_I\phi_i|\hat{H}|\phi_j\rangle
+\langle\phi_i|\hat{H}|\nabla_I\phi_j\rangle
\right], \\
\vec{F}_I^\text{NC} &= \sum_n^{N_\text{occ}}f_n\mathbf{c}_n^\dagger\left[
\mathbf{HS}^{-1}\vec{\mathbf{B}}_I
+\vec{\mathbf{B}}_I^\dagger\mathbf{S}^{-1}\mathbf{H}
+ i\left(
\vec{\mathbf{W}}_I^\dagger - \vec{\mathbf{W}}_I
+ \mathbf{D}^\dagger_I\mathbf{S}^{-1}\vec{\mathbf{B}}_I
- \vec{\mathbf{B}}_I^\dagger\mathbf{S}^{-1}\mathbf{D}_I
\right)\right].
\end{aligned} 
\end{equation}
The first three terms are existing forces in FHI-aims, $\vec F_I^\text{HF}$ are the standard Hellmann-Feynman forces, $\vec F_I^\text{MP}$
are multipole force contributions and $\vec F_I^\text{XC}$ are XC related forces.
$\vec F_I^\text{DBC}$ is the dynamical basis correction term that can be seen as the time-dependent analogue to Pulay forces.
$\vec F_I^\text{NC}$ is the non-adiabatic coupling force with $\mathbf{W}_I=\dot{\mathbf{R}}_I\langle\nabla_I\phi_i|\nabla_I\phi_j\rangle$ and $\mathbf{D}_I=\dot{\vec{R}}_I\cdot\vec{\mathbf{B}}_I$.
\rev{For finite systems, D3 \cite{Grimme2010} and Tkatchenko-Scheffler \cite{ts} dispersion corrections on the forces are available.}
\rev{Ehrenfest dynamics is a mean-field theory and thus disregards 
electron-nuclear dynamical correlations. It can be viewed as a controlled short-time semiclassical 
expansion of the exact Ehrenfest theorem. 
It can perform well for ultrafast, nearly coherent processes where 
the nuclear wave-packet remains sufficiently localised. 
It fails when electronic states have widely different forces
and different possible pathways for the nuclear wave-packet 
would have to be considered.}

In some applications, e.g. in the context of non-adiabatic friction or optical breakthrough in dielectrics, the deviation of the time-propagated wavefunctions from their stationary counterparts can be interpreted as promoting electrons into energetically higher-lying Kohn-Sham states. For instance, the energy lost by a projectile due to electronic friction (i.e. electronic stopping) is spent for the creation of electrons and holes in the target material. 
To obtain the probability $n_{\rm ex}(t)$ of creating such electrons and holes, the time-propagated TDKS wavefunctions $\psi_{n {\bf k}}(t)$ are projected onto a complete set of electronic ground-state wavefunctions $\psi_{n \bf{k},t}(0)$ for the particular atomic geometry at time $t$, 
thus yielding
$$
n_{\rm ex}(t) = \sum_{n n' {\bf k}} \left( \delta_{nn'} - | \langle \psi_{n' {\bf k},ft}(0) | \psi_{n {\bf k}}(t) \rangle |^2 \right) 
$$
This option is available as  well in FHI-aims. 

RT-TDDFT can be performed also with the Nuclear Electronic Orbital (NEO) method\cite{hammes-schiffer_nuclearelectronic_2021}, referred to as ``RT-NEO-TDDFT", such that coupled quantum dynamics of atomic nuclei (usually protons) and electrons can be simulated on an equal footing\cite{xu2023prl,xu2022periodic}. The NEO approach is discussed in Section~6.3.

Finally, we make some comments with respect to other RT-TDDFT implementations. 
For materials simulations, the plane-wave-pseudopotential formalism remains the most popular implementation of electronic structure methods, and this is the case also for RT-TDDFT, as implemented in codes like Qbox\cite{qbox_rt-tddft} and Qb@ll\cite{qball_rt-tddft}. 
Our NAO implementation in the FHI-aims code, due to its all-electron description, has some unique advantages.
For instance, it allows users to model core-electron excitation dynamics in materials. 
At the same time, studying phenomena such as ionization would require careful consideration of the NAO basis set flexibility.

\subsection*{Usability and Tutorials}

A tutorial that describes how to set up RT-TDDFT simulations using FHI-aims is available under \href{https://gitlab.com/FHI-aims-club/tutorials/real-time-time-dependent-dft-in-fhi-aims}{this link}\footnote{\url{https://gitlab.com/FHI-aims-club/tutorials/real-time-time-dependent-dft-in-fhi-aims}}. 
The tutorial provides a brief overview of the theory along with several simple examples. The first section focuses on molecular systems, outlining all the necessary parameters for RT-TDDFT in FHI-aims and explaining how to calculate the absorption spectrum of a single water molecule. The second section addresses the calculation of the optical absorption spectra for periodic systems, using crystalline silicon as the example. Periodic simulations introduce specific challenges that require adjustments to certain parameters, which are thoroughly explained in the tutorial.
With both absorption spectrum examples, two python scripts are provided to assist with data evaluation. The first one, called {\tt eval\_tddft.py}, allows the user to extract observable quantities, such as the externally applied field, the time-dependent dipole moments and current densities, from the output, and to perform Fourier analysis to obtain the spectra. For atomic and molecular systems, determining the precise spectral location of electronic transitions as well as their oscillator strengths is a major concern.  The script {\tt ifit\_abs.py} offers graphical support for this task: by clicking on a peak in the spectrum, an algorithm based on Pad\'e approximants \cite{bruner-2016} outputs the precise peak position. 
\rev{For more refined techniques of extracting excited energy levels, we refer to the recent literature~\cite{Kick_2024}. }

Finally, the last section of the tutorial demonstrates how to perform Ehrenfest dynamics with RT-TDDFT in the FHI-aims code through a simple example of $H_2$ bond dissociation. While the tutorial covers only a few small applications of RT-TDDFT, the implementation allows for many others, such as high-harmonic generation simulations or ion bombardment simulations based on non-adiabatically coupled electron-ion dynamics (Ehrenfest dynamics)\cite{hekele-thesis}. Additional information and example inputs for the optical absorption spectrum of benzene, high-harmonic generation on silicon, and Ehrenfest dynamics of CH$_2$=NH$^+_{2}$ can be found in the appendix of Ph.D. thesis of Ref.~\cite{hekele-thesis}. 
\rev{Moreover, we point out that electronic transport properties in exotic phase of matter, such as a warm, dense hydrogen plasma, have been addressed recently by means of RT-TDDFT simulations \cite{Ramakrishna_2024}.}

\subsection*{Future Plans and Challenges}

Hybrid XC functionals have become increasingly popular in recent years. Currently, hybrid XC functionals in RT-TDDFT are implemented only for molecular (non-periodic) systems.
\rev{Since the Hamiltonian needs to be evaluated in each timestep, efficient methods for calculating exchange integrals are key to the efficiency of hybrid functional RT-TDDFT. 
In contrast to molecular systems, the methods such as the resolution of the identity are crucial for efficiently evaluating exchange integrals for periodic systems, and they are currently being developed in the context of RT-TDDFT.
}

The applications of RT-TDDFT within the FHI-aims code so far have been dealing with systems of light atoms, e.g. first-row elements or silicon (cf. Ref.~\cite{hekele-2021}.  
FHI-aims, being an all-electron method, includes (scalar) relativistic treatment of the core shells of atoms. 
While this treatment (called the zeroth-order approximation ZORA of the Koelling-Harmon scheme) is active and usable also in the present RT-TDDFT implementation, it could lead to stability problems if medium to heavy elements (from Fe onward) shall be treated via RT-TDDFT. 
The highly negative eigen energies of core levels may lead to inaccuracies in the propagation of the phase of the KS wavefunctions in exponential schemes. Similarly, the large spectral range of the all-electron Hamiltonian is a challenge for solving the linear system of equations associated with the implicit Crank-Nicolson propagation scheme. 
As a future development, one may implement a method that uses different schemes to propagate the core and valence states of condensed-matter systems containing heavy elements. 
Moreover, the fully relativistic treatment using four-component spinor wavefunctions (see Section~\ref{Sec:Relativity}) yet awaits its implementation in the RT-TDDFT module. 

For applications that involve excitation of electrons to continuum or near-continuum (e.g. Rydberg) states, the atom-centered basis sets used in FHI-aims obviously poses some limitations. Here, a key point is the accurate description of the density difference between the ground state and the excited state. This could be achieved by introducing some additional unbiased basis functions, and ultimately by an additional real-space grid if required. Such a major code extension could be a long-term project for the future.

\subsection*{Acknowledgements}
This work was partly (Y.K., J.X., C.S.) funded by the National Science Foundation (United States of America) under Award Nos. CHE-1954894 
and partly (J.H.,P.K.) by the Deutsche Forschungsgemeinschaft (DFG, German Research Foundation) Project No. 278162697-CRC1242
\rev{and H.B. was funded by the European Union through the Marie Sklodowska-Curie Grant Agreement no. 101118915 (TIMES).}





\newpage

\section[All-Electron $\Delta$-SCF Calculations in Molecular and Periodic Systems]{All-Electron $\Delta$-Self-Consistent-Field Calculations in Molecular and Periodic Systems}

\sectionauthor[1]{Samuel J. Hall}
\sectionauthor[2]{\textbf{*J. Matthias Kahk}}
\sectionauthor[3]{Benedikt P. Klein}
\sectionauthor[4,5,6]{\textbf{*Reinhard J. Maurer}}
\sectionauthor[7]{Georg S. Michelitsch}
\sectionauthor[4]{Dylan Morgan}
\sectionlastauthor[7]{Karsten Reuter}

\sectionaffil[1]{Helmholtz-Zentrum Berlin für Materialien und Energie GmbH, Berlin, DE 10409, Germany}
\sectionaffil[2]{Institute of Physics, University of Tartu, W. Ostwaldi 1, 50411 Tartu, Estonia}
\sectionaffil[3]{Research Center for Materials Analysis, Korea Basic Science Institute, Daejeon 34133, Republic of Korea}
\sectionaffil[4]{Department of Chemistry, University of Warwick, Gibbet Hill Road, CV4 7AL Coventry, United Kingdom}
\sectionaffil[5]{Department of Physics, University of Warwick, Gibbet Hill Road, CV4 7AL Coventry, United Kingdom}
\sectionaffil[6]{Faculty of Physics, University of Vienna, Vienna A-1090, Austria}
\sectionaffil[7]{Theory Department, Fritz Haber Institute of the Max Planck Society, Faradayweg 4-6, D-14195 Berlin, Germany}

\sectionaffil[*]{Coordinator of this contribution.}




\subsection*{Summary}

In a conventional density functional theory (DFT) calculation, the Kohn-Sham orbitals are populated according to the Aufbau principle. However, it is also possible to perform calculations in which the self-consistent field (SCF) is allowed to converge for an excited electronic configuration \cite{bagus_self-consistent-field_1965}. DFT calculations with non-Aufbau-principle occupation numbers are most commonly used to model core-excited states of materials and molecules, e.g. in theoretical simulations of X-ray Photoelectron Spectroscopy (XPS) and X-ray Absorption Spectroscopy (XAS) \cite{besley_self-consistent-field_2009,hall_characterizing_2023,chong_density-functional_1995,annegarn_combining_2022,uhl_ab_2019,besley_modeling_2021,kahk_predicting_2022,delesma_chemical_2018,bagus_extracting_2018,nishino_identification_2024}. 

As an example, consider the removal of a C 1s core electron from a methane molecule. The electron’s binding energy is defined as:

\begin{equation}
E_\mathrm{B} = E_{N-1,\mathrm{ch}} - E_{N,\mathrm{ground}},
\label{eqn3}
\end{equation}

where $E_\mathrm{B}$ is the binding energy, $E_{N,\mathrm{ground}}$ is the total energy of the $N$-electron ground state, 
and $E_{N-1,\mathrm{ch}}$ is the total energy of the $N-1$ electron final state with a core hole. $E_{N,\mathrm{ground}}$ is straightforward to calculate using ground state DFT. 
For $E_{N-1,\mathrm{ch}}$, one can perform a DFT calculation in which the occupation number of the lowest energy eigenstate in one of the spin channels is fixed at zero throughout the SCF cycle. When both total energies are known, the core electron binding energy is directly obtained as the difference between the two -- this is known as the $\Delta$SCF method. As DFT is a ground-state theory, the formal justification for $\Delta$SCF calculations has been an issue of intense debate. Several works explore the formal theoretical underpinnings of the $\Delta$SCF method \cite{yang_foundation_2024,ziegler_relation_2009,cullen_formulation_2011,ziegler_application_2011,ziegler_implementation_2012,gorling_density-functional_1999,ayers_time-independent_2009}.

FHI-aims allows the user to perform all-electron $\Delta$SCF calculations for both periodic and aperiodic systems, and it contains tools for creating, tracking, and visualizing a localized core hole. Simulations of both charged and neutral excitations from core orbitals, as well as valence excitations are possible - see for example references \cite{klein_nuts_2021,chaudhuri_coexistence_2022,kahk_predicting_2022,kahk_accurate_2019,kahk_core_2021,kahk_combining_2023,taucher_final-state_2020,annegarn_combining_2022,maurer_assessing_2011}.

\subsection*{Current Status of the Implementation}

In FHI-aims, there are currently two ways to set a fixed occupation number for a given Kohn-Sham eigenstate.

In the first method, invoked by the keyword \texttt{deltascf\_projector}, the eigenstate whose occupation number is constrained is identified at the start of the SCF cycle by its index. Then, during subsequent SCF iterations, the maximum overlap method (MOM) \cite{lefebvre-brion_cecam-orsay_2021,gilbert_self-consistent_2008} is used to keep track of that state, even if the energy ordering of the eigenstates were to change.

In the second method, invoked by the keyword \texttt{deltascf\_basis}, the occupation number constraint is applied to the eigenstate that has the largest contribution from a particular basis function specified by the user. This is best used in cases when an eigenstate is clearly dominated by contributions from a single basis function as is often the case for core states.

The keywords \texttt{deltascf\_projector} and \texttt{deltascf\_basis} were introduced into FHI-aims as part of a major overhaul of the occupation-constrained DFT functionality \cite{hall_archer2-ecse04-3_2023}. These new routines replace an older implementation, internally referenced by the keywords \texttt{force\_occupation\_projector} and \\ \texttt{force\_occupation\_basis}, that was not compatible with the use of parallel eigensolvers.
In the new implementation, the application of occupation number constraints is performed with the (pre-)exascale electronic structure interface ELSI \cite{yu_elsi_2018,yu_elsi_2020}. The corresponding routine that creates non-Aufbau occupations can also be used within other codes. As a result of using ELSI, subject to the current limitations listed in Table \ref{Table_KS_methods}, $\Delta$SCF calculations now run at approximately the same speed as regular DFT calculations in FHI-aims.

\begin{table}
    \begin{center}
    \caption{A summary of the currently supported types of $\Delta$SCF calculations in FHI-aims.}
    \begin{tabular}{ c c c c }
        \hline
        Geometry & Kohn-Sham eigensolver & \texttt{deltascf\_projector} & \texttt{deltascf\_basis} \\
        \hline
        \multirow{2}{*}{Aperiodic} & Serial & \ding{51} & \ding{51} \\
         & Parallel & \ding{51} & \ding{51} \\
         \hline
        Periodic & Serial & \ding{51} & \ding{55} \\
        ($\Gamma$-point only) & Parallel & \ding{51} & \ding{55} \\
        \hline
        Periodic & Serial & \ding{51} & \ding{55} \\
        (multiple k-points) & Parallel & \ding{55} & \ding{55} \\
        \hline
    \end{tabular}
    \label{Table_KS_methods}
    \end{center}
\end{table}

\begin{figure}[ht]
    \centering
    \includegraphics[width=8.7cm]{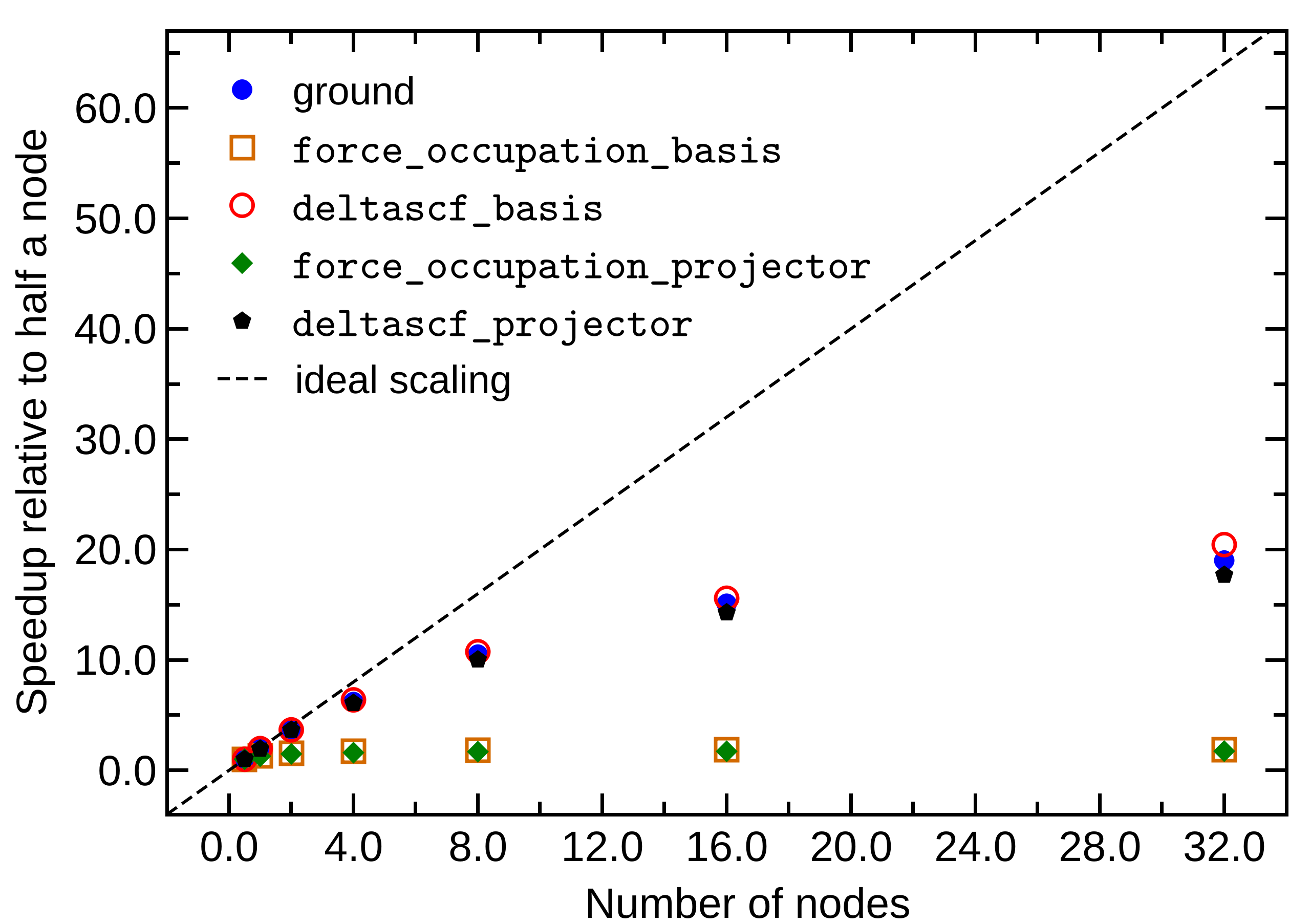}
    \caption{The scalability of the new \texttt{deltascf} routines in FHI-aims, in comparison to regular ground state DFT calculations, and the legacy \texttt{force\_occupation} routines. The data, originally presented in \cite{hall_archer2-ecse04-3_2023}, shows how the average time per SCF step depends on the number of nodes used to run the calculation. The new \texttt{deltascf} routines show similar scalability to regular ground state calculations, whereas with the old implementation, there was almost no improvement in performance when using more than 5 nodes. The test system for the shown scalability calculations was a peroxide-terminated diamond cluster containing 205 atoms - the calculations were run using the PBE functional and the FHI-aims default `intermediate' level basis sets \cite{Perdew1996,Blum2009}. The tests were performed on a machine with 128 CPU cores per node, and speedup was calculated relative to the calculation run on half a node (64 cores). The dashed line indicates an ideal 1:1 speedup with respect to the number of processes.}
    \label{Figure_speedup}
\end{figure}

An illustration of the computational speed-up that the new \texttt{deltascf} routines offer over the previous, \texttt{force\_occupation} routines is provided in Figure \ref{Figure_speedup}. The legacy keywords\\ \texttt{force\_occupation\_projector} and \texttt{force\_occupation\_basis} are now considered obsolete and will be removed from FHI-aims in a future release. 

\subsection*{Usability and Tutorials}
In the following, the workflow for modelling core-excited states in FHI-aims is outlined for each of the \texttt{deltascf} methods described above. A more detailed tutorial on performing $\Delta$SCF calculations in aims is available at \url{https://fhi-aims-club.gitlab.io/tutorials/core-level-with-delta-scf}.

The general workflow consists of three steps: (i) initializing a localized core hole, (ii) running an SCF calculation in which the remaining electrons are allowed to relax in the presence of the core hole, and (iii) verifying the correct localization of the core hole at the end of the SCF calculation.

The procedure for creating a core hole depends on which \texttt{deltascf} method is being used. For \texttt{deltascf\_ projector}, the calculation must be started from an orbital-based restart file (or files). The restart files are typically obtained from a ground state calculation of the same system. To create a core hole on a particular atom, the user must identify the Kohn-Sham eigenstate that best represents the relevant atomic core orbital. In simple cases, this can often be done by examination of the orbital eigenvalues, but tools such as Mulliken analysis \cite{mulliken1955} or visualization using ``cube'' files can also be used for this purpose. When the structure contains multiple equivalent atoms, an additional symmetry-breaking step may be required to create a properly localized core hole. Often, simply using a different basis set on the atom whose core electron is to be removed is sufficient for localizing the core orbitals. Alternatively, for aperiodic systems, Boys localization can be applied to the core eigenstates at the end of the ground state calculation \cite{michelitsch_efficient_2019,michelitsch_ab_2019,foster_canonical_1960,rossi_nuclear_2016}. Finally, as a last resort, localization of the core orbitals can be achieved by obtaining the restart files from a calculation in which the nuclear charge on the chosen atom where the core hole is meant to be introduced is artificially increased by a small amount, e.g. 0.1 e. For \texttt{deltascf\_basis}, the use of restart files is not mandatory, although the user must still ensure that the core hole correctly localizes in systems with equivalent atoms.

Performing a DFT calculation with non-Aufbau-principle occupation numbers is in general no different from a ground state DFT calculation in FHI-aims. When a calculation of a periodic system with a charged unit cell is requested, FHI-aims automatically introduces a compensating uniform background charge. When analyzing the results of periodic calculations, the effect of the spurious interactions between periodic images of the core hole must be taken into account. Due to the highly localized nature of core holes, in homogeneous bulk materials the finite size effect is often well approximated by the first term of the Makov-Payne correction \cite{makov_periodic_1995,kahk_core_2021}. In other cases, systematic convergence testing with respect to unit cell sizes is required \cite{chaudhuri_coexistence_2022}.

Finally, for verifying the correct localization of the core hole, again, Mulliken analysis or direct visualization of the Kohn-Sham orbitals can be used. In systems with a closed shell ground state, Mulliken spin analysis is a particularly convenient tool, as the spin density of a core hole will normally be almost entirely localized onto a single atom.


\subsection*{Future Plans and Challenges}
As noted in Table \ref{Table_KS_methods}, the current implementation of occupation-constrained DFT in FHI-aims has some limitations in calculations of periodic systems. Specifically, at present, \texttt{deltascf\_basis} type constraints cannot be applied in periodic calculations at all, and periodic calculations with more than one k-point that use \texttt{deltascf\_projector} type constraints can presently only be performed using the serial eigensolver. Removing these limitations is the most important short-term goal of the ongoing development work, as calculations with complex periodic structures are often desirable for the interpretation of practical XPS or XAS spectra. Another aim of ongoing work is to remove the limitation that the \texttt{deltascf} keywords cannot be currently used in calculations with a fixed overall spin moment.

In the medium term, our aim is to develop localized numerical basis sets that have been optimized for calculations of atoms with a core hole. To facilitate this, the keyword \texttt{core\_shell\_occ} for the species definition section of the \texttt{control.in} file has been recently introduced. \texttt{core\_shell\_occ} allows the minimal basis functions, that form a part of the FHI-aims default basis sets to be generated for electronic configurations in which the atomic core orbitals are not fully occupied. However, for complete and usable basis sets, additional sets of augmentation functions (analogous to the Tier basis functions used in FHI-aims default basis sets) optimized for atoms with a core hole must also be developed.

In the longer term, there are plans to also integrate the current $\Delta$SCF functionality with the implementation of fully relativistic DFT in FHI-aims \cite{Zhao2021}, to permit calculations of systems with a core hole in a spin-orbit split $p$-, $d$-, or $f$-shell.

\subsection*{Acknowledgements}

This project has received funding from the European Union's Horizon Europe research and innovation programme under grant agreement no. 101131173. Refactorization of the $\Delta$SCF code in FHI-aims was funded under the embedded CSE programme of the ARCHER2 UK National Supercomputing Service (http://www.archer2.ac.uk). RJM acknowledges support through the UKRI Future Leaders Fellowship programme (MR/S016023/1 and MR/X023109/1), and a UKRI frontier research grant (EP/X014088/1). JMK acknowledges support from the European Union’s Horizon Europe project EXANST (contract 101159716), from the Estonian Ministry of Education and Research grant number TK210 and from the Estonian Research Council grant number PSG1037.

\newpage

\chapter{Linear Response Methods}
\label{ChapLinearResponse}




\newpage

\newcommand{\colW}{2.2em} 
\newcommand{\rowH}{\dimexpr\ht\strutbox+\dp\strutbox\relax}
\newcommand{\NAblack}{%
  \begingroup
  \setlength{\fboxsep}{0pt}%
  \colorbox{black}{\makebox[\colW]{\rule{0pt}{\rowH}}}%
  \endgroup
}
\newcolumntype{C}{>{\centering\arraybackslash}m{\colW}}

\section{Density-Functional Perturbation Theory with Numeric Atom-Centered Orbitals}
\label{ChapDFPT}

\sectionauthor[1]{Connor L. Box}
\sectionauthor[1,2,3]{Reinhard J. Maurer}
\sectionauthor[4,5,a]{Honghui Shang}
\sectionauthor[4]{Nathaniel Raimbault}
\sectionauthor[4]{Danilo Simoes Brambila}
\sectionauthor[6]{Matthias Scheffler}
\sectionauthor[7,8]{Volker Blum}
\sectionauthor[6,9,b]{Christian Carbogno}
\sectionlastauthor[10]{\textbf{ *Mariana Rossi}}

\sectionaffil[1]{Department of Chemistry, University of Warwick, Gibbet Hill Road, CV4 7AL Coventry, United Kingdom}
\sectionaffil[2]{Department of Physics, University of Warwick, Gibbet Hill Road, CV4 7AL Coventry, United Kingdom}
\sectionaffil[3]{Faculty of Physics, University of Vienna, Vienna A-1090, Austria}
\sectionaffil[4]{Theory Department (since 1/1/2020: The NOMAD Laboratory), Fritz Haber Institute of the Max Planck Society, Faradayweg 4-6, D-14195 Berlin, Germany}
\sectionaffil[5]{Key Laboratory of Precision and Intelligent Chemistry, University of Science and Technology of China, Hefei, China}
\sectionaffil[6]{The NOMAD Laboratory at the Fritz Haber Institute of the Max Planck Society, Faradayweg 4-6, D-14195 Berlin, Germany}
\sectionaffil[7]{Thomas Lord Department of Mechanical Engineering and Materials Science, Duke University, Durham, NC 27708, USA}
\sectionaffil[8]{Department of Chemistry, Duke University, Durham, NC 27708, USA}
\sectionaffil[9]{Theory Department, Fritz Haber Institute of the Max Planck Society, Faradayweg 4-6, D-14195 Berlin, Germany}
\sectionaffil[10]{Max Planck Institute for the Structure and Dynamics of Matter, 22761 Hamburg, Germany}

\sectionaffil[*]{Coordinator of this contribution.}
\rule[0.25ex]{0.35\linewidth}{0.25pt}

\sectionaffil[a]{{\it Current Address:} Key Laboratory of Precision and Intelligent Chemistry, University of Science and Technology of China, Hefei, China}
\sectionaffil[b]{{\it Current Address:} Theory Department, Fritz Haber Institute of the Max Planck Society, Faradayweg 4-6, D-14195 Berlin, Germany}




\subsection*{Summary}

\setlength{\intextsep}{1.0pt}%
\setlength{\columnsep}{8pt}%

Density-functional perturbation theory (DFPT) extends the framework of DFT to the calculation of response properties of the electronic density. DFPT can give access to a large variety of properties, depending on the type of external stimulus that generates the perturbation~\cite{Baroni2001}. An incomplete list of these properties includes force constant matrices and phonons~\cite{gian+prb1991}, molecular polarizabilities~\cite{maha+pra1980} and dielectric properties of solids~\cite{Baroni2001,gonz+prb1997}, vibrational scattering cross sections~\cite{raim+prm2019, putr+jcp2000}, electron-phonon coupling and non-adiabatic matrix elements~\cite{gius+prb2007, box+es2023}, NMR shifts and J-couplings~\cite{laasner+es2024}.

The implementation of these techniques within a numeric atom-centered orbital (NAO) framework and real-space context, aligned with the FHI-aims code architecture, necessitates specific choices that we outline in the next section. The original implementation, as detailed in papers by H. Shang and coworkers~\cite{shang+cpc2017,shang+njp2018}, employed independent routines for different perturbation scenarios. Recently, this approach has been streamlined by centralizing these routines, resulting in a more robust implementation and improved support for the code. \cite{box+ecse2021}

Like other functionalities of FHI-aims, both periodic and non-periodic evaluations are possible within the same all-electron, NAO infrastructure. For periodic phonons, only limited support (LDA functional, and systems that do not require large memory) is provided in the current implementation. A more comprehensive and very efficient handling of the calculation of phonons with different functionals and larger system sizes is provided through the FHI-aims connection to the FHI-vibes infrastructure~\cite{Knoop2020a}, detailed in another contribution of this Roadmap.

\subsection*{Current Status of the Implementation}
In the following, we summarize the main aspects of the implementation regarding atomic displacement and electric field perturbations. A dedicated section on the implementation for magnetic field perturbations is given in contribution \ref{ChapNMR} of this Roadmap. 

\begin{wrapfigure}{r}{0.4\textwidth}
  \begin{center}
    \includegraphics[width=0.25\textwidth]{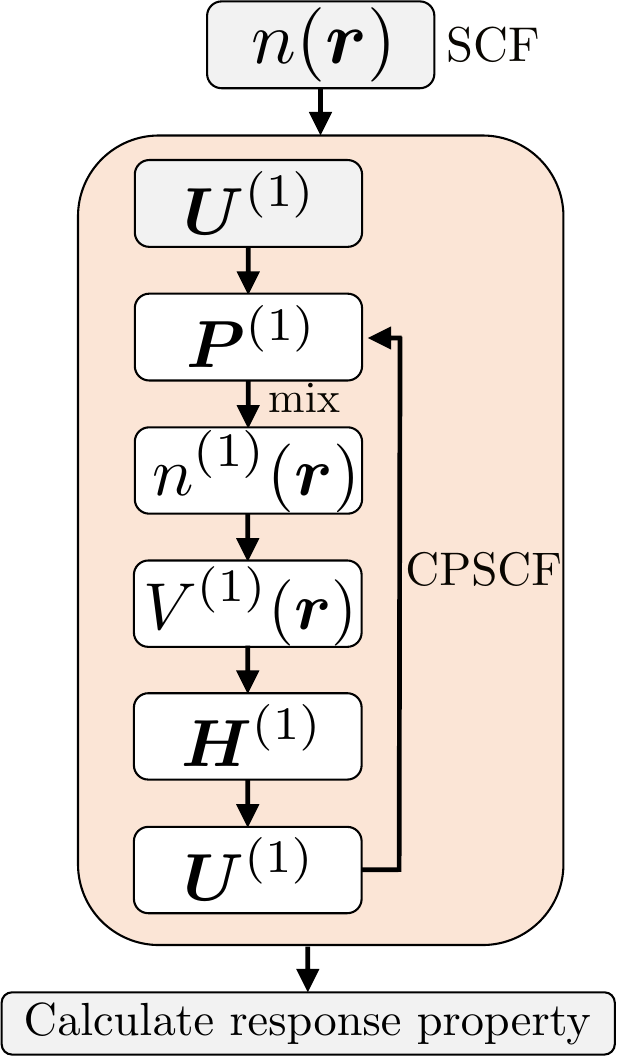}
  \end{center}
  \caption{Self-consistent workflow for response properties in FHI-aims. The ground-state density $n(\bm{r})$ is used for a first guess of the expansion coefficients $\bm{U}^{(1)}$, which enters the response density-matrix $\bm{P}^{(1)}$, which then defines the response density $n^{(1)}$.  $n^{(1)}$ enters the perturbation potential $V^{(1)}$, the Hamiltonian matrix $\bm{H}^{(1)}$ is defined and new expansion coefficients $\bm{U}^{(1)}$ are built. The procedure is evolved self-consistently until the change in $\bm{P}^{(1)}$ falls below a certain threshold. The desired properties are subsequently calculated.} \label{fig:cpscf}
\end{wrapfigure}

In FHI-aims, the Kohn-Sham orbitals $\psi$ are expanded in a local NAO basis (cf. (\ref{Eq:KS-Orbitals})):
\begin{equation}
\psi^{(0)}_l = \sum_i c_{i l}^{(0)} \varphi_i, \label{eq:psi0}
\end{equation}
where in the above notation $l$ is the orbital index, $i$ is an index running over the basis functions, the superscript $(0)$ indicates that this is a ground-state quantity, and $\varphi$ are atom-centered basis functions in finite systems and Bloch-like superpositions of atom-centered basis functions in the periodic case. For the latter case, $\varphi$ also carries a complex phase (see Eqs. 23 and 24 in Ref.~\cite{shang+njp2018}) and the orbitals would depend on $\bm{k}$, which we are not showing to simplify notation. Denoting first-order response quantities with the superscript $(1)$, standard first-order perturbation theory yields the Sternheimer equation~\cite{ster+pr1954} (for each $\bm{k}$ point)
\begin{equation}
(\hat{h}_{KS}^{(0)} - \epsilon_l^{(0)}) | \psi_l^{(1)} \rangle = -(\hat{h}^{(1)}_{KS} - \epsilon_l^{(1)}) | \psi_l^{(0)} \rangle.
\end{equation}
In order to solve it, we can expand the response of the wavefunction analogously to Eq.~\ref{eq:psi0},
\begin{equation}
\psi^{(1)}_l = \sum_i \left[ c^{(1)*}_{il} \varphi_i^{(0)}  + c^{(0)*}_{il} \varphi_i^{(1)}  \right] .\label{eq:psi1}
\end{equation}

For nuclear displacements, the atom-centered basis sets $\varphi$  indeed change upon displacement and thus yield a non-zero $\varphi_i^{(1)}$, whilst in the case of an electric-field perturbation this term is zero and Eq.~\ref{eq:psi1} simplifies to only the first term. While DFPT formalisms concentrate on a self-consistent procedure to determine directly the coefficients $c^{(1)}$, the procedure commonly named coupled perturbed self-consistent field (CPSCF) writes $c^{(1)}_{il} = \sum_{l'} U_{ll'} c_{il'}^{(0)} $ as a linear expansion on the unperturbed coefficients. The new coefficients $U_{ll'}$, where $l$ denotes an occupied orbital and $l'$ an unoccupied one, take the following form
\begin{equation} \label{eq:u1}
 U_{ll'} =  \begin{cases}
     \frac{\sum_{ij}(c_{il'}^{(0)})^{*}H_{ij}^{(1)} c_{jl}^{(0)}}{\epsilon_l^{(0)} - \epsilon_{l'}^{(0)}}, & \text{el. field}\\
    \frac{\sum_{ij}(c_{il'}^{(0)})^{*}H_{ij}^{(1)} c_{jl}^{(0)} - \sum_{ij} (c_{il'}^{(0)})^{*} S_{ij}^{(1)} (\bm{c}^{(0)} \bm{E}^{(0)})_{jl}}{\epsilon_l^{(0)} - \epsilon_{l'}^{(0)}}, & \text{nuclear displ.}
  \end{cases}
\end{equation}
and a self-consistent procedure is also necessary to determine them (see Fig.~\ref{fig:cpscf}). Note that the response of the overlap matrix ($\bm{S}^{(1)}$) is only necessary for nuclear displacement perturbations. The CPSCF formulation, while completely equivalent to DFPT, turns out to be advantageous in FHI-aims because it allows us to make use of existing algorithms for the massively parallel evaluation of matrix elements in this representation. This expansion also allows computing the density response
\begin{eqnarray}
n^{(1)} & = &  \sum_l f(\epsilon_l) \left[\psi^{(1)*}\psi^{(0)} + \psi^{(0)*}\psi^{(1)}\right] = \\ 
 & = & \begin{cases} 
 \sum_{ij} P_{ij}^{(1)} \varphi_{i}^{(0)} \varphi_{j}^{(0)},  &  \text{el. field} \\
 \sum_{ij} [P_{ij}^{(1)} \varphi_{i}^{(0)} \varphi_{j}^{(0)} + P_{ij}^{(0)}(\varphi_{i}^{(1)} \varphi_{j}^{(0)} + \varphi_{i}^{(0)} \varphi_{j}^{(1)}) ] & \text{nuclear displ.}
  \end{cases}
\end{eqnarray}
within a response density-matrix $\bm{P}^{(1)}$ formalism, analogous to the ground-state density evaluation in the code. The full self-consistent procedure is summarized in Fig.~\ref{fig:cpscf}.

\begin{table}[t]
    \centering
    \setlength{\tabcolsep}{6pt}
    \renewcommand{\arraystretch}{1.2}
    \caption{\rev{DFPT Functionality matrix.  Columns represent aperiodic (AP) and periodic (P) implementations across electric field response, electron-phonon coupling, and phonon drivers; ``X'' marks unavailable functionality. Solid-black cells (\NAblack) denote combinations that are not meaningful. The symbol $\bm{q}$ represents the phonon wavevector.}}
\begin{tabular}{|l|C|C|C|C|C|C|}
\hline     & \multicolumn{2}{c|}{Electric} & \multicolumn{2}{c|}{EPC} & \multicolumn{2}{c|}{Phonon} \\
\hline & AP & P & AP & P & AP & P \\
\hline LDA &  &  &  &  &  &  \\
\hline GGA &  &  &  &  &  & X \\
\hline Hybrid  &   & X & X & X & X & X \\
\hline MetaGGA  & X & X & X & X & X & X \\
\hline Frozen core &  &  &  &  &  &  \\
\hline Spin Collinear &  &  &  &  & X & X \\
\hline ScaLAPACK &  &  &  &  &  & X \\
\hline \texttt{collect\_eigenvectors .false.} & X & X & X & X & X & X \\
\hline \texttt{local\_index .true.} & X & X & X & X & X & X \\
\hline Real Eigenvectors &  &  &  &  &  & X \\
\hline Complex Eigenvectors & \NAblack   &  & \NAblack   &  & \NAblack   &  \\
\hline $\bm{q}=0$ &  &  &  &  &  &  \\
\hline $\bm{q}>0$ & \NAblack & \NAblack  & \NAblack & X & \NAblack & X \\
\hline Fractional occupations &  &  &  & X & X & X \\
\hline Dipole correction (surface) & \NAblack   &   & \NAblack   &  & \NAblack   &  \\
\hline
\end{tabular}
\label{tab:functionality-matrix}
\end{table}

\begin{figure}[ht]
    \centering
    \includegraphics[width=\textwidth]{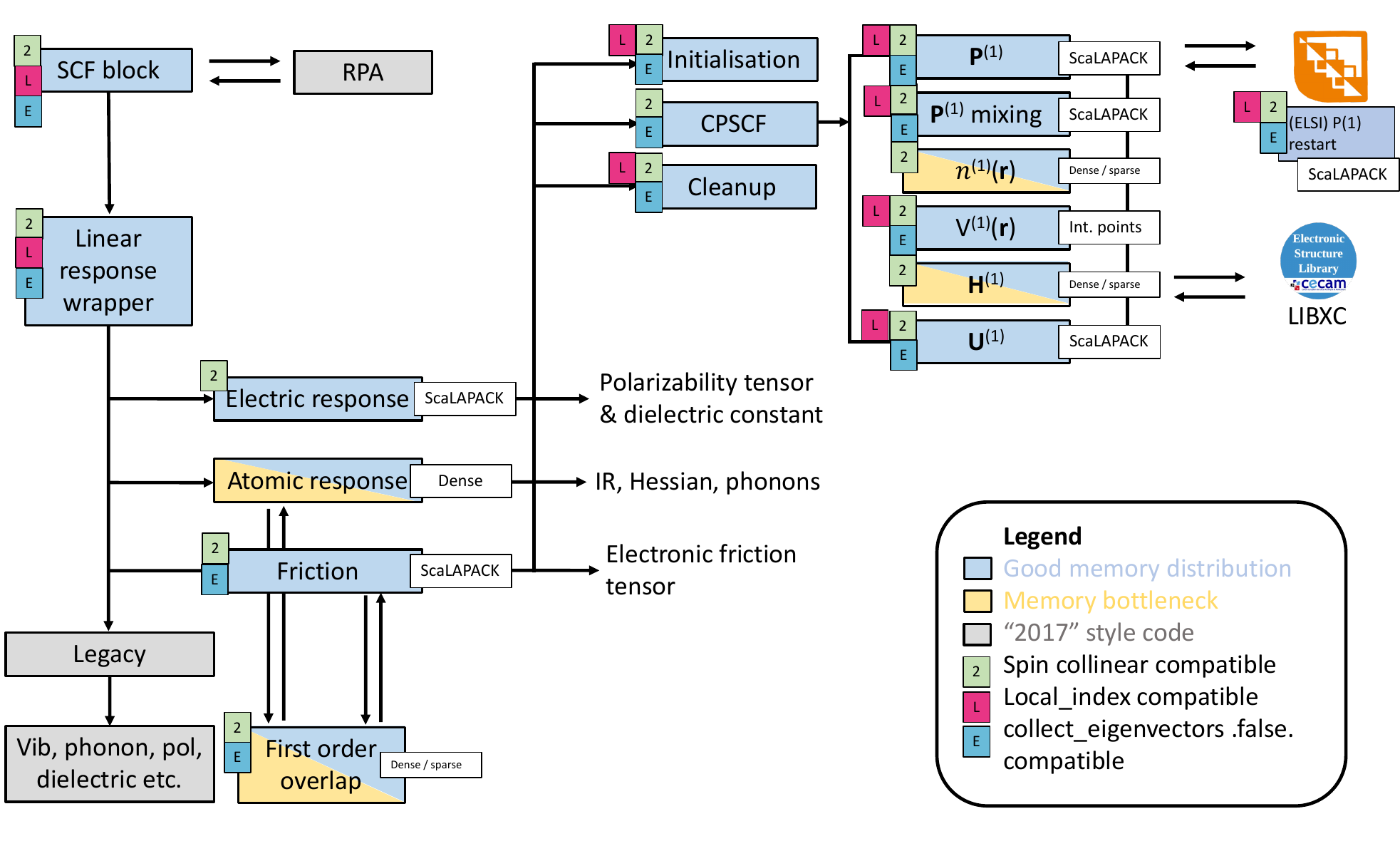}
    \caption{Schematic representation of the current code structure in FHI-aims. ``Dense / sparse'' refers to when the full global dense (aperiodic) or sparse (periodic) response matrices are needed, ``Int. points'' refers to distribution over integration grid points and ``ScaLAPACK'' refers to ScaLAPACK support for distribution of matrices. Mixed coloured regions currently represent a memory bottleneck for aperiodic systems due to the use of full dense matrices.}
    \label{fig:fhiaims-flow}
\end{figure}

\rev{An overview of the currently supported functionality is summarized in Table~\ref{tab:functionality-matrix}, whilst the code layout and support of features for each specific routine is given in Fig.~\ref{fig:fhiaims-flow}. Focusing on Fig.~\ref{fig:fhiaims-flow}, the DFPT implementation in \textsc{FHI-aims} is structured around a central \textit{linear-response wrapper} that is invoked following self-consistent ground-state convergence. This wrapper calls the different response drivers currently available, which include: (i) \textit{electric response} (e.g., polarizabilities and dielectric constants), (ii) \textit{atomic response} (phonon spectra and molecular vibrations), and (iii) \textit{electronic friction} (electron--phonon coupling). These response drivers use the same common DFPT/CPSCF module. The modular design facilitates straightforward extension of the framework: each functionality module adheres to a shared and well-documented interface template, enabling future developments that build upon the same DFPT infrastructure. A natural forthcoming addition is the random-phase approximation (RPA) framework, whose force evaluations depend on response quantities that are provided by this same infrastructure.}

\rev{Recent developments extend the \textit{electric response} and \textit{electronic friction} modules to fractionally occupied systems, including metals, which require specialized handling of the denominator in Eq.~\ref{eq:u1}. Further avenues for optimizations are clear: for example, enabling the \texttt{use\_local\_index} flag reduces memory footprint and can improve parallel efficiency, particularly for large-scale systems, routines that support this are labelled in Fig.~\ref{fig:fhiaims-flow}.  Table~\ref{tab:functionality-matrix} additionally highlights current capabilities such as spin-collinear support (two spin channels) and the use of \texttt{collect\_eigenvectors .false.}, which -- when paired with ScaLAPACK-based matrix operations -- offers significant memory savings for large matrix problems. We note that the cells marked with ``X'' in Table~\ref{tab:functionality-matrix} represent improvements that should be achieved in the near future. In particular, the robust support of hybrid and meta-GGA functionals for the responses is an ongoing effort.}

The central CPSCF module now features interfaces to external open-source libraries such as ELSI~\cite{yu_elsi_2020} and LibXC~\cite{marques+cpc2012}. ELSI enables a restart of the CPSCF routine by parallelised I/O of the first order density matrix and LibXC enables the calculation of the first order Hamiltonian for arbitrary LDA, GGA, and (when supported) hybrid exchange-correlation functionals

Routines that represent significant bottlenecks in memory distribution and parallelisation are shaded yellow in Fig.~\ref{fig:fhiaims-flow}, whilst routines with optimised memory distribution are shaded blue. The half-shading refers to having full dense matrices allocated in the aperiodic case which is memory inefficient when compared to cases that support ScaLAPACK-type distribution, whilst real-space sparse matrices are allocated in the periodic case, which is generally a memory efficient approach. \rev{For the latter case, \texttt{use\_local\_index} is desirable for large scale systems.}

\subsection*{Usability and Tutorials}
All functionality is now accessible through simple keywords in the \texttt{control.in} file, documented in the FHI-aims manual since release 240507. 
Currently, there are keywords that control the central CPSCF shared by all driver routines, these all start with the prefix \texttt{dfpt\_} and control accuracy thresholds, mixing and restart behaviour.
The actual DFPT calculation is triggered by one of the driver-related keywords, currently these are \texttt{electric\_field\_response~DFPT}, \texttt{atomic\_pert\_response~DFPT}, or \texttt{calculate\_friction~DFPT}.

We have recently released online tutorials, which will be continuously updated and augmented, in order to make the functionality more accessible to new users and showcase what the DFPT functionality can achieve. One tutorial employs the electric-field response functionality to calculate non-resonant vibrational Raman spectra of an isolated molecule and a molecular crystal. Notably, this tutorial also showcases the new implementation of the Berry-phase polarization through the use of Wannier orbitals (showcased in another article of this Roadmap) in order to calculate IR spectra of periodic systems. Another tutorial makes use of the electronic friction driver to calculate electron-phonon couplings and vibrational lifetimes including non-adiabatic effects through electronic friction for a molecular overlayer adsorbed on a metallic surface. These effects are discussed in depth in another contribution of this Roadmap. These tutorials can be accessed through \url{https://fhi-aims-club.gitlab.io/tutorials/tutorials-overview/}.

\subsection*{Future Plans and Challenges}
The newly refactored infrastructure~\cite{box+ecse2021} is significantly more compact and easier to maintain than the original. The total number of code lines has been reduced by just over 60\% for the CPSCF code, with the newer interface also including more features. Several existing versions of multiple routines have been consolidated, reducing code duplication and increasing clarity for future developers.
Specific run-mode cases are now only dealt with at the lowest level and high-level developments at the level of the CPSCF cycle are decoupled. In the near future, we plan to extend the support of local dense matrix computations and the distribution of eigenvectors (currently, a copy of the full eigenvector is created on every MPI task for the $\bm{k}$-point(s) it is working on). These changes will improve memory usage and computational efficiency of not just the CPSCF calculations but allow these features to be utilised in the SCF part of the calculation as well, where they are already supported. A scalable implementation of DFPT based on an all-electron NAO basis that performs across various HPC systems has been reported~\cite{Shang2021}. These implementations can address systems up to 200,000 atoms. \rev{The code scalability using the ELSI library support has also been reported for systems of more than 3000 atoms in Ref.~\cite{SHANG2021-2}.} These developments are compatible with FHI-aims.


It is also important to stress the role that machine-learning (ML) plays in the calculation of electronic response properties. While this field is still less advanced than other topics of ML for materials science, FHI-aims is already able to produce the density-response data that is needed to train models that learn the electronic density response directly, or its derived quantities. The success of these frameworks based on FHI-aims data in predicting Raman spectra, dielectric properties of materials, and electronic friction tensors has already been established~\cite{raim+njp2019,zhan+jpcc2020,shan+aipa2021, lewi+jcp2023}. This machinery can massively speed up computations and is expected to have an even larger impact on the calculation of electronic response properties in the coming years.

\subsection*{Acknowledgements}
We acknowledge fruitful discussions with Andrew Logsdail, Ville Havu, and Xinguo Ren. We acknowledge financial support through the UKRI Future Leaders Fellowship programme (MR/S016023/1 and MR/X023109/1), and a UKRI frontier research grant (EP/X014088/1). MS acknowledges support by his TEC1p Advanced Grant (the European Research Council (ERC) Horizon 2020 research and innovation programme, grant agreement No. 740233.

\newpage

\section{Nuclear Magnetic Resonance with Numeric Atom-Centered Basis Sets}
\label{ChapNMR}

\sectionauthor[1,2]{\textbf{*Volker Blum}}
\sectionauthor[1]{Xingtao He}
\sectionauthor[1]{William P. Huhn}
\sectionauthor[1]{Raul Laasner}
\sectionauthor[3]{Iuliia Mandzhieva}
\sectionauthor[1]{Jack Morgenstein}
\sectionauthor[3]{\textbf{*Thomas Theis}}
\sectionlastauthor[3]{Franziska Theiss}

\sectionaffil[1]{Thomas Lord Department of Mechanical Engineering and Materials Science, Duke University, Durham, NC 27708, USA}
\sectionaffil[2]{Department of Chemistry, Duke University, Durham, NC 27708, USA}
\sectionaffil[3]{Department of Chemistry, North Carolina State University, Raleigh, North Carolina 27606, United States}

\sectionaffil[]{(*) Coordinator of this contribution}



\subsection*{Summary}

Magnetic resonance (MR) spectroscopies are disproportionately important in chemistry, physics, materials science and medicine as atomic-scale probes of the structure and dynamics of matter, as well as in the form of non-destructive diagnostic tools, by exerting quantum control over ensembles of spins \cite{Hodgkinson2024}. Key atomic-scale parameters of relevance for MR include nuclear shieldings (i.e., the degree to which the response of an individual nucleus is altered by the presence of a surrounding system), nuclear spin-spin couplings or $J$-couplings (i.e., the effective parameters that govern the quantum-mechanical interaction of two nuclear spins in their surrounding medium), but also molecular magnetizabilities (i.e., the diamagnetic response of the electronic system to an external magnetic field). 

In FHI-aims, shieldings and $J$-couplings for nuclear magnetic resonance (NMR), as well as magnetizabilities are accessible for molecules at the scalar-relativistic level of theory and have been demonstrated for large systems \cite{laasner+es2024}. NAO basis sets capture magnetizabilities and shieldings efficiently. For $J$-couplings, the response at the nucleus is captured by a benchmarked group of specialized NAO basis sets (``NAO-J-n''), which add highly localized Gaussian-type orbitals (GTOs) near the nucleus, following Jensen's GTO-based pc-J-n basis sets \cite{Jensen2006,Jensen2010}. NMR spectra can be interpreted from these observables, either by employing empirical principles or by interfacing with spin dynamics computations. FHI-aims supports the standardized SpinXML output format \cite{Biternas2014}, which allows it to communicate calculated magnetic resonance parameters directly to the SPINACH spin-dynamics package \cite{Hogben2011}.

\subsection*{Current Status of the Implementation}

FHI-aims follows the density-functional perturbation theory (DFPT) formalism \cite{Buehl1999} to compute magnetic resonance observables. The general DFPT formalism is briefly summarized in contribution \ref{ChapDFPT} of this Roadmap. For detailed expressions pertaining to MR as implemented in FHI-aims and for further references, see Ref. \cite{laasner+es2024}). 

The magnetizability $\overleftrightarrow\xi$ of a closed-shell molecule, shielding $\overleftrightarrow\sigma_A$ of a nucleus of $A$, and $J$-coupling $\overleftrightarrow J_{AB}$ of two nuclei of atoms $A$ and $B$ are defined as the following derivatives:
\begin{equation}
  \label{eq:xi_def}
  \overleftrightarrow\xi = - \left. \frac{\partial ^2E}{\partial \bm B^2} \right|_{\bm B = 0},
\end{equation}
\begin{equation}
  \label{eq:shield_def}
  \overleftrightarrow\sigma_A = \left. \frac{\partial ^2 E}{\partial \bm B\partial \bm\mu_A} \right|_{\bm B=0;\bm\mu_A=0},
\end{equation}
\begin{equation}
  \label{eq:J_def}
    \overleftrightarrow J_{AB} = \frac{1}{h} \left. \frac{\partial ^2 E}{\partial \bm I_A\partial \bm I_B} \right|_{\bm I_A=\bm I_B=0} 
    = h\frac{\gamma_A}{2\pi}\frac{\gamma_B}{2\pi} \left. \frac{\partial ^2 E}{\partial \bm\mu_A\partial \bm\mu_B} \right|_{\bm\mu_A=\bm\mu_B=0}.
\end{equation}
The $\overleftrightarrow ~$ notation indicates that each quantity is a $3\times 3$ tensor and depends on molecular orientation, though the trace is frequently used as an average if the molecular orientation is not known. $E$ is the total energy, $\bm B$ is the externally applied magnetic field, $\bm \mu_A$ and $\bm \mu_B$ are nuclear magnetic moments, $\gamma_A$ and $\gamma_B$ their gyromagnetic ratios, and $\bm I_A$ and $\bm I_B$ are the nuclear spins. $h$ is Planck's constant. It is worth noting that $\overleftrightarrow J_{AB}$ is not the total spin-spin tensor but rather only the contribution due to the electronic response. The classical dipole–dipole tensor between the nuclei can be tabulated separately (it does not depend on the electronic system) and must be added to $\overleftrightarrow J_{AB}$ to obtain the total coupling tensor.

FHI-aims implements all three tensor types by DFPT for non-periodic system geometries, tested for over 1,000 atoms \cite{laasner+es2024}. The local-density approximation and generalized-gradient (GGA) density functionals are fully supported. A full self-consistent field cycle without fields or nuclear spins is performed first, followed by a post-processing computation of the desired tensor for each degree of freedom -- a single tensor for $\overleftrightarrow\xi$, $N$ tensors $\overleftrightarrow\sigma_A$ for $N$ nuclei considered, and $N (N-1)/2$ tensors for $\overleftrightarrow J_{AB}$. \rev{The current implementation implements non-relativistic and scalar-relativistic theory, limiting the applicability to light elements, where this level of theory is standard in quantum chemistry. For heavy elements, spin-orbit coupling effects will be necessary but are not yet considered.}

In quantum mechanics, a magnetic field dependence is represented through the vector potential $\bm A$ that generates the field. Since $\bm A = \bm B \times \bm r$, the resulting operator depends on the chosen origin and the perturbed wave function acquires a position-dependent phase factor. The standard solution in the computation of shieldings and magnetizabilities with localized basis functions is to include the gauge as a phase factor in the basis functions (the GIAO approach \cite{Ditchfield1972}). Thus, in FHI-aims, gauge-including NAOs $\varphi_j^\mathrm{GIAO} = e^{-\frac{i}{2} (\bm R_{A(j)}\times\bm r)\bm B} \varphi_j(\bm r)$ are employed. $A(j)$ denotes the atom at which basis function $j$ is centered. Due to the nature of the perturbation, the first-order perturbation of the orbitals is imaginary-valued and the first-order density perturbation associated with shieldings and magnetizabilities is zero. 

In contrast, $J$-couplings do not require the inclusion of a gauge-dependent phase factor. However, the overall perturbation is associated with a non-zero first-order perturbed density. This response of the electronic system to the presence of a nucleus must be represented by an adequate basis set close to the nucleus. By construction, standard NAO basis sets intended for ground-state calculations do not provide this flexibility since the near-nuclear ground state wave function is already well represented by the minimal basis. Therefore, $J$-couplings cannot be computed with any degree of reliability using unmodified basis sets originally constructed for ground-state calculations. Instead, numerically precise $J$-couplings require both additional basis functions very close to the nucleus for stability, as well as dense integration grids very close to the nucleus in order to integrate the near-nuclear perturbation correctly. Therefore, for $J$-couplings, specifically designed basis sets and integration grids must be employed and are available in FHI-aims \cite{laasner+es2024}.

Ref. \cite{laasner+es2024} reports extensive precision benchmarks for molecular magnetizabilities, shieldings, and $J$-couplings computed with FHI-aims, covering light elements in the range $Z$=1-18 and conducted at the DFT-PBE \cite{Perdew1996} level of theory. Importantly, Ref. \cite{laasner+es2024} focused on numerical precision of the implementation, not the physical accuracy of the DFT-PBE functional, while separate and broader benchmark studies \cite{Schattenberg2021,Lehtola2021,Fabian2013} have shown reasonable qualitative accuracy of DFT-PBE for all quantities considered. The central findings of Ref. \cite{laasner+es2024} are:
\begin{itemize}
  \item Magnetizabilities are captured with excellent precision using FHI-aims'   standard ``tier'' basis sets for ground-state DFT \cite{Blum2009}. For example, FHI-aims'     ``tier 2'' basis sets achieve a precision of $\sim$10$^{-30}$ J/T$^2$ for a   set of 25 benchmark molecules, at around 2/3 the size of comparable 
    general-purpose GTO basis sets tested.
  \item For shieldings, either FHI-aims' standard basis sets \cite{Blum2009} or the NAO-VCC-nZ basis sets, originally designed for basis set convergence of correlated methods \cite{Zhang2013}, can be employed. The average precision of shieldings for 25   
    benchmark molecules at the ``tier 2'' level is around 3 ppm, similar to the quality of Jensen's dedicated GTO-based pcS-n basis sets \cite{Jensen2008} and also similar to the NAO-VCC-3 basis set level. More precise shielding calculations require significantly larger, more expensive basis sets; e.g., the NAO-VCC-4 basis sets approach an average precision of $\sim$1~ppm.
  \item For $J$-couplings involving light elements ($Z$=1-18), FHI-aims provides a set of dedicated basis sets ``NAO-J-n'' \cite{laasner+es2024}, constructed by amending the standard NAO-VCC-nZ basis sets by primitive tight GTO functions using exponents from Jensen's specialized pc-J basis sets \cite{Jensen2006,Jensen2010}, used together with dense integration grids. $J$-couplings involving elements H, B, C, N, O, Al, Si, S, and Cl are captured with an average precision level of $\sim$0.3~Hz already at the moderate NAO-J-3 basis set. The elements F and P show somewhat larger absolute deviations, limited to an average precision of $\sim$3~Hz (around 2\% of the overall $J$-couplings targeted for these elements, the magnitudes of which can be quite large).
  \item For calculations of shieldings and $J$-couplings in a single run, the NAO-J-n basis sets can be employed.
  \item Given the larger numerical demands of basis sets and grids for $J$-coupling calculations, it is possible to use the respective NAO-J-n species defaults only on the specific atoms of interest in a structure (e.g., a molecule of interest surrounded by a solvent), while computing the electronic structure of unrelated atoms with standard ground-state basis sets. 
  \item Finally, Jensen's GTO-based, dedicated pcS and pcJ basis sets for shieldings and $J$-couplings, respectively, extend to larger size than FHI-aims' basis sets and can also be used for benchmark precision studies in FHI-aims. 
\end{itemize}
Overall, computing magnetizabilities, shieldings and $J$-couplings for light-element molecules and nanostructures is thus well supported in FHI-aims at the DFT-GGA level of theory (particularly DFT-PBE) -- both using moderately sized NAO basis sets with good precision or larger, dedicated GTO basis sets (pcS or pcJ) for verification. 

\subsection*{Usability and Tutorials}

The DFPT formalism for magnetic resonance spectroscopies in FHI-aims is implemented for local-density and generalized-gradient approximation functionals. Non-relativistic and atomic ZORA scalar-relativistic calculations \cite{Blum2009} are supported. The standard FHI-aims ``tier'' basis sets as well as the NAO-VCC-nZ basis sets are tabulated as species defaults and can be used for magnetizabilities and shieldings. For the NAO-J-n basis sets, dedicated species defaults with stringent numerical settings exist, using FHI-aims' really\_tight numerical defaults as a foundation and increasing the radial density of integration grids by a factor of four compared to the standard really\_tight settings. In FHI-aims' terminology, this choice corresponds to the \texttt{radial\_multiplier} keyword set to a value of 8 \cite{Zhang2013}. Jensen's polarization consistent GTO basis sets (pcS for shieldings \cite{Jensen2008}, pcJ for $J$-couplings \cite{Jensen2006,Jensen2010} and larger, augmented version of the basis sets) are also tabulated as species defaults, with similarly tight settings for grids and for the Hartree potential. It is important to note that the NAO-J-n basis sets and the pcS and pcJ species defaults are thus numerically costly compared to FHI-aims' standard DFT basis sets, for which much lighter numerical settings have been benchmarked.

The actual selection of MR observables to be calculated is made in FHI-aims' input files \texttt{control.in} and \texttt{geometry.in}. At a minimum, the \texttt{magnetic\_response} keyword in \texttt{control.in} must be set to specify which type(s) of response observable(s) are requested. Additionally, the atoms to be included in shielding and $J$-coupling calculations must be identified in \texttt{geometry.in} -- specifically, by placing a \texttt{magnetic\_response} keyword after any selected atom. This is the minimum information needed by the code to instigate a response calculation. The details of the computation and of the DFPT cycle are extensively customizable by further keywords, all of which are documented in a dedicated chapter in FHI-aims' manual. A dedicated tutorial \cite{NMR-Tutorial} explains the necessary steps to compute magnetizabilities, shieldings and $J$-couplings for individual molecules. Furthermore, comments on efficiency and on comparisons to experimentally measured NMR spectra are included.

On their own, computed shielding values and $J$-couplings from static molecular structures can be used to interpret experimental NMR spectra, but directly plotting calculated shielding and $J$-coupling values will usually result in poor agreement with experiment. For rigid molecules in which all $^1$H atoms are equivalent, such as tetramethylsilane (TMS), calculated shielding values from a single point calculation may serve as a reference that is comparable to experiment. However, molecular motion, solvent environments, hydrogen exchange processes and other phenomena mean that the shape of NMR spectra is rarely a direct mapping of calculated shieldings and $J$-couplings. For example, the interaction timeframe of a radio frequency pulse with a nuclear spin is of the order of a microsecond, much longer than timescales associated with nuclear motion and certain exchanges.

\begin{figure}[!ht]
    \centering
    \includegraphics[width=\textwidth]{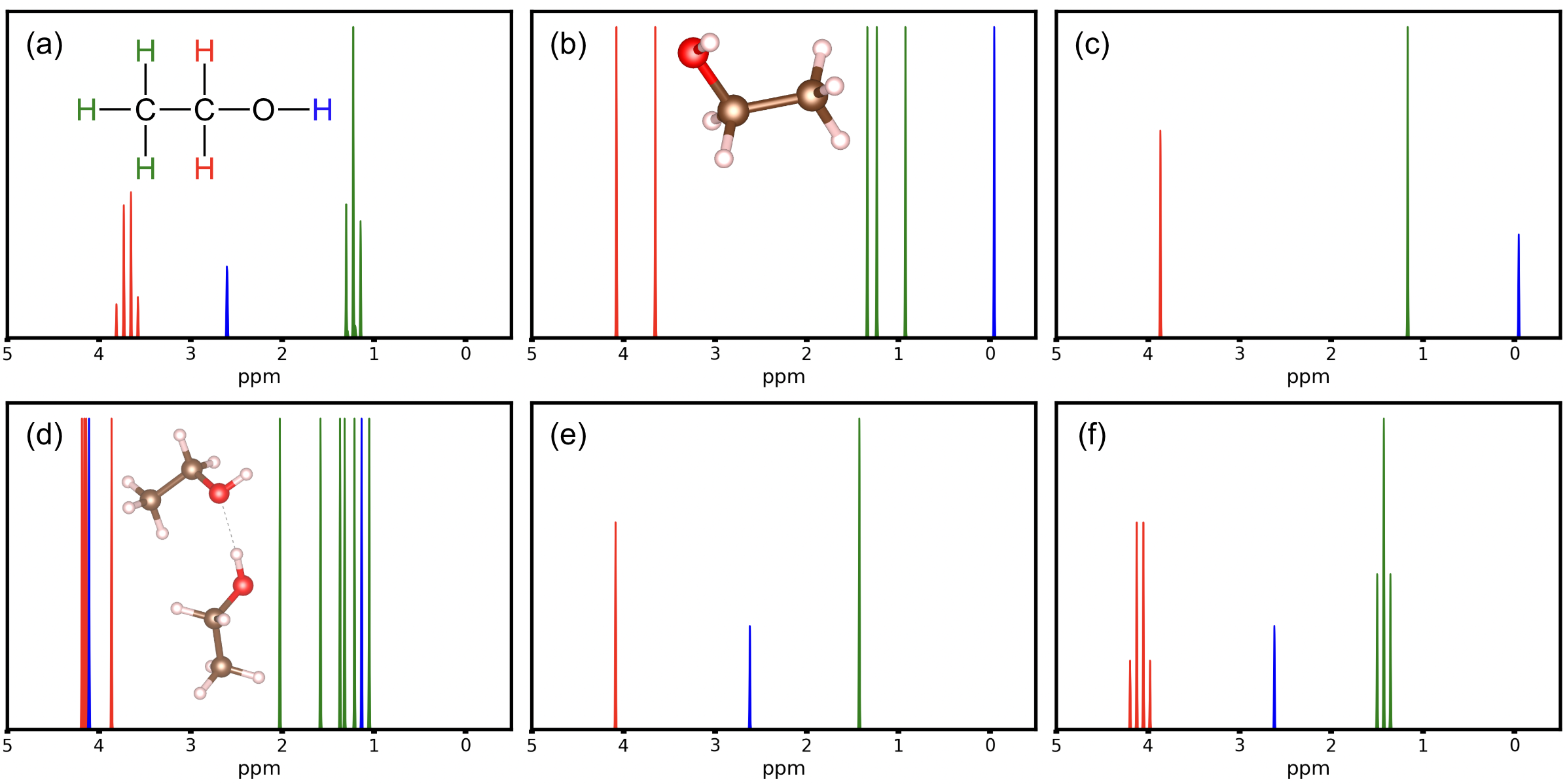}
    \caption{(a) Peaks in the experimental Ethanol $^1$H NMR spectrum at 89.56~MHz for 0.04 ml ethanol in 0.5 ml CDCl$_3$, data taken from \cite{chembook}. Red, blue, and green peaks are tentatively attributed to signals from the CH$_2$ group, OH group, and CH$_3$ group, respectively. (b) DFT-PBE single point calculation of $^1$H chemical shifts with respect to TMS, for a single ethanol molecule (inset). (c) Same data as (b) with signals averaged across the CH$_2$ and CH$_3$ groups in ethanol. (d) Data pertaining to a DFT simulation of the molecular dynamics (see text for details) of a hydrogen-bonded ethanol dimer. Structures were sampled from this ensemble and the NMR response for each atom was averaged across these samples and visualized. (e) Same data as (d) with signals averaged across each of CH$_2$, CH$_3$ and OH. (f) Same data as (e) but including the peak splitting due to $J$-coupling between H atoms in the CH$_2$ and CH$_3$ groups.}
    \label{fig:averaging}
\end{figure}

We exemplify the process of interpreting an experimentally obtained $^1$H NMR spectrum of ethanol, using experimental data obtained from ChemicalBooK \cite{chembook}. First, the experimental spectrum of ethanol itself is far from unique, but instead depends heavily on factors such as solvent used and concentration. Among the five very different spectra included in ChemicalBooK at the time of writing, we focus on a spectrum obtained for 0.04 ml ethanol in 0.5 ml CDCl$_3$ obtained at 89.56 MHz. The spectrum exhibits a set of well defined peaks, schematically redrawn in Figure \ref{fig:averaging}(a). In this instance, the spectrum shows clearly resolved groups of peaks that may be attributable to the CH$_2$ group (red), CH$_3$ group (green) and OH group (blue). The CH$_2$ and CH$_3$ peaks show splittings that can be attributed to interaction between both groups via $J$-couplings.

A first computation of all $^1$H chemical shifts (with respect to TMS, i.e., subtracting the chemical shift obtained in a separate static computation of TMS) associated with each H atom in a static, isolated ethanol molecule (inset) is shown in Figure \ref{fig:averaging}(b) (DFT-PBE, atomic ZORA scalar relativity, NAO-VCC-5Z basis sets \cite{Zhang2013} and numerical settings). In comparison to Figure \ref{fig:averaging}(a), although no $J$-couplings are yet considered in the computed spectrum, there is more than one peak per CH$_2$ and CH$_3$ group, respectively, and the peak associated with the OH group is far off. One obvious measure to achieve physically more appropriate results is to account for averages between the H atoms within the CH$_2$ and CH$_3$ groups, respectively, leading to the spectrum shown in Figure \ref{fig:averaging}(c). In this figure, the peak intensity is scaled to the number of atoms represented by a given peak. The CH$_2$ and CH$_3$ peaks are now reassuringly close to the related peaks in the experimental spectrum, though still without accounting for $J$-couplings. However, the peak associated with OH is far off. Similar observations can be made for spectra in different solvents and under conditions in \cite{chembook} (not shown here), in which the CH$_2$ and CH$_3$ peaks remain roughly in the same locations but the position of the OH peak varies widely. Typically, the OH peak is expected to be impacted by hydrogen bonding and H exchange, depending on the solvent environment.

For the specific conditions in Figure \ref{fig:averaging}(a), a better interpretation of the experimental spectrum can be arrived at by testing specific hypotheses for the ethanol coordination and environment in direct computations. First, each ethanol molecule undergoes molecular motion; second, even in dilute conditions, aggregation of two or more ethanol molecules to form hydrogen bonds could occur. Figure \ref{fig:averaging}(d) tests these hypotheses by averaging over shifts for each H atom individually, obtained from an \textit{ab initio} molecular dynamics (AIMD) trajectory of a hydrogen-bonded ethanol dimer (FHI-aims tight settings, 10~ps, 1~fs AIMD step, stochastic velocity rescaling thermostat \cite{Bussi.2007} at $T$=300~K with 0.5~ps collision time settings) -- note that the signal from the hydrogen-bonded OH group (blue, $\sim$4~ppm) is drastically shifted compared to the non-hydrogen bonded OH (blue, $\sim$1~ppm). Further averaging over chemical $^1$H associated with CH$_3$, CH$_2$ and OH groups in both molecules, respectively, leads to the spectrum shown in Figure \ref{fig:averaging}(e), with peaks in good agreement with Figure \ref{fig:averaging}(a). Finally, the average $J$-coupling between the protons in the CH$_2$ group and those in the CH$_3$ group is computed to be 6.49~Hz for the DFT-PBE functional and NAO-J-5 basis set. Converted to ppm for the experimental frequency used, this amounts to 0.0725~ppm. This $J$-coupling can be applied to the spectrum using multiplicity rules for the interaction between CH$_2$ and CH$_3$, leading to fourfold splitting of the former and threefold splitting of the latter peak in Figure \ref{fig:averaging}(f). This exercise illustrates how mechanistic hypotheses for the behavior of a substance can be tested by comparing to its experimental spectrum, though clearly, experience and significant familiarity with NMR beyond simple static DFT calculations is required. \rev{Furthermore, more general solvation effects are not yet included in the NMR calculations (e.g., by implicit solvation or QM/MM) and represent an important future direction for this work.}

Finally, we briefly mention that a more automated, less heuristic derivation of NMR spectra from first principles is possible using spin dynamics simulation packages such as SPINACH \cite{Hogben2011} and the standardized SpinXML format \cite{Biternas2014} to communicate FHI-aims' computed shieldings and $J$-couplings between the packages. Importantly, SPINACH supports the computation of non-standard NMR spectra such as low-field spectra which do not require expensive superconducting high-field magnets for their collection. In such spectra, shieldings (the separation of which is scaled by the applied magnetic field) collapse into single peaks and $J$-couplings (which reflect absolute interaction strengths, i.e., not relative to an external field) shape the spectra. Some of us have recently published a catalogue of low-field spectra (zero-field and spin-locking induced spectra) for a set of over 200 small molecules \cite{Mandzhieva2025}, calculated using FHI-aims and SPINACH and showing that sensitivity to specific molecules is retained in these spectra, offering potential future avenues for low-cost chemical analyses by magnetic resonance.


\subsection*{Future Plans and Challenges}

Molecular magnetic resonance parameters in FHI-aims are reliable and well benchmarked for light-element molecules using DFT at the GGA level of theory, enabling detailed interpretation of NMR spectra, with a direct connection to spin dynamics calculations that allow for quantitative simulations, e.g. of low-field spectra. 

Beyond the current state of the implementation, significant future improvements are possible. A simple but effective modification would be an approach that contains the need for denser integration grids for $J$-coupling to the near-nuclear region only, a task that is not simple due to the need for consistently spaced grid shells around each atom for other numerical operations in FHI-aims. Obviously, further benchmarks for potentially improved density functionals, such as hybrid DFT, are highly desirable. Indeed, support for magnetic resonance observables from hybrid DFT is partially implemented (for $J$-couplings) but not yet fully tested \cite{laasner+es2024}.

Beyond these technical points, there are multiple desirable extensions of the physical reach of magnetic resonance computations in FHI-aims. Support for periodic systems should be straightforward for $J$-couplings, whereas shieldings and diamagnetic magnetizabilities will require an extension of the GIAO approach to periodic systems, e.g., using a modulated magnetic field \cite{Mauri1996a,Mauri1996b}. For heavier elements, NMR signals will be affected both by relativity (especially spin-orbit coupling \cite{Autschbach2007,Autschbach2008}) and by the actual, finite shape of a nucleus \cite{Autschbach2009}. 

Given high-precision, accurate first-principles data, a host of opportunities exist to advance our understanding and interpretation of NMR-based observables, important in fields as diverse as chemical analysis, geological probes, and medical imaging. Continued development of this functionality in FHI-aims will, over time, enable accurate, tractable ``ground truth'' data across the periodic table.

\subsection*{Acknowledgements}
Acknowledgment is made to the Donors of the American Chemical Society Petroleum Research Fund for partial support of this research.



  


\newcommand{\mat}[1]{\mathbf{\underline{#1}}}
\newcommand{\CODE}[1]{\texttt{#1}\xspace}
\newcommand{\FHIaims}{{FHI-aims}\xspace}
\newcommand{\Vibes}{{FHI-vibes}\xspace}
\newcommand{\Python}{{Python}\xspace}
\newcommand{\controlin}{\CODE{control.in}}
\newcommand{\aimd}{\textit{ai}MD\xspace}
\newcommand{\IFC}{\Phi_{IJ}^{\alpha\beta}}
\renewcommand{\vec}[1]{\mathbf{#1}}
\renewcommand{\t}[1]{\text{#1}}
\newcommand{\CC}[1]{{\textcolor{red}{\bf #1}}}

\newcommand{\ChapVibes}[0]{\ref{ChapVibes}\xspace}
\newcommand{\ChapAIMD}[0]{\ref{ChapMD}\xspace}
\newcommand{\ChapHybrids}[0]{\ref{ChapHybrids}\xspace}
\newcommand{\ChapMLIP}[0]{\ref{ChapMLP_AI}\xspace}
\newcommand{\ChapSergey}[0]{\ref{ChapBSDOS}\xspace}
\newcommand{\ChapFundamentals}[0]{\ref{ChapBasicConcepts}\xspace}
\newcommand{\ChapUnfolding}[0]{\ref{ChapUnfolding}\xspace}
\newcommand{\ChapELPA}[0]{\ref{SecELPA}\xspace}
\newcommand{\ChapElecTrans}[0]{\ref{ChapElecTrans}\xspace}
\newcommand{\ChapHeat}[0]{\ref{ChapHeatTransport}\xspace}
\newcommand{\ChapPIMD}[0]{\ref{ChapQuantumNuclei}\xspace}
\newcommand{\ChapMLDENS}[0]{\ref{ChapML-ES}\xspace}
\newcommand{\ChapDFPT}[0]{\ref{ChapDFPT}\xspace}

\newcommand{\refadd}[1]{\textcolor{red}{{#1}}}
\newcommand{\refrm}[1]{\textcolor{red}{\sout{#1}}}




\newpage

\section[Polarization, Born Effective Charges, and Topological Invariants]{Polarization, Born Effective Charges, and Topological Invariants via a Berry-Phase Approach}
\label{ChapPolarization}

\sectionauthor[1,2,a]{\textbf{*Christian Carbogno}}
\sectionauthor[1,b]{Nikita Rybin}
\sectionauthor[1]{Sara Panahian Jand}
\sectionauthor[1,3]{Alaa Akkoush}
\sectionauthor[1,4]{Carlos Mera Acosta}
\sectionauthor[1,c]{Zhenkun Yuan}
\sectionlastauthor[3]{Mariana Rossi}

\sectionaffil[1]{The NOMAD Laboratory at the Fritz Haber Institute of the Max Planck Society, Faradayweg 4-6, D-14195 Berlin, Germany}
\sectionaffil[2]{Theory Department, Fritz Haber Institute of the Max Planck Society, Faradayweg 4-6, D-14195 Berlin, Germany}
\sectionaffil[3]{Max Planck Institute for the Structure and Dynamics of Matter, 22761 Hamburg, Germany}
\sectionaffil[4]{Center for Natural and Human Sciences, Federal University of ABC, Santo André, SP, Brazil}

\sectionaffil[*]{Coordinator of this contribution.}
\rule[0.25ex]{0.35\linewidth}{0.25pt}

\sectionaffil[a]{{\it Current Address:} Theory Department, Fritz Haber Institute of the Max Planck Society, Faradayweg 4-6, D-14195 Berlin, Germany}
\sectionaffil[b]{{\it Current Address:} Skolkovo Institute of Science and Technology, Bolshoi bulvar 30, build.1, 121205, Moscow, Russia}
\sectionaffil[c]{{\it Current Address:} Thayer School of Engineering, Dartmouth College, Hanover, NH 03755, USA}




\subsection*{Summary}

The Berry connection~$\vec{A}_{mn}(\vec{k})$, the Berry curvature~$\vec{\nabla}\times\vec{A}_{mn}(\vec{k})$, and the Berry phase~$\phi=\oint \vec{A}_{mn}(\vec{k})\cdot d\vec{k}$ are key properties describing the reciprocal-space topology, here 
the connection between two electronic states labeled~$m$ and~$n$. They provide a profound link between the phase of a quantum wave function and macroscopic observables as well as material properties. Most prominently, these quantities are central to our understanding of topological materials and provide a route to classify phases in terms of topology~\cite{Bansil2016}. For instance, all the aforementioned quantities 
enter the definition of the topological-invariant~$\mathbb{Z}_2$ as given by Fu and Kane~\cite{Fu2006}:
\begin{equation}
\label{EQ_Z2}
\mathbb{Z}_2=\frac{1}{2 \pi} \sum_{n}^\t{occ}\left\{\,\oint\limits_{\partial B} \vec{A}_{nn}(\vec{k})\cdot d\vec{k} - \int\limits_B \left[\vec{\nabla}\times\vec{A}_{nn}(\vec{k})\right] \; d^2\vec{k} \,\right\} \;\t{ mod } 2
\end{equation}
Here, $B$ is half the Brillouin zone~(BZ) and $\partial B$ its boundary, the sum runs over all occupied states, and the gauge of~$\vec{A}_{nm}(\vec{k})$ 
is constrained to respect time-reversal symmetry. 

Furthermore, these quantities play a fundamental role for computing the polarization in periodic systems~\cite{Resta1992,KingSmith1993}, or, more precisely, for calculating the polarization density~$\vec{P}$ and its derivatives, the Born effective charges~$\mat{Z}_{I}^*$, via 
\begin{equation}
\label{EQ_pol}
\vec{P} = \frac{1}{V} \left[ \sum_I Z_I \mathbf{R}_I - \frac{eV}{(2 \pi)^3} \sum_n^\t{occ}\oint\limits_\t{BZ} \vec{A}_{nn}(\vec{k}) \; d^3\mathbf{k} \right] \;\; \text{mod } \vec{P}_0 \quad\t{and}\quad Z_{\alpha,I\beta}^*=\frac{V}{e}\frac{\partial P_\alpha}{\partial {R}_{I\beta}} \;.
\end{equation}
Here, $V$ denotes the volume of the unit cell, $\vec{P}_0=\frac{e}{V}(|\vec{a}_1|,|\vec{a}_2|,|\vec{a}_3|)$ the polarization quanta along the lattice vectors~$\vec{a}_\alpha$, $Z_I$ the nuclear charge, $\vec{R}_I$ the nuclear positions, and $e$~the elementary charge. The first term describes the trivial contribution of the bare nuclei, whereas the second term covers the contributions stemming from the electronic states~$\Psi_n(\vec{k})$. 
Besides providing the theoretical foundation for understanding the quantization of adiabatic charge transport~\cite{Thouless1983}, the polarization 
is a key property for describing the electrodynamics in solids,~e.g.,~for modeling light-matter interactions and for studying ferroelectric and piezoelectric effects.

In addition, these quantities play a central role in the assessment of currents, fluxes, magnetization, and, last but not least, in the transformation of delocalized electronic wave functions into a localized Wannier basis~\cite{Blount1962,Marzari1997,Marzari2012}. We refer the interested reader to Ref.~\cite{Vanderbilt2018} for a more thorough discussion of all these effects.

\subsection*{Current Status of the Implementation}

To compute all the aforementioned material properties, the fundamental quantity that needs to be calculated in a first-principles code 
is the Berry connection~\cite{Blount1962}:
\begin{eqnarray}\label{eq:berry_connection}
A_{mn}(\vec{k}) = i \Braket{u_m(\vec{k})| \partial u_n(\vec{k})/\partial \vec{k} } \;.
\end{eqnarray}
Here, $u_l(\vec{k})= \exp(-i\vec{k}\vec{r})\psi_{l}(\vec{k},\vec{r})$ is the lattice-periodic part of the electronic wave function~$\psi_{l}(\vec{k},\vec{r})$ for state~$l$ with wave 
vector~$\vec{k}$ and $\Braket{\cdot|\cdot}$ denotes the scalar product in Hilbert space. Before evaluating this definition,
let us recall that \FHIaims\xspace uses a Bloch-like representation of the wave functions
\begin{eqnarray}
\psi_{l}(\vec{k},\vec{r}) = \sum_{i} c_{i,l}(\vec{k}) \chi_{i}(\vec{k},\vec{r}) \quad \text{with} \quad
\chi_{i}(\vec{k},\vec{r}) = \sum_{N} \exp(i \vec{k}\vec{L}_N) \varphi_{i,N}(\vec{r}-\vec{R}_i - \vec{L}_N) \;,
\end{eqnarray}
as detailed in Contrib.~\ChapFundamentals. Here, $c_{i,l}(\vec{k})$ are the Kohn-Sham expansion coefficients and $\varphi_{i,N}(\vec{r}-\vec{R}_i - \vec{L}_N)$ are the numeric atomic orbitals~(NAOs)
associated with the basis function with index $i$ for the periodic image in the cell~$\vec{L}_N$ of the 
atom located at~$\vec{R}_i$. With that, the Berry connection can be expressed as
\begin{eqnarray}
\label{EQ_connection}
A_{mn}(\vec{k}) & = &
   \overbrace{i \sum_{i,j}  c_{j,m}^*(\vec{k}) \frac{\partial c_{i,n}(\vec{k})}{\partial \vec{k}} S_{ij}(\vec{k})}^{A_{mn}^{(1)}(\vec{k})} \\
&& - \underbrace{\sum_{i,j}  c_{j,m}^*(\vec{k})c_{i,n}(\vec{k}) \left[ {D}_{ij}(\vec{k}) - \vec{R}_i  S_{ij}(\vec{k}) \right]}_{A_{mn}^{(2)}(\vec{k})} \;.\nonumber
\end{eqnarray}
The first term in Eq.~(\ref{EQ_connection}),~i.e.,~$\vec{A}_{mn}^{(1)}(\vec{k})$, denotes the gauge-dependent Berry-connection term, which here includes the overlap matrix~$S_{ij}(\vec{k}) = \sum_{N} e^{i \vec{k} \vec{L}_N}  \Braket{\varphi_{j,0}| \varphi_{i,N}}$ due to the non-orthogonality of the employed NAO basis set. For evaluating the associated contribution to the Berry phase,~i.e.,~the closed-path integrals required for Eq.~(\ref{EQ_Z2}) and Eq.~(\ref{EQ_pol}), the reciprocal-space path is discretized on $K+1$ points~$(\vec{k}_0,\vec{k}_1,\cdots,\vec{k}_K)$, whereby the initial and final point are equivalent with respect to the BZ's periodicity~$\vec{k}_0=\vec{k}_K \text{ mod } \frac{2\pi}{V}$. By 
expressing the $\vec{k}$-derivatives as two-point finite-differences, one obtains~\cite{Marzari1997} 
\begin{equation}
\label{EQ_disc}
\sum_n^\t{occ}\oint\limits_\t{BZ} \vec{A}_{nn}^{(1)}(\vec{k}) \cdot d\vec{k} = - \text{Im}\;\ln\left[ \text{det}\left(\mat{M}_{0,1} \cdot \mat{M}_{1,2} \cdots \mat{M}_{K-2,K-1} \cdot \mat{M}_{K-1,0}\right) \right] \;.
\end{equation}
Note that the Kohn-Sham coefficients~$\mat{c}^\t{occ}(\vec{k})$ used for computing the matrices
$$\mat{M}_{a,b} =  \mat{c}^{\t{occ}^\dag}(\vec{k}_a)  \mat{S}(\vec{k}_a) \mat{c}^\t{occ}(\vec{k}_b)$$ 
entering the above expression only cover the subspace of occupied states.

The second term entering Eq.~(\ref{EQ_connection}),~i.e.,~$\vec{A}_{mn}^{(2)}(\vec{k})$, is gauge-invariant and features the matrix
\begin{eqnarray}
D_{ij}(\vec{k}) & = & - \sum_{N} e^{i \vec{k} \vec{L}_N}  \Braket{\varphi_{j,0}|[\vec{r}-\vec{R}_i-\vec{L}_N]| \varphi_{i,N}} \;,
\end{eqnarray}
which captures the contributions of the NAO basis functions. Since this form exhibits the exact same periodicity as the overlap matrix~$\mat{S}(\vec{k})$, it can be integrated using the real-space routines already present in \FHIaims~\cite{Havu2009,knuth2015all}. Similarly, 
the associated contribution to the Berry-phase required for Eq.~(\ref{EQ_Z2}) and Eq.~(\ref{EQ_pol}) can be computed straightforwardly 
by performing the trace over occupied states and by numerically integrating~$A_{mn}^{(2)}(\vec{k})$ along the exact same path used in Eq.~(\ref{EQ_disc}).

\begin{wrapfigure}{r}{0.55\textwidth}
  \centering
    \includegraphics[width=0.53\textwidth]{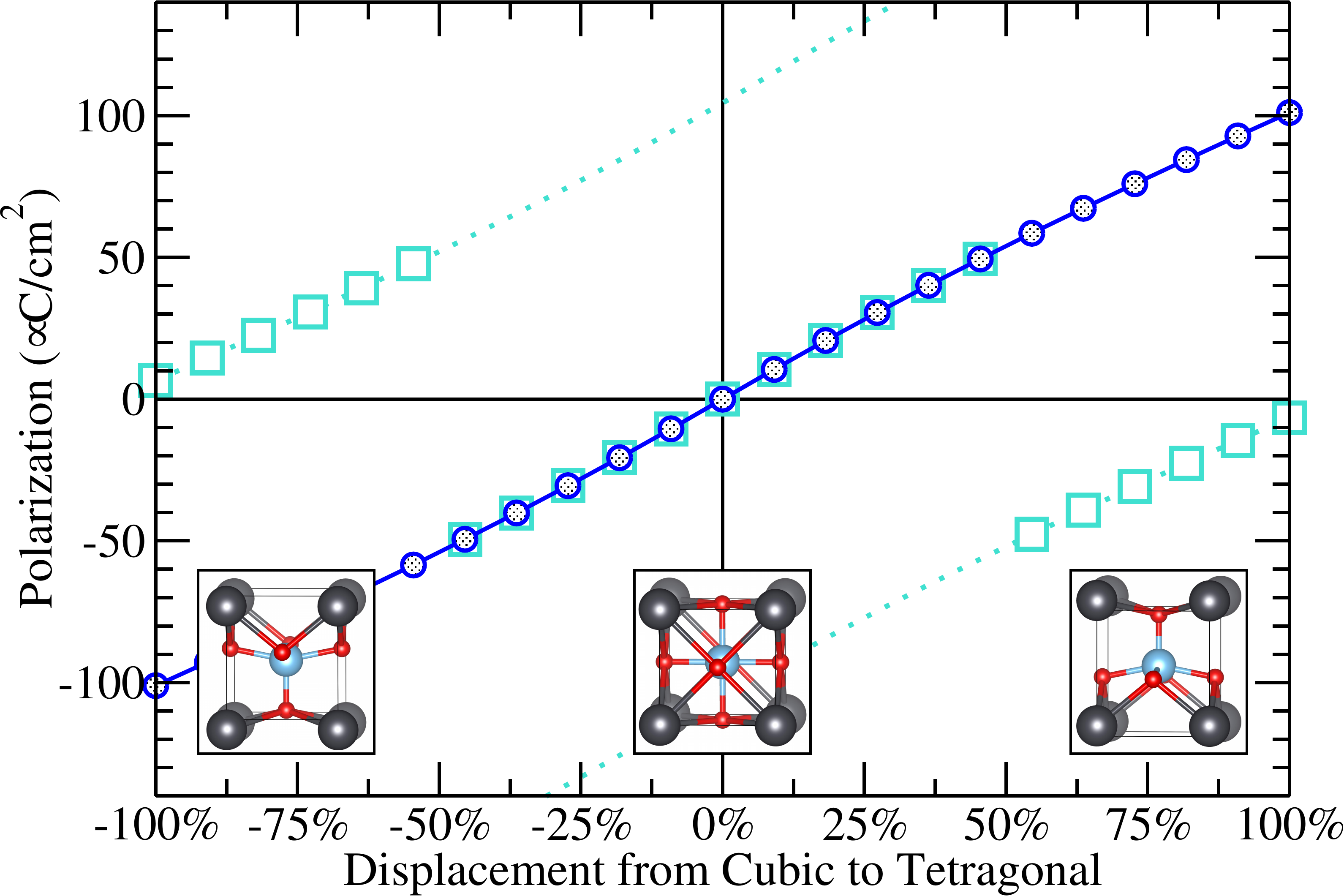}
    \caption{Polarization~($24\times 8 \times 8$ $\vec{k}$-points for the Berry-phase) of PbTiO$_3$~(HSE06, 8$^3$  $\vec{k}$-points) 
    along the minimum-energy path connecting the tetragonal, symmetry-degenerate~$P4mm$ and the cubic, centrosymmetric~$Pm\bar{3}m$ structure.
    Cyan squares denote the bare output, blue circles the ``branch-matched'' polarization, for which the discontinuities associated to the $\text{mod } \vec{P}_0$-operation
    are resolved.}
    \label{FIG_PBTIO3}
\end{wrapfigure}
In the current implementation, the polarization is calculated in the basis of the reciprocal-lattice vectors. To perform the $\vec{k}$-derivative 
along the reciprocal-lattice vector of interest, the closed-loop path is chosen parallel to it. The remaining integrations~$\int d^2\vec{k}$ perpendicular
to this path are performed by splining the Berry phases. As an example for such a calculation, Fig.~\ref{FIG_PBTIO3} shows 
the polarization of PbTiO$_3$, here for smoothly interpolated geometries and lattices between the tetragonal, symmetry-degenerate~$P4mm$ equilibrium structures 
and the cubic, centrosymmetric~$Pm\bar{3}m$ structure. While the latter structure must have a vanishing polarization due to symmetry, the tetragonal configurations 
do not. This highlights that the absolute values of the polarization are meaningless, only relative differences of the polarization matter. To evaluate such differences,
``branch-matching''~\cite{Spaldin2012},~i.e.,~ensuring that the actual polarization values lie on the branch associated with the same multiples of~$\vec{P}_0$ value,
is crucial\footnote{Note that an internal branch matching in \FHIaims\xspace is already performed for splining and integrating over the perpendicular $\vec{k}$-directions.}.

\begin{wrapfigure}{l}{0.55\textwidth}
  \centering
    \includegraphics[width=0.53\textwidth]{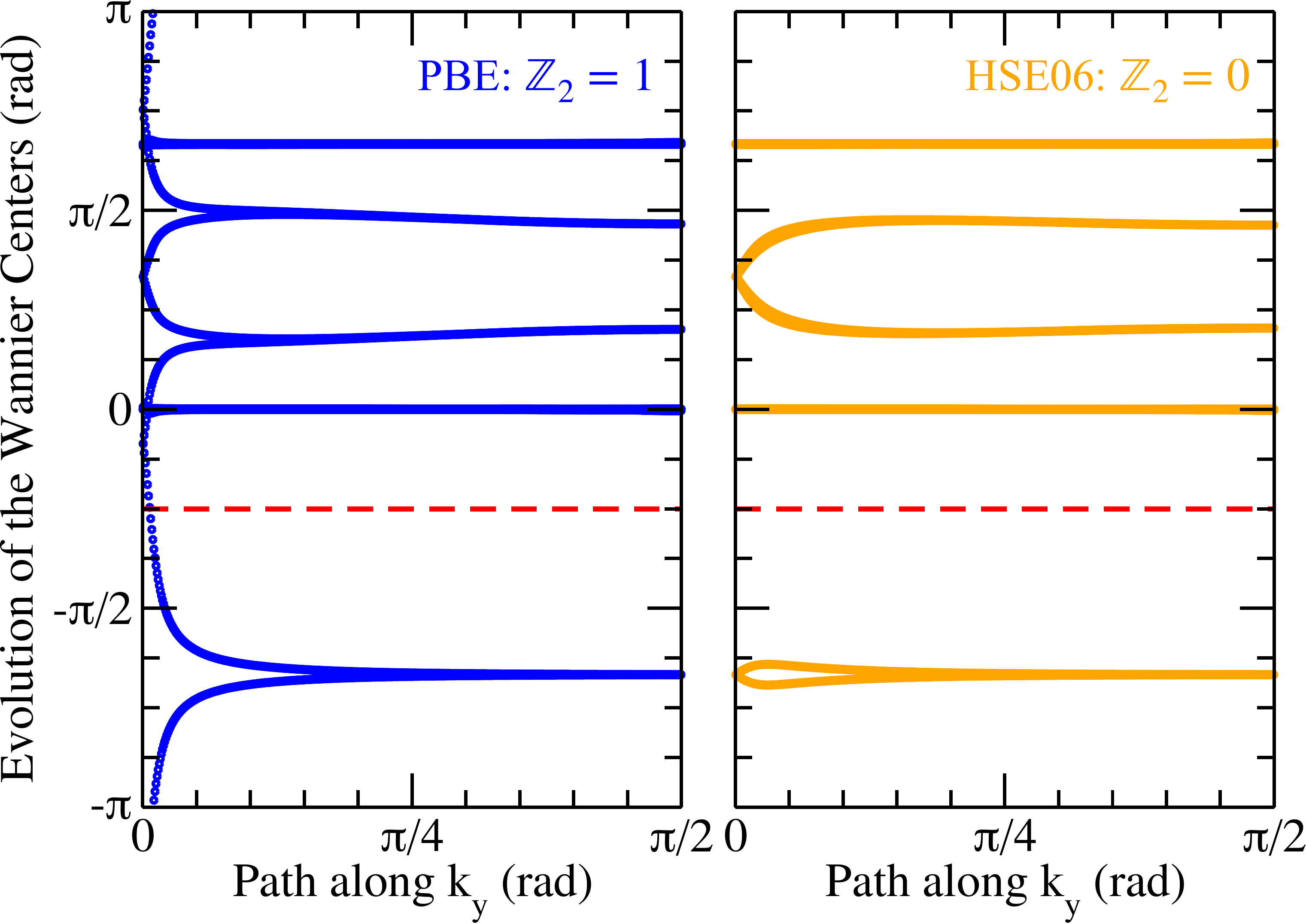}
    \caption{Evolution of the Wannier Centers of Charge for the functionalized 2D-honeycomb structure of GeF$_2$, as computed with PBE~(left) and HSE06~(right) along $\vec{k}_2=k_y$. Both calculations 
    use spin-orbit coupling~\cite{huhn2017}. As indicated by     the red, dashed line, a continuous path along $k_y$ must cross the Wannier Centers an odd~(PBE) viz.~even~(HSE06) 
    number of times, resulting in $\mathbb{Z}_2=1$ and $\mathbb{Z}_2=0$, respectively. This showcases the influence of the exchange-correlation functional on topological invariants~\cite{Si2014,Matthes2016}.}
    \label{FIG_WCC}
\end{wrapfigure}

For the evaluation of the $\mathbb{Z}_2$ invariant, the current implementation follows the formalism proposed in Refs.~\cite{Soluyanov2011,Yu2011}, which is equivalent to the
definition given in Eq.~(\ref{EQ_Z2}), but does not require gauge-fixing. In practice, it requires tracking the evolution of the individual Wannier centers
\begin{equation}
\label{EQ_WCC}
\vec{X}_n(\vec{k}_2) =  \oint\limits_{-\pi}^{\pi} A_{nn}(\vec{k}_1) \cdot d\vec{k}_1 \quad\text{mod}\; 2\pi
\end{equation}
across a path in the BZ described by~$\vec{k}_2$. In practice, one evaluates the line-path integral in Eq.~(\ref{EQ_WCC}) for varying values of~$\vec{k}_2$. Each of these 
integrals is solved for all occupied states as discussed above,~i.e.,~by using $A_{mn}^{(1)}(\vec{k})$ as given by Eq.~(\ref{EQ_disc}) and~$A_{mn}^{(2)}(\vec{k})$ along 
a discretized path~$\vec{k}_1 \perp \vec{k}_2$. The determinant viz. trace is, however, not evaluated. Rather, the obtained matrix is diagonalized 
and the complex phase of the resulting eigenvalues is then tracked, as the example in Fig.~\ref{FIG_WCC} shows. 
If an arbitrary continuous line across the whole $\vec{k}_2$-axis crosses the evolution of the Wannier centers an even number of times,~$\mathbb{Z}_2$ is 0 and otherwise~1.
Similarly, this can be judged by tracking the largest gap between the individual Wannier centers~\cite{Soluyanov2011}.

\subsection*{Usability and Tutorials}

For the evaluation of the polarization, it is sufficient to add one keyword to the \controlin file:
\begin{center}
\texttt{output polarization $\alpha$ $n_1$ $n_2$ $n_3$}
\end{center}
For instance, the line \texttt{output polarization 2 5 10 5} will compute the polarization along the reciprocal-lattice vector number~$2$ using a
grid of~$5\times10\times5$ $\vec{k}$-points in the BZ. As the example highlights, the discretization used along the reciprocal-lattice vector of
interest,~i.e.,~the one along which the closed-path integral in Eq.~(\ref{EQ_disc}) is performed, usually requires denser grids. For convenience,
multiple \texttt{output polarization} statements can be combined and, if all three directions are requested, the code will also report the 
polarization in Cartesian coordinates. For the evaluation of Born effective charges~$\mat{Z}_I$, a Python script~\CODE{BEC.py} is provided 
in the utilities folder of the \FHIaims\xspace distribution to perform the required derivatives, cf.~(\ref{EQ_pol}) via finite differences. Although these functionalities are
rather self-explanatory, a tutorial is provided at \texttt{https://fhi-aims-club.gitlab.io/tutorials/phonons-with-fhi-vibes/}, which also showcases
how the computed Born effective charges can be used to account for long-range dipole interactions in the calculations of phonon spectra in
polar crystals, see~\cite{Baroni2001} and references therein.

Similarly, the evaluation of $\mathbb{Z}_2$ viz.~of the evolution of the Wannier centers of charge only requires to add one keyword to the \controlin file:
\begin{center}
\texttt{output Z2\_invariant $\gamma$ $n_\parallel$ $n_\perp$}
\end{center}
Here, $\gamma$ is an index that encodes which Cartesian planes shall actually be targeted,~e.g.,~$\gamma=1$ implies that Eq.~(\ref{EQ_WCC}) is evaluated along
the first reciprocal-lattice vector using a $\vec{k}$-discretization of $n_\parallel$ points and that this procedure is repeated for $n_\perp$ 
closed-paths that have equidistant~$k_2 \in [0,\pi]$ and~$k_3=0$. The latter scan over $k_2$ can, for instance, be used to discern strong from weak topological
insulators~\cite{Fu2007}.

Eventually, let us note that the implementation supports all exchange-correlation functionals, i.e.,~all semi-local and hybrid functionals, as well as
spin-orbit coupling as described in Ref.~\cite{huhn2017}. Also, given that this functionality targets unit cells and requires rather dense $\vec{k}$-grids,
the parallelization occurs over $\vec{k}$-points using LAPACK for the compute-intense linear-algebra operations.


\subsection*{Future Plans and Challenges}
So far, the existing implementation has proven useful, accurate, and performant for targeting relatively simple materials with few~($<100$) atoms in the unit 
cell.  However, there is increased scientific and technological interest in targeting materials with 
structural or compositional disorder,~e.g.,~for alloyed topological insulators featuring thousands of atoms in the unit cell~\cite{Cao2020}. For such kind 
of systems, the current $\vec{k}$-point-based parallelization strategy  is not efficient \rev{and limited by memory usage, given that each thread stores a local copy of the required matrices in memory}. \removed{Rather,} \rev{To reduce the memory footprint per thread and retain computational efficiency,} support for distributed linear algebra~(ScaLAPACK) is needed and is currently being pursued. \rev{This also requires the implementation of functionality that goes beyond the ScaLAPACK memory and communicator layout currently used in \FHIaims,~e.g.,~for concurrent access to quantities at distinct $\vec{k}$-points, as required for evaluating the matrix elements in the closed-loop integral in Eq.~(\ref{EQ_disc}).}

Furthermore, the described Berry-phase approach does not only give access to polarization, Born effective charges, and topological invariants, but to a multitude 
of other material properties, as described in the introduction. In this context, a systematic interface between the methodologies described in this contributions
and the density-functional perturbation theory implementation described in Contrib.~\ChapDFPT is desirable for the accelerate assessment of response 
properties,~e.g.,~piezoelectric tensors, but also Born effective charges, or other properties pivotal for electrodynamics viz.~light-matter interactions. 
Another route that is being exploited is \removed{the} machine-learning \removed{of} the polarization \removed{for} using systems with a reduced number of atoms 
\rev{to then gain access to } \removed{using a local representation. These 
simulations allow us to treat} much larger unit cells\rev{,~e.g.,~for simulating nuclear dynamics with light-matter coupling} \removed{While learning the polarization requires care because of dealing with a topological quantity~\cite{Xie2022}, our goal is 
to target models that can be used in the context of nuclear dynamics with light-matter coupling, as reported in Refs.}~\cite{Litman2024,Stocco2025}.
\rev{Learning a topological quantity requires special care and several strategies have been proposed in literature for this purpose,~e.g.,~learning the Wannier center positions~\cite{Zhang.2020f5}, learning Born effective charges~\cite{Shimizu.2023,Bergmann.2025god}, or learning multi-valued polarization surfaces~\cite{Stocco2025}.}

Finally, let us emphasize that the implemented infrastructure also makes a transformation to Wannier functions straightforward. While
this functionality would not be particularly useful with \FHIaims\xspace itself, given that the NAOs already provide a localized representation, it would be beneficial
for interfacing to other community codes based on a Wannier representation. For instance, this would give access to all the functionality provided 
by,~e.g.,~Wannier90~\cite{Pizzi2020}, EPW~\cite{Ponce2016}, and Perturbo~\cite{Zhou2021}, and, in turn, enable more systematic, community-wide
benchmarks and collaborations across ``code-boundaries''.

\subsection*{Acknowledgements}
The authors gratefully acknowledge the support of Matthias Scheffler and funding provided by his TEC1p Advanced Grant under the European Research Council (ERC) Horizon 2020 research and innovation programme (Grant Agreement No. 740233).

\newpage

\chapter{Vibrations and Dynamics}
\label{ChapVibrationsDynamics}
\newpage

\section[Phonons, Anharmonicity Quantification, and Strongly Anharmonic Vibrations]{Phonons, Anharmonicity Quantification, and Strongly Anharmonic Vibrations in Solids}
\label{ChapVibes}

\sectionauthor[1,2,a]{\textbf{*Thomas A. R. Purcell}}
\sectionauthor[1,3]{Florian Knoop}
\sectionauthor[1]{Shuo Zhao}
\sectionauthor[1,4,5]{Marcel F. Langer}
\sectionauthor[8]{Marcel H\"ulsberg}
\sectionauthor[1,6]{Daniel T. Speckhard}
\sectionauthor[1]{Maja O. Lenz-Himmer}
\sectionauthor[7,8,b]{J\"org Meyer}
\sectionauthor[1]{\textbf{ *Matthias Scheffler}}
\sectionlastauthor[1,9,c]{\textbf{ *Christian Carbogno}}

\sectionaffil[1]{The NOMAD Laboratory at the Fritz Haber Institute of the Max Planck Society, Faradayweg 4-6, D-14195 Berlin, Germany}
\sectionaffil[2]{Department of Chemistry and Biochemistry, The University of Arizona, Tucson, AZ 85721, USA}
\sectionaffil[3]{Department of Physics, Chemistry and Biology (IFM), Link\"oping University, SE-581 83, Link\"oping, Sweden}
\sectionaffil[4]{Machine Learning Group, Technische Universität Berlin, 10587 Berlin, Germany}
\sectionaffil[5]{Laboratory of Computational Science and Modeling, Institut des Matériaux, École Polytechnique Fédérale de Lausanne, 1015 Lausanne, Switzerland}
\sectionaffil[6]{Physics Department and CSMB, Humboldt-Universität zu Berlin, Zum Großen Windkanal 2, 12489 Berlin, Germany}
\sectionaffil[7]{Leiden Institute of Chemistry, Gorlaeus Laboratories, Leiden University, P.O. Box 9502, 2300 RA Leiden, The Netherlands}
\sectionaffil[8]{Theory Department (since 1/1/2020: The NOMAD Laboratory), Fritz Haber Institute of the Max Planck Society, Faradayweg 4-6, D-14195 Berlin, Germany}
\sectionaffil[9]{Theory Department, Fritz Haber Institute of the Max Planck Society, Faradayweg 4-6, D-14195 Berlin, Germany}

\sectionaffil[*]{Coordinator of this contribution.}
\rule[0.25ex]{0.35\linewidth}{0.25pt}

\sectionaffil[a]{{\it Current Address:} Department of Chemistry and Biochemistry, The University of Arizona, Tucson, AZ 85721, USA}
\sectionaffil[b]{{\it Current Address:} Leiden Institute of Chemistry, Gorlaeus Laboratories, Leiden University, P.O. Box 9502, 2300 RA Leiden, The Netherlands}
\sectionaffil[c]{{\it Current Address:} Theory Department, Fritz Haber Institute of the Max Planck Society, Faradayweg 4-6, D-14195 Berlin, Germany}




\subsection*{Summary}

\begin{figure}
    \centering
    \includegraphics{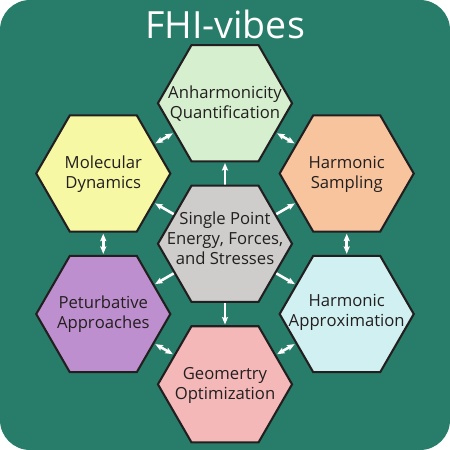}
    \caption{Schematic of how \Vibes can calculate vibrational properties of materials across all orders of anharmonicity.}
    \label{fig:vibes-shematic}
\end{figure}

In condensed-matter physics and materials science, the vibrational modes in a material are the primary drivers of materials properties~\cite{BornHuang}.
At the lowest level of theory, it is common to describe these modes using the {\it harmonic} approximation as a set of uncoupled harmonic oscillators,~i.e.,~phonons~\cite{BornHuang}.
Because all phonon modes are uncoupled, both the classical equations of motion and the quantum Schr\"odinger equation can be solved analytically and thermodynamic quantities such as the harmonic free-energy and heat capacity can be readily computed. 
However, this approximation comes with severe deficiencies and inaccuracies, especially for describing processes at elevated temperatures or featuring corrugated potential-energy surfaces. 
For instance, lattice expansion and vibrational thermal transport cannot be assessed solely within the harmonic approximation, but necessitate the inclusion of additional scattering terms, most prominently those stemming from
 anharmonic effects~\cite{Leibfried1961, Klemens1958, Dove, Fultz2010}. 
Several techniques~\cite{Monacelli2021,Eriksson2019,Hellman2011,Roekeghem2021,TTadano2014} exist that account for such effects at various degrees of approximations. 
Accounting for all orders of anharmonic effects in an unbiased manner, however, typically requires \textit{ab initio} molecular dynamics~(\textit{ai}MD\xspace) simulations.

FHI-vibes is a Python package that aims to facilitate the investigation of harmonic and anharmonic effects with the electronic-structure theory code FHI-aims\xspace. 
It seamlessly bridges between harmonic phonons, approximate treatments of anharmonicity, and a fully anharmonic description of the vibrational dynamics with \textit{ai}MD\xspace~\cite{Knoop2020a}, as illustrated in Figure~\ref{fig:vibes-shematic}. 
To achieve this, it builds on the Atomistic Simulation Environment \xspace (ASE)~\cite{larsen2017ase} to internally represent materials and to interface to first- and second-principles calculators for obtaining energy and forces. 
From here it links to external codes such as spglib~\cite{Togo2024}, phonopy~\cite{Togo2023}, phono3py~\cite{Togo2015}, TDEP~\cite{Knoop2024}, and hiphive~\cite{Eriksson2019} to generate harmonic or anharmonic force-constants. 
Both a command line and a Python API is provided to perform these calculations as well as useful tools for postprocessing the results. 
In particular, this includes an interface to calculate a material's anharmonicity called~$\sigma^\mathrm{A}$~\cite{Knoop2020}, which highlights the advantages of being able to seamlessly process both harmonic and anharmonic dynamics on the same footing. Finally, let us note that these workflows can be run in a high-throughput fashion via FireWorks~\cite{fireworks2015}.

\subsection*{Current Status of the Implementation}
The central quantity of interest is the potential-energy surface~(PES)~$U\left(\{\mathbf{R}\}\right)$, on which the dynamics of the $N$ nuclei with coordinates~$\{\mathbf{R}\}=\{\mathbf{R}_1,\cdots,\mathbf{R}_N\}$ evolve.
In the harmonic approximation, this PES is approximated via the truncated Taylor expansion
\begin{equation}
U\left(\{\mathbf{R}\}\right) \approx U^\text{ha}\left(\{\mathbf{R}\}\right) = U\left(\{\mathbf{R}^0\}\right) + \frac{1}{2}\sum_{I,J,\alpha,\beta}\underbrace{\left.\frac{\partial^2 U\left(\{\mathbf{R}\}\right) }{\partial R_{I\alpha} \partial R_{J\beta}}\right\vert_{\{\mathbf{R}\}=\{\mathbf{R}^0\}}}_{\Phi_{IJ}^{\alpha\beta}} (R_{I\alpha} - R_{I\alpha}^0)( R_{J\beta} - R_{J\beta}^0)\,
\label{EQharmonic}
\end{equation}
around a (local) minimum-energy configuration~$\{\mathbf{R}^0\}$. This already highlights that the dynamics is restricted to oscillations around this specific configuration in the harmonic approximation. The force constants $\Phi_{IJ}^{\alpha\beta}$ characterize the Hessian matrix at this configuration, whereby Greek indices denote Cartesian axes~$x,y,z$. In solid-state physics, it is common to not directly inspect $\Phi_{IJ}^{\alpha\beta}$, but its mass($M$)-weighted Fourier-transform with respect to lattice vectors,~i.e.,~the dynamical matrix~$\mathbf{\underline{D}}(\mathbf{q})$, which is a function of the reciprocal-space vector~$\mathbf{q}$~\cite{BornHuang}. Its eigenvalues~$\omega_s^2(\mathbf{q})$ and eigenvectors characterize the harmonic phonon modes of a material.

The term \textit{anharmonicity} subsumes all effects of the atomic dynamics not captured by the harmonic approximation,~i.e.,~the differences between~$U\left(\{\mathbf{R}\}\right)$ and $U^\text{ha}\left(\{\mathbf{R}\}\right)$. 
When the phonon picture is valid,~i.e.,~in the case of weak anharmonicity as discussed in Contribution~\ref{ChapHeatTransport}, 
these effects can be captured by extending the Taylor expansion in Eq.~(\ref{EQharmonic}) to higher orders,~e.g.,~third- or fourth order~\cite{Eriksson2019,Togo2015}, and/or by effectively incorporating anharmonicity in~$\Phi_{IJ}^{\alpha\beta}$~\cite{KEsfarjani2008}. For instance, this can be achieved by fitting $\Phi_{IJ}^{\alpha\beta}$ to \textit{ai}MD\xspace trajectories, as done in hiphive~\cite{Eriksson2019}. For strongly anharmonic materials, the phonon picture breaks down and \textit{ai}MD\xspace is needed for an accurate description of the dynamics, also see Contribution~\ref{ChapHeatTransport}. Comparing fully anharmonic \textit{ai}MD\xspace with the harmonic approximation is for instance useful to gauge the degree of anharmonicity present in a material at specific thermodynamic conditions.

In the following, a standard workflow for FHI-vibes is described. It is broken down into ordered subsections to highlight the individual components.

\subsubsection{Geometry Optimization}
The goal of the geometry optimization step is to find the minimum-energy structure~$\{\mathbf{R}^0\}$ around which the Taylor expansion in Eq.~(\ref{EQharmonic}) is performed.
Accordingly, $\{\mathbf{R}^0\}$ is the reference geometry around which the vibrational modes are forced to oscillate in the harmonic approximation. FHI-vibes supports geometry optimizations via the algorithms natively implemented
in FHI-aims and via those implemented in ASE. This dual approach allows for a larger flexibility, especially when it comes to enforcing constraints on the relaxation~\cite{Lenz2019}. This is particularly useful when addressing different crystal polymorphs for a given composition.

\subsubsection{Phonons}
Once a minimum-energy structure~$\{\mathbf{R}^0\}$ is identified, the harmonic force constants~$\Phi_{IJ}^{\alpha\beta}$ can be computed using the finite-difference method~\cite{Parlinski1997} as implemented in~phonopy.
Phonopy first generates a list of symmetry-reduced displacements from the reference geometry and then FHI-vibes calculates the forces acting on each atom using FHI-aims.
All relevant data is then stored in a trajectory file and passed back to phonopy to compute $\Phi_{IJ}^{\alpha\beta}$ and all derived properties,~i.e.,~the dynamical matrix, the phonon frequencies
and the associated displacement eigenvectors. Furthermore, FHI-vibes also provides an interface to phonopy's postprocessing tools to calculate and plot properties such as 
phonon bandstructures, densities-of-states~(DOS), and thermodynamic properties such as the vibrational free energy and heat capacity. In this phonon picture, it is assumed
that the harmonic approximation is a valid approximation for the PES.

We note in passing that performing such phonon calculations at different cell volumes enables an approximative treatment of anharmonicity within 
the {\it quasi-harmonic}~approximation~\cite{Biernacki1989,Rasti2019,Rasti2022}. For instance, this allows to model lattice expansion,
as explained in the FHI-vibes tutorials. Further capabilities of FHI-vibes for 
the treatment of anharmonicity at different levels of approximation are covered in detail in Chapter~\ref{ChapHeatTransport} on heat transport.

\subsubsection{\textit{Ab Initio} Molecular Dynamics}
For performing \textit{ai}MD\xspace calculations, FHI-vibes supports the integrators, thermo- and barostats natively implemented in FHI-aims and those in ASE.
As in the case of geometry optimizations, this allows for more flexibility and a wider variety of thermodynamic ensembles to be sampled. 
For the fast initialization and thermalization of \textit{ai}MD\xspace trajectories, information from the harmonic approximation can be used to prepare
a system close to the right thermodynamic equilibrium via harmonic sampling~\cite{West2006}. 
The actual \textit{ai}MD\xspace trajectories including position and velocities are then stored as concatenated JSON strings to include metadata for each calculation. 
This format also facilitates post-processing of the computed \textit{ai}MD\xspace data and several routines are provided to the user for this purpose. 
For instance, (harmonic) force constants~$\Phi_{IJ}^{\alpha\beta}$ can be incorporated in the trajectory file and then subsequently be used to map the trajectory onto harmonic phonon modes.

\subsubsection{Anharmonicity Quantification}
\begin{figure}
    \centering
    \includegraphics[width=0.5\linewidth]{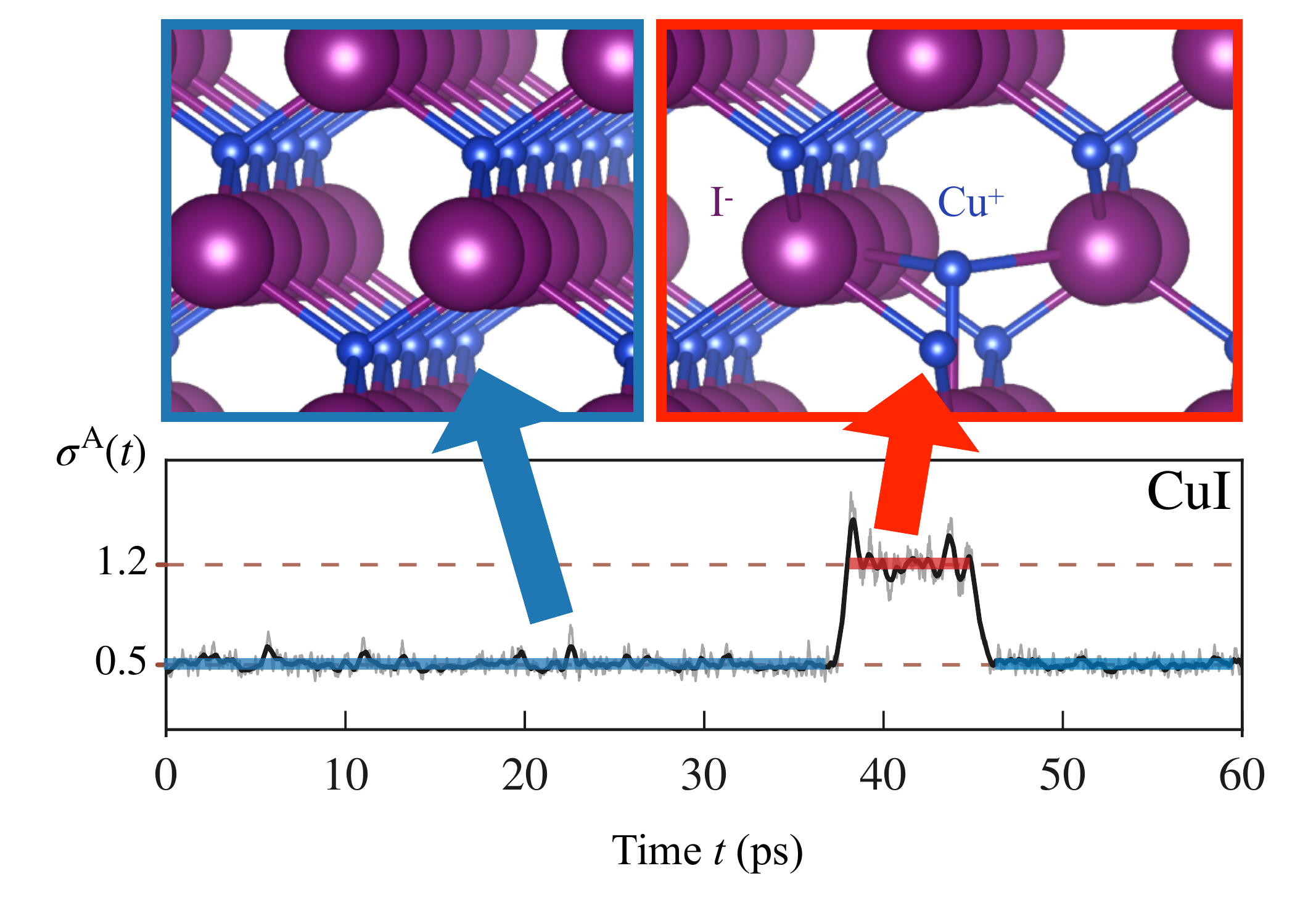}
    \caption{The anharmonicity score per time step for an \textit{ai}MD\xspace trajectory of CuI. The outset structures represent the average positions of the atoms in the CuI supercell over the blue (left) and red (right) highlighted regions of the trajectory. This is reproduced from~\cite{Knoop2023}}
    \label{fig:anharm_time}
\end{figure}

One key advancement enabled by the seamless integration of harmonic models and fully anharmonic \textit{ai}MD\xspace present in FHI-vibes is the development of the anharmonicity measure~$\sigma^\mathrm{A}$~\cite{Knoop2020}.
This measure defines the anharmonicity of a material as
\begin{equation}
\label{eq:anhar}
    \sigma^\mathrm{A}(T) = \sqrt{\frac{\sum_{i,\alpha}\langle \left(F^\mathrm{(2)}_{I,\alpha} - F_{I,\alpha} \right)^2 \rangle}{\sum_{i,\alpha}\langle\left(F_{I,\alpha}\right)^2 \rangle}},
\end{equation}
where $\langle . \rangle$ denotes a thermodynamic ensemble average at temperature~$T$, $F^\mathrm{(2)}_{I,\alpha}$ is the force predicted by the harmonic model and $F_{I,\alpha}$ is the 
fully anharmonic force on atom $I$ and component $\alpha$. 
\rev{In practice, this metric can be used to quantify the relevance of anharmonic effects under different thermodynamic conditions, e.g., for phase-transition detection \cite{Knoop2020}, for estimating thermal conductivities \cite{Knoop2023,purcell2023accelerating}, for analyzing  trends in electronic mobilities \cite{Quan2024}, and even for improving the accuracy and validity of machine-learning potentials for vibrations \cite{kang2025accelerating}.}
Essentially, it normalizes the standard error of the harmonic model to the standard deviation of the DFT forces, and can act as a measure of what fraction of the effects are anharmonic. 
A per-sample definition of $\sigma^\mathrm{A}\left(T, t\right)$ can be defined by replacing the thermodynamic ensemble average with the average force on all atoms for each sample.
This metric has been used to successfully describe the dynamics of thermal transport~\cite{Zeng2024} and to capture, in an adapted formulation, elastic and inelastic thermal transport across a barrier~\cite{Xu2024}.
This measure can also be used to detect when rare events occur in \textit{ai}MD\xspace trajectories, as illustrated in Figure~\ref{fig:anharm_time} for defect formation in CuI~\cite{Knoop2023}.
As mentioned above, FHI-vibes allows to decompose $\sigma^\mathrm{A}$ into atomic contributions and to map it onto the individual phonon modes.
This decomposition allows for determining the role of strongly anharmonic effects for heat transport~(see Contribution~\ref{ChapHeatTransport} and Ref.~\cite{Knoop2023}),
clarifying the origin of anharmonicity in $\beta$-zeolites~\cite{Owain}, and to design machine-learning based descriptors for heat transport~\cite{purcell2023accelerating}. 

\subsection*{Usability and Tutorials}
In this section we will provide a brief tutorial on how to use FHI-vibes, for a complete set of instructions see our documentation here: \href{https://vibes-developers.gitlab.io/vibes/}{https://vibes-developers.gitlab.io/vibes/} 
and the FHI-aims-specific tutorials provided here: \href{https://fhi-aims-club.gitlab.io/tutorials/phonons-with-fhi-vibes/}{https://fhi-aims-club.gitlab.io/tutorials/phonons-with-fhi-vibes/}.

\subsubsection{Setting up a Single Point Calculation}
One of the most important steps in using FHI-vibes is specifying the parameters used for a single-point calculation that is done using a single geometry without optimizing the structure.
No matter what task is being completed FHI-vibes needs to know how to calculate the energy, forces, and stresses for a given material.
Because the interface is built on top of the ASE interface, all supported ASE calculators can be used with FHI-vibes.
An example of what must be defined in this section is shown below.
\begin{verbatim}
[files]
geometries:     geometry.in.???

[calculator]
name:           aims
socketio:       true

[calculator.parameters]
xc:             pw-lda
compute_forces: true

[calculator.kpoints]
density:        3

[calculator.basissets]
Si:             light
\end{verbatim}
All input structures that we want to perform the calculation on are listed in the \texttt{files} section, under either the \texttt{geometry} keyword for a single file or the \texttt{geometries} one for multiple files matching a particular pattern. In the example the ??? signifies the pattern to be matched, any three character string.
The \texttt{calculator} section and corresponding subsections are used to define which ASE calculator object to create, the \texttt{name} keyword mapped to a calculator class, \texttt{socketio} keyword or subsection outlining which socket to use for the SocketIO calculators, the \texttt{parameters} subsection listing all parameters passed to the ASE calculator, and \texttt{basissets} listing all basis set information for the materials.
Because FHI-vibes is designed to work on one or many input structures, the \texttt{k\_grid} is information can be passed both as a parameter, or via a separate section that specifies the desired density of k-points in each crystal direction for consistent calculations.
If no other section is defined in the input file, FHI-vibes will assume that a single-point calculation is requested.

\subsubsection{Running Individual Tasks}
Once the calculation settings are created each individual task can be then be run by creating a new section titled with the task.
The current supported tasks are
\begin{itemize}
    \item relaxation: ASE relaxation
    \item phonopy: 2nd order force constant generator/phonopy calculation
    \item phono3py: 3rd order force constant generator/phono3py calculation
    \item md: Molecular dynamics
\end{itemize}
Additional types of calculations such as harmonic sampling and anharmonicity quantification can be done through utilities and setting up a series of single point calculations.
Each task should have its own input file with tasks chained together through a command line interface.
\rev{For a comparison of the predicted and experimentally measured vibrational properties using these methods we point the reader to Refs.~\cite{Togo2023,Togo2015} and the FHI-vibes tutorials.}

\subsubsection{Using FireWorks for High-Throughput Calculations}
Finally FHI-vibes supports high-throughput calculations using fireworks.
We provide a set of instructions on how to setup a FireWorks job server (LaunchPad) to run these calculations, but importantly a single input file can be used to run a multi-step job.
For these jobs each type of calculation requested should be listed in the input file, with FHI-vibes knowing what order to place each run, e.g. a relaxation should occur before calculating the phonons of a material.
It also adds a set of new sections, e.g. \texttt{statistical\_sampling}, to replace steps that would normally be done by hand.
Additionally a \texttt{use\_aims\_relax} keyword was added to the relaxation step to ensure that if a local FHI-aims relaxation is requested it would not affect the phonon calculations.


\subsection*{Future Plans and Challenges}
From a physical point of view, all fundamental properties to compute key harmonic and anharmonic properties of a solid are implemented in FHI-vibes already, making it a useful and important tool within the FHI-aims software-verse. However, this does not mean that its development has reached an end. Beside maintenance to adapt to updates in the API's of the employed libraries, the main focus is on improving the usability for both common and less common tasks. For example, the inclusion of long-range dipole-dipole interactions~\cite{Gonze1997} is currently being supported and explained in the tutorials, but requires manual calls to phonopy. In this regard, incorporating native support via FHI-vibes is planned for the \rev{near future}. Along these lines, also the computation of the properties needed for computing this long-range effects, namely the Born-effective charges~(see Contribution~\ref{ChapPolarization}) and the dielectric tensor~(see Contribution~\ref{ChapDFPT}) will be incorporated to allow for seamless workflows within FHI-vibes. Similarly, further improvements will be application-driven and revolve around easier access to FHI-vibes' internal routines. For example, our experience in the evaluation of $\sigma^\mathrm{A}$ and of
heat transport has shown that the mapping of the anharmonic motion onto harmonic phonon modes can be pivotal to obtain qualitative
understanding of the dynamics. Accordingly, exposing the relevant routines so that this kind of analysis can be easily used also for other kind of applications is an important task for our users.
\rev{All of these improvements will focus on expanding the types of problems accessible using FHI-vibes}.

Although FHI-vibes was originally developed as an interface for FHI-aims, its MIT licence and the fact that it builds on top of ASE, allows it to be used with all first- and second-principle codes supported by ASE calculators, including those for machine-learned interatomic potentials~(MLIPs, see Contribution~\ref{ChapMLP_AI})\rev{, which would extend the systems accessible by FHI-vibes to those with up to 10,000 atoms.}
This was demonstrated for heat transport calculations in Ref.~\cite{Langer2023}. 
Such MLIP MD simulations are less computationally limited in time- and length-scales compared to \textit{ai}MD\xspace, but still benefit from the features of FHI-vibes when it comes to analyzing and understanding the simulations. 
Naturally, this calls for incorporating training-data creation, training, and testing of MLIPs into FHI-vibes, to allow for a seamless user experience.
In this regard, those results suggests that these plans might not only be beneficial for the usability, but also important for improving the reliability and accuracy of MLIPs.

\subsection*{Acknowledgements}
MS acknowledges support by his TEC1p Advanced Grant (the European Research Council (ERC) Horizon 2020 research and innovation programme, grant agreement No. 740233.

\newpage

\section{Performing Advanced Ab Initio Molecular Dynamics}
\label{ChapMD}

\sectionauthor[1,a]{Yuanyuan Zhou}
\sectionauthor[1]{Herzain I. Rivera-Arrieta}
\sectionauthor[2,3]{Luca M. Ghiringhelli}
\sectionauthor[1,4,b]{Christian Carbogno}
\sectionauthor[1]{Matthias Scheffler}
\sectionlastauthor[5]{\textbf{*Mariana Rossi}}

\sectionaffil[1]{The NOMAD Laboratory at the Fritz Haber Institute of the Max Planck Society, Faradayweg 4-6, D-14195 Berlin, Germany}
\sectionaffil[2] {Department of Materials Science and Engineering, Friedrich-Alexander Universität, Erlangen-Nürnberg, Germany}
\sectionaffil[3] {Erlangen National High Performance Computing Center, Martensstraße 1, Erlangen, 91058, Bayern, Germany}
\sectionaffil[4]{Theory Department, Fritz Haber Institute of the Max Planck Society, Faradayweg 4-6, D-14195 Berlin, Germany}
\sectionaffil[5]{Max Planck Institute for the Structure and Dynamics of Matter, 22761 Hamburg, Germany}

\sectionaffil[*]{Coordinator of this contribution.}
\rule[0.25ex]{0.35\linewidth}{0.25pt}

\sectionaffil[a]{{\it Current Address:} State Key Laboratory of Quantum Functional Materials, Department of Physics, and Guangdong Basic Research Center of Excellence for Quantum Science, Southern University of Science and Technology, Shenzhen 518055, China}
\sectionaffil[b]{{\it Current Address:} Theory Department, Fritz Haber Institute of the Max Planck Society, Faradayweg 4-6, D-14195 Berlin, Germany}



\singlespacing


\subsection*{Summary}
\begin{wrapfigure}{r}{0.45\textwidth}
  \centering
    \includegraphics[width=0.43\textwidth]{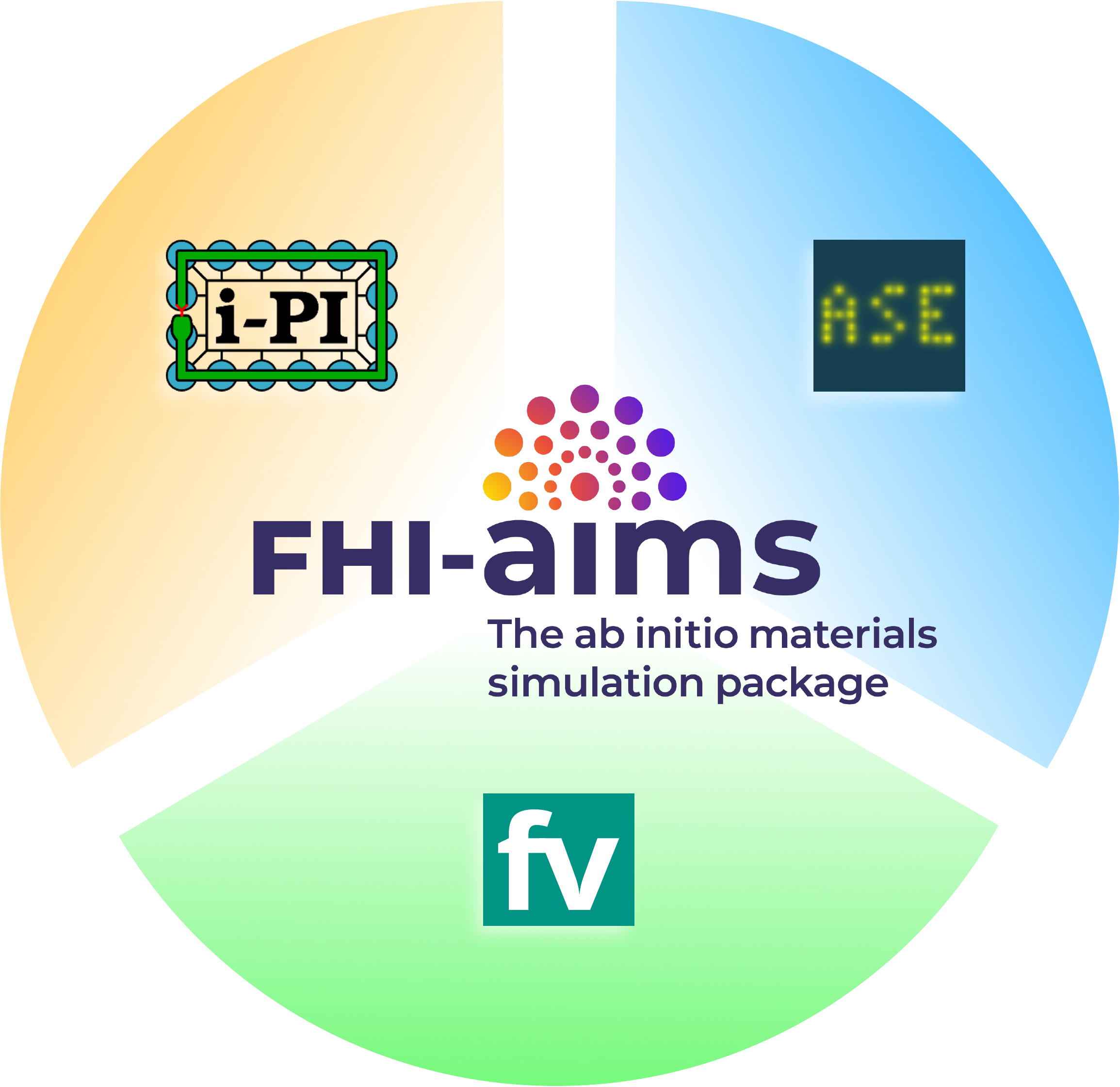}
    \caption{The software packages i-PI~\cite{Litman2024}, ASE~(Atomistic Simulation Environment)~\cite{larsen2017ase}, and \Vibes~\cite{Knoop2020a} provide additionaly \aimd-related
    functionalities beyond what natively supported in \FHIaims.}
    \label{FIG_1_SKETCH}
\end{wrapfigure}
Since its inception~\cite{Car.1985}, {\it ab initio} molecular dynamics~(\aimd) has established itself as one of the most successful numerical techniques in chemistry, solid-state
physics, and materials science. It offers an appealing connection between two prominent fields, namely electronic-structure theory and statistical mechanics. It allows to compute, analyze,
and understand thermodynamic processes and properties without relying on potentially erroneous assumptions for the interatomic interactions. In addition, the electronic-structure that is computed along the whole trajectory enables to assess thermodynamic expectation values of observables that cannot be computed from the nuclear motion alone.

Since its first release~\cite{Blum2009}, \FHIaims\xspace has supported a native implementation of \aimd, as evidenced by early applications that were based on this functionality~\cite{Rossi2010-md, Beret2011-md}. Over time, \FHIaims\xspace has also been interfaced with codes that are able to perform advanced flavors of \aimd. Prominent interfaces that we can mention are i-PI~\cite{Litman2024}, ASE~(Atomistic Simulation Environment)~\cite{larsen2017ase}, and \Vibes~\cite{Knoop2020a}. Several successful applications have built on this feature, as discussed for instance the Roadmap contributions~\ref{ChapVibes}, ~\ref{ChapUnfolding},
~\ref{ChapElecTrans}, ~\ref{ChapHeatTransport}, ~\ref{ChapQuantumNuclei}, and ~\ref{ChapMLP_AI}, as well as a vast array of publications that are too numerous to be singled out. 

In this contribution, we summarize the status of the implementation of \aimd in \FHIaims, the usability of interfaces, and future plans. As a highlight, we discuss the use of \aimd in the context of grand-canonical replica exchange molecular dynamics (REGC) simulations, a method that was first applied with a bespoke interface to \FHIaims~\cite{zhou2019determining}.

\subsection*{Current Status of the Implementation}
The Born-Oppenheimer approximation~\cite{Born.1927} (BOA),~i.e.,~the separation of the dynamics of electronic~$\{\vec{r}\} = \{\vec{r}_1, \vec{r}_2,\cdots\}$ and nuclear degrees of freedom $\{\vec{R}\} = \{\vec{R}_1, \vec{R}_2,\cdots\}$,
is at the very heart of \aimd. The assumption of classical nuclear motion and instantaneous interaction with the ground-state electronic state yields the following equations,
\begin{equation}
\hat{H}_{\{\vec{R}\}}(\{\vec{r}\}) \Psi_{\{\vec{R}\}}(\{\vec{r}\}) = E_{\{\vec{R}\}} \Psi_{\{\vec{R}\}}(\{\vec{r}\}) \quad\text{and}\quad M_I \vec{\ddot{R}}(t) = \vec{\nabla}_I E_{\{\vec{R}\}} \label{eq:eom}
\end{equation}
\begin{wrapfigure}{r}{0.45\textwidth}
  \centering
    \includegraphics[width=0.43\textwidth]{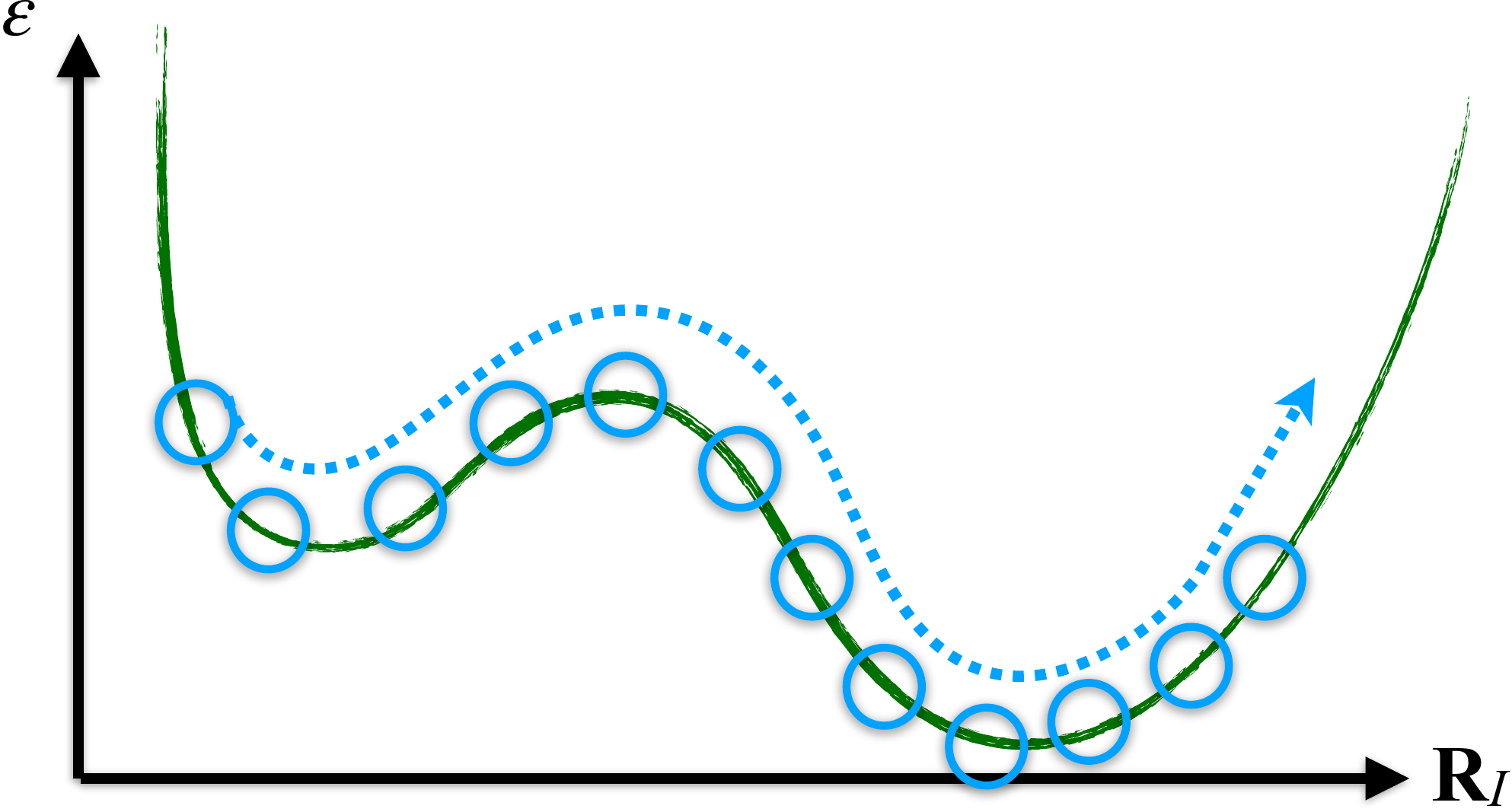}
    \caption{Sketch of an \aimd calculation on a potential-energy surface~$\varepsilon$~(green). By step-wise integrating the equations of motion, one can simulate the dynamics. This requires to solve
    the electronic-structure theory problem in each time step, as indicated by the blue circles.}
    \label{FIG_2_AIMD}
\end{wrapfigure}

Here, the first equation is the electronic-structure theory problem, that depends {\it parametrically} on the instantaneous  nuclear coordinates. Its ground-state solution $E_{\{\vec{R}\}}$
describes the potential-energy surface, on which the nuclei move. Accordingly, it defines the forces acting on the nuclei with masses $M_I$ in the second equation, which is a classical Newtonian equation of motion. 
Related molecular dynamics methods, where the BOA is relaxed, as for example in Ehrenfest dynamics, or where the assumption of classical nuclei is relaxed, or even where both assumptions are relaxed are discussed in contributions~\ref{ChapQuantumNuclei} and ~\ref{SecRTTDDFT}.

The nuclear equations of motion from Eq.~(\ref{eq:eom}) can be step-wise integrated~\cite{Verlet.1967} to simulate the dynamics and obtain time-discretized trajectories~$\{\vec{R}(t_l)\}$. 
At each time step~$t_l$, it is however necessary to solve the associated electronic-structure problem for the Hamiltonian~$H_{\{\vec{R}(t_l)\}}(\{\vec{r}\})$ to obtain the potential and the forces acting on the nuclei,
as sketched in Fig.~\ref{FIG_2_AIMD}. While the straight solution to these equations of motion can only lead to sampling of the microcanonical ensemble (NVE), it has been long recognized~\cite{Frenkel.2002} that coupling the dynamics to thermostats and barostats can lead to sampling the canonical (NVT), isothermal-isobaric (NPT) and the isothermal-isostress (NsT) ensemble. In these ensembles, it is possible to control temperature, pressure and anisotropic stress.
Last but not least, {\it ab initio} replica-exchange molecular dynamics 
can enable the calculation of the $\mu$VT ensemble with variable particle number, but constant chemical potential, as shown below.

\FHIaims\xspace natively supports \rev{basic MD functionality, which can sometimes be advantageous on HPC installations with, e.g., limited or restricted access to (potentially dated) libraries.} 
For more advanced molecular dynamics simulations,
\FHIaims\xspace is linked to external packages. Under the BOA, the interface comes naturally, as it exploits the separation of electronic and nuclear degrees of freedom that underlies \aimd. The external codes take care of solving 
the nuclear equation of motion, while \FHIaims\xspace tackles the electronic-structure problem. Communication between the codes can happen either via input and output file parsing~(e.g., ASE, FHI-PANDA\footnote{\texttt{https://gitlab.com/zhouyuanyuan/fhi-panda}}) or, more efficiently, 
via a UNIX or internet socket communication~(e.g., ASE, i-PI, \Vibes). 

In more detail, the following \rev{functionality is accessible through different implementations:}
\begin{enumerate}
   \item \rev{The native MD functionality of FHI-aims supports NVE calculations with a Velocity-Verlet integrator~\cite{Verlet.1967}, as well as NVT calculations using the Andersen~\cite{Andersen.1980}, Nosé-Hoover~\cite{Hoover.1985}, stochastic velocity rescaling~\cite{Bussi.2007}, and a colored-noise thermostat based on a generalized Langevin ansatz~\cite{Ceriotti.2009}. All necessary infrastructure is provided within the FHI-aims code, so that no additional software package is needed. Related native infrastructure for Ehrenfest dynamics, regarding the coupled motion of electrons and nuclei beyond the BOA are also implemented and discussed in Section~\ref{SecRTTDDFT}.}
    \item ASE, the Atomistic Simulation Environment~\cite{larsen2017ase}, provides a rapid and flexible route to interact with first- and second principles codes via Python. It provides NVE-MD functionality, but also includes support for thermo- and barostats. Besides, it offers a variety of other algorithms to manipulate nuclear coordinates, see contribution~\ref{ChapWorkflows}. \rev{Accordingly, it is particularly useful for rapid prototyping during development and for linking MD calculations to other approaches, c.f. see FHI-vibes below.}
    \item \Vibes~\cite{Knoop2020a} is a Python package to compute vibrational properties of solids from first principles. It can also perform \aimd calculations by using the respective functionality built into ASE. \rev{By linking such aiMD calculations to harmonic and quasi-harmonic descriptions of the dynamics it } offers  a rich set of analysis tools,~e.g.,~to map and compare with lattice-dynamics calculations for quantifying anharmonicity and for computing heat transport coefficients, see contribution~\ref{ChapVibes} and ~\ref{ChapHeatTransport} for more details.
    \item  i-PI~\cite{Litman2024} is a versatile code that is interfaced with a variety of electronic-structure packages, machine-learning potential packages, and empirical-potential packages. \rev{When interfaced with electronic structure codes, it can perform several flavors of nuclear motion, ranging from standard ones such as geometry optimizations, nudged-elastic-band, and harmonic phonons, to advanced ones, such as molecular dynamics with higher-order integrators, NVE, NVT, NPT and NsT dynamics, replica-exchange molecular dynamics, and other enhanced sampling techniques. In particular, its interface with FHI-aims allows transferring not only energies, forces, and virial stresses, but also electronic-structure quantities that can be used in non-adiabatic or light-driven dynamics. For the particular use of i-PI for the simulation of nuclear quantum effects, one of its main strengths, see contribution}~\ref{ChapQuantumNuclei}.
\end{enumerate}
      
      Na\"ively employing \aimd as a phase-space sampling tool is inefficient in general. At times, this inefficiency is so pronounced (systems with multiple minima and large free-energy barriers, rare reactive events, etc.) that a ``vanilla" \aimd sampling is effectively impossible. To address this limitation, many enhanced-sampling methods were introduced. These methods can be classified based on their conceptual 
      approach to overcoming barriers into two main categories~\cite{Frenkel.2002}: (i) incorporating bias terms into the Hamiltonian of the original system, e.g., metadynamics, umbrella sampling, and accelerated molecular dynamics; (ii) generating a generalized ensemble of the original system. Generalized-ensemble simulations involve a broader sampling of the potential energy landscape. 
      Techniques such as simulated tempering, multicanonical sampling, and parallel tempering (also known as replica exchange) fall into this second category.
      These methods are particularly effective in overcoming kinetic barriers. 

      \rev{During the last several decades, \textit{ab initio} atomistic thermodynamics~\cite{ReuterScheffler2003} has been widely and successfully applied to study the thermodynamic stability of surfaces under realistic conditions. This technique requires a prior knowledge of relevant surface phases. Furthermore, in most cases, the vibrational contributions are neglected or at most treated by harmonic approximation. To free from these limitations, an unbiased sampling of the surface phase space is required. In this context, 
    based} on the grand-canonical extension~\cite{yan1999hyper,faller2002multicanonical} of the original parallel-tempering MD scheme~\cite{marinari1992simulated,sugita1999replica}, Zhou, Scheffler and Ghiringhelli~\cite{zhou2019determining,zhou2022ab} have recently developed the grand-canonical replica-exchange (REGC) method coupled to \aimd and implemented it as a wrapper for \FHIaims. Open ensembles (i.e., exchanging both energy and matter with a reservoir) are described at equilibrium by the grand-canonical-ensemble formalism and provide an effective mean to overcome slow-diffusion problems: Thermodynamically possible defect states can be efficiently generated and sampled by means of the atoms' insertion and removal as well as the parallel coupling between different temperatures and different chemical potentials.
    In REGC, an extended ensemble including a set of $S=L\times M$ replicas of the studied system is created. The replicas span $L$ values of temperature and $M$ values of the chemical potential of the selected chemical species. Presently, only one species can be exchanged with a reservoir at given chemical potential, but the extension to multiple species is under development. The partition function of this extended ensemble is the product of the partition functions of the individual $\mu_{m}$VT$_{l}$ ensembles, where $l = 1, 2, \ldots, L$ and $m = 1, 2, \ldots, M$,
\begin{equation}
 Q_\textrm{extended} = \prod_{l=1}^{L}\prod_{m=1}^{M}\frac{e^{\beta_l \mu_m N_{l,m}} V^{N_{l,m}}} {\Lambda_l N_{l,m}!} \int d \bm{R} \, e^{-\beta_{l} E_{\{\vec{R}\}} \left(N_{l,m} \right) }.
\end{equation}
In the equation above, $\bm{R}$ denotes the nuclear coordinates that define the configuration of the system, $\beta = 1 / k_\textrm{B}T$, $\Lambda$ is the thermal wavelength, and  $E_{\{\vec{R}\}}$ is the potential energy of a configuration $\bm{R}$ of the $N$-particle system.
During the REGC simulation, short \aimd trajectories are run for each replica 
in parallel, after which a Metropolis Monte Carlo (MMC) move is attempted. In this MMC move, either a replica exchange or a particle insertion/removal is performed. This decision is based on a random criterion. In the replica-exchange move, a swap of configurations between pairs of replicas at different temperatures and/or chemical potential is tried. This means that an MMC criterion based on the difference in $\beta$ and $\mu$ between the pair of replicas and the difference in energy between the corresponding pair of configurations, accepts or rejects the swap of configurations. The detailed-balance conditions imposes a specific form for the probability of swapping configurations, which is given by Eq.~(6) in Ref.~\cite{zhou2019determining}.
In the particle insertion/removal move, the MMC acceptance criterion is based on the difference between the change in energy upon insertion or removal of a randomly positioned particle and the imposed chemical potential, given by Eqs.~(2) and (3) in Ref.~\cite{zhou2019determining}.

\begin{figure}[ht]
    \centering
    \includegraphics[width=0.5\textwidth]{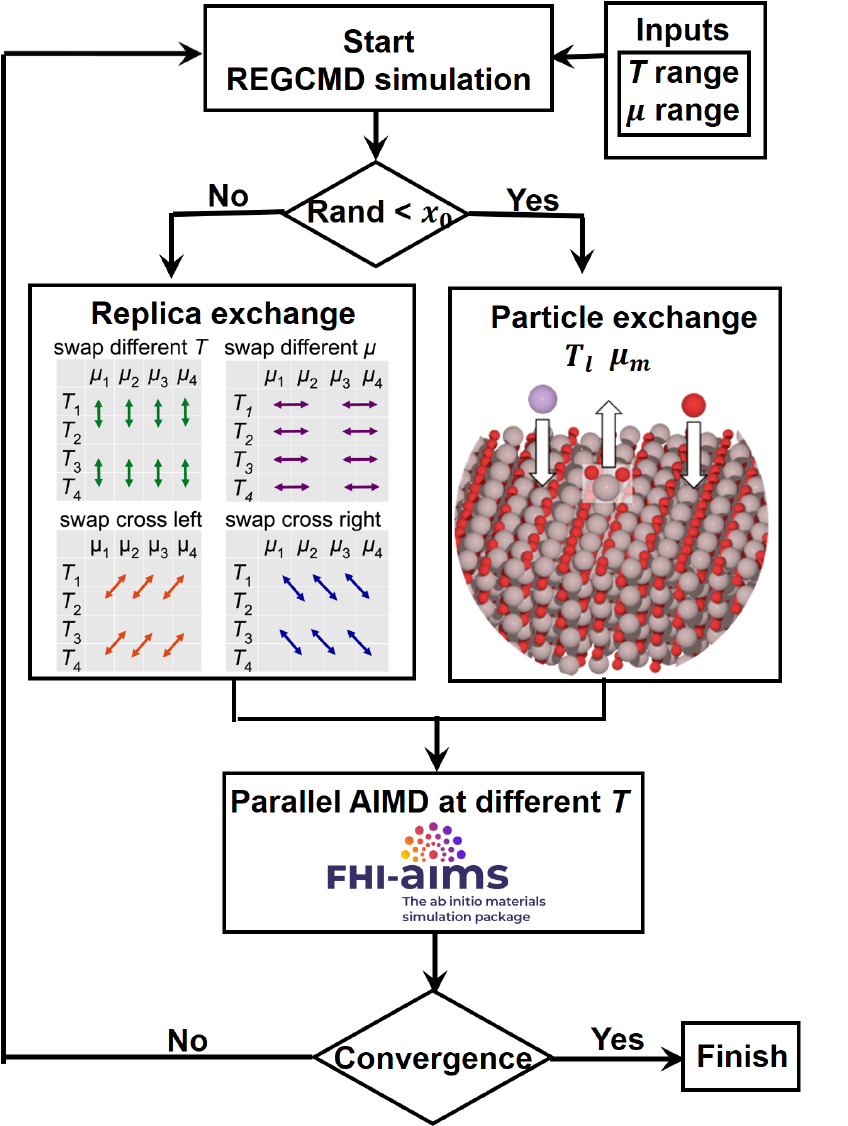}
    \caption{Schematic workflow for the replica-exchange grand-canonical (REGC) \textit{ab initio} molecular dynamics. AIMD: \textit{ab initio} molecular dynamics, which is performed within the \FHIaims code; $T$: temperature; $\mu$: chemical potential. DFT: density-functional theory. $x_0$ is the probability to perform the RE attempt. ($0\leq x_0\leq 1$).}
    \label{fig:REGC}
\end{figure}

The REGC algorithm as described above and sketched in Fig. \ref{fig:REGC}, samples the multi-canonical ensemble at each of the $L$ temperatures and $M$ chemical potentials. In REGC, the interface with \FHIaims is based on output parsing. The configuration at a certain thermodynamic condition ($T_l$, $\mu_m$) for each replica after every particle exchange or replica exchange provides the initial geometry file and the target temperature for FHI-aims to perform \aimd in the canonical ensemble. 
At the end of one iteration of the algorithm, after all the aiMD trajectory have run for the predefined number of steps, the code FHI-PANDA reads all the  configurations and the energies for the next iteration. In order to estimate ensemble averages of observables at any temperature and chemical potential within (and slightly outside) the sampled intervals, one has to apply a re-weighting algorithm, such as the multistate Bennett acceptance ratio (MBAR)~\cite{shirts2008statistically}. In this way, one can construct phase diagrams for given observables~\cite{zhou2019determining}. In particular, if the chosen observable is the heat capacity of the system, one can infer phase boundaries and construct a complete configurational phase diagram without any prior knowledge of the expected phases~\cite{zhou2022ab}. We note that the REGC approach as implemented for \FHIaims is particularly suited to study adsorption on surfaces from the gas phase. In case of bulk systems, the particle insertion MMC move has very low acceptance, leading to poor statistics, unless dedicated biasing mechanisms are implemented.

\subsection*{Usability and Tutorials}
The native implementation of \aimd in FHI-aims is extensively described in section 3.12 of the code manual (year 2024). The central control of the run is accessed via the \texttt{MD\_run} flag. A number of flags defining the time-step, velocity initialization, wave-function extrapolation scheme, and restarting options complete the implementation. Since 2021, the code supports a non-self consistent (NSC) Hartree-potential correction to the Hellmann-Feynman forces of the form~\cite{chan1993accelerating}
\begin{equation}
		\vec{F}_{I}^{(\nu),\text{NSC}}= 
		 \int d\bm{r} \left[ \left(n^{(\nu -1)} (\vec{r})- n_{KS}^{(\nu)} (\vec{r})\right)
			\dfrac{ \partial v_{H}^{(\nu -1)} (\vec{r}; \vec{R}) }{\partial \vec{R}_{I}} \right]
            \label{eq:rossi-force-correction}
\end{equation}
where $\nu$ is the current self-consistent step, $\nu-1$ is the previous step, $n_{\text{KS}}$ is the unmixed Kohn-Sham (KS) density, $n$ is the mixed density, and $v_{H}$ is the Hartree potential. The correction is computed at each SCF step and converges to zero as $n$ approaches the self-consistent solution. In practice, its explicit inclusion allows obtaining very accurate forces with a less-converged density. This implementation, realized by Rivera-Arrieta and Rossi, required a rearrangement of the SCF loop in FHI-aims, in which the density update and mixing step was moved after the solution of the KS equations, instead of being performed at the beginning of the SCF loop. For \aimd simulations, this correction term represented a major improvement in efficiency. In our tests, this correction allows us to obtain energy drifts with a density convergence threshold of $10^{-3}$ $e/a_0^3$ that is comparable to the drift obtained with a $10^{-5}$ $e/a_0^3$ threshold without the correction. In practice, for example for the Zundel cation, this represents a 40\% reduction in the number of SCF cycles in a simulation of 10000 steps. This correction is now automatically included in any force calculation of the code, as it does not incur any appreciable computational overhead.

In regards of the functionality of the code, one important addition compared to the original release is the support of variable-cell \aimd under periodic boundary conditions. This is enabled by the implementation of the stress tensor, which describes the changes of the PES with respect to symmetric strain deformations of the unit cell~\cite{knuth2015all}. Besides enabling the usage of barostats to sample the isothermal-isobaric~(NPT) and the isothermal-isostress~(NsT) ensemble, this also allows to monitor the stress during \aimd runs,~.e.g,~to compute thermal-expansion coefficients~\cite{KnoopDiss}. Furthermore, this gives access to the virials,~i.e.,~the atom-resolved stress contributions, which allow to compute the virial contributions to the heat flux, see contribution~\ref{ChapHeatTransport}.

 The massively parallel REGC algorithm requires no prior knowledge of the phase diagram and takes
 only the potential energy function together with the desired $\mu$ and $T$ ranges as inputs. As shown in Figure \ref{fig:REGC}, during a REGC simulation, each replica is assigned a random number which is compared to a  probability $x_0$ $(0 \leq x_0 \leq 1)$ that determines whether to attempt exchanging a particle with the reservoir or to attempt a replica-exchange move. After the particle/replica-exchange attempt, parallel AIMD runs follow to diffuse the system in the canonical ensemble, i.e., at temperature ${T}_{i}$, with fixed number of particles $N$ and volume $V$ of the system ($NVT$ ensemble). The procedure is then iterated until the convergence of defined quantities is achieved. A 2D (one dimension is $T$ and the other dimension is $\mu$) replica-exchange scheme is performed for the system in contact with a single gas reservoir. Each replica has 3–8 neighbors, compared to 1–2 neighbors in conventional 1D schemes, which improves exchange efficiency. The tutorial for studying a Si$_2$ cluster in contact with gas-phase H$_2$, where the whole workflow of the REGC method is demonstrated, is available \href{https://fhi-aims-club.gitlab.io/tutorials/introduction-of-ab-initio-thermodynamics-and-regc}{under this link.}\footnote{\texttt{https://fhi-aims-club.gitlab.io/tutorials/introduction-of-ab-initio-thermodynamics-and-regc}}\\


\subsection*{Future Plans and Challenges}
With the development of fast and reliable machine-learning interatomic potentials (MLIPs), \aimd simulations have become quickly less popular. The reason is that molecular dynamics with a very similar accuracy as \aimd can be performed at a computational cost that is 4 to 8 orders of magnitude cheaper~\cite{Litman2024} than the underlying electronic-structure method on which the MLIP was trained. Therefore, the system sizes that can be treated with MLIPs are much larger and statistical convergence of thermodynamical observables can be routinely obtained. However \aimd is still useful (and at times still essential), and will likely continue to be for years to come, as MLIP require \textit{ab initio} data to be trained on. In addition, there are systems for which the physics that need to be captured is still challenging for current MLIPs, such as systems with complex charged defects and where electronic spin is particularly relevant. Regarding FHI-aims, a bespoke interface for MLIP active learning based on \aimd will soon be released, which will offer several advantages in terms of data generation and rare-event sampling. Wrappers and workflows around FHI-aims that perform active learning already exist and are described in contribution~\ref{ChapMLP_AI}. It is only natural that the REGC method is also integrated with MLIPs. 

Regarding the REGC algorithm, current and near-future development is dedicated to i) a smart handling of the load unbalance generated by the parallel \aimd run with different number of particles and ii) the extension to multiple species with independent chemical potentials for multi-component phase diagrams, where the challenge is to create an adaptive grid for the sampled temperatures and chemical potentials.

On a more technical note, there are algorithmic improvements related to \aimd that would benefit any situation where the method is used, including data generation. For example, regarding the force-correction term that was discussed in this contribution, one could also compute similar terms for the Hartree-Fock exchange and for the computation of the stress tensor. This implementation would make the convergence of forces from hybrid-functionals and stresses from any functional even faster, making data generation more efficient.


\subsection*{Acknowledgements}
We acknowledge Felix Hanke and J\"urgen Wieferink, who have contributed at fundamental points in the implementation of molecular dynamics in FHI-aims, and Konstantin Lion, who has more recently worked on wave-function propagation schemes. MS acknowledges support by his TEC1p Advanced Grant (the European Research Council (ERC) Horizon 2020 research and innovation programme, grant agreement No. 740233.

\newpage

\section{Electronic Structure Calculations with Quantum Nuclei}
\label{ChapQuantumNuclei}
\sectionauthor[1,2]{\textbf{ *Yosuke Kanai}}
\sectionauthor[3,4,a]{Yair Litman}
\sectionauthor[1]
{\rev{Songrui Liu}}
\sectionauthor[5]{\textbf{ *Mariana Rossi}}
\rev{\sectionauthor[5]{Elia Stocco}}
\sectionauthor[1]{Jianhang Xu}
\sectionauthor[1]
{\rev{Xingchen Zhang}}
\sectionlastauthor[1]{Ruiyi Zhou}

\sectionaffil[1]{Department of Chemistry, University of North Carolina at Chapel Hill, USA}
\sectionaffil[2]{Department of Physics and Astronomy, University of North Carolina at Chapel Hill, USA}
\sectionaffil[3]{Theory Department (since 1/1/2020: The NOMAD Laboratory), Fritz Haber Institute of the Max Planck Society, Faradayweg 4-6, D-14195 Berlin, Germany}
\sectionaffil[4]{Yusuf Hamied Department of Chemistry, Cambridge University, UK}
\sectionaffil[5]{Max Planck Institute for the Structure and Dynamics of Matter, 22761 Hamburg, Germany}

\sectionaffil[*]{Coordinator of this contribution.}
\rule[0.25ex]{0.35\linewidth}{0.25pt}

\sectionaffil[a]{{\it Current Address:} Max Planck Institute for Polymer Research, Mainz, Germany}




\subsection*{Summary}
Electronic structure simulations are most commonly performed within the Born-Oppenheimer approximation (BOA) and under the assumption that nuclei are fixed point charges. However, accounting for quantum nuclear motion often goes beyond a simple correction to such simulations.
Quantum nuclei delocalize, interfere, and tunnel through barriers. These effects strongly impact the thermodynamics and reactive properties of several materials. FHI-aims currently allows atomic nuclei to be treated as quantum-mechanical particles within the BOA, via path-integral methods, or going beyond the BOA through nuclear-electronic orbital (NEO) methods. 

Path-integral molecular dynamics (PIMD) has emerged as a practical \rev{and accurate} methodology to treat nuclear quantum effects, when the BOA holds. Proposed by Rahman and Parrinello~\cite{ParrinelloRahamn1984}, it relies on the molecular dynamics sampling of quantum-mechanical thermodynamical observables  within the imaginary-time path integral (PI) representation of quantum mechanics. Approximations to time-dependent thermodynamic observables can also be obtained from this formalism. A tight integration of FHI-aims with the i-PI code \cite{kapil2019,Litman2024} allows advanced \textit{ab initio} PI simulations to be performed.

The nuclear-electronic orbital method by Hammes-Schiffer and co-workers \cite{webb_multiconfigurational_2002,pavosevic_multicomponent_2020,xu_full-quantum_2020,hammes-schiffer_nuclearelectronic_2021} enables specified atomic nuclei (usually protons) to be treated on an equal footing as electrons in electronic structure calculations. NEO can be combined with various electronic structure methods including DFT \cite{pak_density_2007,chakraborty_development_2008}. In FHI-aims, the NEO method is employed as a practical approach for implementing multicomponent DFT~\rev{\cite{kreibich_multicomponent_2001}}, which allows the atomic nuclei to be modeled as quantum-mechanical particles without the usual BOA.

\subsection*{Current Status of the Implementation}
\begin{figure}[ht]
    \centering
    \includegraphics[width=0.60\textwidth]{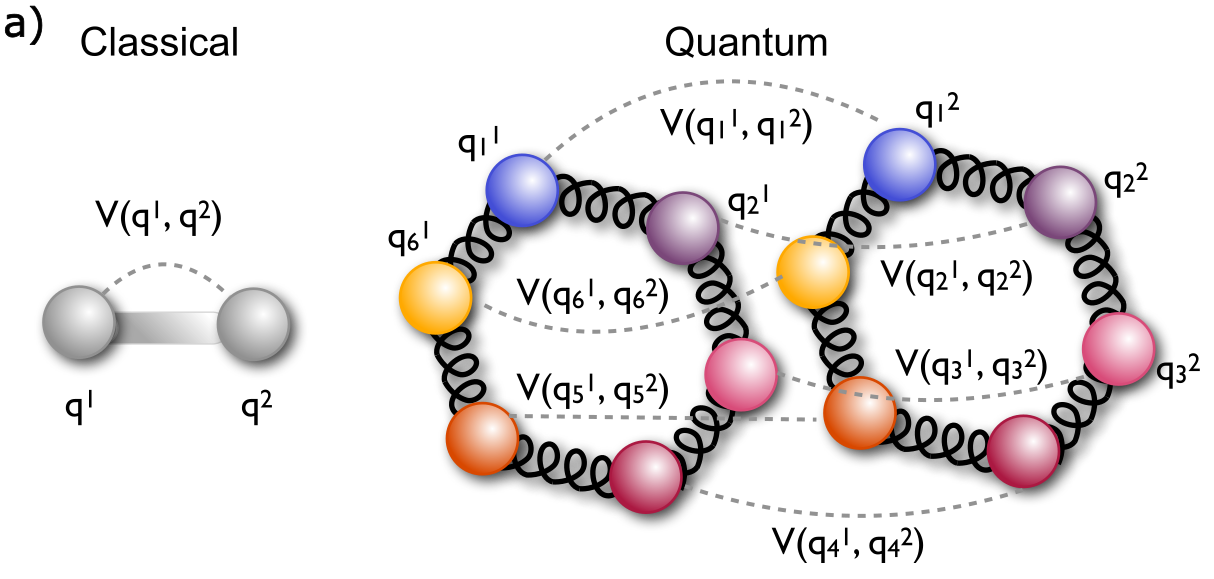}
    \includegraphics[width=0.28\textwidth]{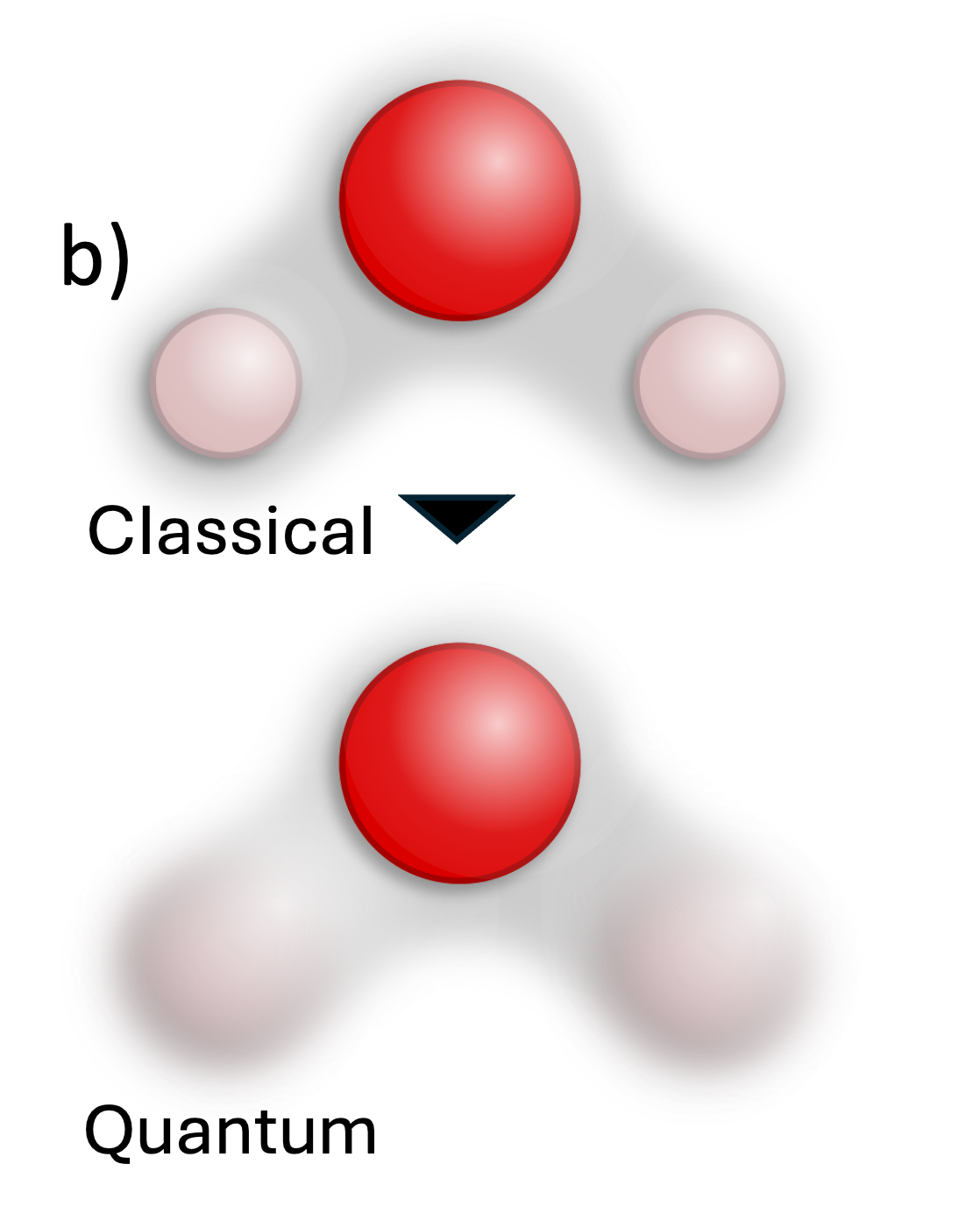}
    \caption{a) A depiction of a classical system of two particles interacting through a potential $V$ and the corresponding quantum path-integral representation of the same system, considering that the particles are distinguishable. Each particle becomes a ring polymer. Each bead in the first particle interacts with the corresponding bead in the second particle through the same potential $V$. Nearest neighbor beads within a ring-polymer are connected by a harmonic spring. b) A water molecule described using the conventional \textit{ab initio} approach and the corresponding nuclear-electronic orbital approach is shown. 
    In the classical picture, both oxygen (red sphere) and hydrogen (pink spheres) atoms are treated as classical nuclei, while electrons (grey cloud) are quantum particles. In the NEO approach, protons (pink cloud) are also treated as quantum particles, at the same level of theory as electrons.}
    \label{fig:method-pics}
\end{figure}

Exploiting the fact that the quantum time propagator is equal to the quantum density matrix $\exp(-\beta\hat{H})$, where $\beta = 1/(k_B T)$, at imaginary time $i\beta \hbar$, one can cast quantum thermodynamics in the Feynman path integral formalism. The canonical partition function $Z = \mathrm{Tr}[\exp^{-\beta \hat{H}}]$ can be calculated by performing a Trotter factorization  of the trace in the position representation, producing, 
\begin{equation}
Z= \lim_{P \to \infty} \left(\frac{1}{2 \pi \hbar }\sqrt{\frac{m}{m'}}\right)^P \int d\bm{R} d\bm{p}
        \exp\left\{-\beta_P  H_P(\bm{p},\bm{R}) \right\}
\end{equation}
where the $m$ is the physical mass of the system, $m'$ are fictitious masses for sampling, $\beta_P = \beta/P$ and $P$ corresponds to the number of identities introduced to factorize the trace. The non-commuting nature of the position operator $\hat{R}$ and momentum operators $\hat{p}$ gives rise to a classical ring-polymer Hamiltonian $H_P$, consisting of $P$ ``beads" (system replicas) connected by harmonic springs. It is given by
\begin{equation}
    H_P = \sum_{j=1}^{P} \left[ \frac{p_j^2}{2m'}+\frac{m \omega_P^2}{2} (R_{j+1}-R_{j})^2 + V[n, R_j] \right], \label{eq:pi-ham}
\end{equation}
where, in \textit{ab initio} PIMD based on DFT, $V[n; R_j]$ is the ground-state Born-Oppenheimer potential calculated at the (fixed) nuclear positions $R_j$ of replica $j$. The spring frequency $\omega_P = P / (\beta \hbar)$ and $R_1 = R_{P+1}$. A pictorial representation of 
 the classical and quantum system represented by ring polymers is given in Fig.~\ref{fig:method-pics}a.
\rev{The definition of the Hamiltonian in Eq.~\ref{eq:pi-ham} allows one to evolve the nuclear positions following Newton's equations of motion and use molecular dynamics techniques (see Section~\ref{ChapMD}) in order to sample the phase space.}

The most advanced PIMD algorithms can reduce the cost of PI simulations by many orders of magnitude, making them almost as cheap as an \textit{ab initio} simulation where nuclei are treated classically. These algorithms involve combinations of higher-order integrators, multiple-time-stepping algorithms, ring-polymer contraction algorithms, and much more~\cite{Marsalek-jcp-2016,Kapil-jcp-2016,Litman_JCP_2018,Poltavsky_JCP_2017, kapil2019}. Because PIMD and related PI methods largely rely on the BOA, it has been deemed most efficient to interface FHI-aims with a code in which a team of developers implements all the most modern algorithmic alternatives: the i-PI code \cite{kapil2019,Litman2024}. This interface is automatically compiled with FHI-aims and using it requires setting only a few keywords in the input file of FHI-aims, as detailed in the next section. FHI-aims provides the BO potential calculated from Kohn-Sham density-functional theory, together with the corresponding forces and stress, and i-PI evolves the equations of motion for the (quantum) nuclei with these quantities. \rev{Advantages of PIMD are that it can be applied to large and complex systems, it naturally captures quantum statistics from all nuclei, and it is exact for distinguishable particles, ensuring the accuracy of the results as long as the BOA holds.}

Approximations to time-correlation functions can be obtained as approximations from this formalism and give access to system response properties such as vibrational spectra, thermal rate constants, heat transport, etc~\cite{Althorpe-ar-2024, Sutherland-jcp-2021}.  
The semiclassical instanton rate theory can also be cast in the same formalism\cite{Miller_JCP_1975,Richardson_JCP_2009}, usually referred to as a ring-polymer instanton rate theory and constitutes a method to calculate nuclear tunnelling splittings and tunnelling rates~\cite{Richardson_Review_2018}.

\textbf{NEO:}  The NEO method is widely used to treat protons quantum-mechanically in electronic structure calculations while any other atomic nuclei can be made quantum in principle. 
It allows us to formulate a practical method for multicomponent DFT. Kohn-Sham (KS) Hamiltonians for electrons and protons are coupled via the classical electrostatic and also the electron-proton correlation functional as in 
\begin{align}
\label{eq:tdks_e}
      \hat{H}^e = -\frac{1}{2}\nabla_{e}^2 
      - \hat{V}_{\text{es}}^{c}(\mathbf{r}^e) - \hat{V}_{\text{es}}^{p}(\mathbf{r}^e) - \hat{V}_{H}^{e}(\mathbf{r}^e) +\frac{\delta E_{XC}^{e}[n ^e]}{\delta n^e}+\frac{\delta E_{epc}[n^e,n^p]}{\delta n^e}
      \\
 \label{eq:tdks_p}
     \hat{H}^p = -\frac{1}{2M^p}\nabla_{p}^2 + 
     \hat{V}_{\text{es}}^{c}(\mathbf{r}^p) + \hat{V}_{H}^{p}(\mathbf{r}^p) + \hat{V}_{\text{es}}^{e}(\mathbf{r}^p) + \frac{\delta E_{XC}^{p}[n ^p]}{\delta n^p}+\frac{\delta E_{epc}[n^e,n^p]}{\delta n^p}
\end{align}
where $\hat{V}_{\text{es}}^{c}$ and $\hat{V}_{\text{es}}^{e/p}$ are the electrostatic potential from classical atomic nuclei and electron/protons, respectively. $\hat{V}_{H}$ is the Hartree potential. $E^e_{XC}$, $E^p_{XC}$, and $E_{epc}$ represent the exchange-correlation (XC) energy of electrons, the XC energy of protons, and the correlation energy between electron and proton, respectively. $\mathbf{r}^e$ and $\mathbf{r}^p$ represent the electron and proton coordinates.
The KS eigenvalue equations thus must be solved together in a self-consistent fashion\rev{, and therefore the electron density is indirectly modified by the proton quantization, as discussed in Ref. \cite{xu2022nuclear}.}
In practice, we approximate $E_{XC}^{p}$ with the Hartree-Fock exchange. It is motivated by the fact that the correlation among protons is negligible in most cases and using the exact exchange makes the protons to be free of the infamous self-interaction error, which would otherwise cause artificial delocalization of the quantum protons.
Additionally, we implemented only $\Gamma$-point sampling of the Brillouin zone for the proton KS orbitals.
For the electron-proton correlation functional, $E_{\text{epc}}$, several approximations have been proposed in literature. Currently, the EPC-17 and EPC-17-2 (LDA equivalent) \cite{yang2017development,brorsen2017multicomponent} are implemented in the FHI-aims code.
For quantum protons, special Gaussian-type orbital basis sets have been developed \cite{yu_development_2020} and FHI-aims code also employs them but using the algorithms described in Ref. \cite{xu2022nuclear}. 

In addition to treating protons quantum-mechanically in equilibrium, NEO method also allows us to study coupled quantum dynamics of protons and electrons via real-time time-dependent density functional theory (RT-TDDFT) and via Ehrenfest method, which are discussed in Section \ref{SecRTTDDFT}. The underlying dynamics of the mixed classical-quantum system is given by the Lagrangian

\begin{equation}
\begin{aligned}
\label{eq:lagrangian}
    L^\text{NEO}(t)
& = \int \text{d}\mathbf{k}\sum_n \bra{\psi^{e}_{n,\mathbf{k}}(t)} i\frac{\partial}{\partial t}+\frac{1}{2}\nabla_{e}^{2} \ket{\psi^{e}_{n,\mathbf{k}}(t)} 
+\sum_n \bra{\psi^{p}_{n}(t)} i\frac{\partial}{\partial t}+\frac{1}{2M^{p}}\nabla_{p}^{2} \ket{\psi^{p}_{n}(t)} \\
& \quad -E_{H}[n^e] -E_{H}[n^p]-E_{XC}^e[n^e]-E_{XC}^p[n^p]\\
    & \quad +
    \int \text{d}\mathbf{r}^e \text{d}\mathbf{r}^p \frac{1}{|\mathbf{r}^e-\mathbf{r}^p|} n^e(\mathbf{r}^e,t)n^p(\mathbf{r}^p,t)-E_{epc}[n^e, n^p] \\
    & \quad + \sum_{I}^{N^c} \frac{1}{2}M_I \dot{\mathbf{R}}_{I}^2(t)
    -\sum_{I<J}^{N^c}\frac{Z_I Z_J}{|\mathbf{R}_{I}(t)-\mathbf{R}_{J}(t) | } \\
    & \quad + \int \text{d}\mathbf{r}^e n^e(\mathbf{r}^e,t) \sum_{I}^{N^c} 
    \frac{Z_I}{|\mathbf{r}^e-\mathbf{R}_I(t)|}
    - \int \text{d}\mathbf{r}^p n^p(\mathbf{r}^p,t) \sum_{I}^{N^c} 
    \frac{Z_I}{|\mathbf{r}^p-\mathbf{R}_I(t)|},
\end{aligned}
\end{equation}
where 
$\mathbf{R}_{I}(t)$ and $Z_I$ are the position coordinates and the charge, respectively, of classical nucleus $I$. 
The upper dot denotes the time derivative, and 
$N^c$ is the total number of classical atomic nuclei.
By applying the variational principle to this NEO action, Euler-Lagrange equation allows us to derive necessary equations of motion (EOM) for quantum (electrons and protons) and classical (non-proton atomic nuclei) degrees of freedom, allowing us to formulate RT-TDDFT and Ehrenfest Dynamics in the context of multicomponent DFT formalism \cite{xu2023prl,xu2024lagrangianformulationnuclearelectronicorbital}. \rev{
The approach is particularly convenient for studying large complex condensed matter because one can select which parts of the system are treated with quantum dynamics and which parts are governed by the classical equations of motion (see Section~\ref{ChapMD}). 
}

\rev{
Additionally, the constrained NEO (cNEO) method~\cite{xu2020constrained} has been also recently implemented in FHI-aims~\cite{liu2025constrained}.
cNEO constrains the expectation value of the quantum nuclear position operator at specific points in space. 
This approach effectively constrains the centers of the quantum nuclear probability densities to designated positions in real space, allowing NEO electronic structure calculations with controlled nuclear localization and providing a straightforward framework to defining nuclear equations of motion for the constrained position and momenta.~\cite{xu2022molecular} 
Its connection to the quantum-corrected effective potential, as usually formulated using quantum-mechanical path integrals~\cite{RamirezLopezCiudad1999PRL}, is discussed in Ref.~\cite{liu2025constrained}.
}


\subsection*{Usability and Tutorials}
\begin{figure}[htb!]
    \centering
    \includegraphics[width=.9\textwidth]{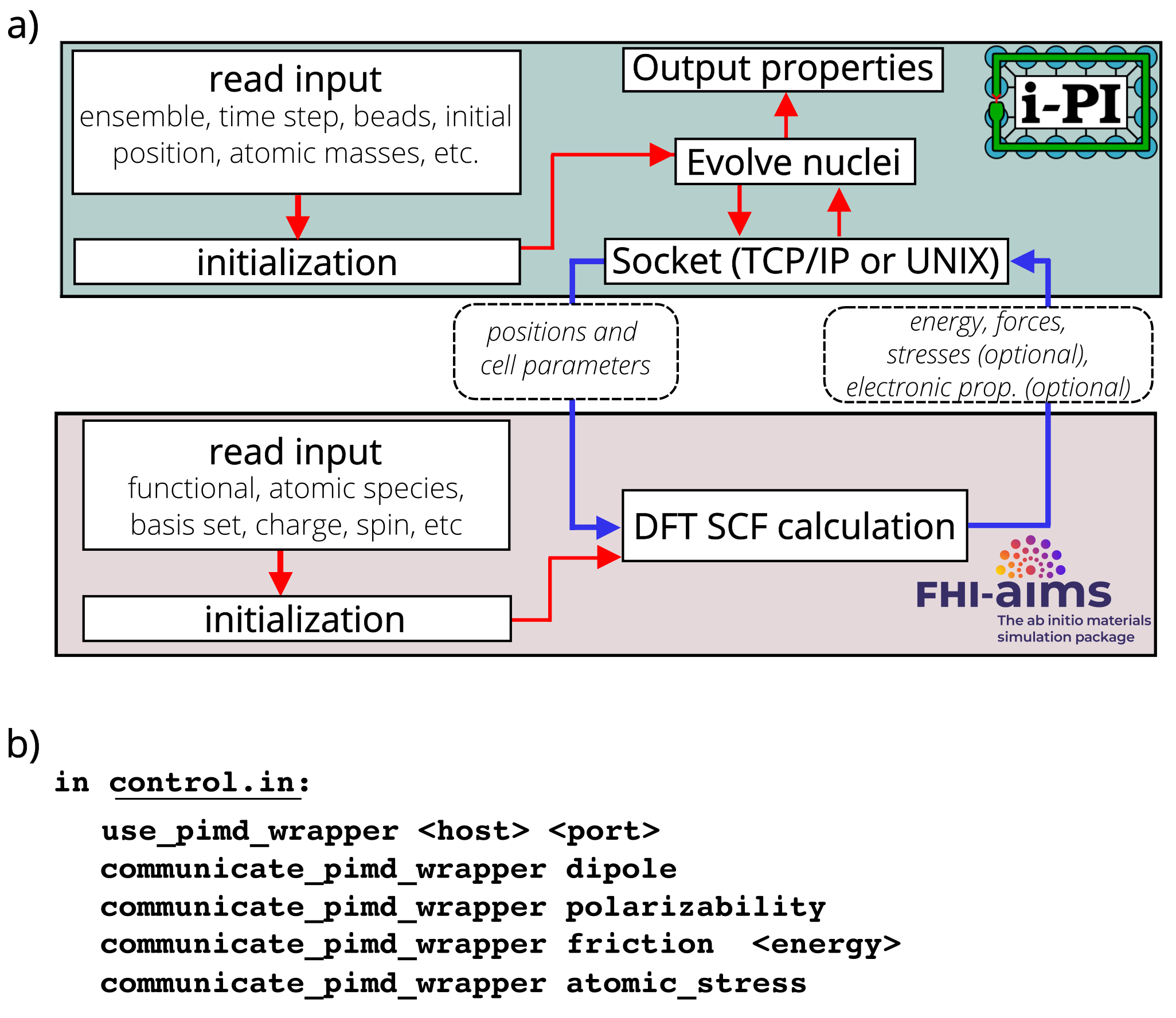}
    \caption{a) A schematic representation of the communication between FHI-aims and i-PI. Once a connection is established, the simulation loop proceeds as follows: i) i-PI sends the nuclear positions and cell parameters to FHI-aims; ii) FHI-aims computes the necessary properties, such as energies and forces, and communicates them back; and iii) i-PI updates the atomic positions using the received data, thus restarting the loop. The red and blue arrows show intra-code and inter-code flow of data, respectively.  b) A snippet of the FHI-aims input file showing the keywords that manage communication with i-PI. In this example, FHI-aims communicates the dipole moment, polarizability, electronic friction tensor, and atomic stress to i-PI. }
    \label{fig:ipi}
\end{figure}

We begin by considering the inclusion of NQEs within the  BOA. For this purpose, a server-client paradigm is adopted, where i-PI takes the role of server and FHI-aims serves as a client (see Fig. \ref{fig:ipi}a). The communication between both codes is established through UNIX or TCP/IP sockets. The latter allows for communication across different nodes as it might be required when running in high-performance computing (HPC) facilities.  The framework enables trivial parallelization over beads by running as many instances of FHI-aims as there are beads in the simulation. As shown in Fig. \ref{fig:ipi}b, the FHI-aims input file requires minimal additions to specify the type and port address of the socket, as well as the communication of properties beyond energy, forces and stresses. At the moment it is possible to communicate polarizabilities, Hirshfeld charges, atomic stresses, dipole moments and electronic friction, some of which can be used by i-PI to determine nuclear dynamics. 

A tutorial that describes how to set up an \textit{ab initio} molecular dynamics simulation using FHI-aims and i-PI is available \href{https://fhi-aims-club.gitlab.io/tutorials/molecular-dynamics-with-i-pi/}{under this link}\footnote{\texttt{https://fhi-aims-club.gitlab.io/tutorials/molecular-dynamics-with-i-pi/}}. The tutorial covers the basics of performing these simulations, such as describing how the codes communicate, setting up the simulation, determining the time-step and assessing the thermostat. It also covers advanced topics like anharmonic free energy calculations and constant pressure simulations. The combination of FHI-aims with i-PI has enabled studies on a wide range of systems and topics including metallic interfaces \cite{Litman_JCP_2018,Fidanyan_AdvThSim_2021,Litman_PRL_2020,Fidanyan_JCP_2023}, molecular crystals \cite{Rossi_PRL_2016,Kapil_PNAS_2022},  nuclear quantum tunnelling in gas-phase molecules \cite{Jelenfi_MolPhys_2023,Litman_JACS_2019,Litman_FarDiss_2020}, quantum dissipative dynamics in metals \cite{Litman_JCP_2022_II,Bridge_JCP_2024}, biomolecules \cite{Rossi_JPCL_2015,Fonseca_JCP_2021} and low-dimensional materials \cite{Jacobs_ElecStruc_2024,Poltavsky_JCP_2017}.

The NEO tutorial, which can be accessed at \href{https://fhi-aims-club.gitlab.io/tutorials/NEO/}{this link}\footnote{\texttt{https://fhi-aims-club.gitlab.io/tutorials/NEO/}}, provides a brief overview of the NEO theory along with several simple examples for running NEO-DFT, RT-NEO-TDDFT simulations in FHI-aims code. 
The first part of the tutorial focuses on standard NEO-DFT calculation, outlining all the necessary parameters and explaining how to calculate the ground state with quantized protons for both isolated systems and extended periodic systems.
It also has an example for performing NEO `geometry optimizations' in which
the centers of proton basis sets are also optimized. For protons, a number of Gaussian-type basis functions have been specifically designed as detailed in Ref. \cite{yu_development_2020}. 
The second part of the tutorial is focused on performing RT-NEO-TDDFT simulation for studying coupled quantum dynamics of electrons and protons. The case of an excited state proton transfer in a small organic molecule is used as an example.

\begin{figure}[ht]
    \centering
    \includegraphics[width=0.9\textwidth]{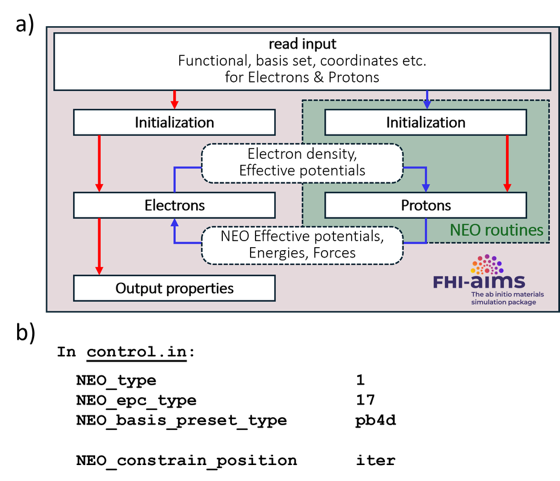}
    \caption{a) A schematic representation of the implementation of the NEO method in FHI-aims. 
    NEO create a quantum subsystem for protons.
    The initialization and SCF cycles are done independently.
    Only necessary information is passed between the electron system and the proton system. The red and blue arrows show intra-system and inter-system flow of data, respectively.
    b) A snippet of the FHI-aims input file showing the keywords for running NEO simulations.    \rev{The keyword in the last line is for cNEO calculations.} }
    \label{fig:code-tutorials NEO}
    
\end{figure}


\subsection*{Future Plans and Challenges}
\textit{Ab initio} PIMD has been extensively and successfully employed in material simulations for decades. However, at best, these simulations carry the same cost as \textit{ab initio} molecular dynamics (AIMD) and often they are at least one order of magnitude more expensive. With the development of the last-generation machine-learning interatomic potentials (MLIPs), the usage of Born-Oppenheimer AIMD and \textit{ab initio} PIMD is becoming restricted as data-generation tool to train or refine these potentials. MLIPs currently offer 5-7 orders of magnitude gain in the speed of force evaluations when compared to DFT, with minimal accuracy loss. 

Nevertheless, it has been repeatedly shown that it is necessary to include data from PIMD simulations to train MLIPs when one is targetting an accurate description of nuclear quantum effects. Therefore, we expect that the necessity of \textit{ab initio} PIMD simulations will continue to exist, and a better integration of i-PI with FHI-aims through MPI communicators is desirable. That would remove the need for socket communication and would make the infrastructure more robust and easier to use in high-performance architectures. Going beyond the BOA with path-integral based approximations is a topic of forefront research~\cite{Ruji2023}. We expect that future methods that follow this approach will directly profit from these implementations.

The NEO method has become an exciting avenue for treating atomic nuclei quantum-mechanically on an equal footing as electrons in electronic structure calculations over the last several years. 
The method is actively developed in the community, and new functionalities will be added also to the FHI-aims code in the near future.
Because quantum nuclei are represented by Gaussian functions centered on a particular point in real space, one needs to ensure that enough proton basis functions are present such that the quantum dynamics of protons can be modeled accurately. 
In order to overcome this inconvenience, the idea of Ehrenfest RT-NEO-TDDFT has been introduced\cite{xu2024lagrangianformulationnuclearelectronicorbital}, and it is under active development. 
Another theoretical challenge is the development of a correlation functional for protons. 
While its effect is insignificant in most molecular/materials systems, such an effect might induce interesting behavior when protons are situated very closely such as under extremely high pressure.

\subsection*{Acknowledgements}
M.R. acknowledges Eszter S. P\'os and Karen Fidanyan for essential contributions to the tutorial regarding the use of FHI-aims with i-PI. 

\newpage

\section{Electron-Phonon Coupling and Electronic Friction}

\sectionauthor[1]{Connor L. Box}
\sectionauthor[2]{ George Trenins}
\sectionauthor[3]{Honghui Shang}
\sectionauthor[2,4,a]{Yair Litman}
\sectionauthor[2]{Mariana Rossi}
\sectionlastauthor[1,5,6]{\textbf{ *Reinhard J. Maurer}}

\sectionaffil[1]{Department of Chemistry, University of Warwick, Gibbet Hill Road, CV4 7AL Coventry, United Kingdom}
\sectionaffil[2]{Max Planck Institute for the Structure and Dynamics of Matter, 22761 Hamburg, Germany}
\sectionaffil[3]{Key Laboratory of Precision and Intelligent Chemistry, University of Science and Technology of China, Hefei, China}
\sectionaffil[4]{Yusuf Hamied Department of Chemistry, Cambridge University, UK}
\sectionaffil[5]{Department of Physics, University of Warwick, Gibbet Hill Road, CV4 7AL Coventry, United Kingdom}
\sectionaffil[6]{Faculty of Physics, University of Vienna, Vienna A-1090, Austria}

\sectionaffil[*]{Coordinator of this contribution.}
\rule[0.25ex]{0.35\linewidth}{0.25pt}

\sectionaffil[a]{{\it Current Address:} Max Planck Institute for Polymer Research, Mainz, Germany}




\subsection*{Summary}
The electron-phonon interaction manifests in many physical properties of materials, such as their thermodynamic, transport and superconducting properties.
Particularly for metals, phonons effectively couple to low-energy electron-hole pair excitations. In these situations, standard molecular dynamics techniques neglect the energy transfer mechanism between lattice vibrations and electronic excitations, and can provide qualitatively incorrect dynamical properties.~\cite{bunermann+sci2015}
When nuclear motion can be considered classical, this energy transfer can be accounted for through molecular dynamics with electronic friction~\cite{head-gordon+jcp1995},  which is evaluated in a Langevin framework:
\begin{equation}
M \ddot{R}_{a \kappa}=-\frac{\partial V(\boldsymbol{R})}{\partial R_{a \kappa}}-\sum_{a^{\prime} \kappa^{\prime}} \Lambda_{a \kappa, a^{\prime} \kappa^{\prime}} \dot{R}_{a^{\prime} \kappa^{\prime}}+\mathcal{R}_{a \kappa}(t),
\end{equation}
where an atom with index $a$, mass $M$ and position $\boldsymbol{R}$ at time $t$ experiences forces from the adiabatic potential energy $V$, a friction force from the electronic friction tensor $\boldsymbol{\Lambda}$ and a random force $\mathcal{R}$. The index $\kappa$ indicates the 3 Cartesian directions $x$, $y$, and $z$. The Markovian friction tensor, $\boldsymbol{\Lambda}$, describes the energy dissipation that arises due to electron-phonon coupling in the quasi-static (zero-frequency) limit. The random force establishes detailed balance between the electronic bath and the atoms.
FHI-aims provides an infrastructure to evaluate the electronic friction tensor, $\Lambda$, and more generally, the electron-phonon-induced vibrational lifetime (or relaxation rates) based on time-dependent perturbation theory for both aperiodic and periodic systems.
The friction module was written by Reinhard J. Maurer, and released in 2016.~\cite{maurer+prb2016} This was re-written by Connor L. Box in 2021, improving parallelism, allowing direct output of electron-phonon coupling matrix elements and completing Honghui Shang's previous efforts in interfacing to the DFPT infrastructure in FHI-aims.~\cite{box+ecse2021,box+es2023} Additional improvements were made by George Trenins in 2024 to improve numerical stability. 

This module has previously supported research efforts to investigate nonadiabatic energy loss of atoms and molecules scattering from metal surfaces~\cite{maurer+prl2017,box+jacsau2021}, the interplay of nonadiabatic effects and quantum effects in hydrogen diffusion in bulk  metals~\cite{Litman_JCP_2022_II}, and the phonon lifetimes of molecular overlayers on metal surfaces.~\cite{maurer+prb2016,box+es2023} Machine learning representations of the FHI-aims-predicted Markovian electronic friction calculated have also been proposed and successfully employed~\cite{zhan+jpcc2020,acefriction}.

\subsection*{Current Status of the Implementation}
For the theory of electron-phonon coupling calculations, the interested reader is referred to Ref~\cite{Giustino2017}, for the specifics of the implementation within FHI-aims, we suggest Ref~\cite{box+es2023}. Considering a phonon of frequency $\omega_{\mathbf{q}\nu}$ and mass $M_{\nu}$, characterized by two quantum numbers, namely the band index, $\nu$,and the momentum vector, $\mathbf{q}$, and electronic states, characterized by band indices $m$, $n$, and momentum $\mathbf{k}$, the EPC matrix element is given by:
\begin{equation}\label{eq:EPCs_SE}
   g_{mn\nu}(\mathbf{k},\mathbf{q}) =  \left(\frac{\hbar}{2M_{\nu}\omega_{\mathbf{q}\nu}}\right)^{1/2} \tilde{g}_{mn\nu}(\mathbf{k},\mathbf{q}),
\end{equation}
with 
\begin{equation}\label{eq:EPCs}
    \tilde{g}_{mn\nu} (\mathbf{k},\mathbf{q}) = \Braket{m\mathbf{k}+\mathbf{q}|\partial_{\mathbf{q}\nu}V|n\mathbf{k}}.
\end{equation}
The EPC matrix elements describe the excitation of an electron from a state $n\mathbf{k}$ to a state $m\mathbf{k}+\mathbf{q}$ by absorption of a phonon $\mathbf{q}\nu$. $V$ in Eq.~\ref{eq:EPCs} is the self-consistent ("screened") effective potential from a Kohn-Sham DFT calculation. These screened EPC elements can be output using ELSI infrastructure for post-processing. They are evaluated in NAO basis and Cartesian framework as:
\begin{equation}\label{eq:local_EPC1_v2}
    \tilde{g}^{a\kappa}_{mn} (\mathbf{k},\mathbf{q}) = \sum_{ij}  \left(C^{j}_{m\mathbf{k}+\mathbf{q}}\right)^* C^{i}_{n\mathbf{k}}  \left( H_{ij}^{a\kappa,(1)}(\mathbf{k},\mathbf{q}) - \epsilon_{n\mathbf{k}}S_{ij}^{a\kappa,L}(\mathbf{k},\mathbf{q}) - \epsilon_{m\mathbf{k}+\mathbf{q}}S_{ij}^{a\kappa,R}(\mathbf{k},\mathbf{q})  \right) ,
\end{equation}
furthermore, an approximation, proposed by Head-Gordan and Tully,~\cite{head-gordon+prb1992} can be applied for the term in response matrices:
\begin{equation}\label{eq:EPC_matrix_HGT}
{G}_{ij}^{a\kappa,\mathrm{HGT}}(\mathbf{k},\mathbf{q}) = \left( {H}_{ij}^{a\kappa,(1)}(\mathbf{k},\mathbf{q}) - \epsilon_F{S}_{ij}^{a\kappa,(1)}(\mathbf{k},\mathbf{q})  \right),
\end{equation}
where $\boldsymbol{H}^{(1)}$ ($\boldsymbol{S}^{(1)}$) is the first order Hamiltonian (overlap) response to the atomic coordinate perturbation, and L(R) superscripts refer to the one-sided left (right) derivatives of the overlap matrix such that $\boldsymbol{S}^{(1)}$ = $\boldsymbol{S}^{L}$ + $\boldsymbol{S}^{R}$.
In FHI-aims, this approximation is applied by default but can be switched off by the user. Its effect on the EPC matrix elements has previously been assessed.~\cite{maurer+prb2016,box+es2023} FHI-aims supports several different variations of the electronic friction tensor, by default, the following expression is used:
\begin{equation}\label{eq:friction_tensor_single_deltab}
    \Lambda_{a\kappa,a'\kappa'}^{\mathbf{q},\mathrm{Ib}}(\hbar\omega) = \pi\hbar \sum_{\sigma mn}\int \frac{d\mathbf{k}}{\Omega_{BZ}}  \tilde{g}^{a\kappa}_{mn} (\mathbf{k},\mathbf{q})   (\tilde{g}^{a'\kappa'}_{mn})^{*} (\mathbf{k},\mathbf{q})   (f_{n\mathbf{k}}-f_{m\mathbf{k}+\mathbf{q}}) \frac{\delta(\epsilon_{m\mathbf{k}+\mathbf{q}}-\epsilon_{n\mathbf{k}}-\hbar\omega)}{\epsilon_{m\mathbf{k}+\mathbf{q}}-\epsilon_{n\mathbf{k}}}.
\end{equation}
The friction tensor can be used in MDEF dynamics directly through on-the-fly evaluation~\cite{maurer+prl2017} or via machine-learning models,~\cite{box+jacsau2021,zhan+jpcc2020} or projected along normal modes (e.g calculated by ASE~\cite{larsen2017ase}) to calculate phonon lifetimes or linewidth broadenings.~\cite{maurer+prb2016,box+es2023}

The current implementation supports ScaLAPACK-type distribution of the first-order response and EPC matrices. The response matrices can be evaluated using finite-difference or DFPT (see Sec. \ref{ChapDFPT}). Collinear spin is also supported. A restart mechanism exists that enables the reading and writing of first-order response matrices to restart electronic friction calculations or to calculate the response atom by atom before evaluating the full friction tensor. The implementation is currently limited to phonon momenta of zero ($\mathbf{q}=0$), consequently, $\mathbf{q}>0$ is available only through post-processing with supercells. However, the code is highly scalable and efficient, enabling the evaluation of EPC for unit cells containing $> 250$ atoms and several $\mathbf{k}$-points. For example, Box et al., employed the code to evaluate the Brillouin-zone averaged vibrational linewidth of the internal stretch vibration of a carbon monoxide layer on a Cu(100) surface.~\cite{box+es2023} 

\rev{The current code has been benchmarked against plane-wave approaches in Ref~\cite{box+es2023}, the interested reader is directed there for a more detailed benchmark and scaling performance.}

\subsection*{Usability and Tutorials}
The aforementioned features have been comprehensively documented in the FHI-aims manual since release 240507. All keywords related to the friction calculations are prefixed with \texttt{friction\_}, with the exception of \texttt{calculate\_friction}. This keyword accepts the options \texttt{numerical\_friction} (for finite-difference evaluation of the electron-phonon coupling matrix elements) or \texttt{DFPT} (for density functional perturbation theory evaluation of the electron-phonon coupling matrix elements) and initiates the friction calculation.

An online tutorial utilizing the friction module is available. This tutorial employs the electronic friction driver to calculate electron-phonon couplings and vibrational lifetimes, including the effect of non-adiabatic effects via electronic friction for a molecule adsorbed on a metallic surface. Figure 1 presents the landing page and a brief description of this tutorial. The tutorials can be accessed at \url{https://fhi-aims-club.gitlab.io/tutorials/tutorials-overview/}.

\begin{figure}[ht]
    \centering
    \includegraphics[width=\textwidth]{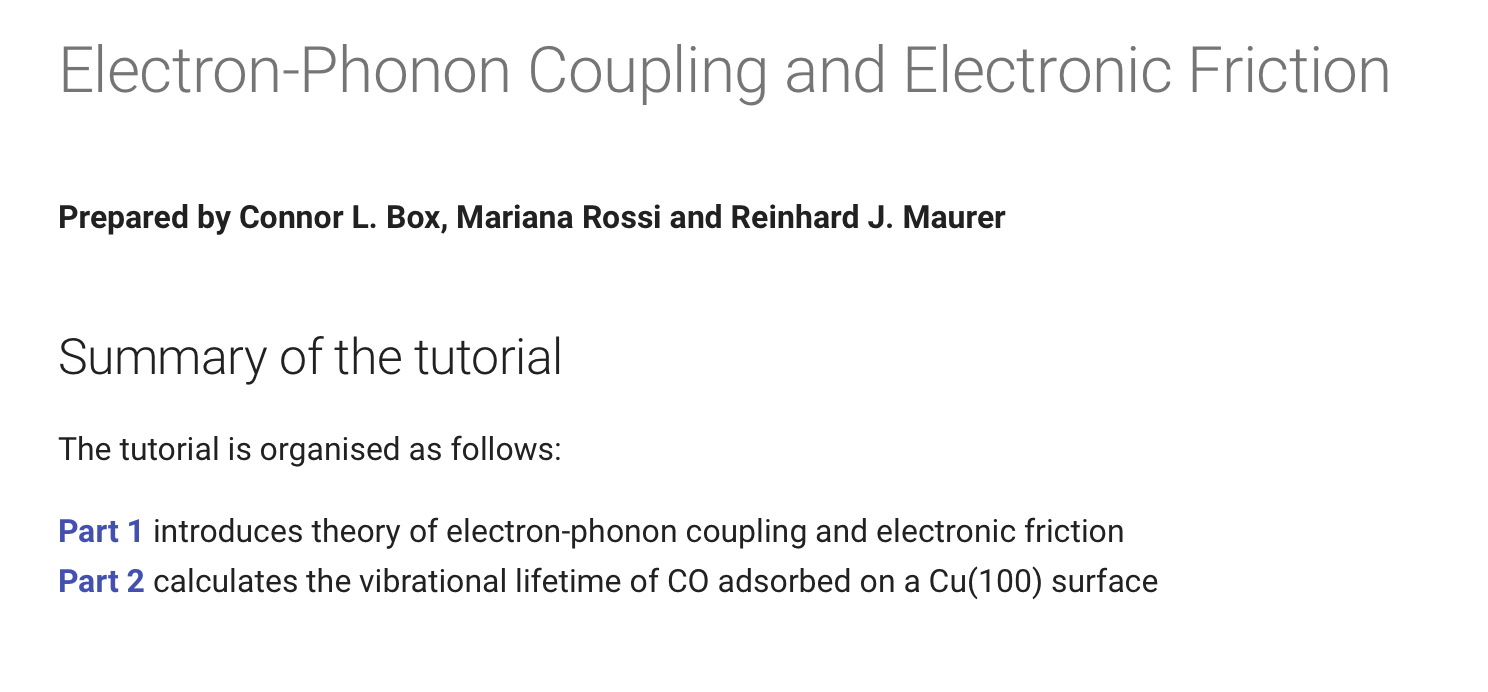}
    \caption{Landing page for electronic friction/electron-phonon coupling tutorial.}
    \label{fig:tutorial}
\end{figure}


\subsection*{Future Plans and Challenges}
Future work focused on expanding the capabilities of the code, will be aimed at expanding the properties that can be calculated, either within FHI-aims or by interfacing to other electron-phonon coupling codes such as Perturbo.~\cite{Zhou2021} For example, the calculation of the Hessian/dynamical matrix to evaluate linewidths requires post processing currently, this can be circumvented by interfacing to existing parts of FHI-aims that calculates these properties. Another example is the evaluation of the Eliashberg function, which is not yet implemented within the friction module. Support for $\mathbf{q}>0$ calculations is desirable, and some existing DFPT routines are available for both reciprocal-space and supercell approaches. 
\rev{Finally,  \texttt{use\_local\_index} is currently not supported, changing this would improve efficiency and memory usage for systems with hundreds of atoms, comparable to the effect for ground-state SCF}.

The combination of the electronic friction formalism with quantum nuclear dynamics is a topic of current research.~\cite{Litman_JCP_2022_II,litman+jcp2022a} A connection of this formalism with instanton theory to describe thermal rates of nuclear tunneling reactions has been derived and implemented in FHI-aims and i-PI.~\cite{ipi3} In the future, we plan extensions of this interface to include MDEF and new algorithms joining path-integral molecular-dynamics approximations to thermal rates and electronic friction.

\subsection*{Acknowledgements}
We acknowledge financial support through the UKRI Future Leaders Fellowship programme (MR/S016023/1and MR/X023109/1), and a UKRI frontier research grant (EP/X014088/1). We acknowledge the contribution of Eszter S. P\'os for her work in the implementation of an interface for the friction tensor output to the i-PI code.

\newpage

\section{Temperature-Dependent Electronic Spectral Functions from Band-Structure Unfolding}
\label{ChapUnfolding}

\sectionauthor[1,2]{\textbf{*Jingkai Quan}}
\sectionauthor[1,3]{Min-Ye Zhang} 
\sectionauthor[1,a]{Nikita Rybin}
\sectionauthor[4,5]{Marios Zacharias} 
\sectionauthor[3,1,b]{Xinguo Ren}
\sectionauthor[6]{Hong Jiang}
\sectionauthor[1]{Matthias Scheffler}
\sectionlastauthor[1,7,c]{\textbf{ *Christian Carbogno}}

\sectionaffil[1]{The NOMAD Laboratory at the Fritz Haber Institute of the Max Planck Society, Faradayweg 4-6, D-14195 Berlin, Germany}
\sectionaffil[2]{Max-Planck Institute for the Structure and Dynamics of Matter, Luruper Chausse 149, 22761, Hamburg, Germany}
\sectionaffil[3]{Institute of Physics, Chinese Academy of Sciences, Beijing 100190, China}
\sectionaffil[4]{Theory Department (since 1/1/2020: The NOMAD Laboratory), Fritz Haber Institute of the Max Planck Society, Faradayweg 4-6, D-14195 Berlin, Germany}
\sectionaffil[5]{Computation-Based Science and Technology Research Center, The Cyprus Institute, Aglantzia, 2121, Cyprus}
\sectionaffil[6]{Beijing National Laboratory for Molecular Sciences,  College of Chemistry and Molecular Engineering, Peking University, 100871 Beijing, China}
\sectionaffil[7]{Theory Department, Fritz Haber Institute of the Max Planck Society, Faradayweg 4-6, D-14195 Berlin, Germany}

\sectionaffil[*]{Coordinator of this contribution.}
\rule[0.25ex]{0.35\linewidth}{0.25pt}

\sectionaffil[a]{{\it Current Address:} Skolkovo Institute of Science and Technology, Bolshoi bulvar 30, build.1, 121205, Moscow, Russia}
\sectionaffil[b]{{\it Current Address:} Institute of Physics, Chinese Academy of Sciences, Beijing, 100190, China}
\sectionaffil[c]{{\it Current Address:} Theory Department, Fritz Haber Institute of the Max Planck Society, Faradayweg 4-6, D-14195 Berlin, Germany}




\subsection*{Summary}
\begin{wrapfigure}{r}{0.45\textwidth}
  \centering
    \includegraphics[width=0.45\textwidth]{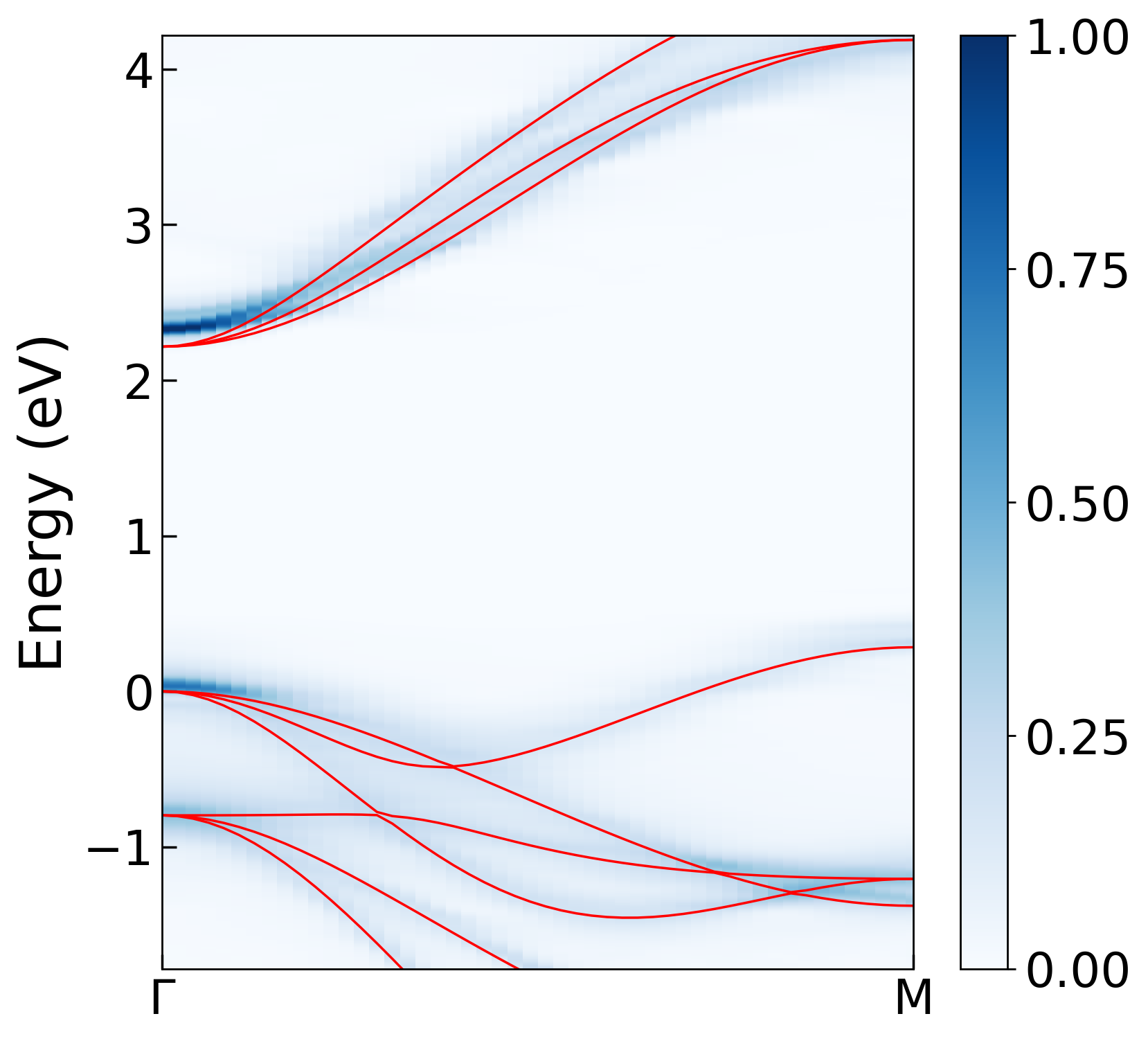}
    \caption{{Electronic spectral function at 300K and the static band structure (red lines) of SrTiO$_3$.}}
    \label{fig:Sketch1}
\end{wrapfigure}

The electronic band structure~$\varepsilon_n(\mathbf{k})$, which describes the periodic dependence of the electronic quantum states~$n$ on the lattice momentum~$\mathbf{k}$ 
in reciprocal space, 
is one of the fundamental concepts in solid-state physics. It is key to our understanding of solid-state devices, since it allows rationalizing the electronic 
properties of periodic materials,~e.g.,~to discern semiconductors with a direct band gap from those with an indirect one. In spite of that, the electronic band 
structure~$\varepsilon_n(\mathbf{k})$ is actually only well-defined for static nuclei,~i.e.,~immobile nuclei fixed at their crystallographic positions. This constitutes a severe approximation that does not even hold 
in the limit of zero Kelvin due to the quantum-nuclear zero-point motion. To account for these thermodynamics effects, the band-structure concept can be generalized
by introducing a temperature-dependent~($T$) spectral-function~$\Braket{A({\bf k}, E)}_T$, as shown in Fig.~\ref{fig:Sketch1}. In this case, the electronic quantum
states at each reciprocal-vector $\mathbf{k}$ are no longer sharp values~$\varepsilon_n(\mathbf{k})$, but are characterized by a finite-width distribution. Several
fundamental physical properties and mechanisms can only be understood and computed when the coupling between nuclear and electronic degrees of freedom is 
accounted for. For instance, this includes the temperature-dependence of key electronic properties such as the band gap~\cite{Giustino2010,E.Cannuccia2012,SPonce2014,Zacharias2020},
optical absorption spectra~\cite{Noffsinger2012,Zacharias2015}, and electronic transport coefficients~\cite{Mustafa2016,Zhou2018,Ponce2020}, also see Contrib.~\ref{ChapElecTrans}.

One possible route to compute spectral functions and the associated observables is many-body perturbation theory, a formally elegant and 
computationally efficient framework for treating electron-phonon coupling that is applicable within and beyond the Born-Oppenheimer approximation~\cite{Giustino2017}. However, both the 
dynamics of the nuclei and that of the electrons are treated approximately in these methods by using the
harmonic approximation for the nuclear degrees of freedom, also see Contribs.~\ref{ChapVibes} and~\ref{ChapHeatTransport}, and expressing the electronic response~(including
some second-order effects~\cite{Giustino2017}) in terms of linear-order electron-phonon coupling elements. These approximations may well fail at elevated temperatures and/or for mobile atoms. For instance, it has been
demonstrated that the complete concept of phonons can break down in systems exhibiting spontaneous defect formation, even if these defects are 
short-lived~\cite{Knoop2023}. To avoid potential inaccuracies from the aforementioned approximations, the electronic spectral function can be equally obtained 
in a non-perturbative fashion within the Born-Oppenheimer approximation, hence capturing higher-order couplings between electronic and vibrational degrees of freedom. 
To this end, {\it ab initio} molecular dynamics~(\textit{ai}MD\xspace) can be run to accurately sample 
the full anharmonic potential-energy surface. The spectral-function is then obtained as the 
thermodynamic average~$\Braket{A({\bf k}, E)}_T$ of the instantaneous electronic-band structures~$\varepsilon_n(\mathbf{k})$ observed during this 
dynamics~\cite{Zacharias2020a, Nery2022, Rybin2023, Quan2026}.
 
\begin{wrapfigure}{r}{0.45\textwidth}
  \centering
    \includegraphics[width=0.450\textwidth]{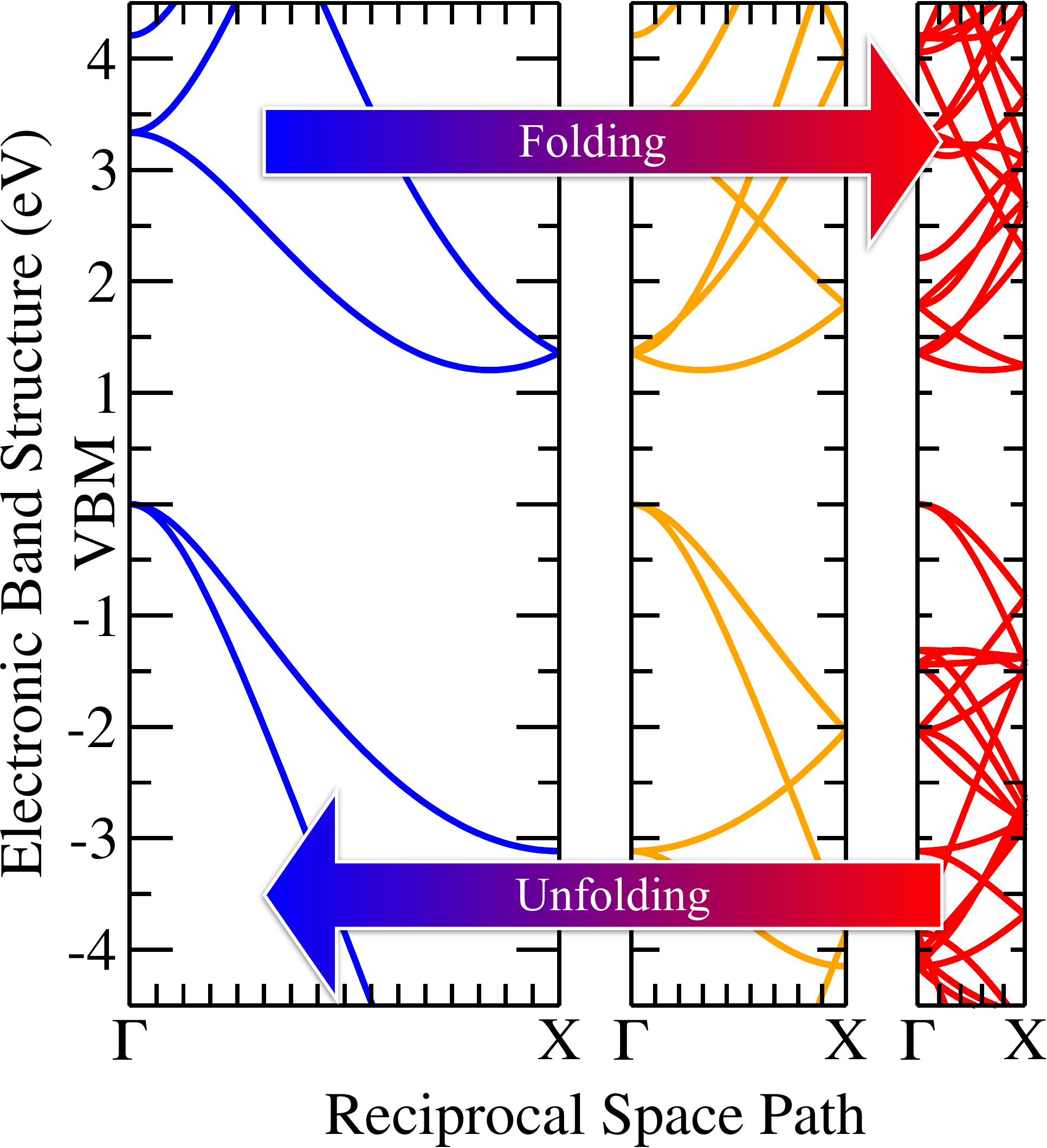}
    \caption{Folding viz. unfolding visualized for the HSE06 band structure of silicon along the high-symmetry $\Gamma\rightarrow X$~path for the primitive unit cell~(left), 
an eight-atom cubic conventional cell~(middle), and a 64-atom cubic supercell~(right).}
    \label{fig:Sketch2}
\end{wrapfigure}
From an electronic-structure point of view, one major hurdle in the implementation of the non-perturbative method arises from the fact that accurately sampling
the dynamics in solids requires extended supercells, see Contrib.~\ref{ChapVibes}. However, the electronic band-structures' topology is fundamentally connected 
to the translational symmetries of a crystal, which results in band-structure folding in supercell electronic-structure theory calculations. In this case, 
electronic states associated to different $\mathbf{k}$-vectors in the first Brillouin zone are mapped onto the same $\mathbf{K}$-vector in the reduced Brillouin zone 
when extended supercells are used in real space, see Fig.~\ref{fig:Sketch2}. Accordingly, the uttermost topological information is lost if the spectral function is computed as simple average
over folded supercell band structures. Rather, it is necessary to first recover the representation in the first Brillouin zone, a process often referred to as 
{\it unfolding}~\cite{Boykin2005,Ku2010,Popescu2012,Allen2013,Zhang2024} and visualized in Fig.~\ref{fig:Sketch2}.
In this contribution, we shortly describe the implementation of the band-structure unfolding technique in the electronic-structure theory package FHI-aims and the
updates made since its original development~\cite{Zacharias2020a}.

\subsection*{Current Status of the Implementation}

The main difficulty in the implementation of band-unfolding techniques in FHI-aims stems from the usage of a non-orthogonal, atomic-centered basis. In this numeric atomic-orbitals~(NAO) basis, 
atom displacements also affect the basis functions, so that changes in the basis set and in the overlap matrix need to be incorporated in the unfolding weight derivation~\cite{Zacharias2020a}. 
We discuss the unfolding technique by establishing a relationship between the electronic structure obtained for a primitive cell~(PC) 
and the one obtained for a supercell~(SC). Notationwise, properties associated to the PC or SC are denoted by using lower- and upper-case letters, 
respectively, and/or PC and SC subscripts when needed for clarity. Accordingly, the PC is characterized by the set of lattice vectors~$\mathbf{\underline{a}} = \left( \mathbf{a}_1, \mathbf{a}_2, \mathbf{a}_3 \right)$ 
and the supercell~(SC) by the lattice vectors~$\mathbf{\underline{A}} = \mathbf{\underline{a}}\cdot\mathbf{\underline{M}}$. Here, the lattice vectors $\bf a$ and $\bf A$ are column vectors, as by FHI-aims convention, and 
the $\mathbf{\underline{M}}$ is a non-singular matrix with integer entries, 
implying that the volume of the SC is  $m=\left\|\text{det}(\mathbf{\underline{M}})\right\|$ times larger than that of the PC. Similarly, the lattice vectors~$\mathbf{\underline{b}}$ of 
the first Brillouin zone~(BZ) associated with the PC and the ones of the reduced BZ associated with the SC are related via~$\mathbf{\underline{B}}=\mathbf{\underline{M}}^{-1}\mathbf{\underline{b}}$.
In turn, the volume of the reduced BZ is $m$ times smaller than that of the first BZ, so that $m$ different $\mathbf{k}$-vectors in the first BZ zone are mapped onto 
one and the same $\mathbf{K}$-vector in the reduced BZ. The unfolding technique reverses this mapping and re-establishes a representation in the first BZ.

To perform the mapping, the projection operator
\begin{equation}
    \mathbf{\underline{P}}_{\bf k} = \ket{{\bf k}}\bra{{\bf k}} 
\label{EQ_proj}
\end{equation}
is used. Here, $\ket{{\bf k}}$ are the eigenvectors with eigenvalue~$\exp(i \mathbf{k} \cdot \mathbf{a})$ that solve the eigenvalue 
problem\footnote{For the sake of clarity, the derivation here assumes non-degenerate eigenvalues. A discussion of the more general case including degeneracy can be found in Ref.~\cite{Quan2026}.}
\begin{equation}
\mathbf{t}\ket{\bf k} = \exp(i \mathbf{k} \cdot \mathbf{a}) \ket{{\bf k}}
\label{EQ_eval}
\end{equation}
 for the translational operators~$\mathbf{t}$ associated to the lattice vectors~$\mathbf{a}$.
With that, it is possible to obtain the weights
\begin{equation} 
 W^{\bf k}_{{\bf K}N} = \Braket{ \Psi_{{\bf K}N} | \mathbf{\underline{P}}_{\bf k} | \Psi_{{\bf K}N}} = |\Braket{{\bf k}|\Psi_{{\bf K}N}}|^2 \,,
\end{equation}
i.e.,~the contribution stemming from the subspace spanned by~$\ket{{\bf k}}$ associated with the translational symmetry of the PC to the SC wave function~$\Psi_{{\bf K}N}$.
For a ``perfect'' SC,~i.e.,~one consisting of identical PC replicas, these weights are either 0 or 1, and one can hence reconstruct the exact band structure in the first BZ 
from a SC calculation. For a disturbed SC, the weights~$W^{\bf k}_{{\bf K}N}$ become fractional, since the SC system is not invariant under PC-translations~$\mathbf{t}$.
Accordingly, the desired remapping can be obtained by summing over all SC wave functions~$\Psi_{{\bf K}N}$ and computing the electronic spectral function
\begin{eqnarray} \label{eq:spectral function}
    A({\bf k}, E) =   \sum_{ {\bf K}N} W^{\bf k}_{{\bf K}N}  \delta (E - E_{{\bf K}N}) \, .
\end{eqnarray}

In practice, the unfolding implementation in FHI-aims start from the representation of the wave function~$\Psi_{{\bf K}N} = \sum_i C_{Ni}(\mathbf{K}) \chi_i(\mathbf{K})$ in the SC,~where
 $C_{Ni}(\mathbf{K})$ denotes a Kohn-Sham expansion coefficient and $\chi_i(\mathbf{K})$ is a Bloch-like basis function build from numeric atomic orbitals. In this SC Bloch-like basis, the algebraic representation of the PC translational operator is then constructed  via $t_{ij} = \Braket{\chi_i(\mathbf{K})|\mathbf{t}|\chi_j(\mathbf{K})}$. 
Let us emphasize that these functions are not orthogonal in real space,~i.e.,~the overlap matrix~$S_{ij}(\mathbf{K}) = \Braket{\chi_i(\mathbf{K})|\chi_j(\mathbf{K})}\neq \delta_{ij}$ is not diagonal. 
Accordingly, Eq.~(\ref{EQ_eval}) becomes a generalized eigenvector  problem in this representation, also see Contrib.~\ref{SecELPA}. Its solution yields the 
eigenvalues~$\exp(i \mathbf{k} \cdot \mathbf{a})$,~i.e.,~those set of $\mathbf{k}$-points in the first BZ that this $\mathbf{K}$-point in the reduced BZ can be mapped to, and 
the set of eigenvectors~$\mathbf{k} = \sum_i F_i({\bf k})\chi_i(\mathbf{K})$ that are needed for constructing the projection operator $\mathbf{\underline{P}}_{\bf k}$ defined in Eq.~(\ref{EQ_proj}). 
With that one obtains the following formula for the weight
\begin{eqnarray} \label{eq:unfold nonorthogonal}
    W^{\bf k}_{{\bf K}N} = |\mathbf{F}^{\dagger}(\mathbf{k}) \mathbf{\underline{S}}_{\bf K}\mathbf{C}_{N}(\mathbf{K}) |^2  = |\mathbf{F}'^{\ \dagger}(\mathbf{k}) \mathbf{C}'_{N}(\mathbf{K}) |^2 \ .
\end{eqnarray}
In the last step, we have introduced the orthogonal representation $\mathbf{C}'_{N}(\mathbf{K}) = \mathbf{\underline{S}}^{1 / 2}_{\bf K} \mathbf{C}_{N}(\mathbf{K})$ and $\mathbf{F}'(\mathbf{k}) = \mathbf{\underline{S}}^{1 / 2}_{\bf K} \mathbf{F}(\mathbf{k})$ via a L{\"o}wedin transformation. Although the square root calculation~$\mathbf{\underline{S}}^{1 / 2}_{\bf K}$ constitutes a computational overhead, the latter orthogonal representation is
advantageous, since this allows to find analytical expressions of the eigenvector expansion coefficients $\mathbf{F}'({\bf k})$ at each $\mathbf{K}$-point in the reduced BZ~\cite{Quan2026}. With that
the numerical solution of the eigenvalue problem in Eq.~(\ref{EQ_eval}) is circumvented, leading to an overall speed-up of the procedure. Eventually, let us note that 
translations in three-dimensional systems form an Abelian group, so that the translations~$\mathbf{t}_\alpha$ associated to different lattice vectors~$\mathbf{a}_\alpha$ can be tackled
consecutively.

\subsection*{Usability and Tutorials}
\begin{wrapfigure}{r}{0.45\textwidth}
  \centering
    \includegraphics[width=0.45\textwidth]{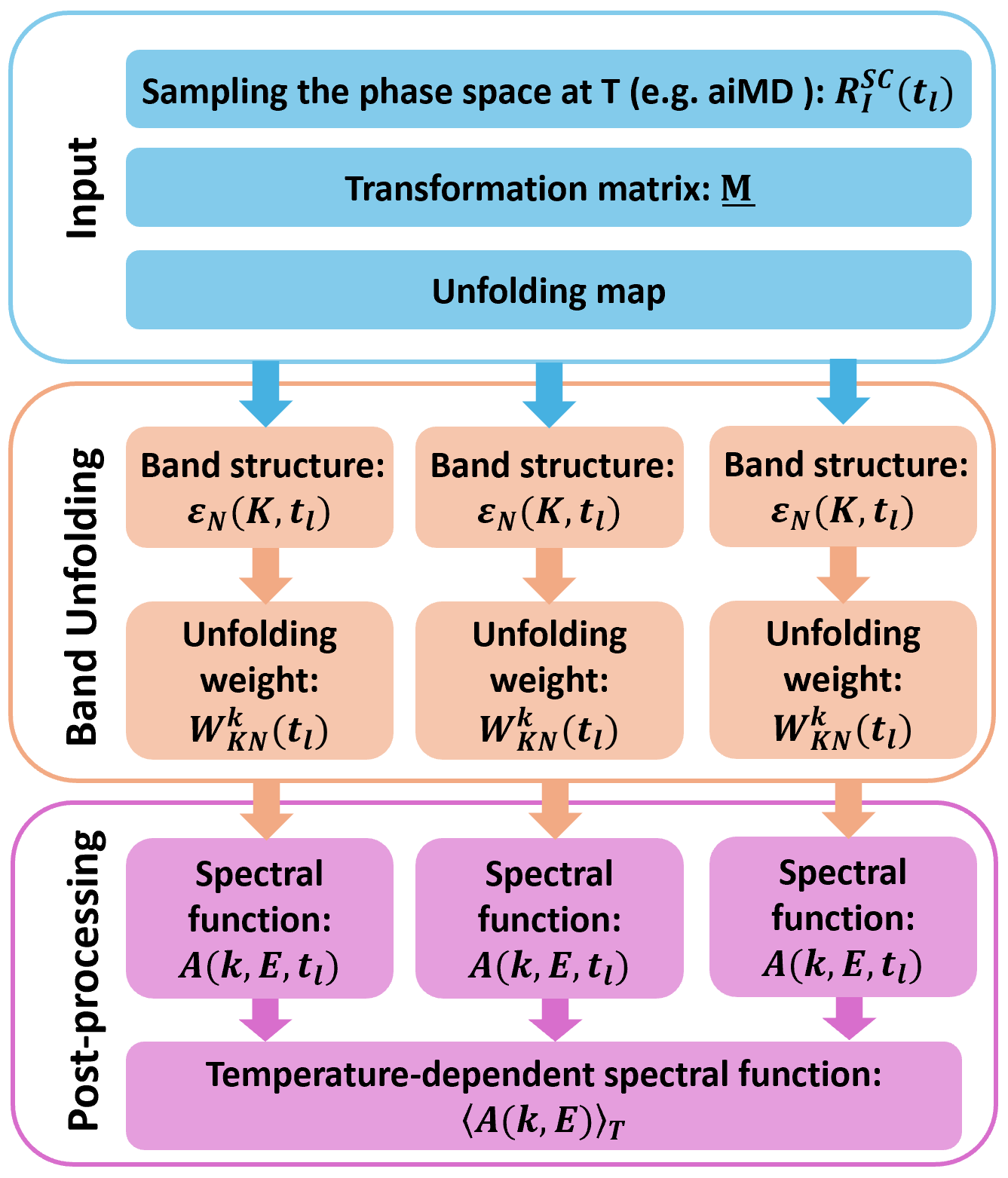}
    \caption{Schematics of a typical workflow used to obtain temperature-dependent spectral functions~$\Braket{A({\bf k}, E)}_T$.}
    \label{FIG_FLOW}
\end{wrapfigure}
The overall workflow for computing temperature-depen\-dent spectral functions~$\Braket{A({\bf k}, E)}_T$ is sketched in Fig.~\ref{FIG_FLOW}. First,
one or multiple \textit{ai}MD\xspace runs are performed in a SC at temperature~$T$ to sample the thermodynamic phase-space. From these trajectories, $L$~single, 
uncorrelated ``samples'',~i.e.,~atomic configurations~$\mathbf{R}_I^\text{SC}(t_l)$ at sufficiently distant times~$t_l$, are extracted and their 
electronic band-structure~$\varepsilon_N(\mathbf{K},t_l)$ is computed at a sufficiently dense~$\mathbf{K}$-grid. In this step, the unfolding routines
are used to map back the band-structure~$\varepsilon_N(\mathbf{K},t_l)$ onto the first BZ,~i.e.,~the weights~$W^{\bf k}_{{\bf K}N}(t_l)$ are computed
and stored on file together with $\varepsilon_N(\mathbf{K},t_l)$. In the last steps, the outputs are post-processed by computing the spectral 
functions~$A({\bf k}, E,t_l)$ of the individual samples, see Eq.~(\ref{eq:spectral function}), and by eventually computing the thermodynamic average
\begin{equation}
\Braket{A({\bf k}, E)}_T = \frac{1}{L}\sum_{l=1}^L A({\bf k}, E,t_l) \;.
\end{equation}
A more detailed description including scripts to pre- and post-process the calculations and to analyze the data are given in the tutorial at~ \url{https://fhi-aims-club.gitlab.io/tutorials/band-unfolding}. In this context, let us note that the method is not strictly restricted to \textit{ai}MD\xspace, but any kind of sampling method can be used 
to explore the relevant phase-space. For instance, path-integral MD, see Contrib.~\ChapPIMD can be used instead, if it is necessary and desirable
to account for quantum-nuclear effects. Along this lines, also more approximative methods,~e.g.,~harmonic sampling as implemented in FHI-vibes~\cite{Knoop2020a} 
and described in Contrib.~\ref{ChapVibes} or MD on machine-learned interatomic potentials as described in Contrib.~\ChapMLIP, can in principle be used. 
In such cases, however, correct spectral functions  can only be obtained if the employed approximations hold, as discussed in the respective contributions.

In more detail, the unfolding procedure in FHI-aims is part of the native band-structure post-processing tool. Accordingly it is invoked by 
 using the keyword {\tt output band} to define the reciprocal-space path for the SC,~i.e.,~the $\mathbf{K}$-points that shall be targeted and
by setting the keyword {\tt bs\_unfolding} in the \texttt{control.in}\xspace file. Additionally, two more input files are required: 
{\tt transformation\_matrix.dat}, where one defines the transformation matrix $\mathbf{\underline{M}}$ between PC lattice vector $\mathbf{\underline{a}}$ and SC lattice vector $\mathbf{\underline{A}}$,
and {\tt unfolding\_map.dat}, describing the mapping between atoms in the PC and the SC that is later used for constructing the translational operator~$\mathbf{t}$. The 
latter is an index map~$I\rightarrow j$ that relates the atomic coordinates in an unperturbed supercell~$\mathbf{R}_I^\text{SC} = \mathbf{R}_j^\text{PC} + \sum_\alpha n_\alpha \mathbf{a}_\alpha \quad n_\alpha \in \mathbb{Z}$ to 
the ones of the atoms~$\mathbf{R}_j^\text{PC}$ in a primitive cell. Upon completion, the unfolding weights for each SC $\bf K$-point are written into separate files named {\tt unfold\_k\_\#\#\#.out}
using the same \texttt{NXY}\xspace format used in the standard band structure output.

\begin{wrapfigure}{r}{0.5\textwidth}
	\centering
	\includegraphics[width=0.45\textwidth]{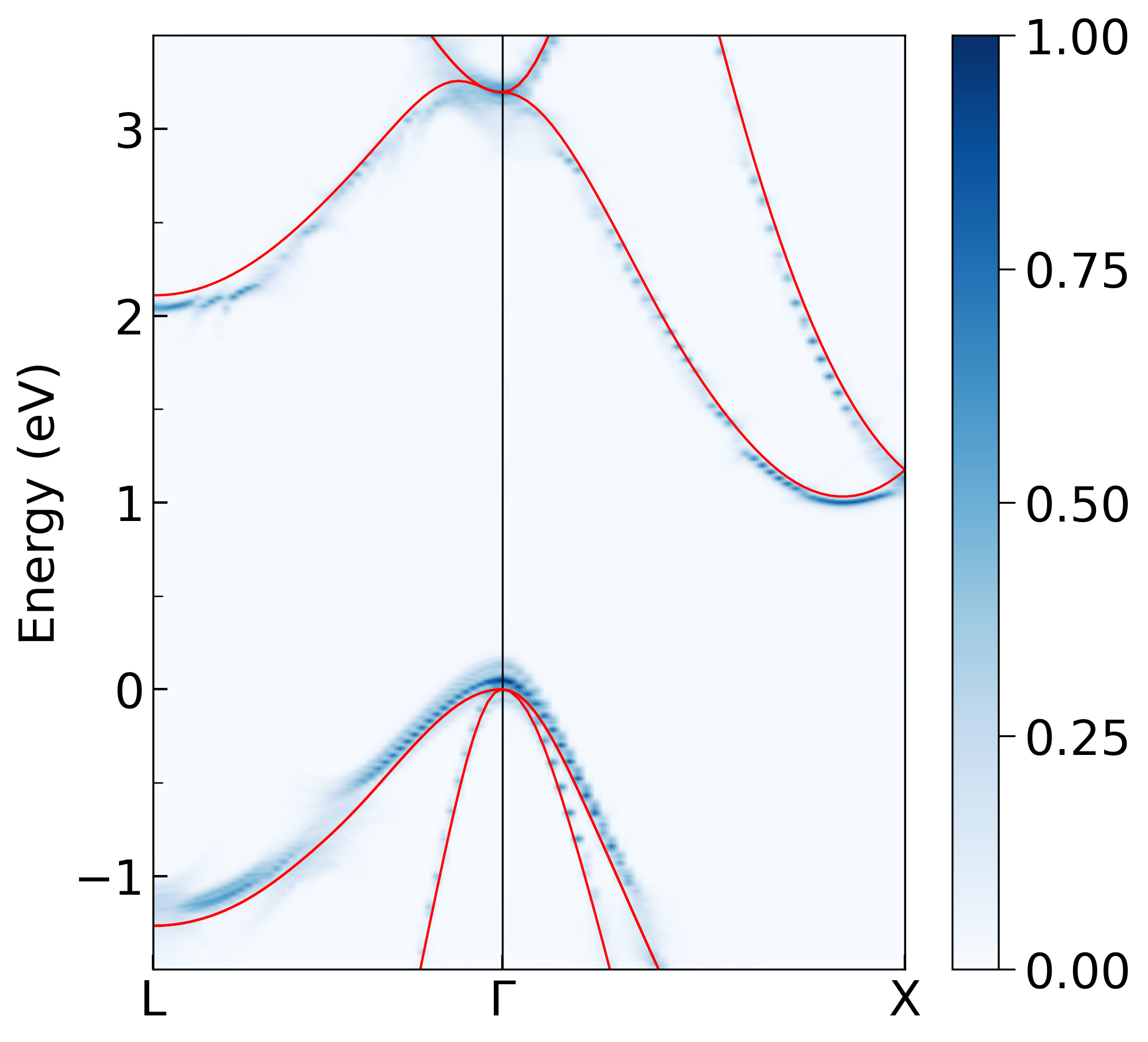}
	\caption{\label{fig:GW} $G_0W_0$ spectral function~(blue) at 300K for 64-atom silicon supercell unfolded onto the first Brillouin zone. The $G_0W_0$ band structure obtained in a PC is shown as comparison in red.}
\end{wrapfigure}
Currently, the band-unfolding implementation in FHI-aims supports all Bravais lattice, all exchange-corre\-lation functionals, and all type of transformation matrices $\mathbf{\underline{M}}$,
including non-diagonal ones typically needed to map primitive fcc and bcc structures to conventional cubic supercells. Both support for LAPACK and ScaLAPACK is implemented,
allowing for a trivially parallel parallelization over $\mathbf{K}$-points in the case of small systems requiring many $\mathbf{K}$-points~(LAPACK) and for a block cyclic distribution 
and distributed linear algebra in the case of large systems with few $\mathbf{K}$-points~({ScaLAPACK). With that, the band unfolding implementation in FHI-aims can routinely handle large 
supercells with small computational overhead. For instance, the unfolding only requires approx.~20\% of the total runtime when computing a 4,096-atom Silicon with one k-point, $> 100,000$ basis 
functions and $> 40,000$ states at the semi-local level of exchange-correlation. 

Eventually, let us note that also {\it electron-electron} coupling,~i.e.,~electronic many-body effects, can play an important role for 
spectral functions~\cite{G.Antonius2014,Karsai2018}. To this end, the current implementation also supports the unfolding of~$G_0W_0$ calculations. In this case, the complex-valued self-energy is treated as a correction to the orbital energy, while the wave function is assumed to remain as in the Kohn-Sham scheme. The unfolded spectral function can then be computed in close analogy to Eq.~(\ref{eq:spectral function}). The required
weights are obtained at the Kohn-Sham level, while the delta-function~$\delta(E-E_{\mathbf{K}N})$ is replaced by the spectral function obtained at the $G_0W_0$ level. By this means, electron correlation is explicitly accounted for. As an example, Fig.~\ref{fig:GW} 
shows the unfolded $G_0W_0$ band structure for silicon from 64-atom supercell configurations.


\subsection*{Future Plans and Challenges}
As described in this contribution, the band-structure unfolding procedures implemented in FHI-aims enable a routine assessment of temperature-dependent spectral functions both at the
density-functional theory and at the $G_0W_0$ level. From
a workflow perspective, the presented methodology is independent on the actual methodologies used (i)~for accurately sampling the thermodynamic phase-space and (ii)~for obtaining
the electronic band-structures~$\varepsilon_N(\mathbf{K})$ in the SC, as detailed above. With that, the availability of (sufficiently accurate) machine-learned interatomic potentials~\cite{Behler2021}\rev{, also see Contrib.~\ChapMLIP,} and 
of (sufficiently accurate) machine-learned electronic-structure theory~\cite{Li2022,Zhang2022,Lewis2021}, \removed{see Contribs.~\ChapMLIP and \ChapMLDENS,} can give access to much larger system sizes\rev{,~i.e.,~up to 10,000 atoms already to date,~c.f.~Contrib.~\ChapMLDENS}. 
Performance-wise, already the current implementation can handle such cases. However, access to such larger system sizes also enables the study of new physical questions,~e.g.,~the influence of point defects, 
of grain boundaries, and of solid-solid interfaces. While the overall theory holds also in such cases, the construction of the translational $\mathbf{t}$ operator will requires case-specific
adaptions to reflect the different types of breaking in translational symmetry.

Furthermore, the above-mentioned machine-learning methods might even help to solve a more fundamental open question with respect to non-adiabatic effects~\cite{Lazzeri2006}, 
which can play a fundamental role for spectral functions, as evidence from many-body perturbation theory shows~\cite{Miglio2020}. So far, however, such kind of effects could
not be studied with the herein presented non-perturbative approach, since it would require prohibitively expensive time-dependent electronic-structure theory
calculations. With the advent of such machine-learning methods, this hurdle may fall and hence allow to study the connection between anharmonic and non-adiabatic effects.

\subsection*{Acknowledgements}
MS acknowledges support by his TEC1p Advanced Grant (the European Research Council (ERC) Horizon 2020 research and innovation programme, grant agreement No. 740233.

\newpage

\chapter{Transport}
\label{ChapTransport}
\newpage

\section{Vibrational Heat Transport in
Weakly, Moderately, and Strongly Anharmonic Solids}
\label{ChapHeatTransport}

\sectionauthor[1,2,a]{\textbf{*Christian Carbogno}}
\sectionauthor[1,3]{Florian Knoop}
\sectionauthor[1,4,b]{\textbf{ *Thomas A.R. Purcell}}
\sectionauthor[1]{Shuo Zhao}
\sectionauthor[1,5,6,c]{Marcel F. Langer}
\sectionauthor[7]{Marcel H\"ulsberg}
\sectionlastauthor[1]{\textbf{ *Matthias Scheffler}}

\sectionaffil[1]{The NOMAD Laboratory at the Fritz Haber Institute of the Max Planck Society, Faradayweg 4-6, D-14195 Berlin, Germany}
\sectionaffil[2]{Theory Department, Fritz Haber Institute of the Max Planck Society, Faradayweg 4-6, D-14195 Berlin, Germany}
\sectionaffil[3]{Department of Physics, Chemistry and Biology (IFM), Link\"oping University, SE-581 83, Link\"oping, Sweden}
\sectionaffil[4]{Department of Chemistry and Biochemistry, The University of Arizona, Tucson, AZ 85721, USA}
\sectionaffil[5]{Machine Learning Group, Technische Universität Berlin, 10587 Berlin, Germany}
\sectionaffil[6]{Laboratory of Computational Science and Modeling, Institut des Matériaux, École Polytechnique Fédérale de Lausanne, 1015 Lausanne, Switzerland}
\sectionaffil[7]{Theory Department (since 1/1/2020: The NOMAD Laboratory), Fritz Haber Institute of the Max Planck Society, Faradayweg 4-6, D-14195 Berlin, Germany}

\sectionaffil[*]{Coordinator of this contribution.}
\rule[0.25ex]{0.35\linewidth}{0.25pt}

\sectionaffil[a]{{\it Current Address:} Theory Department, Fritz Haber Institute of the Max Planck Society, Faradayweg 4-6, D-14195 Berlin, Germany}
\sectionaffil[b]{{\it Current Address:} Department of Chemistry and Biochemistry, The University of Arizona, Tucson, AZ 85721, USA}
\sectionaffil[c]{{\it Current Address:} Laboratory of Computational Science and Modeling, Institut des Matériaux, École Polytechnique Fédérale de Lausanne, 1015 Lausanne, Switzerland}




\subsection*{Summary}

Macroscopic heat transport in materials is characterized by the pressure- and temperature-de\-pen\-dent thermal conductivity tensor~$\mathbf{\kappa}(T,p)$, which relates
heat flux~$\mathbf{J}$ and temperature gradient~$\mathbf{\nabla}T$ in Fourier's law $\mathbf{J}=-\mathbf{\kappa}(T,p)\mathbf{\nabla}T$. For insulators and semiconductors, the dominant 
contribution to~$\mathbf{\kappa}$ stems from the vibrational motion of the atoms~\cite{Ashcroft1976}. However, the harmonic phonon approximation that is commonly used 
for describing the motion in solids, see Contrib.~\ChapVibes, yields an infinite thermal conductivity by construction and is hence insufficient to capture heat transport, as 
first shown by Peierls~\cite{RPeierls1929}. To correctly model and compute thermal conductivities, it is thus quintessential to take into account anharmonic 
effects,~i.e.,~deviations of the potential-energy surface~(PES) from the approximative parabolic PES introduced in Contrib.~\ChapVibes.

From a bird's eye view, two different approaches to compute vibrational thermal conductivities exist and both are supported in \FHIaims. 
While {\it perturbative} methods inherently assume the validity of a harmonic quasi-particle picture, {\it non-perturbative} methods overcome 
this approximation and hence account for strongly anharmonic effects,~e.g.,~defect formation and the exploration of multiple minima on the PES~\cite{Knoop2023}.
In {\it perturbative} methods, the PES is approximated by a truncated Taylor expansion. The terms up to second order,~i.e.,~the {\it harmonic} 
approximation, are used to analytically describe the nuclear motion in terms of phonons with 
frequencies~$\omega_s(\mathbf{q})$, in which $s$~is the mode index and $\mathbf{q}$ the reciprocal-space wave vector, see Contrib.~\ChapVibes. Higher-order terms describing anharmonicity are accounted for
via perturbation theory under the assumption that they only make up for a minor correction of the PES. Conversely, {\it non-perturbative} methods do not make any approximations 
about the PES and hence require an explicit simulation of the nuclear dynamics,~e.g.,~via {\it ab initio} molecular dynamics~(\textit{ai}MD\xspace),~cf.~Contrib.~\ChapAIMD.

\begin{wrapfigure}{r}{0.45\textwidth}
  \centering
    \includegraphics[width=0.45\textwidth]{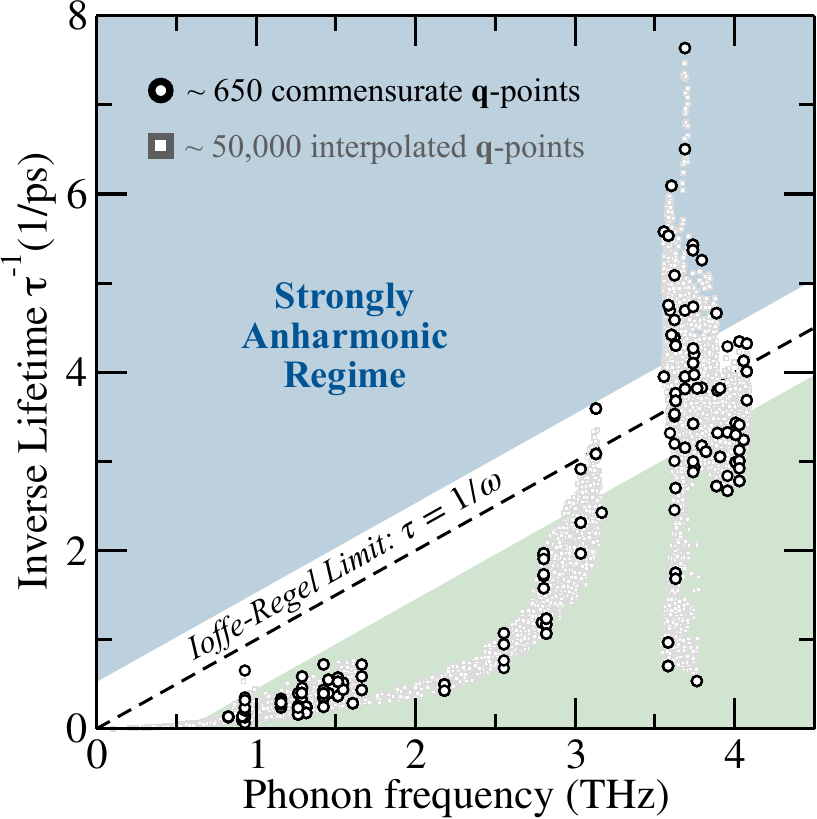}
    \caption{Phonon lifetimes obtained from fully anharmonic {\it ab initio}MD, adapted from Ref.~\cite{Knoop2023a}. The dashed line represents
    the Ioffe-Regel limit that qualitatively separates the quasi-particle regime (light green) from the strongly anharmonic regime.
The quasi-particle picture becomes increasingly questionable when the lifetimes are getting close to $\omega^{-1}$ and breaks
down in the strongly anharmonic regime, for which the Ioffe-Regel limit is violated.}
    \label{fig:Ioffe}
\end{wrapfigure}
The validity of the quasi-particle picture and, with that, of the applicability of a perturbative approach can be rationalized by comparing phonon frequencies~$\omega_s(\mathbf{q})$ and
lifetimes~$\tau_s(\mathbf{q})$ with respect to the Ioffe-Regel limit~\cite{Ioffe1960}, as exemplified in Fig.~\ref{fig:Ioffe}. Perturbative approaches are no longer applicable in 
the overdamped,  strongly anharmonic regime, in which the lifetimes become comparable to or even shorter than a single oscillation period~$\tau_s(\mathbf{q})\ll \omega_s^{-1}(\mathbf{q})$,
because phonons are non longer well defined quasi-particles. Conversely, the perturbative phonon picture holds when the lifetimes span several oscillation periods~$\tau_s(\mathbf{q})\gg \omega_s^{-1}(\mathbf{q})$. Although the Ioffe-Regel
limit does not constitute a strict quantitative rule, it is qualitatively helpful in identifying different transport regimes. Furthermore, similar considerations allow to identify particle-like~(Peierls)
and wave-like~(Wigner) transport regimes within the quasi-particle regime~\cite{Simoncelli2019}.

\subsection*{Current Status of the Implementation}
Both {\it perturbative} approaches  and {\it non-perturbative} approaches are supported in \FHIaims for the calculation of vibrational thermal conductivities, 
whereby FHI-vibes~\cite{Knoop2020a} typically serves as the tool to setup, run, and evaluate the calculations.

{\it Perturbative} approaches start from the harmonic approximation, a second-order Taylor expansion of the PES, see Contrib.~\ChapVibes.
By analytically treating the dynamics in this harmonic potential, quantum-nuclear effects can be accounted for, so that perturbative approaches are applicable also in the low-temperature regime. 
Anharmonic deviations from the harmonic PES are treated as a minor perturbation. 
They are accounted for by determining at least the third-order~\cite{Broido2007}, sometimes also the fourth-order Taylor expansion of the PES~\cite{Feng2016,Ravichandran2018}. In turn, 
this allows one to determine scattering-cross-sections viz.~phonon-lifetimes~$\tau_s(\mathbf{q})$ that are the final ingredient for obtaining $\mathbf{\kappa}$ by solving the Boltzmann-Peierls transport equation. Let us note that
this formalism also allows one to account for isotopic disorder~\cite{Garg2011} as well as glass-like Wigner-transport beyond Peierls' particle-like transport~\cite{Simoncelli2019}.
Similarly, higher-order anharmonic effects can be approximatively accounted for by extracting temperature-dependent force-constants from finite-temperature \textit{ai}MD\xspace~\cite{KEsfarjani2008};
detailed discussions on practical implementations can be found in Refs.~\cite{Hellman2013,TTadano2014,Eriksson2019,Monacelli2021,Roekeghem2021}.

{\it Non-perturbative} approaches rely on explicit simulations of the nuclear motions via classical {\it ab initio} molecular dynamics~(\textit{ai}MD\xspace),~cf.~Contrib.~\ChapAIMD. Then, fully anharmonic thermal conductivities 
can be obtained from non-equilibrium \textit{ai}MD\xspace trajectories,~e.g.,~by explicitly simulating a thermal gradient~\cite{SStackhouse2010}, by imposing a heat flux~\cite{MuellerPlathe1997}, by monitoring 
equilibration from non-equilibrium~\cite{Gibbons2009} or by imposing close-to-equilibrium temperature profiles~\cite{Puligheddu2017}. 
However, such non-equilibrium approaches are typically plagued by poor convergence with respect to simulation time and cell size. For
this reason, it is generally advantageous to compute thermal conductivities from equilibrium \textit{ai}MD\xspace trajectories by leveraging the fluctuation-dissipation 
theorem,~e.g.,~by monitoring energy-density fluctuations~\cite{Cheng2020,Drigo2023} or energy-current fluctuations via the Green-Kubo formalism~\cite{Kubo1957}. 

In the following, we shortly summarize the implementations and workflows available for the perturbative and non-perturbative ansatz.


\subsubsection*{Perturbative Approaches:}
\begin{wrapfigure}{l}{0.4\textwidth}
  \centering
    \includegraphics[width=0.38\textwidth]{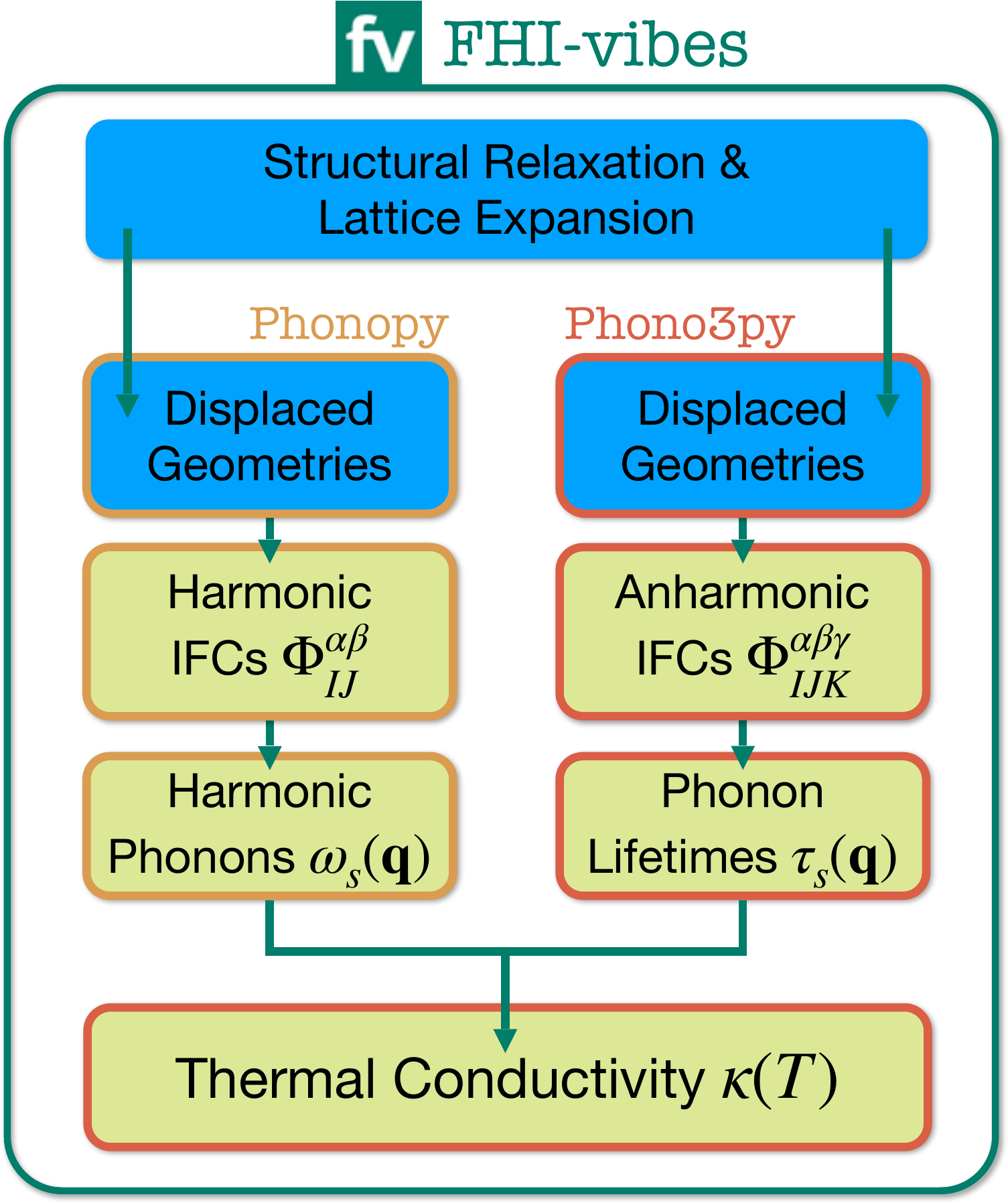}
    \caption{Workflow for computing vibrational thermal conductivities~$\kappa$ using a perturbative approach, for which interatomic force constants~(IFCs) are obtained from finite differences. Blue cells highlight the steps in which \FHIaims is used for electronic-structure theory calculations to obtain,~e.g,~forces.} 
    \label{fig:workflow1}
\end{wrapfigure}
A typical \FHIaims workflow for the calculation of thermal conductivities in a perturbative approaches starts with a geometry relaxation to obtain the equilibrium geometry~$\{\mathbf{R}^0\}$
and a subsequent \textit{harmonic} phonon calculation to obtain the properties of the phonons with frequencies~$\omega_s(\mathbf{q})$ with the Python package FHI-vibes, 
see Contrib.~\ChapVibes for details. Subsequently, third-order derivatives of the PES, the so-called third-order force constants~$\Phi_{IJK}^{\alpha\beta\gamma}$ need to be computed.
They are defined as
\begin{equation}
\Phi_{IJK}^{\alpha\beta\gamma} = \left.\frac{\partial^3 U\left(\{\mathbf{R}\}\right) }{\partial R_{I\alpha} \partial R_{J\beta} \partial R_{K\gamma}}\right\vert_{\{\mathbf{R}\}=\{\mathbf{R}^0\}} \;,
\end{equation}
i.e.,~by extending the Taylor expansion introduced for the definition of the harmonic force constants~$\Phi_{IJ}^{\alpha\beta}$ given in Contrib.~\ChapVibes. As shown in Fig.~\ref{fig:workflow1},
the computation of $\Phi_{IJK}^{\alpha\beta\gamma}$ via finite differences can be orchestrated via FHI-vibes, which (i)~generates geometries with symmetry-reduced displacements using~Phono3py~\cite{Togo2015},
(ii)~evaluates the forces for these structures via~\FHIaims, and (iii)~postprocesses these calculations with~Phono3py to determine the $\Phi_{IJK}^{\alpha\beta\gamma}$.
Eventually, the thermal conductivity is obtained via FHI-vibes viz.~Phono3py by computing the phonon lifetimes~$\tau_s(\mathbf{q})$ using the $\Phi_{IJK}^{\alpha\beta\gamma}$, so to evaluate 
the Boltzmann transport equation. Besides using $\Phi_{IJ}^{\alpha\beta}$ and $\Phi_{IJK}^{\alpha\beta\gamma}$, FHI-vibes also supports the usage of temperature-renormalized force constants, which are obtained via 
linear regression from thermodynamic snapshots generated,~e.g.,~by \textit{ai}MD\xspace. To this end, FHI-vibes offers an interface~(\texttt{vibes utils trajectory 2tdep}\xspace) to the 
temperature-dependent effective potentials method~TDEP described in Refs.~\cite{Hellman2013,Hellman2014,Knoop2024}.

\subsubsection*{Non-Perturbative Approaches:}
For computing thermal conductivities in a {\it non-perturbative} fashion with \FHIaims, the method of choice is the {\it ab initio} Green-Kubo method~\cite{Carbogno2017}. 
In this formalism, it is necessary to run \textit{ai}MD\xspace simulations and to compute the heat flux~$\mathbf{J}(t)$ along the trajectory. The heat-flux implementation in \FHIaims~\cite{Carbogno2017}
supports the calculation of {\it virial} heat fluxes,~i.e.,~the dominant contribution in solids in which mass transport is negligible~\cite{Howell2012}. To this end,
the contributions of each atom~$I$ to the virial~$\mathbf{\underline{\sigma}}_I(t)$ needs to be computed, details regarding the evaluation of these terms can be found in Refs.~\cite{knuth2015all,Carbogno2017}.
In practice, such Green-Kubo calculations are best orchestrated with FHI-vibes, which takes care of running \textit{ai}MD\xspace with \FHIaims and storing all relevant information 
including positions~$\mathbf{R}_I(t)$, velocities~$\dot{\mathbf{R}}_I(t)$, and virials~$\mathbf{\underline{\sigma}}_I(t)$ along the trajectory, as shown in Figure~\ref{fig:workflow2}. 

\begin{wrapfigure}{r}{0.45\textwidth}
  \centering
    \includegraphics[width=0.43\textwidth]{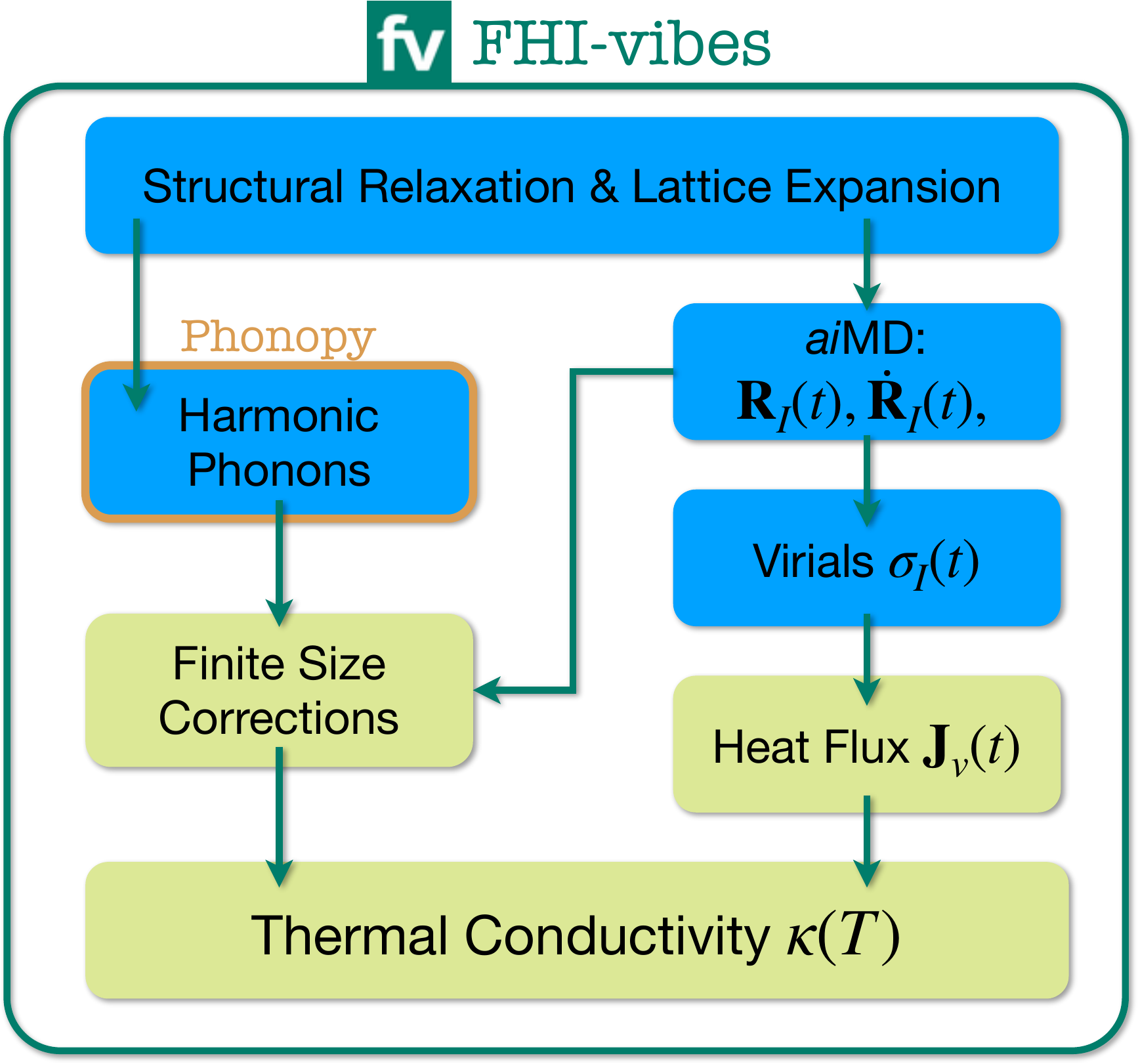}
    \caption{Workflow for computing vibrational thermal conductivities~$\kappa$ via the non-perturbative Green-Kubo approach. Blue cells highlight the steps in which \FHIaims is used for electronic-structure theory calculations to obtain,~e.g,~forces.} 
    \label{fig:workflow2}
\end{wrapfigure}

During postprocessing, FHI-vibes then also takes care to assemble the virial heat flux~$\mathbf{J}_v(t)={V}^{-1} \sum_I \mathbf{\underline{\sigma}}_I(t) \cdot \mathbf{\dot{R}}(t)$, to remove
non-contributing terms~\cite{Ercole2016}, to reduce noise~\cite{Knoop2023a}, and to eventually evaluate the thermal conductivity~$\kappa$ by integrating the heat-flux
autocorrelation function~\cite{Kubo1957}. In the last step, a finite-size correction can be efficiently  applied by mapping the dynamics onto a harmonic model~$\omega_s(\mathbf{q})$,
extracting the associated fully-anharmonic phonon lifetimes~$\tau_s(\mathbf{q})$ from \textit{ai}MD\xspace, and then extrapolating to the bulk viz.~interpolating to the dense~$\mathbf{q}$ limit
in a BTE-type model. This last step requires force constants~$\Phi_{IJ}^{\alpha\beta}$ for the harmonic model, which can be computed at a harmonic level or also extracted from the \textit{ai}MD\xspace via FHI-vibes.

\subsection*{Usability and Tutorials}
From an electronic-structure point of view, {\it perturbative} approaches based on finite differences only require the evaluation of forces, cf.~Fig.~\ref{fig:workflow1}.
Accordingly, they can be used with all levels of theory for which \FHIaims is able to compute forces, notable all semi-local exchange-correlation functionals and 
hybrids, which include a fraction of Fock exchange, see Contrib.~\ChapHybrids. Similarly, the Green-Kubo approach requires the evaluation of forces for performing 
the \textit{ai}MD\xspace simulations and of atomic stresses for evaluating the virial heat flux. For the latter, hybrid functionals are not yet supported, but all necessary ingredients
to evaluate the virials are already implemented also for these cases~\cite{knuth2015all}.

From a practical point of view, the computation of thermal conductivities with perturbative and non-pertur\-bative approaches does, however, require quite complex workflows
and the (system-dependent) convergence of several physical and numerical parameters beyond those important at the electronic-structure theory level. To just name a few, this
includes computing accurately relaxed equilibrium structures within the symmetry group of interest, determining lattice expansion at finite temperatures, and converging the
supercell size used for capturing the interactions relevant for the nuclear dynamics,~i.e.,~for obtaining harmonic and anharmonic force constants~$\Phi_{IJ}^{\alpha\beta}$ and~$\Phi_{IJK}^{\alpha\beta\gamma}$ in
perturbative approaches and for running \textit{ai}MD\xspace simulations in non-perturbative approaches. To guide a novel user through this process, a set of tutorials for the computation
of thermal conductivities with perturbative and non-perturbative methods are provided at~\url{https://gitlab.com/FHI-aims-club/tutorials/thermal-transport-with-fhi-vibes}. 
The tutorials build on the ones provided for the analysis of harmonic 
and anharmonic vibrations with FHI-vibes, cf.~Contrib.~\ChapVibes, and showcase how to compute thermal conductivities using CuI, a strongly anharmonic material~\cite{Knoop2023}, 
as an example. Besides introducing the necessary physical background knowledge, the tutorials provide useful scripts and teach how to use features and functions of FHI-vibes 
to analyze, inspect, and evaluate this kind of calculations.


\subsection*{Future Plans and Challenges}
As described in this contribution, substantial functionality for computing vibrational thermal conductivities~$\kappa(T)$ from first principles is provided within \FHIaims 
and FHI-vibes. This enables to accurately calculate $\kappa(T)$ both for weakly anharmonic materials, but also for moderately and strongly anharmonic cases~\cite{Knoop2023}, for which 
approaches based on the phonon picture are uncertain or even not justified~\cite{Simoncelli2019}. Clearly, further accelerating such kind of calculations to enable a rapid screening of
material space is an important future step. Here, replacing the computationally involved \textit{ai}MD\xspace with simulations performed with MD run with machine-learned interatomic 
potentials~(MLIPs)~\cite{Behler2021}, also see Contrib.~\ChapMLIP, can be extremely beneficial. First steps in this direction have already been taken~\cite{Langer2023,Langer2023Stress,kang2025accelerating}.
To provide accurate $\kappa(T)$ values, it is quintessential that the trained MLIP accurately captures the actuating (strongly) anharmonic effects. This is not guaranteed 
by standard training approaches, but can be facilitated by active-learning schemes~\cite{kang2025accelerating}. Let us note that the infrastructure provided within FHI-vibes 
for thermal-conductivity calculations and for gauging the degree of anharmonicity~\cite{Knoop2020} has also been proven useful in this MLIP context~\cite{Langer2023,Langer2023Stress,kang2025accelerating}. 

Eventually, it is also important to stress that the thermal conductivity of single crystals is certainly a fundamental quantity, but not the only one of interest in materials
science. Here, investigating the role of disorder at different scales, be it mass disorder~\cite{Garg2011}, structural disorder as in glasses~\cite{Simoncelli2019}, or 
interfacial resistance~\cite{Chen2022}, is a very active field of research not yet covered by the implementations so far.

\subsection*{Acknowledgements}
MS acknowledges support by his TEC1p Advanced Grant (the European Research Council (ERC) Horizon 2020 research and innovation programme, grant agreement No. 740233.. FK acknowledges support from the Swedish Research Council (VR) program 2020-04630, and the Swedish e-Science Research Centre (SeRC).

\newcommand{\FF}[1]{{\textcolor{blue}{\bf FF: #1}}}





\newpage

\section{Electrical Transport for Anharmonic Materials}
\label{ChapElecTrans}

\sectionauthor[1,2]{\textbf{*Jingkai Quan}}
\sectionauthor[1]{Florian Fiebig}
\sectionauthor[1,a]{Karsten Rasim}
\sectionauthor[3]{Bjoern Bieniek}
\sectionauthor[1,b]{Yi Yao}
\sectionauthor[4]{Roman Kempt}
\sectionauthor[1,c]{Zhenkun Yuan}
\sectionauthor[1]{\textbf{ *Matthias Scheffler}}
\sectionlastauthor[1,5,d]{\textbf{ *Christian Carbogno}}

\sectionaffil[1]{The NOMAD Laboratory at the Fritz Haber Institute of the Max Planck Society, Faradayweg 4-6, D-14195 Berlin, Germany}
\sectionaffil[2]{Max-Planck Institute for the Structure and Dynamics of Matter, Luruper Chausse 149, 22761, Hamburg, Germany}
\sectionaffil[3]{Theory Department (since 1/1/2020: The NOMAD Laboratory), Fritz Haber Institute of the Max Planck Society, Faradayweg 4-6, D-14195 Berlin, Germany}
\sectionaffil[4]{Technische Universit\"at Dresden, Theoretische Chemie, 01069 Dresden, Germany}
\sectionaffil[5]{Theory Department, Fritz Haber Institute of the Max Planck Society, Faradayweg 4-6, D-14195 Berlin, Germany}

\sectionaffil[*]{Coordinator of this contribution.}
\rule[0.25ex]{0.35\linewidth}{0.25pt}

\sectionaffil[a]{{\it Current Address:} Miele \& Cie. KG, Gütersloh, Germany}
\sectionaffil[b]{{\it Current Address:} Molecular Simulations from First Principles e.V., D-14195 Berlin, Germany}
\sectionaffil[c]{{\it Current Address:} Thayer School of Engineering, Dartmouth College, Hanover, NH 03755, USA}
\sectionaffil[d]{{\it Current Address:} Theory Department, Fritz Haber Institute of the Max Planck Society, Faradayweg 4-6, D-14195 Berlin, Germany}




\subsection*{Summary}

Electrical transport plays a pivotal role for a multitude of scientific and industrial applications.  It critically depends on the chemical and structural details of the materials, its doping, as well as
temperature and pressure. From the conductivity, one can derive crucial thermoelectric quantities for heat transport and energy generation. Similarly, the coupling of mechanical motion and electronic transport gives access to electromechanical properties.
For instance, band-type conductivity is at the very heart of semiconductor technology. While conductivities can be estimated~\cite{Scheidemantel2003}
from electronic band-structures and the associated group velocities, an accurate assessment requires accounting for the coupling
of electronic and nuclear degrees of freedom. 
Commonly, electronic-structure theory and many-body perturbation theory~\cite{Giustino2017} are used to describe electron-phonon coupling and to compute 
band-type mobilities limited by phonon scattering from  first principles~\cite{Mustafa2016,Ponce2018,Ponce2020}. In this approach, the nuclear dynamics 
is approximated by an (effective~\cite{Zhou2018a}) harmonic potential,~i.e.,~phonons, and their first-order coupling to the electronic degrees of freedom. 
In turn, this allows for the calculation of scattering cross-sections that can be fed to transport equations,~e.g.,~the Boltzmann~\cite{Ponce2020} and/or the Wigner formalism~\cite{Cepellotti2021}. 
The perturbative approach works well for harmonic materials~\cite{Ponce2018}, but the aforementioned approximations may fail in anharmonic materials and/or at elevated temperatures. 
For instance, it has been shown that the temperature-dependence of the mobility in the strongly anharmonic perovskites SrTiO$_3$ and BaTiO$_3$ is driven by higher-order anharmonic and electron-vibrational couplings~\cite{Quan2024}. Similarly, thermoelectric materials are known to exhibit strong anharmonicity~\cite{Knoop2020,Knoop2023} that cannot be described within harmonic models and that are
even challenging to describe with machine-learned potentials~\cite{kang2025accelerating}. To capture these effects and to achieve a general, non-perturbative evaluation of mobilities, 
the Kubo-Greenwood formalism~\cite{Greenwood1958} has been implemented in \FHIaims.

\subsection*{Current Status of the Implementation}
\begin{figure}[!htb]
  \centering
  \includegraphics[width=0.9\textwidth]{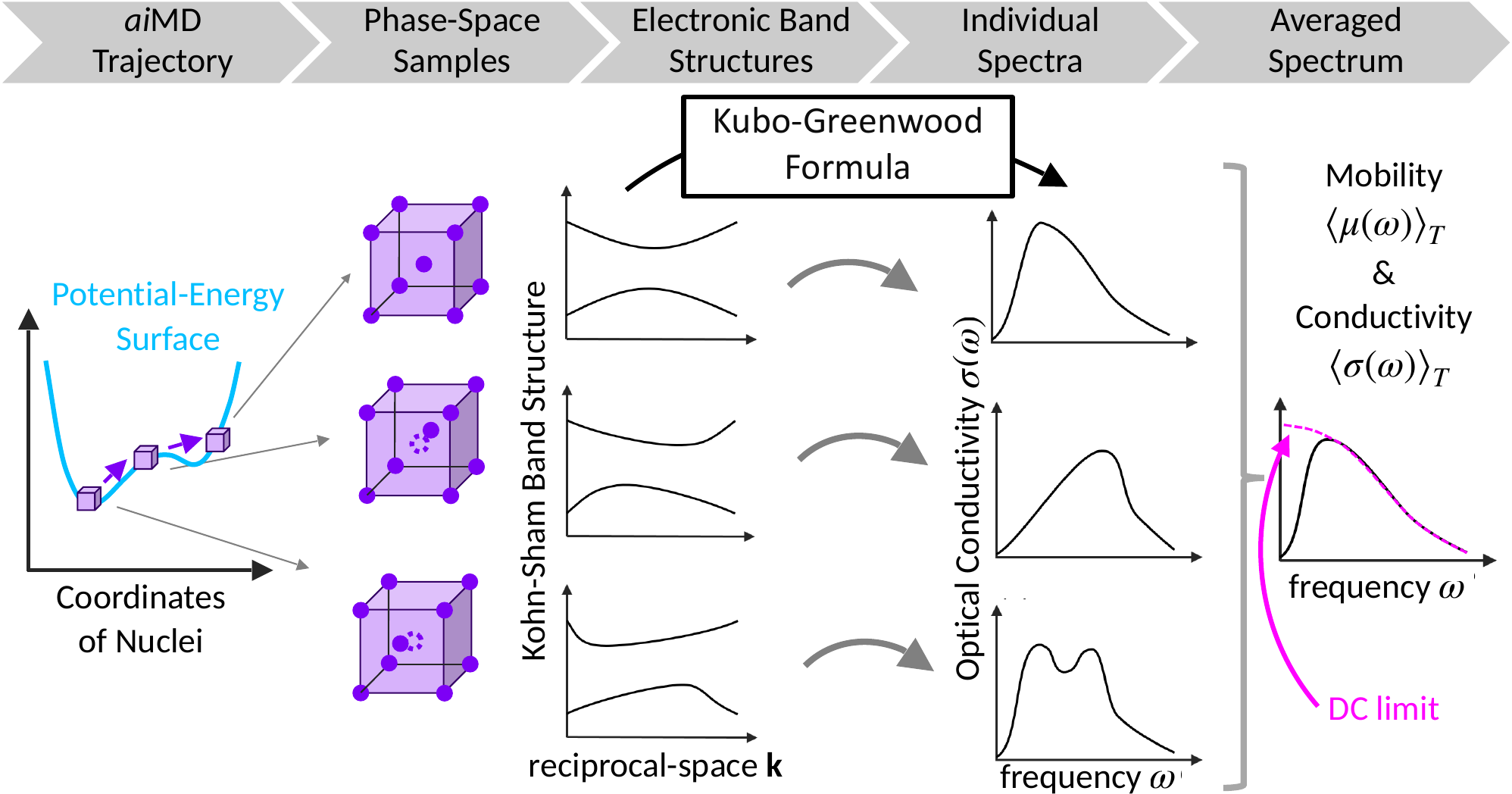}
  \caption{Schematic workflow of a Kubo-Greenwood calculation. For illustration, three snapshots of a bcc crystal structure, in which only the central atom moves, are shown as example. Similarly, the effects on the band structure 
   and on the optical conductivity are exaggerated for the sake of clarity.}
  \label{fig:workflow-electrical-transport}
\end{figure}

A typical workflow for the evaluation of mobilities with the Kubo-Greenwood formalism is shown in Fig.~\ref{fig:workflow-electrical-transport}.
In a first step, an {\it ab initio} molecular-dynamics~(\textit{ai}MD\xspace) simulation is run on the self-consistent potential-energy surface to account for all orders of anharmonic effects,~c.f.~Contrib.~\ChapAIMD. 
In a second step, uncorrelated geometry samples are chosen from this trajectory to cover the phase-space at the thermodynamic conditions of interest.
In a third step, the ground state electronic structure is determined self-consistently for each of these samples, which accounts for all orders of adiabatic electron-vibrational coupling. 
In a fourth step, the obtained electronic-structures are used to compute the frequency-dependent conductivity, typically referred to as optical conductivity~$\sigma(\omega)$ for the individual samples. 
In the final step, the spectra $\sigma(\omega)$ are post-processed to determine the mobilities~$\mu(\omega)$.

By using linear response theory in the Kubo formalism~\cite{Kubo1957} and expressing the band-like current operator in terms of effective, single-particle eigenstates~$\ket{{\textbf k}m}$ one obtains the following Kubo-Greenwood~(KG) formula for the real, diagonal part of the frequency-dependent~($\omega$) optical conductivity for one sample of phase space~\cite{Greenwood1958,Holst2011}:
\begin{equation}
\label{kg-formula}
        \Re(\sigma_{\alpha\alpha}(\omega)) = \frac{2\pi q^2\hbar^2}{m^2V\omega}\sum_{\textbf{k}mn} 
    (\Braket{\textbf{k}n|\nabla_\alpha|\textbf{k}m}\Braket{\textbf{k}m|\nabla_\alpha|\textbf{k}n})
    (f_{\textbf{k}n}-f_{\textbf{k}m}) \delta(\epsilon_{\textbf{k}m}-\epsilon_{\textbf{k}n}-\hbar\omega) \;.
\end{equation}
Here, $\alpha$ denotes a Cartesian axis, $f_{ \textbf{k}m }$ and $\epsilon_{ \textbf{k}m }$ are the occupation number and the energy eigenvalue of eigenstate $\ket{ \textbf{k}m }$, 
$\Braket{\textbf{k}m|\nabla_\alpha|\textbf{k}n}$ are the gradient matrix elements in this basis, and $V$ is the cell volume.  
The electron~$\text{e}$ and hole~$\text{h}$ conductivities~$\sigma_{\text{e/h}}(\omega)$ can be obtained by limiting the summation over~$m$ to states in the conduction and valence band, respectively.
In the \FHIaims implementation, the matrix elements~$\Braket{\textbf{k}m|\nabla_\alpha|\textbf{k}n}$ in 
Kohn-Sham space are obtained via  
\begin{equation}
    \Braket{\textbf{k}m|\nabla_\alpha|\textbf{k}n} = \sum_{ij} [C_{m}^{i}(\textbf{k})]^*C_{n}^j(\textbf{k}) \sum_{\textbf{N}} e^{i\textbf{k}\cdot\bf{T( \textbf{N})}}
\Braket{\phi_{i\bf{0}}(r)|\nabla_\alpha|\phi_{j\textbf{N}}(r)} \ ,
\label{EQmatelem}
\end{equation}
where $C_{n}^j(\textbf{k})$ denotes a Kohn-Sham expansion coefficient, $T(\textbf{N})$ a translation by a combination of lattice vectors, and $\textbf{k}$ a reciprocal-space vector. 
More details on the used notation can be found in Contrib.~\ChapFundamentals. 
The real-space gradient matrix elements 
$\Braket{\phi_{i\bf{0}}(\mathbf{r})|\mathbf{\nabla}|\phi_{j\textbf{N}}(\mathbf{r})}$ in the basis 
of the numeric atom-centered orbitals~$\phi_{j\textbf{N}}(r)$ are computed once using the  
integration scheme discussed in Ref.~\cite{Havu2009} and then stored as sparse matrix. 

\begin{figure}[!htb]
  \centering
    \includegraphics[width=0.9\textwidth]{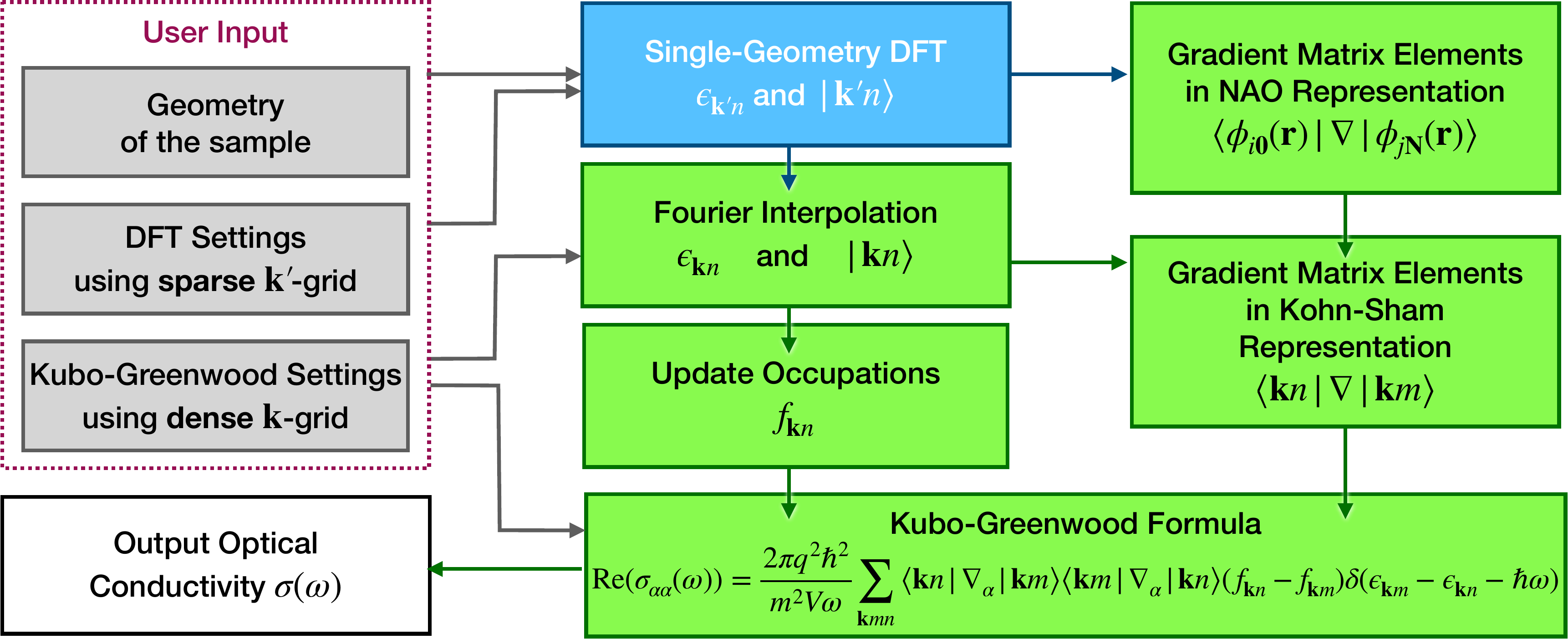}
    \caption{Flowchart for the calculation of the optical conductivity~$\mathbf{\sigma}(\omega)$ for {\it one} sample using the Kubo-Greenwood implementation in \FHIaims. 
     Grey boxes highlight user input, the blue box a ``standard'' \FHIaims calculations, green boxes the steps needed for evaluating Eq.~(\ref{kg-formula}), and the white box the output.}
    \label{fig:Sketch-electrical-transport}
\end{figure}
Achieving numerical convergence of the KG formula is challenging in crystalline materials 
especially in the DC limit~$\omega\rightarrow 0$, since only a small portion of $\textbf{k}$-space contributes to Eq.~(\ref{kg-formula}) due to the fact 
that both $f_{\textbf{k}n}-f_{\textbf{k}m}\neq 0$ and $\epsilon_{\textbf{k}m}-\epsilon_{\textbf{k}n}=\hbar\omega$ need to be fulfilled. 
For the
latter condition, the delta function is numerically resolved by either a Gaussian or Lorentzian broadening function with a finite, but minute 
width~$\eta$ that appropriately represent the limit of vanishing broadness. 

In practice, $\mathbf{k}$-grids in the order of $10^3$-$100^3$ $\mathbf{k}$-points are needed to achieve convergence for 
sufficiently small values of $\eta$, even in extended supercells~\cite{Quan2024,Fiebig2023}. 
Such dense $\textbf{k}$-grids are typically unnecessary for achieving convergence in the self-consistent field cycle~(SCF), since the electronic density and, in turn, the associated real-space Hamiltonian, can typically be converged within a relatively sparse \textbf{k'}-grid during the SCF cycle. 
Due to the locality of the real-space matrix elements, the converged real-space Hamiltonian $\Braket{\phi_{i{\bf 0}}|\mathbf{\underline{H}}|\phi_{j{\bf N}}}$ and overlap matrix $\Braket{\phi_{i{\bf 0}}|\mathbf{\underline{S}}|\phi_{j{\bf N}}}$ can then be Fourier-interpolated to arbitrarily dense \textbf{k}-grids via:
\begin{eqnarray}\label{hk_fourier}
    \Braket{\phi_{i{\textbf k}}|\mathbf{\underline{H}}|\phi_{j{\textbf k}}} & = & \sum_{\textbf{N}} 
    e^{i\textbf{k}\cdot\textbf{T(N)}} \Braket{\phi_{i{\bf 0}}|\mathbf{\underline{H}}|\phi_{j{\bf N}}} \\
    \Braket{\phi_{i{\textbf k}}|\mathbf{\underline{S}}|\phi_{j{\textbf k}}} & = & \sum_{\textbf{N}} 
    e^{i\textbf{k}\cdot\textbf{T(N)}} \Braket{\phi_{i{\bf 0}}|\mathbf{\underline{S}}|\phi_{j{\bf N}}} \;.
\label{sk_fourier}
\end{eqnarray}
As sketched in Fig.~\ref{fig:Sketch-electrical-transport}, this procedure is used in the KG calculation after SCF convergence has been achieved to evaluate Eq.~(\ref{kg-formula}). By this means, the dense $\textbf{k}$-grids required for converging Eq.~(\ref{kg-formula}) can be targeted without incurring into prohibitive numerical costs for the SCF that prevent achieving convergence~\cite{Yuan2022}. 
Note that also the Fermi-level~$\epsilon_\text{F}$ and the electron and hole charge-carrier densities~$n_{\text{e/h}}$ can change when switching to denser $\mathbf{k}$-grids. 
Accordingly, also these quantities are updated in this step, so that the 
optical conductivities~$\sigma_{\text{e/h}}(\omega)$ for \textit{one} sample are obtained within \FHIaims.

Beyond the Kubo-Greenwood method, we have also implemented several semi-empirical methods to quickly estimate properties related to charge transport. This includes the carrier-lattice distance~$d_{\mathrm{c}-1}$~\cite{zhang2023two} that aims at capturing electron–phonon coupling effects. To compute 
\begin{equation} 
d_{\mathrm{c}-1}(\mathrm{CBM} / \mathrm{VBM})=\int_{\mathrm{uc}} d \mathbf{r}\left|\psi_{\mathrm{CBM} / \mathrm{VBM}}(\mathbf{r})\right|^2 \min _\alpha\left\{\left|\mathbf{r}-\mathbf{R}_\alpha\right|\right\},
\end{equation}
we extract the eigenstate density from \FHIaims the at the conduction band minimum (or valence band maximum) and integrate the expression over the whole unit cell~($\mathrm{uc}$). In a similar spirit, conductivities can be estimated by using the interface with the BoltzTraP2~\cite{Boltztrap2} or with the the AICON2~code~\cite{fan2021aicon2}. Here, electronic band-structures are used to determine group velocities; in the AICON2 case, deformation-potential theory is additionally used to approximate relaxation times.

\subsection*{Usability and Tutorials}
The evaluation of optical conductivities~$\sigma(\omega)$ for single samples is directly controlled via keywords in the input file \texttt{control.in}\xspace, most importantly 
{\tt compute\_kubo\_greenwood} for invoking the evaluation of the KG formula and setting numerical parameters such as the broadening~$\eta$. The
Fourier interpolation is controlled via the keyword {\tt dos\_kgrid\_factors}, since the density-of-states routines are used internally for this scope. 
Given that dense, interpolated $\mathbf{k}$-grids are necessary for converging Eq.~(\ref{kg-formula}), the parallelization 
in this routine occurs over $\mathbf{k}$-points with extensive use of LAPACK for linear algebra operations.
This enables good parallel scaling with number of nodes, so that system sizes up to 1,000~atoms can be routinely treated; for even larger systems, extending the implementation to also support distributed ScaLAPACK arrays and {\tt use\_local\_index} in the KG-formula evaluation would be desirable. 
Let us emphasize that the actual SCF cycle predating the KG-formula evaluation can already be run using distributed arrays in ScaLAPACK mode and that
all exchange-correlation~(xc) functionals, including hybrids,~c.f.~Contrib.~\ChapHybrids, are supported.

To eventually obtain thermodynamically relevant conductivities at temperature $T$, 
an ensemble av\-er\-age $\Braket{\sigma(\omega)}_T$ over all samples is performed. 
This step is independent of \FHIaims viz. the electronic-structure theory code and is 
performed as a post-processing procedure. At this stage, also the mobility~$\Braket{\mu}_T = \Braket{\sigma(\omega)}_T / {\Braket{q n}_T}$ is calculated, where $q$ is the carrier's charge and $n$ is the carrier density. 
The DC mobility $\Braket{\mu(0)}_T$ in the thermodynamic bulk limit~$\omega\rightarrow0$ is determined 
by fitting the low-frequency part of $\Braket{\mu(\omega)}_T$ using a Drude function with lifetime $\tau$
\begin{equation}
    \Braket{\mu(\omega)}_T \approx \frac{\mu (0)}{ (\omega \tau)^2 + 1} \;,
\end{equation}
as sketched in the last step in Fig.~\ref{fig:workflow-electrical-transport}. Note that this extrapolation is necessary since
reaching the DC limit would require infinitely large supercells~\cite{Quan2024}, for which the whole electronic-band structure is folded onto the $\Gamma$-point, see Contrib.~\ChapUnfolding.

The conductivity strongly depends on the band-gap through the carrier density~$n$. 
Accordingly, semi-local xc-functionals can overestimate conductivities by orders of magnitude due the notorious band-gap problem. Conversely, mobilities are an intrinsic property that is unaffected by the carrier density in the low-doping regime and are thus reasonably predicted already with semi-local xc-functionals~\cite{Ponce2020}. In this context, constraining the charge-carrier density to a fixed value can be numerically beneficial, as detailed in Ref.~\cite{Quan2024}.

To guide novel user through the workflow of a KG calculation, a tutorial is provided at \href{https://fhi-aims-club.gitlab.io/tutorials/kubo-greenwood-formula}{the following link}\footnote{\texttt{https://fhi-aims-club.gitlab.io/tutorials/kubo-greenwood-formula}}.
Using Al and SrTiO$_3$ as examples, it showcases a typical workflow to calculate the conductivity and mobility with KG formalism, provides scripts for pre- and postprocessing, and explains how the results can be analysed and understood.


\subsection*{Future Plans and Challenges}
The main advantage of the implemented Kubo-Greenwood formalism lies in its ability to account for anharmonic and higher-order coupling effects in the evaluation of mobilities. 
With that, it can be used to effectively assess strongly doped materials, which are for instance common in thermoelectric applications~\cite{Snyder2008}. In this regard,
it is also advantageous that the implemented Kubo-Greenwood formalism can directly deal with high charge-carrier densities and defect scattering  within the supercell approach
and does not require additional {\it ad hoc} models to account for these effects. Besides the electrical conductivity, the Kubo-Greenwood formalism allows to determine electronic contributions to the thermal 
conductivity~$\kappa_{\text{el}}$ and the Seebeck coefficient~$S$ as well.
Thus the exact same code base can be used to calculate charge, heat, and thermoelectrical transport properties of a material. 
The implementation of the respective equations, see Ref.~\cite{Holst2011}, is essentially finished, but requires thorough testing and validation. 
Also in this regard, the implemented formalism can help validating and improving approximations taken in perturbative approaches. 
In turn, a systematic comparison with perturbative approaches can clarify the role of approximations taken in the Kubo-Greenwood formalism, notably using a supercell-approach and the Born-Oppenheimer approximation~\cite{Miglio2020}, which inhibits a systematic inclusion of long-range Fröhlich effects. Combining such studies with the evaluation of temperature-dependent spectral functions, can help including such effects more accurately and more efficiently, see~Contrib.~\ChapUnfolding.

From a numerical point of view, accelerating KG calculations would be desirable. For instance, this would allow to rapidly address weakly or moderately
anharmonic materials, for which the KG approach equally holds, but often becomes numerically impractical. To this end, interfacing to machine-learning 
based approaches is a promising route. Machine-learning approaches for predicting electronic properties such as densities and matrix elements~\cite{Li2022,Zhang2022,Lewis2021} 
can potentially accelerate the evaluation of the KG formula, especially if one can circumvent the computationally dominant eigenvalue problem,see Contrib.~\ChapELPA. 
Similarly, using machine-learned interatomic potentials~\cite{Behler2021} instead of \textit{ai}MD\xspace for the sampling of the thermodynamic phase space can reduce the 
computational cost in the sampling step, see Contrib.~\ChapMLIP. Here, particular care has to be taken to ensure that strongly anharmonic 
effects,~e.g.,~those triggered by spontaneous defect formation~\cite{Knoop2023} are appropriately described by the trained potentials~\cite{kang2025accelerating}.

Eventually, let us emphasize that the developed approach covers band-type conductivity, but not ionic and polaronic conductivities.
Those do not only require different flux operators, but typically also necessitate to simulate even longer time- and length scales~\cite{Rinaldi.2025}.
In this regard, the development of methodologies that cover all these kind of conductivities constitutes a significant challenge for first-principles based modeling~\cite{Merkle2021}.

\subsection*{Acknowledgements}
MS acknowledges support by his TEC1p Advanced Grant (the European Research Council (ERC) Horizon 2020 research and innovation programme, grant agreement No. 740233.
CC and MS acknowledge helpful discussions with Martin French and Ronald Redmer.

\newpage

\section{Molecular Transport}

\sectionauthor[1,2]{\textbf{ *Mar\'ia {Camarasa-G\'omez}}}
\sectionauthor[3]{\textbf{*Daniel {Hernang\'{o}mez-P\'{e}rez}}}
\sectionauthor[4]{Jan {Wilhelm}}
\sectionauthor[5]{Alexej Bagrets}
\sectionlastauthor[4]{\textbf{*Ferdinand Evers}}

\sectionaffil[1]{Departamento de Polímeros y Materiales Avanzados: F\'isica, Qu\'imica y Tecnolog\'ia, Facultad de Qu\'imica, UPV/EHU, Apartado 1072, 20018 Donostia-San Sebasti\'an, Spain}
\sectionaffil[2]{Centro de F\'isica de Materiales (CFM-MPC) CSIC-UPV/EHU, 20018 Donostia-San Sebasti\'an, Spain}
\sectionaffil[3]{CIC nanoGUNE BRTA, Tolosa Hiribidea 76, 20018 San Sebasti\'an, Spain}
\sectionaffil[4]{Institute of Theoretical Physics and Regensburg Center for Ultrafast Nanoscopy (RUN), University of Regensburg, 93040 Regensburg, Germany}
\sectionaffil[5]{\textcolor{black}{Institute of Nanotechnology, Karlsruhe Institute of Technology, 76344 Eggenstein--Leopoldshafen, Germany, and Steinbuch Centre for Computing, Karlsruhe Institute of Technology, 76344 Eggenstein-Leopoldshafen, Germany (\textsl{currently at ITK Engineering GmbH.)}}}

\sectionaffil[*]{Coordinator of this contribution.}

\singlespacing



\subsection*{Summary}

Single-molecule junctions \cite{Su2016chemical, Evers2018, Evers2020, Cuevas2010} --- nanoscale systems where a molecule is connected to metallic electrodes - offer a unique platform for studying charge, spin and energy transport in non-equilibrium many-body quantum systems, with few parallels in other areas of condensed matter physics.
Over the past decades, %
these systems have revealed a wide range of remarkable quantum phenomena, including quantum interference~\cite{Su2016chemical, Camarasa2020}, non-equilibrium spin-crossover \cite{Evers2011b}, diode-like behavior \cite{Elbing2005}, or chiral-induced spin selectivity \cite{Evers2021}, among many others \cite{Evers2018, Evers2020}.
To develop a detailed understanding, 
it turned out essential to have available \textit{ab initio}-based tools for accurately describing quantum transport in such systems \cite{Evers2018, Evers2020}. %
They need to be capable of capturing the intricate electronic structure of molecules, sometimes in the presence of electron-electron or electron-phonon interactions, in out-of-equilibrium environments. %
Such tools are also indispensable for explaining experimentally observed phenomena using parametrized tight-binding models in the context of quantum transport.

While FHI-aims also offers specialized transport routines \cite{Havu2011,Havu2012, Ketolainen2017}, e.g. for chemically functionalized nanotubes or nanotube networks, our focus in this section  is on the AITRANSS package \cite{Arnold2007, Camarasa2024, Bagrets2013, Wilhelm2013} designed for simulations of single-molecule transport.

AITRANSS is an independent post-processing tool that, when combined with FHI-aims, enables the calculation of electronic transport properties, as well as atom-projected density of states, spin properties and the simulation of scanning tunneling microscope images in molecular junctions. Pilot versions of the code extend some of these capabilities to non-linear transport  in the applied bias, with plans to include these features in future releases of the package.

\subsection*{Current Status of the Implementation}

The AITRANSS code implements the non-equilibrium Green's function formalism (NEGF) \cite{Evers2020, Cuevas2010, Jauho2008, DiVentra2008}; AITRANSS can handle closed-shell and spin-polarized contacts~\cite{Bagrets2013} and also spin-orbit coupling \cite{Camarasa2024, Camarasa2021}.

A central element in the calculation of electronic transport is the (ballistic) transmission function, which can be obtained using the trace formula \cite{Wingreen1992}
\begin{equation}\label{eq:trans}
T(E)  = \textnormal{Tr} \left[\hat{\Gamma}_L \hat{G}(E) \Gamma_R \hat{G}^\dagger (E) \right],
\end{equation}
where $\hat{G}(E)$ denotes the Green's function of the extended molecule in the presence of the leads, \textit{i.e.} source and drain. The extended molecule includes  a part of the leads as metal clusters, which are in contact with the molecule; the construction of AITRANSS offers various advantages,  \textit{e.g.}, complete flexibility in the relative orientation of the electrodes. As described elsewhere \cite{Evers2020}, the Green's function can be obtained through  partitioning, 
\begin{equation}\label{eq:green}
 \hat{G} = (E\mathds{1} - \hat{\mathcal{H}}^{\textnormal{KS}} -  \hat{\Sigma})^{-1}.
\end{equation}
Here, $\mathds{1}$ represents the identity operator, $\hat{\mathcal{H}}^ {\textnormal{KS}}$ the Kohn-Sham Hamiltonian and $\hat{\Sigma} = \hat{\Sigma}_{L} + \hat{\Sigma}_{R}$ the self-energy operator of the left ($L$) and right ($R$) leads. Finally, $\hat{\Gamma}_{L,R}$ denotes the anti-Hermitian part of the self-energy operators $\Sigma_{L,R}$.
A distinctive feature of AITRANSS is its computationally efficient implementation of absorbing boundary condition using a  model self-energy approach \cite{Arnold2007, Bagrets2013, Evers2011}. In this scheme, the self-energy is given by
\begin{equation}
    \hat{\Sigma}_{\alpha} = \sum_{\tilde{\mu}, \tilde{\nu} \in \mathcal{S}_{\alpha}}|{\tilde{\mu}}\rangle[\delta \epsilon-i\eta]\delta_{\tilde{\mu}\tilde{\nu}}\langle{\tilde{\nu}}|,
\end{equation}
where $\delta \epsilon$ is the real energy shift, and $\eta$ characterizes the imaginary part, which corresponds to a local, energy-independent, material-specific leakage rate out of the scattering region. 
Note that the self-energy is applied only within a subspace $\mathcal{S}_\alpha$ corresponding to the lead atoms of the extended molecule that are farthest away from the molecule. 

\textit{Calculations with post-processing only.} The simplest  operation mode of AITRANSS is non-self-consistent, see Fig. \ref{fig:f1}, \textit{i.e.} without feedback loop. The process begins with a standard DFT calculation using the optimized geometry of the molecular junction obtained from FHI-aims. It delivers the set of Kohn-Sham energies $\{\epsilon_l\}$ and orbitals 
\begin{equation}
 \psi_l(\mathbf{r}) = \sum_{i  = 1}^{N_\mathcal{B}} c_{il} \varphi_i(\mathbf{r}),
\end{equation}
where $N_\mathcal{B}$ is the number of orbitals, $c_{il}$ are the molecular~orbital coefficients and $\varphi_i$ denotes the FHI-aims basis set (cf. Eq. (\ref{Eq:NAO})). 
These orbitals are used by AITRANSS to reconstruct the Hamiltonian. Because the basis is non-orthogonal, AITRANSS employs the L\"owdin orthogonalization procedure \cite{Lowdin1950} using the overlap matrix, $S$, 
\begin{equation}\label{eq:lowdin}
 \varphi_{\tilde{i}}(\mathbf{r}) = \sum_{j = 1}^{N_\mathcal{B}} S^{-1/2}_{j,i} \varphi_i(\mathbf{r}),
\end{equation}
to orthogonalize the states and reconstruct the Kohn-Sham Hamiltonian in an orthogonal basis,  $\hat{\mathcal{H}}^\textnormal{KS} = S^{1/2} C \bm{\epsilon} C^\dagger S^{1/2}$.  The resulting Hamiltonian is then used for evaluating Eqs. \eqref{eq:trans}-\eqref{eq:green}. 
\begin{figure}
    \centering
    \includegraphics[width=0.5\textwidth]{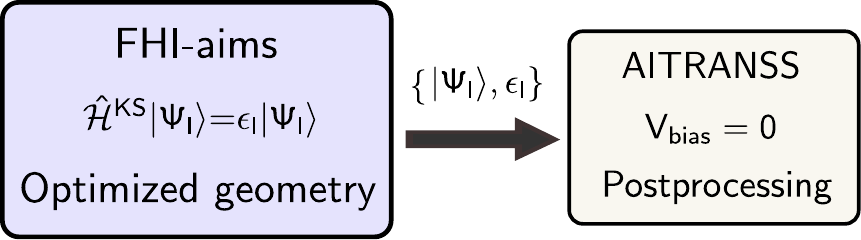}
    \caption{Workflow of the post-processing-only  cycle combining FHI-aims and the AITRANSS package.}
    \label{fig:f1}
\end{figure}

\textit{{Self-consistent calculation procedure.}} In the self-consistent calculation mode, FHI-aims and AITRANSS perform a feedback loop, see Fig. \ref{fig:f2}. 
 As in the post-processing-only variant, the starting point is a standard DFT-calculation for the extended molecule. The role of feedback is to introduce self-consistency in the sense that the DFT-calculation in FHI-aims and the Green's function calculations in AITRANSS refer to the same (non-equilibrium) density matrix $\rho$: during each iteration step the Kohn-Sham Hamiltonian is updated with $\hat{\rho}$, which is constructed using the NEGF \cite{Camarasa2024, Bagrets2013, Camarasa2021}. The feedback-loop establishes, in particular, the Fermi energy, E\textsubscript{F}, at fixed particle number, $N$, and the real part of the self-energy, $\delta \epsilon$. Physically, the cycle ensures correct charge redistribution in the junction,  accounting for the macroscopic nature of the contacts and maintaining charge neutrality while screening the excess charge accumulated at the outermost boundaries of the finite-sized clusters \cite{Arnold2007, Camarasa2024, Bagrets2013, Camarasa2021,  Palacios2023, Naskar2023}.
During the feedback loop, the real part of the self-energy, $\Sigma(\delta \epsilon^\ast)$, is gradually deformed so as to enforce charge-homogeneity on the far sites of the metal cluster, also under finite bias voltage, V\textsubscript{bias}. Physical observables are calculated in the final post-processing step: transmission function, current-voltage characteristics, spin-orbit torques, and more.
%
%
%


\begin{figure}
    \centering
    \includegraphics[width=0.5\textwidth]{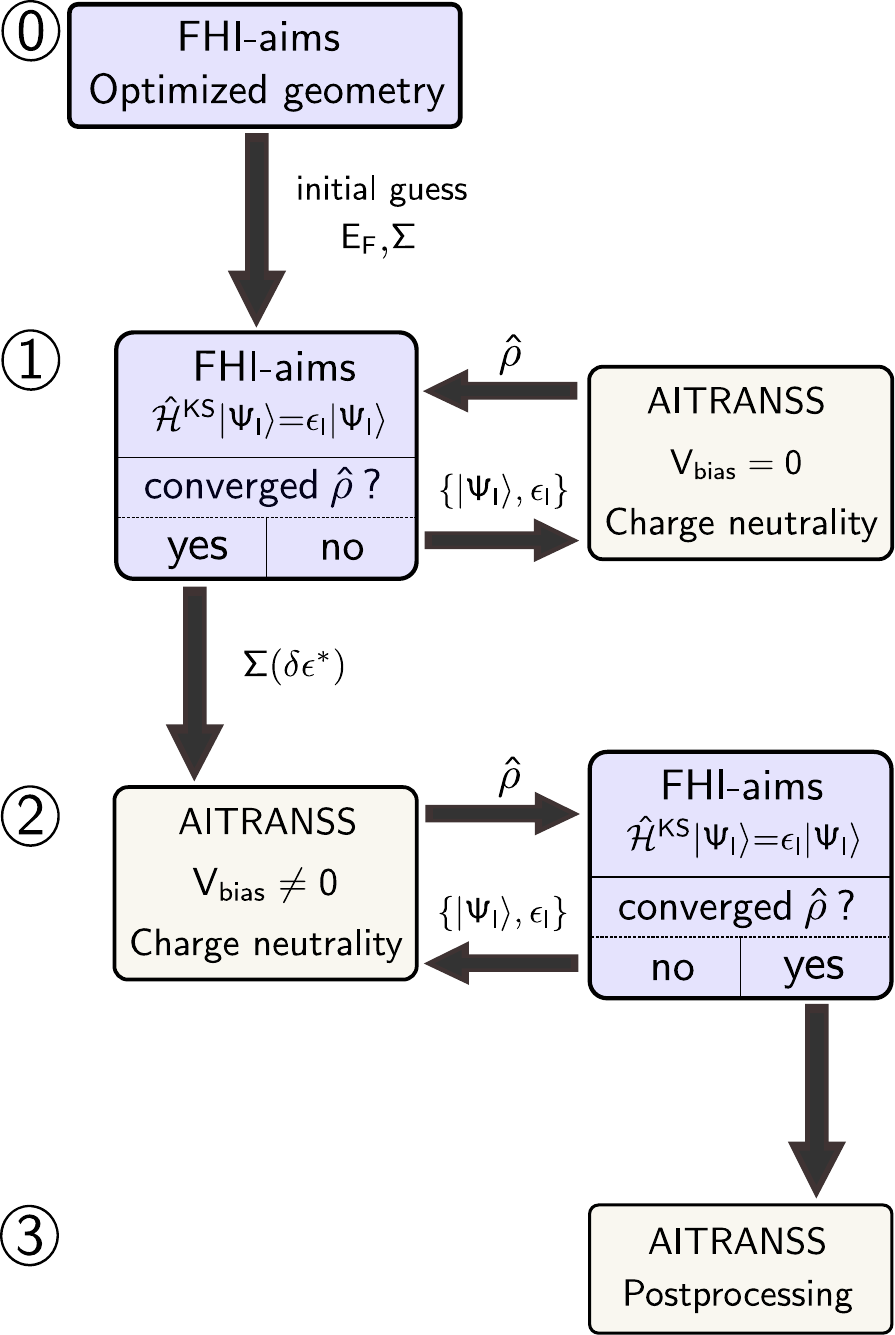}
    \caption{Workflow of the self-consistent cycle combining FHI-aims and the AITRANSS package. The calculation proceeds in four steps: \textcircled{\raisebox{-0.9pt}{0}}, initial preparatory calculation with FHI-aims for the geometry optimization; \textcircled{\raisebox{-0.9pt}{1}}, parametrization of the self-energy and the Fermi energy;  \textcircled{\raisebox{-0.9pt}{2}}, self-consistent loop at finite bias; and  \textcircled{\raisebox{-0.9pt}{3}}, postprocessing step in which observables based on the non-equilibrium density matrix can be computed. Fig. adapted from Fig. 2 in Ref. \cite{Camarasa2024} (CC BY-NC 4.0).}
    \label{fig:f2}
\end{figure}

\subsection*{Usability and Tutorials}

AITRANSS is a project under continuous development since 2002, and is currently centered at Universit\"at Regensburg. The current main developers are Mar\'ia Camarasa-G\'omez, Daniel Hernang\'omez-P\'erez and Ferdinand Evers.
The source code, along with examples and a basic electrode library, is distributed with the FHI-aims package. It can be found in the subdirectory \texttt{external/aitranss}. This directory contains the source code, the script \texttt{tcontrol.aims.x} used to prepare the mandatory \texttt{tcontrol} input file for AITRANSS, a representative electrode library, and a folder with documented examples, including input and output files of FHI-aims and AITRANSS.

Compilation instructions, explanations of the code, and a list of available keywords are provided in the FHI-aims manual.
As an independent package, AITRANSS has its own build system, which has also been integrated into the FHI-aims cmake-based build scheme, as documented in the FHI-aims manual release 240507. The current release of AITRANSS is parallelized using OpenMP directives, enabling the creation of a multithreaded version of the executable.

Without spin-orbit coupling, using the mandatory keyword  \texttt{output aitranss} in the \texttt{control.in} file generates three ASCII files after a successful run of FHI-aims: \texttt{basis.out}, which contains information about the basis functions, \texttt{omat.aims}, which contains the overlap integrals and \texttt{mos.aims} which contains the Kohn-Sham orbitals and energies for the extended molecule. If an open-shell calculation is performed, \texttt{mos.aims} is substituted by two files called \texttt{alpha.aims} and \texttt{beta.aims}. AITRANSS is then run in the same directory where the output files of FHI-aims are located.
To include spin-orbit interaction, the keywords

\texttt{include\_spin\_orbit} \\
\texttt{output soc\_eigenvectors 1} \\
\texttt{output soc\_aitranss}

must also be used, as documented in Ref. \cite{Camarasa2021}.  %
The basis and molecular orbital files are  replaced by \texttt{basis-indices.soc.out
} and \texttt{omat.aims.soc}.
The current implementation of the self-consistent non-equilibrium cycle (Fig. \ref{fig:f2}) is managed by an external shell script and maintains its performance in terms of memory and computational requirements for small molecular junctions. Additional details can be found as well in Ref. \cite{Camarasa2021}.

Finally, as a project in continuous expansion, online tutorials and  updated information can be accessed from the AITRANSS webpage at \href{https://aitranss.ur.de/}{https://aitranss.ur.de/}.


\subsection*{Future Plans and Challenges}
The AITRANSS package is an ongoing project under continuous development, with new features and capabilities to be added in the near future.
From a computational perspective, we will focus on maintaining the package  efficiently. The code is currently being refactored to enhance  readability, making it more compact and easier to manage. Additional examples and tutorials will be added as well.
In addition, we plan to extend the parallelization strategies beyond OpenMP (multithreading) by incorporating MPI parallelization, which will also function independently from the former. Further enhancements include replacing outdated ASCII files with modern formats such as HDF5.

From the functionality perspective, a significant issue, resulting from the limitations of Kohn-Sham transport calculations using semilocal functionals, is the deviation of Kohn-Sham spectral properties from the exact values. We aim to address this issue by employing scissor-like operation techniques based on hybrid functionals and image-charge correction.
Furthermore, we will be expanding the study of transport and dynamical properties in the presence of spin-orbit interactions, for example incorporating spin-orbit effects in scanning tunneling microscope imaging simulations.
We also plan to continue developing the implementations for current-induced forces (both mechanical and spin) and to incorporate interaction with light in the terahertz regime, local currents \cite{Walz2014, Walz2015, Wilhelm2015} as well as phonon effects.
Finally, we plan to work on implementing multiterminal calculations in AITRANSS, particularly in the context of electrochemistry, where electrostatic effects are crucial.


\subsection*{Acknowledgments}
We appreciate pleasant and fruitful discussions with many colleagues over the years: L. Venkataraman, M. S. Inkpen, R. Koryt\'ar,  M. Kamenetska, 
\textcolor{black}{M. Walz, G. Solomon, J. van Ruitenbeek.} \textcolor{black}{Contributions of P. Havu and V. Havu to the development of additional transport routines interfaced with FHI-aims are acknowledged.}
M.C.-G. acknowledges support from the Diputaci\'on Foral de Gipuzkoa through Grant 2024-FELL-000007-01 and from the Gobierno Vasco-UPV/EHU Project No. IT1569-22. D. H.-P. is grateful for funding from the Diputaci\'on Foral de Gipuzkoa through Grants 2023-FELL-000002-01, 2024-FELL-000009-01, and from the Spanish MICIU/AEI/10.13039/501100011033 and FEDER, UE through Project No. PID2023-147324NA-I00. J. W. acknowledges funding by the German Research Foundation (DFG) via the Emmy Noether Programme (Project No. 503985532). F. E. is grateful for support from DFG through the Collaborative Research Center (SFB) 1277 - Project ID 314695032 (subproject A03), from GRK 2905 (project-ID 502572516) and from the State Major Instrumentation Program, INST 89/560-1 (project number 464531296). F. E. and J. W. acknowledge the Gauss Centre for Supercomputing e.V. for providing computational resources on SuperMUC-NG at the Leibniz Supercomputing Centre under the Project ID pn36zo and the computing time provided to them on the high performance computer Noctua 2 at the NHR Center PC2. These are funded by the Federal Ministry of Education and Research and the state governments participating on the basis of the resolutions of the GWK for the national high-performance computing at universities (www.nhr-verein.de/unsere-partner).

\newpage

\section{Computing Diabatic Couplings}

\sectionauthor[1]{Simiam Ghan}
\sectionauthor[2]{\textbf{ *Harald Oberhofer}}
\sectionauthor[3]{Karsten Reuter}
\sectionauthor[4]{Christoph Schober}

\sectionaffil[1]{Catalysis Theory Center, Department of Physics, Technical University of Denmark}
\sectionaffil[2]{Chair for Theoretical Physics VII and Bavarian Center for Battery Technologies, University of Bayreuth, Germany}
\sectionaffil[3]{Theory Department, Fritz Haber Institute of the Max Planck Society, Faradayweg 4-6, D-14195 Berlin, Germany}
\sectionaffil[4]{Department of Computer Science, Deggendorf Institute of Technology, Deggendorf, Germany}

\sectionaffil[*]{Coordinator of this contribution.}



\subsection*{Summary}
The electronic coupling parameter -- also known as the charge transfer integral --  is defined as the Hamiltonian matrix element between two localized \textit{diabatic} states:
\begin{align}
        \text{H}_{ab} = \langle\psi_a|\mathcal{\hat{H}}|\psi_b\rangle.
\end{align}
In molecular systems, the electronic coupling is invoked in the theories of charge transfer\cite{Oberhofer2017}, while on surfaces it is additionally used  in theories of e.g.
chemisorption and catalysis\cite{Hoffmann1988_article, Norskov1995, Newns1969}, scanning-probe microscopy\cite{TersoffHamann1983}, and quantum impurities\cite{Werner2011}. In particular, the electronic coupling is one of the key features determining the efficiency of charge transport in molecular crystals\cite{neef2024}.

Contrary to adiabatic states, which are the delocalized (Kohn-Sham) eigenstates of a system's Hamiltonian, localized diabatic states are nonunique and must be approximated. A large number of diabatization methods have emerged to determine suitable diabatic states and the couplings between them\cite{Oberhofer2017,Voorhis2010}.  
 For FHI-aims, two efficient approaches are currently available: Fragment Orbital DFT (FO-DFT) and Projection-Operator Diabatization (POD2+). Our recent works have demonstrated the use of these methods for both molecular and surface systems.

\subsection*{Current Status of the Implementation}
\rev{Both diabatization methods described below have been shown to work with FHI-aims' native NAO basis as well as the popular Pople\cite{Pople2001} and Dunning\cite{Dunning1989} basis sets. At close distances up to 5\AA{} no difference in accuracy could be detected between a Tier 1  NAO basis set and a triple zeta Gaussian basis, with the former being more efficient due to its smaller size.}
\subsubsection*{Coupling between Molecules}

\subsubsection*{Fragment-Orbital DFT (FO-DFT)}
In FO-DFT the initial and final diabatic states of a charge transfer reaction are constructed from non-interacting fragment densities (see fig.~\ref{fig:fodft}). As a consequence, any interactions between the fragments in the real system are neglected, making FO-DFT a method best suited for weakly interacting donor-acceptor pairs such as found in organic crystals\cite{schober2016virtual, reilly2016report} or physisorbed adsorbates on surfaces.

\begin{figure}[ht]
    \centering
    \includegraphics[width=0.6\textwidth]{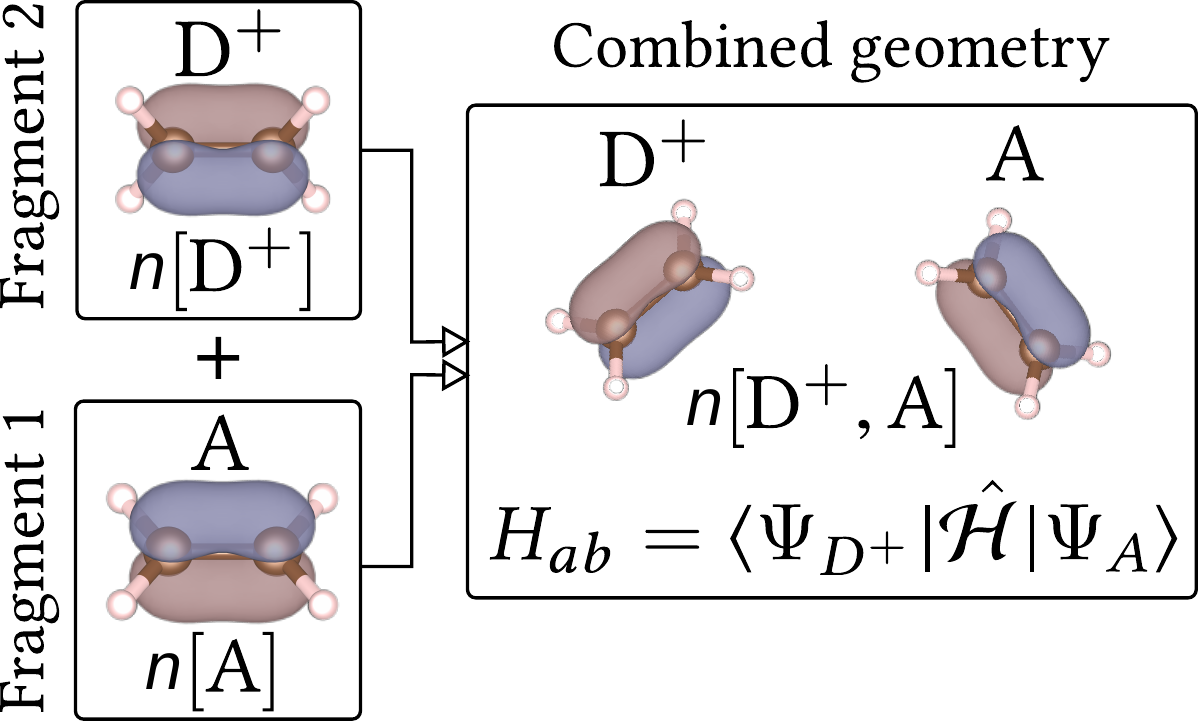}
    \caption{Schematic illustration of the fragment orbital DFT scheme for hole transfer in ethylene, using charged fragments.
    }
    \label{fig:fodft}
\end{figure}

Due to the approximations involved in constructing the fragment and combined system, FO-DFT can be described as a family of methods depending on how the fragment densities and combined Hamiltonian are defined.\cite{schober2016critical} FHI-aims currently implements three FO-DFT flavors:
\begin{itemize}
\item $\mathcal{H}^{2n}@DA$
The original flavor implemented by Senthilkumar et. al\cite{senthilkumar2003charge} with two neutral fragments and $2n$ electrons in the combined Hamiltonian (MRSE HAB11/HAB7:\cite{schober2016critical} -24.6\% / -22.4\%). 
\item $\mathcal{H}^{2n-1}@DA / \mathcal{H}^{2n+1}@D^{-}A^{-}$
This flavor\cite{oberhofer2012revisiting} improves on the original formulation by using the correct number of electrons in the combined Hamiltonian (MRSE HAB11/HAB7:\cite{schober2016critical} -37.7\% / -27.1\%).
\item$\mathcal{H}^{2n-1}@D^+A / \mathcal{H}^{2n+1}@D^-A$
The most accurate flavor of FO-DFT employs charged fragment calculations, most closely mimicking the correct diabatic states in the system (MRSE HAB11/HAB7:\cite{schober2016critical} -22.4\% / -22.9\%). Requires an additional full DFT calculation compared to the previous two schemes, making it computationally more expensive. 
\end{itemize}

\subsubsection*{Projection-Operator Diabatization}

The central idea of Projection-Operator Diabatization (POD) \cite{Kondov2007,Blumberger2017,Ghan2020} lies in a partitioning of the system's Kohn-Sham Hamiltonian into blocks based on the (NAO) basis functions associated with either of the two fragments, labelled a (e.g. surface) and b (e.g. adsorbate). Each block is then diagonalized separately as illustrated in Figure (\ref{fig:matrix}).

\begin{figure}[ht]
    \centering
    \includegraphics[width=0.3\textwidth]{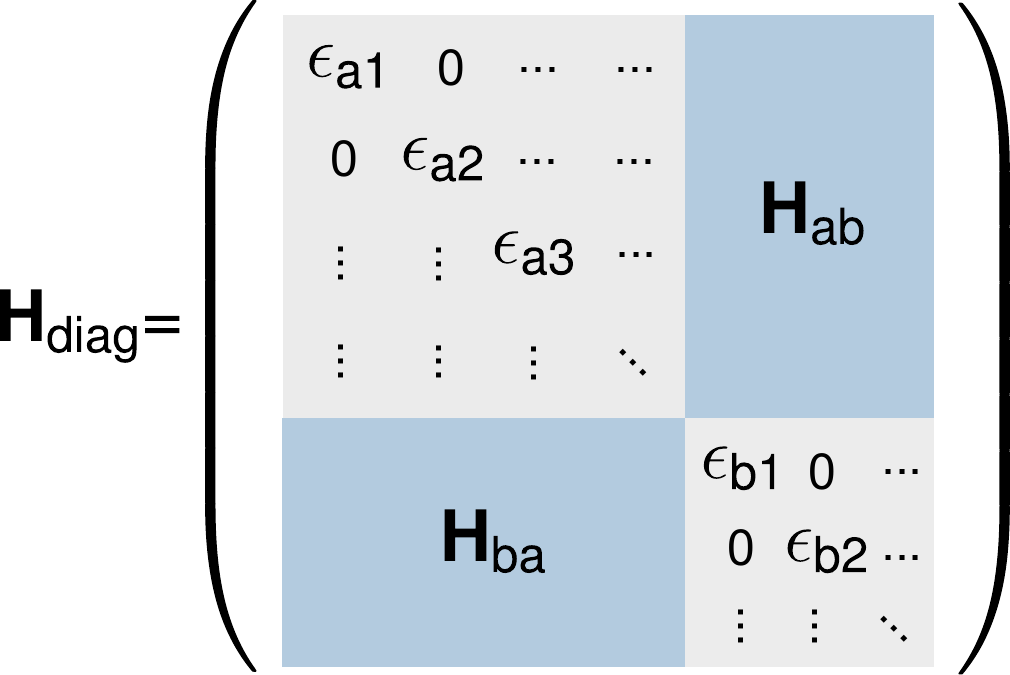}
    \caption{Block-diagonal form of the Hamiltonian.}
    \label{fig:matrix}
\end{figure}

Thereby, diabatic states are approximated as the block eigenstates with energies $\epsilon_{{\rm a},i}$ and $\epsilon_{{\rm b},j}$. First approximations to the coupling between two specific orbitals $\psi_{\text{a}}$ and $\psi_{\text{b}}$ are then simply encoded in the off-diagonal blocks $\mathbf{H}_{\text{ab}}$, in what we refer to as the POD2 method.\cite{Ghan2020} 
Note that the POD2 coupling is not yet comparable to other methods such as FO-DFT, because in general the POD2 states $\psi_{\text{a}}$ and $\psi_{\text{b}}$ are not orthogonal. In the POD2L and POD2GS approaches (referred to as POD2+ below), this is addressed in an additional orthogonalization step based either on L\"owdin-transformation (POD2L) or the Gram-Schmidt method (POD2GS).

Methods based on POD2 diabats may be considered a deductive approach to diabatization, in contrast to the constructive approach of FO-DFT\cite{Voorhis2010}, which builds diabats from the calculation of separated fragments.
Importantly, POD2-based methods allow the diabatic wavefunctions to vary from their frozen FO-DFT counterparts, a result of interactions present in the Hamiltonian. This is found advantageous for e.g. surface systems~\cite{Ghan2023}.
POD2+ methods were found to have an accuracy and numerical performance similar to FO-DFT in the Hab11 benchmark of molecular dimers\cite{Ghan2020}, and improve significantly upon earlier formulations of POD\cite{Kondov2007}.

\subsubsection*{Coupling on Surfaces}

For surface systems such as adsorbates on periodic slabs, couplings can be calculated in the ways described above by performing diabatization seperately at each k-point, i.e.~for each k-points' respective Hamiltonian.  Both POD2+ and FO-DFT-based approaches, and combinations thereof, have in this way been used to calculate couplings for e.g. physisorbed Argon atoms on metal surfaces~\cite{Ghan2023}, illustrated in Figure (\ref{fig:surface}). 

\begin{figure}[ht]
    \centering
    \includegraphics[width=0.8\textwidth]{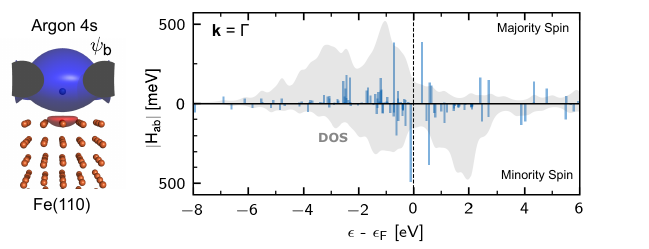}
    \caption{Electronic Couplings of the adsorbed Argon LUMO on Fe(110), using the POD2GS method\cite{Ghan2023}.}
    \label{fig:surface}
\end{figure}

\subsection*{Usability and Tutorials}
\paragraph{FO-DFT} The functionality is fully integrated and documented within FHI-aims. It is available via the \texttt{fo\_dft} tag in the \texttt{control.in} file. A full FO-DFT calculation requires at least 2 invocations of FHI-aims (fragment and combined system). A Python-based package automating the different calculation steps and allowing the re-use of fragment calculations for different geometries is available at \url{https://gitlab.com/schober-ch/aimsutils}.

\paragraph{POD2+} A Python software package for POD2+ methods on surfaces, including tutorial, is available online at \url{https://gitlab.com/simiamghan/bd_utils}.
This is an external post-processing routine which requires as inputs the overlap and converged Hamiltonian matrices of FHI-aims, which can be printed using the keywords \texttt{output hamiltonian\_matrix} and  \texttt{output overlap\_matrix}. Methods are documented in references \cite{Ghan2020,Ghan2023}.


\subsection*{Future Plans and Challenges}
\paragraph{FO-DFT} The calculation of heterogeneous couplings where $H_{ab} \neq H_{ba}$ is not implemented in an automated way and requires explicit fragment geometries. The current implementation also relies on the legacy restart infrastructure in FHI-aims. To further improve the efficiency of the method, future versions will use the more efficient ELSI-based infrastructure.

\paragraph{POD2+} The suitability of the POD2+ diabatization methods for calculating coupling in covalently bound or chemisorbed systems is currently being studied. The POD2-based routines are currently implemented as external post-processing routines with significant storage requirements (FHI-aims must print all matrices to file for parsing). An internal FHI-aims implementation is needed.

\subsection*{Acknowledgements}
HO acknowledges support from the German Science Foundation (DFG) under grant numbers OB 425/9-1 and OB 425/3-2.

\newpage

\chapter{Workflows and AI}
\label{ChapWorkflowsAI}




\newcolumntype{L}[1]{>{\raggedright\let\newline\\\arraybackslash\hspace{0pt}}m{#1}}
\newcolumntype{C}[1]{>{\centering\let\newline\\\arraybackslash\hspace{0pt}}m{#1}}
\newcolumntype{R}[1]{>{\raggedleft\let\newline\\\arraybackslash\hspace{0pt}}m{#1}}

\newpage

\section{Code-Agnostic Materials Science Libraries,\\ Workflow Managers, and Visualization Tools}
\label{ChapWorkflows}
\sectionauthor[1]{Jonas Bj\"ork}
\sectionauthor[2,3]{Volker Blum}
\sectionauthor[4,a]{Felix Hanke}
\sectionauthor[5,1]{Florian Knoop}
\sectionauthor[6]{Alexander Knoll}
\sectionauthor[5,7]{\textbf{*Sebastian Kokott}}
\sectionauthor[8]{Ask Hjorth Larsen}
\sectionauthor[5]{Akhil S. Nair}
\sectionauthor[5,9,b]{Thomas A. R. Purcell}
\sectionauthor[10]{Simon P. Rittmeyer}
\sectionauthor[5]{Matthias Scheffler}
\sectionlastauthor[5,7]{\textbf{*Andrei Sobolev}}

\sectionaffil[1]{Department of Physics, Chemistry and Biology (IFM), Linköping University, SE-581 83, Linköping, Sweden}
\sectionaffil[2]{Thomas Lord Department of Mechanical Engineering and Materials Science, Duke University, Durham, NC 27708, USA}
\sectionaffil[3]{Department of Chemistry, Duke University, Durham, NC 27708, USA}
\sectionaffil[4]{Theory Department (since 1/1/2020: The NOMAD Laboratory), Fritz Haber Institute of the Max Planck Society, Faradayweg 4-6, D-14195 Berlin, Germany}
\sectionaffil[5]{The NOMAD Laboratory at the Fritz Haber Institute of the Max Planck Society, Faradayweg 4-6, D-14195 Berlin, Germany}
\sectionaffil[6]{Research Center Chemical Sciences and Sustainability, Lehrstuhl für Theoretische Chemie II, Ruhr-Universit\"at Bochum, Gesundheitscampus S\"ud 25, 44801 Bochum}
\sectionaffil[7]{Molecular Simulations from First Principles e.V., 14195 Berlin, Germany}
\sectionaffil[8]{CAMD, Department of Physics, Technical University of Denmark, 2800 Kgs. Lyngby, Denmark}
\sectionaffil[9]{Department of Chemistry and Biochemistry, The University of Arizona, Tucson, AZ 85721, USA}
\sectionaffil[10]{Chair for Theoretical Chemistry and Catalysis Research Center, Technische Universit\"at M\"unchen, Lichtenbergstr. 4, 85747 Garching, Germany}

\sectionaffil[*]{Coordinator of this contribution.}
\rule[0.25ex]{0.35\linewidth}{0.25pt}

\sectionaffil[a]{{\it Current Address:} CuspAI Ltd, Cambridge, CB1 2GE, United Kingdom}
\sectionaffil[b]{{\it Current Address:} Department of Chemistry and Biochemistry, The University of Arizona, Tucson, AZ 85721, USA}




\subsection*{Summary}
With routine access to peta-scale resources and the rise of exascale computing, data-centric computational materials science faces significant challenges, such as creating and managing large datasets, visualizing and analyzing diverse data, and assessing the quality of simulation results. To address these issues, the computational materials science community has focused on developing abstractions across all layers of materials simulations --- ranging from representing geometries of a material and computing physical observables to developing a unified language for workflows and providing graphical user interfaces for data visualization. Over the past decade, these concepts have been realized through cross-institutional collaborations that support most existing DFT codes. In this section, we highlight the contributions of the FHI-aims community to these developments or examples that demonstrate the use of code-agnostic frameworks. To avoid any misunderstanding, the use of any of the packages mentioned below should be acknowledged by citing the original publications given by the references in this article.

FHI-aims integrates with the Atomic Simulation Environment (ASE)~\cite{larsen2017ase} and the Python Materials Genomics (pymatgen) library~\cite{ong2013pymatgen}, two widely adopted Python frameworks that offer code-agnostic objects for atomic structures and controlling simulations. To concatenate many calculation steps and automate their execution, several materials science workflow frameworks have been developed, including atomate2~\cite{Ganose}, TaskBlaster~\cite{larsen2025taskblaster}, and AiiDA~\cite{huber2020aiida}, providing support for FHI-aims. Visualization tools such as Vesta~\cite{momma2008vesta}, XCrySDen~\cite{kokalj1999xcrysden}, VMD~\cite{humphrey1996vmd}, jmol~\cite{jmol}, the ASE GUI, and the Graphical Interface for Materials Science (GIMS)~\cite{kokott2021gims} enable rendering three-dimensional geometries from the FHI-aims input format and analyzing structural properties. Additionally, Vesta, jmol, and XCrySDen can visualize volumetric data (such as cube files). GIMS, a browser-based toolbox, also facilitates the generation of input files for basic workflows, along with the analysis and visualization of results derived from output files.

\subsection*{Current Status of the Implementation}
FHI-aims is driven and controlled mainly by the input file interface. To increase efficiency by automation and to standardize the input file generation, tools written in a high-level programming language are usually used. Python has been established as the language of choice to implement such tools. Python is easy to learn due to its straightforward syntax, making scripting and programming accessible even to scientists without programming experience. Further, the vast amount of third-party scientific Python libraries provide tools to solve numerical problems quickly and efficiently. Promoted by the two initiatives, the \href{https://materialsproject.org}{Materials Project} and the \href{https://nomad-lab.eu/nomad-lab/}{NOMAD Project}, two Python software packages have been established in the materials science community: The Python Materials Genomics (pymatgen) library~\cite{ong2013pymatgen} and the Atomic Simulation Environment (ASE)~\cite{larsen2017ase}. These libraries provide code-agnostic classes for representing structures, providing code-specific file I/O for input and output files, and computing physical observables from elemental tasks such as ground state calculations, structure relaxation, or molecular dynamics (MD). Both libraries come with functions that allow creating the two FHI-aims input files (\verb|control.in| and \verb|geometry.in|) and extracting relevant physical quantities, e.g., the total energy, forces, stress, chemical potential, and structure trajectories from output files. In addition, ASE is able to run FHI-aims both externally, through the Aims calculator interface, and interactively via the socket interface (the \verb|SocketIOCalculator| class) adapting infrastructure from the i-PI implementation~\cite{kapil2019}. Through the socket interface, the atomic structure, its forces and, for periodic structures, stress is reported to the calling ASE function after a full self-consistent field cycle (SCF) is completed. In turn, ASE can process the new inputs and return a new structure to FHI-aims; after that, another SCF cycle is evaluated until a user-defined numerical convergence criterion is fulfilled. A full list of features would exceed the scope of the contribution, but can be found in online documentation of \href{https://wiki.fysik.dtu.dk/ase/}{ASE} and \href{https://pymatgen.org}{pymatgen} .

The computational analysis of a material consists of a set of standard workflows. The atomate2 \cite{Ganose} project, based on the pymatgen package, has collected many of these workflows in one framework. The actual execution of the workflows is facilitated by custodian~\cite{ong2013pymatgen}, a job management framework; jobflow~\cite{jobflow2024}, an abstract language to compose and connect workflows; and Fireworks~\cite{fireworks2015}, a framework for executing calculations on high-performance computing (HPC) systems and providing advanced database infrastructure. FHI-aims has now also added support for many of the atomate2 workflows, such as structure optimization, electron band structure calculation with DFT, GW, and other excitation methods, harmonic phonon calculation, computing elastic constants and the equation of state, anharmonicity quantification, and molecular dynamics. Many more features will be added in the near future. Taskblaster~\cite{larsen2025taskblaster} provides a lightweight Python framework for designing and managing workflows and can be extended to support ASE classes and functionality. By using the ASE interface, FHI-aims can be employed to carry out multiple tasks (e.g. structure relaxation, band structure calculation etc.) over a large number of systems. For one of the most advanced workflow managers, AiiDA~\cite{huber2020aiida}, support for FHI-aims was also added recently. AiiDA's core ecosystem is built around high-throughput workflow construction, data provenance, and efficient facilitation of HPC resources. It helps researchers organize, share, and reproduce complex scientific workflows efficiently. All the mentioned workflow managers provide a user-friendly command-line interface for workflow inspection and support for the major job scheduling systems, e.g., SLURM. 

Visualization is an integral part of understanding and analyzing data --- especially the rendering of 3D structure models is an important human consistency check and source of new ideas. Many visualization tools have added support for the FHI-aims input file \verb|geometry.in|. Among them are Vesta~\cite{momma2008vesta}, XCrySDen~\cite{kokalj1999xcrysden}, VMD~\cite{humphrey1996vmd}, jmol~\cite{jmol}, the ASE GUI, and the Graphical Interface for Materials Science (GIMS)~\cite{kokott2021gims}. These visualisation tools allow measuring interatomic distances, the angle of an atom triple, and the torsion angle of two atom pairs, obtaining crystal symmetry information, and many more features.

Web-based applications require no setup and are built around user convenience and data visualization. Examples include the Graphical Interface for Materials Simulations (GIMS)~\cite{kokott2021gims}. GIMS supports FHI-aims, but is easily extendable to any other DFT code, as demonstrated for the \textit{exciting} \cite{gulans2014exciting} code in the current version. GIMS was designed for easy manipulation and visualization of atomic structures, as well as generating corresponding input files. A mechanism of input keyword constraints and consistency checks is implemented to ensure that the results of a calculation with the generated input files are meaningful. GIMS is also able to parse, analyze and plot important FHI-aims output files and provide downloadable figures, such as convergence graphs, band structures, densities of electronic states, and dielectric functions.

\subsection*{Usability and Tutorials}
All of the described Python packages are available on Python Package Index and can be installed with standard Python tools, like \verb|pip|. The majority requires Python 3.10, and the basic functionality is available to the user right after the installation. Nevertheless, further configuration is required following the installation process in order to allow for the integration of the code. The configuration for FHI-aims along with the possible usage scenarios is described in the tutorials on high-throughput calculations \cite{ase-tutorial, atomate2-tutorial, TaskBlasterTutorial, aiida-aims-tutorial}. 

A tutorial specifically written for the use of ASE with FHI-aims is available as part of our tutorial collection \cite{ase-tutorial}. The tutorial introduces the \verb|Atoms| object and the ASE calculators, which are used for the setup of a calculation, demonstrates nudged-elastic band workflow implemented in ASE, as well as showcases the use of the socket interface for a structure optimization. A pragmatic use for the ASE calculator when experimenting with DFT calculations is to compile the complete input files using the \verb|write_input| method. The use of pymatgen is documented as part of a tutorial about atomate2. The \verb|Structure| object is the pymatgen analog to the ASE \verb|Atoms| object. The pymatgen tutorial demonstrates how to create the FHI-aims input files and explains the concept of the pymatgen input sets, which form the basis of the workflows. Eventually, it is explained how the FHI-aims output files can be parsed  with the \verb|AimsOutput| class. It is important to note that the use cases of ASE and pymatgen are by far not limited to the applications in the tutorials and this article: the concepts are widely adopted in several Python packages enabling writing of computational software--agnostic Python code. Such a code greatly reduces the cost of research and is thus of great benefit to the entire community.

Atomate2, being made of several distinct parts, also requires an advanced setup configuration. The workflows are written in the \textit{jobflow} language and can be executed either locally or on the HPC resources using Fireworks or \textit{jobflow-remote} workflow managers. Both workflow managers also use MongoDB to store information about calculation inputs and outputs as well as HPC queue status. \textit{jobflow-remote} does not need access to MongoDB from the HPC. Both workflow managers require that the remote HPC center must allow non-interactive passwordless login. The steps needed to be done to configure and use atomate2 for a simple calculation are outlined in the tutorial \cite{atomate2-tutorial}.

Taskblaster is a workflow tool similar to jobflow, integrated into the ASE ecosystem. Through plugins it allows the execution of multiple tasks (e.g. structural relaxation, bandstructure calculation, postprocessing analysis etc.) via command-line interface on HPC resources. The installation and configuration, as well as the integration of FHI-aims into Taskblaster is demonstrated by the tutorial "FHI-aims with Taskblaster" \cite{TaskBlasterTutorial}.

AiiDA, in its turn, requires PostgreSQL database and RabbitMQ message broker to be installed in a production environment; however, limited functionality is available even without them. AiiDA has a extensive plugin ecosystem --- FHI-aims can be used alongside AiiDA with either ASE or dedicated FHI-aims plugin. The dedicated FHI-aims plugin for AiiDA supports running generic calculations; more specialized workflows will be available in the near future. Several tutorials on working with AiiDA and FHI-aims are available online~\cite{aiida-aims-tutorial, aiida-ase-tutorial}.

All Python package tutorials and their content are summarized in Table~\ref{tab:tutorial-list}.

GIMS, as noted before, does not require any installation. It is a Web application that consists of a collection of apps, such as manipulating atomic structure (Structure Builder), generating input files (Control Generator) and parsing the outputs (Output Analyzer). These apps are called elemental apps. The elemental apps can be chained together to form workflows (Calculation Apps). There are several such apps available for different calculation kinds, including single point, MD, band structure, and GW. Different Calculation Apps have different constraints and consistency checks implemented to ensure the user get the control file with parameters most suitable for their specific calculation. We host a stable release version of GIMS\cite{gims-stable:2024}, as well as a development version \cite{gims-dev:2024} in the Internet. However, also standalone versions of GIMS are released for all major operating systems. It is possible to install GIMS on local machines and laptops. Detailed installation instructions and tutorials for GIMS can be found in the manual \cite{gims-manual}. 

\begin{table}[tb]
    \centering
    \begin{tabular}{L{0.4\linewidth} L{0.5\linewidth}}
       Name & Content  \\\hline
       \href{https://fhi-aims-club.gitlab.io/tutorials/fhi-aims-with-ase/}{FHI-aims with ASE} & - Installation \newline - ASE Atoms and AimsCalculator \newline - Running calculations \newline - Using socket interface \\\hline
       \href{https://workflows-with-atomate2-fhi-aims-club-tutorials-a0acc2efc2fba83.gitlab.io/}{High-throughput workflows with FHI-aims and atomate2}  & - Installation and configuration \newline - Pymatgen support for FHI-aims \newline - Executing workflows locally and remotely \\\hline
       \href{https://fhi-aims-club.gitlab.io/tutorials/fhi-aims-with-taskblaster}{Working in FHI-aims with Taskblaster} & - Installation \newline - Preparing input files \newline - Running workflows \\\hline
       \href{https://fhi-aims-club.gitlab.io/tutorials/fhi-aims-with-aiida/}{Working with FHI-aims and AiiDA} & - Installing AiiDA and FHI-aims plugin \newline - Running calculations \newline - Accessing results \\\hline
       \href{https://fhi-aims-club.gitlab.io/tutorials/fhi-aims-and-aiida/}{Managing Research Data with AiiDA and FHI-aims} & - Using \textit{aiida-ase} plugin to work with FHI-aims \newline - Running single point calculation \newline - Running structure optimization \\\hline
    \end{tabular}
    \caption{The list of available tutorials for the Python infrastructure of FHI-aims}
    \label{tab:tutorial-list}
\end{table}

\subsection*{Future Plans and Challenges}
Maintenance, continuous support, and adding new features form the corners of successful open source projects. While new packages and features are often part of new research, the FHI-aims community will stay dedicated to continuously support and help to maintain open source projects via platforms like the nonprofit organization Molecular Simulations from First Principles e.V. As already indicated above, both major Python computational materials science ecosystems, ASE and pymatgen, require Python functions to generate input files and parse output files. At the moment, each package has its own infrastructure, which complicates their maintenance. A centralized Python package for FHI-aims specific file I/O is currently under development under the name \verb|pyfhiaims|. The package will unify the file I/O functions and provide an API that allows the seamless integration into the ecosystems as a dependency. The big advantage of the centralized \verb|pyfhiaims| package is that changes in the in- and output files of FHI-aims can be faster integrated. Subsequently, more parsers for the various output files, like the band structure and DOS files, will be added. Also a sophisticated API for error handling will be implemented. The error handling is an important part in any workflows to determine a next step in case of a FHI-aims calculation fails. Support for pymatgen and atomate2 was only added recently. More features and integration of FHI-aims in the atomate2 workflows will be added, such as adding anharmonic phonons, magnetic ordering, and defect workflows in atomate2.

\subsection*{Acknowledgements}
MS acknowledges support by his TEC1p Advanced Grant (the European Research Council (ERC) Horizon 2020 research and innovation programme, grant agreement No. 740233.

FK acknowledges support from the Swedish Research Council (VR) program 2020-04630, and the Swedish e-Science Research Centre (SeRC).

\newpage

\section{Workflows for Artificial Intelligence}
\label{ChapMLP_AI}
\sectionauthor[1,2]{J\"org Behler} 
\sectionauthor[3,4,a]{Christian Carbogno} 
\sectionauthor[5]{G{\'a}bor Cs{\'a}nyi} 
\sectionauthor[3]{\textbf{*Lucas Foppa}} 
\sectionauthor[3,b]{Kisung Kang} 
\sectionauthor[6]{Roman Kempt}
\sectionauthor[7]{Marcel F. Langer} 
\sectionauthor[8]{Johannes T. Margraf} 
\sectionauthor[3]{\textbf{*Akhil S. Nair}} 
\sectionauthor[3,9,c]{Thomas A. R. Purcell} 
\sectionauthor[10,11,12,13]{Patrick Rinke} 
\sectionauthor[3]{Matthias Scheffler} 
\sectionauthor[14,15]{Alexandre Tkatchenko} 
\sectionauthor[16]{Milica Todorovi{\'c}} 
\sectionauthor[17]{Oliver T. Unke} 
\sectionlastauthor[18]{Yi Yao} 

\sectionaffil[1]{Lehrstuhl für Theoretische Chemie II, Ruhr-Universit\"at Bochum, D-44780 Bochum, Germany}
\sectionaffil[2]{Research Center Chemical Sciences and Sustainability, Research Alliance Ruhr, D-44780 Bochum, Germany}
\sectionaffil[3]{The NOMAD Laboratory at the Fritz Haber Institute of the Max Planck Society, Faradayweg 4-6, D-14195 Berlin, Germany}
\sectionaffil[4]{Theory Department, Fritz Haber Institute of the Max Planck Society, Faradayweg 4-6, D-14195 Berlin, Germany}
\sectionaffil[5]{Department of Engineering, University of Cambridge, Cambridge CB2 1PZ, U.K.}
\sectionaffil[6]{Technische Universität Dresden, Theoretische Chemie, 01069 Dresden, Germany}
\sectionaffil[7]{Laboratory of Computational Science and Modeling, Institut des Matériaux, École Polytechnique Fédérale de Lausanne, 1015 Lausanne, Switzerland}
\sectionaffil[8]{Bavarian Center for Battery Technology (BayBatt), University of Bayreuth, Weiherstra{\ss}e 26, 95448, Bayreuth, Germany}
\sectionaffil[9]{Department of Chemistry and Biochemistry, The University of Arizona, Tucson, AZ 85721, USA}
\sectionaffil[10]{Department of Applied Physics, Aalto University, P.O. Box 11000, FI-00076 Aalto, Finland}
\sectionaffil[11]{Physics Department, TUM School of Natural Sciences, Technical University of Munich, Garching, Germany}
\sectionaffil[12]{Atomistic Modelling Center, Munich Data Science Institute, Technical University of Munich, Garching, Germany}
\sectionaffil[13]{Munich Center for Machine Learning (MCML), Munich, Germany}
\sectionaffil[14]{Department of Physics and Materials Science, University of Luxembourg, L-1511 Luxembourg, Luxembourg}
\sectionaffil[15]{Institute for Advanced Studies, University of Luxembourg, L-1511 Luxembourg, Luxembourg}
\sectionaffil[16]{Department of Mechanical and Materials Engineering, University of Turku, 20014 Turku, Finland}
\sectionaffil[17]{Google DeepMind, Berlin, Germany}
\sectionaffil[18] {Molecular Simulations from First Principles e.V., D-14195 Berlin, Germany}

\sectionaffil[*]{Coordinator of this contribution.}
\rule[0.25ex]{0.35\linewidth}{0.25pt}

\sectionaffil[a]{{\it Current Address:} Theory Department, Fritz Haber Institute of the Max Planck Society, Faradayweg 4-6, D-14195 Berlin, Germany}
\sectionaffil[b]{{\it Current Address:} Department of Materials Science \& Engineering, Yonsei University, 50 Yonsei-Ro, Seodaemun-Gu, Seoul 03722, Republic of Korea}
\sectionaffil[c]{{\it Current Address:} Department of Chemistry and Biochemistry, The University of Arizona, Tucson, AZ 85721, USA}




\subsection*{Summary}

The efficiency and reliability of artificial-intelligence (AI) models used for physics, chemistry, biophysics, materials science, and engineering depend on the acquisition of sufficient high-quality data. Due to its all-electron, full potential treatment and its scalability to larger systems without precision limitations, FHI-aims provides accurate \textit{ab initio} data from a wide range of computer simulations, such as electronic-structure calculations and molecular dynamics. To leverage the capabilities of AI models, workflows that seamlessly integrate AI tools with FHI-aims are essential. These workflows automate the data acquisition for AI and avoid the calculation of inessential data. Thus, they facilitate the iterative data exchange between AI models and simulations, allowing FHI-aims to be used as a powerful AI-guided calculation engine which puts its emphasis on information richness and not on mere data size. Also, interpretable AI models aid in analyzing the generated data. Finally, AI complements \textit{ab initio} studies as it enables to perform simulations at larger time and length scales. In turn, the AI models must incorporate the physics required for an accurate representation of the \textit{ab initio} data. This contribution highlights workflows developed to integrate FHI-aims with AI and future challenges.

\subsection*{Current Status of the Implementation}

FHI-aims \cite{Blum2009} provides various approaches to the provision and communication of data, which can then be used in workflows for the development of AI models such as machine-learning interatomic potentials (MLIPs). The interface with the Atomic Simulation Environment (\textrm{ASE}) \cite{ase-paper} library provides a framework for the generation or loading of input files, calculation of properties, and storage of calculation outcomes. Interfaces with high-throughput-calculation tools such as atomate2 \cite{Ganose} and Aiida \cite{aiida-aims-tutorial}  also allow rapid acquisition of data for training AI models (see contribution ~\ref{ChapWorkflows}). These interfaces are particularly crucial in workflows where the AI model is retrained iteratively with more data. These active learning (AL) workflows rely on data acquisition strategies informed by the AI model, which ensures that the new data to be collected is relevant. For instance, new data might be acquired when the prediction of the AI model has a high uncertainty or the interesting materials are found \textit{out of distribution} with respect to the training data distribution \cite{bauer2024roadmap}. In the following, we highlight examples of workflows involving FHI-aims and AI.

AL workflows using MLIPs such as \rev{high-dimensional neural network potentials (HDNNP) \cite{P1174}}, Gaussian Approximation Potential (GAP) \cite{deringer2021gaussian} and, more recently, equivariant message passing
neural networks like MACE \cite{batatia2022mace}, have been designed based on training data generated by FHI-aims. Examples of this include workflows for crystal structure prediction \cite{Wengert-2021}, battery materials \cite{Staacke-2021} or catalysis \cite{Stocker-2023,Jung-2023}. In these applications, it proved essential to have large flexibility with respect to simulation types (including global optimization, transition-state searches, molecular dynamics and enhanced sampling methods), and training set selection (e.g., via uncertainty estimation or similarity based search). These requirements have led to the development of the wfl package, a Python toolkit for interatomic potential creation and atomistic simulation workflows that emphasizes modularity and parallelisation over sets of atomic configurations \cite{Gelžinytė-2023,Libatoms}.

Being based on local atomic orbitals, FHI-aims is well suited to build large-scale material models containing defect structures, grain boundaries and stacking faults, which are often challenging for other plane-wave based DFT codes. The potential energy surface of these materials can be sampled with high accuracy in FHI-aims, yielding suitable energies, forces and stresses to train MLIPs. Any ASE-compatible framework to train interatomic potentials can be used, such as DeePMD-kit, which implements the smooth version of the Deep Potential model \cite{han2017deep,wen2022deep}. This approach has been employed to train a Deep Potential model for \ce{PtSe2} nanoflakes up to 8.6 nm \cite{kempt2024edge}.

We note that it is important that the chosen MLIP model is able to describe all the relevant physics, since increasing the amount of data alone does not guarantee a model representing the system correctly. Examples are long-range electrostatic interactions beyond the local cutoff radii often employed in the construction of MLIPs for condensed systems, dispersion interactions, and non-local charge transfer. The latter is crucial in many types of chemical reactions, e.g., if the charge of a molecule is altered by (de)protonation or an atomic oxidation state changes due to electron transfer. For these cases, often non-local approaches like fourth-generation MLIPs may be needed, which take the global structure of the system into account for describing electrostatics~\cite{ko2021}. An alternative solution for small systems is to employ explicit global machine-learning force fields like GDML/BIGMDL \cite{chmiela2023accurate, sauceda2022bigdml} or, more generally, a graph neural network architecture combined with physical models for long-range interactions, such as GEMS \cite{unke2024biomolecular} or SO3LR \cite{frank2024euclidean}. Another package that has been developed linking FHI-aims and MLIPs is the GKX package \cite{GKX} and the FHI-vibes framework (see contribution \ChapVibes). By using GKX, MLIPs trained on high-fidelity data generated with FHI-aims can be used to perform GPU-accelerated MD simulations for systems with thousands of atoms over timescales of nanoseconds. Such simulations can be used, for instance, to obtain converged thermal transport coefficients \cite{Langer2023}.

Despite the growing number of applications of MLIPs, concerns about their reliability arise when they are utilized to predict properties associated with configurations or chemical species that are significantly different from those in the training set \cite{kahle2022quality}. Kang {\it et al.} developed ALMOMD (Active-Learning Machine-Operated Molecular Dynamics ~\cite{kang2025accelerating, almomd:2024}), a Python workflow package interfacing FHI-aims with the MLIP codes NequIP~\cite{Batzner:2022} and so3krates~\cite{Frank:2022}. \texttt{ALMOMD} is designed to effectively train MLIP through an AL scheme with an automated framework that samples unfamiliar data, e.g., rare events, based on uncertainty estimates of MLIP predictions (Figure \ref{Fig:ALMOMD}).

\begin{figure}[ht]
   \centering
   \includegraphics[width=\textwidth]{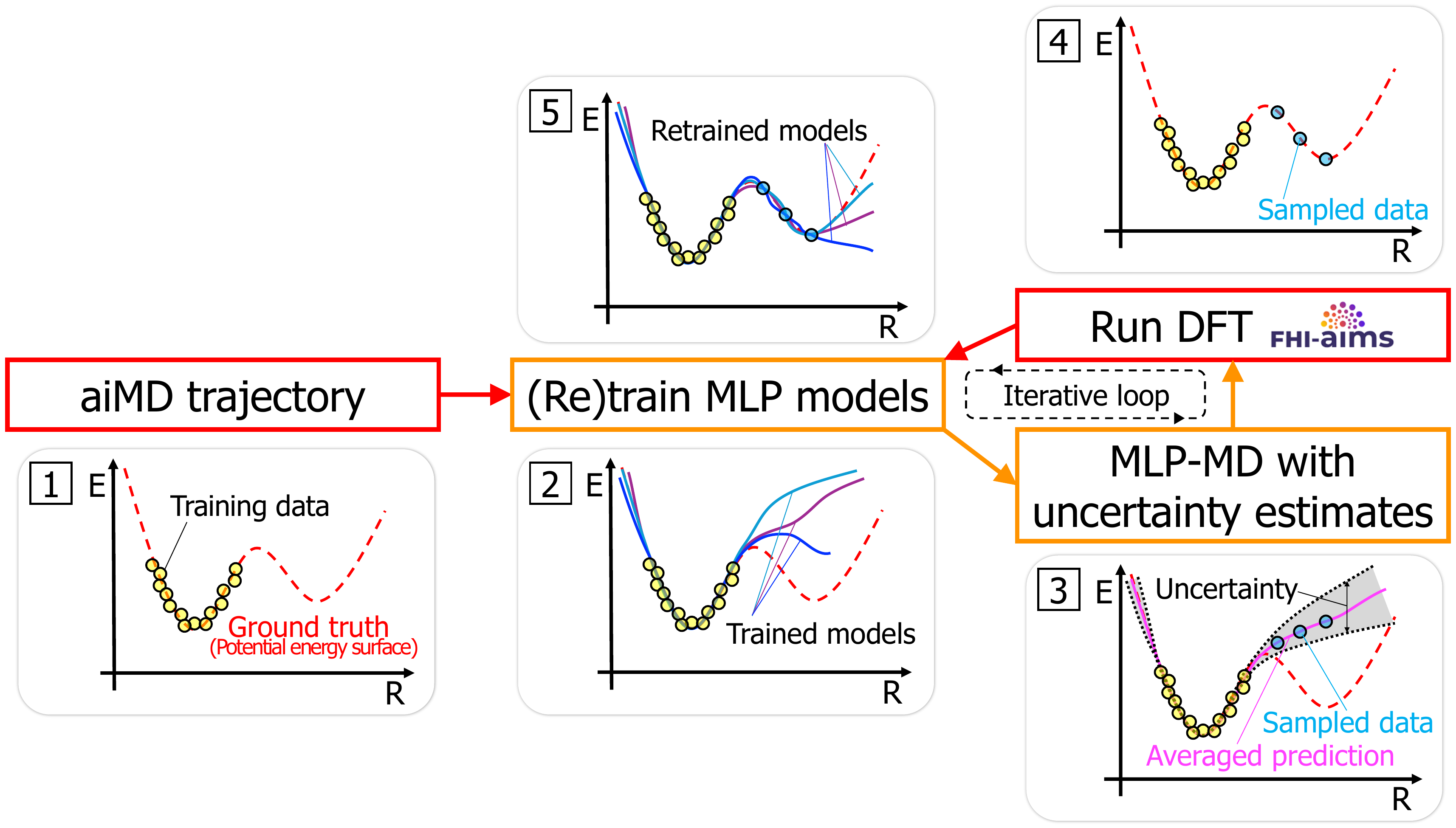}
   \caption{The overall iterative workflow of the ALMOMD. White boxes display indexed sequential steps for exploring the configurational space using MLIP-MD and sampling training data via uncertainty estimates. 
 }
   \label{Fig:ALMOMD}
\end{figure}

While MLIPs hold great promise, many material problems are high-dimensional in nature and involve costly evaluation of an objective function. This can benefit from AL workflows that dramatically reduce sampling, such as those involving Bayesian optimization. The Bayesian Optimization Structure Search (BOSS) algorithm  \cite{todorovic2019, boss} builds probabilistic $N$-dimensional surrogate models for materials energy or property landscapes, then refines them with smart sampling. The strategic acquisition strategy of blending data exploitation with design space exploration ensures fast identification of optimal solutions. BOSS is a Python tool for materials optimization and interoperable with FHI-aims and ASE.  

Finally, high-quality materials data are typically sparse, demanding data-efficient AI approaches. Moreover, interpretability, i.e., the ability to inspect the model's inference is critical, highlighting the importance of using descriptor-based AI methods \cite{oviedo2022interpretable}. Nair {\it et al.} developed an AL workflow integrating the sure-independence screening and specifying operator (SISSO) approach  with FHI-aims \cite{nair2025materials}. SISSO is a symbolic-regression method that utilizes compressed sensing to identify analytical expressions correlated with materials' properties or functions \cite{ouyang2018sisso, Purcell-2023}. It is a data-efficient method and offers better interpretability compared to widely used ML approaches in materials science. SISSO identifies analytical expressions that contain key physicochemical parameters, from many offered ones. The developed workflow utilizes an interface of FHI-aims with the high-throughput utility Taskblaster \cite{larsen2025taskblaster} for executing multiple tasks, such as geometry optimization, band-structure calculation, etc., for a large number of materials. Such workflows achieve efficient data acquisition and mitigate the issue of redundant data \cite{li2023exploiting}, making them suitable for generating information-rich materials databases (e.g., by including materials with high prediction uncertainty) which could be further used to develop AI models with enhanced generalizability.

\subsection*{Usability and Tutorials}

This section illustrates the applications of the workflows with tutorials, that researchers can adapt to their specific projects. The ALMOMD framework is demonstrated in the context of atomistic simulations of strongly anharmonic materials. Incomplete MLIP training often happens due to the absence or insufficiency of data within training regions. For example, the MLIPs may be unable to predict rare dynamical events, like defect creations, that are not included in training data or regularized away in the learning as they are rare. Consequently, it leads to critical deviations in predictions for transport properties. The ALMOMD framework can actively learn these unfamiliar data missed during MLIP training and correct the potential erroneous predictions during molecular dynamics (MD) simulations. It also corrects for fake rare events that an MLIP may produce. 
\begin{wrapfigure}{l}{0.33\textwidth}
  \begin{center}
    \includegraphics[width=0.32\textwidth]{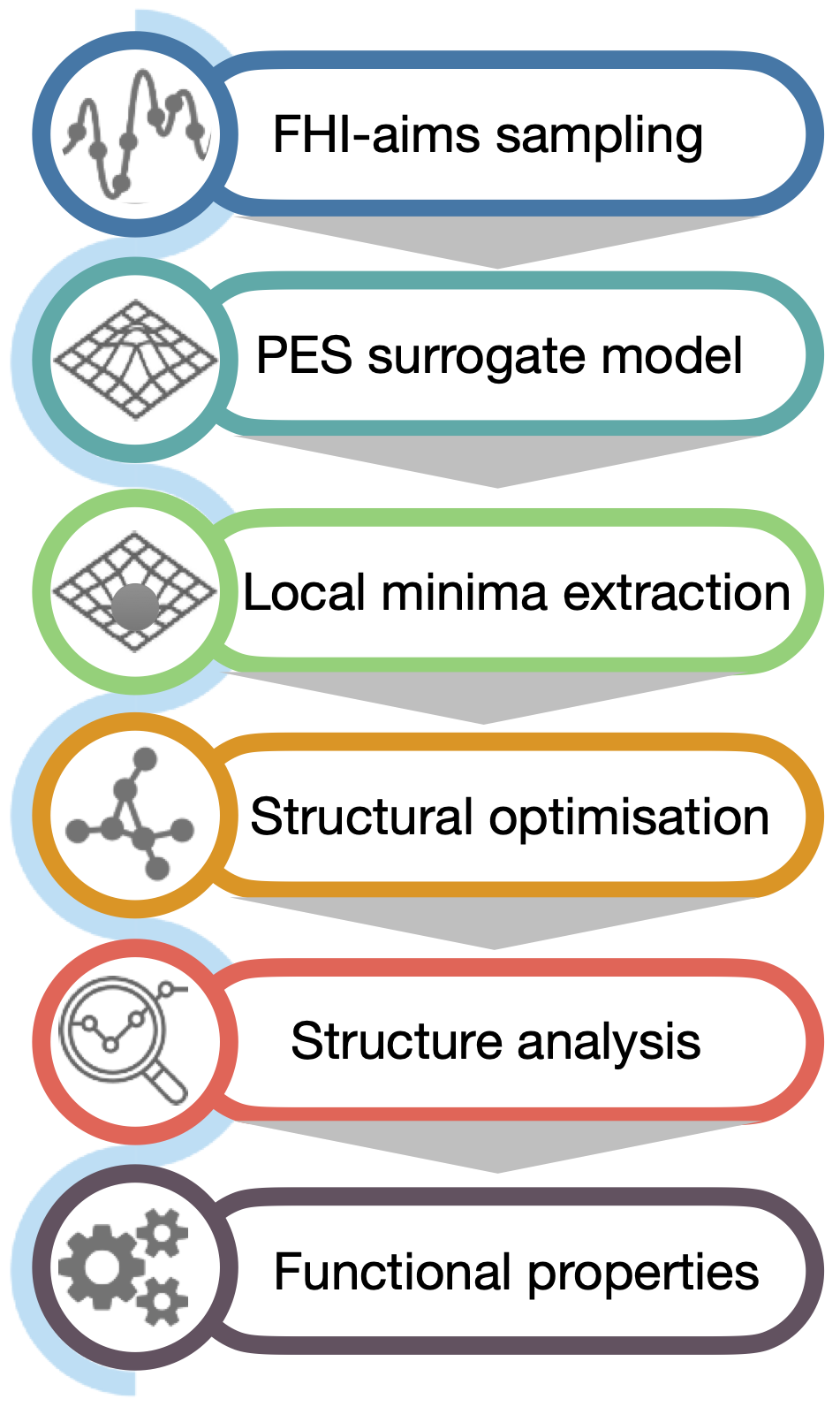}
  \end{center}
  \caption{BOSS AI workflow: from \\surrogate models to optimal solutions.}
  \label{Fig:BOSS}
\end{wrapfigure}
ALMOMD consists of two important steps: exploration and data-sampling. The efficient exploration of configurational space is achieved by explorative MD employing MLIPs (MLIP-MD). Uncertainty estimates serve as a warning indicating when MLIP-MD goes beyond its trained area, and thus, it can identify unfamiliar data that need to be sampled and retrained for MLIPs in subsequent steps. ALMOMD provides the user with an automated workflow environment and online tutorials~\cite{almomd}. 

BOSS was applied to study molecular conformers \cite{fang2021} and surface adsorbates \cite{jarvi2020, jarvi2021}, thin film growth \cite{egger2020}, solid-solid interfaces \cite{fangnon2022} and it can also combine multiple fidelity  simulations. The computation workflow illustrated in Figure \ref{Fig:BOSS} relies on uncertainty-aware and interpretable surrogate model landscapes to extract optimal solution basins, from which structural optimization leads to final structures and associated functional properties.
The BOSS website facilitates adaptations of this workflow to different use cases \cite{boss}, with the code, manual and extensive tutorials available to the research community \cite{boss_code, boss_doc}.

The SISSO-based AL workflow \cite{nair2025materials} is applied to the efficient discovery of acid-stable oxides for water splitting reaction from a large space of candidate materials, i.e., with a reduced number of calculations compared to high-throughput screening (Figure \ref{Fig:SISSO}). Ensembles of SISSO models are used to obtain mean predictions as well as estimates of the prediction uncertainties. This opened the opportunity to use SISSO as a surrogate model in  the aforementioned Bayesian optimization approach. DFT calculations were carried out for these materials by leveraging an efficient implementation of hybrid functionals \cite{kokott2024}. A tutorial demonstrating the workflow can be found at ref \cite{slsisso}. The workflow has also been transformed into a user-friendly web app \cite{webap}, featuring a step-by-step interactive demonstrator that guides users with limited AI experience in utilizing it effectively. Such a workflow reduces the risk of overlooking potentially interesting portions of the materials space that were disregarded in the initial training data and hence enables efficient materials discovery.  

\begin{figure}[ht]
   \centering
   \includegraphics[width=\textwidth]{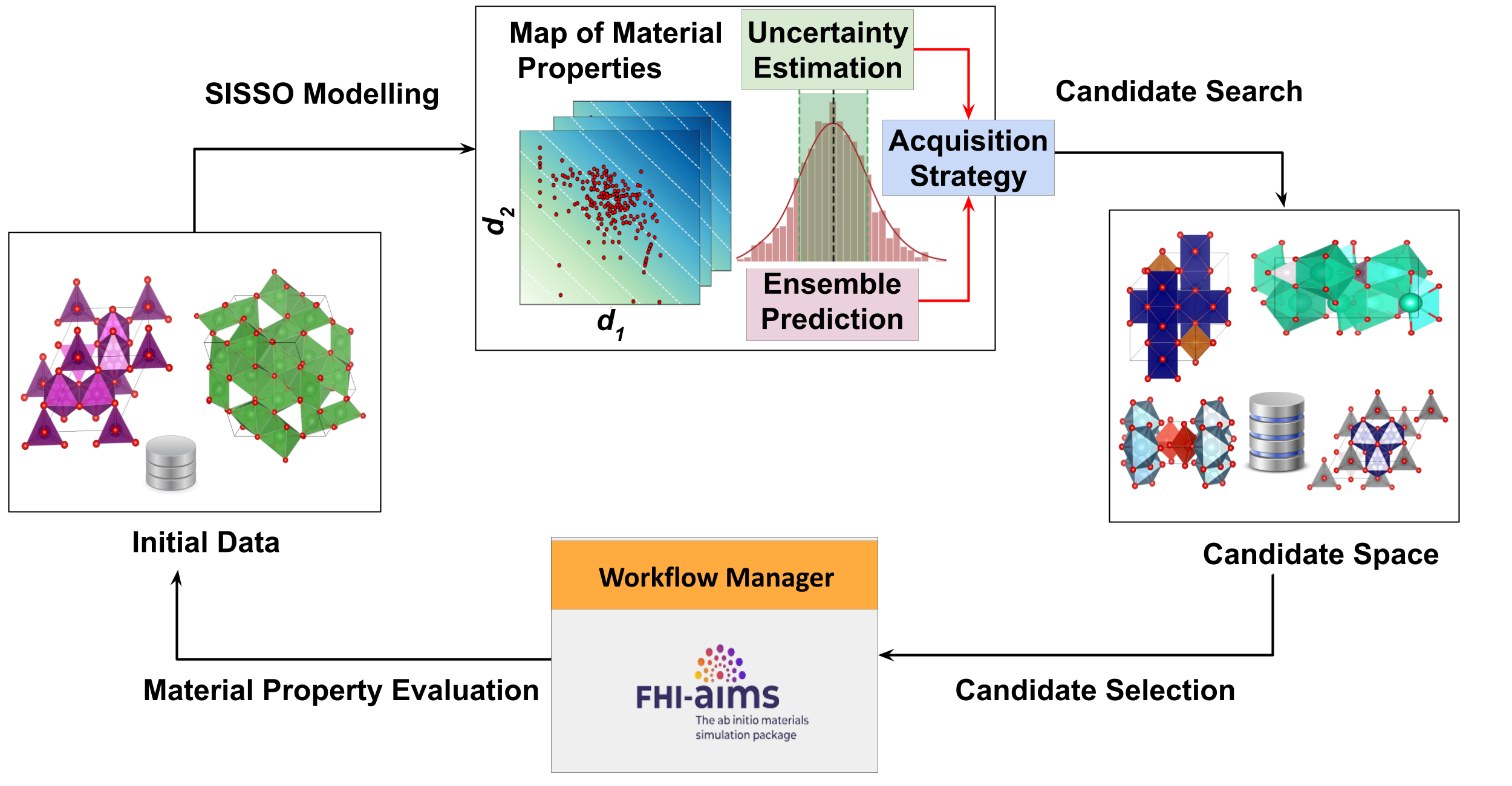}
   \caption{Schematic representation of the workflow integrating SISSO and FHI-aims. $d_1$ and $d_2$ represent the descriptors from the SISSO model constituting a materials map where the initially available data is labeled with red circles. 
 }
   \label{Fig:SISSO}
\end{figure}


\subsection*{Future Plans and Challenges}

Currently, the majority of MLIP-based methods which have been integrated into workflows to guide FHI-aims simulations, function as standalone packages. The next step is to enable direct execution of MLIPs within the FHI-aims environment, creating a more streamlined process that integrates MLIPs into the simulation workflow without the need for separate executions. For AL workflows, obtaining accurate and reliable uncertainty estimates is challenging, as overconfidence could lead to inefficient sampling of the materials or configuration spaces \cite{hwang2024overcoming}. Additionally, for systems like strongly correlated materials, workflows need to be adapted to integrate advancements in beyond-DFT methods, such as \textit{GW} or RPA.

Apart from the use as a calculation engine for data acquisition, AI models could also be used to accelerate FHI-aims calculations or improve their accuracy. For example, promising research directions include AI-based initial guesses for wavefunctions based on learning Kohn-Sham matrices \cite{schutt2019unifying,unke2021se}, the prediction of electron density in 3D space \cite{fu2024recipe} or the development of novel density functionals \cite{snyder2012finding,kirkpatrick2021pushing}. The training data for such approaches could itself be generated by FHI-aims, allowing to improve AI models in a feedback loop. Such methods and their potential applications are discussed in greater detail in the contribution (~\ref{ChapML-ES}).

\subsection*{Acknowledgements}

MS acknowledges support by his TEC1p Advanced Grant (the European Research Council (ERC) Horizon 2020 research and innovation programme, grant agreement No. 740233.




  


\newcommand{\MR}[1]{{\textcolor{purple}{\bf #1}}}




\newpage

\section{Machine Learning the Electronic Structure in a Real-Space and Atom-Centered Framework}
\label{ChapML-ES}
\sectionauthor[1]{Joseph W. Abbott}
\sectionauthor[2]{Michele Ceriotti}
\sectionauthor[2]{Andrea Grisafi}
\sectionauthor[1]{Wei Bin How}
\sectionauthor[3]{Alan Lewis}
\sectionauthor[4]{Andrew J. Logsdail}
\sectionauthor[5]{Zekun Lou}
\sectionauthor[6,7,8]{Reinhard J. Maurer}
\sectionauthor[5]{\textbf{ *Mariana Rossi}}
\sectionauthor[4]{Pavel V. Stishenko}
\sectionauthor[9]{Zechen Tang}
\sectionlastauthor[9,10]{Yong Xu}


\sectionaffil[1]{Institute of Materials, \'Ecole Polytechnique F\'ed\'erale de Lausanne, Rte. Cantonale, Lausanne 1015, Switzerland}
\sectionaffil[2]{Physicochimie des \'Electrolytes et Nanosyst\`emes Interfaciaux, Sorbonne Universit\'e, CNRS, F-75005 Paris, France}
\sectionaffil[3]{Department of Chemistry, University of York, Heslington, York, YO10 5DD, UK}
\sectionaffil[4]{Cardiff Catalysis Institute, School of Chemistry, Cardiff University, Park Place, Cardiff CF10 3AT, United Kingdom}
\sectionaffil[5]{Max Planck Institute for the Structure and Dynamics of Matter, 22761 Hamburg, Germany}
\sectionaffil[6]{Department of Chemistry, University of Warwick, Gibbet Hill Road, CV4 7AL Coventry, United Kingdom}
\sectionaffil[7]{Department of Physics, University of Warwick, Gibbet Hill Road, CV4 7AL Coventry, United Kingdom}
\sectionaffil[8]{Faculty of Physics, University of Vienna, Vienna A-1090, Austria}
\sectionaffil[9]{Department of Physics, Tsinghua University, Beijing 100084, P.R.China}
\sectionaffil[10]{RIKEN Center for Emergent Matter Science (CEMS), Wako, Saitama 351-0198, Japan}

\sectionaffil[*]{Coordinator of this contribution.}




\subsection*{Summary}

Machine-learning (ML) methods are driving a revolution in electronic structure simulations. 
Many \rev{approaches} target the learning and prediction of energies, forces and stresses that are produced by a density functional theory (DFT) calculation, with the goal of training high-accuracy force-fields that can be used to investigate atomic structural and dynamical properties of materials \rev{(see~\ref{ChapMLP_AI})}. 
However, one drawback of such models is that they do not give access to electronic-structure information, contrary to explicit quantum calculations that, in addition to energies and forces, also provide a wealth of derived electronic structure properties. 

A route to access these observables is to train machine learning models to learn a specific property, such as the electronic band gap, dipoles, surface work functions, and many others. 
However, a more \rev{transferable} solution exists in methods that directly target the electronic-density or the Kohn-Sham Hamiltonian (in the framework of DFT). 
When successful, these approaches are advantageous because, with a single ML model, one can recover most electronic structure properties and, in principle, also calculate accurate energies and forces.

The field of ML for electronic structure is admittedly still in its infancy, despite a very fast pace of progress in recent years. Many different approaches have been reported in the literature, based on a diverse set of underlying ML models and basis representations. 
In this contribution, we focus on methods that are directly interfaced to FHI-aims, namely SALTED~\cite{Lewis2021}, rholearn~\cite{rholearn-abbott}, doslearn ~\cite{adaptive-how,rholearn-abbott}, ACEhamiltonians.jl~\cite{aceham, aceham_workflow} and DeepH~\cite{deeph,deephe3,xdeeph}. 

FHI-aims offers unique benefits for ML surrogate models of electronic structure. The numeric atomic-orbital (NAO) basis offers a compact, accurate, and local representation of the electronic structure and yields compact Hamiltonian, overlap, and density matrices. 
The seamless treatment of periodic and non-periodic systems means that molecules and solids can be treated with the same infrastructure, making the training of models easier and more versatile. 
Furthermore, the density fitting (resolution-of-identity) infrastructure that has been developed for hybrid functionals within the NAO basis~\cite{ren2012} provides the ingredients to represent various real-space quantities through a set of basis coefficients with known rotational symmetry properties. Finally, the Atomic Simulation Interface (ASI)~\cite{stishenko2023}, developed in FHI-aims, enables plain C and Python application programming interfaces, which are beneficial for integration with software packages and libraries as they allow in-memory data transfer with minimal latency and storage overheads.


\subsection*{Current Status of the Implementation}

\begin{figure}[ht]
    \centering
    \includegraphics[width=0.8\textwidth]{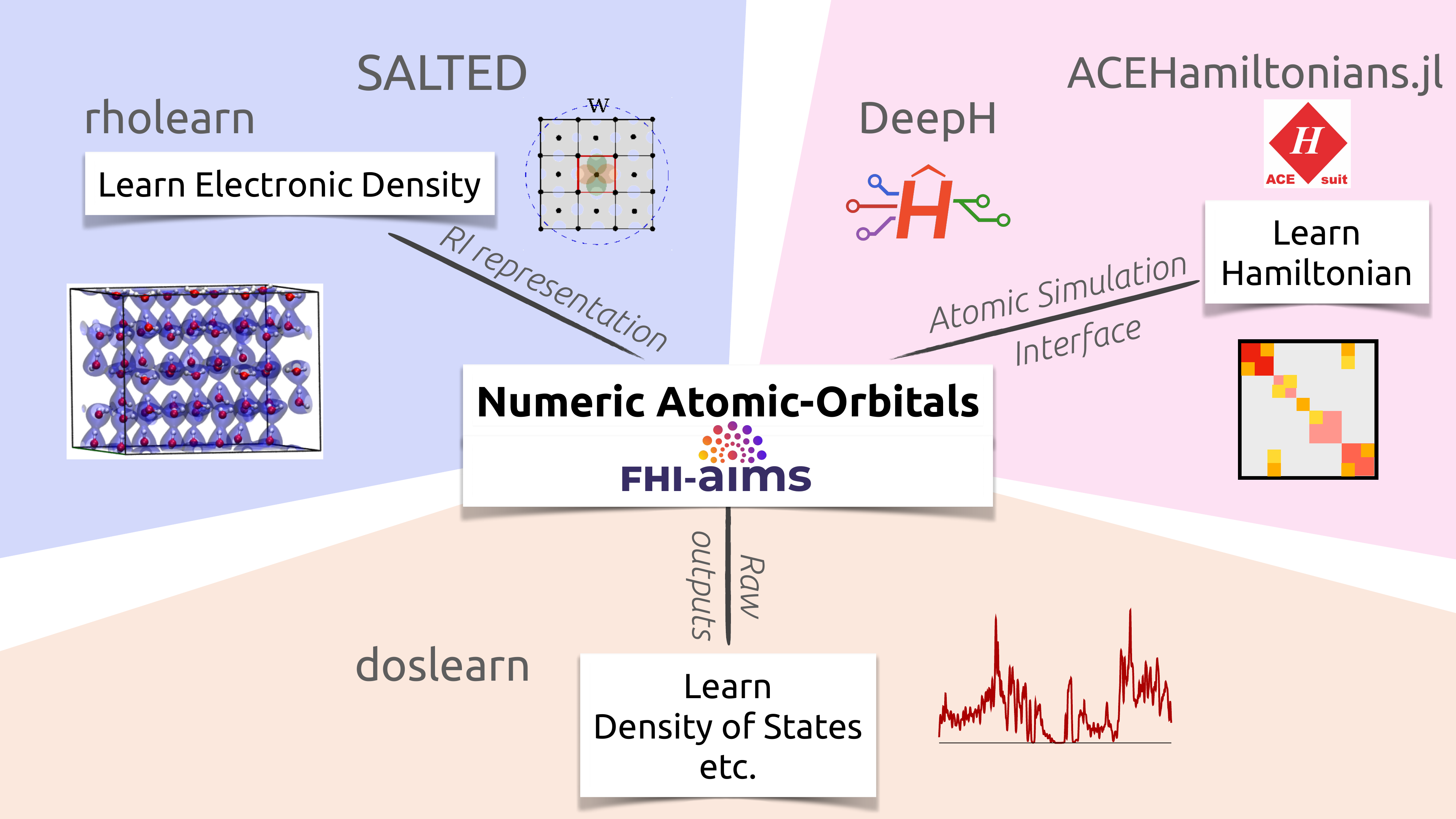}
    \caption{A sketch of the ecosystem of machine-learning (ML) methods for the electronic structure that are interfaced to FHI-aims. The raw outputs of the code, the resolution of the identity (RI) representation of the electronic density, and the atomic simulation interface (ASI) enable diverse types of ML algorithms to be used.}
    \label{fig:ML-Method-Sketch}
\end{figure}
\vspace{0.3em}
Machine-learning methods that interface with FHI-aims and target the electronic structure are, so far, all based on the real-space representation of the Kohn-Sham problem or the density from a quantum mechanical calculation. The challenge of learning these quantities reduces to representing atom-centered local density contributions or intra- and inter-atomic matrix blocks. Because of the common representation of the electronic density $n(\bm{r})$ in isolated and periodic systems, FHI-aims offers a seamless route to learn $n(\bm{r})$ for mixed datasets. 
For the Hamiltonian of periodic systems, once the real-space Hamiltonian in the extended crystal volume is known, the Hamiltonian at arbitrary $k$-points can be evaluated in a straightforward manner, similar to tight-binding techniques.

We start by summarizing methods that target $n(\bm{r})$. SALTED is a machine learning method based on a symmetry-adapted Gaussian process regression.\cite{Lewis2021} Using a resolution-of-the-identity (RI) expansion, $n(\bm{r})$ (from LDA, GGA, or hybrid-functionals) is first expressed as a linear combination of atom-centred basis functions $\phi_{i\sigma}$:
\begin{equation}
    n(\bm{r}) \approx \tilde{n}(\bm{r}) = \sum_{i,\sigma,\mathbf{U}} c_{i\sigma} \phi_{i\sigma}(\bm{r} - \bm{R}_i - \bm{T}(\mathbf{U}))
\end{equation}
where $i$ labels the atom on which the basis function is centered, $\bm{T}$ is a translation vector, and $\sigma=(n\lambda\mu)$ is a compound label describing the radial $R_n$ and spherical $Y_{\lambda\mu}$ part of the basis function. In FHI-aims, these basis functions can be chosen from the auxiliary basis introduced to accelerate the calculation of hybrid functionals.\cite{ren2012} The RI expansion coefficients $c_{i\sigma}\equiv c_\sigma(A_i)$ are then themselves approximated as a linear combination of a similarity measure $k_{\mu \mu^\prime}^\lambda$ between the atomic environment $A_i$ associated with the target coefficient  and each reference atomic environment $M_j$, and their associated regression weights $b_{\mu j}^\lambda$ to be determined using SALTED:
\begin{equation}
     c_\sigma(A_i) = \sum_j \sum_{\mu^\prime \leq \lambda} b_{\mu j}^\lambda k_{\mu \mu^\prime}^\lambda (A_i, M_j).
\end{equation}
These similarity measures are symmetry adapted, being tensors of dimension $2\lambda+1$ that transform as the $\lambda$-th irrep of the O(3) group, since the basis functions and their associated coefficients transform in this way. In SALTED, atom-density based descriptors are computed with the \rev{featomic}~\cite{metatensor-metatomic} library. These can be  smooth-overlap of atomic position (SOAP) descriptors, or descriptors that contain information about long-range electrostatics (\textit{e.g.}, LODE descriptors).

For a full description of the calculation of the regression weight-vector $\bm{b}$, which lies at the core of the symmetry-adapted Gaussian process regression of the SALTED method, the reader is referred to Refs.~\cite{Lewis2021, salted-grisafi}. Here we highlight that, in order to calculate $\mathbf{b}$, two quantities are required for each structure in the training set: $\mathbf{S}$, the overlap matrix of the basis functions $\phi_i$; and the optimal expansion coefficients $\mathbf{c}^\textrm{RI} = \mathbf{S}^{-1}\mathbf{w}$, where $\mathbf{w}$ is a vector containing the projection of the self-consistent electron density onto each basis function $\phi_{i\sigma}$.\footnote{The weights are 
$
    w_{i\sigma} = \sum^{\mathbf{U}_\text{cut}}_\mathbf{U}\braket{\phi_{i\sigma}(\mathbf{U})}{n}_\text{u.c.}.$
 The subscript ``u.c.'' indicates that in FHI-aims the integral is performed only over the ``central'' unit cell in the case of periodic systems, with the sum over $\mathbf{U}$ being truncated to only include unit cells that contain basis functions with some support in this central unit cell. The overlap matrix $\textbf{S}$ is evaluated in a similar way.}  

The calculation of $\mathbf{S}$, $\mathbf{w}$ and 
$\mathbf{c}^{\text{RI}}$ has been implemented in FHI-aims, with the option to write these quantities to files for later use with SALTED to train a ML model. Additionally, SALTED has a number of FHI-aims specific ``helper'' functions to support integration between the two codes. These include functions to calculate the error introduced by the RI approximation, to help users find an accurate set of basis functions $\phi_{i\sigma}$, to convert the data output from FHI-aims into a suitable format to be read by SALTED, and to convert coefficients predicted using SALTED into a format that can be read by FHI-aims. In the latter case, FHI-aims has the capability to read in a set of expansion coefficients $c_{i\sigma}$ during the SCF initialisation, and construct from them $n(\bm{r})$ on the real-space integration grid used in the code. This approach replaces the default choice of the initial electron density (\textit{i.e.}, superposition of atomic densities). In preliminary tests with a SALTED model trained on a subset of around 9000 structures of the QM9 dataset~\cite{qm9} (see Ref.~\cite{salted-grisafi}) we have observed, on average, that this approach leads to a $35$\% reduction in the number of SCF steps needed for convergence.

The infrastructure built for SALTED in FHI-aims, including the  RI representation for $n(\bm{r})$, can be re-used for ML schemes that use a neural network (NN) architecture rather than kernel regression. One such method is rholearn~\cite{rholearn-abbott}.  While the inputs (nuclear coordinates) and outputs (RI coefficients) of the prediction pipeline are the same, there are a few key differences compared to SALTED. 
The \rev{featomic}~\cite{metatensor-metatomic} library  is used to generate equivariant descriptors of arbitrary body order through Clebsch-Gordan products. These descriptor \textit{vectors}, as opposed to the kernels used in SALTED, form the part of the typically (but not necessarily) fixed transformations of the nuclear coordinates into an intermediate representation. Then, the training workflow uses the metatensor-torch library as its backbone. Native torch modules for tensor operations and learning utilities are wrapped to be compatible with TensorMap, the sparse data storage format of the metatensor~\cite{metatensor-metatomic} library. This allows arbitrarily complex equivariance-preserving NNs to be defined, and applied to the equivariant descriptors. Finally, the use of descriptor vectors, the ability to train by minibatch gradient descent, and the use of sparse operations (most notably when evaluating the loss) reduce the memory overhead of rholearn. 

Models that target learning the (Kohn-Sham) Hamiltonian matrix profit from a  more tightly integrated interface with FHI-aims for AI/ML workflows. This interface is realised via ASI \cite{stishenko2023}, which is a plain C interface that provides direct in-memory access to electronic structure quantities within FHI-aims, such as the Hamiltonian, and overlap matrices. It also allows to inject AI/ML predicted matrices into FHI-aims processes at  runtime, providing new opportunities in modular software integration \cite{aceham_workflow}. The integration of ACEhamiltonians.jl and DeepH to FHI-aims use this interface.

ACEhamiltonians.jl~\cite{aceham} uses atom- and bond-centred descriptors based on the Atomic Cluster Expansion (ACE)~\cite{ace-descs} to represent non-orthogonal Hamiltonian matrix blocks that transform equivariantly with respect to the full rotation group. ACEhamiltonians models are analytical and linear, and can produce accurate models with little data. The model can directly be trained on FHI-aims data that is outputted either \textit{via} the keyword \texttt{output\_rs\_matrices} in human readable or HDF5 format, or through the ASI interface. Furthermore, the model is also integrated with FHI-aims through ASI to directly feed predicted quantities such as the Hamiltonian or the density matrix back into FHI-aims, \textit{e.g.}, to provide an improved initial guess of the density matrix for the SCF algorithm. We have also shown that ASI can be used as a bridge to train ACEhamiltonians models on FHI-aims data and then inject them into a different electronic structure code, such as DFTB+, during runtime.~\cite{aceham_workflow}

The deep-learning density functional theory Hamiltonian (DeepH) method~\cite{deeph,deephe3,xdeeph} is an equivariant graph neural network approach for learning and predicting the electronic Hamiltonians for given material structures. Based on the predicted Hamiltonian, various physical properties can also be derived, including band structures, density of states, and the electric susceptibility, among others. Benefiting from the principle of nearsightedness, the neural networks of DeepH can infer the properties of large-scale material structures by learning from small ones. In the original version of DeepH developed in 2021~\cite{deeph}, a local-coordinates-based scheme was proposed to handle the rotational covariance of the DFT Hamiltonian matrix. A subsequent implementation, DeepH-E3~\cite{deephe3}, incorporated an E(3)-equivariant neural network architecture to address the equivariance with respect to the E(3) group, vastly improving the prediction accuracy. Recently, more advanced equivariant tensor product techniques together with transformer architectures have been integrated into the DeepH framework, further enhancing both accuracy and efficiency~\cite{deeph2}. The method is versatile and has also been extended to various deep-learning electronic structure schemes, including magnetic materials~\cite{xdeeph}, density-functional perturbation theory~\cite{deeph-dfpt}, hybrid density functionals~\cite{deeph-hybrid}, among others. The current implementation of the DeepH interface for FHI-aims is based on the ASI package~\cite{stishenko2023}. Structural information, Hamiltonian matrices and overlap matrices are extracted from the ASI plug-ins and converted to DeepH formats. The current interface has been tested on both molecular and periodic systems, in which the open source DeepH-E3 method~\cite{deephe3} reaches sub-meV \rev{mean absolute error in}  Hamiltonian matrix elements across the test systems.

Finally, there are ML methods interfaced with FHI-aims that parse raw outputs of electronic-structure quantities from FHI-aims. doslearn targets the electronic density of states (DOS), employing a locality ansatz in which the total DOS of a structure is built as a sum of atom-centered contributions~\cite{ben_mahmoud2020}. The machine learning pipeline directly takes in FHI-aims outputs, extracts the Kohn-Sham eigenvalues for each structure, and uses scipy to generate cubic Hermite splines of the DOS to support an adaptive energy reference, necessary for  bulk periodic calculations. Model training is performed using the PyTorch package, with the adaptive reference framework supported \textit{via} a modification of the loss function to concurrently optimize the energy reference during model training. Although the current doslearn implementation employs a simple feed forward network using invariant SOAP Power Spectrum descriptors generated from the \rev{featomic}~\cite{metatensor-metatomic} library as inputs, doslearn primarily focuses on the optimization procedure and is model agnostic, allowing it to be easily modified to deploy on any arbitrary model architecture of the user's choice.

\subsection*{Usability and Tutorials}

The usability of the electronic-structure learning techniques described above together with FHI-aims is ensured by specific keywords, with examples and tutorials maintained either natively in the code or in dedicated repositories.

We begin by briefly describing the usability and tutorials of the electronic-density learning approaches.  The SALTED code is written in Python and FORTRAN and it is available through \href{https://github.com/andreagrisafi/SALTED}{a dedicated GitHub repository}\footnote{\texttt{https://github.com/andreagrisafi/SALTED}}. \textit{rholearn} is Python-based and is also available through \href{https://github.com/lab-cosmo/rholearn}{a dedicated GitHub repository}\footnote{\texttt{https://github.com/lab-cosmo/rholearn}}. Their installation and dependencies are described in their respective documentation. 

Briefly, the FHI-aims keywords for \texttt{control.in} that trigger the outputs needed for SALTED and rholearn are:
\begin{itemize}
    \item \texttt{ri\_density\_restart write} will create a file in the working directory containing the coefficients $\textbf{c}^{\textrm{RI}}$.
    \item \texttt{ri\_density\_restart read} will read the expansion coefficients contained in the file previously created (in the same directory) and replace the default initial density with the density obtained from these coefficients.
    \item \texttt{ri\_full\_output}, when used in conjunction with \texttt{ri\_density\_restart write}, will additionally output a file containing the overlap matrix $\textbf{S}$, and the file \texttt{ri\_projections.out} containing the projections vector $\textbf{w}$.
    \item \texttt{ri\_skip\_scf} is used in conjunction with \texttt{ri\_density\_restart write} and allows RI decomposition of a previously-converged density as part of a two-step SCF + RI process. For instance, \texttt{elsi\_restart write} can be used to save a converged solution to the SCF procedure, then \texttt{elsi\_restart read} and \texttt{ri\_skip\_scf} can be used to fit directly to the corresponding converged density.
\end{itemize}

The keywords \texttt{prodbas\_acc}, \texttt{max\_l\_prodbas}, and \texttt{wave\_threshold} are not SALTED or rholearn-specific, but control the construction of the auxiliary basis that is used to expand the density. Specifying these keywords ensures a consistent basis is used, which is especially important if training on mixed periodic and non-periodic datasets.

The SALTED integration with FHI-aims is thoroughly described in \href{https://fhi-aims-club.gitlab.io/tutorials/fhi-aims-with-salted}{an online tutorial}\footnote{\texttt{https://fhi-aims-club.gitlab.io/tutorials/fhi-aims-with-salted}}. The tutorial guides the user on training a SALTED model for the electronic density of a water monomer, and predicting the density of water dimers. The rholearn integration with FHI-aims shares many of the features of the SALTED integration and is also described in \href{https://github.com/lab-cosmo/rholearn/tree/main/example/rholearn-aims-tutorial}{an online tutorial}\footnote{\texttt{https://github.com/lab-cosmo/rholearn/tree/main/example/rholearn-aims-tutorial}}. Data for a subset of the QM7 dataset~\cite{qm7-1,qm7-2} is generated with the \texttt{aims-interface} in rholearn, and a descriptor-based equivariant neural network trained by minibatch gradient descent using a PyTorch- and metatensor~\cite{metatensor-metatomic}-based workflow.

The interfaces for Hamiltonian learning are based on ASI, as mentioned above.
The ASI documentation is available \textit{via} the library website \cite{asi2024}. The Python wrapper for ASI, asi4py, can be installed from the Python Package Index (PyPI) repository. Tutorials on how to train ACEhamiltonians models with example FHI-aims data for bulk Aluminium are provided on GitHub~\cite{aceham-web}. The tutorial covers how to parse and process FHI-aims Hamiltonian data generated with the \texttt{output\_rs\_matrices} keyword into the expected HDF5 database structure. It also covers essential keywords and hyperparameters for model construction, model fitting, and model prediction. ACEhamiltonians directly predicts the real-space representation of the Hamiltonian in the extended crystal volume. Matrices (and therefore eigenvalues and eigenvectors) during prediction can be constructed at arbitrary $\mathbf{k}$-points. 

Documentation on how to use ASI to provide an improved initialisation of the density matrix in FHI-aims during runtime is available in the \texttt{dm} submodule of \href{https://pvst.gitlab.io/asi/ml_dm.html}{asi4py}\footnote{\texttt{https://pvst.gitlab.io/asi/ml\_dm.html}}. This submodule provides a prediction of the density matrix that can reduce the number of required self-consistent-field iterations, when compared to the standard initialization.  The \texttt{ASI\_register\_dm\_init\_callback} function of the ASI API is used to register a user-provided callback function that returns an initial guess of the density matrix at the beginning of the SCF loop. The \texttt{asi4py.dm} module provides a few classes implementing density matrix prediction via ASI; these classes can be used directly, subclassed, or used as inspiration for custom density matrix predictors.

A tutorial for DeepH is available in Zenodo~\cite{aims2deeph-zenodo}, which provides the source code for the interface with FHI-aims, along with demonstrations for generating datasets and two example datasets, each accompanied by configuration files for DeepH-E3 training. The source code for the interface is also distributed alongside FHI-aims. To utilize the interface, FHI-aims should be compiled as a shared library through ASI and the asi4py package should also be installed. The interface can be employed to extract all required data for DeepH training, utilizing the ASI API to access Hamiltonian and overlap matrices. The DeepH-E3 model~\cite{deephe3} has been evaluated on water and graphene datasets provided in the Zenodo repository, achieving mean \rev{absolute} errors of 0.31 and 0.28~meV, respectively, for the Hamiltonian matrix elements in the test data sets. The recently developed version DeepH-2~\cite{deeph2}, which will be soon released, \rev{demonstrates significantly improved accuracy, with mean absolute errors of 0.08 and 0.04 meV for the water and graphene datasets mentioned above, respectively.}

Finally, the doslearn integration with FHI-aims is available in the \href{https://github.com/lab-cosmo/rholearn}{the same repository}\footnote{\texttt{https://github.com/lab-cosmo/rholearn}} as rholearn, with an \href{https://github.com/lab-cosmo/rholearn/tree/main/example/doslearn-aims-tutorial}{an online tutorial} that covers the data processing pipeline and model training of a fully connected network using batch gradient descent. The keyword used to generate the energy eigenvalues necessary for doslearn is: \texttt{output postscf\_eigenvalues}.


\subsection*{Future Plans and Challenges}


The area of ML applied directly to the electronic-structure is growing at a fast pace. Many such methods, including the ones interfaced with FHI-aims discussed in this contribution, show high-accuracy in predicting the electronic-density and Hamiltonian matrix elements. The methods allow the prediction of electronic structure properties of systems containing up to $\sim$10$^5$ atoms when trained on much smaller structures. Efficiency in terms of speed and memory usage can still be bottlenecks, and these are currently being addressed by the community. It is exciting to see that such methods are already able to predict not only the ground-state electronic density and Hamiltonian, but also density linear-responses~\cite{lewi+jcp2023, grisafi-prm-2023, deeph-dfpt}, for example. Remaining challenges in this area regard the proper and efficient handling of equivariant architectures for vector-field prediction. 

It is further interesting to see the extension of methods such as DeepH to unsupervised learning~\cite{deeph-zero}, force fields for magnetic materials~\cite{magnet}, and to train high-accuracy universal models covering over 20 elements in the periodic table and over 10,000 material structures~\cite{deeph-umm}.
For DeepH, despite its $O(N)$ scaling that enables highly efficient computation of large systems, there remain a few post-processing workflows for calculating physical properties that could be computationally expensive for studying large-size material systems. Future advances in efficient algorithms for property calculations, such as sparse-matrix techniques, Green’s function methods, and Wannierization, will significantly enhance the practical applications of deep-learning electronic structure methods, allowing for the computation of a broader range of physical properties in large-scale material simulations.

Finally, to realise the full potential of ML for electronic structure, new infrastructure is necessary to integrate existing software with AI-enabling workflows. 
ML models and workflows to generate and manage data are increasingly available in FHI-aims. Data can be communicated inward and outward in raw or fitted forms, with the latter beneficial for reducing data quantities;  Data can also be communicated to and from FHI-aims \textit{via} file input/output (with minimal overheads), sockets-type interfaces, or through direct interfacing.
Nevertheless, there is an ever growing need for flexible and robust interfaces that allow more user interaction with these calculations, and data extraction. This need motivates further efforts to support workflow connectivity and development of standardised data communication paradigms. For the direct access to in-memory data objects, development plans for ASI include more flexible and extendable interface that supports a wider range of data objects and diverse sparse matrix formats. Long-term aims include applying ASI to develop an infrastructure for integration with a broad range of ML models, which can then accelerate or refine electronic structure calculations.

\subsection*{Acknowledgements}

We acknowledge fruitful discussions with Volker Blum, Ben Hourahine, Thomas Keal, Scott Woodley. We thank Adam McSloy and Connor L. Box for their contributions to Hamiltonian output functionality in FHI-aims.  

AJL and PVS acknowledge funding by the UKRI Future Leaders Fellowship program (MR/T018372/1, MR/Y034279/1). 
AJL, PVS, RJM, and MR acknowledge funding from the ARCHER2 eCSE Programme (eCSE03-10).
RJM acknowledges support through the UKRI Future Leaders Fellowship programme (MR/S016023/1, MR/X023109/1), and a UKRI frontier research grant (EP/X014088/1).
YX and ZT acknowledge funding from the Basic Science Center Project of NSFC (52388201), the Ministry of Science and Technology of China (2023YFA1406400), the National Natural Science Foundation of China (12334003, 12421004, and 12361141826), and the National Science Fund for Distinguished Young Scholars (12025405).




  

\newpage

\chapter{Appendix}

\section{Summary of the Delta-DFT test for the FHI-aims species defaults}

\begin{figure}[ht!]
    \centering
    \includegraphics[width=0.9\textwidth]{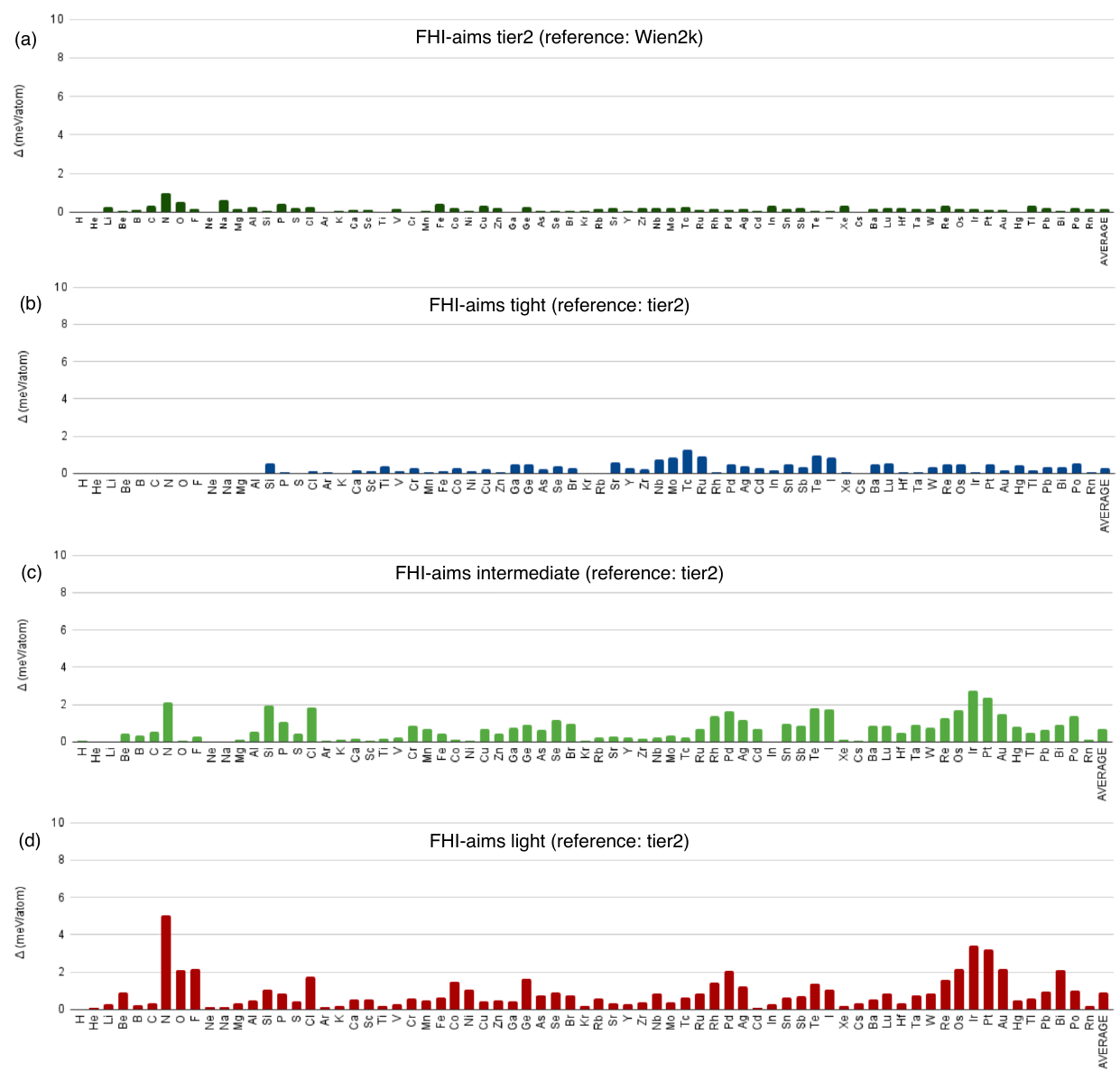}
    \caption{The $\Delta$ gauge \cite{Lejaeghere2016} between (a) WIEN2k and FHI-aims with really tight settings and tier2 basis functions (tier2), (b) FHI-aims tight and tier2 (c) FHI-aims intermediate and tier2, and (d) FHI-aims light and tier2.}
    \label{fig:DeltaTestSummary}
\end{figure}

\newpage

\bibliographystyle{unsrt}
\bibliography{references-clean}

\end{document}